\documentclass[final, 12pt,oneside]{class_diss}

\usepackage{      amsmath} 
\usepackage{     graphicx} 

\usepackage{amsxtra}
\usepackage{amssymb}
\usepackage{amsthm}
\usepackage{latexsym}
\usepackage{styles/deluxetable}

\usepackage{float}
\usepackage{multirow}
\usepackage{cancel}
\usepackage{threeparttable}
\usepackage{relsize}
\usepackage{longtable}

\usepackage[usenames]{color}

\definecolor{  Pink}{rgb}{1.0, 0.5, 0.5}
\definecolor{Maroon}{rgb}{0.8, 0.0, 0.0}

\usepackage[sort&compress]{natbib}
\bibpunct{}{}{,}{s}{}{}
\usepackage{hypernat}

\usepackage[pdftex, plainpages=false, pdfpagelabels]{hyperref}

\hypersetup{
    linktocpage=true,
    colorlinks=true,
    bookmarks=true,
    citecolor=blue,
    urlcolor=red,
    linkcolor=Maroon,
    citebordercolor={1 0 0},
    urlbordercolor={1 0 0},
    linkbordercolor={.7 .8 .8},
    breaklinks=true,
    pdfpagelabels=true,
    }

\begin{document}





\thispagestyle{empty}


\pdfbookmark[0]{Title Page}{PDFTitlePage}

\begin{center}

   \vspace{1cm}


   \Large OBSERVATIONAL CONSTRAINTS ON DARK ENERGY COSMOLOGICAL MODEL PARAMETERS\\

   \vspace{0.5cm}

   by\\

   \vspace{0.5cm}


   MUHAMMAD OMER FAROOQ\\

   \vspace{0.5cm}


   B.Sc., University of Engineering and Technology, Pakistan, 2007 \\
   M.Sc., University of Manchester, UK, 2009 \\

   \vspace{0.65cm}
   \rule{2in}{0.5pt}\\
   \vspace{0.85cm}

   {\Large AN ABSTRACT OF A DISSERTATION}\\

   \vspace{0.5cm}
   submitted in partial fulfillment of the\\
   requirements for the degree\\

   \vspace{0.5cm}


   {\Large DOCTOR OF PHILOSOPHY}\\
   \vspace{0.5cm}


   Department Of Physics\\
   College of Arts and Sciences\\

   \vspace{0.5cm}
   {\Large KANSAS STATE UNIVERSITY}\\
   Manhattan, Kansas\\


   2013\\
   \vspace{1cm}

\end{center}

\begin{abstract}
   \setcounter{page}{-1}
   \pdfbookmark[0]{Abstract}{PDFAbstractPage}

\pagestyle{empty}
\vspace{1cm}
\setlength{\baselineskip}{0.8cm}



The expansion rate of the Universe changes with time, initially slowing (decelerating)
when the universe was matter dominated, because of the mutual gravitational attraction of 
all the matter in it, and more recently speeding up (accelerating). A number of 
cosmological observations now strongly support the idea that the
Universe is spatially flat (provided the dark energy density is at least approximately
time independent) and is currently undergoing an accelerated cosmological 
expansion. A majority of cosmologists consider ``dark energy" to be
the cause of this observed accelerated cosmological expansion.

The ``standard" model of cosmology is the spatially-flat $\Lambda$CDM model.
Although most predictions of the $\Lambda$CDM model are reasonably 
consistent with measurements, the $\Lambda$CDM model has some curious
features. To overcome these difficulties, different Dark
Energy models have been proposed. Two of these models, the XCDM parametrization and the slow rolling
scalar field model $\phi$CDM, along with ``standard" $\Lambda$CDM, with the generalization of XCDM and 
$\phi$CDM in non-flat spatial geometries are considered here and observational data are used to constrain
their parameter sets.

In this thesis, we start with a overview of the general theory of relativity, Friedmann's equations, and distance measures in cosmology.
In the following chapters we explain how we can constrain the three above mentioned
cosmological models using three data sets: measurements of the Hubble parameter $H(z)$,
Supernova (SN) apparent magnitudes, and the baryonic acoustic oscillations (BAO) peak length scale, as functions of redshift $z$. We then 
discuss constraints on the deceleration-acceleration transition redshift $z_{\rm da}$ 
using unbinned and binned $H(z)$ data. 
Finally, we incorporate the spatial curvature in the XCDM and $\phi$CDM model and determine observational constraints on the parameters of these
expanded models.

   \vfill
\end{abstract}



\newpage


\thispagestyle{empty}


\begin{center}

   \vspace{1cm}


   \Large OBSERVATIONAL CONSTRAINTS ON DARK ENERGY COSMOLOGICAL MODEL PARAMETERS\\

   \vspace{0.5cm}

   by\\

   \vspace{0.5cm}


   MUHAMMAD OMER FAROOQ\\

   \vspace{0.5cm}


   B.Sc., University of Engineering and Technology, Pakistan, 2007 \\
   M.Sc., University of Manchester, UK, 2009 \\

   \vspace{0.65cm}
   \rule{2in}{0.5pt}\\
   \vspace{0.85cm}

   {\Large A DISSERTATION}\\

   \vspace{0.5cm}
   submitted in partial fulfillment of the\\
   requirements for the degree\\

   \vspace{0.5cm}


   {\Large DOCTOR OF PHILOSOPHY}\\
   \vspace{0.5cm}


   Department Of Physics\\
   College of Arts and Sciences\\

   \vspace{0.5cm}
   {\Large KANSAS STATE UNIVERSITY}\\
   Manhattan, Kansas\\


   2013\\
   \vspace{1cm}

\end{center}

\begin{flushleft}
   \hspace{10cm}Approved by:\\
   \vspace{ 1cm}
   \hspace{10cm}Major Professor\\


   \hspace{10cm}Dr. Bharat Ratra\\
\end{flushleft}





\newpage

\thispagestyle{empty}

\begin{center}

{\bf \Huge Copyright}

\vspace{1cm}


   \Large MUHAMMAD OMER FAROOQ\\

   \vspace{0.5cm}


   2013\\

   \vspace{0.5cm}

\end{center}


\begin{abstract}



    \vfill

\end{abstract}


\newpage
\pagenumbering{roman}


\setcounter{page}{6}


\phantomsection

\tableofcontents
\listoffigures
\listoftables




\newpage
\begin{center}
{\bf \Huge Acknowledgments}
\end{center}
\vspace{1cm}
\setlength{\baselineskip}{0.8cm}



Finishing a PhD in physics is truly a marathon event, and it would have not been possible for me to 
finish this long and tedious journey without the help of countless people that were around me in last
four years. Some of them were not with me physically during this PhD, but they are in my heart all the time. 
I must first express my gratitude towards my advisor Bharat Vishnu Ratra. He is an outstanding 
mentor and teacher. His leadership, support, guidance, attention to detail, hard work, patience and kindness have set 
an example I hope to match some day. Thank you for your help and guidance.
I would like to thank Larry Weaver from whom I have learned a lot. He taught me more than I could read in any book on physics. 
He is always in his office and very welcoming. I would like to thank my roommate, my colleague and co-author 
of a couple of my papers, Data Mania, who helped me a lot in understanding cosmology. He is the best roommate 
one can have. Thanks Data. I also want to say thanks to my fellow graduate students Shawn Westmoreland 
and Mikhail Makouski for having very helpful discussions with me during my research. I want to thank 
my sister Saima Farooq and my friend Arjun Nepal and his family for being an excellent support to 
me during my stay here in Manhattan. Thanks to Foram Madiyar and Misty Long for being good friends to me. I want to thank all of my students whom make me think about physics 
problems more deeply, and their questions deepen my understanding of physics.
I am extremely fortunate to have an enthusiastic and talented proofreaders in the form of Shawn Westmoreland, Max Goering, Sara Crandall and Levi Delissa. 
I want to say special thanks to them for spending lot of time in pointing out lots of inevitable mistakes and typos and places 
where my presentation didn't sparkle quite as much as I thought it did. Their sharp eyes and hard work did much 
to make this thesis better. Any remaining errors or omissions are obviously the sole responsibility of mine.
Special thanks to Sara Crandall co-author of one of my paper. I enjoyed working with her alot. 
Thanks to Daniel Nelson, my undergraduate student for letting me use his computer for some of the calculations during
my work at Kansas State University. I want to give a very special thanks, though this 
word ``thanks" is not enough for my best friend May Ebbeni, for teaching me most 
of physics and being one of the best friends one can ever have. You helped make 4 years of my life in graduate school 
more fun and interesting. Thanks alot May. Finally, and most importantly, this work would not have been possible
without the endless support and encouragement from my parents. I dedicate this thesis to them. 
I will always remember the wonderful time that I spend here 
with my friends, graduate student and teachers in Physics department of Kansas State Univerity. 

This work was supported in part by DOE grant DEFG03-99EP41093 and NSF grant AST-1109275.

\phantomsection
\addcontentsline{toc}{chapter}{Acknowledgements}



\newpage
\begin{center}
{\bf \Huge Dedication}
\end{center}
\vspace{1cm}
\setlength{\baselineskip}{0.8cm}



To my parents Farooq Ahmad Uppal and Abida Bano who trust in me in all respect more than I do myself.

\phantomsection
\addcontentsline{toc}{chapter}{Dedication}




\newpage
\pagenumbering{arabic}
\setcounter{page}{1}




\cleardoublepage


\chapter{Introduction}
\label{Chapter1}

Cosmology is the study of the Universe, or cosmos, regarded as a whole. Physical cosmology 
is the scholarly and scientific study of the origin, evolution, structure, dynamics, and
ultimate fate of the Universe, as well as the natural laws that keep it in order. The study of
cosmology is fueled by the curiosity of wanting to know more about the Universe in which we are living and wanting to find 
answers to some fundamental questions like \textit{Where do we come from? What are we? Where
are we going? Does the Universe have a beginning? Will the Universe have an end? Is the Universe
infinite? How did we get here? Are we special?} Cosmology grapples with these questions by describing the 
past, explaining the present, and predicting the future of the Universe. In this chapter, we will 
summarize the basics and fundamentals of Einstein general theory of relativity and Friedmann's equations 
undoubtedly the most important equation in cosmology. After that we will solve Friedmann's equation in different cases.

\section{Basics and Fundamentals}
The structure of any theory in science is based on some fundamental axioms, often summarized generally by a 
genius, based on a lot of observations. These are the axioms on which the theory 
depends, which are sometimes incompletely experimentally tested, but, if one does not believe in these
axioms then one does not believe in the theory. 
The theory, which is generally a mathematical formula, will make some predictions that one has to develop 
experimental set-up to check. If the theory make predictions that are consistent with the experimental results
within the uncertainty of the experimental error, this means that the theory can be trusted, until we have some experiment
which gives a contradictory result. In this case it is said that the theory needs modifications. After
finding a number of experimental results consistent with the predictions of the theory, the theory is accepted
by the scientific community and one can make deductions from it. On the other hand, if the theory gives 
results that disagree with experiments, then the theory is wrong, independent of who developed the theory, how
beautiful the theory is, and how smart the person who developed it is.    

Standard, cosmology is based on a fundamental axiom, the {\bf cosmological
principle}. This principle states that Universe is homogeneous and isotropic in space on a sufficiently 
large scale (roughly $100$ Mpc or more).\footnote{The parsec (symbol: pc) is a unit of 
length used in astronomy, equal to about 30.9 trillion kilometers (19.2 trillion miles) or 3.26 light-years. 
The parsec is equal to the length of the adjacent side of an imaginary right triangle in space. 
The two dimensions on which this triangle is based are the angle (which is defined as 1 arcsecond), 
and the opposite side (which is defined as 1 astronomical unit, which is the average distance from the Earth 
to the Sun). Using these two measurements, along with the rules of trigonometry, the length of the 
adjacent side (the parsec) can be found. 1 Mpc is $10^6$ pc.} This simply means that there is no special location
for any observer, in any part of cosmos, and the large-scale picture of the Universe will look the 
same from any point is space.

Going back to history, Newtonian mechanics is an approximation which works quite well for most of our “earthly” needs, at
least when the velocity $v \ll c$ where $c$ is the speed of light.\footnote{Speed of light can be taken to be $3 \times 10^8$ m s$^{-1}$.} 
A more general theory was developed by Albert Einstein. The basic differences and
analogies between Newtonian and Einsteinian physics are presented in Table\ (\ref{table:N&E}).


\begin{threeparttable}
\centering
\begin{tabular}{|p{7.5cm}|p{7.5cm}|}
\hline \hline 
\hspace{1.3 cm}\textbf{Newtonian Mechanics.} & \hspace{1.3 cm}\textbf{Einsteinian Mechanics.} \\
\hline\hline
Absolute time and absolute space. & Dynamical spacetime, one entity.\\
[0.2cm]
\hline
Galilean invariance of space (simultaneity). &  Lorentz invariance of spacetime (time dilation, length contraction, no simultaneity).\\
[0.2cm]
\hline
Existence of preferred inertial frames (at rest or moving with constant velocity w.r.t. absolute space). & No preferred frames (physics is the same everywhere).\\
[0.6cm]
\hline
Infinite speed of light $c$ (instantaneous action at a distance). & Finite and fixed speed of light $c$ (nothing physical can propagate faster than c).\\
[0.2cm]
\hline
There is no upper limit on the speed with which mass can travel. & There is a upper limit of speed with which mass can travel, $c$.\\
[0.2cm]
\hline
Gravity is a force. & Gravity is a distortion of the fabric of spacetime. \\
[0.2cm]
\hline
Newton’s Second Law: & Geodesic equation: \\
[0.2cm]
$\frac{d^2 x^i}{dt^2}=-\delta^{ij}\frac{\partial \Phi}{\partial x^j}$.&$\frac{d^2 x^{\nu}}{d\lambda^2}=-\Gamma^{\nu}_{\gamma\delta}\frac{dx^\gamma}{d\lambda}\frac{dx^\delta}{d\lambda}$.\\
[0.2cm]
\hline
Poisson equation: & Einstein’s field equation:\\
[0.2cm]
${\nabla}^2\Phi(x)=4\pi G \rho_m$&$G_{\mu \nu}=8 \pi G T_{\mu \nu}$\\
[0.2cm]
\hline
Mass produces a field $\Phi$ causing a force on the other mass $m$ given by:& Spacetime is curved and mass particles move along curved geodesics defined by metric:\\
$\vec{F}=-\vec{\nabla}\Phi$. & $ds^2=g_{\mu \nu}(x)dx^\mu dx^\nu$.\\
[0.2cm]  
\hline
Absolute space acts on matter but is not acted upon (Newton's interpretation of the bucket experiment\tnote{a} ).\tnote{b}  & Spacetime acts on matter and is acted upon by matter (Einstein's field equation).\\
[0.2cm] 
\hline \hline 
\end{tabular}
\caption{\rm Differences and analogies between Newtonian and Einsteinian mechanics.}
\label{table:N&E}
\begin{tablenotes}
\item[a]{Interested readers can read more about it in the Principia \cite{newton1687philosophiae} or on Wikipedia.}
\item[b]{Special thanks to Shawn Westmoreland.}
\end{tablenotes}
\end{threeparttable}

\vspace{0.8 cm}
Newtonian mechanics quickly runs into phenomenon that it cannot explain:

\begin{itemize} 
    \item[\checkmark] Why do all observers measure the same speed of light $c$ (in a vacuum), as demonstrated by the Michelson-Morley experiment?

    \item[\checkmark] Why don't Maxwell's equations respect Galilean invariance?\footnote{For a detail proof see ``\textit{On the Galilean non-invariance of classical electromagnetism}" Preti \textit{et al.}.\cite{0143-0807-30-2-017} This is an excellent read.} 

    \item[\checkmark] Why do all bodies experience the same gravitational acceleration regardless of their mass? Why are the inertial and gravitational mass the same (as measured experimentally)?

    \item[\checkmark] Why does the perihelion of the orbit of Mercury not behave as required by Newton's equations?
\end{itemize}

To answer some of these questions, Einstein proposed his theory of Special Relativity in 1905, in which he
introduced some revolutionary concepts:

\begin{itemize} 
     \item[\checkmark] ``Abolished" absolute time --- introduced 4-dimensional spacetime as an inseparable entity.
     \item[\checkmark] However, the 4-dimensional spacetime considered in special relativity is still flat Minkowski spacetime.\footnote{This is discussed in detail later in this chapter.}
     \item[\checkmark] Finite and fixed speed of light $c$, independent of the observer.
     \item[\checkmark] Established the equivalence between energy and mass.     
     \item[\checkmark] Prohibition on any particle with non-zero rest mass to move with speed $v\geq c$.
\end{itemize}

Einstein’s theory of General Relativity, which he proposed in 1915, continued the revolution by adding the following ideas to the intellectual data base of humanity:

\begin{itemize} 
     \item[\checkmark] \textit{Equivalence principle}: Established the equivalence between inertial and gravitational mass.\footnote{It can also be stated as: There is no way of distinguishing between the effects on an observer of a uniform gravitational field and of constant acceleration. This is the fundamental axiom of general relativity.}
     \item[\checkmark] \textit{Cosmological principle}: Our position is “as mundane as it can be” (on large spatial scales, the Universe is homogeneous and isotropic).
     \item[\checkmark] \textit{Relativity}: The laws of physics are the same everywhere.
     \item[\checkmark] \textit{New definition of gravity}: Gravity is not a force but the distortion of the structure of spacetime as caused by the presence of matter and energy. The paths followed by matter and energy in spacetime are governed by the structure of spacetime. This great feedback loop is described by Einstein’s field equation. In the beautiful words of John Wheeler\footnote{He popularized the terms ``black hole", ``quantum foam", and ``wormhole." He with his two students Kip Thorne and Charles Misner wrote the book ``\textit{Gravitation}", which is  known as the `bible' of general relativity or `MTW'.\cite{kipthorne}} ``Mass-energy tells space-time how to curve, Curved space-time tells mass-energy how to move." So the 4-dimensional spacetime considered in general relativity is no longer flat (no longer Minkowskian).

\end{itemize}

After establishing general relativity as the way to describe the Universe and learning its mathematical formalism, 
we will finally embark on a journey of expressing, mathematically, the world around us on larger
scales, and physically interpreting the implications of this and relating this to the observations.
Many of the phenomena for which we now have overwhelming evidence, for --- the big bang, the expanding and accelerated Universe,
the cosmic microwave background (CMB) radiation, black holes, among others --- were first predicted from
Einstein's field equation. Therefore, it is the mathematics that hold the keys to unlocking
the mysteries of the Universe. So let us begin by reviewing the required mathematical ideas.

\section{Mathematical Background}

\subsection{Notation}

In this section we will develop some basic mathematical notations needed for 
general relativity.

\textbf{4-vectors}: \: $(t,x,y,z) \equiv (x^0,x^1,x^2,x^3)=x^\mu$.

\vspace{3mm}

\textbf{Conventions for indices}: 

\begin{itemize} 

       \item[$\star$] Roman letters $(i, j, k, l, m, n)$ run from 1 to 3;

       \item[$\star$] Greek letters $(\alpha,~\beta,~\gamma,~\delta,~\mu,~\nu,~\eta,~\xi)$ run from 0 to 3.
\end{itemize} 

\textbf{Einstein summation}: (summation over repeated indices): $\mu'^{\alpha}=\sum\limits_{\beta=0}^{3}{\frac{\partial x'^{\alpha}}{\partial x^{\beta}} \mu^{\beta} \equiv \frac{\partial x'^{\alpha}}{\partial x^{\beta}}\mu^{\beta}}$. Under change of coordinates $x^\beta \rightarrow x'^\beta.$

\textbf{Contravariant vector}: (index is a superscript) transforms as $A'^{\alpha}=\frac{\partial x'^{\alpha}}{\partial x^{\beta}} A^{\beta}$

\textbf{Covariant vector}: (index is a subscript) transforms as $A'_{\alpha}=\frac{\partial x^{\beta}}{\partial x'^{\alpha}} A_{\beta}$

\textbf{Tensors}: These are objects with multiple indices.

\begin{equation} 
\rm First~~rank~~tensor~~(one~~index):
\left\{
       \begin{array}{lr}
	    \vspace{6pt}
           \rm {contravariant}: \textit{A}'^{\alpha}=\frac{\partial x'^{\alpha}}{\partial x^{\beta}} \textit{A}^{\beta}. \hspace{1.6 cm} \\
	    \vspace{6pt}
           \rm {covariant}: \textit{A}'_{\alpha}=\frac{\partial x^{\beta}}{\partial x'^{\alpha}} \textit{A}_{\beta}.\hspace{1.6 cm} \\
       \end{array}
\right.
\end{equation}

\begin{equation} 
\rm Second~~rank~~Tensor~~(two~~indices):
\left\{
       \begin{array}{lr}

	    \vspace{6pt}

           \rm {contravariant}: \textit{A}'^{\alpha \beta}=\frac{\partial x'^{\alpha}}{\partial x^{\xi}}\frac{\partial x'^{\beta}}{\partial x^{\nu}} \textit{A}^{\xi \nu}.\\

	    \vspace{6pt}

           \rm {covariant}:  \textit{A}'_{\alpha \beta}=\frac{\partial x^{\alpha}}{\partial x'^{\xi}}\frac{\partial x^{\beta}}{\partial x'^{\nu}} \textit{A}_{\xi \nu}.\\

	    \vspace{6pt}

           \rm {mixed}: \textit{A}'^{\alpha}_{\beta}=\frac{\partial x'^{\alpha}}{\partial x^{\xi}} \frac{\partial x^{\nu}}{\partial x'^{\beta}} \textit{A}_{\nu}^{\xi}.\\

       \end{array}
\right.
\end{equation}

\begin{equation} 
\rm N^{th}~~rank~~Tensor~~(\textit{N}~~indices):
\left\{
      \begin{array}{lr}

	    \vspace{6pt}

            \rm {mixed}: \textit{A}'^{\alpha_1...\alpha_s}_{\alpha_{s+1}...\alpha_{N}}=
	    \frac{\partial x'^{\alpha_1}}{\partial x^{\beta_1}}... \frac{\partial x'^{\alpha_s}}{\partial x^{\beta_s}} 
            \frac{\partial x^{\alpha_{s+1}}}{\partial x'^{\beta_{s+1}}}... \frac{\partial x^{\alpha_N}}{\partial x'^{\beta_N}} 
            \textit{A}_{\beta_{s+1}...\beta_{N}}^{\beta_1...\beta_s}.\\

      \end{array}
\right.
\end{equation}

\textbf{Tensor Operations}: 

\begin{itemize} 

       \item[$\star$] Addition: $A^{\alpha \beta}_{\mu \nu}+B^{\alpha \beta}_{\mu \nu}=C^{\alpha \beta}_{\mu \nu}$

       \item[$\star$] Subtraction: $A^{\alpha \beta}_{\mu \nu}-B^{\alpha \beta}_{\mu \nu}=D^{\alpha \beta}_{\mu \nu}$

       \item[$\star$] Tensor Product:  $A^{\alpha \beta}_{\mu \nu}~B^{\gamma \delta}_{\eta \xi}=F^{\alpha \beta \gamma \delta}_{\mu \nu \eta \xi}$

       \item[$\star$] Contraction: $A^{\alpha \psi}_{\psi \gamma}=H^{\alpha}_{\gamma}$ \hspace{7 mm} (summed over $\psi$)

       \item[$\star$] Inner Product: $A^{\alpha \beta}_{\mu \nu}~B^{\nu \gamma}_{\delta \eta}=P^{\alpha \beta \nu \gamma}_{\mu \nu \delta \eta}= K^{\alpha \beta \gamma}_{\mu \delta \eta}$

\end{itemize} 

\textbf{Why Tensors are Important}:
 
When the equations of motion are written in tensor form, they are invariant under some appropriately-defined transformations. For example:

\begin{itemize} 

       \item[$\star$] Newtonian Mechanics: 3~-~vectors $(x,y,z)=(x^1,x^2,x^3)$ are invariant under Galilean transformations.

       \item[$\star$] Special Relativity:  4~-~vectors $(t,x,y,z)=(x^0,x^1,x^2,x^3)$ are invariant under Lorentz transformations.

       \item[$\star$] General Relativity:  4~-~vectors $(t,x,y,z)=(x^0,x^1,x^2,x^3)$ are invariant under general coordinate transformations.

\end{itemize}

\textbf{Scalar}: They are \textbf{invariant}, which means they are the same in all coordinate systems.

\subsection{Metric Tensor}

\subsubsection{Flat Euclidean space} 

Our everyday experience has taught us to think in terms of a flat space metric
(Euclidean), where parallel lines never cross and the sum of the interior angles of a triangle is $180^0$. In this case, the
invariant line element of space in Cartesian coordinates $(x,y,z)=(x^1,x^2,x^3)$ is:
\begin{equation} 
ds^2=(dx^1)^2+(dx^2)^2+(dx^3)^2,
\end{equation} 
and space is flat. An equivalent way of writing the above metric is:
\begin{equation} 
ds^2=\delta_{ij}dx^idx^j,
\end{equation} 
where $\delta_{\alpha \beta}$ is the Kronecker delta function defined as: 
\begin{equation}
\delta_{\alpha \beta}= \left\{
     \begin{array}{cc}
       1 & \mathrm{for}\ \alpha=\beta, \\
       0 & \mathrm{for}\ \alpha \neq \beta. \\
       \end{array}
   \right.
\end{equation}
Therefore, the
Euclidean space metric tensor for Cartesian coordinates is:
\begin{equation}
\delta_{ij}=\left( {\begin{array}{ccc}

1 & 0 & 0 \\

0 & 1 & 0 \\

0 & 0 & 1 \\
\end{array} } \right).
\end{equation}
An invariant line element in an arbitrary coordinate system in flat space, can be written in terms of
Cartesian coordinates (via change of variables):
\begin{equation}
ds^2=\delta_{ij}dx^idx^j=\delta_{ij}\left(\frac{\partial x^i}{\partial x'^k}dx'^k\right)\left(\frac{\partial x^j}{\partial x'^l}dx'^l\right)=
\delta_{ij}\frac{\partial x^i}{\partial x'^k}\frac{\partial x^j}{\partial x'^l}dx'^kdx'^l\equiv p_{kl}dx'^kdx'^l.
\label{eq:ds2FES}
\end{equation}
where $p_{kl}$ is the metric of the new coordinate system.

Since the line element is invariant under the interchange of $dx'$ and $dx$, we may, without loss of generality, take the metric tensor to be symmetric in general relativity. Furthermore, isotropy and homogeneity (as in the flat Euclidean
space) implies that the metric tensor in such a space will necessarily be \textit{diagonal}.

\begin{figure}[h]
\centering
    \includegraphics[height=3.5in]{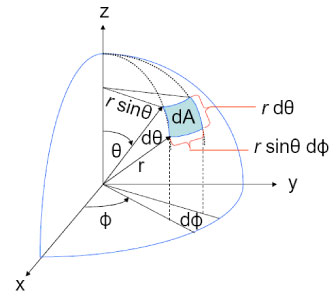}
    \caption{Spherical coordinate system. Figure shows $r=$ constant surface, hence, $dr=0$ on the surface.}
    \label{fig:spherical}
\end{figure}

Consider an example of spherical coordinates $(r,\theta,\phi)$, Fig.\ (\ref{fig:spherical}), where we are at the center of the spherical coordinate system.
As we look out into the ``cosmos," the flat space part of the metric (line element) is given by the following line element:\cite{griffithED}
\begin{eqnarray}
ds^2=dr^2+r^2\left(d\theta^2+\mathrm{sin^2}\theta d\phi^2\right),
\end{eqnarray}
where $\theta$ is now measured from the north pole and is $\pi$ at the south pole. It is useful to
abbreviate the term between parenthesis as:
\begin{eqnarray}
d\Omega^2=d\theta^2+\mathrm{sin^2}\theta d\phi^2,
\label{eq:angular}
\end{eqnarray}
because it is a measure of angle on the sky of the observer. Since the Universe is
isotropic, the angle between two galaxies as we see it is, in fact, the true angle from our
vantage point. The expansion of the Universe (which we will discuss later) does not change this angle
[we will explain this in Sec.\ (\ref{sec:Hubble law})]. Therefore
we need only $d\Omega$. So, for flat space, the line element is:
\begin{eqnarray}
ds^2=dr^2+r^2d\Omega^2.
\label{eq:ds2FES2}
\end{eqnarray}

\subsubsection{Flat Minkowski Spacetime} 

We can now generalize the interval to 4-dimensional flat spacetime $(x^0,x^1 ,x^2 ,x^3)$:
\begin{equation} 
ds^2=-(dx^0)^2+(dx^1)^2+(dx^2)^2+(dx^3)^2,
\label{eq:ds2FMS1}
\end{equation}
which can be written in compact notation as:
\begin{equation} 
ds^2=\eta_{\alpha \beta}dx^{\alpha}dx^{\beta},
\label{eq:ds2FMS}
\end{equation} 
where $\eta_{\alpha \beta}$ is the Minkowski (flat) spacetime metric tensor:
\begin{equation} 
\label{eq:Minkowskimetric}
\eta_{\alpha \beta}=
            \left( {\begin{array}{cccc}
-1 & 0 & 0 & 0 \\

0 & 1 & 0 & 0 \\

0 & 0 & 1 & 0 \\

0 & 0 & 0 & 1 \\
                \end{array} } \right).
\end{equation} 
Again, \textbf{isotropy and homogeneity} of spacetime leads to a diagonal metric tensor.

\subsubsection{Curved Three-Dimensional Space}

For a general (possibly curved) covariant spacetime metric tensor $g_{\alpha \beta}$, the
invariant line element is given by
\begin{equation} 
ds^2=g_{\alpha \beta}dx^{\alpha}dx^{\beta}.
\label{eq:ds2CST}
\end{equation} 
The contravariant spacetime metric tensor $g^{\alpha \beta}$ is the inverse of the covariant tensor $g_{\alpha \beta}$:
\begin{equation}
g^{\alpha \beta}g_{\beta \gamma}=\delta^{\alpha}_{\gamma}.
\end{equation}
This implies that whenever the metric tensor is diagonal:
\begin{equation}
g^{\alpha \beta}=(g_{\alpha \beta})^{-1}.
\end{equation}
One can take inner products of tensors with the metric tensor, thus lowering or raising indices:
\begin{equation}
A_{\alpha \beta}=g_{\alpha \mu}A^{\mu}_{\beta}, \hspace{7 mm} A^{\alpha \beta}=g^{\alpha \mu}A_{\mu}^{\beta}.
\end{equation}

For the spatial part of $g_{\alpha \beta}$, as proven by Robertson and Walker, the only alternative line elements, 
beside Eq.\ (\ref{eq:ds2FES2}), that obey both isotropy and homogeneity is:
\begin{eqnarray}
ds^2=dr^2+S_k(r)^2d\Omega^2,
\label{eq:ds2CST2}
\end{eqnarray}
where the function $S_k(r)$ is a function of space curvature given by:
\begin{equation}
S_k(r)= \left\{
     \begin{array}{cc}
       \frac{1}{\sqrt{k}}\ \mathrm{sin}\big(\sqrt{k}\ r\big) & \ \ \ \ \mathrm{for}\ k>0 \\
       r & \ \ \ \ \mathrm{for}\ k=0 \\
       \frac{1}{\sqrt{-k}}\ \mathrm{sinh}\big(\sqrt{-k}\ r\big) & \ \ \ \ \mathrm{for}\ k<0 \\
     \end{array},
   \right.
\label{eq:Sk}
\end{equation}
where the central $k=0$ case is given in Eq.\ (\ref{eq:ds2FES2}).
This means that the circumference of a sphere around us with radius $r$, for $k\neq 0$,
is no longer equal to $C=2\pi r$, but is smaller for $k > 0$ and larger for $k < 0$. Also,
the surface area of that sphere is no longer $S = 4\pi r^2$, but is smaller for $k > 0$
and larger for $k < 0$. For small $r$ (to be precise, for $r \ll |{k}|^{−1/2}$ ) the deviation from
$C = 2\pi r$ and $S = 4\pi r^2$ is small, but as $r$ approaches  $|{k}|^{−1/2}$ the deviation can
become very large. This can be checked by writing the Taylor's series expansion of Eq.\ (\ref{eq:Sk}).
This is very similar to the 2-dimensional example of the Earth’s surface.


If we stand on the North
Pole, and use $r$ as the distance from us along the sphere (i.e. the longitudinal distance) from the north pole
and $d\phi$ as the 2-dimensional version of $d\Omega$, then the circumference of a circle at $r = 10000$ km
(i.e. a circle that is the equator in this case) is “just” 40000 km instead of $2\pi \times10000 =
62831$ km, i.e. a factor of 0.63 smaller than it would be on a flat surface.

The constant $k$ is the curvature constant. We can also define a ``radius of curvature", as:
\begin{eqnarray}
R_{\mathrm{curvature}}=|k|^{-1/2},
\end{eqnarray}
which, for our 2-dimensional example of the Earth’s surface, is the radius of the Earth. In our 3-dimensional
Universe it is the radius of a 3-dimensional ``surface" of a 4-dimensional ``sphere"
in 4-dimensional space.

Note that the expression given in Eq.\ (\ref{eq:ds2CST2}) is not the only possible way of writing the 
metric in curved space. For instance, if we switch to a very commonly used parametrization in which we
change the radial coordinate from $r$ to $x \equiv S_{k}(r)$, then from Eq.\ (\ref{eq:Sk}):
\begin{equation}
r= \left\{
     \begin{array}{cc}
       \vspace{0.2cm}
       \frac{1}{\sqrt{k}}\ \mathrm{sin^{-1}}\left(\sqrt{k}\ x\right) & \ \ \ \ \mathrm{for}\ k>0 \\
       \vspace{0.2cm}
       x & \ \ \ \ \mathrm{for}\ k=0 \\
       \vspace{0.2cm}
       \frac{1}{\sqrt{-k}}\ \mathrm{sinh^{-1}}\Big(\sqrt{-k}\ x\Big) & \ \ \ \ \mathrm{for}\ k<0 \\
     \end{array},
   \right.
\label{eq:Sk2}
\end{equation}
which implies that:
\begin{equation}
dr= \left\{
     \begin{array}{cc}
       \vspace{0.2cm}
       \frac{1}{ \sqrt{k}}\left(\frac{ \sqrt{k}}{\sqrt{1-\left(\sqrt{k}x\right)^2}}dx\right) & \ \ \ \ \mathrm{for}\ k>0 \\
       \vspace{0.2cm}
       dx & \ \ \ \ \mathrm{for}\ k=0 \\
       \vspace{0.2cm}
       \frac{1}{ \sqrt{-k}}\left(\frac{ \sqrt{-k}}{\sqrt{1+\left(\sqrt{-k}x\right)^2}}dx\right) & \ \ \ \ \mathrm{for}\ k<0 \\
     \end{array}.
   \right.
\label{eq:Sk2.5}
\end{equation}
By squaring both sides of Eq.\ (\ref{eq:Sk2.5}) we get:
\begin{equation}
dr^2= \left\{
     \begin{array}{cc}
       \vspace{0.2cm}
       \frac{1}{1-kx^2}dx^2& \ \ \ \ \mathrm{for}\ k>0 \\
       \vspace{0.2cm}
       dx^2 & \ \ \ \ \mathrm{for}\ k=0 \\
       \vspace{0.2cm}
       \frac{1}{1-kx^2}dx^2 & \ \ \ \ \mathrm{for}\ k<0 \\
     \end{array}.
   \right.
\label{eq:Sk3}
\end{equation}
Then, the metric for homogeneous, isotropic, 3-dimensional space can be written as
\begin{eqnarray}
ds^2=\frac{dx^2}{1-kx^2}+x^2d\Omega^2,
\end{eqnarray}
which we can rewrite, by changing the name of the variable from $x$ to $r$, as
\begin{eqnarray}
ds^2&=&\frac{dr^2}{1-kr^2}+r^2d\Omega^2,\\
\Rightarrow ds^2&=&\frac{dr^2}{1-kr^2}+r^2\left(d\theta^2+\mathrm{sin^2}\theta d\phi^2\right).
\end{eqnarray}
Note that this metric is different only in the way we choose our coordinate $r$; it is not different
in any physical way from Eq. (\ref{eq:ds2CST2}).

\subsubsection{Expanding flat spacetime.}
\ \ \ \ The metric tensor for a flat, homogeneous, and isotropic spacetime, which is expanding or 
contracting spatially  with  scale factor $a(t)$, is obtained from the Minkowski metric by scaling the spatial
coordinates by $a^2(t)$:
\begin{equation} 
g_{\alpha \beta}=
            \left( {\begin{array}{cccc}
-1 & 0 & 0 & 0 \\

0 & a^2(t) & 0 & 0 \\

0 & 0 & a^2(t) & 0 \\

0 & 0 & 0 & a^2(t) \\
                \end{array} } \right).
\label{eq:FLRW}
\end{equation} 
Then the metric takes the form:
\begin{eqnarray}
ds^2=-dt^2+a^2(t)\left[dr^2+r^2d\Omega^2\right].
\end{eqnarray}
In Cartesian coordinates:
\begin{eqnarray}
ds^2=-dt^2+a^2(t)d{x}^2,
\end{eqnarray}
where $dx$ is known as the coordinate or comoving infinitesimal distance while $a dx$ is the physical or proper infinitesimal distance. 
\subsubsection{Expanding curved spacetime (Friedmann-Lema\^{i}tre-Robertson-Walker metric tensor).}

\ \ \ \ In cosmology, a common zeroth order approximation is to slice spacetime into
spacelike slices which are exactly homogeneous and isotropic. This means that there
exists a coordinate system in which the constant $t$ hypersurfaces are homogeneous
and isotropic. The proper time $t$, which labels the hypersurfaces, is called the cosmic
time.

There is evidence that the Universe is indeed statistically homogeneous (all places
look the same) and isotropic (all directions look the same) on scales larger than few
100 Mpc, as we noted at the beginning of the chapter. 
This does not prove that the Universe is well described by a model
which is exactly homogeneous and isotropic, but it does motivate us to use it as a first
approximation. We shall see that this approximation, is in fact, quite good, and at
early times it is excellent, as the Universe was then more homogeneous and isotropic.\footnote{Over
the age of the Universe, due to gravitational attraction, masses clustered together to form 
galaxies, voids (vacant spaces), clusters etc, increasing the spatial inhomogeneity of the matter distribution.}

Since the spacetime is spatially homogeneous and isotropic, its curvature is the
same at all points in space, but can vary in time. It can be shown that the spacetime metric of curved expanding space
can be written (by a suitable choice of the coordinates) in the form:\cite{liddle2000cosmological,
weinberg1972gravitation, kolb1990early, padmanabhan2000theoretical}
\begin{eqnarray}
ds^2=-dt^2+a^2(t)\left[\frac{dr^2}{1-kr^2}+r^2d\theta^2+r^2\mathrm{sin^2}\theta d\phi^2\right].
\label{eq:FRLWM}
\end{eqnarray}
An alternative form, in Cartesian as opposed to spherical coordinates, is
\begin{eqnarray}
ds^2=-dt^2+a^2(t)\frac{1}{(1+\frac{k}{4}r^2)^2}\delta_{ij}dx^idx^j.
\end{eqnarray}
In either form, this is called the \textit{Robertson-Walker} (RW) metric, sometimes the
\textit{Friedmann-Robertson-Walker} (FRW) metric or the \textit{Friedmann-Lema\^{i}tre-Robertson-Walker} (FLRW) metric.\footnote{The most commonly used term is the FRW metric. However, some authors prefer to make the distinction between the geometry (with the names Robertson and Walker attached) and the equations of motion (endowed with the name Friedmann and sometimes also Lema\^{i}tre).} It was first derived by
Friedmann in 1922 and then more generally by Robertson in late 1920s and early 1930's and Walker in
1935. Note that both form of the metric has the same amount
of symmetry as the spacetime itself: the metrics are isotropic, and homogeneous.
The full symmetry of the spacetime is usually not apparent in the metric itself, even
though all physical quantities calculated from the metric display the symmetry. The
time coordinate $t$ is the \textit{cosmic time.} Here $k$ is a constant, related to curvature of
space (not spacetime) and $a(t)$ is a function of time which governs how the Universe
expands (or contracts). In Eq.\ (\ref{eq:FRLWM}), the coordinates $r$, $\theta$, and $\phi$
are known as comoving coordinates. A
freely moving particle is at rest in these coordinates.
Equation\ (\ref{eq:FRLWM}) is a purely kinematic statement. In this
problem, the dynamics are associated with the scale factor $a(t)$. 
The Einstein equations allow us to determine the scale
factor $a(t)$ provided the matter content of the Universe is specified.

\subsection{Covariant Derivative}
 Consider a vector $\vec{A}$ given in terms of its components along the basis vectors $\hat{e}_{\alpha}$ as:
\begin{equation} 
\vec{A}=A^{\alpha}\hat{e}_{\alpha}.
\end{equation} 
Differentiating the vector $\vec{A}$ using the standard Leibniz rule for the differentiation 
of the product of functions $(fg)'=f'g+fg'$, we get:
\begin{equation} 
\frac{\partial \vec{A}}{\partial x^\alpha}=\frac{\partial}{\partial x^\alpha}\Big(A^\beta \hat{e}_\beta \Big)=
\frac{\partial A^\beta}{\partial x^\alpha}\hat{e}_\beta+A^\beta \frac{\partial \hat{e}_\beta}{\partial x^\alpha}.
\label{eq:leibnizrule}
\end{equation} 
In flat Cartesian coordinates the basis vectors are constant, so the last term in the Eq.\ (\ref{eq:leibnizrule})
vanishes. However, this is not the case in general curved spaces. In general, the derivative in the
last term will not vanish, and it will itself be given in terms of the original basis vectors:
\begin{equation}
\frac{\partial \hat{e}_\beta}{\partial x^\alpha} \equiv \Gamma^{\nu}_{\alpha \beta}\hat{e}_\nu,
\end{equation} 
where $\Gamma^{\nu}_{\alpha \beta}$ is called Christoffel symbol. It is given in terms of the metric tensor $g_{\mu \nu}$ as 
(see MWT \cite{kipthorne}):
\begin{equation}
\Gamma^{\nu}_{\alpha \beta}\equiv \frac{1}{2}g^{\nu \gamma}\Big(g_{\alpha\gamma,\beta}+g_{\gamma\beta,\alpha}-g_{\alpha\beta,\gamma}\Big).
\label{eq:christoffel}
\end{equation}
Here it is important to note that Christoffel symbols are not tensors.

Taking the curvature of the ambient manifold into account when taking derivatives of a scalar field $\phi$, a vector $A^{\alpha}$, or a
co-vector $A_{\alpha}$ will yield covariant derivatives:
\begin{equation}
\partial_{;\mu}\phi \equiv \partial_{,\mu}\phi,
\hspace{9 mm}
A_{\alpha;\beta}\equiv A_{\alpha, \beta}-\Gamma^{\nu}_{\alpha \beta}A_{\nu},
\hspace{9 mm}
A^{\alpha}_{;\beta}\equiv A^{\alpha}_{, \beta}+\Gamma^{\nu}_{\alpha \beta}A^{\nu},
\end{equation}
where we have used the short hand notation $\partial_{,\mu}\phi \equiv \frac{\partial \phi}{\partial x^\mu}$, 
$A_{\alpha,\beta} \equiv \frac{\partial A_{\alpha}}{\partial x^{\beta}}$ and 
$A^{\alpha}_{,\beta} \equiv \frac{\partial A^{\alpha}}{\partial x^{\beta}}$.
Other covariant derivatives of second rank contravariant and covariant tensor are defined as
\begin{eqnarray}
\nabla_{\rho}A^{\mu \nu}\equiv A^{\mu \nu}_{;\rho}\equiv A^{\mu \nu}_{,\rho}+\Gamma^{\mu}_{\rho \alpha}A^{\alpha \nu}+\Gamma^{\nu}_{\rho \beta}A^{\mu \beta}, \\
\nabla_{\rho}A_{\mu \nu}\equiv A_{\mu \nu;\rho}\equiv A_{\mu \nu,\rho}-\Gamma^{\alpha}_{\rho \mu}A_{\alpha \nu}-\Gamma^{\beta}_{\rho \nu}A_{\mu \beta},
\label{eq:covderoften}
\end{eqnarray}
respectively. The covariant derivative of mixed tensor is defined as
\begin{eqnarray}
\nabla_{\rho}A^{\mu}_{\nu}\equiv A^{\mu}_{\nu;\rho}\equiv A^{\mu}_{\nu,\rho}+\Gamma^{\mu}_{\rho \alpha}A^{\alpha}_{\nu}-\Gamma^{\beta}_{\rho \nu}A^{\mu}_{\beta},
\label{eq:covderofmixten}
\end{eqnarray}
where $A^{\mu \nu}_{,\rho}=\frac{\partial  A^{\mu \nu}}{\partial x^{\rho}}$, $A_{\mu \nu, \rho}=\frac{\partial  A_{\mu \nu}}{\partial x^{\rho}}$,
and $A^{\mu}_{\nu,\rho}=\frac{\partial  A^{\mu}_{\nu}}{\partial x^{\rho}}$.

For vector $A^{\alpha}$, and co-vector $A_{\alpha}$, defined along a curve $x^\beta = x^\beta(s)$, the covariant derivative along this
curve are
\begin{equation}
\frac{D A^\alpha}{Ds}\equiv \frac{d A^\alpha}{ds}+\Gamma^{\alpha}_{\beta\gamma}\frac{dx^\gamma}{ds}A^{\beta},
\hspace{7 mm}
\frac{D A_\alpha}{Ds}\equiv \frac{d A_\alpha}{ds}-\Gamma^{\beta}_{\alpha\gamma}\frac{dx^\gamma}{ds}A_{\beta},
\end{equation}

The covariant derivative in a curved spacetime is the analog to the ordinary derivative in Cartesian coordinates
in flat spacetime.

\subsubsection{Principle of General Covariance}
This principle states that all tensor equations valid in Special Relativity will also be valid
in General Relativity if:
\begin{itemize} 

      \item[$\star$] The Minkowski metric $\eta_{\alpha \beta}$ is replaced by a general curved metric $g_{\alpha \beta}$.
	\begin{eqnarray}
	ds^2=\eta_{\alpha \beta}dx^{\alpha}dx^{\beta} &\Rightarrow& ds^2=g_{\alpha \beta}dx^{\alpha}dx^{\beta}\\
        \eta_{\alpha \beta} u^{\alpha}u^{\beta}=-1   &\Rightarrow& g_{\alpha \beta} u^{\alpha}u^{\beta}=-1,
	\end{eqnarray}
      \item[$\star$] All the partial derivatives are replaced by covariant derivatives; in simple language the commas in the equations will be replaced by semicolon (, $\rightarrow$ ;). E.g.,
\begin{eqnarray}
 T^{\alpha \beta}_{,\beta}=0  &\Rightarrow&  T^{\alpha \beta}_{;\beta}=0.
\end{eqnarray}
 \end{itemize} 

\subsection{Geodesic Equation}
\label{Subsec:Geodesic Equation}
The fundamental axiom on which Newtonian mechanics is based is Newton's second Law, which
states that when a net force $\vec{F}$ acts on a body of mass $m$ it produces acceleration $\vec a$ in the direction of the net force such that
\begin{equation}
m\frac{d^2\vec{x}}{dt^2}=\vec{F}=-\vec{\nabla}\Phi \hspace{3 mm} \Rightarrow \hspace{3 mm} \frac{d^2\vec{x}}{dt^2}=-\frac{1}{m}\vec{\nabla}\Phi,
\end{equation}
where $\Phi$ is the scalar potential field, $d/dt$ is the time derivative and $\vec{\nabla}\Phi$ is the gradient of the scalar potential.
In the absence of forces acting on a body, Newton's second Law reduces to Newton's first Law 
(which means the first Law is the special case of second Law),
\begin{equation}
\frac{d^2\vec{x}}{dt^2}=-\frac{1}{m}\vec{\nabla}\Phi=0.
\label{eq: NFL}
\end{equation}
In flat Euclidean space and flat Minkowski spacetime, this also leads to straight lines.

It is a fundamental assumption of general relativity that, in curved spacetime, free particles (i.e., particles
feeling no non-gravitational effects) follow paths that extremize their proper interval $ds$. Such paths
are called  geodesics. Therefore, generalizing Newton's Laws of motion of a particle in the absence
of forces, Eq. (\ref{eq: NFL}), to a general curved spacetime metric, leads to the geodesic equation.

\textbf{Derivation of geodesic equation.}

We derive the geodesic equation using the variational principle (Lagrange's
equation).

Suppose the points $x^i$ lie on a curve parametrized by the parameter $\psi$, i.e.,
\begin{equation}
x^{\alpha}\equiv x^{\alpha}(\psi), \hspace{10 mm} dx^{\alpha}=\frac{dx^{\alpha}}{d\psi}d\psi,
\end{equation}
and the distance between two points A and B is denoted by $S_{AB}$ and is given by
\begin{equation}
S_{AB}=\int\limits_{A}^{B}ds=\int\limits_{A}^{B}\frac{ds}{d\psi}d\psi=
\mathlarger{\mathlarger{\int}}\limits_{A}^{B}\sqrt{g_{\alpha \beta}\frac{dx^{\alpha}}{d\psi}\frac{dx^{\beta}}{d\psi}}d\psi.
\label{eq:distance}
\end{equation}
The shortest path between the points A and B is called the geodesic, and it is found by extremizing
(minimizing) the path $S_{AB}$. This is done using the standard tools of variational calculus and leads to the
Lagrange equations; we review this technique next.


Consider a functional $G(x)$ of the form:
\begin{equation}
G(x)\equiv\mathlarger{\mathlarger{\int}}\limits^{B}_{A} L\left(\lambda, x, \frac{dx}{d\lambda}\right)d\lambda.
\label{eq:functional}
\end{equation}
Note that here our curve is parametrized by $\lambda$.
Let $x=\rm X(\lambda)$ be the curve extremizing $G(x)$ (this is what we are looking for). Then a nearby curve passing through A and B can be
parametrized as $x=\rm X(\lambda)+\varepsilon \eta(\lambda)$, such that $\eta(A)=\eta(B)=0$. To extremize Eq.\ (\ref{eq:functional}), we
have to require $\frac{dG}{d\varepsilon}\Big |_{\varepsilon=0}=0$. This means
\begin{figure}[h]
\begin{center}
    \includegraphics[height=3.5in]{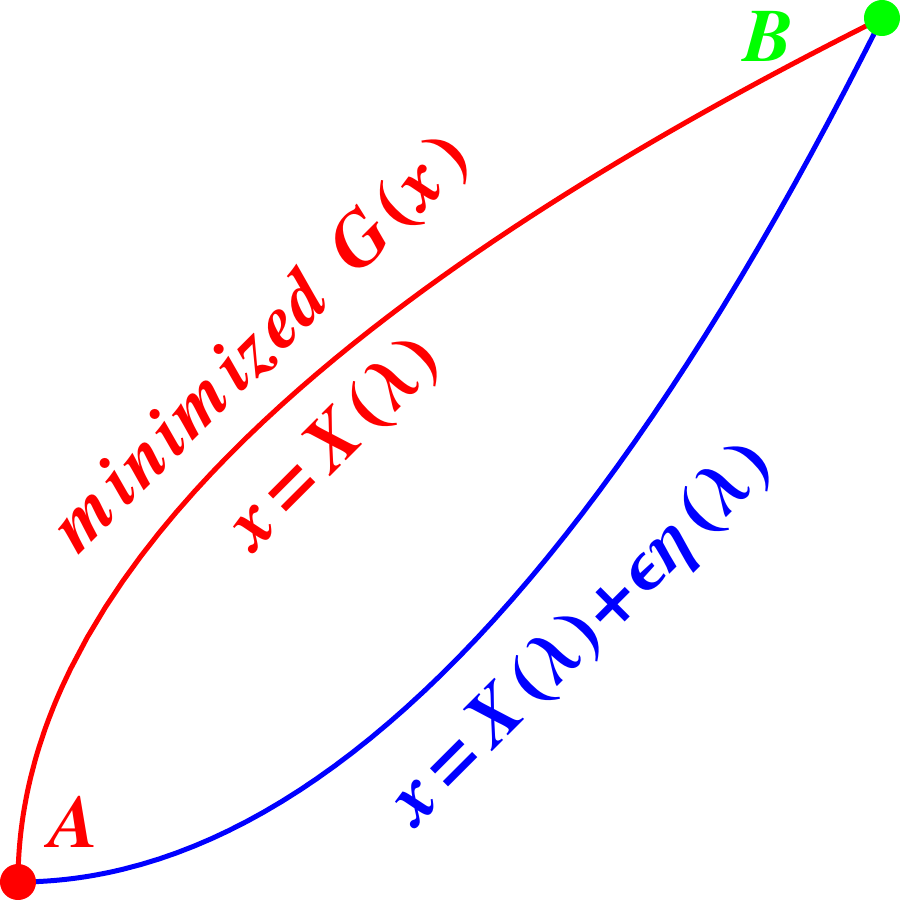}
    \caption{Variational Principle.}
    \label{fig:variational}
\end{center}
\end{figure}
\begin{eqnarray}
\frac{dG}{d \varepsilon}\bigg|_{\varepsilon=0} &=& \int \limits^{A}_{B} \left[\ \ \frac{\partial L}{\partial x}\frac{dx}{d\varepsilon}+
\frac{\partial L}{\partial \dot x}\frac{d \dot x}{d \varepsilon}+
\cancelto{0}{\frac{\partial L}{\partial \lambda}\frac{d \lambda}{d\varepsilon}}\ \ \ \ \right]d\lambda=0, 
\hspace{5 mm}	\left(\because \ \frac{d\lambda}{d \varepsilon}=0 \right) \nonumber \\
&=& \int \limits^{B}_{A} \left(\frac{\partial L}{\partial x}\eta+\frac{\partial L}{\partial \dot x}\dot\eta\right)d\lambda, 
\hspace{34 mm}	\left(\text{where } \dot x\equiv \frac{dx}{d\lambda},\ \dot \eta\equiv \frac{d\eta}{d\lambda}\right)  \nonumber \\
 &=& \int \limits^{B}_{A} \frac{\partial L}{\partial x}\eta~d\lambda +\int \limits^{B}_{A} \frac{\partial L}{\partial \dot x}\dot\eta ~d\lambda,    \nonumber \\
 &=& \int \limits^{B}_{A} \frac{\partial L}{\partial x}\eta~d\lambda +\frac{\partial L}{\partial \dot x} \eta \bigg|_{A}^{B}-\int 
\limits_{A}^{B}\eta \frac{d}{d\lambda}\frac{\partial L}{\partial \dot x}~d\lambda, \hspace{6 mm} \left(\text{here we integrated by parts}\right) \nonumber \\
&=& \int \limits^{B}_{A} \frac{\partial L}{\partial x}\eta~d\lambda +\frac{\partial L}{\partial \dot x}\Bigg |^{B}_{A} \left[\eta(B)-\eta(A)\right]-\int 
\limits_{A}^{B}\eta \frac{d}{d\lambda}\frac{\partial L}{\partial \dot x}~d\lambda, \nonumber \\
&=& \int \limits_{A}^{B} \eta \left[\frac{\partial L}{\partial x}-\frac{d}{d\lambda}\frac{\partial L}{\partial \dot x}\right]d\lambda=0. 
\hspace{10 mm} \left(\text{remember}\ \eta(A)=\eta(B)=0\right). 
\end{eqnarray}
But the function $\eta$ is arbitrary, so in order to have $\frac{dG}{d \varepsilon}\big|_{\varepsilon=0}=0$, 
the square brackets in the integrand must vanish, and so we arrive at Lagrange's equation
\begin{equation}
\frac{\partial L}{\partial x}-\frac{d}{d\lambda}\frac{\partial L}{\partial \dot x}=0.
\end{equation}
This can be extended to any number of phase-space coordinates as follows
\begin{equation}
\frac{\partial L}{\partial x^{\alpha}}-\frac{d}{d\lambda}\frac{\partial L}{\partial \dot x^{\alpha}}=0.
\label{eq:lagrange equation}
\end{equation}

\textbf{Euler-Lagrange equation in Scaler field theory}\cite{mandl2010quantum}

In the case of field theory in curved spacetime, the action $S(\Omega)$ for a scaler filed $\phi$ in an arbitrary region $\Omega$ of four-dimensional spacetime is given by:
\begin{eqnarray}
S(\Omega)=\int \limits_{\Omega}\sqrt{-g}\ \mathcal{L}(\phi,\partial_\alpha \phi)\mathrm{d^4}x,
\label{eq:EL}
\end{eqnarray}
where the Lagrangian density $\mathcal{L}(\phi,\partial_\alpha \phi)$ depends on the scalar field, and the first derivatives of the scalar 
field with respect to the coordinates only. This is not the most general case possible, but it covers all the theories in this work. 
Here $g$ is the determinant of the metric $g_{\alpha \beta}$ and $\mathrm{d^4}x$ stands for the four-dimensional volume
element $\mathrm{d}x^0\mathrm{d^3}\bf{x}$.

Now we postulate that the equations of motion (i.e., the field equations), are obtained from the variational 
principle.\cite{mandl2010quantum} For any region $\Omega$, we consider variations of the fields
\begin{eqnarray}
\phi(x)\rightarrow \phi(x)+\delta \phi (x),
\end{eqnarray}
which vanish on the surface $\Gamma(\Omega)$ bounding the region $\Omega$:
\begin{eqnarray}
\delta \phi (x)=0 \hspace{5 mm} \mathrm{on}\ \ \ \Gamma(\Omega).
\end{eqnarray}
Let's take the variation $\delta S(\Omega)$. We get
\begin{eqnarray}
\delta S(\Omega)=\int \limits_{\Omega}\left(\frac{\partial \left(\sqrt{-g}\mathcal{L}\right)}{\partial \phi}\delta \phi+
\frac{\partial \left(\sqrt{-g}\mathcal{L}\right)}{\partial(\partial_\alpha \phi)}\delta (\partial_{\alpha}\phi) \right)\mathrm{d^4}x,
\label{eq:ftl1}
\end{eqnarray}
but
\begin{eqnarray}
\frac{\partial\left(\sqrt{-g}\mathcal{L}\right)}{\partial \left(\partial_{\alpha} \phi\right)}\ \delta \left(\partial_{\alpha}\phi\right)=
\frac{\partial}{\partial x^\alpha}\left[\frac{\partial \left(\sqrt{-g}\mathcal{L}\right)}{\partial \left(\partial_{\alpha}\phi\right)}\delta \phi\right]-
\frac{\partial}{\partial x^\alpha}\left[\frac{\partial \left(\sqrt{-g}\mathcal{L}\right)}{\partial \left(\partial_{\alpha}\phi\right)}\right]\delta \phi,
\label{eq:ftl2}
\end{eqnarray}
and using Eq.\ (\ref{eq:ftl2}) in Eq.\ (\ref{eq:ftl1}), we find:
\begin{eqnarray}
\delta S(\Omega)=\int \limits_{\Omega}\left(\frac{\partial \left(\sqrt{-g}\mathcal{L}\right)}{\partial \phi}-
\frac{\partial}{\partial x^\alpha}\left[\frac{\partial \left(\sqrt{-g}\mathcal{L}\right)}{\partial \left(\partial_{\alpha}\phi\right)}\right]\right)\delta \phi\ \mathrm{d}^4x+
\int \limits_{\Omega}\frac{\partial}{\partial x^\alpha}\left[\frac{\partial \left(\sqrt{-g}\mathcal{L}\right)}{\partial(\partial_\alpha \phi)}\delta \phi \right]\mathrm{d}^4x. \hspace{5 mm}
\label{eq:ftl3}
\end{eqnarray}
The last term in Eq.\ (\ref{eq:ftl3}) can be converted to a surface integral over the surface $\Gamma(\Omega)$
using Gauss's divergence theorem in four dimensions. Since $\delta \phi=0$ on $\Gamma(\Omega)$, this surface
integral vanishes. If $\delta S(\Omega)$ is to vanish for arbitrary regions $\Omega$ and arbitrary variations $\delta \phi$,
Eq.\ (\ref{eq:ftl3}) leads to the Euler-Lagrange equation
\begin{eqnarray}
\frac{\partial \left(\sqrt{-g}\mathcal{L}\right)}{\partial \phi}-
\frac{\partial}{\partial x^\alpha}\left(\frac{\partial \left(\sqrt{-g}\mathcal{L}\right)}{\partial (\partial_\alpha \phi)}\right)=0.
\label{eq:ELE1}
\end{eqnarray}
This is the equation of motion of the field. We will use this equation in Chapter\ (\ref{Chapter3}), 
when we derive the equation of motion of a scalar field $\phi$, see Eq.\ (\ref{eq:SF3}). \\

\textbf{Geodesic equation, continued}

After this interesting derivation, we now focus again on the geodesic equation. We can now
apply the Lagrange equation to the Lagrangian given in Eq.\ (\ref{eq:distance}). 
Using $L=\sqrt{g_{\gamma \delta} \dot{x}^{\gamma}\dot{x}^{\delta}}$, is more traditional, but it leads us into mathematical ambiguity. Since
squaring and scaling the Lagrangian will not effect the equation of motion,\footnote{Here Lagrangian is invariant hence in this particular case any function of Lagrangian will give us same equation of motion. But in general we cannot do that.} we will use: 
\begin{equation}
L=\frac{1}{2}g_{\gamma \delta} \dot{x}^{\gamma}\dot{x}^{\delta},
\label{eq:lagrangian}
\end{equation}
because it is easy to derive equation of motion from here.
After substituting Eq.\ (\ref{eq:lagrangian}) into Eq.\ (\ref{eq:lagrange equation}) we have:
\begin{eqnarray}
\frac{d}{dx^\alpha}\left[\frac{1}{2}\ g_{\gamma \delta}\dot x^{\gamma} \dot x^{\delta}\right]-\frac{d}{d \lambda}\left[\frac{\partial}{\partial \dot x^\alpha}\left(\frac{1}{2}\ g_{\gamma \delta}\dot x^{\gamma} \dot x^{\delta}\right)\right]&=&0, \nonumber \\
 \nonumber \\
\Rightarrow \frac{1}{2}\ \frac{d}{dx^\alpha} \left(g_{\gamma \delta}\right)\dot x^{\gamma} \dot x^{\delta}-
\frac{d}{d \lambda}\left[\frac{1}{2}g_{\gamma \delta}\left(\dot x^\gamma \delta^{\delta}_{\alpha}+\dot x^\delta \delta^{\gamma}_{\alpha}\right)\right]&=&0, \nonumber \\
\nonumber \\
\Rightarrow \frac{1}{2}\ g_{\gamma \delta,\alpha}\dot x^{\gamma} \dot x^{\delta}-\frac{d}{d\lambda}\left[\frac{1}{2}\ \left(g_{\gamma\alpha}\dot x^\gamma+g_{\alpha\delta}\dot x^\alpha\right)\right]&=&0, \nonumber \\
\nonumber \\
\Rightarrow \frac{1}{2}\ g_{\gamma \delta,\alpha}\dot x^{\gamma} \dot x^{\delta}-\frac{d}{d\lambda}\left[\frac{1}{\cancel{2}}\ \left(\cancel{2}g_{\gamma\alpha}\dot x^\gamma\right)\right]&=&0,
\end{eqnarray}
where we have used the fact that $\gamma$, and $\alpha$ are dummy indices and $g_{\gamma \delta,\alpha}\equiv\frac{\partial g_{\gamma \delta}}{\partial x^{\alpha}}$. Then we will get
\begin{eqnarray}
\frac{1}{2}g_{\gamma \delta, \alpha} \dot{x}^{\gamma}\dot{x}^{\delta}-\frac{d}{d\lambda}\left(g_{\gamma \alpha} \dot{x}^\gamma \right)&=&0, \nonumber \\
\Rightarrow\ \ \ \frac{1}{2}g_{\gamma \delta, \alpha} \dot{x}^{\gamma}\dot{x}^{\delta}-\dot{x}^\gamma\frac{d}{d\lambda}(g_{\gamma \alpha})-g_{\gamma \alpha} \ddot{x}^{\gamma}&=&0,
\end{eqnarray}
From the chain rule
\begin{equation}
\frac{d}{d \lambda}g_{\gamma \alpha}=\frac{\partial g_{\gamma \alpha}}{\partial x^\delta}\cdot \frac{dx^\delta}{d\lambda}=\frac{\partial g_{\gamma \alpha}}{\partial x^\delta}\dot x^{\delta}=g_{\gamma \alpha,\delta}\dot x^{\delta}, 
\hspace{10 mm} \left(\text{Here $\dot \kappa \equiv \frac{d \kappa}{d \lambda}$}\right) 
\end{equation}
So we find,
\begin{eqnarray}
\frac{1}{2}g_{\gamma \delta, \alpha}\dot x^{\gamma} \dot x^{\delta}-g_{\gamma \alpha, \delta}\dot x^{\delta}\dot x^{\gamma}-g_{\gamma \alpha}\ddot x^{\gamma}&=&0, \nonumber \\
\Rightarrow 
\left(\frac{1}{2}g_{\gamma \delta, \alpha}-g_{\gamma \alpha, \delta}\right)\dot x^{\gamma} \dot x^{\delta}-g_{\gamma \alpha}\ddot x^{\gamma}&=&0.
\label{eq:geodesic1}
\end{eqnarray}
To simplify we multiply Eq.\ (\ref{eq:geodesic1}) by $g^{\nu \alpha}$ and get
\begin{equation}
g^{\nu \alpha}\left(\frac{1}{2}g_{\gamma \delta, \alpha}-g_{\gamma \alpha, \delta}\right)\dot x^{\gamma} \dot x^{\delta}-\ddot x^{\nu}=0.
\label{eq:geodesic2}
\end{equation}
Writing this equation like Newton's second Law, it takes the form
\begin{eqnarray}
\ddot x^{\nu}&=&-g^{\nu \alpha}\left(g_{\gamma \alpha, \delta}-\frac{1}{2}g_{\gamma \delta, \alpha}\right)\dot x^{\gamma} \dot x^{\delta}, \nonumber \\
&=&-\frac{1}{2}g^{\nu \alpha}\left(2g_{\gamma \alpha, \delta}-g_{\gamma \delta, \alpha}\right)\dot x^{\gamma} \dot x^{\delta},
\label{eq:geodesic3}
\end{eqnarray}
or using the symmetry of the metric $g_{\mu \nu}$, we can write Eq.\ (\ref{eq:geodesic3}) 
in terms of the Christoffel symbol $\Gamma^{\nu}_{\gamma \delta}$ which is defined in Eq.\ (\ref{eq:christoffel}):
\begin{equation}
\ddot x^{\nu}=-\Gamma^{\nu}_{\gamma \delta}\dot x^{\gamma} \dot x^{\delta}.
\label{eq:geodesic4}
\end{equation}
More commonly it is written as:
\begin{equation}
\frac{d^2 x^\nu}{d \lambda^2} +\Gamma^{\nu}_{\gamma\delta}\frac{dx^\gamma}{d\lambda}\frac{dx^\delta}{d\lambda}=0.
\label{eq:geodesic5}
\end{equation}
Note that here we have used $g_{\gamma \alpha, \delta}\dot x^{\gamma} \dot x^{\delta}=
g_{\alpha \delta, \gamma}\dot x^{\gamma} \dot x^{\delta}$ [see Eq.\ (\ref{eq:geodesic3}) and Eq.\ (\ref{eq:christoffel})].
In Euclidean space and Minkowski spacetime, $g_{\alpha \beta}$ is diagonal and constant, so its derivatives, and
consequently the Christoffel symbol, vanish, thus leaving us with the equation of motion for a straight line, as it must.
Another advantage for using the Lagrangian in the form given in Eq.\ (\ref{eq:lagrangian}) is that solving the
Lagrange equation in (\ref{eq:lagrange equation}) in each coordinate yields the differential equation of the same form as
the geodesic equation in (\ref{eq:geodesic5}). The Christoffel symbols can then simply be read off.

\textbf{Simple examples:}
\begin{description} 
    \item[First: Geodesics on the surface of a sphere] \hfill \\
     In spherical polar coordinates the vector line element is
\begin{equation}
d\vec {r}=dr ~\hat{a}_r+r~d\theta ~\hat{a}_\theta+r~\mathrm{sin}(\theta)~d\phi ~\hat {a}_\phi.
\end{equation}
Without loss of generality, we may take the sphere to be of
unit radius. The length of a path from A to B between two fixed points $\theta_1$ and $\theta_2$ is given by:
\begin{eqnarray}
s&=&\int \limits_{\theta_1}^{\theta_2}ds, \nonumber \\
&=&\int \limits_{\theta_1}^{\theta_2}\sqrt{d\theta^2+\mathrm {sin^2}\theta ~d\phi^2}\ d\theta, \hspace{10mm} (\mathrm{since ~~~} dr=0)\nonumber \\
&=&\int \limits_{\theta_1}^{\theta_2}\sqrt{1+\mathrm {sin^2}\theta ~\left(\frac{d\phi}{d\theta}\right)^2}\ d\theta, \nonumber \\
&=&\int \limits_{\theta_1}^{\theta_2}\sqrt{1+\mathrm {sin^2}\theta ~\phi'^2}\ d\theta, \hspace{10mm} \left(\mathrm{where}~~\phi'\equiv \frac{d\phi}{d\theta}\right).
\end{eqnarray}
Therefore, one can use $L(\theta, \phi, \phi')=(1+\mathrm{sin^2}\theta~\phi'^2)^{1/2}$ to compute the trajectories between
$\theta_1$ and $\theta_2$ which have the shortest distance. Such trajectories are called geodesics and will
play a significant role in the following discussion of general relativity. In this case, the Euler-Lagrange equation (\ref{eq:lagrange equation}) takes the form:
\begin{eqnarray}
\frac{dL}{d\phi}-\frac{d}{d\theta}\left(\frac{\partial L}{\partial \phi'}\right)=0. \nonumber
\end{eqnarray}
Since $L$ does not depend upon $\phi$, hence Euler-Lagrange equation becomes
\begin{eqnarray}
\frac{d}{d\theta}\left(\frac{\partial}{\partial \phi'}\sqrt{1+\mathrm {sin^2}\theta~\phi'^2} \right)=0,
\end{eqnarray}
so that
\begin{eqnarray}
\frac{\mathrm{sin^2}\theta~\phi'}{\sqrt{1+\mathrm{sin^2}\theta~\phi'^2}}=c.
\end{eqnarray}
Rewriting, we have,
\begin{eqnarray}
\phi'=\frac{c}{\mathrm{sin}\theta~\sqrt{\mathrm{sin^2}\theta-c^2}}, 
\end{eqnarray}
and integrating with respect to $\theta$ gives
\begin{eqnarray}
\phi=\int \frac{c}{\mathrm{sin}\theta~\sqrt{\mathrm{sin^2}\theta-c^2}}d\theta.
\end{eqnarray}
To do the integral, use the substitution $u=\mathrm{cot}\theta$, 
so $du=-\mathrm{cosec^2}\theta d\theta \Rightarrow d\theta=-\mathrm{sin^2}\theta du \Rightarrow d\theta=-\frac{du}{1+u^2}$,
where $\mathrm{sin}\theta=\frac{1}{\sqrt{1+u^2}}$, which gives
\begin{eqnarray}
\phi&=&\mathlarger{\mathlarger{\mathlarger{\int}}} \frac{c}{\frac{1}{\sqrt{1+u^2}} \left(\frac{1}{1+u^2}-c^2\right)^{1/2}}\left(\frac{-du}{\left(\sqrt{1+u^2}\right)^2}\right), \nonumber \\
    &=&\mathlarger{\int} \frac{-c}{\sqrt{1-c^2 (1+u^2)}}du, \nonumber \\
    &=&-\mathlarger{\int} \frac{1}{\sqrt{a^2-u^2}}du, \hspace{10mm} \left(\mathrm{where,}~ac=\sqrt{1-c^2}\right) \nonumber \\
    &=&\mathrm{cos^{-1}}\left(\frac{u}{a}\right)+\phi_0, 
\end{eqnarray}
where $\phi_0$ is the constant of integration. Hence, the geodesic path is given by:
\begin{eqnarray}
\mathrm{cot}\theta=a \mathrm{cos}(\phi-\phi_0),
\end{eqnarray}
and the arbitrary constants $a$ and $\phi_0$ must be found using the end-points. This is a 
great circle path.
    \item[Second: The brachistochrone (shortest time) problem] \hfill \\
     Two fixed points, $P$ and $Q$, are connected by a smooth wire lying in the vertical plane
      that contains $P$ and $Q$, see Fig.\ (\ref{fig:brachistochrone}). A particle is released from rest at $P$ and slides, under uniform
      gravity, along the wire to $Q$. Let's calculate the shape the wire should be so that the transfer of the particle is
      completed in the shortest time. 
\begin{figure}[h]
\centering
    \includegraphics[height=2.5in]{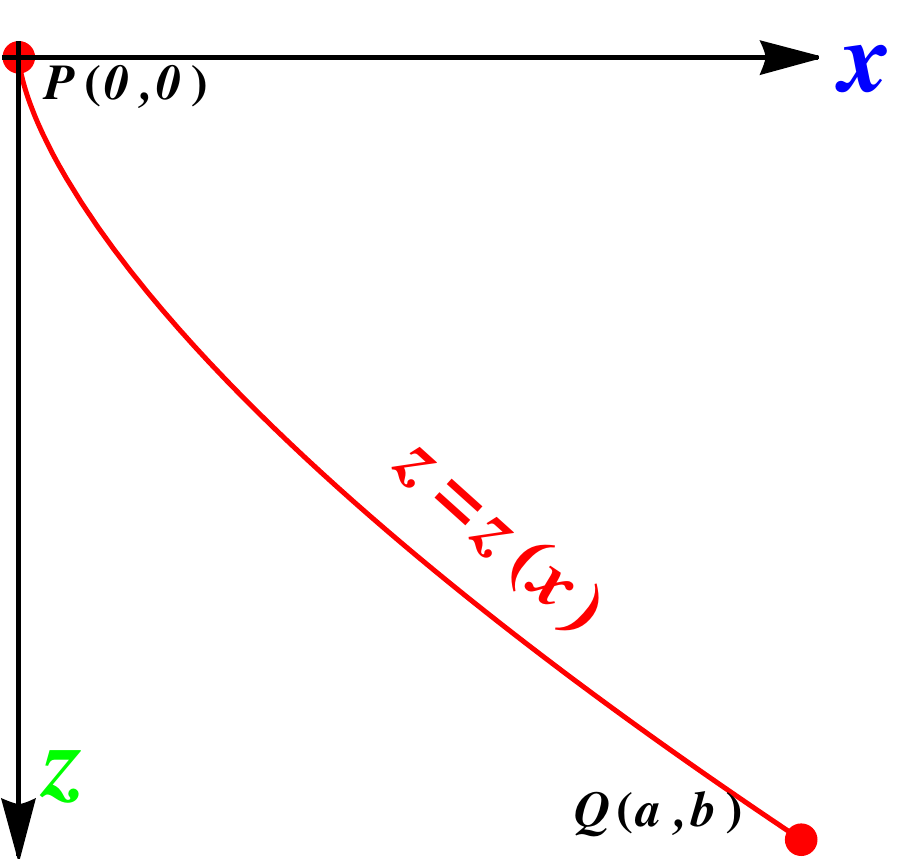}
    \caption{Brachistochrone problem, curve of fastest descent. It is the curve between the two points $P(0,0)$ and $Q(a,b)$ that is covered in the least time by a point particle that starts at $P(0,0)$ with zero kinetic energy and is constrained to move along the curve ``Brachistochrone" to $Q(a,b)$, under the action of only constant gravity and assuming no friction.}
    \label{fig:brachistochrone}
\end{figure}

Let's start by setting up the co-ordinate system. Suppose that the wire 
lies in the $(x, z)$-plane with $Pz$ pointing vertically downwards,
$P$ at the origin, and $Q$ at the point $(a, b)$. Let the shape of the wire be given by
the curve $z = z(x)$. Then, since the particle is released from rest when $z = 0$, energy
conservation implies that the speed of the particle when its downward displacement
is $z$ is given as $v(z)=\sqrt{2gz}$. Then: 
\begin{eqnarray}
&           & v(z)=\sqrt{2gz}, \nonumber \\
&\Rightarrow& v^2(z)=2gz, \hspace{44 mm} \left(\mathrm{squaring\ both\ sides}\right) \nonumber \\
&\Rightarrow& v_x^2+v_z^2=2gz,  \nonumber \\
&\Rightarrow& \left(\frac{dx}{dt}\right)^2+\left(\frac{dz}{dt}\right)^2=2gz, \nonumber \\
&\Rightarrow& \left(\frac{dx}{dt}\right)^2\left[1+\frac{dz}{dx}\right]^2=2gz, \nonumber \\
&\Rightarrow& (2g)^{-1/2}\frac{\left[1+\dot z^2\right]}{z^{1/2}}dx=dt \hspace{15 mm} \left(\mathrm{defining}\ \dot z=\frac{dz}{dx}\right). 
\end{eqnarray}
The time, $T$, taken for the particle to complete the transfer is therefore
\begin{eqnarray}
T\left(z(x)\right)=(2g)^{-1/2} \int \limits_{0}^{a}\frac{(1+\dot z^2)^{1/2}}{z^{1/2}}dx.
\end{eqnarray}
The problem is to find the function $z(x)$, satisfying the end
conditions $z(0) = 0$, $z(a) = b$, and that minimizes $T$ [see Fig.\ (\ref{fig:brachistochrone})]. 
If $x=x^*$ minimizes $T$ , then it must make $T$ stationary and so be an extremized 
$T$. Since $x$ is not explicitly present in this functional, we can use the integrated form of Eq.\ (\ref{eq:lagrange equation}), $\dot z\frac{\partial L}{\partial \dot z}-L=0$.
Substituting $L(x,z,\dot z)=\left(1+\dot z^2\right)^{1/2}/z^{1/2}$ and simplifying we obtain:
\begin{eqnarray}
z\left(1+\dot z^2\right)=2C,
\end{eqnarray}
where $C$ is a positive constant (the constant of integration is called $2C$ for convenience in latter calculations). This
equation can be rearranged as
\begin{eqnarray}
\dot z=\pm \left(\frac{2C-z}{z}\right)^{1/2},
\end{eqnarray}
a pair of first order separable ODEs. Integration gives:
\begin{eqnarray}
x=\pm \int \left(\frac{z}{2C-z}\right)^{1/2}dz.
\end{eqnarray}
To perform the integral, we use the substitution $z=C(1-\mathrm{cos}\ \psi)$, in which case
\begin{eqnarray}
x&=&\pm \int \left(\frac{1-\mathrm{cos}\ \psi}{1+\mathrm{cos}\ \psi}\right)^{1/2}\mathrm{sin}\ \psi\ d\psi \nonumber \\
&=&\pm C(\psi-\mathrm{sin}\ \psi)+D,
\end{eqnarray}
where $D$ is a constant of integration. Hence the parametric form of $z(x)$ that minimize $T(z)$ is
\begin{eqnarray}
x=\pm C(\psi-\mathrm{sin}\ \psi)+D, \hspace{10mm} z=C(1-\mathrm{cos}\ \psi).
\end{eqnarray}
Here, $C$ and $D$ are two constants that can be fixed by fixing the two points $P$ and $Q$.
\end{description}

\subsection{Relating the Geodesic Equation with Newtonian Gravity}
To see that the geodesic equation\ (\ref{eq:geodesic5}) describes the motion of a particle in a more general theory of gravity than Newton's gravity, we recover Newton's second Law of motion as an approximation of Eq.\ (\ref{eq:geodesic5}). In order to do this, we make following three 
approximations.
\begin{description} 
    \item[First] \hfill \\
	      The first approximation is the slow motion, non-relativistic ($v\ll c$) one. In this case
              $\frac{dx^i}{d\lambda} \ll \frac{dx^0}{d\lambda}$, $\left(\mathrm{or}\ \frac{dx^i}{d\lambda} \ll 1\right)$,\footnote{Since:
	      $\Delta x^i=v \Delta x^0$, but $v \ll c$, $\Rightarrow$ $\Delta x^i \ll \Delta x^0$, $\Rightarrow$ 
	      $\frac{\Delta x^i}{\Delta \lambda} \ll \frac{\Delta x^0}{\Delta \lambda}$.} and the 
              geodesic equation Eq.\ (\ref{eq:geodesic5}) will take the form (keeping only $\gamma=\delta=0$ terms): 
\begin{eqnarray}
\centering
 \frac{d^2 x^\nu}{d \lambda^2} +\Gamma^{\nu}_{00}\frac{dx^0}{d\lambda}\frac{dx^0}{d\lambda}&=&0, \nonumber \\
 \Rightarrow\ \ \ \frac{d^2 x^\nu}{d \lambda^2} +\Gamma^{\nu}_{00}\left(\frac{dx^0}{d\lambda}\right)^2&=&0.
\label{eq:NfromG1}
\end{eqnarray}
    \item[Second] \hfill \\
		  The second approximation for the metric tensor of spacetime is the weak field approximation (nearly flat spacetime). In this case, we write
\begin{equation}
g_{\mu \nu}(x)=\eta_{\mu \nu}+h_{\mu \nu}(x),
\label{eq:NfromG2}
\end{equation}
where $\eta_{\mu \nu}$ is the Minkowski (flat) spacetime metric tensor, and $h_{\mu \nu}(x)$ is a small perturbation $\left(|h_{\mu \nu}(x)|\ll 1\right)$ that depends only on position. 

Since the components of the correction tensor $h_{\mu \nu}$ are small, we can use 
$\eta_{\mu \nu}$, and its inverse to raise and lower indices without including any $h_{\mu \nu}$ term. We will need $g^{\mu \nu}(x) \approx \eta^{\mu \sigma}-h^{\mu \sigma}$, justified as follows, 
\begin{eqnarray}
g_{\mu \nu}g^{\mu \sigma}&=& g_{\mu \nu}\left(\eta^{\mu \sigma}-h^{\mu \sigma}\right), \nonumber \\
&=&\left(\eta_{\mu \nu}+h_{\mu \nu}\right)\left(\eta^{\mu \sigma}-h^{\mu \sigma}\right), \nonumber \\
&=&\eta_{\mu \nu}\eta^{\mu \sigma}-\eta_{\mu \nu}h^{\mu \sigma}+h_{\mu \nu}\eta^{\mu \sigma}-\underbrace{h_{\mu \nu}h^{\mu \sigma}}_{O[h^2]\rightarrow \mathrm{Neglect}}, \nonumber \\
&\approxeq &\delta_{\nu}^{\sigma}-\eta_{\mu \nu}h^{\mu \sigma}+h_{\mu \nu}\eta^{\mu \sigma}.
\label{eq:NfromG2.1}
\end{eqnarray}
Let's consider the last term of Eq.\ (\ref{eq:NfromG2.1}):
\begin{eqnarray}
h_{\mu \nu}\eta^{\mu \sigma}&=&\left(\eta_{\nu \rho}h_{\mu}^{\rho}\right)\left(\eta^{\sigma \lambda}\eta_{\lambda}^{\mu}\right), \nonumber \\
&=&\eta_{\nu \rho}\eta^{\sigma \lambda}h_{\mu}^{\rho}\eta_{\lambda}^{\mu}, \nonumber \\
&=&\eta_{\nu \rho}\eta^{\sigma \lambda}\left(\eta_{\alpha \mu} h^{\alpha \rho}\right)\left(\eta^{\alpha \mu}\eta_{\alpha \lambda}\right), \nonumber \\
&=&\eta_{\nu \rho}\eta^{\sigma \lambda}\underbrace{\left(\eta_{\alpha \mu}\eta^{\alpha \mu} \right)}_{=1}\left(h^{\alpha \rho}\eta_{\alpha \lambda}\right), \nonumber \\
&=&\eta_{\nu \rho}\underbrace{\left(\eta^{\sigma \alpha}\eta_{\alpha \lambda}\right)}_{=\delta_{\alpha}^{\sigma}}h^{\alpha \rho}, \nonumber \\
&=&\eta_{\nu \rho}\delta_{\alpha}^{\sigma}h^{\alpha \rho}, \nonumber \\
&=&\eta_{\nu \rho}h^{\sigma \rho}, \nonumber \\
&=&\eta_{\nu \mu}h^{\mu \sigma}.
\label{eq:NfromG2.2}
\end{eqnarray}
Hence, from Eq.\ (\ref{eq:NfromG2.1}) and Eq.\ (\ref{eq:NfromG2.2}) we can see $g_{\mu \nu}\left(\eta^{\mu \sigma}-h^{\mu \sigma}\right)=\delta_{\nu}^{\sigma}$:
\begin{equation}
\Rightarrow g^{\mu \nu}(x)=\eta^{\mu \nu}-h^{\mu \nu}(x).
\label{eq:NfromG3}
\end{equation}
    \item[Third] \hfill \\
		  The third approximation is the time independence of the perturbation. We will consider $h_{\mu \nu}(x)$ to be the time independent, i.e.
\begin{equation}
\partial_0{h_{\mu \nu}(x)}=\frac{\partial h_{\mu \nu}(x)}{\partial x^0}=\frac{\partial h_{\mu \nu}(x)}{\partial t}=0.
\label{eq:NfromG4}
\end{equation}
\end{description}
We now compute the Christoffel symbol in Eq.\ (\ref{eq:NfromG1}) under the second and third approximation and the definition in Eq.\ (\ref{eq:christoffel}) we have:
\begin{eqnarray}
\Gamma^{\nu}_{00}&\equiv& \frac{1}{2}g^{\nu \gamma}\Big(g_{0 \gamma,0}+g_{\gamma 0,0}-g_{00,\gamma}\Big).
\label{eq:NfromG5}
\end{eqnarray}
Because of the stationary field approximation $\left(g_{\alpha \beta,0}=0\right)$, this becomes
\begin{eqnarray}
\Gamma^{\nu}_{00}&=& -\frac{1}{2}g^{\nu \alpha}g_{00,\alpha} = -\frac{1}{2}g^{\nu i}g_{00,i}.
\label{eq:NfromG6}
\end{eqnarray}
From Eq.\ (\ref{eq:NfromG2})
\begin{eqnarray}
g_{\alpha \beta, i}(x)=\eta_{\alpha \beta, i}+h_{\alpha \beta, i}(x)=h_{\alpha \beta, i}(x). \hspace{5 mm} \left(\mathrm{since}\ \eta_{\alpha \beta, i}=0\right)
\label{eq:NfromG7}
\end{eqnarray}
Eqs.\ (\ref{eq:NfromG3}), (\ref{eq:NfromG5}), and (\ref{eq:NfromG7}), lead to:
\begin{eqnarray}
\Gamma^{\nu}_{00}&=&-\frac{1}{2}g^{\nu i}h_{00,i} \nonumber \\
&=&-\frac{1}{2}\left(\eta^{\nu i}-h^{\nu i}\right)h_{00,i} \nonumber \\
&=&-\frac{1}{2}\eta^{\nu i}h_{00,i}+\underbrace{\frac{1}{2}h^{\nu i}h_{00,i}}_{O^2(h)\rightarrow \mathrm{Neglect~it}}.
\label{eq:NfromG8}
\end{eqnarray}
Since $\Gamma_{00}^{0}=0$, Eqs.\ (\ref{eq:NfromG1}) and (\ref{eq:NfromG8}), gives the geodesic equation as
\begin{eqnarray}
 \frac{d^2 x^j}{d \lambda^2}&=&\frac{1}{2}\eta^{ji}h_{00,i}\left(\frac{dx^0}{d\lambda}\right)^2.
\label{eq:NfromG9}
\end{eqnarray}
Now
\begin{eqnarray}
\frac{d x^j}{d \lambda}&=&\frac{d x^j}{dt}\frac{dt}{d \lambda}, \nonumber \\
\frac{d^2 x^j}{d \lambda^2}&=&\frac{d}{d \lambda}\left(\frac{d x^j}{dt}\frac{dt}{d \lambda}\right)=\frac{d}{d t}\left(\frac{d x^j}{dt}\frac{dt}{d \lambda}\right)\frac{dt}{d \lambda}=\frac{d^2 x^j}{dt^2}\left(\frac{dt}{d \lambda}\right)^2.
\label{eq:NfromG10}
\end{eqnarray}
So the geodesic equation becomes
\begin{eqnarray}
\frac{d^2 x^j}{dt^2}=\frac{1}{2}\eta^{ji}h_{00,i}.
\label{eq:NfromG11}
\end{eqnarray}
Recalling that $x^j=(x,y,z)$, and expressing it in vector format, we arrive at
\begin{eqnarray}
\frac{d^2 \vec{x}}{dt^2}&=&\frac{1}{2}\vec \nabla h_{00}(\vec x)=\vec \nabla \left(\frac{h_{00}(\vec x)}{2}\right).
\label{eq:NfromG12}
\end{eqnarray}
When we compare this to Newton's second Law
\begin{eqnarray}
\frac{d^2 \vec{x}}{dt^2}&=&-\vec \nabla \Phi,
\label{eq:NfromG13}
\end{eqnarray}
we see that
\begin{eqnarray}
h_{00}&=&-2\Phi.
\label{eq:NfromG14}
\end{eqnarray}
Hence $g_{00}=\eta_{00}+h_{00}=-1-2\Phi$. In a spherically symmetric situation $\Phi=-\frac{GM}{r}$, so
\begin{eqnarray}
g_{00}&=&-\left(1-\frac{2GM}{r}\right), 
\label{eq:NfromG15}
\end{eqnarray}
where $G$ is the Newtonian gravitational constant. Thus, if mass $M$ is small or we are far away from mass $M$ then $g_{00}\approx -1$ 
like in Minkowski spacetime, and hence time and proper time are indistinguishable in the Newtonian limit.
Equation\ (\ref{eq:NfromG15}) quantifies how mass curves
spacetime in the Newtonian approximation. Hence the geodesic equation is the general relativity equivalent of Newton's Laws.
This describes how a particle moves in a curved spacetime, like Newton's second Law describes how a particle moves under the action of a force.

\section{Einstein's Field Equation}

Einstein's field equation, the general relativity generalization 
of Poisson's equation for gravity, is a set of 10 equations that governs gravity. In Albert Einstein's 
general theory of relativity, which describes the fundamental interaction of gravitation as a result of 
spacetime being curved by matter and energy.\cite{peebles1993principles} First published by Einstein in 1915 as a tensor equation, 
the Einstein field equation equates local spacetime curvature (expressed by the Einstein tensor $G_{\mu \nu}$) to the local energy and 
momentum within that spacetime (expressed by the stress-energy tensor $T_{\mu \nu}$).

Similar to the way that electromagnetic fields are determined from the source charges and currents through Maxwell's 
equations, Einstein's field equations are used to determine the spacetime geometry resulting from the presence of mass-energy 
and linear momentum (sources), that is, they determine the metric tensor of spacetime for a given arrangement of 
stress-energy in the spacetime. The relation between the metric tensor and the Einstein tensor 
allows the Einstein field equation to be written as a set of non-linear partial differential equations. 
The solutions of the Einstein field equation are the components of the metric tensor. The inertial trajectories (geodesics)
of particles and radiation in the resulting geometries are then calculated using the geodesic 
equation.\cite{Hartle}

\subsection{Riemann Tensor, Ricci Tensor, Ricci Scalar, Einstein Tensor}

\subsubsection{Riemann (curvature) tensor} 
The Riemann curvature tensor, or Riemann-Christoffel tensor, is the most common tensor used to describe the curvature of Riemannian manifolds. 
It associates a tensor to each point of a Riemannian manifold (i.e., it is a tensor field) that measures the extent to which 
the metric tensor is not locally isometric to a Euclidean flat space and so specifies the geometrical properties
of spacetime.
More precisely, the Riemann tensor governs the evolution of a vector on a displacement parallel propagated along a geodesic.\cite{Loveridge:2004pp}
It is defined in terms of Christoffel symbols as
\begin{eqnarray}
R^{\alpha}_{\beta \gamma \delta} \equiv \Gamma^{\alpha}_{\beta \delta, \gamma}-\Gamma^{\alpha}_{\beta \gamma, \delta}+
\Gamma^{\nu}_{\beta \delta} \Gamma^{\alpha}_{\nu \gamma}-\Gamma^{\nu}_{\beta \gamma} \Gamma^{\alpha}_{\nu \delta},
\label{eq:Riemanntensor}
\end{eqnarray}
where $\Gamma^{\alpha}_{\beta \delta, \gamma}\equiv \frac{\partial \Gamma^{\alpha}_{\beta \delta}}{\partial x^{\gamma}}$.
The spacetime is considered flat if the Riemann tensor vanishes everywhere.
The Riemann tensor can also be written directly in terms of the spacetime metric
\begin{eqnarray}
R_{\alpha \beta \gamma \delta} \equiv \frac{1}{2}\left(g_{\beta \gamma, \alpha \delta}+g_{\alpha \delta, \beta \gamma}
-g_{\beta \delta, \alpha \gamma}-g_{\alpha \gamma, \beta \delta}\right)+g_{\mu \nu}\Gamma^{\nu}_{\alpha \gamma}
\Gamma^{\mu}_{\beta \delta}-g_{\mu \nu}\Gamma^{\nu}_{\alpha \delta}\Gamma^{\mu}_{\beta \gamma}.
\end{eqnarray}
The Riemann tensor has the following symmetries.
\subsubsection{Skew symmetry} 
\begin{eqnarray}
R_{\alpha \beta \gamma \delta}=-R_{\beta \alpha \gamma \delta}=-R_{\alpha \beta \delta \gamma}.
\end{eqnarray}
\textbf{Interchange symmetry:} 
\begin{eqnarray}
R_{\alpha \beta \gamma \delta}=R_{\gamma \delta \alpha \beta},
\end{eqnarray}
\subsubsection{First Bianchi identity} 
\begin{eqnarray}
R_{\alpha \beta \gamma \delta}+R_{\alpha \gamma \delta \beta}+R_{\alpha \delta \beta \gamma}=0,
\end{eqnarray}
which is often written as: 
\begin{eqnarray}
R_{\alpha [\beta \gamma \delta]}=0.
\end{eqnarray}
Because of the symmetries above, the Riemann tensor in 4-dimensional spacetime has only 20
independent components out of $4^4=256$. The general rule for computing the number of independent components
in an $N$-dimensional spacetime is $N^2\left(N^2-1\right)/12$.\cite{Sean}\\
\subsubsection{Second Bianchi identity}
\begin{eqnarray}
R_{\alpha \beta \gamma \delta;\nu}+R_{\alpha \beta \nu \gamma;\delta}+R_{\alpha \beta \delta \nu;\gamma}=0,
\end{eqnarray}
which can be written as
\begin{eqnarray}
R_{\alpha \beta [\gamma \delta;\nu]}=0,
\end{eqnarray}
Raising two indices we have,
\begin{eqnarray}
R^{\alpha \beta}_{[\gamma \delta;\nu]}=0.
\end{eqnarray}
Setting $\gamma=\alpha$ and $\delta=\beta$, we get
\begin{eqnarray}
\hspace{-2 cm}  R^{\alpha \beta}_{[\alpha \beta;\nu]}&=&0, \nonumber \\
\Rightarrow\ \  R^{\alpha \beta}_{\alpha \beta;\nu}+R^{\alpha \beta}_{\beta \nu;\alpha}+R^{\alpha \beta}_{\nu \alpha; \beta}&=&0, \nonumber \\
\Rightarrow\ \  R^{\alpha \beta}_{\alpha \beta;\nu}-R^{\alpha \beta}_{\nu \beta;\alpha}+R^{\alpha \beta}_{\alpha \nu; \beta}&=&0, \nonumber \\
\Rightarrow\ \  \mathcal{R}_{;\nu}-R^{\alpha}_{\nu;\alpha}+R^{\beta}_{\nu; \beta}&=&0, \nonumber \\
\Rightarrow\ \  2R^{\alpha}_{\nu;\alpha}&=& \mathcal{R}_{;\nu},
\end{eqnarray}
which after raising the index is,
\begin{eqnarray}
R^{\alpha \beta}_{; \alpha}=\frac{1}{2}g^{\alpha \beta}\mathcal{R}_{;\alpha}.
\label{eq:bianchi}
\end{eqnarray}
Here $\mathcal{R}$ is Ricci scalar, discuss below. We will use Eq.\ (\ref{eq:bianchi}) in Sec.\ (\ref{sec:derivation of friedmann}).
\subsubsection{Ricci tensor}
The Ricci tensor, or the Ricci curvature tensor, governs the evolution of a small volume parallel propagated along a geodesic.\cite{Loveridge:2004pp} 
It is obtained from the Riemann tensor by contracting over two of the indices,
\begin{eqnarray}
R_{\alpha \beta}\equiv R^{\gamma}_{\alpha \gamma \beta}.
\end{eqnarray}
It is symmetric, which means that it has at most 10 independent components out of $4\times 4=16$. 
For the case of vacuum we will see later, the field equation is $R_{\mu \nu}=0$.
\subsubsection{Ricci scalar}
The Ricci scalar $\mathcal{R}$ is obtained by contracting the Ricci tensor over the remaining two indices and is denoted by :
\begin{eqnarray}
\mathcal{R}\equiv g^{\alpha \beta}R_{\alpha \beta}=R^{\alpha}_{\alpha}.
\end{eqnarray}
\subsubsection{Einstein tensor}
The Einstein tensor is defined in terms of the Ricci tensor and Ricci scalar as
\begin{eqnarray}
G_{\alpha \beta}\equiv R_{\alpha \beta}-\frac{1}{2}g_{\alpha \beta}\mathcal{R}.
\end{eqnarray}
One can use Eq.\ (\ref{eq:bianchi}) to derive a very important property of the Einstein tensor, $G_{\alpha \beta; \alpha}=0$.

\subsection{Energy-Momentum Tensor}

In the Newtonian approximation, the gravitational field is directly proportional to mass. In general relativity, 
mass is just one of several sources of spacetime curvature. The energy-momentum (stress-energy) tensor, denoted by $T^{\mu \nu}$, 
includes all possible forms of sources (energy) that
can curve spacetime, and it describes the density and flow of the 4-momentum $(-E, p_\mathrm x,p_\mathrm y,p_\mathrm z,)$.\cite{kipthorne}
In simple terms, the stress-energy tensor quantifies all the stuff that causes spacetime to curve, and thus to the gravitational field. 
More rigorously, the components $T^{\mu \nu}$ of the stress-energy tensor is the flux of the $\mu$ component of the four momentum crossing
the surface of constant $x^{\nu}$. A surface of constant $x^{\nu}$ is simply a 3-plane perpendicular to the $x^{\nu}$-axis. Hence, 
the stress-energy tensor is the flux of a 4-momentum across a surface of a constant coordinate. In other words, the stress-energy tensor 
describes the density of energy and momentum and the flux of energy and momentum in a region. Since, under the mass-energy equivalence 
principle, we can convert mass units to energy units and vice-versa, then the stress-energy tensor can describe all the mass 
and energy in a given region of spacetime. In simple layman's language, the stress-energy tensor represents everything that gravitates.

The stress-energy tensor, being a tensor of rank two in four-dimensional spacetime, has sixteen components that can 
be written as a 4$\times$4 matrix, and has the following structure in an orthonormal basis


\begin{equation} 
T^{\mu \nu}=
            \left( {\begin{array}{cccc}
\textcolor{blue}{T^{00}} & \textcolor{red}{T^{01}} & \textcolor{red}{T^{02}} & \textcolor{red}{T^{03}} \\

\textcolor{magenta}{T^{10}} & \textcolor{green}{T^{11}} & \textcolor{cyan}{T^{12}} & \textcolor{cyan}{T^{13}} \\

\textcolor{magenta}{T^{20}} & \textcolor{cyan}{T^{21}} & \textcolor{green}{T^{22}} & \textcolor{cyan}{T^{23}} \\

\textcolor{magenta}{T^{30}} & \textcolor{cyan}{T^{31}} & \textcolor{cyan}{T^{32}} & \textcolor{green}{T^{33}} \\
                \end{array} } \right).
\end{equation} 

\begin{itemize} 

      \item[1.] Here \textcolor{blue}{$T^{00}=T^{tt}$}, represents the energy flow $p^0$, crossing a hypersurface of constant time $(x^0=t)$.
		A hypersurface of constant time is the volume. Hence, $T^{00}$ is the energy density.
      \item[2.] \textcolor{red}{$T^{0i}$} represents the flow (flux) of energy in the $x^i$ direction.
      \item[3.] \textcolor{magenta}{$T^{i0}$} represents the $i-$component of the momentum density.
      \item[4.] \textcolor{blue}{$T^{ij}$} ($i \neq j$) represents the flow of the $i-$component of momentum in the $j-$direction 
		(shear stress, i.e., stress applied tangential to the region).
      \item[5.] \textcolor{green}{$T^{ii}$} represents the components of normal stress, or stress applied perpendicular to the region (normal 
		stress is another term for pressure).
 \end{itemize} 
Note that the components \textcolor{blue}{$T^{00}$}, \textcolor{magenta}{$T^{10}$}, \textcolor{magenta}{$T^{20}$} and
\textcolor{magenta}{$T^{30}$} are interpreted as densities. A density is what you get when you measure the 
flux of 4-momentum across a 3-surface of constant time, which means the instantaneous value of 4-momentum flux is density. 

To develop more insight of the energy-momentum tensor, let's consider the example of a cylinder with a piston that is 
fixed and cannot move, hence the volume of the cylinder is fixed. Initially we consider that the pressure of the air 
inside and outside is the same, which means that no net force acts on the walls of the cylinder. If we now heat the cylinder, 
the temperature of the air inside will increase according to the ideal gas law. The mass of the cylinder will remain the same, 
but the energy content inside the box has increased due to the extra kinetic energy given to the molecules of the air 
inside the cylinder during the process of heating. This will increase the time-time component of the stress-energy tensor \textcolor{blue}{$T^{00}$} 
and consequently increase the spacetime curvature around the cylinder. It is against our intuition that just by 
increasing the pressure inside the cylinder it will make it heavier and cause spacetime to curve more around the cylinder. 
This is because, in daily life, the contribution of increased pressure and kinetic energy on the gravitational effect of 
the cylinder is negligible, as compared to the mass contribution, and our intuition is developed on the basis of daily life 
experience. Hence, it is our intuition that needs to be blamed, and we should consider it wrong. On larger scales, such as 
the sun, pressure and temperature contribute significantly to the gravitational field.   

We can see that the stress-energy tensor neatly quantifies all static and dynamic properties of a region of spacetime, 
from mass to momentum to temperature to pressure to shear stress. This is the only mathematical quantity that we want
to know about a particular region of space to find its gravitational effects. 

A very good proof that stress-energy tensor is symmetric $\left(T^{\mu \nu}=T^{\nu \mu}\right)$, is given in 
Gravitation\cite{kipthorne} on page 141. Hence, it has only 10 independent components. The conservation equation 
(which incorporates both energy and momentum conservation in a general metric)\footnote{Note here that the Einstein summation convention is adopted, in which repeated upper and lower indices are implicitly summed over.} 
\begin{eqnarray}
T^{\mu \nu}_{;\nu}=0.
\label{eq:energyconservation} 
\end{eqnarray}
In the limit of flat spacetime (Minkowski metric), the covariant derivative reduced to an ordinary derivative and we will get:
\begin{eqnarray}
\frac{\partial T^{\mu \nu}}{\partial x^\nu}=0. 
\end{eqnarray}
For $\mu=0=t$, one finds the continuity equation for energy conservation as:
\begin{eqnarray}
\frac{\partial T^{tt}}{\partial t}+\frac{\partial T^{it}}{\partial x^i}= \frac{\partial \varepsilon}{\partial t}+\vec {\nabla} \cdot \vec {J}=0, 
\end{eqnarray}
where $\varepsilon=T^{tt}$ and $J^i=T^{it}$ are the energy and momentum densities respectively. 

We now consider three types of energy-momentum tensors frequently
used in GR: classical vacuum, dust and perfect fluid.

\textbf{Vacuum:} This is the simplest possible stress-energy tensor in which all the values are zero:
\begin{eqnarray}
T^{\mu \nu}=0.
\end{eqnarray}
This tensor represents a region of space in which there is no matter, energy, or fields. This is not just at a 
given instant, but over the entire period of time in which we're interested in. Nothing exists 
in this region, and nothing happens in this region.

So one might assume that in a region where the stress-energy tensor is zero, the gravitational field
must also necessarily be zero. There's nothing there to gravitate, so it follows naturally that there 
can be no gravitation. In fact, it's not that simple. For details see van Nieuwenhove.\cite{van2007vacuum}

\textbf{Dust:} Imagine a time-dependent distribution of identical, massive, non-interacting, electrically neutral particles. 
In general relativity, such a distribution is called a dust. Let's break down what this means. 

\begin{itemize} 

      \item[$\bullet$] \textbf{Time-dependent:} The distribution of particles in dust is not a constant; that is to say, the particles may be in motion. The overall configuration you see when you look at the dust depends on the time at which you look at it, so the dust is said to be time-dependent.

      \item[$\bullet$] \textbf{Identical:} The particles that make up dust are all exactly the same; they don't differ from each other in any way.
      \item[$\bullet$] \textbf{Massive:} Each particle in dust has some rest mass. Because the particles are all identical, their rest masses must also be identical. We'll call the rest mass of an individual particle $m_0$. 

      \item[$\bullet$] \textbf{Non-interacting:} The particles don't interact with each other in any way; they don't collide, and they don't attract or repel each other. This is, of course, an idealization; since the particles have mass $m_0$, they must at least interact with each other gravitationally, if not in other ways. However, we're constructing our model in such a way that gravitational effects between the individual particles are so small they can be neglected. Either the individual particles are very tiny, the average distance between them is very large, or may be both. 

      \item[$\bullet$] \textbf{Electrically neutral:} In addition to the obvious electrostatic effect of two charged particles either attracting or repelling each other, violating our ``non-interacting" assumption, allowing the particles to be both charged and in motion would introduce electrodynamic effects that would have to be included in the stress-energy tensor. We prefer to ignore these effects for the sake of simplicity, so by definition, the particles in dust are all electrically neutral. 
 \end{itemize} 
To fully describe the dust we need to write its energy-momentum tensor, which is given by
\begin{eqnarray}
T^{\mu \nu}=\rho u^{\mu}u^{\nu}.
\end{eqnarray}
For a comoving observer, the 4-velocity is given by $\vec u = (1,0,0,0)$, so the stress-energy tensor
reduces to
\begin{equation} 
T^{\mu \nu}=
            \left( {\begin{array}{cccc}
\rho & 0 & 0 & 0 \\

0 & 0 & 0 & 0 \\

0 & 0 & 0 & 0 \\

0 & 0 & 0 & 0 \\
                \end{array} } \right).
\end{equation} 
Dust is an approximation to the content of the Universe at later times, when radiation is negligible.

\textbf{Perfect fluid:} It is a fluid that has no heat conduction or viscosity. It is fully parametrized by its
mass density $\rho$ and the pressure $P$ . The stress-energy tensor is given by
\begin{eqnarray}
T^{\mu \nu}=\left(\rho+P\right) u^{\mu}u^{\nu}+Pg^{\mu \nu}.
\end{eqnarray}
For a comoving observer, the 4-velocity is $\vec u = (1,0,0,0)$, so the stress-energy tensor
reduces to
\begin{equation} 
T^{\mu \nu}=
            \left( {\begin{array}{cccc}
\rho & 0 & 0 & 0 \\

0 & P & 0 & 0 \\

0 & 0 & P & 0 \\

0 & 0 & 0 & P \\
                \end{array} } \right).
\label{eq:dusttmunu}
\end{equation} 
In the limit as $P\rightarrow 0$, the perfect fluid approximation reduces to that of dust.
A perfect fluid can be used an approximation to the components of the Universe at earlier times, when radiation dominates.

\subsection{Conservation Laws and Energy Evolution}
\label{sec:energyconservation}

Stress-Energy conservation, Eq.\ (\ref{eq:energyconservation}), can be used to determine 
how components of the energy-momentum tensor evolve with time. Considering the special case of perfect 
fluid, and using Eq.\ (\ref{eq:dusttmunu}) and Eq. (\ref{eq:Minkowskimetric}), the
mixed energy-momentum tensor is
\begin{eqnarray}
T^{\mu}_{\nu}=\left( {\begin{array}{cccc}
-\rho & 0 & 0 & 0 \\

0 & P & 0 & 0 \\

0 & 0 & P & 0 \\

0 & 0 & 0 & P \\
                \end{array} } \right),
\label{eq:mixedtmunu}
\end{eqnarray}
and the four conservation equations using Eq.\ (\ref{eq:covderofmixten}) are
\begin{eqnarray}
T^{\mu}_{\nu;\mu}\equiv \frac{\partial T^{\mu}_{\nu}}{\partial x^{\mu}}+\Gamma^{\mu}_{\alpha \mu}T^{\alpha}_{\nu}-\Gamma^{\alpha}_{\nu \mu}T^{\mu}_{\alpha}=0.
\end{eqnarray}

Consider the $\nu=0$ equations
\begin{eqnarray}
\frac{\partial T^{\mu}_{0}}{\partial x^{\mu}}+\Gamma^{\mu}_{\alpha \mu}T^{\alpha}_{0}-\Gamma^{\alpha}_{0 \mu}T^{\mu}_{\alpha}=0.
\label{eq:conservation}
\end{eqnarray} 
As a consequence of the isotropy, assumed as the fundamental principle of cosmology, all non-diagonal terms of $T^{\mu \nu}$ vanish
(i.e., $T^{\mu \nu}=0$ if $\mu\neq \nu$), hence $T^{i}_{0}=0$. This means that $\mu=0$ in the first term
of Eq.\ (\ref{eq:conservation}) and $\alpha=0$ in the second term, and Eq.\ (\ref{eq:conservation}) becomes
\begin{eqnarray}
\frac{\partial T^{0}_{0}}{\partial x^{0}}+\Gamma^{\mu}_{0 \mu}T^{0}_{0}-\Gamma^{\alpha}_{0 \mu}T^{\mu}_{\alpha}=0.
\label{eq:conservation2}
\end{eqnarray} 
From Eq.\ (\ref{eq:mixedtmunu}), $T^{0}_{0}=-\rho$, so we have
\begin{eqnarray}
\frac{\partial \rho}{\partial t}+\Gamma^{\mu}_{0 \mu}\left(\rho+T^{\mu}_{\alpha}\right)&=&0.
\label{eq:conservation3}
\end{eqnarray}
Now, we need to calculate the Christoffel symbol $\Gamma^{\mu}_{0 \mu}$. For simplicity we consider expanding flat space-time with the flat Friedman-Lema\^{i}tre-Robertson-Walker metric tensor given by
Eq.\ (\ref{eq:FLRW}):
%
%
%
The Christoffel symbols are given by Eq.\ (\ref{eq:christoffel}). To start, let's put $\beta=0$ in Eq.\ (\ref{eq:covderoften}) to get $\Gamma^{\alpha}_{0 \mu}$:
\begin{equation}
\Gamma^{\alpha}_{0 \mu} = \frac{1}{2}g^{\alpha \gamma}\Big(g_{\gamma 0, \mu}+g_{\mu \gamma, 0}-g_{0 \mu ,\gamma}\Big).
\label{eq:christoffel3}
\end{equation}
But $g_{\gamma 0}$ and $g_{0 \mu}$ are constants and hence $g_{\gamma 0,\mu}=g_{0 \mu,\gamma}=0$
\begin{eqnarray}
\Gamma^{\alpha}_{0 \mu}=\frac{1}{2}g^{\alpha \gamma}g_{\mu \gamma,0}.
\end{eqnarray}
Let's figure out its components:
\begin{itemize} 

      \item[1.] when $\alpha=\mu=0$:
		\begin{equation}
		      \Gamma^{0}_{00}=\frac{1}{2}g^{0 \gamma}g_{0\gamma,0}=0. \hspace{7 mm} \left(\because \ g_{0\gamma,0}=0\right)
		\end{equation}
      \item[2.] when $\alpha=\mu=i$, where $i\in \left\{1, 2, 3\right\}$:
		\begin{eqnarray}
		       \Gamma^{i}_{0i}&=&\frac{1}{2}g^{i \gamma}g_{i \gamma,0}, \nonumber \\
		       &=&\frac{1}{2}\left(g^{i1}g_{i1,0}+g^{i2}g_{i2,0}+g^{i3}g_{i3,0}\right), \nonumber \\
		       &=&\frac{1}{2}\left(g^{i1}+g^{i2}+g^{i3}\right)\left(g_{i1,0}\right), \hspace{7 mm} \left(\because \ g_{\mu \nu} \ \mathrm{is\ diagonal}\right)\nonumber \\
		       &=&\frac{1}{2}\left(\frac{1}{a^2}+\frac{1}{a^2}+\frac{1}{a^2}\right) (2a\dot a), \nonumber \\
		       &=&3\frac{\dot a}{a}.
		\end{eqnarray}
      \item[3.] when $\alpha=i$, \textit{$\mu=j$}\ and \ $i \neq j$:
		\begin{equation}
		      \Gamma^{j}_{0i}=\frac{1}{2}g^{j \gamma}g_{i\gamma,0}=0.
		\end{equation}
 \end{itemize}
Hence, we can conclude:
\begin{eqnarray}
\Gamma^{\mu}_{0 \mu}=\Gamma^{i}_{0i}=3\frac{\dot a}{a}.
\end{eqnarray}

So the conservation laws in an expanding Universe, Eq.\ (\ref{eq:conservation3}), takes the form
\begin{eqnarray}
\centering
&           & \hspace{5 mm}-\frac{\partial \rho}{\partial t}-3\frac{\dot a}{a}\rho-\frac{\dot a }{a} T^{i}_{i}=0, \nonumber \\
&& \hspace{5 mm} \frac{\partial \rho}{\partial t}+3\frac{\dot a}{a}\rho+\frac{\dot a }{a} \left(3P\right)=0, \nonumber \\
&& \hspace{5 mm} \frac{\partial \rho}{\partial t}=-3\frac{\dot a}{a}\left(\rho+P\right).
\label{eq:energycon}
\end{eqnarray}
Here $a(t)$ is the cosmological scale factor and an over-dot denotes a derivative with respect to cosmological time.

For a perfect fluid the equation-of-state, which is the relationship between pressure $P$ and the density $\rho$ of the fluid is: 
\begin{eqnarray}
P=\rho\ \omega.
\end{eqnarray}
Using the equation of state, Eq.\ (\ref{eq:energycon}) takes the form:
\begin{eqnarray}
\frac{\partial \rho}{\partial t}&=&-3\ \rho \left(\frac{\dot a}{a}\right)\left(1+ \omega\right).
\label{eq:diffcon}
\end{eqnarray}
This is separable first order ordinary differential equation with the solution
\begin{equation}
\rho (t)=\rho_0\left(\frac{a_0}{a}\right)^{3(1+\omega)}.
\label{eq:density}
\end{equation} 
Here $\rho_0$ is the density of matter at scale factor $a_0$, which we take to be the value today. We can assume
$a_0=1$ without loss of generality. 

Equation\ (\ref{eq:density}) describes the evolution of a particular kind of species whose time-dependent energy
density is $\rho$, with equation-of-state parameter $\omega$. Now let's consider different kinds of energy densities and their 
evolution according to Eq.\ (\ref{eq:density}):
\begin{itemize} 

       \item[$\bullet$] For cold matter (dust), we have zero pressure, $P_m=0$, so $\omega_m=0$ and
		      \begin{equation}
		      \rho_m (t) \propto \frac{1}{a^3}. 
		      \end{equation}

		      This should come as no surprise, because the total amount of matter is conserved, and the volume of the 
		      Universe goes as $V\propto a^3$, hence, $\rho \propto \frac{1}{V} \propto a^{-3}$. 

	\item[$\bullet$] For radiation $P_r=\frac{1}{3}\rho_r$, so $\omega_r=\frac{P_r}{\rho_r}=\frac{1}{3}$. This means that:
		\begin{equation}
		\rho_r (t) \propto \frac{1}{a^4}. 
		\end{equation}
		This too should not surprise us --- since radiation density is directly
		proportional to the energy per particle and inversely proportional to the total volume, \text{i.e.}, 
		$\rho_r \propto \frac{\hbar \nu}{V} \propto \frac{\hbar}{\lambda V} \propto a^{-4}$, because 
		$\lambda \propto a$. The last part states that the energy per particle decreases as
		the Universe expands.

	\item[$\bullet$] For the case of curvature where $\omega_k=-\frac{1}{3}$, using Eq.\ (\ref{eq:density}) we find
		    \begin{equation}
		    \rho_k (t) \propto \frac{1}{a^2}, 
		    \end{equation}
	  where $\rho_{k}$ is the time-dependent  energy density corresponding to the space curvature of the Universe.

\end{itemize} 
Now, we are ready to postulate Einstein's field equation. 

\subsection{Einstein's Field Equation}

Just as  Maxwell's equations govern the electric and magnetic field response to electric charges and current (sources),
Einstein's field equations describe how the metric is governed by energy and momentum (sources). 

The general relativity must describe both parts of the dynamical picture. \cite{Sean}
\begin{itemize}
	    \item[1.] How gravity affects the motion of particles (i.e., how force is applied on the particle, when it is placed  in the gravitational field).
	    \item[2.] How the gravitational field is generated by a source of mass energy.
\end{itemize}

The first part is described by the geodesic equation, derived in Sec.\ (\ref{Subsec:Geodesic Equation}):
\begin{equation}
\frac{d^2 x^\nu}{d \lambda^2} +\Gamma^{\nu}_{\gamma\delta}\frac{dx^\gamma}{d\lambda}\frac{dx^\delta}{d\lambda}=0,
\label{eq:geodesicEFQ}
\end{equation}
which is analogous to Newton's second law of motion $\vec F=m\vec a$.

The second part requires finding the analog of the Poisson equation
\begin{eqnarray}
\nabla^2 \Phi(\vec x)=4 \pi G \rho(\vec x),
\label{eq:poissoneq} 
\end{eqnarray}
which specifies how matter (or energy in general relativity) curves spacetime. Here $\nabla^2=\delta^{ij}\partial_i \partial_j$
is the Laplacian in space and $\rho$ is the mass density [the explicit form of $\phi=-GM/r$ is the solution of Eq. (\ref{eq:poissoneq}) for
the case of a spherically symmetric mass distribution].

In classical Newtonian gravity, gravitational effects are produced by the mass at rest.
In modified Newtonian gravity, which we can call special relativity, we learned that
rest mass is also a form of energy; thus special relativity put mass and energy on equal footing.
Extending this idea, one should expect that in general relativity, all sources of both
energy and momentum contribute in generating spacetime curvature. On the left hand side of Eq.\ 
(\ref{eq:poissoneq}) we have a second order differential operator acting on the gravitational potential
and on the right hand side we have the measure of mass density. The relativistic generalization of
the Poisson equation should be the relationship between tensors. The tensor generalization of
mass density can be $T^{\mu \nu}$. This means that we consider the
stress-energy tensor $T^{\mu \nu}$ as the source of spacetime curvature (with an unknown scaling
factor), in the same sense that the mass density $\rho$ is the source for the potential $\Phi$ in 
Newtonian gravity. Hence, the right hand side of the general relativity analog 
of the Poisson equation should be $\kappa T^{\mu \nu}$ (where $\kappa$ is an unknown constant to be 
determined latter.) 

As far as the left hand side of general relativity analog of the Poisson equation, we have seen earlier in
Eq.\ (\ref{eq:NfromG15}), the spacetime metric in the Newtonian limit is modified by a term that is 
proportional to $\Phi$. Extending this idea, the general relativity counterpart of $\nabla^2 \Phi(\vec x)$
would contain terms having the second derivative of the metric tensor. Something along the lines of:
\begin{eqnarray}
\Big[\nabla^2\ g \Big]_{\mu \nu} = \kappa T^{\mu \nu}.
\label{eq:shauneq}
\end{eqnarray}
But of course we want it to be completely tensorial and the left-hand side of Eq.\ (\ref{eq:shauneq}) is not a 
tensor. It is just simplistic notation that indicates we need something on the left hand side that should have 
the second derivative of the metric.

The Riemann tensor $R_{\alpha \beta \gamma \delta}$,
and consequently its contractions, the Ricci tensor $R_{\alpha \beta}=R^{\gamma}_{\alpha \gamma \beta}$, and the
Ricci scalar $\mathcal{R}=R^{\alpha}_{\alpha}$, contain the second derivative of the metric and therefore is a candidate for the
left hand side of Einstein's field equations. 

Following this line of thought, Einstein originally suggested that the field equations might be
\begin{equation}
R_{\mu \nu}=\kappa T_{\mu \nu},
\end{equation}  
but one can see directly that this can not be correct. While the conservation of energy and 
momentum require $T^{\mu \nu}_{;\mu}=0$, the same in general is not true for the Ricci tensor $R^{\mu \nu}_{;\mu} \neq 0$.
However Einstein's tensor, $G_{\mu \nu}=R_{\mu \nu}-\frac{1}{2}g_{\mu \nu}\mathcal{R}$, which is a combination 
of the Ricci tensor and scalar, satisfies the divergence-less condition $\nabla^\mu G_{\mu \nu}=0$, see 
Eq.\ (\ref{eq:bianchi}). Therefore, Einstein's field equation becomes
\begin{equation}
G_{\mu \nu} \equiv R_{\mu \nu}-\frac{1}{2}g_{\mu \nu} \mathcal {R}= \kappa T_{\mu \nu}.
\label{eq:EFE1}
\end{equation}
This equation satisfies all of the obvious requirements: the right-hand side of is a covariant expression of the
energy and momentum density in the form of a symmetric and conserved tensor, while the left-hand side is also a
symmetric and conserved tensor constructed from the first and second derivatives of the metric tensor and the metric
tensor itself. The only issue that remains is to fix the constant $\kappa$. 
By matching Einstein's equation in the Newtonian limit to the Poisson equation, the constant $\kappa$ was found to 
$8 \pi G$,\footnote{For a detailed derivation see Carroll\cite{Sean} page numbers 155-159.} where $G$ 
is the Newtonian gravitational constant. Then Eq.\ (\ref{eq:EFE1}) takes the form
\begin{equation}
G_{\mu \nu} \equiv R_{\mu \nu}-\frac{1}{2}g_{\mu \nu} \mathcal {R}= 8 \pi G T_{\mu \nu}.
\label{eq:EFE2}
\end{equation}

Summarizing, Einstein's equations for the gravitational field came from requiring that the equations
of motion were generally covariant under coordinate transformations and reduced to the
Newtonian form in weak stationary gravitational fields. The field equation relates the Ricci tensor, that is
made up of second derivatives of the metric tensor, to the Ricci scalar formed by contracting
the Ricci tensor, and to the energy-momentum content of the Universe. Remember, this is not the proof of the field equations.

\section{Hubble's Law and Redshift of Distant Galaxies}
\label{sec:Hubble law}

In 1912 Slipher discovred most nearby galaxies were moving away and measured their velocities $v$, but he did not know they were galaxies. Hubble\cite{Hubble15031929} showed they were galaxies in 1925-26 and measured distance $r$ and showed in 1929 
\begin{eqnarray}
 v \propto r,
\label{eq:HL1}
\end{eqnarray}
This phenomenon was observed as a redshift of a galaxy's spectrum. This redshift appeared to 
have a larger value for fainter, presumably farther away, galaxies. Hence, the farther a galaxy, 
the faster it is receding from Earth. Consider two galaxies separated by the distance $r$, then
the relative velocity $v \propto r$, this is called Hubble's law and mathematically given
as
\begin{eqnarray}
\vec v=H_0 \vec r,
\label{eq:HL1}
\end{eqnarray} 
where, $H_0$ is Hubble constant that relates the distance of the galaxy (for the earth based observers 
this is conveniently taken to be the distance from our galaxy) to its recession velocity.
The Hubble constant $H_0$ is one of the most important numbers in cosmology, because it may be used to 
estimate the size of the observable Universe and its age. It indicates the rate at which the Universe is expanding. 
Since more generally a related law holds at all times, $\vec v(t)=H(t) \vec r(t)$ where Eq.\ (\ref{eq:HL1}) $H(t)$ is the Hubble
parameter whose current value is the Hubble constant $H(t_0)=H_0$.

The units of the Hubble constant (or parameter) are kilometers per second per megaparsec. In other words, for 
each megaparsec of distance, the velocity of a distant object appears to increase by some value. 
For example, if the Hubble constant was determined to be 65 kms$^{-1}$ Mpc$^{-1}$, a galaxy at 10 Mpc would 
have a redshift corresponding to a radial recession velocity of 650 kms$^{-1}$. 

The reasonable current summary value of Hubble constant is $H\pm \sigma_{H_0}=68\pm 2.8$ kms$^{-1}$ Mpc$^{-1}$. \cite{Chen2011a, chen03}

The discovery of Hubble's Law marked the commencement of the era of quantitative cosmology
in which theories of the Universe could be subject to observational test.
Since the days of Hubble, advances in technology have enabled
astronomers to measure the light from increasingly deeper space and more ancient time, and our ideas of the entire
history of the expanding universe have been gradually converging into a unified picture called the Big Bang---model.

At first glance, it looks like Hubble's law is a violation of cosmological principle, because
all galaxies are moving away from us, which might seen to put us in a special location in the Universe. In fact, 
what we see here in our Galaxy exactly what you would be expected in a Universe which is 
undergoing homogeneous and isotropic expansion. We see distant galaxies moving away from us; but 
observers in any other galaxy would also see distant galaxies moving away from them.

Let's define some terminologies:

Physical or proper distance: The actual distance between the two points in space is called the physical distance.
It is denoted by $\vec r(t)$.

Comoving or coordinate distance: This is just the label of the point in space and it is independent of time. It is
denoted by $\vec x$

The physical and comoving distance: are related through
\begin{eqnarray}
\vec r(t)=a(t)\ \vec x,
\label{eq:HL2}
\end{eqnarray}
where $a(t)$ is called the scale factor, a time dependent scalar that describes the the expansion (or contraction)
of the Universe. Scale factor $a(t)$ is a dimensionless function of time that carries important information about the cosmological expansion of Universe. The current 
value of the scale factor is denoted by $a(t_0)=a_0$ and its value is often set to 1. The evolution of the scale factor is governed by general relativity. Differentiating Eq.\ (\ref{eq:HL2}) with respect to the time we get
\begin{eqnarray}
\vec v(t)=\dot a(t)\ \vec x,\\
\Rightarrow \frac{\dot a(t)}{a(t)} a(t)\ \vec x=\frac{\dot a(t)}{a(t)} \vec r(t),
\label{eq:HL2.1}
\end{eqnarray}
and comparing to Eq.\ (\ref{eq:HL1}) we can write the Hubble parameter $H$ in terms of the scale factor
$a(t)$, rewriting  Eq.\ (\ref{eq:HL2}) as:
\begin{eqnarray}
H=\frac{\dot a(t)}{a(t)},
\label{eq:HL3}
\end{eqnarray}
where the over-dot represents a time derivative.
Current evidence suggests that the expansion rate of the Universe is accelerating, which means that the 
second derivative of the scale factor $\ddot{a}(t)$ is positive, or equivalently that the first derivative $\dot{a}(t)$ is increasing over time.

When the galaxies move relative to us, we observe the change in the wavelength of the light emitted by those galaxies. To describe this it is convenient to define a
redshift denoted by $z$. The redshift is a 
dimensionless quantity defined as the change in the wavelength of the light divided by the rest wavelength of the light:
\begin{eqnarray}
z=\frac{\lambda_{o}-\lambda_{e}}{\lambda_{e}},
\label{eq:HL4}
\end{eqnarray}
where $\lambda_{e}$ is wavelength of the emitted wave, and $\lambda_{o}$ is the wavelength of observed wave.

Redshift $z$ is an observable which can be related with the mathematical construct $a(t)$ through:\footnote{For the proof of Eq.\ (\ref{eq:HL5}) see Ryden.\cite{ryden2003introduction}}
\begin{eqnarray}
1+z=\frac{a(t_0)}{a(t)}=\frac{1}{a(t)},
\label{eq:HL5}
\end{eqnarray}
Here we have used the usual convention that $a(t_0) =a_0= 1$.
 
Thus, if we observe a galaxy with a redshift $z = 2$, we are observing it
as it was when the Universe had a scale factor $a(t_e)=1/3$, where $t_e$ 
is the cosmological time when photon was emitted from the galaxy. This means that we are observing it 
at the time the Universe was $1/3$ of its present size.\footnote{This statement is strictly true if we consider uniform expansion of Universe over whole cosmic history} The redshift
we observe for a distant object depends only on the relative scale factors
at the time of emission and the time of observation. It does not depend on
how the transition between $a(t_e)$ and $a(t_0)$ was made. It does not matter if
the expansion was gradual or abrupt; it does not matter if the transition was
monotonic or oscillatory. All that matters is the scale factors at the time of
emission and the time of observation.

\section{Metric of the Universe}

The assumption of homogeneity of the standard model requires the Universe to have the same
curvature everywhere (just like the 2-dimensional surface of a sphere has the same curvature everywhere.)
Thus, we must investigate $3$-dimensional curved spaces that are 
homogeneous. Let us first consider this
question in two dimensions, where visualization is easier,
and then generalize to three spatial dimensions.

\subsection{Homogeneous, 2-Dimensional Spaces}

In two dimensions there are three independent possibilities for homogeneous, isotropic spaces:

\begin{itemize} 

       \item[1.] Flat Euclidean space.

       \item[2.] A sphere of constant (positive) curvature.

       \item[3.] An hyperboloid of constant (negative) curvature.

\end{itemize}

\begin{figure}[h!]
\centering
    \includegraphics[height=2.3in]{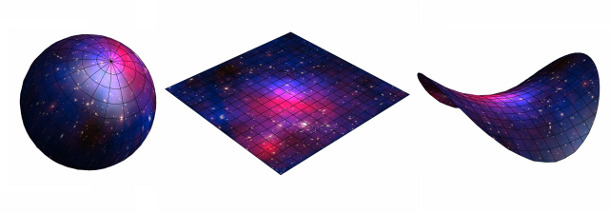}
\caption{2-dimensional examples of positive, zero and negative curvature universes (from left to right respectively). The zero and negative ones show only a finite part of the space, which extend to infinity in all directions.}
\label{fig:pfn}
\end{figure}

These three possibilities are illustrated in Fig.\ (\ref{fig:pfn}). Constant negative 
curvature surfaces cannot be embedded in 3-D Euclidean space. The saddle-like open 
surface only approximates constant negative curvature near its center. In each case the 
corresponding space has neither a special point nor a special direction. Thus, these are 
2-dimensional spaces with underlying metrics consistent with the cosmological principle.

Let us examine the 2-sphere as a representative example: We can think of 
2-spheres in terms of a 2-dimensional surface embedded in a 3-dimensional space. The 
center of the sphere is not the part of the space: it is inside the surface. Since a metric 
defines intrinsic properties of a space that should be independent of any additional embedding 
dimensions, thus technically it should be possible to express the metric of the 2-sphere 
in terms of only two coordinates. Now in terms of coordinates $\vec x=(x_1,x_2,x_3)$ in the embedding
3-dimensional flat euclidean space, the surface of sphere of radius $D$ is the collection of points that 
satisfies
\begin{eqnarray}
(x^1)^2+(x^2)^2+(x^3)^2=D^2,
\label{eq:2D-1.1}
\end{eqnarray}
so differentiating and simplifying we get
\begin{eqnarray}
\Rightarrow (dx^3)^2=\frac{(x^1dx^1+x^2dx^2)^2}{D^2-(x^1)^2-(x^2)^2}.
\label{eq:2D-1}
\end{eqnarray}
The line elements for the distance between two points $(\vec x, \vec x+d\vec x)$ on the 2-sphere is
\begin{eqnarray}
dl^2&=&(dx^1)^2+(dx^2)^2+(dx^3)^2.
\label{eq:2D-2}
\end{eqnarray}
Using this relation and Eq.\ (\ref{eq:2D-1}) we may rewrite this as 
\begin{eqnarray}
dl^2&=&(dx^1)^2+(dx^2)^2+\frac{(x^1dx^1+x^2dx^2)^2}{D^2-(x^1)^2-(x^2)^2}.
\label{eq:2D-2}
\end{eqnarray}
This line element describes distances on the 2-dimensional surface of the sphere and depends on only two coordinates 
($D$ is a constant). Distances are specified entirely by coordinates intrinsic 
to the 2D surface, independent of third embedding dimension. Let us now generalize 
this discussion to the 3-dimensional homogeneous space.

\subsection{Homogeneous, 3-Dimensional Spaces}

Consider a 3-dimensional sphere embedded in a 4-dimensional Euclidean hyperspace
\begin{eqnarray}
\left(x^1\right)^2+\left(x^2\right)^2+\left(x^3\right)^2+\left(x^4\right)^2=R^2,
\label{eq:MU1}
\end{eqnarray}
where $R$ is the radius of the 3-dimensional sphere. The distance between two neighboring points in the 4-dimensional space is given by
\begin{eqnarray}
\left(dx^1\right)^2+\left(dx^2\right)^2+\left(dx^3\right)^2+\left(dx^4\right)^2=dl^2.
\label{eq:MU2}
\end{eqnarray}
Differentiating Eq.\ (\ref{eq:MU1}) and solving for $dx^4$ , we obtain\footnote{We are using the Einstein summation convention notation.}
\begin{eqnarray}
x^4dx^4&=&=-x^1dx^1-x^2dx^2-x^3dx^3, \nonumber \\
dx^4&=&-\frac{x^i dx^i}{\sqrt{R^2-x^i x^i}}. \hspace{10 mm} \left(\mathrm {recall}\ i \in \left\{1,2,3\right\}\right)
\label{eq:MU3}
\end{eqnarray}
Then Eq.\ (\ref{eq:MU2}) can be rewritten as
\begin{eqnarray}
dl^2=\left(dx^1\right)^2+\left(dx^2\right)^2+\left(dx^3\right)^2+\frac{\left(x^i dx^i\right)^2}{R^2-x^i x^i}.
\label{eq:MU4}
\end{eqnarray}
In spherical coordinates
\begin{eqnarray}
x^1&=&r\ \mathrm{sin}\theta\ \mathrm{cos}\phi, \nonumber \\
x^2&=&r\ \mathrm{sin}\theta\ \mathrm{sin}\phi, \\
x^3&=&r\ \mathrm{cos}\theta.\nonumber 
\label{eq:MU5}
\end{eqnarray}
Using Eq.\ (\ref{eq:MU5}), and very simple algebra and differential calculus, we find
\begin{eqnarray}
x^ix^i&=&\left(dx^1\right)^2+\left(dx^2\right)^2+\left(dx^3\right)^2=r^2, \nonumber \\
x^idx^i&=&rdr, \\
dx^idx^i&=&dr^2+r^2d\theta^2+\left(r\ \mathrm{sin}\theta\right)^2d\phi^2.\nonumber 
\label{eq:MU6}
\end{eqnarray}
Substituting Eqs.\ (\ref{eq:MU6}) in Eq.\ (\ref{eq:MU4}), we can express the differential length element $dl^2$ in terms of
polar coordinates and $R$ as
\begin{eqnarray}
dl^2&=&\frac{dr^2}{1-\left(\frac{r}{R}\right)^2}+r^2d\theta^2+\left(r\ \mathrm{sin}\theta\right)^2d\phi^2. 
\label{eq:MU7}
\end{eqnarray}

The 3-sphere (with $R^2 > 0$) has the following important properties:
 
\begin{enumerate}

       \item It corresponds to a homogeneous, isotropic space that is closed and bounded.

       \item It has great circles as geodesics.

       \item It is a space of constant positive curvature.

       \item An ant dropped onto the surface of an otherwise featureless 3-sphere would find that:
       \begin{enumerate}

       \item No point or direction appears any different from any other.

       \item The shortest distance between any two points corresponds to a segment of a great circle.

       \item A sufficiently long journey in a fixed direction would return one to the starting point.

       \item The total area of the sphere was finite.
       \end{enumerate}
\end{enumerate}

We can also have a negatively curved 3-dimensional space (a ``saddle") where in the above we have to replace $R^2$ with $-R^2$. Then
\begin{eqnarray}
\left(x^1\right)^2+\left(x^2\right)^2+\left(x^3\right)^2+\left(x^4\right)^2=-R^2, \nonumber\\
\Rightarrow\ \ \ dl^2=\frac{dr^2}{1+\left(\frac{r}{R}\right)^2}+r^2d\theta^2+\left(r\ \mathrm{sin}\theta\right)^2d\phi^2.
\label{eq:3d-1}
\end{eqnarray}
This negatively curved 3-dimensional space have the following important properties:

\begin{enumerate}

       \item It is a  homogeneous and isotropic, space that is unbounded and infinite.

       \item It has hyperbolas as geodesics.

       \item It is a space of constant negative curvature.

       \item An ant dropped onto the this surface would find that:
       \begin{enumerate}

       \item No point or direction appears any different from any other.

       \item The shortest distance between any two points corresponds to a segment of a hyperbola.

       \item The ant would never return to the starting point by continuing an infinite distance in a fixed direction.

       \item The ant would find that the volume of the space is infinite.

       \end{enumerate}
\end{enumerate}

Finally the 3-dimensional flat Euclidean case is the $R\rightarrow \infty$ limit of either of the above space. It has
\begin{eqnarray}
dl^2=dr^2+r^2 d\Omega^2=dr^2+r^2 d\theta^2+r^2 \mathrm{sin^2}\theta d\phi^2.
\end{eqnarray}
This space has the following properties:
\begin{enumerate}

       \item It is homogeneous and isotropic.

       \item It is of infinite extent, with straight lines as geodesics.

       \item Obviously, this space corresponds to the limit of zero spatial curvature.

       \item The volume of this space is infinite, and a straight path in one direction will never return to the starting point.

\end{enumerate}

In summary, $dl^2$ is given by
\begin{eqnarray}
dl^2&=&\frac{dr^2}{1-k\ r^2}+r^2d\theta^2+\left(r\ \mathrm{sin}\theta\right)^2d\phi^2, 
\label{eq:MU8}
\end{eqnarray}
where $k = 1/R^2$ can be positive, zero, negative and the corresponding spacetime interval is  
\begin{eqnarray}
ds^2&=&-dt^2+\frac{dr^2}{1-k\ r^2}+r^2d\theta^2+\left(r\ \mathrm{sin}\theta\right)^2d\phi^2. 
\label{eq:MU9}
\end{eqnarray}
Where $k$ negative, zero, and positive correspond to an infinitely open or negatively curved Universe, 
an infinitely flat Universe 
and a finite closed Universe respectively. 

The metric in Eq.\ (\ref{eq:MU9}) is the metric of stationary curved spacetime. As a special case we can put $k=0$,
and we will recover the Minkowski flat spacetime metric, which applies only within the context
of special relativity, so called, because it deals with the special case in which
spacetime is not curved by the presence of mass and energy. Without any
gravitational effects, Minkowski spacetime is flat and static.
Using Eq.\ (\ref{eq:angular}), we can write the above metric as:
\begin{eqnarray}
ds^2&=&-dt^2+\frac{dr^2}{1-k\ r^2}+r^2d\Omega^2. 
\label{eq:MU10}
\end{eqnarray}
When gravity is added, however, the permissible spacetimes are more interesting.
With the assumption of homogeneity of space, the spatial components of the metric tensor can still be 
time dependent. The generic line element which meets these conditions is:
\begin{eqnarray}
ds^2&=&-dt^2+a^2(t)\left[\frac{dr^2}{1-k\ r^2}+r^2d\Omega^2\right]. 
\label{eq:MU10}
\end{eqnarray}
The term in square bracket in above equation is the spatial part of the metric that does not depend on time; all the 
time dependence is in the function $a^2(t)$, where $a(t)$ is the scale factor. Note that the substitutions
\begin{eqnarray}
k \rightarrow \frac{k}{|k|}, \nonumber \\
r \rightarrow \sqrt{|k|}r, \\
a \rightarrow \frac{a}{\sqrt{|k|}}, \nonumber
\end{eqnarray}
leaves Eq.\ (\ref{eq:MU10}) invariant.
Therefore, the only relevant parameter is $k/|k|$, which can have only three discrete values: $-1,\ 0,\ \mathrm{and}\ 1$, corresponding to the hyperbolic, flat, 
and spherical Universe respectively. Also $t$ is the comoving cosmic time\footnote{It is the time measured by an observer who sees the surrounding Universe expand uniformly.} and $d\Omega^2$ is the line element of the unit 2-sphere. Here
it is important to note that locally we actually see around us the distribution of matter in the form of stars, galaxies,
galaxy clusters and so on. The local spacetime structure around such objects is certainly different from what is given 
by the above line element. This above line element is meant to describe the spacetime of cosmology on much large scale, scale of
about 100 Mpc and larger, where the cosmological principle is valid.  

The above Friedmann-Lema\^{i}tre-Robertson-Walker (FLRW) spacetime line element may be expressed in an alternative form by introducing the 4-dimensional generalization of polar
angles
\begin{eqnarray}
w&=&R\ \mathrm{cos}\chi,\\
x&=&R\ \mathrm{sin}\chi \mathrm{sin}\theta \mathrm{cos}\phi,\\
y&=&R\ \mathrm{sin}\chi \mathrm{sin}\theta \mathrm{sin}\phi,\\
z&=&R\ \mathrm{sin}\chi \mathrm{cos}\theta,
\end{eqnarray}
with the ranges $0\leq \phi \leq 2\pi$, $0\leq \theta \leq \pi$, and $0\leq \chi \leq \pi$ for the case of spherical
geometry and with the substitution $w=iw$, $\chi=i\chi$, and $R=iR$ for hyperbolic geometry. 

The Friedmann-Lema\^{i}tre-Robertson-Walker (FLRW) spacetime line element may be written as:

\begin{eqnarray}
ds^2=-dt^2 + a^2(t)\left\{
      \begin{array}{lc}
           d\chi^2+\mathrm{sin^2}\chi \left(d\theta^2+\mathrm{sin^2}\theta d\phi^2\right) & \ \ \ \ \mathrm{(closed)} \\
           d\chi^2+\chi^2 \left(d\theta^2+\mathrm{sin^2}\theta d\phi^2\right) & \ \ \ \ \mathrm{(flat)} \\
           d\chi^2+\mathrm{sinh^2}\chi \left(d\theta^2+\mathrm{sin^2}\theta d\phi^2\right) & \ \ \ \ \mathrm{(open)} \\
     \end{array},
   \right.
\label{eq:4dploarmetric}
\end{eqnarray} 
which is related to
\begin{eqnarray}
ds^2&=&-dt^2+a^2(t)\left[\frac{dr^2}{1-k\ r^2}+r^2d\theta^2+r^2\mathrm{sin^2}\theta d\phi^2\right].,
\end{eqnarray}
by the change of variables:
\begin{eqnarray}
r=\left\{
      \begin{array}{lc}
           \mathrm{sin}\chi  & \ \ \ \ \mathrm{(closed)} \\
           \chi  & \ \ \ \ \mathrm{(flat)} \\
           \mathrm{sinh^2}\chi  & \ \ \ \ \mathrm{(open)} \\
     \end{array}.
   \right.
\label{eq:variablechange}
\end{eqnarray} 
Here it is important to take note of the following points:
\begin{enumerate}

       \item The derivation of the FLRW metric was purely geometrical, subject to the constraints of homogeneity.

       \item No dynamical considerations enter explicitly into its formulation.

       \item Of course, dynamics are implicit, to the extent that the overall dynamical structure of the Universe must be consistent with the cosmological principle that was used to construct the metric.
\end{enumerate}
The Friedmann-Lema\^{i}tre-Robertson-Walker (FLRW) the line element may be expressed in matrix form as
\begin{eqnarray}
ds^2&=&g_{\mu \nu} dx^\mu dx^\nu,\nonumber \\
&=&\left(
      \begin{array}{cccc}
           dt & dr & d\theta & d\phi \\
     \end{array}
   \right)
\left(
      \begin{array}{cccc}
           -1     &     0     &     0    &     0 \\
	    0     &\frac{a^2(t)}{1-kr^2}& 0  & 0 \\
	    0     &     0     & a^2(t)r^2 &    0 \\
	    0     &     0     &     0    & r^2\mathrm{sin^2}\theta    \\
     \end{array}
   \right)\left(
      \begin{array}{c}
           dt \\ 
           dr \\
           d\theta \\
           d\phi \\
     \end{array}
   \right)
\end{eqnarray}
Thus, the FLRW metric is diagonal, with non zero covariant elements
\begin{eqnarray}
g_{00}=-1, \hspace{0.5 cm} g_{11}=\frac{a^2}{1-kr^2}, \hspace{0.5 cm} g_{22}=a^2r^2, \hspace{0.5 cm} g_{33}=a^2 r^2\mathrm{sin^2}\theta,
\end{eqnarray}
and the corresponding contravariant components are
\begin{eqnarray}
g^{00}=-1, \hspace{0.5 cm} g^{11}=\frac{1-kr^2}{a^2}, \hspace{0.5 cm} g^{22}=\frac{1}{a^2r^2}, \hspace{0.5 cm} g^{33}=\frac{1}{a^2 r^2\mathrm{sin^2}\theta}.
\end{eqnarray}

\section{Derivation of Friedmann's Equations from General Relativity}
\label{sec:derivation of friedmann}

Though we can derive Friedmann's equations of motion which are the backbone of cosmology almost entirely by using Newtonian Mechanics 
(interested readers can see, Liddle,\cite{liddle2003introduction} Ryden,\cite{ryden2003introduction} or
 Raine and Thomas.\cite {raine2001introduction}), here we derive them from Einstein's general relativity field
equations.

Considering the most general Friedmann-Lemaitre-Robertson-Walker (FLWR) metric given in Eq. (\ref{eq:MU10}), the only
non-zero elements and first derivative of the metric are

\begin{eqnarray}
g_{00}&=&-1, \hspace{20 mm} g^{00}=-1,\nonumber \\
g_{11}&=&\frac{a^2}{1-kr^2}, \hspace{11 mm}g^{11}=\frac{1-kr^2}{a^2},\nonumber \\
g_{22}&=&a^2r^2, \hspace{18 mm} g^{22}=\frac{1}{a^2r^2},\nonumber \\
g_{33}&=&a^2r^2\mathrm{sin^2}\theta, \hspace{9 mm} g^{33}=\frac{1}{a^2r^2\mathrm{sin^2}\theta},\nonumber \\
g_{11,0}&=&\frac{2a\dot a}{1-kr^2}, \nonumber \\
g_{11,1}&=&\frac{2kra^2}{(1-kr^2)^2}, \nonumber \\
g_{22,0}&=&2a \dot a\ r^2,\nonumber \\
g_{22,1}&=&2a^2 r,\nonumber \\
g_{33,0}&=&2a\dot a r^2 \ \mathrm{sin^2}\theta,\nonumber \\
g_{33,1}&=&2a^2r \ \mathrm{sin^2}\theta,\nonumber \\
g_{33,2}&=&2a^2r^2 \mathrm{sin}\theta \ \mathrm{cos}\theta.
\label{eq:metricandderivatives}
\end{eqnarray} 
We now compute the 
Christoffel symbols
\begin{eqnarray}
\Gamma^{\gamma}_{\alpha \beta}=\frac{1}{2}g^{\gamma \delta}\Big(g_{\delta \alpha, \beta}+g_{\beta \delta, \alpha}-g_{\alpha \beta, \delta}\Big).
\label{eq:chr}
\end{eqnarray} 
Let's compute term by term:
\begin{itemize} 

       \item[\textbf {a)}]  Time-Time Terms

We put $\gamma=0$, hence also have to set $\delta=0$, to ensure $g^{\gamma \delta}$ is non-zero,
\begin{eqnarray}
\Gamma^{0}_{\alpha \beta}=\frac{1}{2}g^{00}\Big(g_{0\alpha,\beta}+g_{\beta 0,\alpha}-g_{\alpha \beta,0}\Big)=-\frac{1}{2}g^{00}g_{\alpha \beta,0}=\frac{1}{2}g_{\alpha \beta,0}.
\end{eqnarray}
Here we use Eqs.\ (\ref{eq:metricandderivatives}) and the fact that $g_{0 \alpha, \beta}=g_{\beta 0, \alpha}=0$. From this we have
\begin{eqnarray}
\Gamma^{0}_{0 0}&=&\Gamma^{0}_{\alpha 0}=\Gamma^{0}_{0 \beta}=0, \\
\Gamma^{0}_{ij}&=&\frac{1}{2}g_{ij,0}. \hspace{15 mm} \left(\mathrm{where ~~}i,j\ \in \left\{ 1, 2, 3\right\}\right)
\end{eqnarray}
\begin{eqnarray}
\Rightarrow \Gamma^{0}_{12}=\Gamma^{0}_{13}=\Gamma^{0}_{21}=\Gamma^{0}_{23}=\Gamma^{0}_{31}=\Gamma^{0}_{32}=0,
\end{eqnarray}
and the $i=j$ terms are
\begin{eqnarray}
\Gamma^{0}_{11}&=&\frac{1}{2} \left(\frac{2 a \dot a}{1-kr^2}\right)=\frac{a \dot a}{1-kr^2}, \\
\Gamma^{0}_{22}&=&\frac{1}{2}\left(2 a \dot a r^2\right)=a\dot a r^2, \\
\Gamma^{0}_{33}&=&\frac{1}{2}\left(2a \dot a r^2 \mathrm{sin^2}\theta\right)=a\dot a r^2 \mathrm{sin^2}\theta.
\end{eqnarray}
       \item[\textbf {b)}] Space-Time Terms

Let $\alpha=0$ and $\gamma=i$ in Eq.\ (\ref{eq:chr}), we will get
\begin{eqnarray}
\Gamma^{i}_{0 \beta}=\frac{1}{2}g^{ij}\big(g_{j0,\beta}+g_{\beta j, 0}-g_{0 \beta, j}\Big).
\end{eqnarray}
The first and third term in the parentheses are zero (because our metric is diagonal), the second term is
also zero unless $\beta=j$, in which case:
\begin{eqnarray}
\Gamma^{i}_{0 \beta}&=&\frac{1}{2}g^{ij}g_{ij,0}\delta^{i}_{\beta}, \nonumber \\
&=&\frac{1}{2}g^{ij}\left(\frac{2a \dot a}{a^2}g_{ij}\right)\delta^{i}_{\beta}, \hspace{10 mm} \left(\because\ g_{ij,0}=\frac{2a \dot a}{a^2}g_{ij},\ \ \mathrm{from\ Eq.\ (\ref{eq:metricandderivatives})} \right) \nonumber \\
&=&\frac{\dot a}{a}g^{ij}g_{ij}\delta^{i}_{\beta}, \nonumber \\
&=&\frac{\dot a}{a}\delta^{i}_{\beta}.
\label{eq:FtGRgamma}
\end{eqnarray}

       \item[\textbf {c)}] Space-Space Terms

We need to use $\alpha=j, \ \beta=k, \mathrm{and}\ \gamma=i$ in Eq.\ (\ref{eq:chr}),
\begin{eqnarray}
\Gamma^{i}_{j k}=\frac{1}{2}g^{i l}\Big(g_{l j, i}+g_{k l, j}-g_{j k, l}\Big),
\label{eq:chr1}
\end{eqnarray}
then the required Christoffel symbols will be:
\begin{eqnarray}
\Gamma^{1}_{11}&=&\frac{1}{2}g^{11}\big(g_{11,1}+\cancel{g_{11,1}}-\cancel{g_{11,1}}\big)=\frac{1}{2}g^{11}g_{11,1}=\frac{kr}{1-kr^2},\nonumber \\
\Gamma^{1}_{22}&=&\frac{1}{2}g^{11}\big(g_{12,2}+g_{21,2}-g_{22,1}\big)=-r(1-kr^2),\nonumber \\
\Gamma^{1}_{33}&=&\frac{1}{2}g^{11}\big(g_{13,3}+g_{31,3}-g_{33,1}\big)=-r(1-kr^2)\mathrm{sin^2}\theta,\nonumber \\
\Gamma^{2}_{33}&=&\frac{1}{2}g^{22}\big(g_{23,3}+g_{32,3}-g_{33,2}\big)=-\mathrm{sin}\theta \mathrm{cos}\theta,\nonumber \\
\Gamma^{2}_{12}&=&\Gamma^{2}_{21}=\frac{1}{2}g^{22}\big(g_{22,1}+g_{12,2}-g_{21,2}\big)=\frac{1}{r},\nonumber \\
\Gamma^{3}_{13}&=&\Gamma^{3}_{11}=\frac{1}{2}g^{33}\big(g_{33,1}+g_{13,3}-g_{31,3}\big)=\frac{1}{r},\nonumber \\
\Gamma^{3}_{23}&=&\Gamma^{3}_{22}=\frac{1}{2}g^{33}\big(g_{33,2}+g_{23,3}-g_{32,3}\big)=\mathrm{cot}\theta.
\label{eq:allgamma}
\end{eqnarray}
All other terms are zero.
\end{itemize} 
Now  we can compute the Ricci tensor, using Eq.\ (\ref{eq:Riemanntensor}),
\begin{eqnarray}
R^{\gamma}_{\alpha \gamma \beta}=R_{\alpha \beta}=\Gamma^{\gamma}_{\alpha \beta,\gamma}-\Gamma^{\gamma}_{\gamma \alpha, \beta}+
\Gamma^{\gamma}_{\gamma \lambda}\Gamma^{\lambda}_{\beta \alpha}-\Gamma^{\gamma}_{\beta \lambda}\Gamma^{\lambda}_{\gamma \alpha}.
\end{eqnarray}
The $R_{00}$ component is
\begin{eqnarray}
R_{0 0}=\Gamma^{\gamma}_{0 0,\gamma}-\Gamma^{\gamma}_{\gamma 0, 0}+
\Gamma^{\gamma}_{\gamma \lambda}\Gamma^{\lambda}_{0 0}-\Gamma^{\gamma}_{0 \lambda}\Gamma^{\lambda}_{\gamma 0}.
\end{eqnarray}
It is very easy to see that the first and the third term are zero because $\Gamma^{\gamma}_{00}=0$ from Eq.\ (\ref{eq:allgamma}).
Hence we have
\begin{eqnarray}
R_{00}=-\Gamma^{\gamma}_{\gamma 0, 0}-\Gamma^{\gamma}_{0 \lambda}\Gamma^{\lambda}_{\gamma 0}.
\label{eq:R00}
\end{eqnarray}
The first term is
\begin{eqnarray}
\Gamma^{\gamma}_{\gamma 0, 0}&=&\frac{d}{dx^0}\left[\Gamma^{\gamma}_{\gamma 0}\right], \nonumber \\
&=&\frac{d}{dx^0}\left(\frac{\dot a}{a}\delta^{\gamma}_{\gamma}\right), \hspace{10 mm} \Big(\mathrm{using\ Eq.\ (\ref{eq:FtGRgamma})}\Big) \nonumber \\
&=&\frac{d}{dx^0}\left(3\frac{\dot a}{a}\right), \nonumber \\
&=&3\left(\frac{a \ddot a-\dot a^2}{a^2}\right).
\label{eq:ft}
\end{eqnarray}
Now, the second term on the right hand side of Eq.\ (\ref{eq:R00}) is zero if $\gamma=0$ so
\begin{eqnarray}
\Gamma^{\gamma}_{0 \lambda}\Gamma^{\lambda}_{\gamma 0} 
&=& \Gamma^{i}_{0 j}\Gamma^{j}_{i 0} \nonumber \\
&=& \left(\frac{\dot a}{a}\right)\delta^{i}_{j} \left(\frac{\dot a}{a}\right)\delta^{j}_{i}, \hspace{10 mm} \Big(\mathrm{using\ Eq.\ (\ref{eq:FtGRgamma})}\Big) \nonumber \\
&=& \left(\frac{\dot a}{a}\right)^2 \underbrace{\delta^{i}_{j}\ \delta^{j}_{i}}_{=3}, \nonumber \\
&=& 3 \left(\frac{\dot a}{a}\right)^2.
\label{eq:st}
\end{eqnarray}
Now Eq.\ (\ref{eq:R00}) with Eq.\ (\ref{eq:ft}) and Eq.\ (\ref{eq:st}) results
\begin{eqnarray}
R_{00}&=&-3\left(\frac{a \ddot a-\dot a^2}{a^2}\right)-3 \left(\frac{\dot a}{a}\right)^2=-3\frac{\ddot a}{a}. 
\label{eq:R002} 
\end{eqnarray}
Hence we can raise an index to get
\begin{eqnarray}
R^{0}_{0}=g^{00}R_{00}=-R_{00}=3\ \frac{\ddot a}{a}.
\label{eq:R00f}
\end{eqnarray}

Along the same lines, straightforward, yet tedious, computation yields the other non-zero components of the Ricci tensor,
\begin{eqnarray}
R_{11}&=&\frac{a \ddot a +2\dot a^2+2k}{1-kr^2}, \nonumber \\
R_{22}&=&r^2 \left(a \ddot a+2\dot a^2+2k\right), \nonumber \\
R_{22}&=&r^2 \left(a \ddot a+2\dot a^2+2k\right)\mathrm{sin^2}\theta.
\end{eqnarray}
These can be written in a compact form after raising the index,
\begin{eqnarray}
R^{0}_{i}&=&0, \nonumber \\
R^{i}_{j}&=&\frac{1}{a^2}\left(a\ddot a+2\dot a^2+2k\right)\delta^{i}_{j}.
\label{eq:Rijf}
\end{eqnarray}

The Universe is not empty, so we are interested in non-vacuum solutions to Einstein's equations. 
We will choose to model the matter and energy in the Universe by perfect fluids. We discussed 
perfect fluids earlier. They are defined as fluids which are isotropic in their 
rest frame. The energy-momentum tensor for a perfect fluid can be written as
\begin{eqnarray}
T_{\alpha \beta}=\left(\rho+P\right)u_{\alpha}u_{\beta}+Pg_{\alpha \beta}.
\label{eq:perfectgas}
\end{eqnarray}
Raising the index of the stress-energy tensor we have
\begin{eqnarray}
T^{\alpha}_{\beta}&=&\left(\rho+P\right)u^{\alpha}u_{\beta}+P\delta^{\alpha}_{\beta}.
\label{eq:setud}
\end{eqnarray}
Here $\rho$ and $P$ are the energy density and pressure as measured in the rest frame, and $u^\nu$ is the 
4-velocity of the fluid. A fluid that is isotropic in some frame must lead to a metric isotropic 
in that frame. That is, the fluid must be at rest in comoving coordinates. The final 4-velocity is then
\begin{eqnarray}
u^\nu=(1,0,0,0),
\end{eqnarray}
and lowering the index gives
\begin{eqnarray}
u_\nu=(-1,0,0,0). 
\end{eqnarray}
Now, we shall have need for the contraction over indices $\alpha$ and $\beta$ of Eq.\ (\ref{eq:setud}),
\begin{eqnarray}
T \equiv T^{\alpha}_{\alpha}=-\left(\rho+P\right)+4P=3P-\rho.
\label{eq:contractedT} 
\end{eqnarray} 

Raising an index of Einstein's field equations gives
\begin{eqnarray}
R^{\alpha}_{\beta}-\frac{1}{2} \delta^{\alpha}_{\beta} \mathcal{R}&=& 8 \pi G T^{\alpha}_{\beta}.
\label{eq:eeud}
\end{eqnarray}
Contracting over indices $\alpha$ and $\beta$ of Eq.\ (\ref{eq:eeud}) we have
\begin{eqnarray}
-\mathcal{R}=8\pi GT, \hspace{10 mm} \Big(\mathrm{where}\ T\equiv T^{\alpha}_{\alpha},\ \delta^{\alpha}_{\alpha}=4,\ \mathrm{and}\ R^{\alpha}_{\alpha}=\mathcal{R}.\Big)
\end{eqnarray}
Hence Einstein's field equation can also be written as 
\begin{eqnarray}
R^{\alpha}_{\beta}=8\pi G \left(T^{\alpha}_{\beta}-\frac{1}{2}\delta^{\alpha}_{\beta}T\right).
\label{eq:ricciud}
\end{eqnarray}
Using Eq.\ (\ref{eq:setud}) and Eq.\ (\ref{eq:contractedT}) in Eq.\ (\ref{eq:ricciud}), we will find
\begin{eqnarray}
R^{\alpha}_{\beta}&=&8\pi G \left[\left(\rho+P\right)u^{\alpha}u_{\beta}+P \delta^{\alpha}_{\beta}-\frac{1}{2}\delta^{\alpha}_{\beta}\left(3P-\rho\right)\right], \nonumber \\
&=& 8 \pi G \left[\left(\rho+P\right)u^{\alpha}u_{\beta}+\frac{1}{2}\delta^{\alpha}_{\beta}\left(\rho-P\right)\right].
\label{eq:ricciud1}
\end{eqnarray}
The $R^{0}_{0}$ component of Eq.\ (\ref{eq:ricciud1}) is
\begin{eqnarray}
R^{0}_{0}&=& 8 \pi G \left[-\left(\rho+P\right)+\frac{1}{2}\left(\rho-P\right)\right] \nonumber \\
&=&8 \pi G \left[-\frac{1}{2}\left(\rho+3P\right)\right]=-4 \pi G \left(\rho+3P\right).
\end{eqnarray}
Comparing this with Eq.\ (\ref{eq:R00f}), leads to standard acceleration equation as:
\begin{eqnarray}
\frac{\ddot a}{a}=-\frac{4 \pi G}{3}\left(\rho+3P\right).
\label{eq:accelerationeq}
\end{eqnarray}
Similarly, comparing the $R^{i}_{i}$ component of Eq.\ (\ref{eq:ricciud1}) with Eq.\ (\ref{eq:Rijf}), leads to the other standard Friedman equation as follows
\begin{eqnarray}
\frac{1}{a^2}\left(a\ddot a+2 \dot a^2+2k\right)&=&8 \pi G \left[0+\frac{1}{2}\left(\rho-P\right)\right], \nonumber \\
\Rightarrow \left(\frac{\ddot a}{a}\right)+2\left(\frac{\dot a}{a}\right)^2+2\frac{k}{a^2}&=&4 \pi G (\rho -P), \nonumber \\
\Rightarrow \left(\frac{\dot a}{a}\right)^2&=&\frac{8 \pi G}{3}\rho-\frac{k}{a^2}.
\label{eq:Friedmanneq}
\end{eqnarray}
Here we have used Eq.\ (\ref{eq:accelerationeq}) in the last step. By combining this with the conservation of energy equation, that we already derived in Sec.\ (\ref{sec:energyconservation}), 
and the equation of state, we obtain a closed system of Friedmann equations.

\section{Solutions of Friedmann's Equations}

Given the closed set of cosmological equations of motion, which gives the relationship between the scale factor $a(t)$, the energy density $\rho(t)$ ($c=1$, mass and energy have the same units), and pressure $P$, for open, flat and closed Universe models (as denoted by three discrete values of $k=-1,\ 0,$ and, $1$ respectively), we can solve these equations in various cases and obtain the form of the scale factor $a(t)$ as a function of time. Let's start with the definition of the deceleration parameter. 

\subsection{Deceleration Parameter}

From the definition  of Hubble parameter $H$ in Eq.\ (\ref{eq:HL3}) we have
\begin{eqnarray}
\dot H&=&\frac{a\ddot a-\dot a^2}{a^2}=-H^2+\frac{\ddot a}{a}=-H^2\left(1-\frac{\ddot a}{H^2 a}\right)\equiv -H^2(1+q),
\label{eq:qfactor1}
\end{eqnarray}
where the dimensionless deceleration parameter $q$ is defined as
\begin{eqnarray}
q \equiv -\frac{\ddot a}{H^2 a}.
\label{eq:qfactor2}
\end{eqnarray} 
The present value of all time-dependent quantities are denoted by a subscript of $0$ for example, the present value of the deceleration parameter is
denoted by $q_0$ and is
\begin{eqnarray}
q_0=-\frac{1}{H_0^2}\left(\frac{\ddot a}{a}\right)_0.
\label{eq:qfactor3}
\end{eqnarray}
Note the choice of sign in defining $q_0$. Since $H_0^2$ and $a$ are strictly positive numbers, $q_0<0$ for
$\ddot a >0$ and vice-versa, which means a negative value of $q_0$ represents an accelerating cosmological expansion.

\subsection{Curvature-Dominated Universe ($k \neq 0$, $q_0=0$, $\rho_i=0$)}

The simplest (but not interesting) Universe is one that is completely empty, no matter, no energy, no radiation, etc.
For such Universe, the Friedmann equation takes the form
\begin{eqnarray}
\dot a^2=-k.
\label{eq:CD1}
\end{eqnarray}
This equation has two solutions. The first is $\dot a=0$ and $k=0$ has the solutions $a=$ constant which is the spatially-flat static Minkowski spacetime. The second one is governed by
\begin{eqnarray}
\dot a=\pm \sqrt{-k},
\label{eq:CD2}
\end{eqnarray} 
which is physically consistent only when $k=-1$ (since $\dot a$ cannot be complex). A Universe that is positively curved and empty is not allowed by Friedmann's equations.
In the negative curvature case, solving the differential equation Eq.\ (\ref{eq:CD2}), gives: 
\begin{eqnarray}
a(t) \propto t.
\label{eq:CD3}
\end{eqnarray}
This is negative space-curvature Milne spacetime. In the language of Newtonian mechanics, in the absence of any gravitational force (as in this case) the relative 
velocity of any two points in space is constant, which leads to the scale factor $a(t)$ being a linear function of time
in an empty Universe. From the second Friedmann equation, we get
\begin{eqnarray}
\hspace{2cm} \frac{\ddot a}{a}=0.   \hspace{1cm} (\mathrm{since}\ \rho=p=\Lambda=0)
\label{eq:CD4}
\end{eqnarray}
This means that an empty Universe, which has to be negatively curved, should expand with zero acceleration. Also from 
Eq.\ (\ref{eq:qfactor2}) we see:
\begin{eqnarray}
-H^2q=0,
\label{eq:CD5}
\end{eqnarray}
since $H \neq 0$ this implies that $q=0$. 

Note that in an empty Universe since $H=\dot a/a=t^{-1}$, $t_0$, the age of the Universe is equal to the Hubble time,
\begin{eqnarray}
t_0=H_0^{-1},
\label{eq:CD6}
\end{eqnarray} 
because there is nothing 
to speed up or slow down expansion.


\subsection{Spatially-Flat Single-Component Universe (${k=0,\ q_0=(1+3\omega)/2}$)}

In a spatially-flat Universe, $k=0$, the Friedmann equation is
\begin{eqnarray}
\left(\frac{\dot a}{a}\right)^2=\frac{8 \pi G}{3} \rho,
\label{eq:SFU1}
\end{eqnarray} 
where $\rho$ is the time dependent energy density of the single type of matter present in Universe. From the energy evolution equation [see Eq.\ (\ref{eq:density}) with the
general assumption of $a_0=1$];
\begin{eqnarray}
\rho(t)=\rho_0 a^{-3(1+\omega)},
\label{eq:SFU2}
\end{eqnarray}
where $a(t)$ is the scale factor, and we have assumed an ideal fluid of time independent equation of state parameter $\omega$. Hence, Eq. (\ref{eq:SFU1}) takes the form
\begin{eqnarray}
\left(\frac{\dot a}{a}\right)^2=\frac{8 \pi G}{3} \rho_0 a^{-3(1+\omega)}.
\label{eq:SFU3}
\end{eqnarray}
To find how the scale factor $a(t)$ evolves in this kind of Universe, we have to solve the above differential equation.
In order to do that, let's guess a power law solution $a \propto t^p$, then the left and right hand side of Eq.\ (\ref{eq:SFU3}) evolve with time as
\begin{eqnarray}
\hspace{30 mm} \left(\frac{\dot a}{a}\right)^2\propto t^{-2}, \hspace{5 mm} \frac{8 \pi G}{3} \rho_0 a^{-3(1+\omega)} \propto t^{-3p(1+\omega)}.
\label{eq:SFU4}
\end{eqnarray}
Matching the powers of $t$ we get
\begin{eqnarray}
p&=&\frac{2}{3(1+\omega)},
\label{eq:SFU5}
\end{eqnarray}
with the restriction of $\omega \neq -1$. Hence:
\begin{eqnarray}
a(t)\propto t^{2/3(1+\omega)}.
\label{eq:SFU6}
\end{eqnarray}
If $\omega \neq -1/3$, $\ddot a \neq 0$ and this particular Universe is undergoing accelerated or decelerated expansion. From the second 
Friedmann's equation
\begin{eqnarray}
\frac{\ddot a}{a}=-\frac{4 \pi G}{3}\rho \left(1+3 \omega \right). 
\label{eq:SFU7}
\end{eqnarray}
In terms of deceleration parameter, $\ddot a/a=-qH^2$, this equation gives
\begin{eqnarray}
\rho&=&\frac{3H^2}{4 \pi G (1+3\omega)}q,
\label{eq:SFU8}
\end{eqnarray}
which when substituted in the first Friedmann equation $H^2=8\pi G\rho/3$, gives
\begin{eqnarray}
H^2\left(1-\frac{2}{1+3\omega}q\right)&=&0.
\label{eq:SFU10}
\end{eqnarray}
Since $H \neq 0$, 
\begin{eqnarray}
q=\frac{1}{2}\left(1+3\omega\right).
\end{eqnarray}
Hence $q$ of the spatially-flat Universe is $\left(1+3\omega\right)/2$.\footnote{Incorporating space-curvature $k$, the deceleration parameter is where $h$ is the Hubble constant multiple of 100 km s$^{-1}$ Mpc$^{-1}$,
\begin{eqnarray}
q=\frac{1}{2}\left(1+3\omega\right)\left(1+\frac{k}{a^2H^2}\right),
\end{eqnarray}
so, if $k=1,\ q>\frac{1}{2}\left(1+3\omega\right)$ and
if $k=-1,\  q<\frac{1}{2}\left(1+3\omega\right)$.
}

We now define the critical density of the Universe, 
which is the density of the matter when $q=1/2$ (spatially-flat case) with $\omega=0$ or pressure less matter.\footnote{This is called the Einstein-de Sitter spacetime or model.} From Eq.\ (\ref{eq:SFU8}),
\begin{eqnarray}
\rho_{cr}=\frac{3H^2}{8 \pi G}.
\label{eq:SFU11}
\end{eqnarray}
\textit{This is the density needed to yield a spatially-flat Universe}. Numerically this is 
\begin{eqnarray}
\rho_{cr}=1.88\times  10^{-29}\ h^2\mathrm{g~ cm^{-3}},
\label{eq:SFU12}
\end{eqnarray}
The deceleration parameter $q$ relates the density of the Universe $\rho$ 
to the critical density $\rho_{cr}$ through
\begin{eqnarray}
q=\frac{\rho}{2\rho_{cr}}\left(1+3\omega\right).
\label{eq:SFU13}
\end{eqnarray}

From Eq.\ (\ref{eq:SFU6}) $H={2}/({3(1+\omega)t})$, so
the age of the Universe, $t_0$, in terms of the Hubble constant, is
\begin{eqnarray}
t_0=\frac{2}{3(1+\omega)} H^{-1}_{0}.
\label{eq:SFU15}
\end{eqnarray}
If $\omega<-1/3$ then the age of the single-component-dominated spatially-flat Universe is larger than the 
Hubble time $1/H_0$, and if $\omega>-1/3$ the age is less than the 
Hubble time. 

Now let's consider two special cases of the spatially-flat Universe.

\begin{itemize} 

       \item[\textbf {a)-}]  \textbf{Non-Relativistic-Matter-Dominated Universe ($\boldsymbol{k=0,\ P=0}$)}

			      Non-Relativistic matter (dust) does not exert pressure, hence $P=0$ and also $\omega=0$. 
			      From Eqs.\ (\ref{eq:density}) and (\ref{eq:SFU6}) we get 
			      \begin{eqnarray}
			      \rho_m (t)=\rho_{m0}a^{-3}(t), 
			      \label{eq:SFU16}
			      \end{eqnarray}
			      and integrating Friedmann differential equation we find
			      \begin{eqnarray}
			      a(t)=(6\pi G \rho_{m0} a^3_0)^{1/3}t^{2/3},
			      \label{eq:scalefof matter}
			      \end{eqnarray}
			      where constant of integration has been chosen so that, $a(t=0)=0$. This Universe is called Einstein-de Sitter spacetime or model.

			      Since $\rho_{m} a^3=$ constant  
			      we can put $\rho_{m}a^3=\rho_{m0}a^3_0$ in Eq.\ (\ref{eq:scalefof matter}) to obtain
			      \begin{eqnarray}
			      \rho_{m}(t)=\frac{1}{6\pi G t^2},
			      \end{eqnarray}
			      for the density as a function of time.  
			      The age of the Universe can be computed from Eq.\ (\ref{eq:SFU15}) as
			      \begin{eqnarray}
			      t_0=\frac{2}{3H_0}. 
			      \label{eq:SFU17}
			      \end{eqnarray}
			      Also, the deceleration parameter for Einstein-de Sitter Universe is
			      \begin{eqnarray}
			      q=\frac{1}{2}.			     
			      \end{eqnarray}

       \item[\textbf {b)-}]  \textbf{Relativistic-Matter-Dominated Universe ($k=0$, $P=\rho/3$)}

			      Radiation is often modeled by the perfect fluid approximation with $P=\rho/3$ or $\omega=1/3$. Using this 
			      value of $\omega$ in Eqs.\ (\ref{eq:density}) and (\ref{eq:SFU6}), we will
			      \begin{eqnarray}
			      \rho_r (t)=\rho_{r0}a^{-4}(t),
			      \label{eq:SFU18}
			      \end{eqnarray}
			      and integrating the Friedmann equation we find
			      \begin{eqnarray}
			      a(t)=\left(\frac{32\pi G}{3} \rho_{r0} a^4_0\right)^{1/4}t^{1/2},
			      \label{eq:scalefof rad}
			      \end{eqnarray}
			      where the constant of integration has been chosen so that, $a(t=0)=0$.

			      Since $\rho_{r} a^4=$constant  
			      we can put $\rho_{r}a^4=\rho_{r0}a^4_0$ in Eq.\ (\ref{eq:scalefof rad}) to obtain
			      \begin{eqnarray}
			      \rho_{r}(t)=\frac{3}{32\pi G t^2},
			      \end{eqnarray}
			      for the density as a function of time.  
			      The age of the Universe can be computed from Eq.\ (\ref{eq:SFU15}) as
			      \begin{eqnarray}
			      t_0=\frac{1}{2H_0}. 
			      \label{eq:SFU19}
			      \end{eqnarray}
			      Also, the deceleration parameter for the spatially-flat relativistic-matter-dominated Universe is
			      \begin{eqnarray}
			      q=1.			     
			      \end{eqnarray}

\end{itemize}

\subsection{Multi-Component Universes}

Now let's consider different cases when the Universe is dominated by two energy density sources at a time. Our basic equations are Friedmann's equations,
so we start with the general form of Friedmann's equations,
\begin{eqnarray}
\left(\frac{\dot a}{a}\right)^2=\frac{8 \pi G}{3}\sum_{i}\rho_i-\frac{k}{a^2},
\label{eq:MCU1}
\end{eqnarray}
\begin{eqnarray}
\frac{\ddot a}{a}=-\frac{4 \pi G}{3}\sum_{i} \left(\rho_i+3P_i\right).
\label{eq:MCU2}
\end{eqnarray}
Here $\rho_i$ is the time-dependent matter (energy) density contribution of the $i^{th}$ component, and $P_i$ is the corresponding 
pressure exerted by that particular component. The equations of state are:
\begin{eqnarray}
P_i=P_i(\rho_i)=\omega_i \rho_i,
\label{eq:MCU3}
\end{eqnarray} 
where $\omega_i$ is dimensionless equation-of-state-parameter of the $i^{th}$ component of matter.

We know our Universe contains matter for which $\rho_m \propto a^{-3}$, and radiation for which $\rho_r \propto a^{-4}$. Current
evidence supports the presence of a cosmological constant with mass density $\rho_{\Lambda} =\rho_{\Lambda0}=\ $constant.
It is certainly possible that the Universe contains other components as well, but we will consider only the above-mentioned one.\footnote{In this Chapter, 
while we introduce the cosmological constant $\Lambda$ here, we will not consider solutions that involve constant or time-variable dark energy until Chapter (\ref{Chapter3}).}  
The relative density parameter for the $i^{th}$ component of the Universe is:
\begin{equation}
\Omega_{i}\equiv \frac{\rho_{i}}{\rho_{cr}}, \hspace{1cm} \rho_{cr}\equiv \frac{3H^2}{8 \pi G}.
\label{eq:MCU4}
\end{equation}
The present values of the density parameter of $i^{th}$ component of Universe is:
\begin{equation}
\Omega_{i0}\equiv \frac{\rho_{i0}}{\rho_{cr}},\ \ \ \ ~~~~~~\mathrm{so}\ \ \ \ ~~~~~~~\sum_{i} \Omega_{i0}=1.
\label{eq:MCU5}
\end{equation}
Now to be more specific let's say that $\Omega_{m0}$, $\Omega_{r0}$, 
$\Omega_{k0}$, and $\Omega_{\Lambda 0}$ are non-relativistic, relativistic, 
curvature, and cosmological constant density parameters at present time, respectively, then
\begin{equation}
\Omega_{m0}+\Omega_{r0}+\Omega_{k0}+\Omega_{\Lambda 0}=1,
\label{eq:MCU6}
\end{equation}
where
\begin{equation}
   \Omega_{m0} = \frac{8\pi G}{3H_0^2} \rho_{m0}, \ \
   \Omega_{r0} = \frac{8\pi G}{3H_0^2} \rho_{r0}, \ \
   \Omega_{k0}=\frac{-k}{(H_0 a_0)^2}, \ \ 
   \Omega_{\Lambda 0} = \frac{\Lambda}{3H_0^2}.
\label{eq:MCU7}
\end{equation}

Hence, the Friedmann's equation takes the form:
\begin{eqnarray}
\left(\frac{\dot a}{a}\right)^2&=&H^{2}_{0}\left[\Omega_{r0}a^{-4}+\Omega_{m0}a^{-3}+\Omega_{\Lambda 0}+\Omega_{k0}a^{-2}\right], \nonumber \\
&=&H^{2}_{0}\left[\Omega_{r0}a^{-4}+\Omega_{m0}a^{-3}+\Omega_{\Lambda 0}+(1-\Omega_{0})a^{-2}\right].
\label{eq:MCU8}
\end{eqnarray}
Here
\begin{eqnarray}
\Omega_0=\Omega_{r0}+\Omega_{m0}+\Omega_{\Lambda 0},
\label{eq:sum}
\end{eqnarray}
will determine the value of this source curvature. If $\Omega_0<1$ then $k=-1$, if $\Omega_0=1$ then
$k=0$, and if $\Omega_0>1$ then $k=+1$. Recent analyses of the observable data indicates that the space-curvature term 
$\frac{1-\Omega_0}{a^2}$ is probably small but not necessarily zero but very small.\cite{Komatsu2009,farooq5}

Let's consider the two special cases, the non-flat Universe dominated by non relativistic matter, and the non-flat Universe dominated by relativistic matter 
and compute the scale factor $a(t)$. We will see that, it is not possible to 
get the scale factor $a(t)$ as an explicit function of time $t$, and, we will instead derive parametric solutions.

\begin{itemize} 

       \item[\textbf {a)}]    \textbf{Curvature and Non-Relativistic-Matter-Dominated Universe (${k \neq 0}$, ${P=0}$, ${q_0 \neq 1/2}$)}

			      In this case (when $\Omega_{r0}=0$ and $\Omega_{\Lambda0}=0\ \mathrm{so} \ \Omega_{m0}=\Omega_{0}$) the Friedmann equation.\ (\ref{eq:MCU8}) takes the form:
			      \begin{eqnarray}
			      \left(\frac{\dot a}{a}\right)^2=H^{2}_{0}\left[\frac{\Omega_0}{a^3}+\frac{1-\Omega_0}{a^2}\right],
			      \label{eq:MCU9}
			      \end{eqnarray}
			      \begin{eqnarray}
			      \Rightarrow \frac{da}{dt}&=&H_{0}\left[\frac{\Omega_0}{a}+(1-\Omega_0)\right]^{1/2}, \nonumber \\
			      \Rightarrow H_0 t&=&\int \limits_{0}^{a}\frac{da'} {\left[\frac{\Omega_0}{a'}+(1-\Omega_0)\right]^{1/2}}.
			      \label{eq:MCU10}
			      \end{eqnarray}
			      There are two cases:

			      \textbf{Case 1. Open Universe (${\Omega_0<1}$) with ${k=-1}$, ${q_0<1/2}$}

			      With the substitution,
			      \begin{eqnarray}
			      a'(\eta)=\frac{\Omega_0}{2\left(1-\Omega_0\right)}\left(\mathrm{cosh}\eta-1\right), \ \ \
			      \ \  \ \ \mathrm{d}a'(\eta)=\frac{\Omega_0}{2\left(1-\Omega_0\right)}\mathrm{sinh}\eta\ \mathrm{d}\eta,
			      \label{eq:MCU11}
			      \end{eqnarray}
			      the indefinite integral in Eq.\ (\ref{eq:MCU10})becomes
			      \begin{eqnarray}
			      H_0 t&=&\frac{\Omega_0}{2\left(1-\Omega_0\right)^{3/2}} \int\frac{\mathrm{sinh}\eta} {\left[\frac{2}{\mathrm{cosh}\eta-1}+1\right]^{1/2}}\ \mathrm{d}\eta,  \nonumber \\
			      &=&\frac{\Omega_0}{2\left(1-\Omega_0\right)^{3/2}} \int\frac{\displaystyle \sqrt{\mathrm{cosh}\eta-1}\ \mathrm{sinh}\eta} {\displaystyle \sqrt{\mathrm{cosh}\eta+1}}\ \mathrm{d}\eta, \nonumber \\
			      &=&\frac{\Omega_0}{2\left(1-\Omega_0\right)^{3/2}} \int \left(\mathrm{cosh}\eta-1\right)\ \mathrm{d}\eta, \hspace{7 mm} \left(\because \mathrm{sinh}\eta=\displaystyle\sqrt{\left(\mathrm{cosh}\eta-1\right)\left(\mathrm{cosh}\eta+1\right)}\right)\nonumber \\
			      &=&\frac{\Omega_0}{2\left(1-\Omega_0\right)^{3/2}} \left(\mathrm{sinh}\eta-\eta\right)+ c_1.
			      \label{eq:MCU12}
			      \end{eqnarray}
			      Here $c_1$ is constant of integration. Using the initial condition $a(t=0)=0$ leads $c_1=0$.
			      Hence, the parametric dependence of the scale factor $a(t)$ on time in this Universe is
			      \begin{eqnarray}
			      a(\eta)&=&\frac{\Omega_0}{2\left(1-\Omega_0\right)} \left(\mathrm{cosh}\eta-1\right), \\
			      t(\eta)&=&\frac{\Omega_0}{2 H_0\left(1-\Omega_0\right)^{3/2}} \left(\mathrm{sinh}\eta-\eta\right).
			      \label{eq:MCU13}
			      \end{eqnarray}
			      In terms of the deceleration parameter $q_0$, using Eq.\ (\ref{eq:SFU13}) (which can be read as $q_0=\Omega_0/2$), these lead to:  
			      \begin{eqnarray}
			      a(\eta)&=&\frac{q_0}{\left(1-2q_0\right)} \left(\mathrm{cosh}\eta-1\right), \\
			      t(\eta)&=&\frac{q_0}{ H_0\left(1-2q_0\right)^{3/2}} \left(\mathrm{sinh}\eta-\eta\right).
			      \label{eq:MCU14}
			      \end{eqnarray}

			      \textbf{Case 2. Closed Universe (${\Omega_0>1}$) with ${k=1}$, ${q_0>1/2}$}

			      With the substitution,
			      \begin{eqnarray}
			      a'(\theta)=\frac{\Omega_0}{2\left(\Omega_0-1\right)}\left(1-\mathrm{cos}\theta\right), \ \ \
			      \ \ \ \ \mathrm{d}a'(\theta)=\frac{\Omega_0}{2\left(\Omega_0-1\right)}\mathrm{sin}\theta\ \mathrm{d}\theta
			      \label{eq:MCU15}
			      \end{eqnarray}
			      the indefinite integral in Eq.\ (\ref{eq:MCU10}) becomes
			      \begin{eqnarray}
			      H_0 t&=&\frac{\Omega_0}{2\left(\Omega_0-1\right)^{3/2}} \int\frac{\mathrm{sin}\theta} {\left[\frac{2}{1-\mathrm{cos}\theta}-1\right]^{1/2}}\ \mathrm{d}\theta, \nonumber \\
			      &=&\frac{\Omega_0}{2\left(\Omega_0-1\right)^{3/2}} \int\frac{\displaystyle \sqrt{1-\mathrm{cos}\theta}\ \mathrm{sin}\theta} {\displaystyle \sqrt{1+\mathrm{cos}\theta}}\ \mathrm{d}\theta, \nonumber \\
			      &=&\frac{\Omega_0}{2\left(\Omega_0-1\right)^{3/2}} \int \left(1-\mathrm{cos}\theta \right)\ \mathrm{d}\theta, \hspace{7 mm} \left(\because \mathrm{sin}\theta=\displaystyle\sqrt{\left(1-\mathrm{cos}\theta\right)\left(1+\mathrm{cos}\theta\right)}\right)\nonumber \\
			      &=&\frac{\Omega_0}{2\left(\Omega_0-1\right)^{3/2}} \left(\theta-\mathrm{sin}\theta\right)+ c_2.
			      \label{eq:MCU16}
			      \end{eqnarray}
			      Here $c_2$ is constant of integration. Using the initial condition $a(t=0)=0$ leads $c_2=0$.
			      Hence, the parametric dependence of the scale factor $a(t)$ on time in this Universe is
			      \begin{eqnarray}
			      a(\theta)&=&\frac{\Omega_0}{2\left(\Omega_0-1\right)} \left(1-\mathrm{cos}\theta\right), \\
			      t(\theta)&=&\frac{\Omega_0}{2 H_0\left(\Omega_0-1\right)^{3/2}} \left(\theta-\mathrm{sin}\theta\right).
			      \label{eq:MCU17}
			      \end{eqnarray}
			      In terms of the deceleration parameter $q_0$, we have
			      \begin{eqnarray}
			      a(\theta)&=&\frac{q_0}{\left(2q_0-1\right)} \left(1-\mathrm{cos}\theta\right), \\
			      t(\theta)&=&\frac{q_0}{ H_0\left(2q_0-1\right)^{3/2}} \left(\theta-\mathrm{sin}\theta\right).
			      \label{eq:MCU18}
			      \end{eqnarray}

       \item[\textbf {b)}]  \textbf{Curvature and Relativistic-Matter-Dominated Universe (${k \neq 0}$, ${P=\rho/3}$, ${q_0 \neq 1}$)}

In this case (when $\Omega_{m0}=0$ and $\Omega_{\Lambda0}=0\ \mathrm{so}\ \Omega_{r0}=\Omega_{0}$) the Friedmann Eq.\ (\ref{eq:MCU1}) takes the form:
\begin{eqnarray}
\left(\frac{\dot a}{a}\right)^2=H^{2}_{0}\left[\frac{\Omega_0}{a^4}+\frac{1-\Omega_0}{a^2}\right],
\label{eq:MCU19}
\end{eqnarray}
\begin{eqnarray}
\Rightarrow \frac{da}{dt}&=&H_{0}\left[\frac{\Omega_0}{a^2}+(1-\Omega_0)\right]^{1/2}, \nonumber \\
\Rightarrow H_0 t&=&\int \limits_{0}^{a}\frac{da'} {\left[\frac{\Omega_0}{a'^2}+(1-\Omega_0)\right]^{1/2}}.
\label{eq:MCU20}
\end{eqnarray}
There are two cases:

\textbf{Case 1. Open Universe ($\Omega_0<1$) with ${k=-1}$, ${q_0<1}$}

With the substitution,
\begin{eqnarray}
a'(\eta)=\sqrt{\frac{\Omega_0}{1-\Omega_0}}\ \mathrm{sinh}\eta, \ \ \
\ \ \ \ \ \mathrm{d}a'(\eta)=\sqrt{\frac{\Omega_0}{1-\Omega_0}}\ \mathrm{cosh}\eta\ \mathrm{d}\eta,
\label{eq:MCU21}
\end{eqnarray}
the indefinite integral in Eq.\ (\ref{eq:MCU20}) becomes
\begin{eqnarray}
H_0 t&=&\frac{\sqrt\Omega_0}{1-\Omega_0} \int\frac{\mathrm{cosh}\eta} {\left[\frac{1}{\mathrm{sinh^2}\eta}+1\right]^{1/2}}\ \mathrm{d}\eta, \nonumber \\
\ \ \ \ &=&\frac{\sqrt\Omega_0}{1-\Omega_0} \mathrm{cosh}\eta+c_3,
\label{eq:MCU22}
\end{eqnarray}
where $c_3$ is constant of integration. Using the initial condition $a(t=0)=0$ leads $c_3=-\frac{\sqrt{\Omega_{0}}}{1-\Omega_{0}}$.
Hence, the parametric dependence of the scale factor $a(t)$ on time in this Universe is
\begin{eqnarray}
a(\eta)&=&\sqrt{\frac{\Omega_0}{1-\Omega_0}}\ \mathrm{sinh}\eta \\
t(\eta)&=&\frac{\sqrt{\Omega_0}}{H_0\left(1-\Omega_0\right)} \left(\mathrm{cosh}\eta-1\right). 
\label{eq:MCU23}
\end{eqnarray}
In terms of the deceleration parameter $q_0$, using Eq.\ (\ref{eq:SFU13}) (which can be read as $q_0=\Omega_{r0}=\Omega_0$), these lead to  
\begin{eqnarray}
a(\eta)&=&\sqrt{\frac{q_0}{1-q_0}} \left(\mathrm{sinh}\eta\right), \\
t(\eta)&=&\frac{\sqrt{q_0}}{ H_0\left(1-q_0\right)} \left(\mathrm{cosh}\eta-1\right).
\label{eq:MCU24}
\end{eqnarray}
\\

\textbf{Case 2: Closed Universe (${\Omega_0>1}$) with ${k=1}$, ${q_0>1}$ :-}

With the substitution,
\begin{eqnarray}
a'(\theta)=\sqrt{\frac{\Omega_0}{\Omega_0-1}}\ \mathrm{sin}\theta, \ \ \
\ \  \ \ \mathrm{d}a'(\theta)=\sqrt{\frac{\Omega_0}{\Omega_0-1}}\ \mathrm{cos}\theta\ \mathrm{d}\theta.
\label{eq:MCU25}
\end{eqnarray}
the indefinite integral in Eq.\ (\ref{eq:MCU10}) becomes
\begin{eqnarray}
H_0 t&=&\frac{\sqrt\Omega_0}{\Omega_0-1} \int\frac{\mathrm{cos}\theta} {\left[\frac{1}{\mathrm{sin^2}\theta}+1\right]^{1/2}}\ \mathrm{d}\theta, \nonumber \\
\ \\ \ \ &=&-\frac{\sqrt\Omega_0}{\left(\Omega_0-1\right)} \mathrm{cos}\theta+c_4,
\label{eq:MCU26}
\end{eqnarray}
where $c_4$ is constant of integration. Using the initial condition $a(t=0)=0$ leads to $c_4=\frac{\sqrt{\Omega_{0}}}{\Omega_{0}-1}$.
Hence, the parametric dependence of the scale factor $a(t)$ on time in this Universe is
\begin{eqnarray}
a(\theta)&=&\sqrt{\frac{\Omega_0}{\Omega_0-1}}\ \mathrm{sin}\theta, \\
t(\theta)&=&\frac{\sqrt\Omega_0}{H_0\left(\Omega_0-1\right)} \left(1-\mathrm{cos}\theta \right).
\label{eq:MCU27}
\end{eqnarray}
In terms of the deceleration parameter $q_0$, we have
\begin{eqnarray}
a(\theta)&=&\sqrt{\frac{q_0}{q_0-1}}\ \mathrm{sin}\theta, \\
t(\theta)&=&\frac{\sqrt{q_0}}{H_0\left(q_0-1\right)} \left(1-\mathrm{cos}\theta\right).
\label{eq:MCU28}
\end{eqnarray}

\end{itemize}

\begin{figure}[h]
\centering
    \includegraphics[height=5.5in]{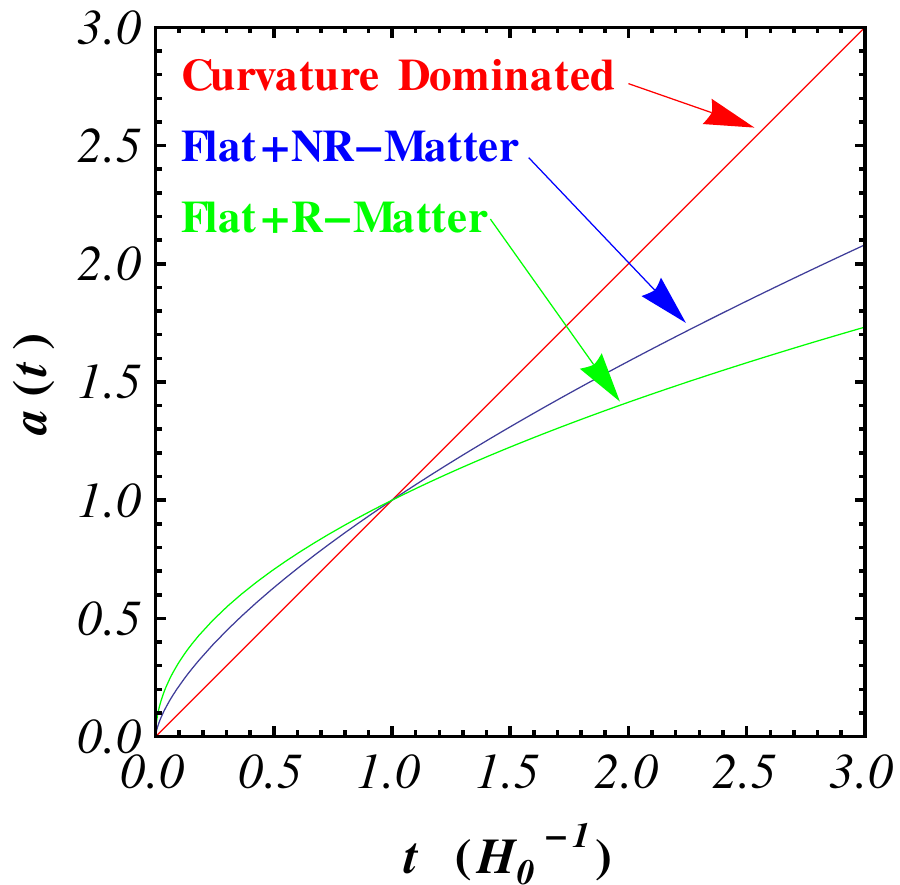}
\caption{This figure shows the evolution of scale factor $a(t)$  as a function of cosmic time (in units of Hubble time) of single component Universes. The red line is for the Universe that is only curvature dominated (remember that according to Friedmann's equation, a Universe with curvature only should be open, so $k=-1$) and scale factor behaves like $a(t)\propto t$. The scale factor as a function of time for the non-relativistic (NR) matter dominated Universe is shown as the blue curve. Mathematically $a(t)\propto t^{2/3}$. The relativistic (R) matter dominated Universe will expand as $a(t)\propto t^{1/2}$ and is shown as the green curve. All three Universes expand forever with no finite maximum value of scale factor.}
\label{fig:1.5}
\end{figure}

\begin{figure}[h]
\centering
    \includegraphics[height=5.5in]{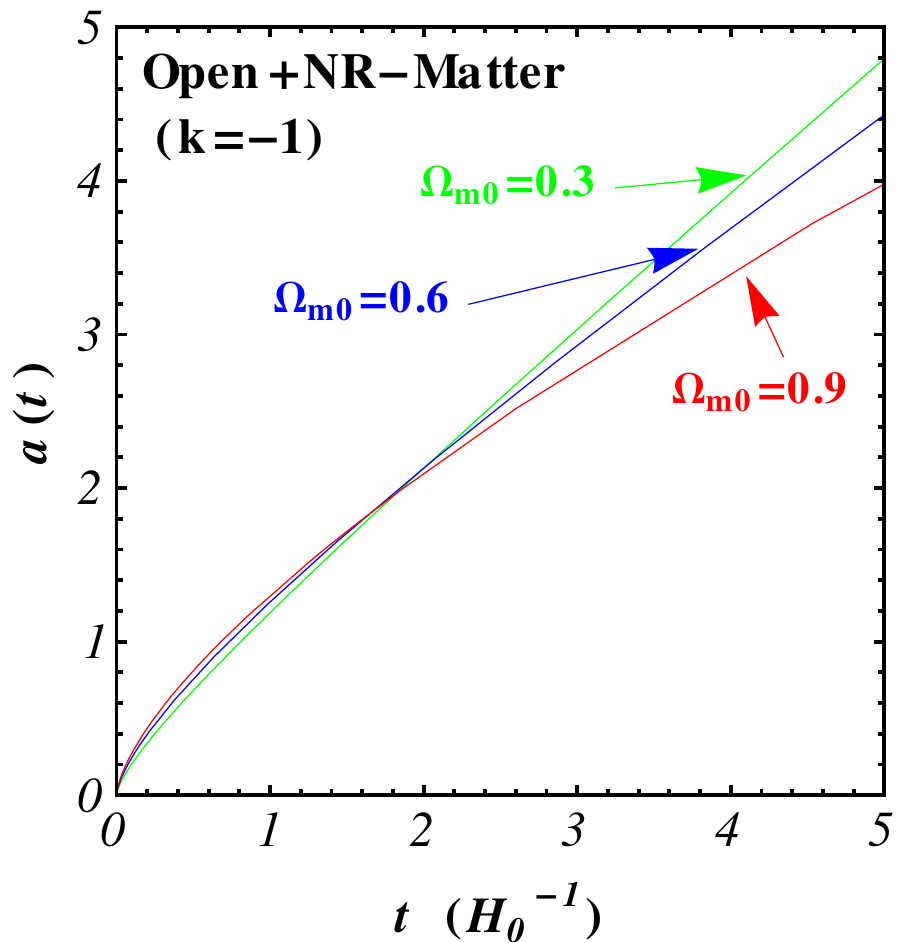}
\caption{This figure shows the evolution of scale factor $a(t)$ as a function of cosmic time (in units of Hubble time) of open, non-relativistic (NR)  matter dominated Universes, for three different values of $\Omega_{m0}$. It is clear from the plot that a larger value of $\Omega_{m0}$ decreases the rate of expansion of the Universe, but if $\Omega_{m0}<1$ the Universes is not only open, but will expand forever. All three Universes in this case will expand forever with no finite maximum values of scale factor. $\Omega_{m0}=0.3$ is the most realistic model of our Universe among the three models presented in this figure.}
\label{fig:1.6}
\end{figure}

\begin{figure}[h]
\centering
    \includegraphics[height=5.5in]{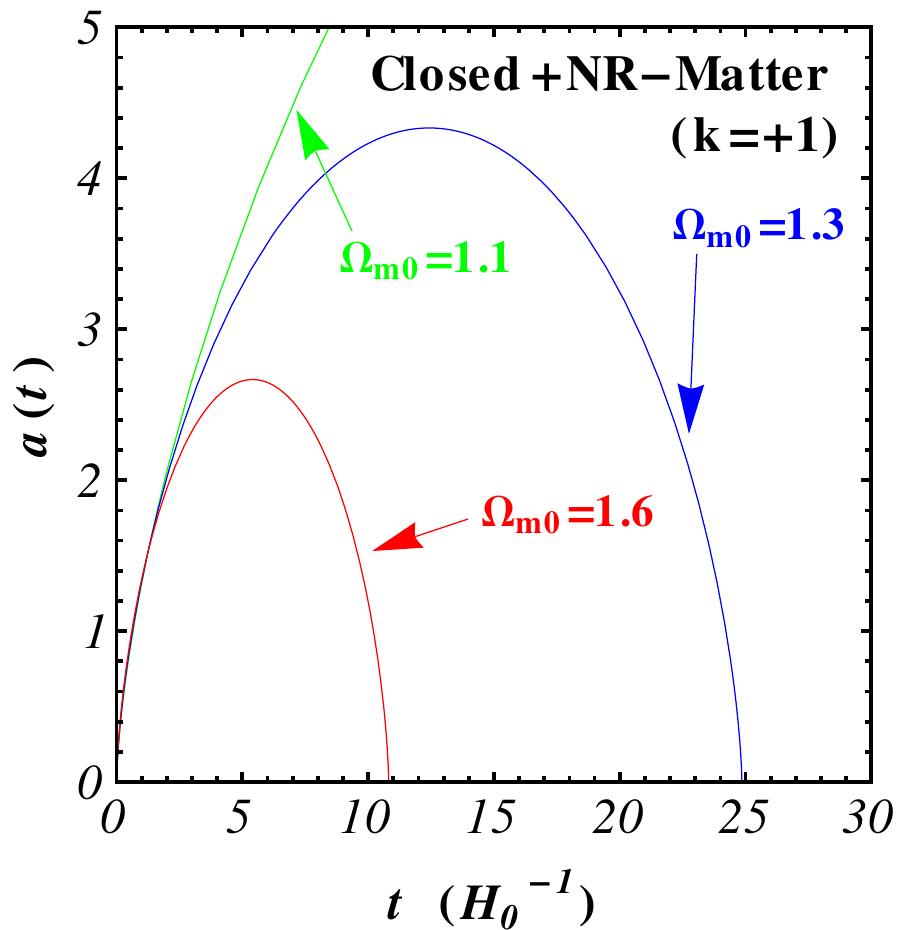}
\caption{This figure shows the evolution of scale factor $a(t)$ as a function of cosmic time (in units of Hubble time) of closed, non-relativistic (NR) matter dominated Universes, for three different values of $\Omega_{m0}$. It is clear from the plots that a larger value of $\Omega_{m0}$ decreases the rate of expansion of the Universe. These Universes reach to a maximum scale factor of $a=a_{\mathrm{max}}=\frac{\Omega_{m0}}{2\left(\Omega_{m0}-1\right)}$ and then collapse to a big crunch. The larger the matter density, the faster the Universe will collapse (no surprise since collapse is due to the gravitational attraction between the matter in the Universe). All three Universes will eventually recollapse. The green curve that corresponds to the Universe having $\Omega_{m0}=1.1$, appears to indicate that it will continue expanding forever but if we increase the $t$ range of the plot, it too will come back to a big crunch, like the red and blue curves, but after a longer time.}
\label{fig:1.7}
\end{figure}

\begin{figure}[h]
\centering
    \includegraphics[height=5.5in]{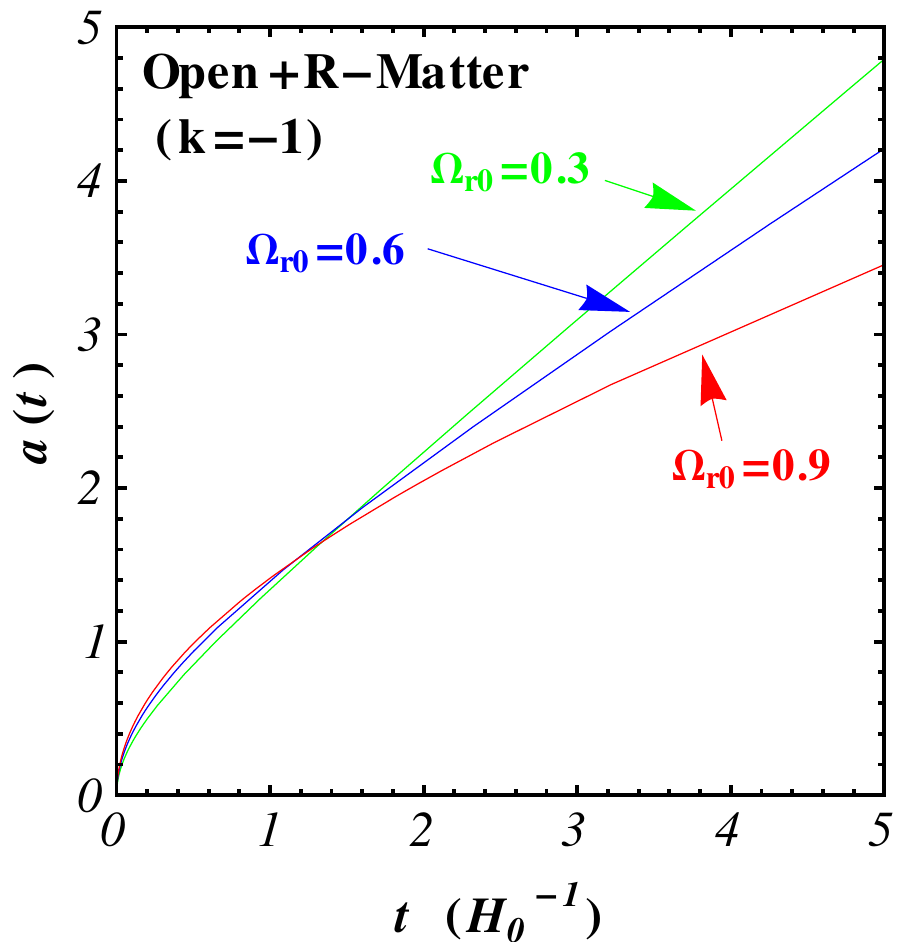}
\caption{This figure shows the evolution of scale factor $a(t)$ as a function of cosmic time (in units of Hubble time) of open, relativistic (R) matter dominated Universes, for three different values of $\Omega_{r0}$. It is clear that a larger value of $\Omega_{r0}$ decreases the rate of expansion of the Universe, more than for the non-relativistic matter dominated case [see Fig.\ (\ref{fig:1.6})], but if $\Omega_{r0}<1$ the Universe is not only open, but will expand forever. All three Universe in this case will expand forever with no finite value of scale factor.}
\label{fig:1.8}
\end{figure}

\begin{figure}[h]
\centering
    \includegraphics[height=5.5in]{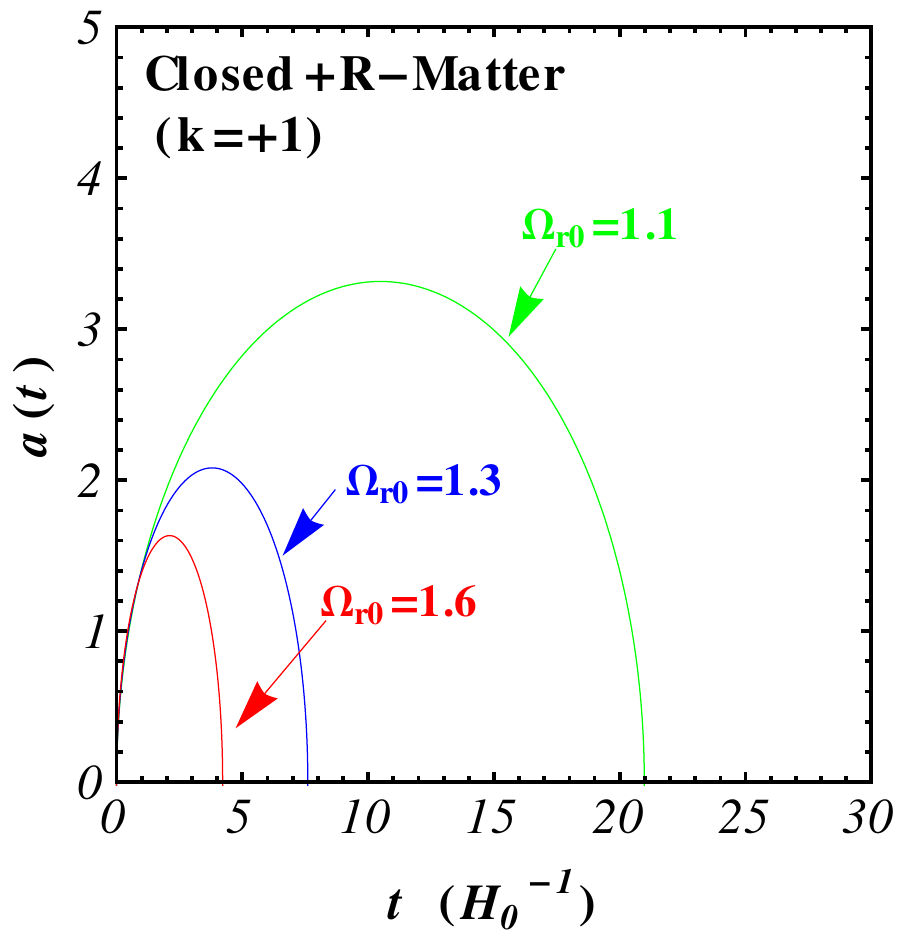}
\caption{This figure shows the evolution of scale factor $a(t)$ as a function of cosmic time (in units of Hubble time) of closed, relativistic (R) matter dominated Universes, for three different values of $\Omega_{r0}$. It is clear that a larger value of $\Omega_{r0}$ decreases the rate of expansion of the Universe. These Universes reach maximum scale factor of $a=a_{\mathrm{max}}=\sqrt{\frac{\Omega_{r0}}{\Omega_{r0}-1}}$ and then collapse to a big crunch. The larger the radiation density faster the Universe will collapse. All three Universes will eventually recollapse.}
\label{fig:1.9}
\end{figure}

A summary of the scale factor $a(t)$ for the various models is given in the Table\ (\ref{table:Solution of Friedman}), and the plots in Fig.\ (\ref{fig:1.5})---(\ref{fig:1.9}) illustrate the evolution of the scale factor as a function of time for the different cases.

\begin{threeparttable}
\centering
\begin{tabular}{|p{5cm}|p{10cm}|}
\hline \hline 
\hspace{1.5 cm}\textbf{Universe} & \hspace{3.5 cm}\textbf{Properties} \\
\hline\hline
\vspace{0.5 mm} 
Curvature-dominated & \vspace{0.5 mm} $k \neq 0$, (only negative curvature possible),\ \  $\rho_i=0$ $\forall\ i$, \\
Universe \vspace{0.5 mm} & $q_0=0$,\ \ $a(t) \propto t$,\ \  $t_0=H^{-1}_{0}$,\ \  $\ddot a(t)=0$ $\forall\ t$.\vspace{0.5 mm} \\
\hline
\vspace{0.5 mm} 
Spatially-flat Universe with single component(equation-of-state-parameter $\omega$)  &  \vspace{0.5 mm} $k=0$, $\rho \neq 0$,\ \ $q_0=\frac{1}{2}$,\ \ 
$a(t)\propto t^{2/3(1+\omega)}$,\ \ $t_0=\frac{2}{3(1+\omega)}H^{-1}_{0}$, $\ddot a(t)<0$ (we are ignoring dark energy),\ $q_0=\frac{1}{2}(1+3\omega)$\\
\vspace{0.5 mm} 
a) Non-relativistic-matter dominated & \vspace{0.5 mm} $P=0$,\hspace{0.2 cm} $\omega=0$,\hspace{0.2 cm} $a(t)\propto t^{2/3}$,\hspace{0.2 cm} $t_0=\frac{2}{3}H^{-1}_{0}$,\ $q_0=\frac{1}{2}$\\
b) Relativistic-matter-dominated \vspace{0.5 mm} & $P=\frac{1}{3}\rho$,\hspace{0.2 cm} $\omega=\frac{1}{3}$,\hspace{0.2 cm} $a(t)\propto t^{1/2}$,\hspace{0.2 cm} $t_0=\frac{1}{2}H^{-1}_{0}$,\ $q_0=\frac{1}{2}$ \\
\hline
\multicolumn{2}{|c|}{\textbf{Two-component Universes}\tnote{a}} \\
\hline
\vspace{0.5 mm}
a) Open $+$ non-relativistic-matter-dominated & \vspace{0.5 mm}\hspace{1 cm} $k=-1$,\hspace{0.2 cm} $P=0$,\hspace{0.2 cm} $\Omega_0<1$,\hspace{0.2 cm} $q_0<\frac{1}{2}$.\\
   & \hspace{1.9 cm}$a(\eta)=\frac{\Omega_0}{2(1-\Omega_0)}\left(\mathrm{cosh}\eta-1\right)$, \\

\vspace{0.5 mm}    & \vspace{0.5 mm}\hspace{1.8 cm} $t(\eta)=\frac{\Omega_0}{2H_0(1-\Omega_0)^{3/2}}\left(\mathrm{sinh}\eta-\eta\right)$.\\
\vspace{0.5 mm} 
b) Closed $+$ non-Relativistic-matter-dominated \vspace{0.5 mm} & \vspace{1 mm}\hspace{1 cm} $k=1$,\hspace{0.2 cm} $P=0$,\hspace{0.2 cm} $\Omega_0>1$,\hspace{0.2 cm} $q_0>\frac{1}{2}$.\\
   & \hspace{1.9 cm}$a(\theta)=\frac{\Omega_0}{2(\Omega_0-1)}\left(1-\mathrm{cos}\theta\right)$, \\

\vspace{0.5 mm}    & \vspace{0.5 mm}\hspace{1.8 cm} $t(\theta)=\frac{\Omega_0}{2H_0(\Omega_0-1)^{3/2}}\left(\theta-\mathrm{sin}\theta\right)$.\vspace{2 mm}\\
\hline
\vspace{0.5 mm} 
c) Open $+$ relativistic-matter-dominated & \vspace{0.5 mm}\hspace{1 cm} $k=-1$,\hspace{0.2 cm} $P=\frac{1}{3}\rho$,\hspace{0.2 cm} $\Omega_0<1$,\hspace{0.2 cm} $q_0<1$.\\
   & \hspace{1.9 cm}$a(\eta)=\sqrt{\frac{\Omega_0}{1-\Omega_0}}\ \mathrm{sinh}\eta$, \\

\vspace{0.5 mm}    & \vspace{0.5 mm}\hspace{1.8 cm} $t(\eta)=\frac{\sqrt{\Omega_0}}{H_0(1-\Omega_0)}\left(\mathrm{cosh}\eta-1\right)$.\\
\vspace{0.5 mm} 
d) Closed $+$ relativistic-matter-dominated \vspace{0.5 mm} & \vspace{0.5 mm}\hspace{1 cm} $k=1$,\hspace{0.2 cm} $P=\frac{1}{3}\rho$,\hspace{0.2 cm} $\Omega_0>1$,\hspace{0.2 cm} $q_0>1$.\\
   & \hspace{1.9 cm}$a(\theta)=\sqrt{\frac{\Omega_0}{\Omega_0-1}}\ \mathrm{sin}\theta$, \\

\vspace{0.5 mm}    & \vspace{0.5 mm}\hspace{1.8 cm} $t(\theta)=\frac{\sqrt{\Omega_0}}{H_0(\Omega_0-1)}\left(1-\mathrm{cos}\theta\right)$.\vspace{2 mm}\\
\hline\hline
\end{tabular}
\caption{\rm{Solutions of Friedmann's equations in various cases (with out dark energy)}.}
\label{table:Solution of Friedman}
\begin{tablenotes}
\item[a]{Special thanks to Sara Crandall and Max Goering, for checking these results.}
\end{tablenotes}
\end{threeparttable}

\cleardoublepage


\chapter{Distance Measures in Cosmology}
\label{Chapter2}

In order to discuss and derive observational constraints on dark energy, it is important to
introduce distance measures that are directly related to observations in the FLRW spacetime of Eq.\ (\ref{eq:FRLWM}).

In cosmology there are different ways to specify the distances between two points in space, is because different techniques can be used to define and measure the distance between two points! In addition in the expanding Universe the physical distance between co-moving
objects changes and Earth-based observers look back in time as they look deep
into space. We will define several different kinds of distances here. The commonality between all these different distances is that they are the measure 
of the separation between events on the \textit{radial null} trajectories i.e., the trajectories of photons which terminate at the observer. In fact, a large part of the 
evidence for the existence of dark energy\footnote{We will discuss some of, evidence for dark energy in detail in Chapter\ (\ref{Chapter3}).} comes from the measurements of cosmological distances. Let's start with the 4-dimensional
\begin{eqnarray}
ds^2 &=& g_{\mu \nu}dx^{\mu}dx^{\nu}, \nonumber \\
&=&-\mathrm{d}t^2+a^2(t)\mathrm{d}\sigma^2, \nonumber \\
&=&-\mathrm{d}t^2+a^2(t)\left[\mathrm{d}r^2+r^2\mathrm{d}\Omega^2\right], \nonumber \\
&=&-\mathrm{d}t^2+a^2(t)\left[\mathrm{d}r^2+r^2\mathrm{d}\theta^2+r^2\mathrm{sin^2}\theta \mathrm{d}\phi^2\right],
\label{eq:interval}
\end{eqnarray}
where d$\sigma$ is the comoving spatial line element.
\section{Comoving or Coordinate Distance}

\begin{figure}[h]
\centering
    \includegraphics[height=4.5in]{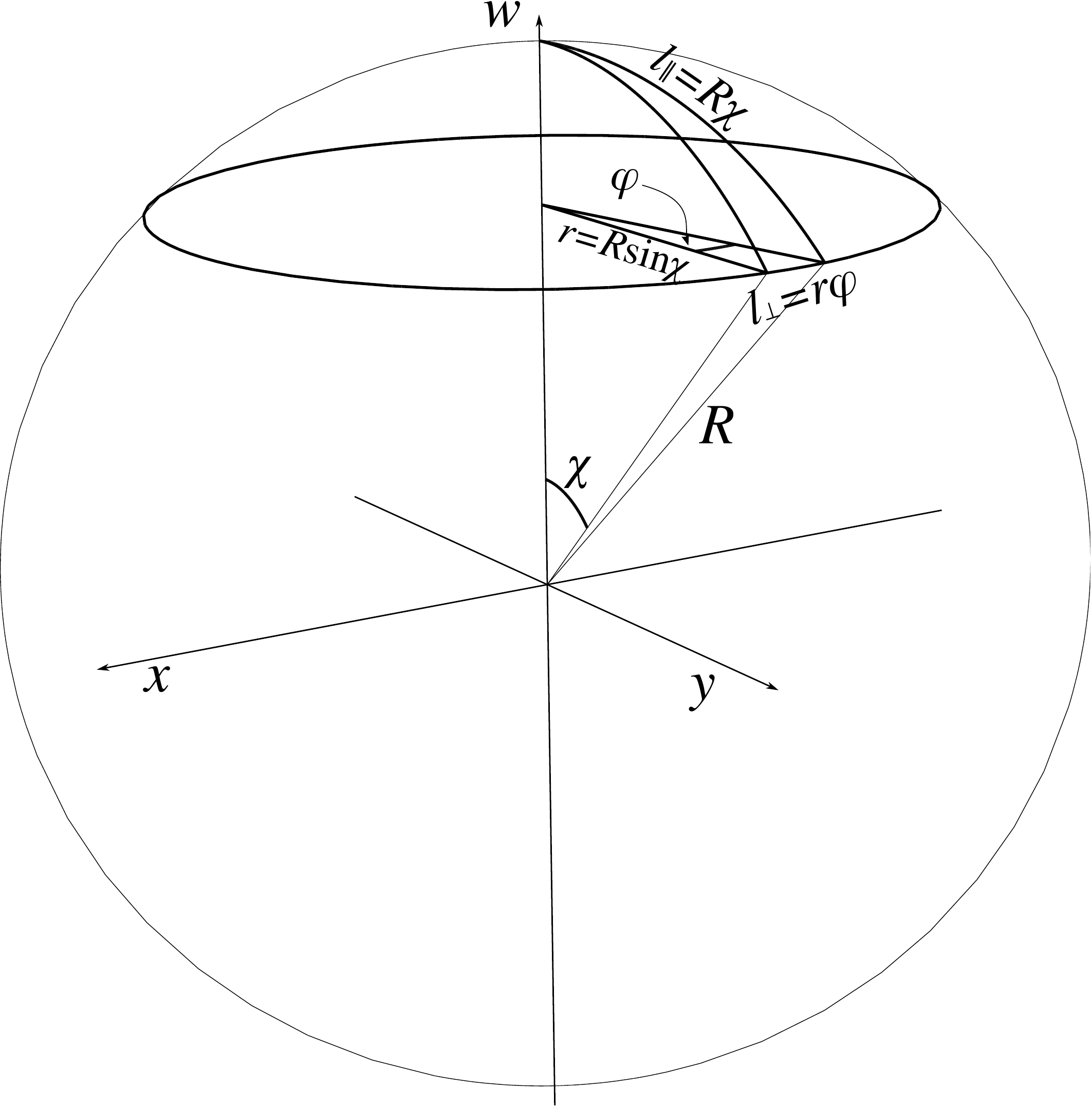}
    \caption{Two-dimensional FLRW geometry. This figure is taken from the Master's thesis of Data Mania.\cite{datathesis}}
    \label{fig:distance}
\end{figure}

Let's first discuss the comoving or coordinate distance which we denote by d$_{co}$. The co-moving distance d$_{co}$ 
between two neighboring objects in the cosmos is the distance between them that remain constant with epoch
if the two objects are moving only with the Hubble flow.\footnote{The motion of astronomical objects solely due to the 
expansion of Universe is known as the Hubble flow.} This is the intuitive way of defining distance, but it is not directly
measurable. In simple language, it is the distance between the two objects 
measured by a ruler at the time they were simultaneously observed (called proper or physical distance), divided 
by the ratio of the scale factor of the Universe then to now. So it is proper distance multiplied by $a_0/a=(1+z)$. The
total line-of-sight comoving distance along a null geodesic can be found by setting $\mathrm{d}\Omega=0$, and
$\mathrm{d}s=0$ in Eq.\ (\ref{eq:interval}), then
\begin{eqnarray}
\mathrm{d}_{co}(z)=R\ \chi=\int\limits^{t_0}_{t_e}\frac{\mathrm{d}t'}{a(t')}=\int\limits^{a_0}_{a_e}\frac{\mathrm{d}a}{a\dot a}=\frac{1}{a_0H_0}\int\limits^{z}_{0}\frac{\mathrm{d}z'}{E(z')},
\label{eq:comoving}
\end{eqnarray}
where $t_e$ and $a_e$ is the cosmological time and scale factor at the time of emission of the photon from the source, similarly
$t_0$ and $a_0$ are the values of cosmological time and scale factor at the time of observation and 
$z_e$ is the redshift of the source at the time of emission of the photon. According to convention 
$a_0=1$. $E(z)\equiv H(z)/H_0$ is the dimensionless Hubble 
parameter and $R\chi$, shown in the Fig.\ ({\ref{fig:distance}}), is a comoving distance. $dt'$ is the physical distance
(remember $c$=1) traveled by photon in time $dt'$, but dividing the physical distance by the scale factor $a(t')$ we get 
the comoving distance, therefore $R \chi$ can be interpreted as the total, integrated comoving distance between the emitter and the 
observer. If the space is flat, this comoving distance is just the difference in the radial coordinates.
Although it is integrated over time for any two distinct galaxies moving with the Hubble flow,
its value remains constant with time. It is just  like the labels on a stretchable ruler: while the 
distance between two marks increases when the ruler is stretched,the labels remain the same.

The function $\int\limits^{z_e}_{0}\frac{\mathrm{d}z}{E(z)}$, can be expanded around $z_e=0$,\footnote{
\vspace{-0.5cm}
\begin{eqnarray}
&&f(z_e)=\int\limits^{z_e}_{0}\frac{\mathrm{d}z}{E(z)}=f(0)+f'(0)z_e+\frac{f''(0)}{2!}z_e^2+\frac{f'''(0)}{3!}z_e^3+\frac{f^{\mathrm{(iv)}}(0)}{4!}z_e^2+\cdots, \\
&&f(0)=0, \hspace{0.5cm} f'(z_e)=\frac{1}{E(z_e)} \hspace{0.25cm}\Rightarrow \hspace{0.25cm} f'(0)=\frac{1}{E(0)}=\frac{1}{1}=1,  \nonumber \\
&&f''(z_e)=-\frac{E'(z_e)}{E^2(z_e)}\hspace{0.25cm}\Rightarrow \hspace{0.25cm} f''(0)=-\frac{E'(0)}{E^2(0)}=-E'(0), \nonumber \\
&&f'''(z_e)=-\frac{E^2(z_e)E''(z_e)-2E'^2(z_e)E(z_e)}{E^4(z_e)}\hspace{0.25cm}\Rightarrow \hspace{0.25cm} f'''(0)=-\frac{E''(0)-2E'^2(0)}{1}=2E'^2(0)-E''(0).\nonumber
\end{eqnarray}
}
\begin{eqnarray}
\int\limits^{z_e}_{0}\frac{\mathrm{d}z}{E(z)}\ \approx\ z_e-\frac{E'(0)}{2}z_e^2+\frac{1}{6}\left[2E'^2(0)+E''(0)\right]z_e^3+ O(z_e^4)+\cdots,
\label{eq:comoving2}
\end{eqnarray}
where a prime represents a derivative with respect to $z$. 

For redshift $z$ much smaller than unity the comoving distance is approximately given from Eq.\ (\ref{eq:comoving}),
\begin{eqnarray}
\mathrm{d}_{co}(z_e) \approx \frac{1}{a_0 H_0}z_e, \hspace{2 cm} (\mathrm{for}\ z_e \ll 1)
\label{eq:comoving3}
\end{eqnarray}
Since $z \approx v$,\footnote{For low redshift $z=\frac{v}{c}$ but in our convention $c=1$.} for a very small $z$, where $v$ is the speed of the source, we find
\begin{eqnarray}
\mathrm{d}_{co} \approx \frac{1}{a_0 H_0}v,
\label{eq:comoving3}
\end{eqnarray}
which is the Hubble law.

\subsection{Some Insight: Start with 2-Dimensional Expansion}

\begin{figure}[h!]
\centering
    \includegraphics[height=3.0 in]{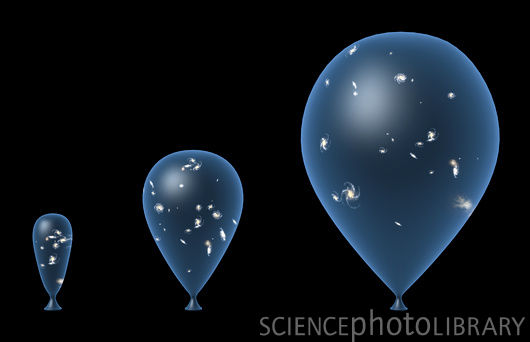}
    \caption{Hubble expansion in two spatial dimensions.}
    \label{fig:balloon}
\end{figure}

Let's think about the homogeneous, isotropic expansion of a Universe with two 
spatial dimension (e.g., a 2-sphere embedded in three dimensional Euclidean space). We place dots 
(galaxies) on a balloon and allow it to expand see Fig.\ (\ref{fig:balloon}).

\begin{enumerate}

       \item The spherical coordinates $(\theta,\phi)$ remain the same but the distance between points on the balloon changes in proportional to the scale factor $a(t)$. For concreteness think of the surface of earth, defined by latitudes and longitudes. Expand the size of globe (of course in imagination) by a factor of 2. The actual distance between two cities also increases by a factor of 2, but the $(\theta,\phi)$ coordinates of the two cities will remain unchanged.

       \item An observer attached to a comoving coordinate (comoving or fundamental observer) sees all other points receding from him/her and sees a homogeneous, isotropic Universe.

       \item An observer not comoving does not see an isotropic Universe.

       \item The receding points maintain their comoving coordinates.

\end{enumerate} 

\subsubsection{Some misconceptions}
Although the balloon analogy is useful, one must guard against misconceptions that it can generate.
\begin{enumerate}

       \item The surface that is expanding is two dimensional; the ``center" of the balloon is in the third dimension and is not part of the surface, which has no center.

       \item The balloon is expanded by the pressure difference between the inside and the outside, but the Universe is not being expanded by pressure. 

       \item Pressure couples to gravity in the Einstein equation, so the addition of (positive) pressure to the Universe would slow, not increase, the expansion rate see Eq.\ (\ref{eq:accelerationeq}).

       \item If the dots on the balloon represent galaxies, they too will expand. But real galaxies do not expand due to general Hubble expansion because they are gravitationally bound objects. We can make a better analogy by gluing solid objects (like 10 cent coins) to the surface of the balloon to represent galaxies, so that they do not expand when the balloon expands.
\end{enumerate} 

\subsection{3-Dimensional Expansion}

Now we can extend our understanding to 3-dimensional space expansion.
\begin{enumerate}

       \item The coordinates $(r,\theta, \phi)$ of Friedmann-Lema\^{i}tre-Robertson-Walker metric are comoving coordinates.

       \item As the Universe expands (to the degree that peculiar motion relative to the Hubble flow can be ignored) comoving observer keeps the same coordinates $(r,\theta, \phi)$, and only the FLRW scale factor $a(t)$ changes with time.

       \item Galaxies recede from us, but, if we are comoving observers, the receding galaxies maintain their comoving coordinates and the recession is described entirely by the time dependence of the scale factor $a(t)$.

       \item Peculiar velocities will change the comoving coordinates, but these effects are small on large scales where the cosmological principle and therefore the FLRW metric is valid.
\end{enumerate} 

In some sense the comoving distance is the fundamental distance measure in cosmology, all others are expressed in terms of it.

\section{Physical Distance}

It is the actual proper distance, denoted by $r_p$, between the two objects in the cosmos that can be measured by a physical ruler.
It is related to the co-moving distance d$_{co}(z)$ through
\begin{eqnarray}
r_p(z)=a(t)\mathrm{d}_{co}(z),
\label{eq:pd1}
\end{eqnarray} 
where $a(t)$ is the scale factor of the Universe. At the present epoch
\begin{eqnarray}
r_p(z=0)=a(t_0)\mathrm{d}_{co}(z).
\label{eq:pd2}
\end{eqnarray} 
Since $z \approx v$ for small values of $z$ and using Eq.\ (\ref{eq:comoving3}), we have
\begin{eqnarray}
r_p(z) \approx \frac{1}{H_0}v, \hspace{2 cm} \mathrm{or}\ v\approx H_0r.
\label{eq:pd3}
\end{eqnarray}
which means that Hubble's law is satisfied. $v \propto z$ is only valid at small $z$. It is just a consequence of uniform expansion Hubble's law  that $v=H_0 r$ is exactly always valid.

\section{Transverse Comoving Distance}

The comoving distance between two events at the same redshift or distance, but separated on the sky by some small angle $\phi$ 
is $r \phi$ where the transverse coordinate or comoving distance is denoted by $r$. In Fig.\ (\ref{fig:distance}) it is $r=R\mathrm{sin}\chi$, so considering Eq.\ (\ref{eq:comoving}) we have
\begin{equation}
r(z)= \left\{
      \begin{array}{lc}
           \frac{1}{\sqrt{k}}\mathrm{sin}\left(\frac{\sqrt{k}}{a_0 H_0}\int\limits^{z}_{0}\frac{dz'}{E(z')}\right) & \ \ \ \ \mathrm{for}\ k>0 \\
           \frac{1}{a_0 H_0}\int\limits^{z}_{0}\frac{dz'}{E(z')} & \ \ \ \ \mathrm{for}\ k=0 \\
           \frac{1}{\sqrt{-k}}\mathrm{sinh}\left(\frac{\sqrt{-k}}{a_0 H_0}\int\limits^{z}_{0}\frac{dz'}{E(z')}\right) & \ \ \ \ \mathrm{for}\ k<0 \\
     \end{array},
   \right.
\label{eq:coordinatedistance1}
\end{equation}
where the trigonometric functions account for the curvature of space. 
Space curvature is not coordinate independent; a change of coordinates can make space flat. The
only coordinate independent curvature is spacetime curvature, which is related to the local mass-energy density 
or really the stress-energy tensor.\cite{Hogg:1999ad} 
In terms of the curvature density parameter $\Omega_{k0}$, the coordinate distance can be written as
\begin{equation}
r(z)= \left\{
      \begin{array}{lc}
           \frac{1}{a_0 H_0\sqrt{\Omega_{k0}}}\mathrm{sinh}\left(\sqrt{\Omega_{k0}}\int\limits^{z}_{0}\frac{dz'}{E(z')}\right) & \ \ \ \ \mathrm{for}\ \Omega_{k0}>0 \\
           \frac{1}{a_0 H_0}\int\limits^{z}_{0}\frac{dz'}{E(z')} & \ \ \ \ \mathrm{for}\ \Omega_{k0}=0 \\
           \frac{1}{a_0 H_0\sqrt{-\Omega_{k0}}}\mathrm{sin}\left(\sqrt{-\Omega_{k0}}\int\limits^{z}_{0}\frac{dz'}{E(z')}\right) & \ \ \ \ \mathrm{for}\ \Omega_{k0}<0 \\
     \end{array}.
   \right.
\label{eq:coordinatedistance3}
\end{equation}

For $\Omega_\Lambda=0$, there is an analytical expression for the
coordinate distance:\cite{peebles1993principles, weinberg1972gravitation}
\begin{eqnarray}
r(z)=\frac{2\left[2-\Omega_{m0}(1-z)-(2-\Omega_{m0})\sqrt{1+\Omega_{m0}z}\ \right]}{a_0H_0\Omega^{2}_{m0}(1+z)}.
\label{eq:coordinatedistance2}
\end{eqnarray}

We stress here that coordinate distance $\boldsymbol r$ is not the proper distance --- it is 
the proper distance divided by the ratio of the scale factor.

\section{Dimensionless Coordinate Distance}

It is useful to define the dimensionless coordinate distance denoted by $y(z)\equiv a_0H_0d_m(z)$,
\begin{equation}
y(z)= \left\{
      \begin{array}{lc}
           \frac{a_0H_0}{\sqrt{k}}\mathrm{sin}\left(\frac{\sqrt{k}}{a_0 H_0}\int\limits^{z}_{0}\frac{dz'}{E(z')}\right) & \ \ \ \ \mathrm{for}\ k>0 \\
           \int\limits^{z}_{0}\frac{dz'}{E(z')} & \ \ \ \ \mathrm{for}\ k=0 \\
           \frac{a_0H_0}{\sqrt{-k}}\mathrm{sinh}\left(\frac{\sqrt{-k}}{a_0 H_0}\int\limits^{z}_{0}\frac{dz'}{E(z')}\right) & \ \ \ \ \mathrm{for}\ k<0 \\
     \end{array}.
   \right.
\label{eq:dlcd1}
\end{equation}
In terms of the curvature density parameter $\Omega_{k0}$, it is given by
\begin{equation}
y(z)= \left\{
      \begin{array}{lc}
           \frac{1}{\sqrt{\Omega_{k0}}}\mathrm{sinh}\left(\sqrt{\Omega_{k0}}\int\limits^{z}_{0}\frac{dz'}{E(z')}\right) & \ \ \ \ \mathrm{for}\ \Omega_{k0}>0 \\
           \int\limits^{z}_{0}\frac{dz'}{E(z')} & \ \ \ \ \mathrm{for}\ \Omega_{k0}=0 \\
           \frac{1}{\sqrt{-\Omega_{k0}}}\mathrm{sin}\left(\sqrt{-\Omega_{k0}}\int\limits^{z}_{0}\frac{dz'}{E(z')}\right) & \ \ \ \ \mathrm{for}\ \Omega_{k0}<0 \\
     \end{array}.
   \right.
\label{eq:coordinatedistance3}
\end{equation}

\section{Angular Diameter Distance}

The angular diameter distance, $d_A$, is the ratio of an object's physical transverse 
size, denoted by $l_\perp$, to its angular size (in radians). See Fig.\ (\ref{fig:distance}). Mathematically,
\begin{eqnarray}
d_A(z)=\frac{l_\perp}{\phi}=\frac{ar(z)\phi}{\phi}=ar(z)=\frac{a_0 r(z)}{1+z}.
\label{eq:ada1}
\end{eqnarray}
In terms of dimensionless coordinate distance it is given as,
\begin{eqnarray}
d_A(z)=\frac{y(z)}{H_0(1+z)}.
\label{eq:ada3}
\end{eqnarray}
This distance is often used when discussing CMB anisotropies observations.

\section{Luminosity Distance}

The luminosity distance denoted by $d_L$ is defined by the relations between the bolometric
(i.e. integrated over all frequencies) flux $\mathcal{F}$ and the bolometric absolute luminosity $L$ (defined as the total power radiated in watts). It
is a measure of how far an object of known luminosity $L$ 
is that produces the luminosity flux of $\mathcal{F}$, assuming the inverse-square law, 
\begin{eqnarray}
\mathcal{F}&=&\frac{L}{4 \pi d^{2}_{L}}\ , \\
\Rightarrow \hspace{0.5 cm} d_{L}&=&\sqrt{\frac{L}{4 \pi \mathcal{F}}}\ .
\label{eq:ldq1}
\end{eqnarray}
Note that the observed luminosity (that we will denote as $L_0$), which is luminosity detected at d$_{co}=0$ ($z=0$),
is different from the absolute luminosity $L$ of the source (emitted at a comoving distance d$_{co}$ with redshift $z$).
Thus $d_L$ can be found from measurements of $\mathcal{F}$, provided $L$ is known; the trouble is that generally, it is not.
However, there are certain classes of objects for which we do know $L$, to some accuracy, and all higher levels of the 
cosmic distance ladder are based on these ``standard candles".

The flux $\mathcal{F}$ is defined by $\mathcal{F}=L_0/S$, where $S$ is the surface area over which that flux is spread
and it is equal to $S=4\pi (a_0 r)^2$ (area of the sphere at $z=0$). Then from Eq.\ (\ref{eq:ldq1}):
\begin{eqnarray}
\mathcal{F}=\frac{L}{4 \pi d^{2}_{L}}=\frac{L_0}{4 \pi (a_0 r)^2}, \\
\Rightarrow \hspace{0.5 cm} d^{2}_{L}= (a_0r)^2\frac{L}{L_0}.
\label{eq:ldq2}
\end{eqnarray} 
Now, we need to find the ratio $\frac{L}{L_0}$ in terms of known quantities.

If energy $\Delta E_e$ is emitted in time interval $\Delta t_e$ from a distinct object (with redshift $z$),
then the absolute luminosity is
\begin{eqnarray}
 L=\frac{\Delta E_e}{\Delta t_e}.
\label{eq:ldq3}
\end{eqnarray}
The flux of this energy observed at $z=0$, denoted by $L_0$, is
\begin{eqnarray}
 L_0=\frac{\Delta E_0}{\Delta t_0},
\label{eq:ldq4}
\end{eqnarray}
where $\Delta E_0$ is the energy of light detected during time interval $\Delta t_0$. From Planck's theory, the energy of the
photon is inversely proportional to it's wavelength $\lambda$ $\left(E={hc}/{\lambda}\right)$, hence,
\begin{eqnarray}
\frac{\Delta E_e}{\Delta E_0}=\frac{\lambda_0}{\lambda_e}=1+z,
\label{eq:ldq5}
\end{eqnarray}
where $\lambda_e$ and $\lambda_0$ are the wavelengths of the light at the point of emission (at redshift $z$) and 
detection (at $z=0$) respectively, and we have used Eq.\ (\ref{eq:HL5}) in the second step,
Also, since $c=\frac{\lambda}{\Delta t}$,
\begin{eqnarray}
\frac{\lambda_e}{\Delta t_e}=\frac{\lambda_0}{\Delta t_0},
\label{eq:ldq7}
\end{eqnarray} 

\begin{eqnarray}
\Rightarrow \hspace{0.5 cm} \frac{\lambda_0}{\lambda_e}=\frac{\Delta t_0}{\Delta t_e}=1+z,
\label{eq:ldq8}
\end{eqnarray} 
and so,
\begin{eqnarray}
\frac{L_e}{L_0}&=&\frac{\Delta E_e}{\Delta t_e} \cdot \frac{\Delta t_0}{\Delta E_0}=\frac{\Delta E_e}{\Delta E_0} \cdot \frac{\Delta t_0}{\Delta t_e}, \nonumber \\
&=&(1+z)\cdot(1+z). \hspace{2 cm} (\mathrm{using}\  \mathrm{Eq.}\ (\ref{eq:ldq7})-(\ref{eq:ldq8}))
\label{eq:ldq9}
\end{eqnarray}
Hence, from Eq.\ (\ref{eq:ldq2}),
\begin{eqnarray}
d^{2}_{L}&=&(a_0r)^2(1+z)^2.
\label{eq:ldq10}
\end{eqnarray}
Taking the square root on both sides we will get:
\begin{eqnarray}
d_{L}&=&(a_0r)(1+z).
\label{eq:ldq11}
\end{eqnarray}
In terms of the dimensionless coordinate distance, the luminosity distance is,
\begin{eqnarray}
d_{L}&=&\frac{1}{H_0}y(z)(1+z),
\label{eq:ldq12}
\end{eqnarray}
so the luminosity distance can be expressed using Eq.\ (\ref{eq:coordinatedistance3}) in terms of curvature density parameter as
\begin{equation}
d_L(z)= \left\{
      \begin{array}{lc}
           \frac{(1+z)}{H_0\sqrt{\Omega_{k0}}}\mathrm{sinh}\left(\sqrt{\Omega_{k0}}\int\limits^{z}_{0}\frac{dz'}{E(z')}\right) & \ \ \ \ \mathrm{for}\ \Omega_{k0}>0 \\
           \frac{1+z}{H_0}\int\limits^{z}_{0}\frac{dz'}{E(z')} & \ \ \ \ \mathrm{for}\ \Omega_{k0}=0 \\
           \frac{(1+z)}{H_0\sqrt{-\Omega_{k0}}}\mathrm{sin}\left(\sqrt{-\Omega_{k0}}\int\limits^{z}_{0}\frac{dz'}{E(z')}\right) & \ \ \ \ \mathrm{for}\ \Omega_{k0}<0 \\
     \end{array}
   \right.
\label{eq:ldq13}
\end{equation}
and using Eq.\ (\ref{eq:dlcd1})in terms of curvature parameter as,
\begin{equation}
d_L(z)= \left\{
      \begin{array}{lc}
           \frac{a_0(1+z)}{\sqrt{k}}\mathrm{sin}\left(\frac{\sqrt{k}}{a_0H_0}\int\limits^{z}_{0}\frac{dz'}{E(z')}\right) & \ \ \ \ \mathrm{for}\ \ k>0 \\
           \frac{1+z}{H_0}\int\limits^{z}_{0}\frac{dz'}{E(z')} & \ \ \ \ \mathrm{for}\ \ k=0 \\
           \frac{a_0(1+z)}{\sqrt{-k}}\mathrm{sinh}\left(\frac{\sqrt{-k}}{a_0H_0}\int\limits^{z}_{0}\frac{dz'}{E(z')}\right) & \ \ \ \ \mathrm{for}\ \ k<0 \\
     \end{array}
   \right.
\label{eq:lumdis}
\end{equation}

Now let's compute the approximate  expression for the luminosity distance when $z \ll 1$, that we will use in Chapter\ (\ref{Chapter3}) where we will calculate the general expression of luminosity distance as a function of $z \ll 1$ in non-flat Universe having dark energy. In order
to do this we read the Maclaurin series expansion of the integral $\int\limits^{z}_{0}\frac{dz'}{E(z')}$,\footnote{$E(0)=1.$}
\begin{eqnarray}
\int\limits^{z}_{0}\frac{dz'}{E(z')}\  \ = \ z-\frac{E'(0)}{2}z^2+\frac{1}{6}\left[2E'(0)^2-E''(0)\right]z^3+O(z^4).
\label{eq:app1}
\end{eqnarray}
Now using the expansion of sinh($x$)$= x+\frac{x^3}{6}+O(x^5)$, we get from Eq.\ (\ref{eq:ldq13}), 
\begin{eqnarray}
d_L(z)\  &=&\ \frac{(1+z)}{H_0 \sqrt{\Omega_{k0}}}\ \mathrm{sinh}\left[\sqrt{\Omega_{k0}}\left(z-\frac{E'(0)}{2}z^2+\frac{1}{6}\left[2E'(0)^2-E''(0)\right]z^3+O(z^4)\right)\right], \nonumber \\
\nonumber \\
&=&\ \frac{(1+z)}{H_0 \sqrt{\Omega_{k0}}}\ \left[\sqrt{\Omega_{k0}}\left(z-\frac{E'(0)}{2}z^2+\frac{1}{6}\left[2E'(0)^2-E''(0)\right]z^3\right)\vphantom{\frac{(\Omega_{k0})^{3/2}z^3}{6}} \right.\nonumber \\ 
&&\hspace{2.2 cm}+\ \left. \frac{(\Omega_{k0})^{3/2}z^3}{6}+\dots \right], \nonumber \\
\nonumber \\
&\approx& \frac{(1+z)}{H_0 \cancel{\sqrt{\Omega_{k0}}}}\ \ \cancel{\sqrt{\Omega_{k0}}}\left[z-\frac{E'(0)}{2}z^2+\frac{1}{6}\left[2E'(0)^2-E''(0)+\Omega_{k0}\right]z^3\right], \nonumber \\
\nonumber \\
&\approx&\frac{1}{H_0}\left[z+z^2-\frac{E'(0)}{2}z^2-\frac{E'(0)}{2}z^3+\frac{1}{6}\left(2E'(0)^2-E''(0)+\Omega_{k0}\right)z^3\right]. \nonumber \\
\nonumber \\
&=&\frac{1}{H_0}\left[z+\left(1-\frac{E'(0)}{2}\right)z^2+\frac{1}{6}\left(2E'(0)^2-3E'(0)-E''(0)+\Omega_{k0}\right)z^3\right].
\label{eq:app2}
\end{eqnarray}
Reintroducing the speed of light is $c$, the approximate luminosity distance for $z \ll 1$ is,
\begin{eqnarray}\hspace{-1 cm}
d_L(z) &\approx& \frac{c}{H_0}\left[z+\left(1-\frac{E'(0)}{2}\right)z^2+\frac{1}{6}\left(2E'(0)^2-3E'(0)-E''(0)+\Omega_{k0}\right)z^3\right].
\label{eq:app3}
\end{eqnarray}

We can relate the angular diameter distance $d_A$ and the luminosity distance using Eq.\ (\ref{eq:ada3}), and (\ref{eq:ldq12}),
\begin{eqnarray}
d_{A}(z)&=&\frac{y(z)}{H_0(1+z)}=\frac{d_L(z)}{(1+z)^2}.
\label{eq:app4}
\end{eqnarray}

\section{Distance Modulus}
\label{sec:Distance Modulus}
The apparent magnitude, $m$, of an astronomical object is defined by the ratio of the apparent flux of
the object to some reference flux,
\begin{eqnarray}
m=-2.5\ \mathrm{log_{10}}\left(\frac{\mathcal{F}}{\mathcal{F}_{\mathrm{ref}}}\right).
\label{eq:dm1}
\end{eqnarray}
The absolute magnitude, $M$, is defined as the apparent magnitude the object would have if it were 10 pc away.
The difference between them is known as distance modulus and can be expressed as: 
\begin{eqnarray}
\mu\equiv m-M&=&-2.5\ \mathrm{log_{10}}\left(\frac{\mathcal{F}}{\mathcal{F}_{\mathrm{ref}}}\right)-(-2.5)\ \mathrm{log_{10}}\left(\frac{\mathcal{F}_{\mathrm{10\ pc}}}{\mathcal{F}_{\mathrm{ref}}}\right), \nonumber \\
&=&-2.5\ \mathrm{log_{10}}\left(\frac{\mathcal{F}}{\mathcal{F}_{\mathrm{10\ pc}}}\right),\nonumber \\
&=&-2.5\ \mathrm{log_{10}}\left(\frac{L}{4\pi d_L^2}\cdot \frac{4 \pi (10\ \mathrm{pc})^2}{L}\right),\nonumber \\
&=&-2.5\ \mathrm{log_{10}}\left(\frac{10\ \mathrm{pc}}{d_L}\right)^2,\nonumber \\
&=&5\ \mathrm{log_{10}}\left(\frac{d_L}{\mathrm{10\ pc}}\right).
\label{eq:dm2}
\end{eqnarray}
Let's write this distance modulus in terms of dimensionless coordinate distance $y(z)$, as a function of the dimensionless Hubble parameter, using Eq.\ (\ref{eq:ldq12}),\footnote{In Eq.\ (\ref{eq:ldq12}) we were working in units where $c=1$, but we now bring $c$ back for computational simplicity. In this case Eq.\ (\ref{eq:ldq12}) takes the form,
\begin{eqnarray}
d_{L}(z)&=&\frac{c}{H_0}y(z)(1+z).
\label{eq:dm3}
\end{eqnarray}
}
\begin{eqnarray}
\mu= 5\ \mathrm{log_{10}}\left(\frac{d_L}{\mathrm{10\ pc}}\right)=5\ \mathrm{log_{10}}\left[\frac{c}{H_0}y(z)(1+z)\frac{1}{\mathrm{10\ pc}}\right].
\label{eq:dm4}
\end{eqnarray}
Now, with the value of $c=3 \times 10^5$ km s$^{-1}$, and $H_0=h$($100$ km s$^{-1}$ Mpc$^{-1}$), we get
\begin{eqnarray}
\mu(z,h)&=& 5\ \mathrm{log_{10}}\left[\frac{3\times 10^5\ \mathrm{km\ s^{-1}}}{\frac{100h\ \mathrm{km\ s^{-1}}}{10^6\ \mathrm{pc}}}y(z)(1+z)\frac{1}{10\ \mathrm{pc}}\right],\nonumber \\
\nonumber \\
&=& 5\ \mathrm{log_{10}}\left[\frac{3\times 10^8}{h}y(z)(1+z)\right],\nonumber \\
\nonumber \\
&=& 5\ \mathrm{log_{10}}\Big[3000y(z)(1+z)\Big]+25- 5\ \mathrm{log_{10}}(h).
\label{eq:dm5}
\end{eqnarray}
This is the equation we will use in the supernova (SN) analyses in Chapter\ (\ref{Chapter5}).


\cleardoublepage


\chapter{Dark Energy and Dark Energy Models}
\label{Chapter3}

In this chapter we will discuss two commonly used dark energy models, and one dark energy parametrization, but prior to that 
we discuss the observational evidence for most mysterious type od substance i.e., dark energy.

\section{Observational Evidence for Dark Energy}

From the last two decades, evidence for the most striking results in modern 
cosmology has been steadily growing, namely the existence of 
a cosmological constant or dark energy, which is believed to be the cause of 
the current accelerated expansion of the Universe\footnote{
Some cosmologists instead view these observations as an indication 
that general relativity needs to be modified on these large 
length scales. For recent reviews of modified gravity see \cite{Tsujikawa2010},
\cite{Bolotin2011}, \cite{Capozziello2011}, \cite{Starkman2011}, and
references therein. These theories reduces to GR on the scale of our galaxy where 
GR predictions are exactly consistent with observations. In this thesis we assume that general relativity 
provides an adequate description of gravitation on cosmological scales.}
as first observed by Perlmutter \textit{et al.},\cite{Perlmutter99} and Riess \textit{et al.}\cite{Riess98} 
Although it may not have come as such a surprise to a few theorists, 
who were at that time considering the interplay between a number 
of different types of observations,\cite{Krauss,Peebles&Ratra1988,peebles84,Ratra&Peebles1988} for the majority 
it came as something of a bombshell. The Universe is not only expanding, 
it is expanding with acceleration. The results first published by 
Perlmutter \textit{et al.},\cite{Perlmutter99} and Riess \textit{et al.}\cite{Riess98} 
where the one that caused a major change in the way we have started thinking about the universe.

There is a key problem that we have to explain, 
and it is fair to say it has yet to be understood. 
The value of the energy density stored in the cosmological constant or dark energy today (the assumed cause of accelerated expansion), 
has to be of order of the critical density, 
namely $\rho_{\Lambda} \sim 10^{-3}\,{\rm eV}^4$. 
It is unfortunate that no reasonable explanation exists as to why a true 
cosmological constant should be at this scale. There are reasons to think it should naturally 
be much larger. Typically, since it is conventionally associated 
with the energy of the vacuum in quantum field theory, we expect it to 
have a magnitude of order the typical energy scale of early Universe phase transitions. 
Even the QCD scale, it would imply a much larger value, 
$\rho_{\Lambda} \sim 10^{-3}\,{\rm GeV}^4$. 
The question then remains, why does $\Lambda$ have the value that it has today?\footnote{We 
will discuss this more in sec.\ (\ref{sec:problems in LCDM})}

In order to explain the current accelerated expansion of the universe, 
we require an exotic substance dubbed ``dark energy'' 
with an equation-of-state parameter satisfying $\omega<-1/3$, as mentioned in Eq.\ (\ref{eq:SFU6}).
Newtonian gravity cannot account for the accelerated 
expansion, it allows only decelerated cosmological expansion.\footnote{In Newtonian gravity there is no concept of dark energy, all matter gravitates and 
hence accelerated expansion is out of the question.} Let's consider a homogeneous sphere with radius $a$ and energy density $\rho$. 
The Newton's equation of motion for a point particle with mass $m$
on this sphere is
\begin{eqnarray}
F&=&-\frac{GMm}{a^2},\nonumber \\
m \ddot{a} &=&-\frac{Gm}{a^2} \left(\frac{4\pi a^3 \rho}{3} \right)\ , \hspace{1 cm} \left(\mathrm{Here},\ M=\rho V= \frac{4 \pi a^3 \rho}{3}\right)\nonumber \\
\Rightarrow \ \ \
\frac{\ddot{a}}{a} &=&
-\frac{4\pi G}{3}\rho\,.
\label{Newtoneq}
\end{eqnarray}
The difference compared to the Einstein's equation (\ref{eq:Friedmanneq})
is the absence of the pressure term, $P$. 
This appears in Einstein's
equation because of relativistic effects.
The condition $\omega<-1/3$ means that we essentially require 
a large negative pressure in order to give rise to an accelerated expansion.

Now let's consider mathematically how to accommodate energy with $\omega<-1/3$ in Friedmann's equations.
From Eq.\ (\ref{eq:energycon}), the energy density $\rho$ is constant with respect to 
cosmological time $t$ for a fluid which has $\omega=-1$. In this case, in a spatially-flat model, the Hubble rate,
$\dot a/a$ in Eq.\ (\ref{eq:Friedmanneq}) also is constant. This leads
to the exponential evolution of the scale factor with respect to time, 
\begin{eqnarray}
a(t) \propto e^{Ht}.
\end{eqnarray}
This is the de Sitter spacetime or Universe. As we will see in a
later section, this exponential expansion also arises in the Einstein equations in the case of the cosmological constant $\Lambda$ dominant Universe.

We now discuss some of the observational facts in the support of the existence of dark energy. These include:

\begin{itemize} 
     \item The age of the Universe compared to that of the oldest star in combination with estimate of $H_0$.
     \item Supernovae apparent magnitude observations (SNeIa).
     \item Cosmic microwave background (CMB) anisotropy observations in combination with the estimation of $\Omega_{m0}$, the non relativistic matter density parameter.
     \item Baryonic acoustic oscillation (BAO) peak length scale measurements.
     \item Hubble parameter measurements.    
     \item Large-scale-structure (LSS) observations.
 \end{itemize} 

Even in the last two decades of the $20^{th}$ century, there were some observations that the
age of a cold dark matter (CDM) dominated Universe  with $\Omega_{m0}=1$ was smaller than the age of the oldest stars
in globular clusters in our Milky Way galaxy. Adding dark energy and so reducing the amount of CDM, 
can address this apparent inconsistency by increasing the age
of the Universe. The stronger evidence supporting accelerated cosmological expansion 
(and consequently the presence of dark energy along with that of CDM) came from measurements of the luminosity distance
of type Ia Supernovae (SNeIa). CMB observations also favor the presence of the dark energy, yet the
constraints obtained from CMB alone on dark energy are very weak. The BAO measurements are another 
independent source of support for the idea of the dark energy, as are $H(z)$ measurements. The last idea that we will
discuss is that that power spectrum density irregularities is also consistent with the existence of dark
energy. We now discuss some of these observations in a little more detail.

\subsection{Age of the Universe}

By looking at Table.\ (\ref{table:Solution of Friedman}) in which the age of the Universe $t_0$ is given in terms
of the inverse of the Hubble constant $H^{-1}_{0}$ for single-component Universes, it is easy to see that 
dimensionally $H_0^{-1}$ sets the scale for $t_0$ and we expect this to be the case in multiple components Universes, like ours, also. 
In this Section we try to make a more
precise guess (estimate of $H_0$ by incorporating factors that we have previously ignored) and then will 
compare this to $t_0$ the age of the oldest known stars. We now consider dark energy, thus with the time dependent dark energy density $\rho_{DE}(t)$ and using Eq.\ (\ref{eq:diffcon})  write
\begin{eqnarray}
\dot \rho_{DE}(t)=3\left(\frac{\dot a}{a}\right)\Big(\rho_{DE}(t)+P_{DE}(t)\Big).
\label{eq:AOU1.1}
\end{eqnarray}
We assume a dark energy equation-of-state-parameter $\omega_{DE}$ which
is time dependent so,
\begin{eqnarray}
P_{DE}(t)=\omega_{DE}(t)\ \rho_{DE}(t).
\label{eq:AOU1.2}
\end{eqnarray}
Equation\ (\ref{eq:AOU1.1}) then takes the form
\begin{eqnarray}
\frac{\dot \rho_{DE}(t)}{\rho_{DE}(t)}=3\left(\frac{\dot a}{a}\right)\Big(1+\omega_{DE}(t)\Big).
\label{eq:AOU1.3}
\end{eqnarray}
or, 
\begin{eqnarray}
\frac{d \rho_{DE}}{\rho_{DE}}=3 \left(1+\omega_{DE}\right) \frac{d a}{a}.
\label{eq:AOU1.4}
\end{eqnarray}
Integrating from some time in the past to the present when $a_0=1$,
\begin{eqnarray}
\mathlarger{\int} \frac{d \rho_{DE}}{\rho_{DE}}=\int\limits^{1}_{a}3 \left(1+\omega_{DE}\right) \frac{d a'}{a'}+\mathrm{ln}\rho_{DE,0}\ ,
\label{eq:AOU1.5}
\end{eqnarray}
where ln$\rho_{DE,0}$ is the constant of integration. Substituting:
\begin{eqnarray}
a'=\frac{1}{1+z}, \hspace{1cm} \ \ \ \ \   \hspace{1cm} da'=-\frac{1}{(1+z)^2}dz,
\label{eq:AOU1.6}
\end{eqnarray}
\begin{eqnarray}
\Rightarrow\ \ \  \mathlarger{\int} \frac{d \rho_{DE}}{\rho_{DE}}&=&\int\limits^{0}_{z}-3 \frac{\left(1+\omega_{DE}(z')\right)}{1+z'}dz'+\mathrm{ln}\rho_{DE,0}\ ,\\
\Rightarrow\ \ \  \mathrm{ln}\rho_{DE}&=&\int\limits^{z}_{0} \frac{3\left(1+\omega_{DE}(z')\right)}{1+z'}dz'+\mathrm{ln}\rho_{DE,0},
\label{eq:AOU1.7}
\end{eqnarray}
Thus we will get redshift $z$ dependent dark energy density
\begin{eqnarray}
\rho_{DE}(z)=\rho_{DE,0}\ \mathrm{exp}\left[\int\limits^{z}_{0} \frac{3\left(1+\omega_{DE}(z')\right)}{1+z'}dz'\right].
\label{eq:AOU1.8}
\end{eqnarray}
Which can also be written by introducing a average $\langle \omega_{DE} \rangle$ as:
\begin{eqnarray}
\rho_{DE}(z)=\rho_{DE,0}(1+z)^{3(1+\langle \omega_{DE} \rangle)}.
\label{eq:AOU1}
\end{eqnarray}
Where $\langle \omega_{DE} \rangle$ is defined as:\footnote{Can be calculated by comparing Eqs.\ (\ref{eq:AOU1.8}) and (\ref{eq:AOU1})}
\begin{eqnarray}
\langle \omega_{DE} \rangle = \frac{1}{\mathrm{ln}(1+z)}\int\limits^{z}_{0}\frac{\omega_{DE}(z')}{1+z'}dz'.
\label{eq:AOU1.9}
\end{eqnarray}

If we will take into account other components like radiation, non-relativistic matter,
and space curvature, then Eq.\ (\ref{eq:MCU1}) is of the from:
\begin{eqnarray}
H^2=\frac{8\pi G}{3}\left(\rho_r+\rho_m+\rho_{DE}\right)-\frac{k}{a^2}.
\label{eq:AOU1.10}
\end{eqnarray}
Present epoch density parameters, defined in Eq.\ (\ref{eq:MCU7}), obey 
\begin{equation}
\Omega_{m0}+\Omega_{r0}+\Omega_{k0}+\Omega_{DE,0}=1,
\label{eq:AOU1.11}
\end{equation}
so we can rewrite Eq.\ (\ref{eq:AOU1.10}) in the form, $H^2(z)=H_0^2E^2(z)$,
Where $E(z) \equiv H(z)/H_0$ is
\begin{eqnarray}
E^2(z)&=&\left[\Omega_{r0}(1+z)^4+\Omega_{m0}(1+z)^3 \hspace{3 cm} \vphantom{\left\{\int \limits^{z}_{0}\frac{3(1+\omega_{DE})}{1+z'}\mathrm{d}z'\right\}} \right.\nonumber \\
&&    \left.+\ \Omega_{DE,0}\ \mathrm{exp}\left\{3\int \limits^{z}_{0}\frac{(1+\omega_{DE}(z'))}{1+z'}\mathrm{d}z'\right\} +\Omega_{k0}(1+z)^2\right].
\label{eq:AOU3}
\end{eqnarray}

Since $H={\dot a}/{a}$, $a={1}/({1+z})$ $\Rightarrow$ $H=-\dot z/{(1+z)}$, the age of the Universe can be expressed as
\begin{eqnarray}
t_0=\frac{1}{H_0}\int\limits^{\infty}_{0} \frac{dz}{E(z)(1+z)}.
\label{eq:AOU4}
\end{eqnarray}
This integral is dominated by the terms at low redshift. Since $\Omega_{r0}$ is of the order of
$10^{-4}$, for detail see,\cite{ryden2003introduction,raine2001introduction,
liddle2000cosmological,weinberg1972gravitation,kolb1990early,padmanabhan2000theoretical}
radiation will play an important role only at high redshifts $z\gtrsim 1000$. Hence for the sake of
simplicity we ignore radiation when evaluating Eq.\ (\ref{eq:AOU4}) then Eq.\ (\ref{eq:AOU1.11}) reduced to $\Omega_{m0}+\Omega_{DE,0}+\Omega_{K0}=1$. 

Let's consider the special case of time independent $\omega_{DE}=-1$ 
(Einstein's cosmological constant). Then the age of the 
Universe is
\begin{eqnarray}
t_0=\frac{1}{H_0}\mathlarger{\mathlarger{\int}}\limits_{0}^{\infty} \frac{dz}{(1+z)\Big[\Omega_{m0}(1+z)^3+ \Omega_{DE,0}+\Omega_{K0}(1+z)^2\Big]^{1/2}}.
\label{eq:AOU5}
\end{eqnarray}
Further simplifying $t_0$ the flat case (with $\Omega_{K0}=0$, and so $\Omega_{DE,0}=1-\Omega_{m0}$), and substituting
\begin{eqnarray}
(1+z)^3=\frac{1-\Omega_{m0}}{\Omega_{m0}}\ \mathrm{sinh^2}\beta, \hspace{1 cm} \mathrm{d}z=\frac{2}{3}\left(\frac{1-\Omega_{m0}}{\Omega_{m0}}\right)^{1/3}\mathrm{sinh^{-1/3}}\beta\ \mathrm{cosh}\beta\ \mathrm{d}\beta,
\label{eq:AOU6.1}
\end{eqnarray}
Then the integral Eq.\ (\ref{eq:AOU5}) becomes,
\begin{eqnarray}
t_0=\frac{2}{3H_0\sqrt{1-\Omega_{m0}}}\int\limits_{\mathrm{sinh^{-1}}\left[\sqrt{\frac{\Omega_{m0}}{1-\Omega_{m0}}}\ \right]}^{\infty}\mathrm{cosech}\beta\ \mathrm{d}\beta,
\label{eq:AOU6.2}
\end{eqnarray}
which can be easily integrated, and in terms of $\Omega_{m0}$ gives,
\begin{eqnarray}
t_0=\frac{1}{3H_0\sqrt{1-\Omega_{m0}}}\mathrm{ln} \Bigg(\frac{1+\sqrt{1-\Omega_{m0}}}{1-\sqrt{1-\Omega_{m0}}}\Bigg).
\label{eq:AOU7}
\end{eqnarray}
In the limit $\Omega_{DE,0}\rightarrow 0$, we have our old result for Einstein-de Sitter model derived in Eq.\ (\ref{eq:SFU17}):\footnote{It is fun to use L H\^{o}pital's rule here.}
\begin{eqnarray}
t_0=\frac{2}{3H_0}. 
\label{eq:AOU8}
\end{eqnarray}
Using the then-favored dimensionless Hubble parameter value $h=0.72 \pm 0.08$ (today $h=0.68 \pm 0.028$ is probably a batter estimate see \cite{Gott2001, 2003PASP..115.1143C})
the age of the Universe in the absence of the cosmological constant is in the range $8.2\ \mathrm {Gyr}\ \leqslant t_0\leqslant 10.2\ \mathrm {Gyr}$
($9.2\ \mathrm {Gyr}\leqslant t_0\leqslant 10.2\ \mathrm {Gyr}$). So this is the theoretical value we have. On the other hand, Carretta \textit{et al.}
\cite{Carretta2000} estimated the age of a globular cluster\footnote{A spherical collection of the dust known stars that orbits a galactic core as a satellite. 
Globular clusters are very tightly bound by gravity, which gives them their spherical shapes and relatively high stellar densities toward 
their centers.} in the Milky Way galaxy to be $12.9 \pm 2.9$ Gyr. Jimenez \textit{et al.} \cite{jimenez1996ages} calculated the value of 
$13.5 \pm 2$ Gyr, whereas Hansen \textit{et al.}\cite{Hansen} constrained the age of the globular cluster Messier 4 (M$4$) to be $12.7 \pm 0.7$ Gyr 
by using the method of the white dwarf cooling sequence. Overall, observations of the ages of the globular clusters are larger than about
11 Gyr. Consequently, this leads to some doubt about the numerical age of the Universe results obtained in 
the Einstein-de Sitter model from Eq.\ (\ref{eq:AOU8}).

One of the ways to address this problem is to take a cosmological constant (or dark energy with an equation-of-state-parameter 
$\omega_{DE}\approx-1$) into account. From Eq.\ (\ref{eq:AOU7}) it is clear that $t_0$ increases with the decrease of $\Omega_{m0}$. 
The limit $\Omega_{m0}\rightarrow 0$ results in $t_0 \rightarrow \infty$.  

\begin{figure}[h]
\centering
    \includegraphics[height=4.5in]{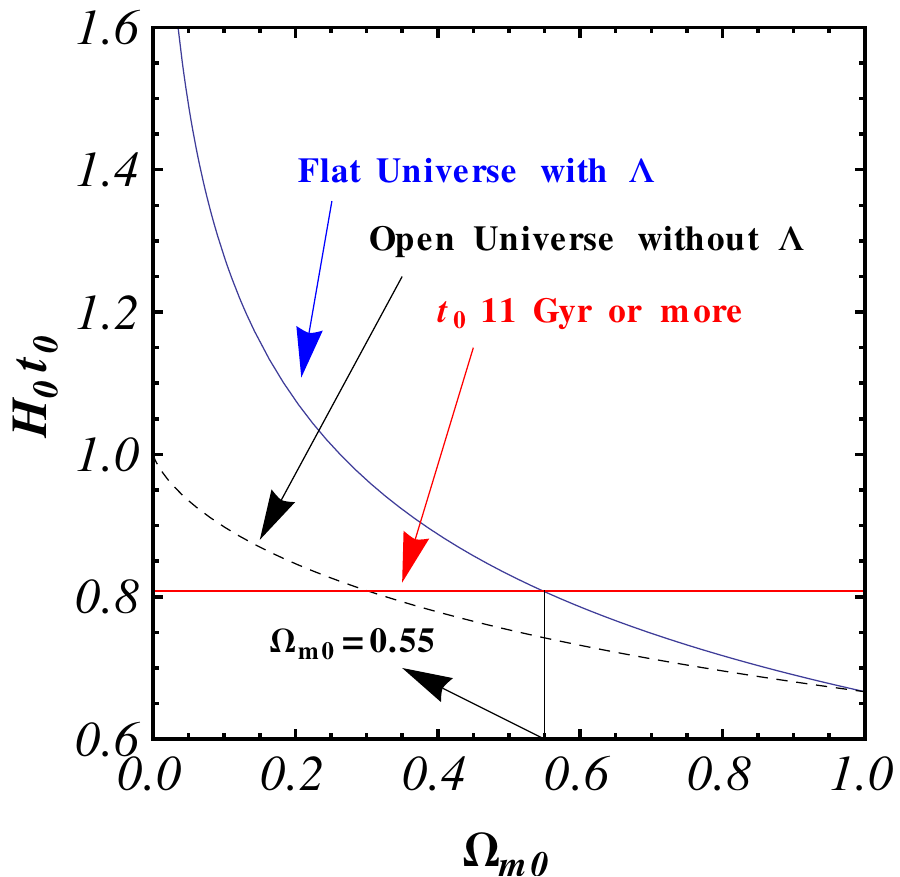}
\caption{The cosmic age $t_0$ in terms of $H^{-1}_{0}$ verses $\Omega_{m0}$. The thin-solid blue curve
describes a flat Universe in the presence of the cosmological constant $\Lambda$ with the constraints
$\Omega_{m0}+\Omega_{DE,0}=1$. The dashed black curve corresponds to an open Universe without the 
cosmological constant $\Lambda$. The red horizontal line is a minimum age of the Universe allowed to form
the oldest globular cluster ($>$11 Gyr), here we suppose $H_0=72\pm8$ km s$^{-1}$ Mpc$^{-1}$. Here the intersection 
of the blue line and red line is at $\Omega_{m0}=0.55$, which gives the constraint that $\Omega_{m0}<0.55$ if the 
age of the spatially-flat with dark energy Universe has to be more than the age of the oldest globular clusters.}
\label{fig:AgeofUniverse}
\end{figure}
Figure\ (\ref{fig:AgeofUniverse}) shows the cosmic age predicted by Eq.\ (\ref{eq:AOU7}) as a function of
$\Omega_{m0}$. It shows that if the $t_0$ must be greter than 11 Gyr then $\Omega_{m0} < 0.55$. 
The WMAP 5-year estimates on the cosmic age (assuming the $\Lambda$CDM model) is 
$t_0=13.73\pm0.12$ Gyr.\cite{Komatsu2009} Under this bound we find that the density parameter of non-relativistic matter
is constrained to be $0.245<\Omega_{m0}<0.261$ for $h=0.72$. When $h=0.68$ then $0.27 < \Omega_{m0} < 0.29$.\cite{Farooq20131}

We only considered the flat case and we showed that the way to increase the cosmic age of Universe is by incorporating dark energy
(the cosmological constant), but what if we will consider an open Universe, i.e. let $\Omega_{K0}>0$? To address this issue we have to integrate
Eq.\ (\ref{eq:AOU5}) without incorporating dark energy, with positive curvature density ($\Omega_{K0}>0$). Hence, 
$\Omega_{m0}+\Omega_{K0}=1$. substituting 
\begin{eqnarray}
1+z=\frac{1-\Omega_{m0}}{\Omega_{m0}}\ \mathrm{sinh^2}\beta, \hspace{1 cm} \mathrm{d}z=2\left(\frac{1-\Omega_{m0}}{\Omega_{m0}}\right)\mathrm{sinh}\beta\ \mathrm{cosh}\beta\ \mathrm{d}\beta,
\label{eq:AOU9}
\end{eqnarray} 
in Eq.\ (\ref{eq:AOU5})the equation becomes
\begin{eqnarray}
t_0=\frac{2\ \Omega_{m0}}{H_0\left(1-\Omega_{m0}\right)^{3/2}}\int\limits_{\mathrm{sinh^{-1}}\left[\sqrt{\frac{\Omega_{m0}}{1-\Omega_{m0}}}\ \right]}^{\infty}\mathrm{cosech^3}\beta\ \mathrm{d}\beta,
\label{eq:AOU10}
\end{eqnarray}
which is easily integrated and in terms of $\Omega_{m0}$, we get
\begin{eqnarray}
t_0=\frac{1}{H_0(1-\Omega_{m0})}\left[1+\frac{\Omega_{m0}}{2\sqrt{1-\Omega_{m0}}}\mathrm{ln} \Bigg(\frac{1-\sqrt{1-\Omega_{m0}}}{1-\sqrt{1+\Omega_{m0}}}\Bigg)\right].
\label{eq:AOU11}
\end{eqnarray}
The dashed black lines in Fig.\ (\ref{fig:AgeofUniverse}) show the age of the open Universe without incorporating
dark energy. In the limit $\Omega_{m0}\rightarrow 1$, we recover the value in Eq.\ (\ref{eq:AOU8}) in the flat Universe.
On the other hand, in the limit $\Omega_{m0}\rightarrow 0$, we have $t_0\rightarrow H_{0}^{-1}$. So we can conclude that the
cosmic age of the open Universe does not become as large as in the case of flat Universe with dark energy (cosmological
constant). Since the curvature $|\Omega_{K0}|$ has been constrained to be much smaller then unity from WMAP 
measurements,\cite{Komatsu2009} it is almost impossible to satisfy the condition $t_0>11$ Gyr for $h=0.72 \pm 0.08$
in the open Universe without dark energy.

The above discussion shows that the existence of dark energy can resolve the cosmic age problem.

\subsection{Supernovae Apparent Magnitude Observations (SNeIa)} 

Persuasive evidence for the current cosmic acceleration, and hence for  dark energy, came from observations of the distance modulus of distant supernova of type Ia (SNIa).\footnote{A supernova, denoted by SN, is a very energetic explosion of a massive super-giant  star. Supernovae (plural of supernova denoted by SNe) are extremely luminous and cause a burst of radiation that, most of the time, lighten an entire galaxy for several weeks to months. During this short time, a supernova can radiate as much as $10^{44}$ J of energy which is equal to the energy radiated by sun in 10 Gyr ($10^{44}$ J=1 foe, a unit of energy used to measure the energy of supernovae. The word is an acronym derived from the phrase [ten to the power of] fifty-one ergs.)} Two independent groups working on observations of supernovae [Riess \textit{et al.} named the High-Redshift Supernovae Search Team (HSST)\cite{Riess98} and Perlmutter \textit{et al.} named the Supernovae Cosmology Project (SCP)\cite{Perlmutter99}] reported late time cosmic acceleration. By 1998 Riess \textit{et al.} has studied 16 high-redshift SNIa along with the 34 nearby supernovae, while by 1999 the SCP group had studied 42 SNIa in the redshift range $0.18 \leq z \leq 0.83$.

The supernova explosion, in which large amounts of energy are released as electromagnetic radiation to interstellar space, can be triggered in two ways. In both ways, it is gravity that gives a SN its energy.
\begin{enumerate} 
    \item By the abrupt reignition of nuclear fusion in a compact star.\footnote{Compact star is the term used to refer collectively to white dwarfs, neutron star, and black holes.} The compact star may accumulate sufficient material from it's surroundings, either by a merger or through accretion, to raise its core temperature and ignite nuclear fusion, completely disrupting the star.    

    \item By the collapse of the core of the massive star. Mass flows into the core of the star by the continued formation of iron from nuclear fusion. Once the core has gained so much mass that it cannot withstand its own weight (gravity), the core implodes. This implosion can some time be halted by neutron degeneracy pressure (neutrons form by the fusion of electrons and protons),  depending upon the mass of the star core. If the mass of the star core is higher then the Chandrasekhar limit ($1.38 M_{\odot}$),\cite{Chandrasekhar1931ApJ....74...81C} then even neutrons fail to stop the implosion. When collapse is abruptly stopped by a neutron flux, matter bounces off the hard iron core, thus turning the implosion into an explosion. This produces shock waves that move up as an expanding shell of gas and dust called a \textit{supernova remnant}.    
\end{enumerate}

\textbf{Classification of supernovae}

Supernovae can be classified according to their light curves\footnote{Light curve is a graph of SN apparent magnitude as a function of time.} and the absorption line of different chemical elements that appear in their spectra.

The first element considered for classification of SNe is obviously the first element in the periodic table, and the most common element in the Universe i.e. hydrogen. If a supernova spectrum contains lines of hydrogen (Balmer series in the visual part of the electromagnetic spectrum) it is classified as a Type II supernova, otherwise it is a Type I supernovae. In each of these two types one can have further subdivisions according to the presence of lines from other elements or the shape of the light curve. If the spectrum of Type I supernovae, contain a singly ionized silicon line at 615 nm, it is called Type Ia, generally abbreviated as SNeIa. It is the most common type of supernova in the cosmos. If in the spectrum of a Type I supernova there is a strong non-ionized helium (He) line at 587.6 nm present, it is called Type Ib SN, and in the case when He lines are not present then that type of supernova is classified as Type Ic. Similarly Type II supernovae are also subdivided on the basis of their light curves (for detail see\cite{2001ASSL..264..199C,Ferreras:2002xk}).

A Type Ia supernova occur when the mass of the white dwarf in a binary system exceeds the Chandrasekhar limit ($1.38 M_{\odot}$)\cite{Chandrasekhar1931ApJ....74...81C} by absorption of gas from other stars and its surroundings. Due to the fact the absolute luminosity of a Type Ia SN at peak brightness can be determined from how fast the explosion dims, Type Ia SN are standardizeable candles and the distance of a SNIa can be determined by measuring its observed (apparent) luminosity. Thus SNIa are very bright `standard candles', which makes then very useful candidates for measuring the geometry of the Universe.

It is conventional to use apparent magnitude $m$ from Eq.\ (\ref{eq:dm2}) as a measure of the brightness of stars observed from earth. To understand this quantity more deeply, let's consider two stars $A$ and $B$ whose apparent fluxes are $\mathcal{F}_A$ and $\mathcal{F}_B$ with corresponding apparent magnitudes $m_A$ and $m_B$ given as
\begin{eqnarray}
m_A=-2.5\ \mathrm{log_{10}}\left(\frac{\mathcal{F}_A}{\mathcal{F}_{\mathrm{ref}}}\right), \hspace{1 cm} m_B=-2.5\ \mathrm{log_{10}}\left(\frac{\mathcal{F}_B}{\mathcal{F}_{\mathrm{ref}}}\right).
\label{eq:du1}
\end{eqnarray}
By subtracting we get
\begin{eqnarray}
 m_A-m_B=-2.5\ \mathrm{log_{10}}\left(\frac{\mathcal{F}_A}{\mathcal{F}_{B}}\right).
\label{eq:du2}
\end{eqnarray}
This equation means that if $m_A=1$ and $m_B=3.5$, then star $A$ is 10 times brighter than the star $B$.\footnote{According to definition of apparent magnitude, the brighter the star the smaller its apparent magnitude. Apparent magnitude of the Sun and Moon are $m_{\odot}=-26.5$ and $m_{\mathrm{moon}}=-12.6$ respectively.} As discussed in Sec.\ (\ref{sec:Distance Modulus}) the Absolute Magnitude $M$ of an object in terms of an apparent magnitude $m$ and luminosity distance $d_L$ is
\begin{eqnarray}
\mu=m-M= 5\ \mathrm{log_{10}}\left(\frac{d_L}{10 \mathrm{pc}}\right),
\label{eq:du3}
\end{eqnarray}
where $\mu$ is the distance modulus. In words, the absolute magnitude corresponds to the apparent magnitude the object would have if it were located at the luminosity distance $d_L=10\ \mathrm{pc}$ from the observer.

If the distance is expressed in Mega-parsecs the relation is
\begin{eqnarray}
m-M=5\ \mathrm{log_{10}}d_L+25.
\label{eq:du4}
\end{eqnarray}
It has been observed experimentally that the absolute magnitude of SNIa is around $M=-19$ at the peak of the brightness. 

If we consider two SNeIa, $A$ and $B$, whose apparent magnitudes are $m_A$ and $m_B$ with the respective luminosity distances $d_{L_A}$ and $d_{L_B}$, then by using Eqs.\ (\ref{eq:du2}) and (\ref{eq:ldq1}), we can relate apparent magnitude with the luminosity distance as follows through
\begin{eqnarray}
m_A-m_B=5\ \mathrm{log_{10}}\left(\frac{d_{L_A}}{d_{L_B}}\right).
\label{eq:du5}
\end{eqnarray}
The luminosity distance $d_L(z)$ can be obtained from Eq. (\ref{eq:du4}) by observing the apparent magnitude $m$, because the peak absolute magnitude $M$ is same for any standardized SNIa under the assumption of standard candles. 

The redshift $z$ of the corresponding SNIa can be found by measuring the wavelength, $\lambda$, of the light, using Eqs.\ (\ref{eq:HL4}) and (\ref{eq:HL5}). The observations of many SNIa provides the dependence of distance modulus $\mu$ or luminosity distance $d_L$ on redshift $z$. Comparing the observational data with the theoretical distance, Eq.\ (\ref{eq:ldq13}), it is possible to infer the expansion history of Universe.

\begin{figure}[h!]
\centering
    \includegraphics[height=5.4in]{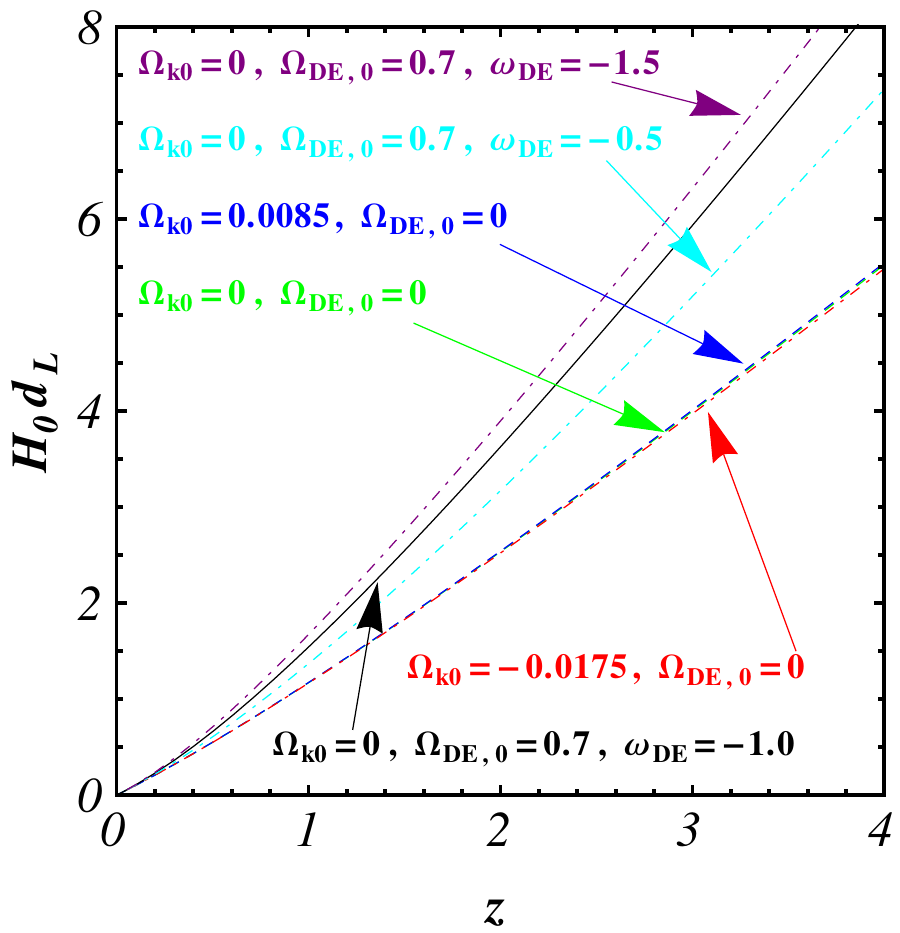}
\caption{The luminosity distance $d_L$ versus the redshift $z$ for six cases: (a) A flat Universe without dark energy (green dashed line, not very clearly visible since it lies between the blue dashed and red dot-dashed lines), (b) An open Universe ($\Omega_{k0}=0.0085$) without dark energy (blue dashed line), (c) A closed Universe ($\Omega_{k0}= - 0.0175$) without dark energy (red dot-dashed line), (d) A flat Universe  with cosmological constant  with $\Omega_{DE,0}=0.7$ and $\omega_{DE}=-1$ (black solid line), (e) A flat Universe with dark energy whose equation of state parameter $\omega_{DE}=-0.5$ and density $\Omega_{DE}=0.7$ (cyan dot-dashed line), (f) A flat Universe with dark energy whose equation of state parameter $\omega_{DE}=-1.5$ and density $\Omega_{DE}=0.7$ (cyan dot-dashed line). The presence of dark energy leads to a larger luminosity distance relative to the case without it. In the open Universe the luminosity distance also gets larger than that in the flat Universe. Also, dark energy with smaller $\omega_{DE}$ and larger $\Omega_{DE}$ leads to a larger luminosity distance.}
\label{fig:luminositydistance}
\end{figure}

Let's consider the special case in which the Universe is dominated by non-relativistic matter and dark energy with equation of state parameter $\omega_{DE}(z)$. In this case,\footnote{Here we are neglecting the contribution of radiation, $\Omega_{r0}=0$.} Eq.\ (\ref{eq:AOU3}) will take the form
\begin{eqnarray}
H(z)=H_0\left[\Omega_{m0}(1+z)^3 +\ \Omega_{DE,0}\ \mathrm{exp}\left\{\int \limits^{z}_{0}\frac{3\big[1+\omega_{DE}(z')\big]}{1+z'}\mathrm{d}z'\right\}\hspace{1 cm}  \right.\nonumber \\
\nonumber \\
     \left. \vphantom{\left\{\int \limits^{z}_{0}\frac{3(1+\omega_{DE})}{1+z'}\mathrm{d}z'\right\}}+\left(1-\Omega_{DE}-\Omega_{m0}\right)(1+z)^2\right]^{1/2},
\label{eq:du6}
\end{eqnarray}
and the luminosity distance for $z \ll 1$ using the approximation (\ref{eq:app3}) is given by
\begin{eqnarray}
d_L(z)=\frac{1}{H_0}\left[z+\frac{1}{4}\left(1-3\omega_{DE}\Omega_{DE,0}+\Omega_{k0}\right)z^2+O(z^3)\right].
\label{eq:du7}
\end{eqnarray}
When we consider a flat Universe without dark energy ($\Omega_{DE,0}=0$), then the luminosity distance for $z \ll 1$ is
\begin{eqnarray}
d_L(z)=\frac{1}{H_0}\left[z+\frac{1}{4}z^2+O(z^3)\right].
\label{eq:du8}
\end{eqnarray}
Figure\ (\ref{fig:luminositydistance}) shows the plot of luminosity distance $H_0 d_L$ vs redshift $z$ for different Universes.
This plot shows that incorporating dark energy (black solid line) leads to a larger luminosity distance relative to the case when dark energy is neglected (red, green and blue plots). Also, it is clear from Fig.\ (\ref{fig:luminositydistance}) that for smaller $\omega_{DE}$ and for larger $\Omega_{DE}$ the luminosity distance becomes larger significant. Here a point to note is that an open Universe $\Omega_{k0}>0$ ($k<0$) has a larger  luminosity distance than a flat Einstein-de Sitter Universe.\footnote{The green dashed line is not very visible at this resolution. It is inbetween the blue dashed (open with out dark energy) and red dot-dashed (closed with out dark energy) lines as expected.} But since the curvature of the Universe is constrained very close to flat\cite{Komatsu2009} ($-0.0175<\Omega_{k0}<0.0085$) from WMAP 5 year data, it is very difficult to increase the luminosity distance in an open Universe without dark energy as compared to a flat Universe with dark energy.

We now consider some examples of the observational apparent luminosity verses redshift for high redshift Type Ia supernovae ($0.2 <z < 0.8$),\cite{Riess98, Perlmutter99} \footnote{Infact Riess and Perlmutter also used the previously known low-redshift data as well.}, to illustrate how luminosity distance is determined.

In Table (\ref{tab:lowredshift}) we consider the apparent magnitude data in low redshift range. For low $z$ the luminosity distance is $d_L \approx {z}/{H_0}$ from Eq.\ (\ref{eq:app3}). Considering a Hubble constant,  $H_0 = 70$ km s$^{-1}$ Mpc$^{-1}$ leads to the absolute magnitude of both SNeIa being approximately $-19$ as discussed above. 
\begin{table}[htb]
\begin{center}
\begin{threeparttable}[b]
\caption{Illustrate SNeIa Apparent Magnitude Data at Low $z$}
\vspace{1mm}
\begin{tabular}{ccccc}
\hline\hline
\multirow{3}{*}{Name of SN}  &  \multirow{2}{*}{Redshift} & Apparent           &     Luminosity      & Absolute           \\
			       &           		    &   Magnitude        &  Distance\tnote{a}  &  Magnitude\tnote{b} \\
                               &    $z$    		    &       $m$          &     $H_0d_L$        &        $M$          \\ 
\hline

1990O			       &  0.03     		    &     16.26		&       0.03	       &       -19.28	     \\
\hline

1992bg			       &  0.036   		    &     16.66		&       0.036	      &       -19.29	      \\
\hline\hline
\end{tabular}
\begin{tablenotes}
\item[a]{Here redshift $z \ll 1$ so $d_L H_0=z$ (considering $c=1$).} 
\item[b]{Absolute magnitude is computed from Eq.\ (\ref{eq:dm2}) in the small $z$ approximation thus $M=m-5\ \mathrm{log_{10}}\left(\frac{z}{10\  \mathrm{pc}\  H_0}\right)$. Here we assume $H_0 = 70$ km s$^{-1}$ Mpc$^{-1}$.}
\end{tablenotes}
\label{tab:lowredshift}
\end{threeparttable}
\end{center}
\end{table}
Let's consider 2 high redshift SNeIa data points from Perlmutter \textit{et al.}\cite{Perlmutter99} shown in the Table\ (\ref{tab:highredshift}).
\begin{table}[htb]
\begin{center}
\begin{threeparttable}[b]
\caption{Illustrate SNeIa Apparent Magnitude Data at High $z$\cite{Perlmutter99}}
\vspace{1mm}
\begin{tabular}{cccc}
\hline\hline
\multirow{3}{*}{Name of SN}\hspace{7mm}  &  \multirow{2}{*}{Redshift}\hspace{7mm} & Apparent \hspace{7mm}          &     Luminosity          \\
			       &           		    &   Magnitude \hspace{7mm}       &  Distance\tnote{a}      \\
                               &    $z$    \hspace{7mm}		    &       $m$    \hspace{7mm}      &     $H_0d_L$           \\ 
\hline

1997R	\hspace{7mm}		       &  0.657    \hspace{7mm} 		    &     23.83	\hspace{7mm}	&       0.920	       	    \\
\hline

1995ck \hspace{7mm}	       &  0.656   	\hspace{7mm}	    &     23.57	\hspace{7mm}	&       0.817	     	   \\
\hline\hline
\end{tabular}
\begin{tablenotes}
\item[a]{To calculate the luminosity distance, we assumed an absolute magnitude $M=-19.15$. Note that in this case the small redshift approximation is not valid hence we use Eq.\ (\ref{eq:du4}) for the computation of the luminosity distance.} 
\end{tablenotes}
\label{tab:highredshift}
\end{threeparttable}
\end{center}
\end{table}
Here, luminosity distance is calculated by assuming absolute magnitude $M=-19.15$. Also for large redshift we cannot use approximate luminosity distance ${z}/{H_0}$, instead we use the more general equation (\ref{eq:du4}).

Let's consider a flat Universe ($\Omega_{k0}=0$) with dark energy whose equation of state parameter is $\omega_{DE} \approx -1$ (i.e. a cosmological constant). Then from Eq.\ (\ref{eq:MCU8}):
\begin{eqnarray}
E(z)=\sqrt{\Omega_{m0}(1+z)^3+\Omega_{DE,0}}.
\label{eq:du9}
\end{eqnarray}
In this case the luminosity distance is given by
\begin{eqnarray}
d_L(z)=\frac{(1+z)}{H_0}\mathlarger{\mathlarger{\mathlarger{\int}}}\limits^{z}_{0}\frac{dz'}{\Big[\Omega_{m0}(1+z')^3+\Omega_{DE,0}\Big]^{1/2}},
\label{eq:du10}
\end{eqnarray}
which needs to be evaluated numerically. In order to satisfy observational data $H_0d_L(z=0.657)=0.920$ for $1997$R we require that $\Omega_{DE,0}=0.7$. Similarly we get $\Omega_{DE,0}=0.38$ from the $1995$ck data. Both indicate the requirement of dark energy to make an agreement between the theoretical predictions and the observations. 

Of course, two data points are not enough to conclude that of dark energy is required and that the current cosmological expansion is accelerating. Using 570 SNeIa data points,\cite{suzuki2012} Farooq and Ratra,\cite{Farooq:2012ev} put constraints on $\omega_{DE}$ (considering a time-independent equation of state parameter) which came out to be $-0.97 <\omega_{DE}< -1.03$ ($1\sigma$ interval) and the $1\sigma$ constraints on non-relativistic  matter density was found to be $0.24 <\Omega_{m0} < 0.34$.

All this discussion leads to the conclusion that dark energy can provide a good fit to the observational data.

\subsection{Baryonic Acoustic Oscillation (BAO) Measurements} 

Before the recombination epoch, baryons are tightly coupled to photons so sound waves oscillations will be imprinted in the baryon perturbations as well in CMB temperature anisotropies. Eisenstein $et\ al.$ \cite{Eisenstein2005} were the first, in 2005, to report on baryon acoustic oscillation peak in the large-scale correlation function measured from a spectroscopic sample of 46,748 luminous red galaxies observed by the Sloan Digital Sky Survey (SDSS). The detection of BAOs provides another independent technique for probing dark energy.

The sound horizon at which baryons were released from the Compton drag of photons determines the location of the baryon acoustic oscillation peak length scale. This epoch, called the  drag epoch, occurs at the redshift $z_{d}$ and the length scale is
\begin{eqnarray}
r_s(z_d)=\int\limits^{\eta_{\mathrm{drag}}}_{0}c_s(\eta)\ \mathrm{d}\eta,
\label{eq:bao1}
\end{eqnarray}
where $c_s$ is the speed of sound and $\eta$ is conformal time. We emphasize that the drag epoch is not the same that as recombination epoch at which the photons were release from the electrons. 

Using the fitting formula of $z_d$ by Eisenstein and Hu,\cite{Eisenstein1998} $z_d$ and $r_s(z_d)$ are constrained (from WMAP 5 year data\cite{Komatsu2009}) to be $z_d \approx 1020.5 \pm 1.6$ and $r_s(z_d)=153.3 \pm 2.0$ Mpc.

The observed angular and redshift distributions of galaxies can be characterized by a power spectrum $P(k_{\bot}, k_{\|})$ in redshift $z$ space. Here $k_{\bot}$ and $k_{\|}$ are the wave numbers parallel and perpendicular to the line of sight respectively. In principle, we can measure the following two ratios\cite{amendola2010dark,Masatoshi2009}
\begin{eqnarray}
\theta_s(z)=\frac{r_s(z_d)}{(1+z)d_A(z)}, \hspace{1 cm} \delta z_s(z)=\frac{r_s(z_d)H(z)}{c},
\label{eq:bao2}
\end{eqnarray}
where $c$ is the speed of light reintroduced for clarity. Here $\theta_s(z)$  characterizes the angle distribution orthogonal to the line of sight, while $\delta z_s$  corresponds to the oscillation along the line of sight.

Most current BAO observations cannot measure both $\theta_s(z)$ and $\delta z_s(z)$ independently, but form a spherically averaged spectrum, it is possible to determine a combined distance scale ratio given as\cite{Eisenstein2005}
\begin{eqnarray}
\left[\theta_s^2(z)\delta z_s(z)\right]^{1/3}=\frac{r_s(z_d)}{\Big[(1+z)^2d_A^2(z)c/H(z)\Big]^{1/3}},
\label{eq:bao3}
\end{eqnarray}
or an effective distance ratio
\begin{eqnarray}
D_V(z)=\left[(1+z)^2 d_A^2(z)\frac{c}{H(z)}\right]^{1/3}.
\label{eq:bao4}
\end{eqnarray}
In 2005, Eisenstein \text{et al.}\cite{Eisenstein2005} obtained $D_V(z)=1370 \pm 64$ Mpc at $z=0.68$.

In 2007, Percival \text{et al.},\cite{Percival2007} measured the effective distance ratio defined by\footnote{In literature this is sometimes denoted by $d_z$.\cite{Percival2007,Percival2010}}
\begin{eqnarray}
r_{\mathrm{BAO}}(z)=\frac{r_s(z_d)}{D_V(z)},
\label{eq:bao5}
\end{eqnarray}
at two different redshifts, $r_{\mathrm{BAO}}(z=0.2)=0.1980\pm 0.0058$ and $r_{\mathrm{BAO}}(z=0.35)=0.1094\pm 0.0033$. This is based on data from the 2-degree Field (2-dF) Galaxy redshift survey.
\begin{figure}[h]
\centering
    \includegraphics[height=5.0in]{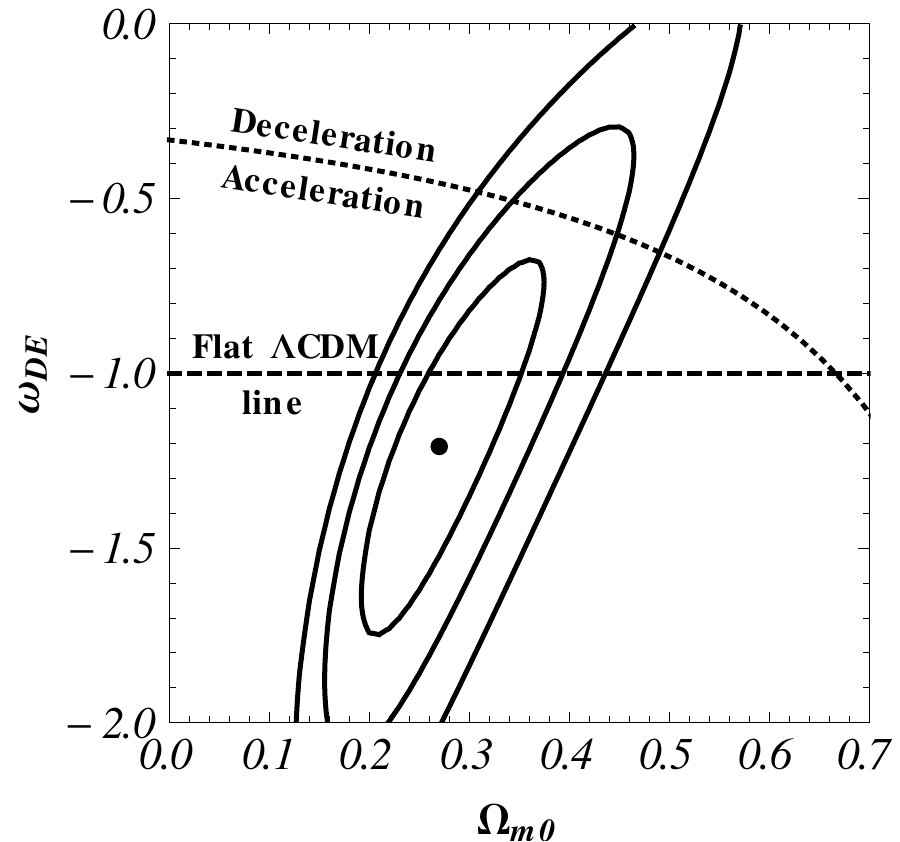}
\caption{$1\sigma$, $2\sigma$, and $3\sigma$ constraint contours leaving $\omega_{DE}$ as free parameter. The dashed horizontal lines at $\omega_{DE}=-1$ correspond to spatially-flat $\Lambda$CDM models and the curved dotted lines demarcate zero-acceleration models. The constraints contours are obtained using the 6 BAO data given in Table\ (\ref{tab:BAODATA}) the plot is taken from Farooq \textit{et al.}\cite{Farooq:2012ev} The method to obtain these contours is explained in Chapter\ (\ref{Chapter5}). The solid dot is best fit point located at $(\Omega_{m0},\omega_{X})=(0.27,-1.21)$. The corresponding $\chi^2_{\mathrm{min}}=5.5$.}
\label{fig:bao}
\end{figure}
In 2010, Percival \text{et al.}\cite{Percival2010} used SDSS data to determine
$r_{\mathrm{BAO}}(z=0.2)=0.1905\pm 0.0061$ and $r_{\mathrm{BAO}}(z=0.35)=0.1097\pm 0.0036$. In 2011, using 6dF Galaxy Survey (6dFGS) measurements, Beutler \text{et al.}\cite{Beutler11} found $r_{\mathrm{BAO}}(z=0.106)=0.3360\pm 0.015$.

Then in 2011, Black \text{et al.}\cite{blake11} considered the acoustic parameter introduced by Eisenstein \text{et al.},\cite{Eisenstein2005}
\begin{eqnarray}
A(z) \equiv \frac{100 D_V(z)\sqrt{\Omega_m h^2}}{cz},
\label{eq:bao6}
\end{eqnarray}
and based on the data from WiggleZ Dark Energy Survey found $A(z)$ at three different redshifts [given in the Table\ (3) of Blake \cite{blake11}]: $A(z=0.44)=0.474 \pm 0.034$, $A(z=0.6)=0.442 \pm 0.020$, and $A(z=0.44)=0.424 \pm 0.021$. These are summarized in Table\ (\ref{tab:BAODATA}).

Using the data from Percival \textit{et al.}\cite{Percival2010}, Blake \textit{et al.}\cite{blake11}, and Beutler \textit{et al.}\cite{Beutler11}, Farooq \textit{et al.}\cite{Farooq:2012ev} found the constraints on the $\omega_{DE}$ shown in the Fig.\ (\ref{fig:bao}). Figure\ (\ref{fig:bao}) shows the constraints obtained using BAO observational data for the general but time independent equation of state parameters $\omega_{DE}$ for dark energy, also provides evidence for a current epoch cosmological expansion that is accelerating, and thus support for the existence of dark energy.

\section{Spatially-Flat $\Lambda$CDM Model (Standard Cosmological Model)}

Most currently available cosmological data are largely consistent with a spatially-flat, cosmological constant ($\Lambda$) dominated model with $\Omega_{\Lambda}\approx 0.7$ and with the rest of the energy being non-relativistic (baryonic and cold dark) matter $\Omega_{m0} \approx 0.3$, where baryonic matter makes up only 5\% of the Universe ($\Omega_{b}\approx 0.05$). The cosmological constant can be considered to be an ideal fluid with equation-of-state parameter $\omega_{\Lambda}=-1$, hence from Eq.\ (\ref{eq:energycon}) its energy density does not change with time ($\rho_{\Lambda}=\rho_{\Lambda 0} \ \ \forall$ t). This spatially-flat $\Lambda$CDM model\cite{peebles84} is often referred to as the ``standard model".
In this model the Hubble parameter, from the Friedmann equation (\ref{eq:MCU8}), obeys
\begin{eqnarray}
H(z,H_0,\textbf{p})&=&H_{0}\left[\Omega_{m0}(1+z)^3+(1-\Omega_{m0})\right]^{1/2}.
\label{eq:SFLCDMMHz1}
\end{eqnarray}
Here  we have neglected the radiation term\footnote{Radiation energy density $\rho_{\mathrm{rad}}\propto a^{-4}$ from Eq.\ (\ref{eq:energycon}), hences its contribution dies quickly, we usually do not account for it in calculations.} and set $1-\Omega_0=0$ for this spatially-flat case so $\Omega_{\Lambda}=1-\Omega_{m0}$. This is one parameter model, with $\textbf{p}=\Omega_{m0}$ being the parameter characterize the model, and $H_0$ is the current value of Hubble parameter.

Although the spatially-flat $\Lambda$CDM model is a reasonable fit to most observations, the data is not yet precise enough to rule out other models of dark energy.

\section{$\Lambda$CDM with Non-Zero Curvature}

This is a generalization of the standard model of cosmology within the frame work of general relatively. Here we also consider spatial curvature $k$ with present value density parameter denoted by $\Omega_{k0}$, as defined in Eq.\ (\ref{eq:MCU7}). In this case the Friedmann equation (\ref{eq:MCU8}) for the evolution of the Hubble parameter is
\begin{eqnarray}
H(z)=H_0 \left[\Omega_{m0}(1+z)^3+\Omega_{\Lambda}+(1-\Omega_{m0}-\Omega_{\Lambda})(1+z)^2\right]^{1/2},
\label{eq:LCDMHz1}
\end{eqnarray}
where, since $\Omega_{m0}+\Omega_{k0}+\Omega_{\Lambda}=1$, we have replaced $\Omega_{k0}$ by $1-\Omega_{m0}-\Omega_{\Lambda}$ and this is with parameters $\textbf{p}=(\Omega_{m0},\Omega_{\Lambda})$.

\subsection{No Big Bang Region in the ${\Lambda}$CDM Model}
\label{subsection:NBBL}
Returning to the general case of models with a combination of energy in the vacuum $(\Lambda)$ and normal components, we have to distinguish three cases. For models that start from a big bang (in which case radiation dominates at early times), the Universe will either recollapse or expand forever. The latter outcome becomes more likely for low densities of matter and radiation and high vacuum density. It is however also possible to have models in which there is no big bang: the Universe was collapsing in the distant past, but was slowed by the repulsion of a positive $\Lambda$ term and underwent a ``bounce'' to reach its present state of expansion. 

We can compute an analytical formula $\Omega_{\Lambda}(\Omega_{m0})$ which separates models with a big bang from those without a big bang. It is given as follows:

\begin{equation}
\label{eq:nobigbang1}
\Omega_{\Lambda}(\Omega_{m0}) \geq \left\{
     \begin{array}{lc}
        4\ \Omega_{m0}\ \mathrm{cosh^3} \left[\frac{1}{3}\mathrm{cosh^{-1}\left(\frac{1-\Omega_{m0}}{\Omega_{m0}}\right)}\right], & 0 \leq \Omega_{m0} \leq \frac{1}{2}\\
															     &					  \\
        4\ \Omega_{m0}\ \mathrm{cos^3} \left[\frac{1}{3}\mathrm{cos^{-1}\left(\frac{1-\Omega_{m0}}{\Omega_{m0}}\right)}\right], & \frac{1}{2} \leq \Omega_{m0} \\
     \end{array}.
   \right.
\vspace{1cm}
\end{equation}
\textbf{The derivation follows}

The Friedman equations, which are derived from Einstein's theory of general relativity in Chapter\ (\ref{Chapter1}), are
\begin{eqnarray}
\left(\frac{\dot a}{a}\right)^2=\frac{8 \pi G}{3}\rho+\frac{\Lambda}{3}-\frac{k}{a^2},
\label{eq:nobigbang2}
\end{eqnarray}
\begin{eqnarray}
\frac{\ddot a}{a}=-\frac{4 \pi G}{3}\left(\rho+3p\right)+\frac{\Lambda}{3}.
\label{eq:nobigbang3}
\end{eqnarray}
Equation\ (\ref{eq:nobigbang2}) can be rewritten as:
\begin{eqnarray}
\frac{k}{a^2}&=&H^2(t)\left[\Omega_m(t)-1\right]+\frac{\Lambda}{3}=H^2(t)\left[\Omega_m(t)-1+\frac{\Lambda}{3H^2(t)}\right].
\label{eq:nobigbang4.1}
\end{eqnarray}
At the present time $a=a_0=1$ and $H=H_0$, hence the above equation takes the form:
\begin{eqnarray}
k= H_0^2\left[\Omega_{m0}-1+\frac{\Lambda}{3H_0^2}\right].
\label{eq:nobigbang4}
\end{eqnarray}
A Universe with a large value of $\Omega_{\Lambda} \equiv {\Lambda}/{3H^{2}_{0}}$ (say $\Omega_{\Lambda_c}$), loses the big bang in the beginning and we instead obtain an Eddington-Lema\^itre (EL) model, asymptotic to Einstein's static model in the infinite past.\cite{Felten1986} For $\Omega_{\Lambda} \geq \Omega_{\Lambda_c}$, we obtain models that collapse from infinity, reach a minimum value $a(t)$ and then expand again. These Universes are sometimes called ``catenary Universes". Here it is important to note that the catenary Universes, like the re-collapsing models with large negative $\Lambda$, are time symmetric about their extrema in $a$, where $\dot a(t)=0$. This is expected, because the Friedman Eq.\ (\ref{eq:nobigbang2}) is manifestly time symmetric: if $a(t)$ is a solution, $a(-t)$ is also.

\begin{figure}[h]
\centering
    \includegraphics[height=4.2in]{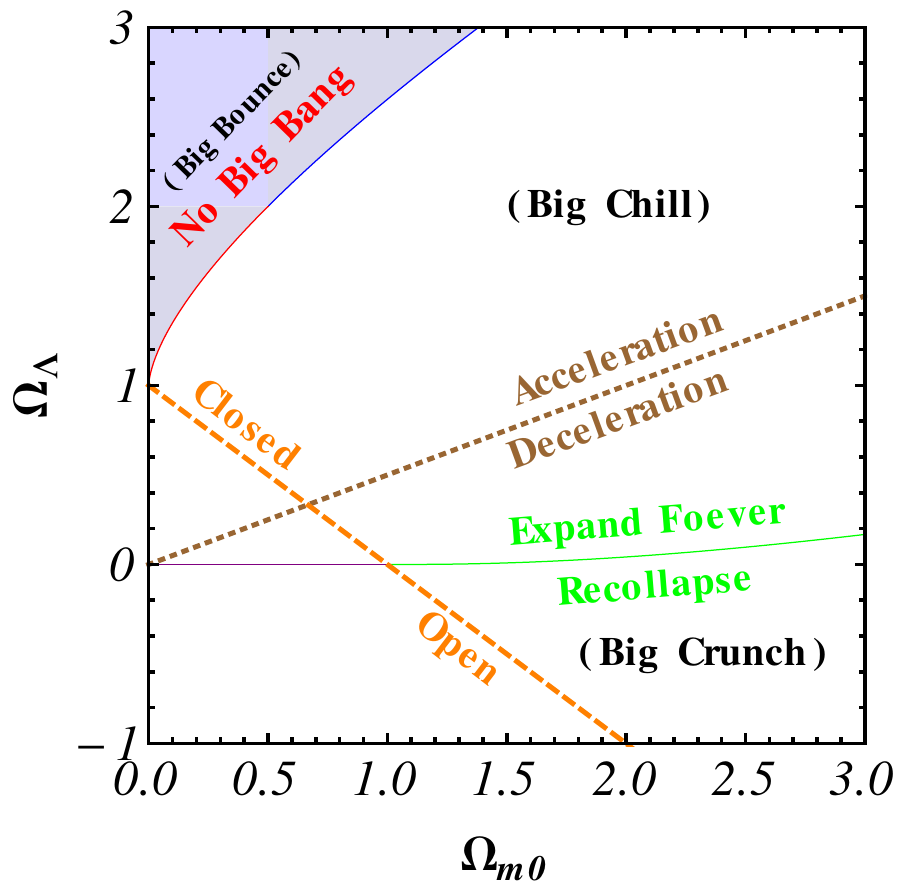}
\caption{Different regions of $\Omega_{m0}-\Omega_{\Lambda}$ plane of the two-parameter $\Lambda$CDM model represent different evolutionary cosmological histories. The brown dotted line [which is the plot of Eq.\ (\ref{eq:flza6})] separates accelerating and decelerating Universes. The Orange dashed line [which is the plot of Eq.\ (\ref{eq:flza1})] demarcates closed and open Universes. The gray portion of the parameter plane above the red and the blue curves at the left top [the curves are the plot of Eq.\ (\ref{eq:nobigbang20})] represent Universes with no big bang back in time. These Universes are called ``Big Bounce" Universes because they started contracting from a non zero size, reached a minimum size, and then start expanding. The purple line and the green curve [which are the plot of Eq.\ (\ref{eq:cofe13})] distinguish between the Universes that will expand forever those that will recollapse. Universes that will recollapse lie below the purple-green curve ``Big Crunch", while Universe above the purple-green-line are in the ``Big Chill", in the region labeled.}
\label{fig:ConditionofLCDM}
\end{figure}

It is of interest to obtain an analytical expression for the critical value $\Omega_{\Lambda}$ at which the Universe no longer starts from a big bang. To do this we note that the critical (EL) model is asymptotic to a static Einstein model in the infinite past, so that in that state it must satisfy $\dot a=\ddot a=p=0$,\footnote{This is the way no big bang equation is derived by Felten and Isaacman.\cite{Felten1986} We can reach the same result by requiring just $\dot a=p=0$.} then $\Omega_\Lambda=\Omega_{\Lambda_c}$, $\Lambda=\Lambda_c$ and $a=a_c$ with these conditions, using Eqs.\ (\ref{eq:nobigbang2}) and (\ref{eq:nobigbang3}), we can require:
\begin{eqnarray}
\left(\frac{\dot a}{a}\right)^2_c+2\left(\frac{\ddot a}{a}\right)_c&=&0,
\label{eq:nobigbang5}
\end{eqnarray}
\begin{eqnarray}
\Rightarrow \cancel{\frac{8\pi G}{3}\rho_c}+\frac{\Lambda_c}{3}-\frac{k}{a^{2}_{c}}-\cancel{\frac{8\pi G}{3}\rho_c}+\frac{2\Lambda_c}{3}=0,
\label{eq:nobigbang6}
\end{eqnarray}
this can be simplified as:
\begin{eqnarray}
\Lambda_c=\frac{k}{a^{2}_{c}}.
\label{eq:nobigbang7}
\end{eqnarray} 
From Eq.\ (\ref{eq:nobigbang3}) at these conditions leads to:\footnote{Here $\rho_c(t)$ means the value of the non-relativistic matter density in the Universe at the time when $\dot a=\ddot a=0$. This should not be confused this with the critical density of the Universe, which is the energy density required to make the Universe spatially flat.}
\begin{eqnarray}
\Lambda_c&=&4\pi G \rho_c(t)=\frac{3}{2}\left(\frac{8\pi G}{3}\rho_0\right)\frac{1}{a^{3}_{c}}.
\label{eq:nobigbang8}
\end{eqnarray} 
Defining,
\begin{eqnarray}
\mathcal{C} \equiv \frac{8\pi G}{3}\rho_0 =H^{2}_{0}\Omega_{m0},
\label{eq:nobigbang9}
\end{eqnarray} 
we can write Eq.\ (\ref{eq:nobigbang8}) as
\begin{eqnarray}
\Lambda_c=\frac{3}{2}\ \mathcal{C}a^{-3}_{c}.
\label{eq:nobigbang10}
\end{eqnarray} 
Comparing Eqs.\ (\ref{eq:nobigbang7}) and (\ref{eq:nobigbang10}) we have
\begin{eqnarray}
\Lambda_c=k\ a^{-2}_{c}=\frac{3}{2}\ \mathcal{C}a^{-3}_{c}.
\label{eq:nobigbang11}
\end{eqnarray} 
Solving Eq.\ (\ref{eq:nobigbang11}) for $a_c$ we find
\begin{eqnarray}
a_c=\frac{3\mathcal{C}}{2k}.
\label{eq:nobigbang12}
\end{eqnarray} 
Plugging the value of $a_c$ back in Eq.\ (\ref{eq:nobigbang10}) and simplifying,
\begin{eqnarray}
\Lambda_c=\frac{4}{9}\mathcal{C}^{-2}k^3,
\label{eq:nobigbang13}
\end{eqnarray}
using the definition of $\mathcal{C}$ from Eq.\ (\ref{eq:nobigbang9}) leads to
\begin{eqnarray}
\Lambda_c=\frac{4}{9}\left(\Omega_{m0}H^{2}_{0}\right)^{-2}k^3,
\label{eq:nobigbang14}
\end{eqnarray}
and dividing both sides by $\left(\Omega_{m0}H^{2}_{0}\right)$ gives
\begin{eqnarray}
\frac{\Lambda_c}{\left(\Omega_{m0}H^{2}_{0}\right)}=\frac{4}{9}\left(\frac{k}{\Omega_{m0}H^{2}_{0}}\right)^{3},
\label{eq:nobigbang15}
\end{eqnarray}
Plugging the value of $k$ from Eq.\ (\ref{eq:nobigbang4}) in to Eq.\ (\ref{eq:nobigbang15}) results in:
\begin{eqnarray}
\frac{\Lambda_c}{\left(\Omega_{m0}H^{2}_{0}\right)}=\frac{4}{9}\left[\frac{\cancel{H_0^2}\left(\Omega_{m0}-1+\frac{\Lambda}{3H_0^2}\right)}{\Omega_{m0}\cancel{H^{2}_{0}}}\right]^{3}.
\label{eq:nobigbang16}
\end{eqnarray}
and simplifying this leads to
\begin{eqnarray}
\left(\frac{\Lambda_c}{12\Omega_{m0}H^{2}_{0}}\right)^{1/3}=\frac{1}{3}\left[\frac{\Omega_{m0}-1}{\Omega_{m0}}+4\ \frac{\Lambda_c}{12\Omega_{m0}H^{2}_{0}}\right].
\label{eq:nobigbang17}
\end{eqnarray}
Defining
\begin{eqnarray}
x\equiv \left(\frac{\Lambda_c}{12\Omega_{m0}H^{2}_{0}}\right)^{1/3},
\label{eq:nobigbang18}
\end{eqnarray}
the above equation takes the from:
\begin{eqnarray}
x^3-\frac{3}{4}x+\frac{1}{4}\left(\frac{\Omega_{m0}-1}{\Omega_{m0}}\right)=0.
\label{eq:nobigbang19}
\end{eqnarray}
This is a standard cubic equation whose, solutions are discussed in Appendix (\ref{Appendix:C}). 
Using Eqs.\ (\ref{eq:cubic9}) and (\ref{eq:cubic13}) the value of $x$ is
\begin{equation}
\label{eq:nobigbang20}
x= \left\{
     \begin{array}{lc}
        \mathrm{cosh} \left[\frac{1}{3}\mathrm{cosh^{-1}\left(\frac{1-\Omega_{m0}}{\Omega_{m0}}\right)}\right], & 0 \leq \Omega_{m0} \leq \frac{1}{2}\\
															     &					  \\
        \mathrm{cos} \left[\frac{1}{3}\mathrm{cos^{-1}\left(\frac{1-\Omega_{m0}}{\Omega_{m0}}\right)}\right], & \frac{1}{2} \leq \Omega_{m0} \\
     \end{array}.
   \right.
\end{equation}
We can write these no big bang conditions in terms of $\Omega_{\Lambda_c}$, using Eq.\ (\ref{eq:nobigbang18}), as
\begin{equation}
\label{eq:nobigbang20}
\Omega_{\Lambda}(\Omega_{m0}) \geq \Omega_{\Lambda_c}=\left\{
     \begin{array}{lc}
        4\ \Omega_{m0}\ \mathrm{cosh^3} \left[\frac{1}{3}\mathrm{cosh^{-1}\left(\frac{1-\Omega_{m0}}{\Omega_{m0}}\right)}\right], & 0 \leq \Omega_{m0} \leq \frac{1}{2}\\
															     &					  \\
        4\ \Omega_{m0}\ \mathrm{cos^3} \left[\frac{1}{3}\mathrm{cos^{-1}\left(\frac{1-\Omega_{m0}}{\Omega_{m0}}\right)}\right], & \frac{1}{2} \leq \Omega_{m0} \\
     \end{array}.
   \right.
\end{equation}
These are plotted (in red and blue) in Fig.\ (\ref{fig:ConditionofLCDM})

\subsection{Condition for Recollapse in the ${\Lambda}$CDM Model}

As $a\rightarrow 0$, in the past space curvature and vacuum energy is negligible and the $\Lambda$CDM model will behave as an Einstein-de Sitter model until the radiation become important. As $a \rightarrow \infty$, in the future space curvature and the matter energy becomes negligible and the $\Lambda$CDM model will asymptote to the de Sitter model, unless the scale factor never reaches $\infty$ because the $\Lambda$CDM model Universes re-collapses at a finite time. Two possibilities for the evolution of $\Lambda$CDM model are:
\begin{itemize} 
     \item[$\star$] $\Omega_{\Lambda} < 0$: always decelerates and recollapses [as explained in Section\ (\ref{subsection:NBBL})] as vacuum energy is always going to dominate.
     \item[$\star$] $\Omega_{\Lambda} \geq 0$: recollapse is possible if $\Omega_{m0}$ is large enough to halt the expansion before $\Omega_{\Lambda}$ dominates.
\end{itemize}

To determine the dividing line between perpetual expansion and eventual recollapse, note that collapse requires that the Hubble parameter to pass through $0$ as it changes from positive to negative, so at turn around
\begin{eqnarray}
H^2(t_t)=0=\frac{8\pi G}{3}\ \left[\rho_{m0}\ a^{-3}_{t}+\rho_{\Lambda,0}+\rho_{k0}\ a^{-2}_{t}\right],
\label{eq:cofe1}
\end{eqnarray}
where $t_t$ and $a_t$ is the time and the scalar factor at turn around. Dividing by $H^{2}_{0}$ and using $\Omega_{k0}=1-\Omega_{m0}-\Omega_{\Lambda}$, and rearranging, we obtain
\begin{eqnarray}
\Omega_{\Lambda}\ a^{3}_{t} + \left(1-\Omega_{m0}-\Omega_{\Lambda}\right)\ a^{2}_{t}+\Omega_{m0}=0.
\label{eq:cofe2}
\end{eqnarray}
But what we really care about is not really $a_t$, but the range of $\Omega_{m0}$ and $\Omega_{\Lambda}$ for which there is at least one real solution of Eq.\ (\ref{eq:cofe2}). 

Let's find the condition for Eq.\ (\ref{eq:cofe2}) to have one positive real solution. Equation\ (\ref{eq:cofe2}) can be written as
\begin{eqnarray}
a^{3}_{t} + \frac{1-\Omega_{m0}-\Omega_{\Lambda}}{\Omega_{\Lambda}}\ a^{2}_{t}+\frac{\Omega_{m0}}{\Omega_{\Lambda}}=0.
\label{eq:cofe3}
\end{eqnarray}
Comparing this with Eq.\ (\ref{eq:cubic4}), which is standard from of cubic equation, we define the parameters
\begin{eqnarray}
x=a_{t}, \hspace{1cm} P=\frac{1-\Omega_{m0}-\Omega_{\Lambda}}{\Omega_{\Lambda}},\hspace{1cm} Q=\frac{\Omega_{m0}}{\Omega_{\Lambda}}.
\label{eq:cofe3}
\end{eqnarray}

Consider the cubic polynomial
\begin{eqnarray}
y(x)=x^3+Px+Q.
\label{eq:cofe4}
\end{eqnarray}

If $Q<0$, this means that the plot of $y(x)$ will have the negative $y$-intercept and since the co-efficient of $x^3$ is positive the corresponding cubic equation $y(x)=0$ will have at least one positive root.

If $Q>0$ (then we need $P<0$ so that the curve will have negative slope when it intersects the $y$-axis; this means $\Omega_{m0}\geq 1$) then the cubic equation will have at least one negative real solution (from the same argument as above). In this case to make sure that the cubic equation will have at least are positive solution as well, we have to find an $x_0>0$ such that $\frac{dy}{dx}\big|_{x_0}=0$ and $y(x_0)=0$. Thus:
\begin{eqnarray}
\frac{dy}{dx}&=&3x^2+P=0.\nonumber \\
\Rightarrow x_0&=&+\sqrt{\frac{-P}{3}}, \hspace{1cm} \left(\mathrm{we\ will\ consider\ only\ positive\ root}\right)
\label{eq:cofe5}
\end{eqnarray}
and $y(x_0)=0$ leads to:
\begin{eqnarray}
\Rightarrow \left(-\frac{P}{3}\right)^{3/2}+P\left(-\frac{P}{3}\right)^{1/2}+Q=0,
\label{eq:cofe6}
\end{eqnarray}
\begin{eqnarray}
\Rightarrow -4P^3=27Q^2.
\label{eq:cofe7}
\end{eqnarray}
Substituting the values of $P$ and $Q$ from Eq.\ (\ref{eq:cofe3}) we find:
\begin{eqnarray}
\left(1-\Omega_{m0}-\Omega_{\Lambda}\right)^3=-\frac{27}{4}\Omega_{m0}^2\Omega_{\Lambda}.
\label{eq:cofe8}
\end{eqnarray} 
Under the given condition $\Omega_{m0}\geq 1$, we can write:
\begin{eqnarray}
\left(\frac{1-\Omega_{m0}-\Omega_{\Lambda}}{3\Omega_{m0}}\right)^3=-\frac{\Omega_{\Lambda}}{4\Omega_{m0}}.
\label{eq:cofe9}
\end{eqnarray} 
\begin{eqnarray}
\Rightarrow \frac{1-\Omega_{m0}}{3\Omega_{m0}}-\frac{4}{3}\left(\frac{\Omega_{\Lambda}}{4\Omega_{m0}}\right)=-\left(\frac{\Omega_{\Lambda}}{4\Omega_{m0}}\right)^{1/3}.
\label{eq:cofe10}
\end{eqnarray} 
Defining
\begin{eqnarray}
x\equiv \frac{4}{3}\left(\frac{\Omega_{\Lambda}}{4\Omega_{m0}}\right)^{1/3},
\label{eq:cofe11}
\end{eqnarray} 
we can rewrite Eq.\ (\ref{eq:cofe10}) as
\begin{eqnarray}
 x^3-\frac{3}{4}x-\frac{1}{4}\left(\frac{1-\Omega_{m0}}{\Omega_{m0}}\right)=0.
\label{eq:cofe12}
\end{eqnarray}
Using the solution of standard cubic equation from Appendix (\ref{Appendix:C}), the solution of this equation is
\begin{eqnarray}
x=\mathrm{cos}\left[\frac{1}{3}\mathrm{cos^{-1}}\left(\frac{1-\Omega_{m0}}{\Omega_{m0}}\right)+\frac{4\pi}{3}\right],
\label{eq:cofe13}
\end{eqnarray}
and in terms of the parameters of the $\Lambda$CDM model we have
\begin{eqnarray}
\Omega_{\Lambda}=4\ \Omega_{m0}\ \mathrm{cos^{3}}\left[\frac{1}{3}\mathrm{cos^{-1}}\left(\frac{1-\Omega_{m0}}{\Omega_{m0}}\right)+\frac{4\pi}{3}\right].
\label{eq:cofe13}
\end{eqnarray}
The plot of this equation is shown (green curve) in Fig.\ (\ref{fig:ConditionofLCDM})

\subsection{Flat ${\Lambda}$CDM and the Zero Acceleration Line for General ${\Lambda}$CDM}

Since $\Omega_{k0}=1-\Omega_{m0}-\Omega_{\Lambda}$, for the flat case when $\Omega_{k0}=0$ we have
\begin{eqnarray}
\Omega_{\Lambda}=1-\Omega_{m0}.
\label{eq:flza1}
\end{eqnarray}
The plot of this equation is shown (orange dashed line) in Fig.\ (\ref{fig:ConditionofLCDM}).

To derive zero acceleration line, consider the acceleration equation with
only non-relativistic matter and a cosmological constant
\begin{eqnarray}
\frac{\ddot a}{a}=-\frac{4\pi G}{3}\Big[\rho_m\left(1+3\omega_m\right)+\rho_\Lambda\left(1+3\omega_\Lambda\right)\Big].
\label{eq:flza3}
\end{eqnarray}
But as discussed before $\omega_m=0$, and $\omega_{\Lambda}=-1$, hence:
\begin{eqnarray}
\frac{\ddot a}{a}=-\frac{4\pi G}{3}\left[\rho_m-2\rho_\Lambda\right]=-2H_0^2\left[\frac{\Omega_{m0}}{a^3}-2\Omega_\Lambda\right].
\label{eq:flza4}
\end{eqnarray}
For the zero acceleration today we set $\ddot a=0$, and $a=a_0=1$, which gives
\begin{eqnarray}
\hspace{2 cm}\Omega_\Lambda=\frac{1}{2}\Omega_{m0}. \hspace{2 cm} \left[\mathrm{since}\ H_0 \neq 0\right]
\label{eq:flza6}
\end{eqnarray}
The plot of this equation is shown (brown dotted line) in Fig.\ (\ref{fig:ConditionofLCDM})

\section{Potential Problems with $\Lambda$CDM}
\label{sec:problems in LCDM}

The $\Lambda$CDM cosmological model is a well defined, simple, classically consistent, and predictive model which is largely consistent with most of the current cosmological observations. Despite of these successes there are some theoretical and observational reasons to go beyond the standard cosmological model.
\subsection{Theoretical Puzzles}
\subsubsection{Cosmological Constant Puzzle}

The cosmological constant is difficult to motivate from fundamental physics. Since the cosmological constant is equivalent to a constant energy density, it likely receives contributions from many sources. A problematic source is zero point vacuum energy. For example, the standard quantum mechanical harmonic oscillator (or any other bound system) has a energy proportional to the frequency, given by $\hbar\omega /2$ even in the absence of excitations. The zero point energy of some field of mass $m$ with momentum $k$ and frequency $\omega$ is given by $E=\hbar \omega/2=\hbar\sqrt{k^2+m^2}/2$. Summing over the zero point energies of this field up to the cut off scale $k_{\mathrm{max}}$ we obtain for the vacuum energy density
\begin{eqnarray}
\rho_{\mathrm{vac}}=\mathlarger{\mathlarger{\int}}\limits^{k_{\mathrm{max}}}_{0}\frac{\hbar}{2}\sqrt{k^2+m^2}\frac{d^3k}{(2\pi)^3}.
\label{eq:PLCDM1}
\end{eqnarray}     
Since the integral is dominated by short wavelength modes with $k \gg m$, we find
\begin{eqnarray}
\rho_{\mathrm{vac}}\approx \mathlarger{\mathlarger{\int}}\limits^{k_{\mathrm{max}}}_{0}\frac{\hbar}{2} k \frac{4 \pi k^2 dk}{(2\pi)^3}=\frac{k_{\mathrm{max}}^4}{16 \pi^2}.
\label{eq:PLCDM2}
\end{eqnarray}
Quantum field theory is expected to break down at the Planck scale of around $10^{19}$ GeV. If we use this as the cutoff limit in Eq.\ (\ref{eq:PLCDM2}), we get a huge number
 $\rho_{\mathrm{vac}}\approx 10^{74}\ \mathrm{GeV^{4}}$ that exceeds the observed dark energy density ($\rho_{DE,0}\approx 10^{-44}\ \mathrm{GeV^{4}}$)\cite{Weinberg89} by 120 orders of magnitude.\cite{2010arXiv1004.1493T}

In a super-symmetric model every boson has a fermion of equal mass as a super-symmetric partner and the vacuum energies of these partners cancel. Super-symmetry (SUSY), if existent, is believed to be broken at an energy of roughly 1 TeV or so. If we cut off the upper integration limit in Eq.\ (\ref{eq:PLCDM2}) at the energy of SUSY breaking we will still get a difference of around 46 orders of magnitude. This discrepancy between the small measured value of the cosmological constant and the much larger theoretically “expected” values of vacuum energy is known as the ``smallness" problem.\cite{Weinberg89}

One potential explanation of the smallness problem is based on anthropic arguments. In
string theory, multiple vacuum states with all possible values of vacuum energy are possible.
Different causally disconnected patches of the Universe spontaneously choose vacuum states
that are independent of each other. If the Universe is infinite there will always be parts of it
that have a given value, no matter how unlikely, of the vacuum energy and we just happen
to live in one of those regions with a very small value of vacuum energy 
density.\cite{Susskind:2003kw,Bousso:2012dk}
\subsubsection{Coincidence Puzzle}

Another interesting fact that is difficult to explain in the $\Lambda$CDM model is that current energy density of non-relativistic matter $\rho_{m0}$ and that of the cosmological constant $\rho_{\Lambda}$ have a comparable magnitudes, but their relative scaling $\rho_{\Lambda}/\rho_{m}\propto a^3$, implies that $\Lambda$ was completely negligible in the cosmological past and will absolutely dominate in the future. Summarizing, the radiation energy density redshift as $a^{-4}$ hence the Universe was dominated by radiation at very early times, the Universe then became dominated by non-relativistic (baryonic and cold dark) matter (since $\rho_{m} \propto a^{-3}$), and very recently\cite{Farooq:2013hq} the Universe became $\Lambda$ dominated. In the future dark energy will be the only component driving cosmic expansion. If we consider cosmological constant to be set as an initial condition in the very early Universe (when the non-relativistic energy density was very high as compared to the current value of non-relativistic energy density), it seems very unlikely that $\Lambda$ should have a value comparable to that of matter at the present cosmological epoch, when galaxies and other large scale structures have formed. If the cosmological constant was a couple of orders of magnitude higher than what is now observed, the Universe would be empty of large-scale-structure. But if it were just a couple of orders of magnitude smaller $\Lambda$ would be hardly detectable.

The lack of an explanation for why dark energy has the same order as non-relativistic matter density at the present epoch is known as the coincidence problem. One potential resolution of this is again the anthropic principle.\cite{Susskind:2003kw}   

\subsection{Potential Observational Problems\cite{SixPuzzles}}

There are a number of observations that are not consistent with the standard $\Lambda$CDM model of cosmology. It is not yet clear if all of these measurements are definitive. More observational work is needed to settle the issue, some are given here:

\begin{description}
  \item[Large-Scale Velocity Flows] \hfill \\
           $\Lambda$CDM predicts a significantly smaller amplitude and scale of velocity flows than what some observations seems to indicate.
  \item[Brightness of Type Ia Supernovae (SNIa)] \hfill \\
            $\Lambda$CDM predicts fainter SNIa at high redshift then some that are observed.
  \item[Emptiness of Voids] \hfill \\
           $\Lambda$CDM predicts more dwarf or irregular galaxies in the voids than are observed.
  \item[Profiles of Cluster Haloes] \hfill \\
            $\Lambda$CDM predicts shallow, low concentration and less dense profiles, in contrast to the observations which indicate denser high concentration cluster haloes.
  \item[Profiles of Galaxy Haloes] \hfill \\
           $\Lambda$CDM predicts galaxy halo mass profiles with more dense cores and low outer density while lensing and dynamical observations indicate a central core of constant density and a flattish high dark-mass density outer profile.
  \item[Sizable Population of Disk Galaxies] \hfill \\
            $\Lambda$CDM predicts a smaller fraction of disk galaxies than is observed due to recent mergers that are expected to disrupt cold rotationally-supported disks. 
\end{description}

It is interesting the most of the above-mentioned problems are apparently related to the dark matter. On the scale of clusters dark energy would play a significant role in the stability of some of these systems especially if the dark energy was coupled to dark matter through a new, non-gravitational force.

At the moment these discrepancies between observations and theoretical predictions do not carry great weight in the overall picture. The ΛCDM model is in general a good statistical fit to the combined data. These inconsistencies might go away, or could become more pressing, when new high-quality data become available.\cite{ladothesis}

\section{XCDM Parameterization}

The cosmological constant time-independent dark energy can be thought as a spatially homogeneous fluid with equation-of-state parameter $\omega_{\Lambda}=p_\Lambda/\rho_\Lambda=-1$, where $\rho_{\Lambda}$ and $p_\Lambda$ are the fluid energy density and pressure respectively. There is not yet strong observational evidence for the time-independence of dark energy. Hence, we can model dark energy as a spatially-homogeneous $X-$fluid with equation-of-state parameter $\omega_X=p_X/\rho_X<-1/3$, an arbitrary constant, where $\rho_{X}$ and $p_X$ are the energy density and pressure of the $X-$fluid respectively. When $\omega_X=-1$ the XCDM parametrization reduces to $\Lambda$CDM. 
In the more general, $\omega_X<-1/3$ case, the Hubble parameter evolves according to
\begin{eqnarray}
H(z)=H_0 \left[\Omega_{m0}(1+z)^3+\Omega_{X0}(1+z)^{3(1+\omega_X)}+\Omega_{k0}(1+z)^2\right]^{1/2}.
\label{eq:XCDMHz1}
\end{eqnarray}
Since $\Omega_{m0}+\Omega_{k0}+\Omega_{X0}=1$, hence we can write Eq.\ (\ref{eq:XCDMHz1}) explicitly in terms of the three free parameters as 
\begin{eqnarray}
H(z,H_0,\textbf{p})=H_0 \left[\Omega_{m0}(1+z)^3+(1-\Omega_{m0}-\Omega_{k0})(1+z)^{3(1+\omega_X)}+\Omega_{k0}(1+z)^2\right]^{1/2}.
\label{eq:XCDMHz2}
\end{eqnarray}
Here $\textbf{p}=(\Omega_{m0},\omega_X,\Omega_{k0})$ are the parameters of this model. In the special case when we consider $\Omega_{k0}=0$ (flat space) then in this spatially-flat XCDM model $H(z)$ evolves as
\begin{eqnarray}
H(z,H_0,\textbf{p})=H_0 \left[\Omega_{m0}(1+z)^3+(1-\Omega_{m0}-\Omega_{k0})(1+z)^{3(1+\omega_X)}\right]^{1/2}.
\label{eq:XCDMHz3}
\end{eqnarray}
In this case the flat XCDM model parameters are $\textbf{p}=(\Omega_{m0},\omega_X)$. 

\subsection{No Big Bang Surface in the Parameter Space of the Non-Flat XCDM Parametrization}
\label{subsection:NBBLinXCDM}

In the XCDM model, the Friedmann and acceleration equations are
\begin{eqnarray}
\left(\frac{\dot a}{a}\right)^2=\frac{8 \pi G}{3}\Big[\rho_{m0}a^{-3}+\rho_{X0}a^{-3(1+\omega_X)}\Big]-\frac{k}{a^2},
\label{eq:XCDMNBB1}
\end{eqnarray}
\begin{eqnarray}
\left(\frac{\ddot a}{a}\right)=-\frac{4 \pi G}{3}\Big[\rho_{m0}a^{-3}+\rho_{X0}a^{-3(1+\omega_X)}\left(1+3\omega_X\right)\Big].
\label{eq:XCDMNBB2}
\end{eqnarray}
By applying the condition of no big bang as we did in Section\ (\ref{subsection:NBBL}), i.e., setting $\dot a=\ddot a=0$, to define $a=a_c$ we get,
\begin{eqnarray}
\left(\frac{\dot a}{a}\right)^2\Bigg|_c+2\left(\frac{\ddot a}{a}\right)\Bigg|_c=-\frac{k}{a_c^2}-\frac{8 \pi G}{3}\Big[3~ \omega_X\rho_{X0}a^{-3(1+\omega_X)}_c\Big]=0.
\label{eq:XCDMNBB3}
\end{eqnarray}
Solving for $a_c$ we find
\begin{eqnarray}
a_c=\left[-\frac{8\pi G \omega_X \rho_{X0}}{k}\right]^{1/(1+3\omega_X)}.
\label{eq:XCDMNBB4}
\end{eqnarray}
Using $\ddot a=0$ and $a=a_c$ in Eq.\ (\ref{eq:XCDMNBB2}) leads to
\begin{eqnarray}
\rho_{m0}=-\rho_{X0}\left(1+3\omega_X\right)a^{-3\omega_X}_{c}.
\label{eq:XCDMNBB5}
\end{eqnarray}
Since $\Omega_{m0}=\frac{8\pi G\rho_{m0}}{3H^{2}_{0}}$, $\Omega_{X0}=\frac{8\pi G\rho_{X0}}{3H^{2}_{0}}$, and $\Omega_{k0}=-\frac{k}{H^{2}_{0}}$, we can write this as
\begin{eqnarray}
\Omega_{m0}=\Omega_{X0}\left(1+3\omega_X\right)\left[\frac{\Omega_{X0}}{\Omega_{m0}}\omega_X\right]^{-\frac{3\omega_X}{1+3\omega_X}},
\label{eq:XCDMNBB6}
\end{eqnarray}
and since $\Omega_{X0}=1-\Omega_{m0}-\Omega_{k0}$, the implicit condition for no big bang in the XCDM parametrization is
\begin{eqnarray}
\Omega_{m0}+\left(1-\Omega_{m0}-\Omega_{k0}\right)\left(1+3\omega_X\right)\left[\frac{\left(1-\Omega_{m0}-\Omega_{k0}\right)}{\Omega_{m0}}\omega_X\right]^{-\frac{3\omega_X}{1+3\omega_X}} \leq 0.
\label{eq:XCDMNBB7}
\end{eqnarray}

\begin{figure}[h]
\centering
    \includegraphics[height=5.5in]{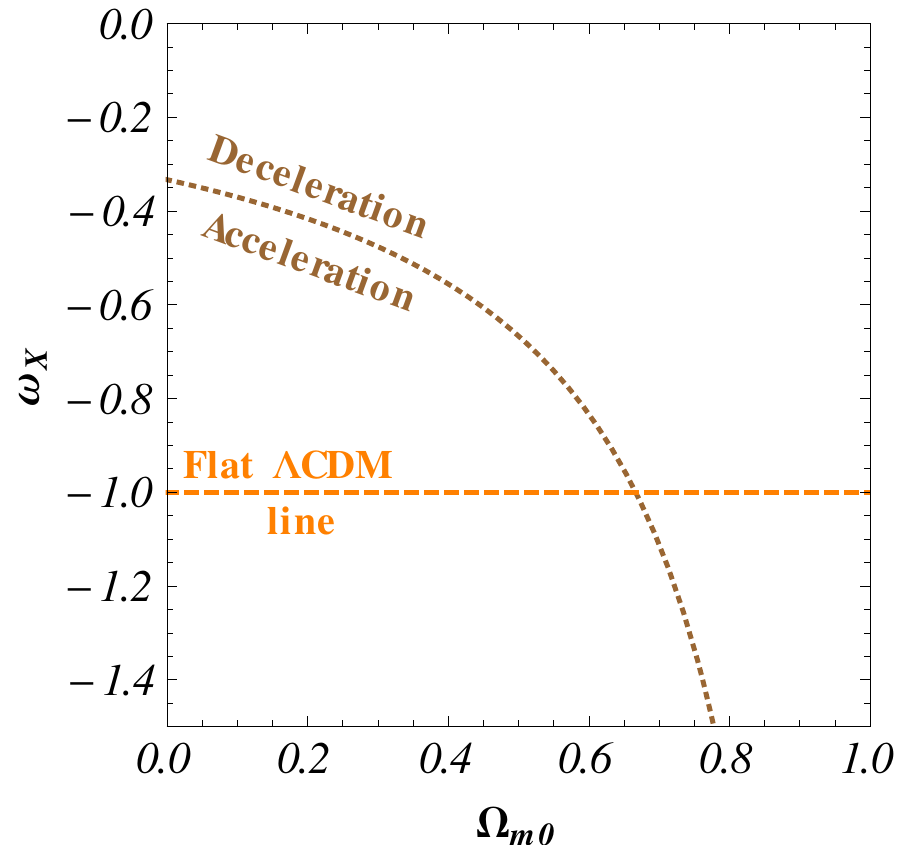}
\caption{Different regions of $\Omega_{m0}-\omega_{X}$ plane represent different cosmological behavior in spatially-flat XCDM parametrization. The brown dotted line [which is the plot of Eq.\ (\ref{eq:zacxcdm4})] separates accelerating from decelerating Universes, the orange dashed line [which is the plot of $\omega_X=-1$] shows the flat $\Lambda$CDM model.}
\label{fig:ConditionofXCDM}
\end{figure}

\subsection{Zero Acceleration Condition in the XCDM Parametrization}

In the XCDM parametrization, the acceleration equation (\ref{eq:XCDMNBB2}) in terms of density parameters, is
\begin{eqnarray}
\frac{\ddot a}{a}=-\frac{1}{2}H_0^2 \left[\frac{\Omega_{m0}}{a^3}+\frac{\Omega_{X0}}{a^{3(1+\omega_X)}}\left(1+3\omega_{X}\right)\right].
\label{eq:zacxcdm2}
\end{eqnarray}
For zero cosmological acceleration at the present time we set $a=a_0=1$, $\ddot a=0$ at present time we set $a=a_0=1$ in Eq.\ (\ref{eq:zacxcdm2}), and get 
\begin{eqnarray}
\Omega_{m0}+\Omega_{X0}\left(1+3\omega_X\right)=0.
\label{eq:zacxcdm3}
\end{eqnarray}
Using $\Omega_{X0}=\left(1-\Omega_{m0}\right)$, in spatially-flat XCDM parametrization we can write the above equation as
\begin{eqnarray}
\omega_{X}=\frac{1}{3\left(\Omega_{m0}-1\right)}.
\label{eq:zacxcdm4}
\end{eqnarray}

\section{Scalar Field Dark Energy Models}

The cosmological constant is equivalent a fluid with constant equation-of-state parameter $\omega_{\Lambda}=-1$. In the XCDM parametrization the time-independent equation-of-state parameter $\omega_{X}=p_X/\rho_X$ can take any value ($<-1/3$ to make sure that the cosmic expansion accelerate which is now observationally established to high confidence). Observations which constrain the value of $\omega$ today to be close to that of a cosmological constant, actually tell us little about the time evolution of equation-of-state parameter of dark energy at early times. At very early times, it is widely thought that a scalar field drive inflation. Scalar fields naturally arise in particle physics and string theory and these might be candidates for dark energy. A wide variety of scalar-field dark energy models have been proposed (for a summery see Samushia\cite{ladothesis}).

The kind of scalar field dark energy model we will discuss is also known as Quintessence. It is described by a scalar field $\phi$ minimally coupled to gravity (but with particular potentials that leads to late time inflation-like accelerating cosmological expansion). The action of quintessence is 
\begin{eqnarray}
S=\int \sqrt{-g}\left[\frac{m^{2}_{p}}{16 \pi}\left(\frac{1}{2}g^{\mu \nu}\partial_{\mu}\phi\partial_{\nu}\phi-V(\phi)-\mathcal{R}\right)+\mathcal{L}_m\right]d^4x,
\label{eq:SF1}
\end{eqnarray} 
where $m_p=1/\sqrt{G}$ is the Plank mass with $G$ being Newton's gravitational constant, $\phi$ is a classical scalar field whose stress-energy tensor act like time-variable $\Lambda$, with potential energy density $V(\phi)$, $\mathcal{R}$ is the Ricci scalar curvature, and $\mathcal{L}_m$ is Lagrangian density for ordinary matter and radiation. The action Eq.\ (\ref{eq:SF1}) can be arrived at by writing down the Einstein-Hilbert action with the term $2\Lambda$ replaced by the Lagrangian density for scalar field $\phi$. In this sense one has a modification of General Relativity in which cosmological `constant' can vary. 

The Lagrangian density for the $\phi$ field is given as:
\begin{eqnarray}
\mathcal{L}_{\phi}=\frac{m^{2}_{p}}{16 \pi}\left[\frac{1}{2}g^{\mu \nu}\partial_{\mu}\phi\partial_{\nu}\phi-V(\phi)\right].
\label{eq:SF2}
\end{eqnarray} 
In a FLRW spacetime, the variation of this action with respect to $\phi$ gives equation of motion of $\phi$ field\footnote{For the proof See Appendix \ref{Appendix:A}}
\begin{eqnarray}
\ddot\phi+3\left(\frac{\dot a}{a}\right)\dot \phi+\frac{dV}{d\phi}=0.
\label{eq:SF3}
\end{eqnarray} 

The energy-momentum tensor of the $\phi$ field is determined by varying $\mathcal{L}_\phi$ from Eq.\ (\ref{eq:SF1})
\begin{eqnarray}
T_{\mu \nu}=-2\frac{\delta\mathcal{L}_\phi}{\delta g^{\mu \nu}}+\mathcal{L}_\phi g_{\mu \nu},
\label{eq:SF4}
\end{eqnarray}
and as derived in Appendix (\ref{Appendix:B}),
\begin{eqnarray}
 T_{\mu \nu}=\frac{m^{2}_{p}}{16 \pi}\left[\partial_{\mu}\phi \partial_{\nu}\phi-g_{\mu \nu}\left(\frac{1}{2}g^{\alpha \beta}\partial_{\alpha}\phi \partial_{\beta}\phi+V(\phi)\right) \right].
\label{eq:SF5}
\end{eqnarray}
The scalar field energy density
\begin{eqnarray}
\rho_{\phi}=\frac{m^{2}_{p}}{16 \pi}\left[\frac{1}{2}\dot \phi^2+V(\phi)\right],
\label{eq:SF6}
\end{eqnarray}
and scalar field pressure
\begin{eqnarray}
p_{\phi}=\frac{m^{2}_{p}}{16 \pi}\left[\frac{1}{2}\dot \phi^2-V(\phi)\right].
\label{eq:SF7}
\end{eqnarray}
Then the equation-of-state parameter for $\phi$ field is
\begin{eqnarray}
\omega_{\phi}=\frac{p_{\phi}}{\rho_{\phi}}=\frac{\frac{1}{2}\dot \phi^2-V(\phi)}{\frac{1}{2}\dot \phi^2+V(\phi)},
\label{eq:SF8}
\end{eqnarray}

Using Eq.\ (\ref{eq:accelerationeq}) and (\ref{eq:SFU1}), and the above expressions for the scalar field energy density and pressure, we get the Friedmann equation
\begin{eqnarray}
H^2&=&\frac{1}{6}\left[\frac{1}{2}\dot\phi^2+V(\phi)\right],
\label{eq:SF9}
\end{eqnarray}
\begin{eqnarray}
\frac{\ddot a}{a}&=&-\frac{1}{6}\left[\dot\phi^2+V(\phi)\right].
\label{eq:SF10}
\end{eqnarray}

\subsection{Ratra-Peebles Scalar Field Model}

In 1988 Ratra and Peebles proposed a scalar field model\cite{Peebles&Ratra1988} known as $\phi$CDM and in order to alleviate two problems of the standard $\Lambda$CDM model of cosmology. These are the coincidence and energy scale problems (discussed in Sec.\ (\ref{sec:problems in LCDM}). The energy problem is that the very small spatial curvature of the Universe at the present epoch in some models of inflation\cite{Guth1981} requires a cosmological constant of very small energy scale $E_{\Lambda}$\cite{Peebles&Ratra1988,shawnthesis}
\begin{eqnarray}
E_{\Lambda}=\left[\frac{3(1-\Omega)H^{2}_{0}\hbar^3 c^5}{8\pi G}\right]^{1/4},
\label{eq:RP1}
\end{eqnarray}
where $\Omega=8\pi G\rho_m/(3 H^{2}_{0})$, $\hbar$ is Planck's constant, $H_0$ is the present value of Hubble parameter, $c$ is the speed of light and $G$ is the universal gravitational constant. Using the values of the parameters obtained by Planck Collaboration\cite{Planckdata} ($\Omega=0.315$, and $H_0=67.3$ km s$^{-1}$ Mpc$^{-1}$), we get $E_{\Lambda}=2.2$ meV, a very low energy scale that is difficult to understand theoretically. 

In order to alleviate these problems Peebles and Ratra\cite{Peebles&Ratra1988} postulated a model with scalar field potential energy density of the form $V(\phi)=\frac{1}{2}\kappa m^{2}_{p}\phi^{-\alpha}$, where $\alpha$ is a non-negative constant and $\kappa$ depends on $\alpha$ through 
\begin{eqnarray}
\kappa=\frac{8}{3}\left(\frac{\alpha+4}{\alpha+2}\right)\left[\frac{2}{3}\alpha(\alpha+2)\right]^{-\alpha/2},
\label{eq:RP10}
\end{eqnarray}
Then Eq.\ (\ref{eq:SF3}) gives the equation of motion of the $\phi$ field as
\begin{eqnarray}
\ddot \phi+3\left(\frac{\dot a}{a}\right)\dot \phi-\frac{1}{2}\kappa m^{2}_{p}\phi^{-(\alpha+1)}=0,
\label{eq:RP2}
\end{eqnarray}
and the corresponding energy density, pressure, and the equation-of-state parameter for the $\phi$ field are, from Eq.\ (\ref{eq:SF6})---(\ref{eq:SF8})
\begin{eqnarray}
\rho_{\phi}=\frac{m^{2}_{p}}{32 \pi}\left[\dot\phi^2+\kappa m^{2}_{p} \phi^{-\alpha}\right],
\label{eq:RP3}
\end{eqnarray}
\begin{eqnarray}
p_{\phi}=\frac{m^{2}_{p}}{32 \pi}\left[\dot\phi^2-\kappa m^{2}_{p} \phi^{-\alpha}\right],
\label{eq:RP4}
\end{eqnarray}
\begin{eqnarray}
\omega_{\phi}=\frac{\dot\phi^2-\kappa m^{2}_{p} \phi^{-\alpha}}{\dot\phi^2+\kappa m^{2}_{p} \phi^{-\alpha}}.
\label{eq:RP5}
\end{eqnarray}
In the $\phi$CDM model the Hubble parameter evolves according to
\begin{eqnarray}
H(z,H_0,\textbf{p})=H_0\left[\Omega_{m0}(1+z)^3+\Omega_{\phi}(z,\alpha)+\Omega_{k0}(1+z)^2\right]^{1/2},
\label{eq:RP6}
\end{eqnarray}
where
\begin{eqnarray}
\Omega_{\phi}=\frac{1}{12H^{2}_{0}}\left[\dot\phi^2+\kappa m^{2}_{p} \phi^{-\alpha}\right]
\label{eq:RP7}
\end{eqnarray}
here $\textbf{p}=(\Omega_{m0},\alpha,\Omega_{k0})$ are the parameters of this model. In the special case when we consider $\Omega_{k0}=0$ (spatially flat) then $H(z)$ in the spatially-flat $\phi$CDM model obey
\begin{eqnarray}
H(z,H_0,\textbf{p})=H_0\left[\Omega_{m0}(1+z)^3+\Omega_{\phi}(z,\alpha)\right]^{1/2},
\label{eq:RP8}
\end{eqnarray}
In this case the parameters of the flat $\phi$CDM model are $\textbf{p}=(\Omega_{m0},\alpha)$. 

Equations\ (\ref{eq:RP2}), (\ref{eq:RP6}), and (\ref{eq:RP7}) form the coupled system of partial differential equations that can be solved using the initial conditions described in Peebles and Ratra,\cite{Peebles&Ratra1988} that for $a\ll a_0$, $\rho_\phi \ll \rho_m$ (this is non-relativistic matter dominated epoch), then with Einstein-de Sitter solution
\begin{eqnarray}
a(t \ll t_0) \propto t^{2/3}, \hspace{1cm} \mathrm{and}\hspace{1cm} \phi(t \ll t_0)\propto t^{2/(2+\alpha)}.
\label{eq:RP9}
\end{eqnarray}
so the initial value of $\phi-$field is taken to be the solution of Eq.\ (\ref{eq:RP2}),
\begin{eqnarray}
\phi(t \ll t_0)=\left[\frac{2}{3}\alpha (\alpha+2)\right]^{1/2}\left(\frac{a}{a_1}\right)^{3/(\alpha+2)},
\label{eq:RP11}
\end{eqnarray}
where $a=a_1$ is the epoch at which $\rho_\phi \approx \rho_m$. Using these initial conditions one can solve the coupled differential equations numerically to get $H(z)$ in Eq.\ (\ref{eq:RP8}) for this model.

\subsection{Zero Acceleration Condition in the ${\phi}$CDM model}

For $\phi$CDM model, the acceleration equation is given as:
\begin{eqnarray}
\frac{\ddot a}{a}=-\frac{1}{2}H_0^2\left[\frac{\Omega_{m0}}{a^3}+\Omega_{\phi}\left(1+3\omega_{\phi}\right)\right].
\label{eq:zacpcdm2}
\end{eqnarray}
For zero acceleration at present time require $\dot a=0$ at $a=a_0=1$, and so 
\begin{eqnarray}
\Omega_{m0}+\Omega_{\phi, 0}\left(1+3\omega_{\phi}\right)=0.
\label{eq:zacpcdm3}
\end{eqnarray}
Using $\Omega_{\phi,0}=\left(1-\Omega_{m0}\right)$ we can write this as:
\begin{eqnarray}
\omega_{\phi}=\frac{1}{3\left(\Omega_{m0}-1\right)}.
\label{eq:zacpcdm4}
\end{eqnarray}
Solving Eqs. (\ref{eq:RP2}), (\ref{eq:RP4}) numerically using the initial condition of Eq.\ (\ref{eq:RP11}) we can find the set of parameter values that leads to zero cosmic acceleration in this model.

\subsection{Generalized Relation for ${\kappa({\alpha})}$ for $a(t)\propto t^n$ Initial Condition:}

The two constants $\alpha$ and $\kappa$ in the potential function $V(\phi)$ are not independent. They depends on each other and their relationship is dependent on the initial conditions considered. 

Let's write all the equations needed to derive the required relation. We first write the equation of motion of scalar field from Eq.\ (\ref{eq:RP2})
\begin{eqnarray}
\ddot \phi+3\left(\frac{\dot a}{a}\right)\dot\phi-\frac{1}{2}\kappa \alpha m^{2}_{p}\phi^{-(\alpha+1)}=0.
\label{eq:AK1}
\end{eqnarray}
Friedmann's equation in flat case $(k=0)$ including $\phi$ field energy density $\rho_\phi$ from Eq.\ (\ref{eq:Friedmanneq}) as
\begin{eqnarray}
\left(\frac{\dot a}{a}\right)^2=\frac{8 \pi}{3 m^{2}_{p}}(\rho_m+\rho_\phi),
\label{eq:AK2}
\end{eqnarray}
and finally equation of scalar field energy density in terms of $\phi$ field and its derivatives, from Eq.\ (\ref{eq:RP3}):
\begin{eqnarray}
\rho_{\phi}=\frac{m^{2}_{p}}{32 \pi}\left[\dot\phi^2+\kappa m^{2}_{p} \phi^{-\alpha}\right],
\label{eq:AK3}
\end{eqnarray}
There are a coupled differential equations, to find the solution of this system we need to have initial conditions. If we will consider that in very early times $(a \ll a_0)$, the Universe was dominated by a single component (other than scalar field $\phi$), then we can assume the exponential solutions for $a(t)$ as well as for $\phi(t)$. Let say, the the scale factor $a(t)$ and scalar field $\phi$ is of the form
\begin{eqnarray}
a(t)=a_1t^n, \hspace{2cm} \phi(t)=At^p,
\label{eq:AK4}
\end{eqnarray}
then first derivatives of $a(t)$ and $\phi(t)$ with respect to time will be
\begin{eqnarray}
\dot a=a_1nt^{n-1}, \hspace{2cm} \dot\phi=Apt^{p-1},
\label{eq:AK5}
\end{eqnarray}
\begin{eqnarray}
\Rightarrow \frac{\dot a}{a}=nt^{-1}, \hspace{2cm} \ddot\phi=Ap(p-1)t^{p-2}.
\label{eq:AK6}
\end{eqnarray}
Now plugging back Eqs.\ (\ref{eq:AK4})---(\ref{eq:AK6}) in Eq.\ (\ref{eq:AK1}) we get
\begin{eqnarray}
\Big[Ap(p-1)+3nAp\Big]t^{p-2}-\left[\frac{1}{2}\kappa \alpha m^{2}_{p} A^{-(\alpha+1)}\right]t^{-p(\alpha+1)}=0.
\label{eq:AK7}
\end{eqnarray}
Comparing the powers of $t$ from both sides of Eq.\ (\ref{eq:AK7}), we can get the value of $p$ as:
\begin{eqnarray}
p-2 &=&-p(\alpha+1),\nonumber \\
\Rightarrow p&=&\frac{2}{2+\alpha}.
\label{eq:AK8}
\end{eqnarray}
Now, let's calculate the coefficient. For that using Eqs.\ (\ref{eq:AK7}) and (\ref{eq:AK8}) we get
\begin{eqnarray}
\left[Ap(p-1)+3nAp-\frac{1}{2}\kappa \alpha m^{2}_{p} A^{-(\alpha+1)}\right]t^{-(\alpha+1)}=0,
\label{eq:AK9}
\end{eqnarray}
since, $t$ is never zero hence the coefficient should be equal to zero
\begin{eqnarray}
Ap(p-1)+3nAp-\frac{1}{2}\kappa \alpha m^{2}_{p} A^{-(\alpha+1)}=0,
\label{eq:AK10}
\end{eqnarray}
solving for $A$ in-terms of $\kappa$, $\alpha$, $n$, and $p$ we get:
\begin{eqnarray}
A^{\alpha+2}=\frac{\kappa \alpha m^{2}_{p}/2}{p(p-1)+3np}.
\label{eq:AK11}
\end{eqnarray}
Using Eq.\ (\ref{eq:AK8}), we can eliminate $p$ from Eq.\ (\ref{eq:AK11})
\begin{eqnarray}
A^{\alpha+2}=\frac{(\alpha+2)^2}{2\big[6n+3n\alpha-\alpha\big]}\ \frac{1}{2}\ \kappa \alpha m^{2}_{p}.
\label{eq:AK12}
\end{eqnarray}
Here $A$ is $\kappa$ dependent which is still unknown ($\alpha$ is model parameter), so till now we have two unknowns $\kappa$, and $A$, hence we need another equation. Let's use  Eqs.\ (\ref{eq:AK4})---(\ref{eq:AK6}) in Eq.\ (\ref{eq:AK3}) we will get
\begin{eqnarray}
\rho_{\phi}&=&\frac{m^{2}_{p}}{32\pi}\left[\left(Apt^{p-1}\right)^2+\kappa m^{2}_{p}\left(At^p\right)^{-\alpha}\right], \nonumber \\
&=&\frac{m^{2}_{p}}{32\pi}A^2\left[p^2+\kappa m^{2}_{p}A^{-(\alpha+2)}\right]t^{\beta}.
\label{eq:AK13}
\end{eqnarray}
Here we have defined $\beta$ as
\begin{eqnarray}
\beta \equiv -\alpha p=2p-2=-\frac{2\alpha}{\alpha+2}.
\label{eq:AK14}
\end{eqnarray}
Using Eq.\ (\ref{eq:AK8}) in Eq.\ (\ref{eq:AK13}) gives us $\alpha$ and $A$ (which contains $\kappa$) dependent coefficient of scalar field density as:
\begin{eqnarray}
\rho_{\phi}&=&\frac{m^{2}_{p}}{32\pi}A^2\left[\frac{4}{(\alpha+2)^2}+\frac{4\big[6n+3n\alpha-\alpha\big]}{\alpha (\alpha+2)^2}\right]t^{\beta}, \nonumber \\
&=&\frac{m^{2}_{p}}{8\pi}\ \frac{A^2}{\alpha(\alpha+2)}\ 3n\ t^{\beta}.
\label{eq:AK15}
\end{eqnarray} 
Now lets put Eqs.\ (\ref{eq:AK4})---(\ref{eq:AK6}) in Eq.\ (\ref{eq:AK2}) we will get
\begin{eqnarray}
\Big(\frac{n}{t}\Big)^2=\frac{8\pi}{3m^{2}_{p}}\ \rho,
\label{eq:AK16}
\end{eqnarray}
here $\rho$ is the density of that single component that we want to assume was dominated in the very early times in the Universe. Also if we suppose that the density of that particular type of matter is $\rho_1$ at $a=a_1$, and its equation-of-state parameter $\omega$ then takes the form
\begin{eqnarray}
\rho (a)=\rho_1\left(\frac{a_1}{a}\right)^{3(1+\omega)}, 
\label{eq:AK17}
\end{eqnarray}
but $3(1+\omega)=2/n$.\footnote{For detail proof of this see page 41 of Shawn Westmoreland master Thesis\cite{shawnthesis}} Hence from Eq.\ (\ref{eq:AK16})
\begin{eqnarray}
\frac{1}{t^2}=\frac{8\pi}{3n^2m^{2}_{p}}\ \rho_1\ \left(\frac{a_1}{a}\right)^{2/n}. 
\label{eq:AK18}
\end{eqnarray}
Plugging Eq.\ (\ref{eq:AK18}) in Eq.\ (\ref{eq:AK15})
\begin{eqnarray}
\rho_{\phi}&=&\frac{m^{2}_{p}}{8\pi}\ \frac{A^2}{\alpha(\alpha+2)}\ 3n\  \left(\frac{1}{t^2}\right)^{-\beta/2}, \nonumber\\
&=&\frac{m^{2}_{p}}{8\pi}\ \frac{A^2}{\alpha(\alpha+2)}\ 3n\  \left[\frac{8\pi}{3n^2m^{2}_{p}}\ \rho_1\ \left(\frac{a_1}{a}\right)^{2/n}\right]^{-\beta/2},
\label{eq:AK19}
\end{eqnarray}
but $\beta/2=\alpha/(\alpha+2)$thud, from Eq.\ (\ref{eq:AK14})
\begin{eqnarray}
\rho_{\phi}&=&\frac{m^{2}_{p}}{8\pi}\ \frac{A^2}{\alpha(\alpha+2)}\ 3n\  \left[\frac{8\pi}{3n^2m^{2}_{p}}\ \rho_1\ \left(\frac{a_1}{a}\right)^{2/n}\right]^{\alpha/(\alpha+2)}.
\label{eq:AK19}
\end{eqnarray}
Put $a=a_1 \Rightarrow \rho_{\phi}=\rho_1$, the above equation looks like
\begin{eqnarray}
\rho_{1}&=&\frac{m^{2}_{p}}{8\pi}\ \frac{A^2}{\alpha(\alpha+2)}\ 3n\  \left[\frac{8\pi}{3n^2m^{2}_{p}}\right]^{\alpha/(\alpha+2)} \rho_1^{\alpha/(\alpha+2)}.
\label{eq:AK20}
\end{eqnarray}
\begin{eqnarray}
\Rightarrow \rho_{1}^{2/(2+\alpha)}&=&\frac{m^{2}_{p}}{8\pi}\ \frac{A^2}{\alpha(\alpha+2)}\ 3n\  \left[\frac{8\pi}{3n^2m^{2}_{p}}\right]^{\alpha/(\alpha+2)}.
\label{eq:AK20}
\end{eqnarray}
Calculating $A^2$ from Eq.\ (\ref{eq:AK20})
\begin{eqnarray}
A^2= \frac{8\pi}{m^{2}_{p}}\left[\frac{3n^2m^{2}_{p}}{8\pi}\right]^{\alpha/(\alpha+2)}\frac{\alpha(\alpha+2)}{3n}\rho_{1}^{2/(2+\alpha)}.
\label{eq:AK21}
\end{eqnarray}
From Eq.\ (\ref{eq:AK12}) we can write:
\begin{eqnarray}
A^{2}=\left[\frac{(\alpha+2)^2}{2\big[6n+3n\alpha-\alpha\big]}\ \frac{1}{2}\ \kappa \alpha m^{2}_{p}\right]^{2/(2+\alpha)}.
\label{eq:AK22}
\end{eqnarray} 
Comparing Eq.\ (\ref{eq:AK21}) with Eq.\ (\ref{eq:AK22}) we will get:
\begin{eqnarray}
\left[\frac{(\alpha+2)^2}{2\big[6n+3n\alpha-\alpha\big]}\ \frac{1}{2}\ \kappa \alpha m^{2}_{p}\right]^{2/(2+\alpha)}&=&\frac{8\pi}{m^{2}_{p}}\left[\frac{3n^2m^{2}_{p}}{8\pi}\right]^{\alpha/(\alpha+2)}\frac{\alpha(\alpha+2)}{3n}\rho_{1}^{2/(2+\alpha)}, \nonumber \\
\nonumber \\
\Rightarrow\hspace{1 cm} \frac{(\alpha+2)^2}{2\big[6n+3n\alpha-\alpha\big]}\ \frac{1}{2}\ \kappa \alpha m^{2}_{p}&=&\left(\frac{8\pi}{m^{2}_{p}}\right)^{(\alpha+2)/2}\left[\frac{3n^2m^{2}_{p}}{8\pi}\right]^{\alpha/2}\left(\frac{\alpha(\alpha+2)}{3n}\right)^{(\alpha+2)/2}\rho_1, \nonumber \\
\nonumber \\
\Rightarrow\hspace{1 cm} \frac{(\alpha+2)^2}{2\big[6n+3n\alpha-\alpha\big]}\ \frac{1}{2}\ \kappa \alpha m^{2}_{p}&=&\left[\frac{\cancel{8\pi}}{\cancel{m^{2}_{p}}}\frac{\cancel{3}n^{\cancel{2}}\cancel{m^{2}_{p}}}{\cancel{8\pi}}\frac{\alpha(\alpha+2)}{\cancel{3}\cancel{n}}\right]^{\alpha/2}\frac{8\pi}{m^{2}_{p}}\frac{\alpha(\alpha+2)}{3n}\rho_1, \nonumber \\
\nonumber \\
\Rightarrow\hspace{1 cm} \frac{(\alpha+2)^{\cancel{2}}\cancel{\alpha} m^{2}_{p}}{4\big[6n+3n\alpha-\alpha\big]}\ \kappa &=&\big[n\alpha(\alpha+2)\big]^{\alpha/2}\frac{8\pi}{m^{2}_{p}}\frac{\cancel{\alpha}\cancel{(\alpha+2)}}{3n}\rho_1.
\label{eq:AK23}
\end{eqnarray} 
Solving this for $\kappa$ will results in
\begin{eqnarray}
\kappa=\frac{32\pi}{3n m^{4}_{p}}\left(\frac{6n+3n\alpha-\alpha}{\alpha+2}\right)\big[n\alpha(\alpha+2)\big]^{\alpha/2}\rho_1.
\label{eq:AK24}
\end{eqnarray} 
which is the relation between $\alpha$, $\kappa$, and $\rho_1$. Now, lets calculate the initial value of the scalar filed in terms of $\alpha$. Hence, using Eq.\ (\ref{eq:AK24}) in Eq.\ (\ref{eq:AK12}) we will get
\begin{eqnarray}
A^{\alpha+2}&=&\frac{(\alpha+2)^{\cancel{2}}}{2\big[\cancel{6n+3n\alpha-\alpha}\big]}
\left[\frac{32\pi}{3n m^{4}_{p}}\left(\frac{\cancel{6n+3n\alpha-\alpha}}{\cancel{\alpha+2}}\right)\big[n\alpha(\alpha+2)\big]^{\alpha/2}\rho_1\right], \nonumber \\
\nonumber \\
&=&\frac{8\pi \alpha (\alpha+2)}{3n m^{2}_{p}}\rho_1\big[n\alpha(\alpha+2)\big]^{\alpha/2}
\label{eq:AK25}
\end{eqnarray} 
Using Eq.\ (\ref{eq:AK4}), we can write
\begin{eqnarray}
\phi^{\alpha+2}=A^{\alpha+2}t^2, \hspace{2cm} \left(\mathrm{Since}\ p=\frac{2}{\alpha+2}\right)
\label{eq:AK26}
\end{eqnarray} 
Using Eq.\ (\ref{eq:AK25}) and Eq.\ (\ref{eq:AK18}) in Eq.\ (\ref{eq:AK26})
\begin{eqnarray}
\phi^{\alpha+2}=\frac{\cancel{8\pi}\alpha(\alpha+2)}{\cancel{3}\cancel{n}\cancel{m^{2}_{p}}}\cancel{\rho_1}\big[n\alpha(\alpha+2)\big]^{\alpha/2}\frac{\cancel{3}n^{\cancel{2}}\cancel{m^{2}_{p}}}{\cancel{8\pi}\cancel{\rho_1}}\left(\frac{a}{a_1}\right)^{2/n},
\label{eq:AK27}
\end{eqnarray}
which after simplification leads to the final result of:
\begin{eqnarray}
\phi=\big[n\alpha(\alpha+2)\big]^{1/2}\left(\frac{a}{a_1}\right)^{2/n(\alpha+2)},
\label{eq:AK28}
\end{eqnarray}
which is the initial value of the scalar field we have to take as a function of model parameter depending on type of the matter (considered) which dominates the early Universe.
Putting Eq.\ (\ref{eq:AK28}) in Eq.\ (\ref{eq:AK1}) leads to the general $\alpha$ $\kappa$ relation as\footnote{Thanks to Shawn Westmoreland for useful discussions.}
\begin{eqnarray}
\kappa=\frac{4n}{m^{2}_{p}}\left[\frac{6n+3n\alpha-\alpha}{\alpha+2}\right]\big[n\alpha(\alpha+2)\big]^{\alpha/2}.
\label{eq:AK29}
\end{eqnarray}


\cleardoublepage


\chapter{Data Analysis Techniques}
\label{Chapter4}

Generally we are interested in observables $X_{i,obs}$ measured at redshift $z_i$ (or in redshift bins of width $\Delta z_i$). $X_{i,obs}$ could be, e.g., the Hubble parameter, the luminosity distance, the angular diameter distance, or any other observable quantity. Let's consider theoretical model predicted $X_{th}(z,\textbf{p},H_0)$\footnote{$H_0$ is nuisance parameter discussed latter in this Chapter. We will suppress $H_0$ in this section.} which gives the values of the same quantity at the redshift $z_i$, which we will denote $X_{i,th}$. Here \textbf{p} are the parameters of the model considered. The parameters for the models we studied are listed in Table\ (\ref{table:parameters})

\begin{center}
\begin{threeparttable}
\centering
\begin{tabular}{|p{5cm}|p{5cm}|}
\hline \hline 
\hspace{0.2 cm}\textbf{Cosmological Model} & \hspace{1.0 cm}\textbf{Parameters} \\
\hline\hline
\vspace{0.5 mm} \hspace{0.8 cm}
Non-flat $\Lambda$CDM & \vspace{0.5 mm}\hspace{1.5 cm}$\Omega_{m0}$, $\Omega_{\Lambda}$\\
\hline
\vspace{0.5 mm} \hspace{1.2 cm}
Flat XCDM & \vspace{0.5 mm}\hspace{1.5 cm}$\Omega_{m0}$, $\omega_{X}$\\
\hline
\vspace{0.5 mm} \hspace{1.2 cm}
Flat $\phi$CDM & \vspace{0.5 mm}\hspace{1.5 cm}$\Omega_{m0}$, $\alpha$\\
\hline
\vspace{0.5 mm} \hspace{0.8 cm}
Non-flat XCDM & \vspace{0.5 mm}\hspace{1.0 cm}$\Omega_{m0}$, $\omega_{X}$, $\Omega_{k0}$ \\
\hline
\vspace{0.5 mm} \hspace{0.8 cm}
Non-flat $\phi$CDM & \vspace{0.5 mm}\hspace{1.1 cm}$\Omega_{m0}$, $\alpha$, $\Omega_{k0}$ \\
\hline\hline
\end{tabular}
\caption{\rm{Dark energy models and their parameters.}}
\label{table:parameters}
\end{threeparttable}
\end{center}

For given values of model parameters \textbf{p} we can compute the corresponding theoretical value of observable $X_{i,th}(\textbf{p})$. We will find the best-fit parameters $\mathbf{p_0}$
for which the theoretical model predictions $X_{th}$ best match the observations $X_{i,obs}$ according to some measure. We will also construct the confidence level intervals that are likely to cover the true value of parameters with a specified probability.\footnote{Those probabilities are taken to be 68.27\%, 95.45\%, and 99.73\%, the 1$\sigma$, 2$\sigma$, and  3$\sigma$ limits for a Gaussian distribution.}  

We will compare confidence contours of different models and also see if the data favors one model over the other, as well as quantify the degree of discrepancy between the models and observed data.

\section{${\chi^2}$ and Likelihood Function} 

Consider $N$ independent measurements $X_{i,obs}$ at known redshifts $z_i$ with known standard deviations $\sigma_i$. The theoretical model considered is $X_{th}(\textbf{p})$. Then $\chi^2(\textbf{p})$ as a function of model parameters is defined as:
\begin{eqnarray}
\chi^2(\textbf{p})=\sum\limits_{i=1}^{N}\frac{\big[X_{th}(z_i,\textbf{p})-X_{i,obs}\big]^2}{\sigma_i^2}.
\label{eq:chi2-1}
\end{eqnarray}
$\chi^2(\textbf{p})$ quantifies the discrepancy between theoretical predictions and observations at a particular value of the parameters $\textbf{p}$. 

Small values of $\chi^2$ indicates a good fit, similarly a large value of $\chi^2$ corresponds to a large difference between the theoretical prediction and the observed data. The set of parameters $\mathbf{p_0}$ that minimizes $\chi^2$ are called best-fit parameters, and $\chi^2$ defines the least square estimator for the general case even when the observed data is not Gaussian distributed as long as the measurements are independent. If they are not independent but rather have a covariance matrix $V_{ij}=\mathrm{cov}[X_i,X_j]$, then the $\chi^2(\mathbf{p})$ is
\begin{eqnarray}
\chi^2(\textbf{p})=\big[\boldsymbol{X_{th}(\mathbf{p})}-\boldsymbol{X_{i,obs}}\big]^{T}\ V^{-1}\ \big[\boldsymbol{X_{th}(\mathbf{p})}-\boldsymbol{X_{i,obs}}\big],
\label{eq:chi2-2}
\end{eqnarray}
where $\boldsymbol{X_{i,obs}}$ is the vector of measurements, $\boldsymbol{X_{th}(\mathbf{p})}$ is the corresponding vector of the predicted values from theory [understood as column vector in Eq.\ (\ref{eq:chi2-2})], and the superscript $T$ denotes the transposed (i.e., row) vector. 

We can define the corresponding likelihood function $\mathcal{L}(\mathbf{p})$ as
\begin{eqnarray} 
\mathcal{L}(\mathbf{p})=\mathrm{exp}\left[-\frac{1}{2}\chi^2(\mathbf{p})\right]=\mathrm{exp}\left[-\frac{1}{2}\big[\boldsymbol{X_{th}(\mathbf{p})}-\boldsymbol{X_{i,obs}}\big]^{T}\ V^{-1}\ \big[\boldsymbol{X_{th}(\mathbf{p})}-\boldsymbol{X_{i,obs}}\big]\right].
\label{eq:lik-1}
\end{eqnarray}
The likelihood function maximize at the same set of parameters $\mathbf{p_0}$ at which $\chi^2(\mathbf{p})$ minimizes. If the measurements are independent and Gaussian distributed with mean $X_{i,obs}$ and variance $\sigma_i$, then the best-fit values of parameters are an unbiased estimator of their true values. Values of the parameters that result in a higher value of the likelihood function are more likely to be true parameters.

If the models considered have two parameters [e.g., flat XCDM, or for $\phi$CDM, see Table\ (\ref{table:parameters})] then we define $1\sigma$, $2\sigma$, and $3\sigma$ confidence intervals as two-dimensional parameter sets bounded by $\chi^2(\mathbf{p})=\chi^2(\mathbf{p_0})+2.3$, $\chi^2(\mathbf{p})=\chi^2(\mathbf{p_0})+6.17$, and $\chi^2(\mathbf{p})=\chi^2(\mathbf{p_0})+11.8$ respectively.

\section{Nuisance Parameter}

It is very common that the theoretical models $X_{th}$ depend not only on the parameter of interest $\mathbf{p}$, but also on a nuisance parameter $\mathbf{\nu}$, whose value is known with limited accuracy. An example of a nuisance parameter in dark energy models is the Hubble constant $H_0$. It is often possible to assume some estimated prior distribution for $\mathbf{\nu}$, $\pi(\mathbf{\nu})$. Often a Gaussian probability density function with a mean value of $\mathbf{\nu_0}$ and variance $\sigma_{\mathbf{\nu}}$ provides a reasonable model. Then, we can build posterior likelihood function that will depend only on $\mathbf{p}$,
\begin{eqnarray}
\mathcal{L}(\mathbf{p})=\int\mathcal{L}(\mathbf{p},\mathbf{\nu})\pi(\mathbf{\nu})d\mathbf{\nu}.
\label{eq:lik-2}
\end{eqnarray}
Considering $\pi(\mathbf{\nu})$ to be a Gaussian, we have
\begin{eqnarray}
\mathcal{L}(\mathbf{p})=\frac{1}{\sqrt{2\pi\sigma_{\mathbf{\nu}}^2}}\mathlarger{\mathlarger{\int}}\mathcal{L}(\mathbf{p},\mathbf{\nu})\mathrm{exp}\left[-\frac{(\mathbf{\nu}-\mathbf{\nu_0})^2}{2\sigma_{\mathbf{\nu}}^2}\right]d\mathbf{\nu}.
\label{eq:lik-3}
\end{eqnarray}
where
\begin{eqnarray}
\mathcal{L}(\mathbf{p},\mathbf{\nu})=\mathrm{exp}\left[-\frac{1}{2}\chi^2(\mathbf{p,\mathbf{\nu}})\right],
\label{eq:lik-5}
\end{eqnarray}
is a prior likelihood. By maximizing $\mathcal{L}(\mathbf{p})$, or equivalently minimizing $\tilde{\chi}^2(\mathbf{p})\equiv -2\mathrm{ln}\mathcal{L}(\mathbf{p})$, we can estimate the best-fit point and calculate $1\sigma$, $2\sigma$, and $3\sigma$ contours as describe above, using $\tilde{\chi}^2(\mathbf{p})$ this time. 

The two frequently used priors on the Hubble constant are 68 $\pm$ 2.8 km s$^{-1}$ Mpc$^{-1}$ and 73.8 $\pm$ 2.4 km s$^{-1}$ Mpc$^{-1}$.  The first is from a median 
statistics analysis updated from Gott \textit{et al.},\citep{Gott2001} of 553 measurements of $H_0$ \citep{Chen2011a}; this estimate has been remarkably stable for over a decade now. \citep{Gott2001, chen03} The second value is the most precise recent one, based on HST measurements of Riess \textit{et al.}\citep{Riess2011} Other recent measurements are not inconsistent with at least one of the two values we use as a prior see, e.g., Freedman \textit{et al.},\cite{Freedman2012} Tammann \&\ Reindl,\cite{Tammann2012} and Sorce \textit{et al.}\cite{Sorce2012} 

\section{Constraints on Individual Cosmological Parameters}

One-dimensional confidence limits and best-fit values can be computed for individual cosmological parameters, say $p_1$ and $p_2$ of two parameter models. We take the two-dimensional likelihood function $\mathcal{L}(\mathbf{p}) \equiv \mathcal{L}(p_1,p_2)$ from Eq.\ (\ref{eq:lik-3}) and integrate it with respect to the other parameter with a (in this work) flat prior
\begin{eqnarray}
\mathcal{L}(p_1)=\int\mathcal{L}(\mathbf{p})dp_2=\int\limits_{\mathrm{all\ }p_2}\mathcal{L}(p_1,p_2)dp_2,
\label{eq:lik-6}
\end{eqnarray}
\begin{eqnarray}
\mathcal{L}(p_2)=\int\mathcal{L}(\mathbf{p})dp_1=\int\limits_{\mathrm{all\ }p_1}\mathcal{L}(p_1,p_2)dp_1.
\label{eq:lik-7}
\end{eqnarray}
For each parameter we find the best-fit value that maximizes the corresponding one-dimensional likelihood function. We also find $1\sigma$ and $2\sigma$ confidence intervals $\left[\ p_{1_L},p_{1_H}\right]$ and $\left[\ p_{2_L},p_{2_H}\right]$ such that
\begin{eqnarray}
r_1=\frac{\int\limits^{p_{1_H}}_{p_{1_L}}\mathcal{L}(p_1)dp_1}{\int\limits_{\mathrm{all\ space}}\mathcal{L}(p_1)dp_1},
\label{eq:r1}
\end{eqnarray}
\begin{eqnarray}
r_2=\frac{\int\limits^{p_{2_H}}_{p_{2_L}}\mathcal{L}(p_2)dp_2}{\int\limits_{\mathrm{all\ space}}\mathcal{L}(p_2)dp_2},
\label{eq:r2}
\end{eqnarray}
where $r_1$, and $r_2$ equal to $0.6827$ and $0.9545$, respectively.


\cleardoublepage


\chapter{Hubble Parameter Measurements Constraints}
\label{Chapter5}

This chapter is based on Farooq \textit{et al.}\cite{Farooq:2012ev}, and Farooq \& Ratra\cite{Farooq20131}
\\

\section{Dark Energy Models}
As discussed in Chapter (\ref{Chapter3}), in $\Lambda$CDM model Hubble's parameter evlove as:
\begin{equation}
\label{eq:LCDMFM}
H^2(z; H_0, \textbf{p}) = H_0^2 \left[\Omega_{m0} (1+z)^3 + \Omega_{\Lambda}
                          + (1-\Omega_{m0}-\Omega_\Lambda) (1+z)^2 \right],
\end{equation}
where $H_0$ is the value of the Hubble parameter at present, $\Omega_{m0}$, and $\Omega_{\Lambda}$ are the non-relativistic matter and dark energy density parameters, and $\mathbf{p}=(\Omega_{m0},\Omega_{\Lambda})$ are the parameters of this model.

For flat XCDM parameterization:
\begin{equation}
   H^2(z; H_0, \textbf{p}) = H_0^2 \left[\Omega_{m0}(1+z)^3 + 
      (1 - \Omega_{m0}) (1+z)^{3(1+\omega_{\rm X})}\right], 
\end{equation}
where $\mathbf{p}=(\Omega_{m0},\omega_{X})$ are the parameters of this model.

In the spatially flat $\phi$CDM model, the Hubble parameter is:
\begin{equation}
\label{eq:phicdmfriedman}
   H^2(z; H_0, \textbf{p}) = H_0^2 \left[\Omega_{m0}(1+z)^3+\Omega_\phi (z,\alpha)\right],
\end{equation}
with scalar field energy density given by:
\begin{equation}
   \rho_{\phi}=\frac{m_{p}^{2}}{16\pi}\left(\frac{1}{2}\dot\phi^2+\kappa m_{p}^{2} \phi^{-\alpha}\right).
\end{equation}
In this case $\mathbf{p}=(\Omega_{m0},\alpha)$ are the parameters of this model.

\section{Constraints from the $H(z)$ Data}

We use 21 independent $H(z)$ data points from  Simon \textit{et al.}\cite{simon05}, Gazta\^{n}aga \textit{et al.}\cite{gaztanaga09}, Stern \textit{et al.}\cite{Stern2010}, and Moresco \textit{et al.}\cite{moresco12}., listed in Table\ (\ref{tab:Hz1}), to constrain cosmological model
parameters. The observational data consist of measurements of the 
Hubble parameter $H_{\rm obs}(z_i)$ at redshifts $z_i$, with the 
corresponding one standard deviation uncertainties $\sigma_i$.

To constrain cosmological parameters $\textbf{p}$ of the 
models of interest we compute the $\chi_{H}^2$ function defined as:\footnote{Since the covariance matrix is diagonal in $H(z)$ case, hence we used Eq.\ (\ref{eq:chi2-1}).} 
\begin{equation}
\label{eq:chi2H} 
\chi_{H}^2 (H_0, \textbf{p}) =
\sum_{i=1}^{21}\frac{[H_{\rm th} (z_i; H_0, \textbf{p})-H_{\rm
obs}(z_i)]^2}{\sigma^2_i}.
\end{equation}
where $H_{\rm th} (z_i; H_0, \textbf{p})$ is the model-predicted
value of the Hubble parameter at the redshift $z_i$. As defined in Eq.\ (\ref{eq:comoving}), 
$ H_{\rm th} (z_i; H_0, \textbf{p}) \equiv H_0 E(z;\textbf{p})$, so
from Eq. ({\ref{eq:chi2H}}) we can write:
\begin{equation}
\label{eq:chi2Hs} 
\chi_{H}^2 (H_0, \textbf{p}) =
H_0^2 \sum_{i=1}^{21}\frac{E^2(z_i; \textbf{p})}{\sigma^2_i}
-2H_0 \sum_{i=1}^{21}\frac{H_{\rm obs}(z_i)E(z_i; \textbf{p})}{\sigma^2_i}
+\sum_{i=1}^{21}\frac{H^2_{\rm obs}(z_i)}{\sigma^2_i}.
\end{equation}

$\chi_{H}^2$ depends on the model parameters $\textbf{p}$ as well as 
on the nuisance parameter $H_0$, whose value is not known exactly.
We assume that the distribution of $H_0$ is a Gaussian with one 
standard deviation width  $\sigma_{H_0}$ and mean $\Bar{H_0}$. We
can then build the posterior likelihood function $\mathcal{L}_{H}(\textbf{p})$
that depends only on the $\textbf{p}$ by integrating the product 
of exp$(-\chi_H^2 /2)$ and the $H_0$ prior likelihood function 
exp$[-(H_0-\bar H_0)^2/(2\sigma^2_{H_0})]$ see e.g., Ganaga \textit{et al.}: \cite{Ganga1997}
\begin{equation}
\label{eq:likely}
\mathcal{L}_{H}(\textbf{p})=\frac{1}{\sqrt{2\pi \sigma^2_{H_0}}}
   \int \limits_0^\infty e^{-\chi_H^2(H_0,\textbf{p})/2} 
   e^{-(H_0-\bar H_0)^2/(2\sigma^2_{H_0})} dH_0.
\end{equation}
Defining:
\begin{equation}
\alpha=\frac{1}{\sigma_{H_0}^2}+\sum_{i=1}^{21}\frac{E^2(z_i; \textbf{p})}{\sigma^2_i},~
\beta=\frac{\bar H_0}{\sigma_{H_0}^2}+\sum_{i=1}^{21}\frac{H_{\rm obs}(z_i)E(z_i; \textbf{p})}{\sigma^2_i},~
\gamma=\frac{\bar H_0^2}{\sigma_{H_0}^2}+\sum_{i=1}^{21}\frac{H^2_{\rm obs}(z_i)}{\sigma^2_i}, 
\end{equation}
the integral can be expressed in terms of $\alpha$, $\beta$, and $\gamma$ as:
\begin{eqnarray}
\mathcal{L}_{H}(\textbf{p})=\frac{1}{\sqrt{2\pi \sigma^2_{H_0}}}
   \int \limits_0^\infty e^{-\frac{1}{2}\left(\alpha H^{2}_{0}-2\beta H_0+\gamma\right)} dH_0.
\end{eqnarray}
Completing the square in the exponent we get:
\begin{eqnarray}
\mathcal{L}_{H}(\textbf{p})&=&\frac{1}{\sqrt{2\pi \sigma^2_{H_0}}}e^{-\frac{1}{2}\gamma}
   \int \limits_0^\infty e^{-\frac{1}{2}\alpha\left(H^{2}_{0}-2\frac{\beta}{\alpha} H_0+\frac{\beta^2}{\alpha}-\frac{\beta^2}{\alpha}\right)} dH_0.\nonumber \\
\nonumber \\
&=& \frac{1}{\sqrt{2\pi \sigma^2_{H_0}}}\mathrm{exp}\left[-\frac{1}{2}\left(\gamma-\frac{\beta^2}{\alpha}\right)\right]
   \int \limits_0^\infty e^{-\frac{1}{2}\alpha\big[H_{0}-\frac{\beta}{\alpha}\big]^2} dH_0.
\end{eqnarray}
Substituting the following in the above integral:

\begin{figure}[t]
\centering
    \includegraphics[height=3.0in]{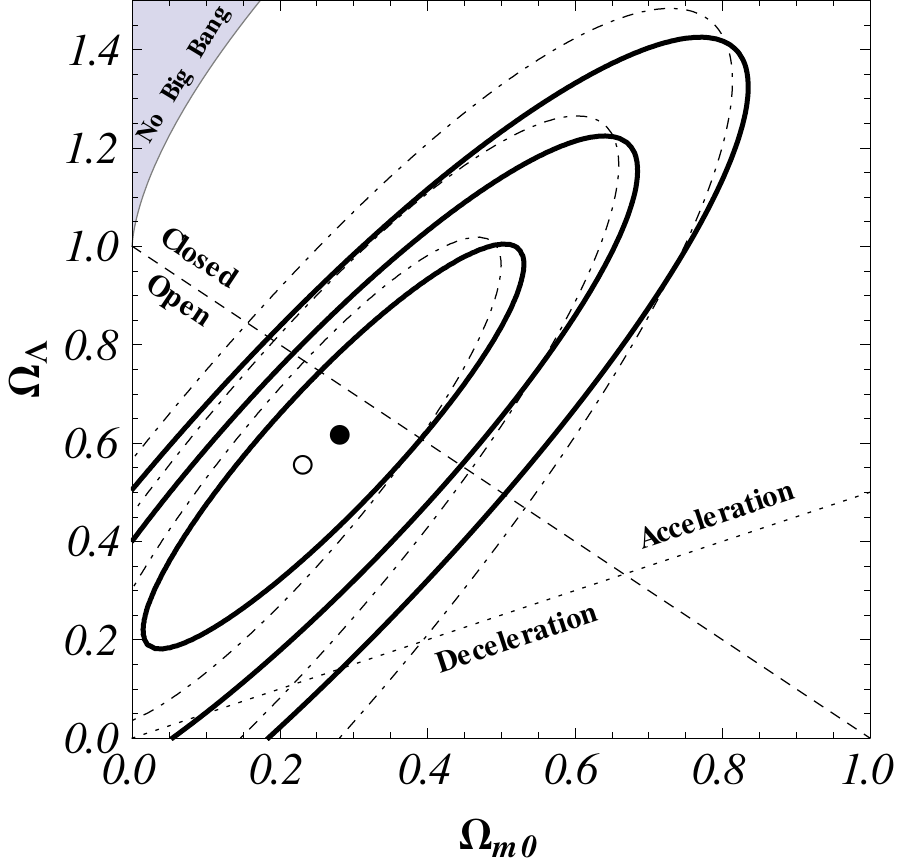}
    \includegraphics[height=3.0in]{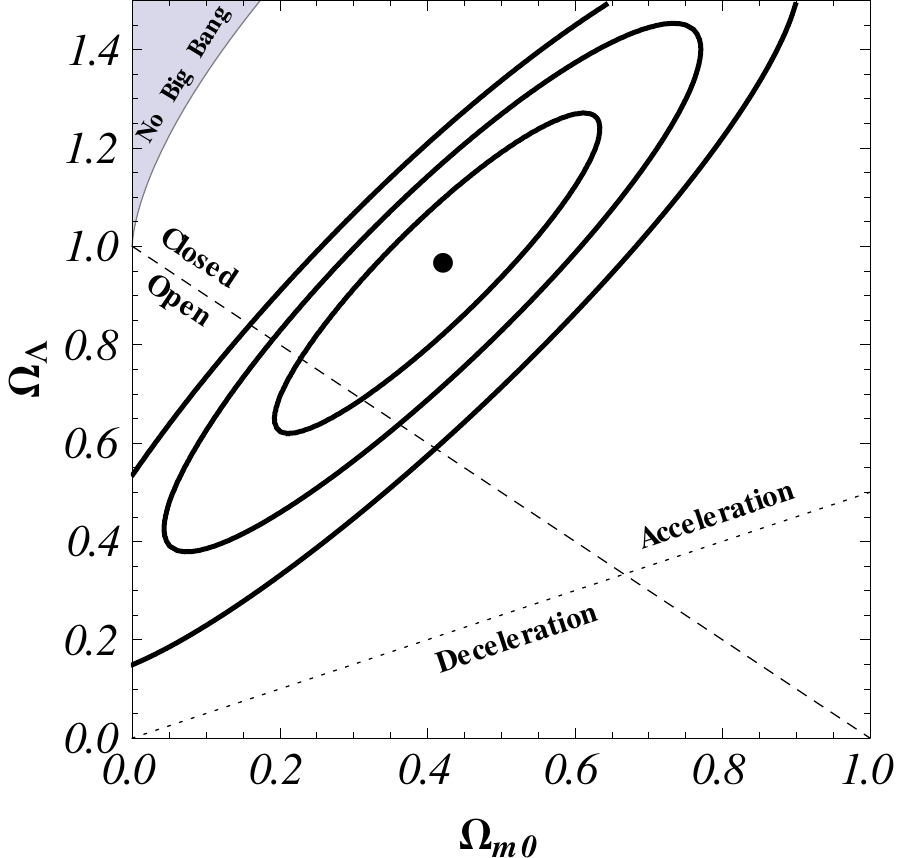}
\caption{Solid lines shows 1$\sigma$, 2$\sigma$, and 3$\sigma$ constraint contours for 
the $\Lambda$CDM model from the $H(z)$ data. The left panel is for the 
$H_0 = 68 \pm 2.8$ km s$^{-1}$ Mpc$^{-1}$ prior and the right 
panel is for the $H_0 = 73.8 \pm 2.4$ km s$^{-1}$ Mpc$^{-1}$ one.
Thin dot-dashed lines in the left panel are 1$\sigma$, 2$\sigma$, and 3$\sigma$ 
contours reproduced from, Yun \& Ratra\cite{Chen2011b} where the prior is 
$H_0 = 68 \pm 3.5$ km s$^{-1}$ Mpc$^{-1}$; the empty circle is the 
corresponding  best-fit point.
The dashed diagonal lines correspond to spatially-flat models, the 
dotted lines demarcate zero-acceleration models, and the shaded area 
in the upper left-hand corners are the region  for which there is no 
big bang. The filled black circles correspond to best-fit points. For
quantitative details see Table (\ref{tab:results-1}).}
\label{fig:LCDM_Hz21}
\end{figure}

\begin{figure}[t]
\centering
    \includegraphics[height=2.9in]{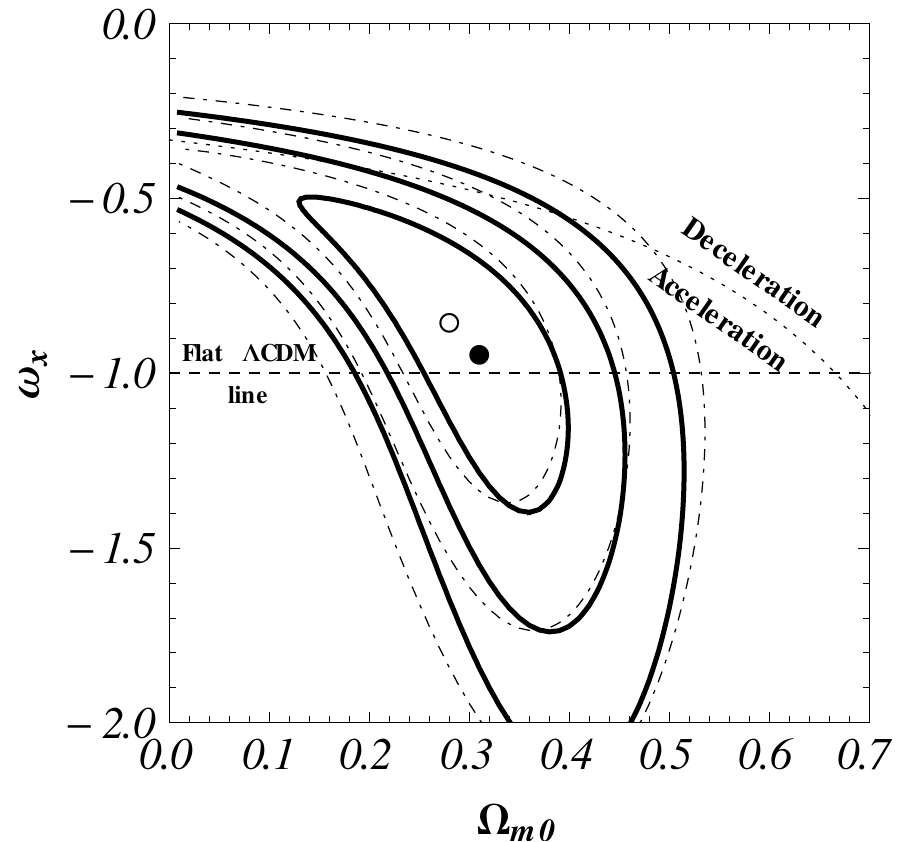}
    \includegraphics[height=2.9in]{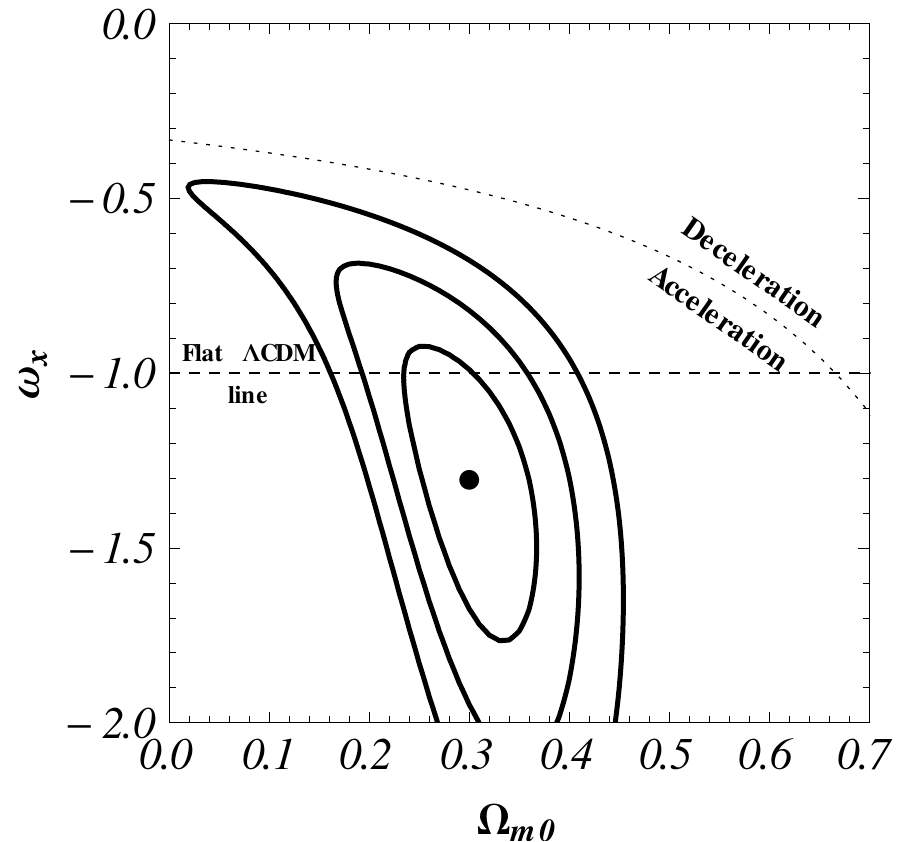}
\caption{Solid lines shows 1$\sigma$, 2$\sigma$, and 3$\sigma$ constraint contours for the XCDM
parametrization from the $H(z)$ data. The left panel is for the 
$H_0 = 68 \pm 2.8$ km s$^{-1}$ Mpc$^{-1}$ prior and the right 
panel is for the $H_0 = 73.8 \pm 2.4$ km s$^{-1}$ Mpc$^{-1}$ one.
Thin dot-dashed lines in the left panel are 1$\sigma$, 2$\sigma$, and 3$\sigma$ 
contours reproduced from Yun \& Ratra\cite{Chen2011b}, where the prior is 
$H_0 = 68 \pm 3.5$ km s$^{-1}$ Mpc$^{-1}$; the empty circle is the 
corresponding best-fit point.
The dashed horizontal lines at $\omega_{\rm X} = -1$ correspond to 
spatially-flat $\Lambda$CDM models and the curved dotted lines demarcate 
zero-acceleration models. The filled black circles correspond to best-fit
points. For quantitative details see Table (\ref{tab:results-1}).}
\label{fig:XCDM_Hz21}
\end{figure}

\begin{figure}[t]
\centering
    \includegraphics[height=3.0in]{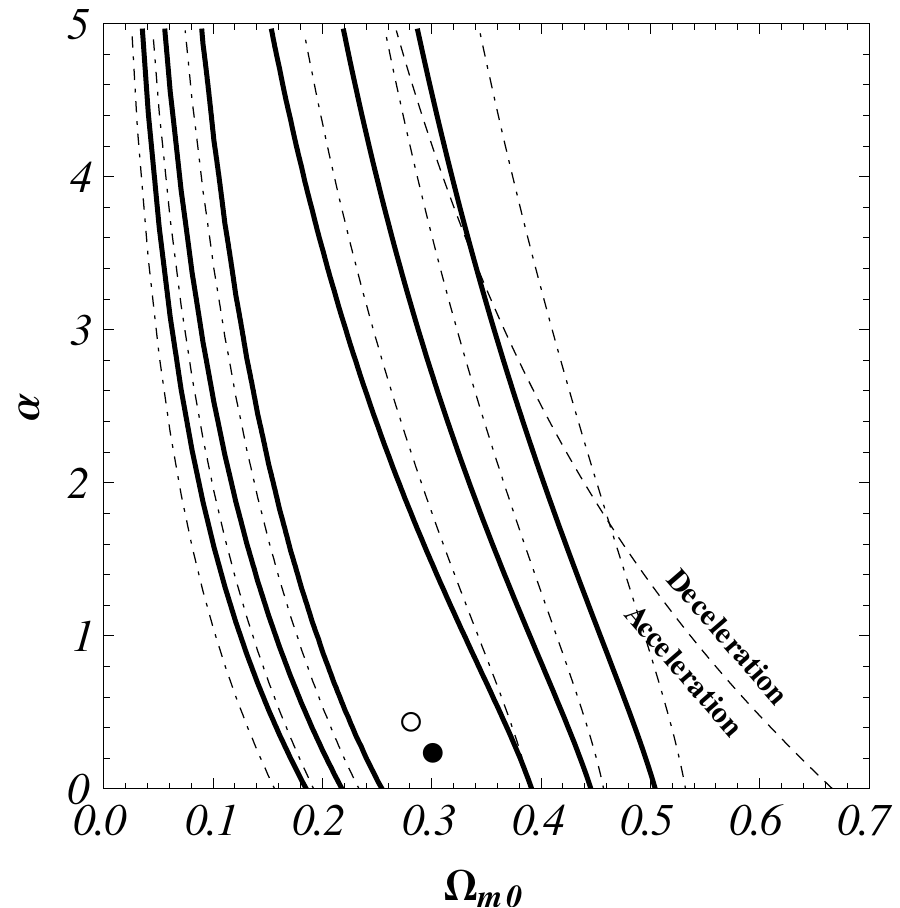}
    \includegraphics[height=3.0in]{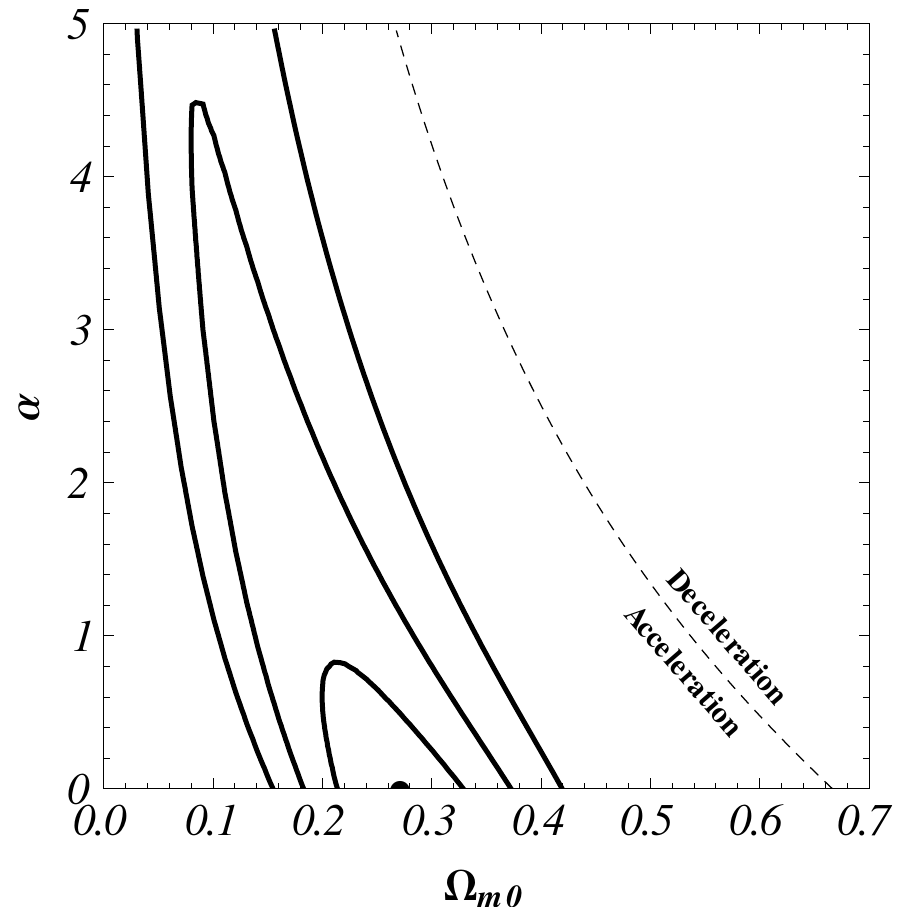}
\caption{Solid lines shows 1$\sigma$, 2$\sigma$, and 3$\sigma$ constraint contours for 
the $\phi$CDM model from the $H(z)$ data. The left panel is for the 
$H_0 = 68 \pm 2.8$ km s$^{-1}$ Mpc$^{-1}$ prior and the right 
panel is for the $H_0 = 73.8 \pm 2.4$ km s$^{-1}$ Mpc$^{-1}$ one.
Thin dot-dashed lines in the left panel are 1$\sigma$, 2$\sigma$, and 3$\sigma$ 
contours reproduced from Yun \& Ratra\cite{Chen2011b}, where the prior is 
$H_0 = 68 \pm 3.5$ km s$^{-1}$ Mpc$^{-1}$; the empty circle is 
the corresponding best-fit point.
The horizontal axes at $\alpha = 0$ correspond to spatially-flat 
$\Lambda$CDM models and the curved dotted lines demarcate 
zero-acceleration models. The filled black circles correspond to best-fit
points. For quantitative details see Table (\ref{tab:results-1}).}
\label{fig:phiCDM_Hz21}
\end{figure}

\begin{eqnarray}
t=\sqrt{\frac{1}{2}\alpha}\left[H_0-\frac{\beta}{\alpha}\right], \hspace{1cm} &\Rightarrow& \hspace{1cm} dH_0=\sqrt{\frac{2}{\alpha}}dt,\nonumber \\
\mathrm{When,} \hspace{2cm} H_0=0 \hspace{0.5cm} &\Rightarrow& \hspace{0.5cm} t=-\frac{\beta}{\sqrt{2\alpha}},\\
\mathrm{When,} \hspace{1.7cm} H_0\rightarrow\infty \hspace{0.5cm} &\Rightarrow& \hspace{0.5cm} t\rightarrow\infty. \nonumber
\end{eqnarray}
Thus we can write the posterior likelihood function $\mathcal{L}_{H}(\textbf{p})$ as:
\begin{eqnarray}
\mathcal{L}_{H}(\textbf{p})&=& \frac{1}{\sqrt{\cancel{2}\pi \sigma^2_{H_0}}}\mathrm{exp}\left[-\frac{1}{2}\left(\gamma-\frac{\beta^2}{\alpha}\right)\right]
   \int \limits^{\infty}_{-\beta/ \sqrt{2 \alpha}} e^{-t^2} \left(\frac{\sqrt{\cancel 2}}{\alpha}\right) dt, \nonumber \\
\nonumber \\
&=& \frac{1}{\sqrt{\alpha\pi \sigma^2_{H_0}}}\mathrm{exp}\left[-\frac{1}{2}\left(\gamma-\frac{\beta^2}{\alpha}\right)\right]
   \left[\int \limits^{\beta/ \sqrt{2 \alpha}}_0 e^{-t^2} dt+\int \limits^{\infty}_0 e^{-t^2} dt\right].
\end{eqnarray}
 Using the definition of error function,\footnote{$
{\rm erf}(x)={\frac{2}{\sqrt{\pi}}}\int \limits_{0}^{x}{{e^{-t^2}dt}}$.} and the very commonly known result, that the second integral in square brackets of above equation is $\sqrt{\pi}/2$, we get:
\begin{equation}
   \mathcal{L}_H(\textbf{p})=\frac{1}{2 \sqrt{\alpha~ \sigma_{H_0}^2}} 
   \exp \left[-\frac{1}{2}\left({\gamma}-\frac{\beta^2}{\alpha}\right)\right]
   \left[ 1 + \mathrm{erf}\left({\frac{\beta }{\sqrt{2\alpha}}}\right)\right].
\end{equation}
We maximize the likelihood $\mathcal{L}_H(\textbf{p})$, 
or equivalently minimize $\chi_H^2(\textbf{p}) = -2 
\mathrm{ln}\mathcal{L}_{H}(\textbf{p})$, with respect to the 
parameters $\textbf{p}$ to find the best-fit parameter values 
$\mathbf{p_0}$. In the models we consider $\chi_H^2$ depends on 
two parameters. We define 1$\sigma$, 2$\sigma$, and 3$\sigma$ 
confidence intervals as two-dimensional parameter sets bounded by 
$\chi_H^2(\textbf{p}) = \chi_H^2(\mathbf{p_0})+2.3,~\chi_H^2(\textbf{p}) = 
\chi_H^2(\mathbf{p_0})+6.17$, and $\chi_H^2(\textbf{p}) = 
\chi_H^2(\mathbf{p_0})+11.8$, respectively.

Even though the precision of measurements of the Hubble constant have
greatly improved over the last decade, the concomitant improvement 
in the precision of other cosmological measurements means that in some
cases the Hubble constant uncertainty still significantly affects 
cosmological parameter estimation. For a recent example see Calabrese 
\textit{et al.}\cite{calabrese12} The values of $\bar{H_0}\pm\sigma_{H_0}$ 
that we use in this paper are 68 $\pm$ 2.8 km s$^{-1}$ Mpc$^{-1}$ and 
73.8 $\pm$ 2.4 km s$^{-1}$ Mpc$^{-1}$. The first is from a median 
statistics analysis Gott \textit{et al.}\citep{Gott2001} of 553 measurements of $H_0$
Chen \textit{et al.};\citep{Chen2011a} this estimate has been remarkably stable for over 
a decade now \citep{Gott2001, chen03}. The second value is the most
precise recent one, based on HST measurements Riess \textit{et al.}\citep{Riess2011}. Other 
recent measurements are not inconsistent with at least one of the two 
values we use as a prior see, e.g., Freedman \textit{et al.},\cite{Freedman2012} Sorce \textit{et al.},\cite{Sorce2012} and 
Tammann \text{et al.}\cite{Tammann2012}

\begin{table}[htb]
\begin{center}
\begin{tabular}{ccccccc} 
\hline\hline 
\multirow{2}{*}{Model and prior} &  
\multicolumn{2}{c}{$H(z)$}
&\multicolumn{2}{c}{SNeIa}
&\multicolumn{2}{c}{BAO}\\
{} & $\chi^2_{\rm min}$ & B.F.P & $\chi^2_{\rm min}$ & B.F.P & $\chi^2_{\rm min}$ & B.F.P\\ 
\hline \hline 
$\Lambda$CDM & \multirow{2}{*}{$14.6$} & $\Omega_{m0}=0.28$ & \multirow{4}{*}{545} & \multirow{2}{*}{$\Omega_{m0}=0.29$} & \multirow{4}{*}{5.5} & \multirow{2}{*}{$\Omega_{m0}=0.27$} \\ 
$h = 0.68 \pm 0.028$ &  & $\Omega_{\Lambda}=0.62$ &  & {} &  & {} \\ 
\cline{1-3}

$\Lambda$CDM & \multirow{2}{*}{$14.6$} & $\Omega_{m0}=0.42$ & \multirow{4}{*}{} & \multirow{2}{*}{$\Omega_{\Lambda}$=0.69} & \multirow{2}{*}{} & \multirow{2}{*}{$\Omega_{\Lambda}$=0.87} \\ 
$h = 0.738 \pm 0.024$ &  & $\Omega_{\Lambda}=0.97$ &  & \multirow{4}{*}{} &  & \multirow{4}{*}{}\\ 
\hline
\hline

XCDM & \multirow{2}{*}{$14.6$} & $\Omega_{m0}=0.31$ & \multirow{4}{*}{545} & \multirow{2}{*}{$\Omega_{m0}=0.29$} & \multirow{4}{*}{$5.5$} & \multirow{2}{*}{$\Omega_{m0}=0.27$} \\ 
$h = 0.68 \pm 0.028$ &  & $\omega_{X}=-0.94$ &  & {} &  & {} \\
\cline{1-3}

XCDM & \multirow{2}{*}{$14.6$} & $\Omega_{m0}=0.30$ & \multirow{4}{*}{} & \multirow{2}{*}{$\omega_{X}=-0.99$} & \multirow{2}{*}{} & \multirow{2}{*}{$\omega_{X}=-1.21$} \\ 
$h = 0.738 \pm 0.024$ &  & $\omega_{X}=-1.3$ &  & \multirow{4}{*}{} &  & \multirow{4}{*}{}\\ 
\hline 
\hline

$\phi$CDM & \multirow{2}{*}{$14.6$} & $\Omega_{m0}=0.30$ & \multirow{4}{*}{545} & \multirow{2}{*}{$\Omega_{m0}=0.27$} & \multirow{4}{*}{5.9} & \multirow{2}{*}{$\Omega_{m0}=0.30$} \\ 
$h = 0.68 \pm 0.028$ &  & $\alpha=0.25$ &  & {} &  & {} \\
\cline{1-3}

$\phi$CDM & \multirow{2}{*}{$15.6$} & $\Omega_{m0}=0.27$ & \multirow{4}{*}{} & \multirow{2}{*}{$\alpha =0.20$} & \multirow{2}{*}{} & \multirow{2}{*}{$\alpha =0.00$} \\ 
$h = 0.738 \pm 0.024$ &  & $\alpha =0.00$ &  & \multirow{4}{*}{} &  & \multirow{4}{*}{}\\ 
\hline
\hline  
\end{tabular}

\caption{The minimum value of $\chi^2$ and the corresponding best-fit 
points (B.F.P) that maximize the likelihood for the three individual
data sets. The SNIa values are for the case including systematic errors. 
Ignoring SNIa systematic errors, for the $\Lambda$CDM model
$\chi_{SN}^2(\mathbf{p_0})=562$, at $(\Omega_{m0}, \Omega_\Lambda) = (0.28,0.73)$; for 
the XCDM case $\chi_{SN}^2(\mathbf{p_0})=562$ at $(\Omega_{m0}, 
\omega_{\rm X}) = (0.28,-1.01)$; and for the $\phi$CDM model 
$\chi_{SN}^2(\mathbf{p_0})=562$, at $(\Omega_{m0}, \alpha) = (0.27,0.05)$.}
\label{tab:results-1}
\end{center}
\end{table}
Figures (\ref{fig:LCDM_Hz21})---(\ref{fig:phiCDM_Hz21}) show the constraints
from the $H(z)$ data for the three dark energy models we consider, and
for the two different $H_0$ priors. Table\ (\ref{tab:results-1}) lists 
the best fit parameter values. Comparing these plots with Figs.\ 
1---3 of Yun \& Ratra,\cite{Chen2011b} whose 1$\sigma$, 2$\sigma$ and 3$\sigma$ constraint 
contours are reproduced here as dot-dashed lines in the left panels of 
Figs (\ref{fig:LCDM_Hz21})---(\ref{fig:phiCDM_Hz21}), we see that the 
contours derived from the new data are more constraining, 
by about a standard deviation, because of the 8 new, more
precise, data points used here taken from Moresco \textit{et al.}\cite{moresco12} On comparing the 
left and right panels in these three figures, we see that the constraint
contours are quite sensitive to the value of $H_0$ used, as well as to 
the uncertainty associated with the Hubble constant measurement. 

\section{Constraints from the SNIa Data}
\label{SNeIa}

While the $H(z)$ data provide tight constraints on a linear
combination of cosmological parameters, the very elongated constraint
contours of Figs.\ (\ref{fig:LCDM_Hz21})---(\ref{fig:phiCDM_Hz21})
imply that these data alone cannot significantly discriminate
between cosmological models. To tighten the constraints we must add other 
data to the mix.

The second set of data that we use are the Type Ia supernova data 
from the Suzuki \textit{et al.}\cite{suzuki2012}
Union2.1 compilation of 580 SNIa distance modulus $\mu_{\rm obs}(z_i)$ 
measurements at measured redshifts $z_i$ (covering the redshift range
of 0.015 to 1.414) with associated one standard deviation  
uncertainties $\sigma_i$. The predicted distance modulus from Eq.\ ({\ref{eq:dm5}}) is: 
\begin{equation}
\label{eq:distance modulus theoratical}
\mu_{\rm th} (z_i; H_0, \textbf{p}) = \underbrace{5 ~ \mathrm{log}_{10}\left(3000 ~ y(z) (1+z)\right) +25}_{=\mu_{0}} ~ -\ 5 ~ \mathrm{log}_{10}(h),
\end{equation} 
where $H_0 = 100 h$ km s$^{-1}$ Mpc$^{-1}$ and 
$y(z)$ is the dimensionless coordinate distance given in Eq.\ (\ref{eq:dlcd1}) as:
\begin{equation}
y(z)= \left\{
      \begin{array}{lc}
           \frac{a_0H_0}{\sqrt{k}}\mathrm{sin}\left(\frac{\sqrt{k}}{a_0 H_0}\int\limits^{z}_{0}\frac{dz'}{E(z')}\right) & \ \ \ \ \mathrm{for}\ k>0 \\
           \int\limits^{z}_{0}\frac{dz'}{E(z')} & \ \ \ \ \mathrm{for}\ k=0 \\
           \frac{a_0H_0}{\sqrt{-k}}\mathrm{sinh}\left(\frac{\sqrt{-k}}{a_0 H_0}\int\limits^{z}_{0}\frac{dz'}{E(z')}\right) & \ \ \ \ \mathrm{for}\ k<0 \\
     \end{array}.
   \right.
\label{eq:Angular size distance}
\end{equation}
As the SNIa distance modulus measurements $\mu_{\rm obs}$ are correlated, 
$\chi^2$ is defined as: 
\begin{equation}
\label{eq:chiSN-1}
\chi_{SN}^{2}(h,\textbf{p})=\Delta\boldsymbol\mu^T~{\mathcal{C}}^{-1}~
\Delta\boldsymbol{\mu}.
\end{equation}
Here $\Delta \boldsymbol\mu $ is a vector of differences 
$\Delta{\mu_i}= \mu_{\rm th}(z_i;H_0,\textbf{p}) -\mu_{\rm obs}(z_i)$, 
and $\mathcal{C}^{-1}$ is the inverse of the 580 by 580 Union 2.1 compilation 
covariance matrix. In index notation: 
\begin{equation}
\label{eq:chiSN-2}
\chi^{2}_{SN}(h,\textbf{p})=\sum_{\alpha, \beta}\left[\mu_0 -5 
    \mathrm{log}_{10} h - \mu_{\rm obs} \right]_\alpha
    (\mathcal{C}^{-1})_{\alpha \beta}
    \left[\mu_0 - 5 \mathrm{log}_{10} h - \mu_{\rm obs}
    \right]_\beta.
\end{equation}
Let's simplify this:
\begin{eqnarray}
\chi^{2}_{SN}(h,\textbf{p})&=&\sum\limits_{\alpha \beta}\left[\mu_0-\mu_{\mathrm{obs}}\right]_{\alpha}(\mathcal{C}^{-1})_{\alpha \beta}\left[\mu_0-\mu_{\mathrm{obs}}\right]_{\beta} -\big(5 \mathrm{log}_{10} h\big)\left[\sum\limits_{\alpha \beta}\left[\mu_0-\mu_{\mathrm{obs}}\right]_{\alpha}(\mathcal{C}^{-1})_{\alpha \beta}\right]\nonumber\\
&&-\big(5 \mathrm{log}_{10} h\big)\left[(\mathcal{C}^{-1})_{\alpha \beta}\sum\limits_{\alpha \beta}\left[\mu_0-\mu_{\mathrm{obs}}\right]_{\beta}\right]+\big(5 \mathrm{log}_{10} h\big)^2\sum\limits_{\alpha \beta}(\mathcal{C}^{-1})_{\alpha \beta},
\end{eqnarray}
since the the covariance matrix is symmetric hence the second and the third terms in above equation will just add, an we will get:
\begin{eqnarray}
\label{eq:SNchi1}
\chi^{2}_{SN}(h,\textbf{p})&=&\sum\limits_{\alpha \beta}\left[\mu_0-\mu_{\mathrm{obs}}\right]_{\alpha}(\mathcal{C}^{-1})_{\alpha \beta}\left[\mu_0-\mu_{\mathrm{obs}}\right]_{\beta} -\big(10 \mathrm{log}_{10} h\big)\left[(\mathcal{C}^{-1})_{\alpha \beta}\sum\limits_{\alpha \beta}\left[\mu_0-\mu_{\mathrm{obs}}\right]_{\beta}\right]\nonumber\\
&&+\big(5 \mathrm{log}_{10} h\big)^2\sum\limits_{\alpha \beta}(\mathcal{C}^{-1})_{\alpha \beta},
\end{eqnarray}
Defining:
\begin{equation}
\label{eq:chiSN-3a}
\begin{array}{lr}
A(\textbf{p})= \sum\limits_{\alpha, \beta} (\mu_{0}-\mu_{\rm obs})_{\alpha}\ 
    (\mathcal{C}^{-1})_{\alpha \beta}\ (\mu_{0}-\mu_{\rm obs})_{\beta}\\
B(\textbf{p})= \sum\limits_\alpha (\mu_{0}-\mu_{\rm obs})_\alpha\sum\limits_\beta 
    (\mathcal{C}^{-1})_{\alpha \beta} \\
C= \sum\limits_{\alpha,\beta} (\mathcal{C}^{-1})_{\alpha \beta}.
\end{array}
\end{equation}
then Eq.\ (\ref{eq:SNchi1}) takes the form:
\begin{equation}
\label{eq:chiSN-3}
\chi^{2}_{SN}(h,\textbf{p})=A(\textbf{p}) - 
10 B(\textbf{p}) \mathrm{log}_{10}(h)+ 25 C [\mathrm{log}_{10}(h)]^2
\end{equation}
The corresponding likelihood function, when considering a flat $H_0$ prior,
is:
\begin{eqnarray}
\label{eq:SNchi2}
\hspace{-1cm}
 \mathcal{L}_{SN}(\textbf{p}) &=&  \int\limits_{0}^{\infty} {e^{-\chi^2_{SN}(h,\textbf{p})/2} dh}, \nonumber\\
&=&\mathrm{exp}\left[-\frac{1}{2}A(\textbf{p})\right]\mathlarger{\mathlarger{\int}}\limits^{\infty}_{0}\mathrm{exp}\left[-\frac{25C}{2} (\mathrm{log}_{10}h)^2+5 B(\textbf{p})(\mathrm{log}_{10}h)\right]dh,
\end{eqnarray}
completing the square in the exponent results in:
\begin{eqnarray}
\label{eq:SNchi3}
\hspace{-1cm}
 \mathcal{L}_{SN}(\textbf{p}) &=&\mathrm{exp}\left[-\frac{1}{2}\left(A(\textbf{p})-\frac{B^2(\textbf{p})}{C}\right)\right]\mathlarger{\mathlarger{\int}}\limits^{\infty}_{0}\mathrm{exp}\left[-\frac{25C}{2}\left(\mathrm{log}_{10}h-\frac{B(\textbf{p})}{5C}\right)^2 \right]dh.  
\end{eqnarray}
Using the property of logarithm that $\mathrm{log_{10}}x=\frac{\mathrm{ln}x}{\mathrm{ln}10}$, we get:
\begin{eqnarray}
\label{eq:SNchi4}
\hspace{-1cm}
 \mathcal{L}_{SN}(\textbf{p}) &=&\mathrm{exp}\left[-\frac{1}{2}\left(A(\textbf{p})-\frac{B^2(\textbf{p})}{C}\right)\right]\mathlarger{\mathlarger{\int}}\limits^{\infty}_{0}\mathrm{exp}\left[-\frac{25C}{2}\left(\frac{\mathrm{ln}h}{\mathrm{ln}10}-\frac{B(\textbf{p})}{5C}\right)^2 \right]dh,  
\end{eqnarray}
simplifying we get:
\begin{eqnarray}
\label{eq:SNchi5}
\hspace{-1.2cm}
 \mathcal{L}_{SN}(\textbf{p})\hspace{-0.2cm}&=&\hspace{-0.2cm}\mathrm{exp}\left[-\frac{1}{2}\left(A(\textbf{p})-\frac{B^2(\textbf{p})}{C}\right)\right]\mathlarger{\mathlarger{\int}}\limits^{\infty}_{0}\mathrm{exp}\left[-\frac{25C}{2\left(\mathrm{ln}10\right)^2}\left(\mathrm{ln}h-\frac{B(\textbf{p})(\mathrm{ln}10)}{5C}\right)^2 \right]dh,  
\end{eqnarray}
 Defining: 
\begin{equation}
\delta = \frac{25C}{2\mathrm{(ln10)^2}}~, 
 ~~~\varepsilon = \frac{B(\textbf{p})\mathrm{ln10}}{5C},
 \nonumber
\end{equation}
\begin{eqnarray}
\label{eq:SNchi6}
 \mathcal{L}_{SN}(\textbf{p})\hspace{-0.2cm}&=&\hspace{-0.2cm}\mathrm{exp}\left[-\frac{1}{2}\left(A(\textbf{p})-\frac{B^2(\textbf{p})}{C}\right)\right]\mathlarger{\mathlarger{\int}}\limits^{\infty}_{0}\mathrm{exp}\Big[-\delta\big(\mathrm{ln}h-\varepsilon\big)^2 \Big]dh.
\end{eqnarray}
Substituting:
\begin{eqnarray}
\label{eq:SNchi7}
\mathrm{ln}h-\varepsilon=y, \hspace{0.5cm} \Rightarrow \hspace{0.5cm} h=e^{y+\varepsilon},
\end{eqnarray}
\begin{eqnarray}
\label{eq:SNchi8}
\hspace{-1cm} \Rightarrow \hspace{1cm} \frac{dh}{h}=dy,  \hspace{0.5cm} \Rightarrow \hspace{0.5cm} dh=e^{y+\varepsilon}dy.
\end{eqnarray}
Using Eqs.\ (\ref{eq:SNchi7}) and (\ref{eq:SNchi8}), it is easy to see that:
\begin{eqnarray}
\label{eq:SNchi9}
&\mathrm{When:}& \hspace{0.5cm} h\rightarrow 0, \hspace{1cm} y\rightarrow -\infty,\\
&\mathrm{When:}& \hspace{0.5cm} h\rightarrow \infty, \hspace{1cm} y\rightarrow +\infty.
\end{eqnarray}
Thus the likelihood function takes the form:
\begin{eqnarray}
\label{eq:SNchi10}
\mathcal{L}_{SN}(\textbf{p})\hspace{-0.2cm}&=&\hspace{-0.2cm}\mathrm{exp}\left[-\frac{1}{2}\left(A(\textbf{p})-\frac{B^2(\textbf{p})}{C}\right)\right]\int\limits^{\infty}_{-\infty} e^{-\delta y^2} e^{y+\varepsilon} dy, \nonumber\\
&=&\hspace{-0.2cm}\mathrm{exp}\left[-\frac{1}{2}\left(A(\textbf{p})-\frac{B^2(\textbf{p})}{C}-2\varepsilon\right)\right]\int\limits^{\infty}_{-\infty} e^{-\delta\left( y^2-\frac{1}{\delta}y\right)}dy.
\end{eqnarray}
Completing the square in the exponent we will get:
\begin{eqnarray}
\label{eq:SNchi11}
\mathcal{L}_{SN}(\textbf{p})\hspace{-0.2cm}&=&\hspace{-0.2cm}\mathrm{exp}\left[-\frac{1}{2}\left(A(\textbf{p})-\frac{B^2(\textbf{p})}{C}-2\varepsilon\right)\right]\int\limits^{\infty}_{-\infty} e^{-\delta\left(y^2-2\frac{1}{2\delta}y+\frac{1}{4 \delta^2}-\frac{1}{4 \delta^2}\right)}dy, \nonumber\\
&=&\hspace{-0.2cm}\mathrm{exp}\left[-\frac{1}{2}\left(A(\textbf{p})-\frac{B^2(\textbf{p})}{C}-2\varepsilon-\frac{1}{2\delta}\right)\right]\int\limits^{\infty}_{-\infty} e^{-\delta\left(y-\frac{1}{2\delta}\right)^2}dy.
\end{eqnarray}
This is very standard integral called Gaussian integral.\footnote{We can find this integral by converting in to polar coordinates. Calling it I, we have:
\begin{eqnarray}
I^2=\int\limits^{\infty}_{-\infty}\int\limits^{\infty}_{-\infty}e^{-\delta(x^2+y^2)}dxdy=\int\limits^{2\pi}_{0}\int\limits^{\infty}_{0}e^{-\delta r^2}rdrd\theta=\frac{\pi}{\delta}=\int\limits^{2\pi}_{0}\int\limits^{\infty}_{0}e^{-t}\frac{1}{2\delta}dtd\theta=\frac{\cancel{2}\pi}{\cancel{2}\delta}\int\limits^{\infty}_{0}e^{-t}dt=\frac{\pi}{\delta}.
\end{eqnarray}
}
\begin{figure}[h!]
\centering
    \includegraphics[height=3.0in]{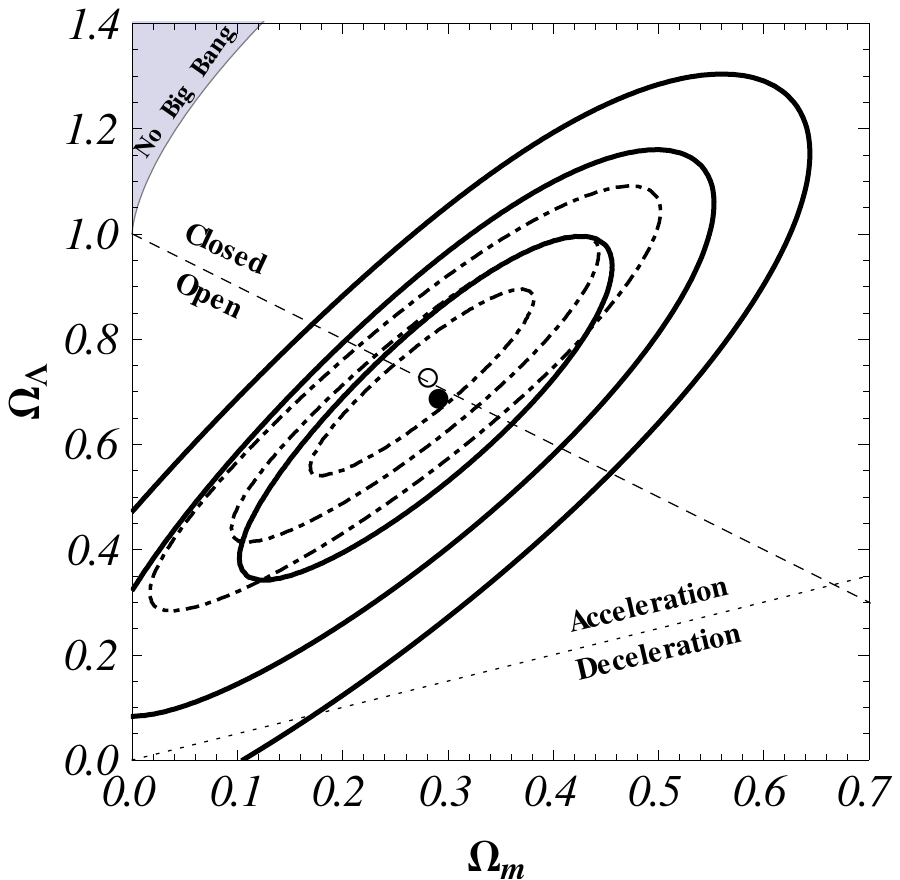}
    \includegraphics[height=3.0in]{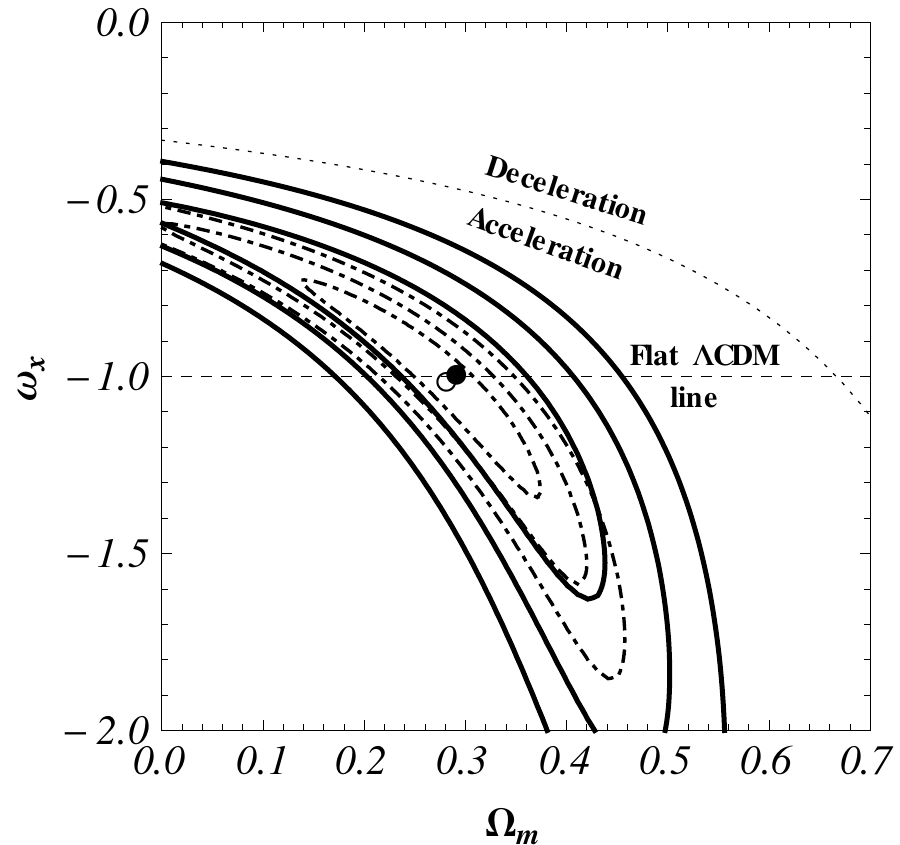}
    \includegraphics[height=3.0in]{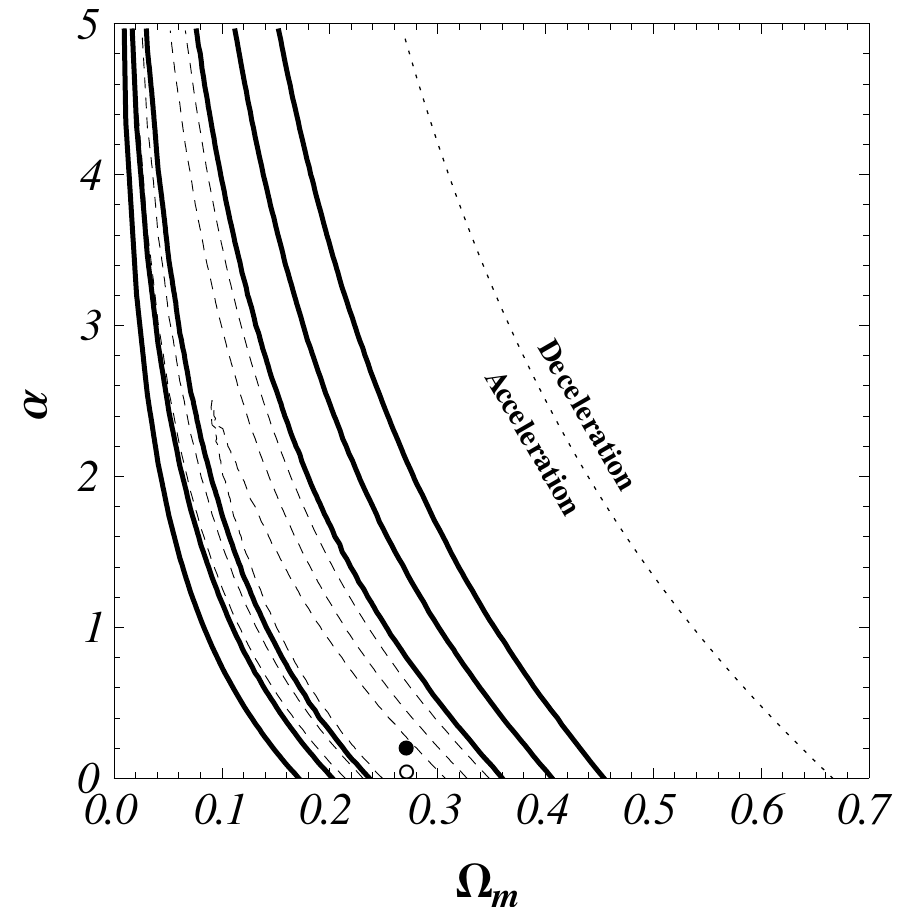}
\caption{Thick solid (dot-dashed) lines are 1$\sigma$, 2$\sigma$, and 3$\sigma$ constraint contours 
from SNIa data with (without) systematic errors. Filled (open) circles 
demarcate likelihood maxima for the case of data with (without) systematic 
errors. The top left plot is for the $\Lambda$CDM model, the top right 
plot is for the XCDM parametrization, and the bottom one is for the 
$\phi$CDM model. For quantitative details see Table\ (\ref{tab:results-1}).}
\label{fig:SNEIA-lxCDM}
\end{figure}
The above integral will come out to be:
\begin{eqnarray}
\int\limits_{-\infty}^{\infty}e^{-\delta\left(y-\frac{1}{2\delta}\right)^2}dy=\sqrt{\frac{\pi}{\delta}}.
\end{eqnarray}
Hence, the likelihood function will become:\footnote{Please note the typo in Eq.\ (26) of Farooq \textit{et al.}\cite{Farooq:2012ev} where we put $\delta^2$ by mistake actually it should be just $\delta$. The correct likelihood function is given in Eq.\ (\ref{eq:chiSN-4}), though it does not effect the calculations since it is constant and neglected anyway.}
\begin{eqnarray}
\label{eq:chiSN-4}
 \mathcal{L}_{SN}(\textbf{p}) =  \sqrt{\frac{\pi}{\delta}} 
{\rm exp}\left[-\frac{1}{2}\left(A(\textbf{p})-\frac{B^2(\textbf{p})}{C}
-2\varepsilon - \frac{1}{2\delta}\right)\right].
\end{eqnarray}
The $h$-independent: 
\begin{equation}
\label{eq:chiSN-5}
\chi^2_{SN}(\textbf{p}) =  -2~\mathrm{ln} \mathcal{L}_{SN}(\textbf{p}) 
 = A(\textbf{p})-\frac{B^2(\textbf{p})}{C} - 
   \frac{2 \mathrm{ln}(10)}{5C}B(\textbf{p})- Q,
\end{equation}
where $Q$ is a constant that does not depend on the model parameters $\textbf{p}$:
\begin{equation}
\label{eq:G}
Q=\frac{2(\mathrm{ln}10)^4}{625~C^2}+2~\mathrm{ln}\left(\frac{2\pi(\mathrm{ln}10)^2}{25~C}\right), \nonumber
\end{equation}
and so can be ignored.
We minimize $\chi_{SN}^2(\textbf{p})$ with respect to the model parameters 
$\textbf{p}$ to find the best-fit parameter values $\mathbf{p_0}$ and
constraint contours.

Figure (\ref{fig:SNEIA-lxCDM}) shows constraints from the SNIa data on 
the three dark energy models we consider here. For the $\Lambda$CDM 
model and the XCDM parametrization the constraints shown in Fig.\ 
(\ref{fig:SNEIA-lxCDM}) are in very good agreement with those in Figs.\ 
5 and 6 of Suzuki \textit{et al.}\cite{suzuki2012} The $\phi$CDM model SNIa data constraints 
shown in Fig.\ (\ref{fig:SNEIA-lxCDM}) have not previously been computed. 
Comparing the SNIa 
constraints of Fig.\ (\ref{fig:SNEIA-lxCDM}) to those which follow 
from the $H(z)$ data, Figs.\ (\ref{fig:LCDM_Hz21})---(\ref{fig:phiCDM_Hz21}),
it is clear that SNIa data provide tighter constraints on the  
$\Lambda$CDM model. For the XCDM case both SNIa data and $H(z)$ data
provide approximately similar constraints, while the SNIa constraints
are somewhat more restrictive than the $H(z)$ ones for the $\phi$CDM
model. However, in general, the SNIa constraints are not very significantly 
more restrictive than the $H(z)$ constraints, which is a remarkable
result. It is also reassuring that both data favor approximately 
similar regions of parameters space, for all three models we consider.
However, given that the degeneracy in parameter space is similar for the
$H(z)$ and SNIa data, a joint analysis of just these two data sets is 
unlikely to greatly improve the constraints.  

\section{Constraints from the BAO Data}
\label{BAO}
In an attempt to further tighten the cosmological parameter constraints,
we now include BAO data in the analysis. To constrain cosmological 
parameters using 
BAO data we follow the procedure of Blake \textit{et al.}\cite{blake11} To derive the BAO
constraints we make use of the distance parameter $D_V (z)$, a 
combination of the angular diameter distance and the Hubble parameter, 
given by:
\begin{equation}
\label{eq:D_V}
D_V(z)= \left[(1 + z)^2 d_A(z)^2 \frac{c~z}{H(z)}\right]^{1/3}.
\end{equation}
Here $d_A(z)$ is the angular diameter distance from Eq.\ (\ref{eq:ada3}):
\begin{equation}
\label{eq:D_A}
d_A(z)=\frac{y(z)}{H_0(1+z)},
\end{equation}
where $y(z)$ is the dimensionless coordinate distance given in Eq.\ 
({\ref{eq:Angular size distance}}).

\begin{figure}[h!]
\centering
    \includegraphics[height=3.0in]{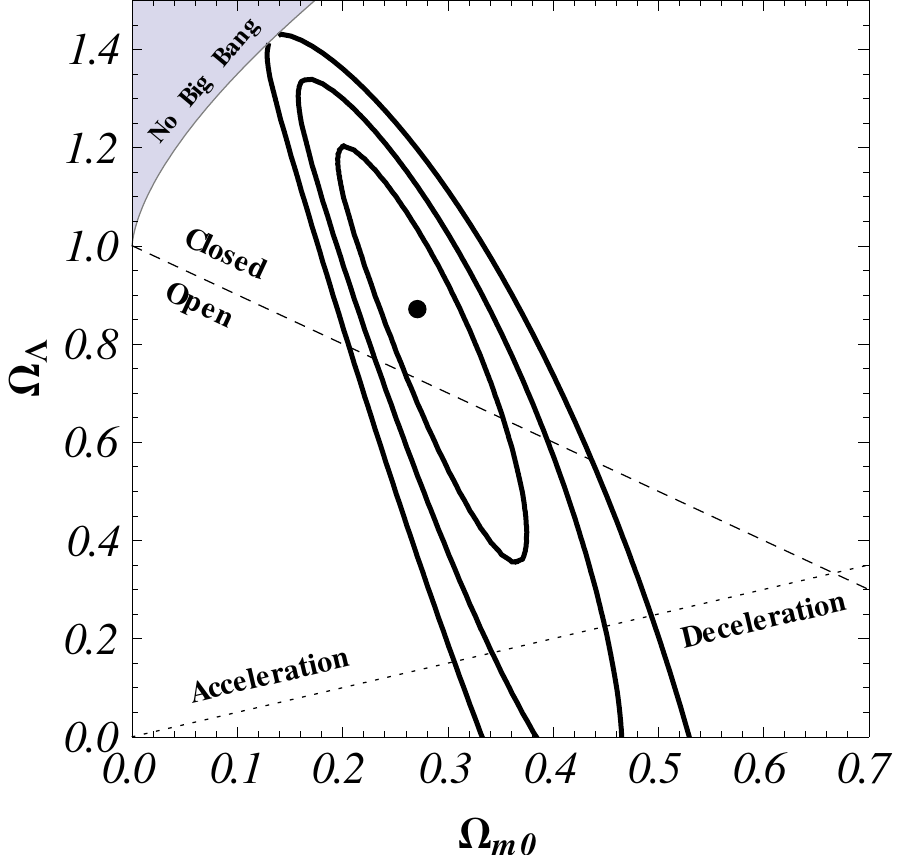}
    \includegraphics[height=3.0in]{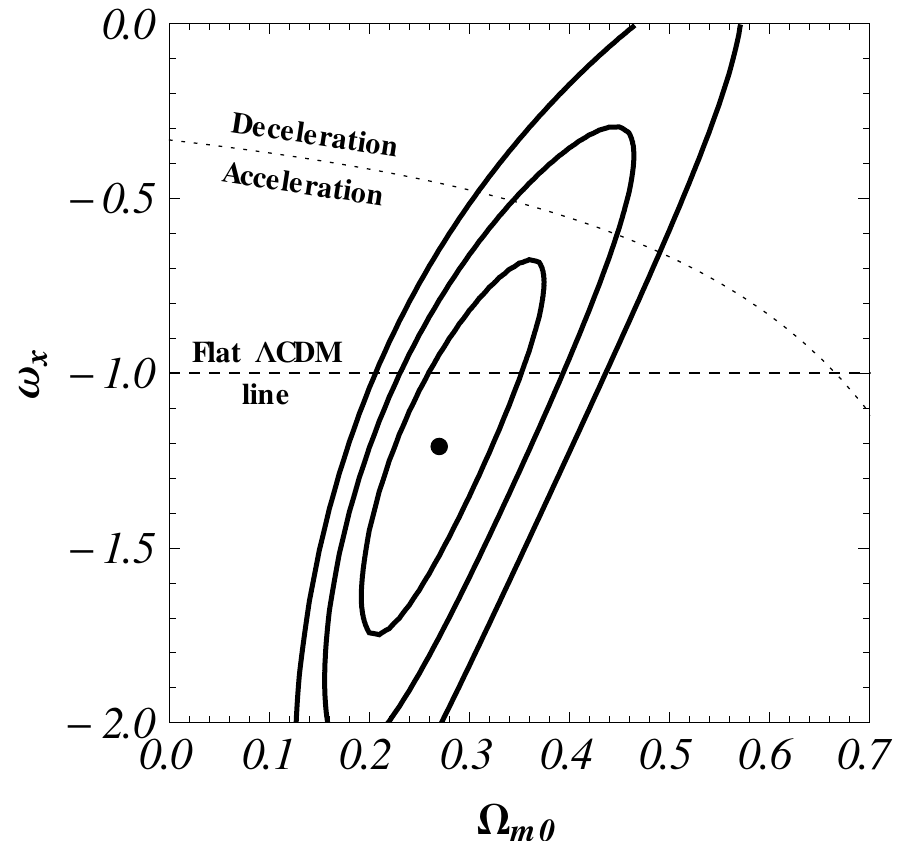}
    \includegraphics[height=3.0in]{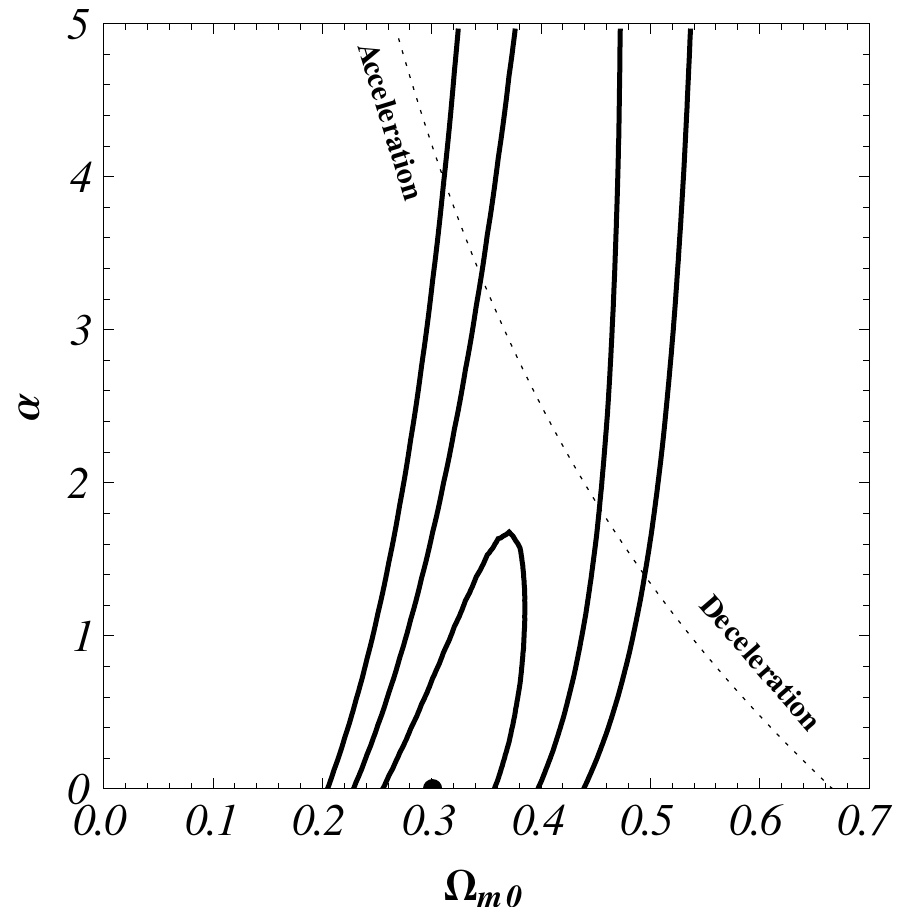}
\caption{1$\sigma$, 2$\sigma$, and 3$\sigma$ constraint contours from the BAO data. Filled 
circles denote likelihood maxima. The top left plot is for the 
$\Lambda$CDM model, the top right one is for the XCDM parametrization, 
and the bottom plot is for the $\phi$CDM model. For quantitative 
details see Table (\ref{tab:results-1}).}
\label{fig:allonlybao}
\end{figure}
We use measurements of the acoustic parameter $A(z)$ from Blake \textit{et al},\cite{blake11} 
where the theoretically-predicted $A_{\rm th}(z)$ is given in Eq.\ (5) 
of Eisenstein \textit{et al.}:\cite{Eisenstein2005}
\begin{equation}
\label{eq:A_z}
A_{\rm th}(z)=\frac{100~D_V(z)~\sqrt{\Omega_mh^2}}{z}.
\end{equation}
Using Eqs.\ ({\ref{eq:D_V}})---({\ref{eq:A_z}}) we have: 
\begin{equation}
\label{eq:A_z2}
A_{\rm th}(z)=\sqrt{\Omega_m} \left[\frac{y^2(z)}{z^2 E(z)}\right]^{1/3},
\end{equation}
which is $h$ independent and where $E(z)$ is defined in 
Chapter\ (\ref{Chapter2}). 

Using the WiggleZ $A_{\rm obs}(z)$ data from Table (3) of Blake \textit{et al},\cite{blake11}
we compute:
\begin{equation}
\chi_{A_z}^2(\textbf{p}) = \Delta{\textbf{A}}^T ({\rm C}_{A_z})^{-1} 
    \Delta{\textbf{A}}.
\end{equation}
Here $\Delta {\textbf{A}}$ is a vector consisting of differences
 $\Delta {A_i} = A_{\rm th}(z_i;\textbf{p}) - A_{\rm obs}(z_i)$ and 
$({\rm C}_{Az})^{-1}$ is the inverse of the 3 by 3 covariance matrix 
given in Table (3) of Blake \textit{et al}.\cite{blake11} 

We also use the 6dFGS and SDSS data, three measurements from Beutler \textit{et al},\cite{Beutler11} 
Percival \textit{et al},\cite{Percival2010}, listed in Blake \textit{et al}.\cite{blake11}
In this case the distilled parameter: 
\begin{equation}
\label{eq:dz}
d_{\rm th}(z)=\frac{r_s(z_d)}{D_V(z)},
\end{equation}
where $r_s(z_d)$ is the sound horizon at the drag epoch, is given in 
Eq.\ (6) of Eisenstein \textit{et al.}\cite{Eisenstein1998} The correlation coefficients for this case 
are also given in Table (3) of Blake \textit{et al.}\cite{blake11} Using the covariance 
matrix we define:
\begin{equation}
\label{eq:chiBAOdz}
\chi_{d_z}^2(h,\textbf{p}) = \Delta{\textbf{d}}^T ({\rm C}_{d_z})^{-1}
   \Delta{\textbf{d}}
\end{equation} 
where $\Delta {\textbf{d}}$ is a vector consisting of differences
$\Delta {d_i} = d_{\rm th}(z_i;h,\textbf{p}) - d_{\rm obs}(z_i)$ and 
${\rm C}_{d_z}$ is the the covariance matrix Blake \textit{et al.}.\cite{blake11}
We then marginalize over a flat prior for $H_0$ to get:
\begin{equation}
\label{eq:chiBAOdz2}
\chi_{d_z}^2(\textbf{p}) = -2~\mathrm{ln}\left[
\int^\infty_0{e^{-\chi_{d_z}^2(h,\textbf{p})/2}dh}\right].
\end{equation}

Since $\chi_{A_z}^2(\textbf{p})$ and $\chi_{d_z}^2(\textbf{p})$ correspond
to independent data, the combined BAO data:
\begin{equation}
\chi_{BAO}^2(\textbf{p})=\chi_{A_z}^2(\textbf{p})+ \chi_{d_z}^2(\textbf{p}) .
\end{equation}
We can maximize the likelihood by minimizing $\chi_{BAO}^2(\textbf{p})$ 
with respect to the model parameters $\textbf{p}$ to get best-fit
parameter values $\mathbf{p_0}$ and constraint contours. Figure 
(\ref{fig:allonlybao}) show the constraints from the BAO data on the 
three dark energy models we consider here. The XCDM parametrization 
constraints shown in this figure are in good agreement with those
shown in Fig.\ 13 of Blake \textit{et al.}\cite{blake11} The constraints shown in the 
other two panels of Fig.\ (\ref{fig:allonlybao}) have not previously 
been computed. Comparing to the $H(z)$ and SNIa constraint 
contours of Figs.\ (\ref{fig:LCDM_Hz21})---(\ref{fig:SNEIA-lxCDM}), we see 
that the BAO contours are also very elongated, although largely orthogonal 
to the $H(z)$ and SNIa ones. Consequently, a joint analysis of these data
will result in significantly tighter constraints than those derived
using any one of these data sets.  

\section{Joint Constraints}
\label{Joint}
To constrain cosmological parameters from a joint analysis of the $H(z)$,
SNIa, and BAO data we compute:  
\begin{equation}
\chi^2(\textbf{p}) = \chi_{H}^2(\textbf{p}) + \chi_{SN}^2(\textbf{p})
      + \chi_{BAO}^2(\textbf{p})
\end{equation}
for each of the three cosmological models considered here. We minimize 
$\chi^2(\textbf{p})$ with respect to model parameters $\textbf{p}$ to 
get best-fit parameter values $\mathbf{p_0}$ and constraint contours.

\begin{figure}[h]
\centering
    \includegraphics[height=3.0in]{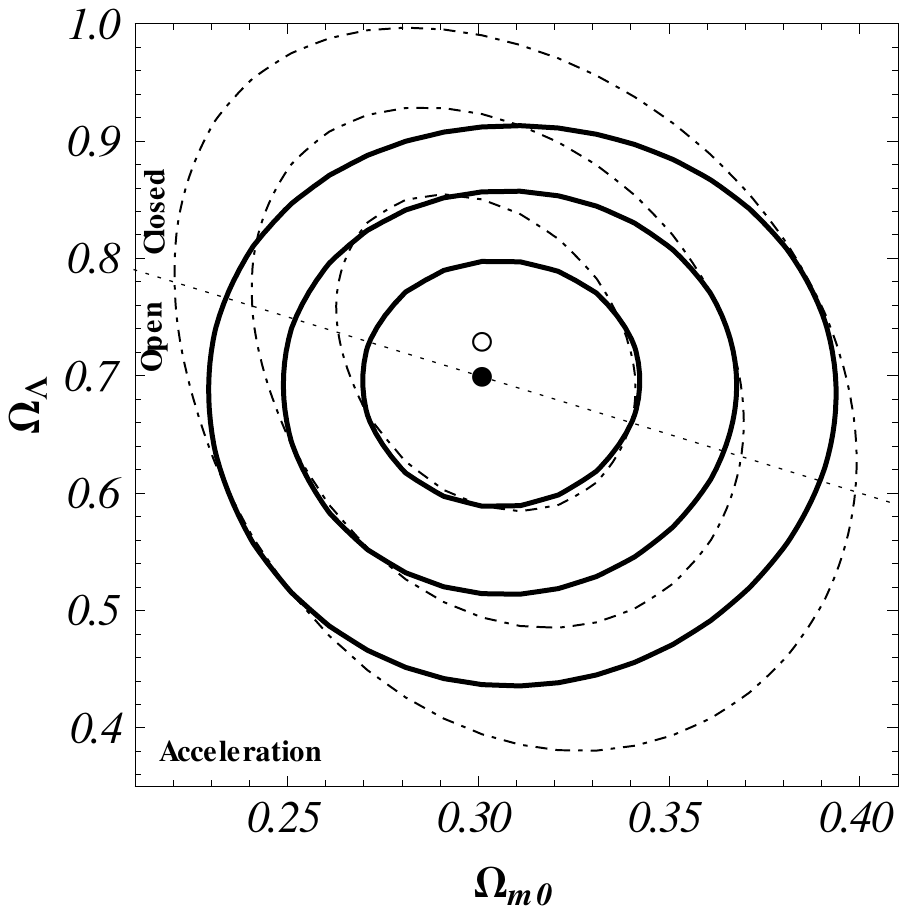}
    \includegraphics[height=3.0in]{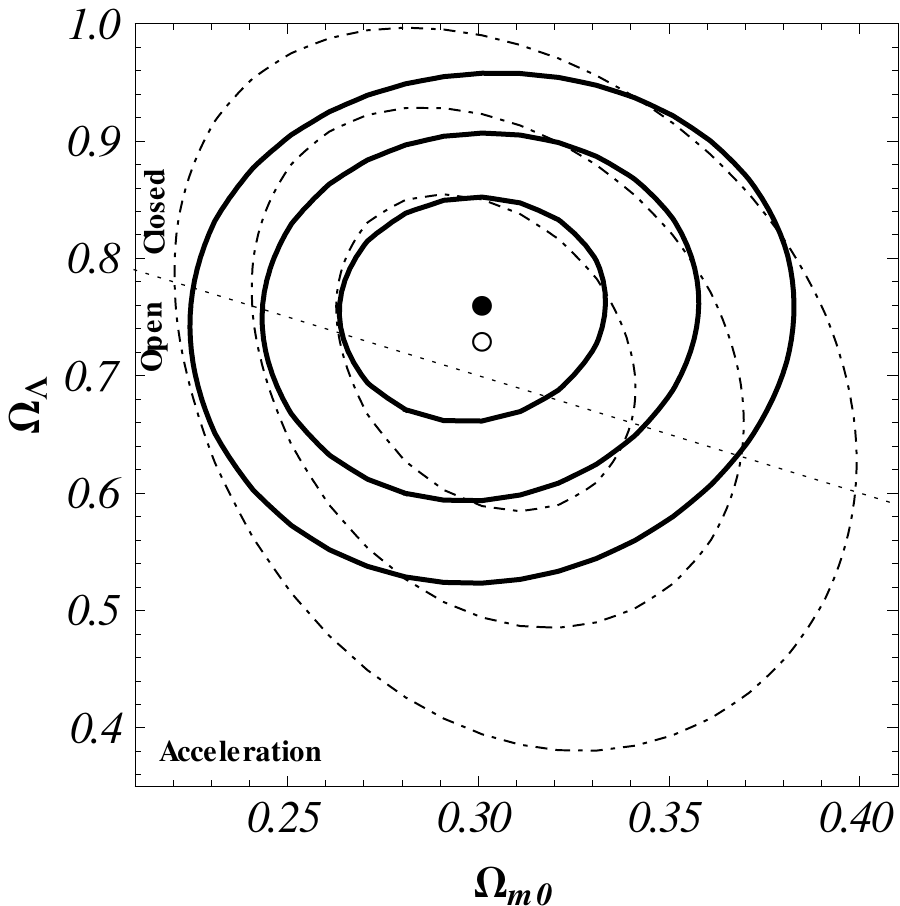}
\caption{Thick solid (dot-dashed) lines are 1$\sigma$, 2$\sigma$, and 3$\sigma$ constraint
contours for the $\Lambda$CDM model from a joint analysis of the BAO
and SNIa (with systematic errors) data, with (without) the $H(z)$ data. 
The full (empty) circle marks the best-fit point determined from the 
joint analysis with (without) the $H(z)$ data. The dotted sloping line 
corresponds to spatially-flat $\Lambda$CDM models. In the left panel 
we use the $H_0$ = 68 $\pm$ 2.8 km s$^{-1}$ Mpc$^{-1}$ prior while 
the right panel is for the $H_0$ = 73.8 $\pm$ 2.4 km s$^{-1}$ Mpc$^{-1}$ 
case. For quantitative details see Table (\ref{tab:results-2}).}
\label{fig:LCDM_com}
\end{figure}

\begin{figure}[h]
\centering
    \includegraphics[height=3.0in]{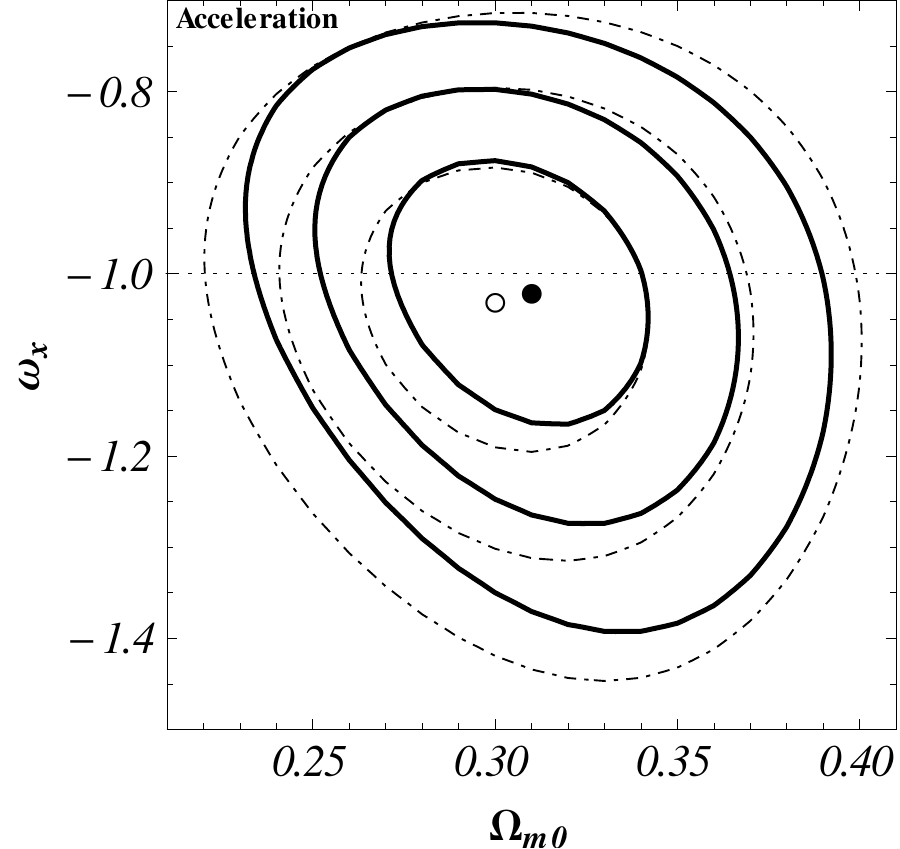}
    \includegraphics[height=3.0in]{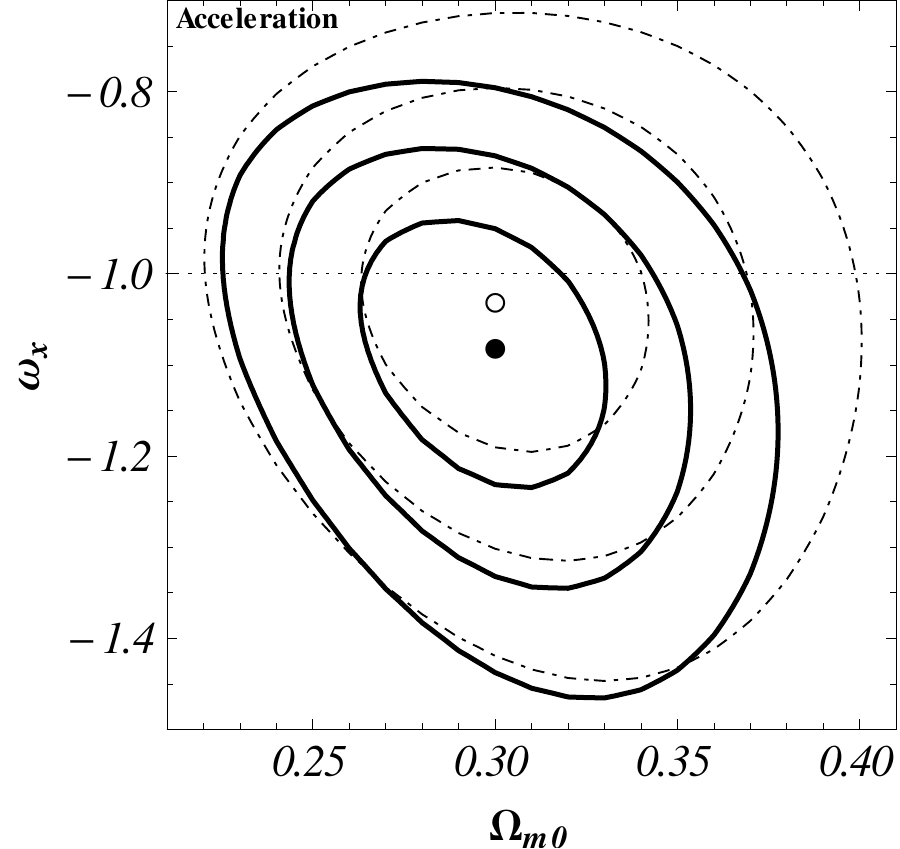}
\caption{Thick solid (dot-dashed) lines are 1$\sigma$, 2$\sigma$, and 3$\sigma$ constraint
contours for the XCDM parametrization from a joint analysis of the BAO
and SNIa (with systematic errors) data, with (without) the $H(z)$ data. 
The full (empty) circle marks the best-fit point determined from the 
joint analysis with (without) the $H(z)$ data. The dotted horizontal 
line at $\omega_{\rm X} =-1$  corresponds to spatially-flat $\Lambda$CDM 
models. In the left panel we use the $H_0$ = 68 $\pm$ 2.8 km s$^{-1}$ 
Mpc$^{-1}$ prior while the right panel is for the $H_0$ = 73.8 
$\pm$ 2.4 km s$^{-1}$ Mpc$^{-1}$ case. For quantitative details see 
Table (\ref{tab:results-2}).}
\label{fig:XCDM_com}
\end{figure}

\begin{figure}[h]
\centering
    \includegraphics[height=3.0in]{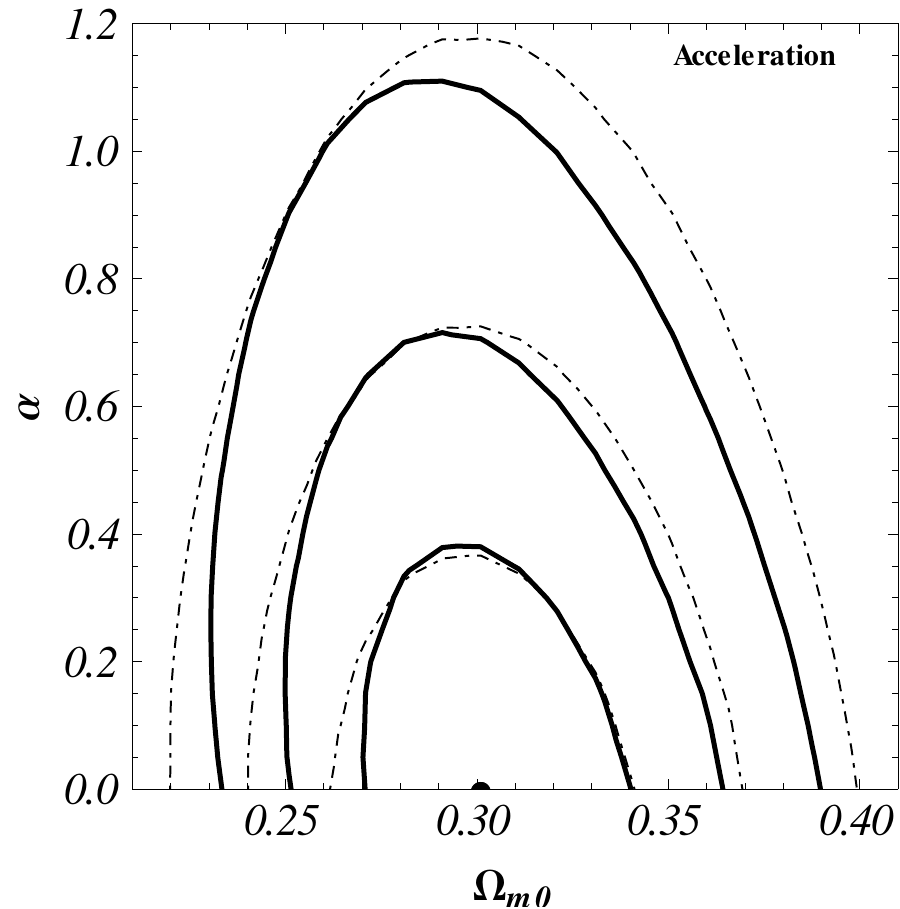}
    \includegraphics[height=3.0in]{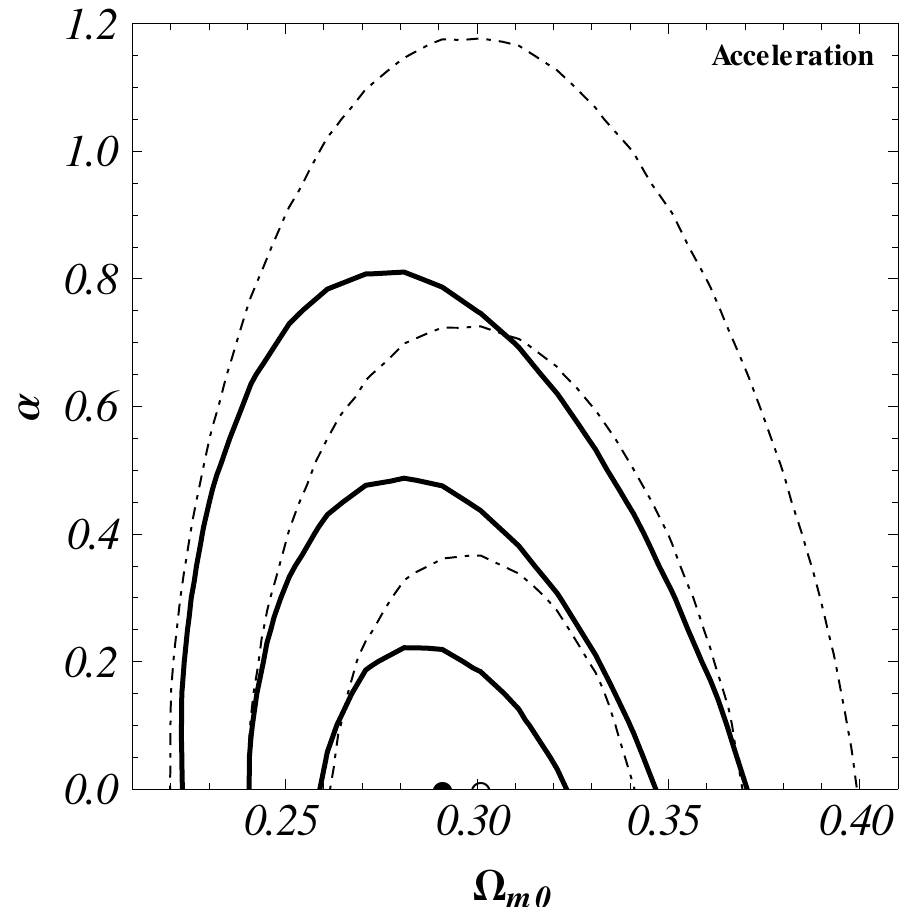}
\caption{Thick solid (dot-dashed) lines are 1$\sigma$, 2$\sigma$, and 3$\sigma$ constraint
contours for the $\phi$CDM model from a joint analysis of the BAO
and SNIa (with systematic errors) data, with (without) the $H(z)$ 
data. The full (empty) circle marks the best-fit point determined
from the joint analysis with (without) the $H(z)$ data (in the left
panel the full and empty circles overlap). The $\alpha = 0$
horizontal axes correspond to spatially-flat $\Lambda$CDM models.
In the left panel we use the $H_0$ = 68 $\pm$ 2.8 km s$^{-1}$ Mpc$^{-1}$ 
prior while the right panel is for the $H_0$ = 73.8 $\pm$ 2.4 km s$^{-1}$ 
Mpc$^{-1}$ case. For quantitative details see Table (\ref{tab:results-2}).}
\label{fig:phiCDM_com}
\end{figure}

\begin{table}[htb]
\begin{center}
\begin{tabular}{ccccccc} 
\hline\hline 
\multirow{2}{*}{Model and prior} &  
\multicolumn{2}{c}{$H(z)$+BAO}
&\multicolumn{2}{c}{$H(z)$+SNIa+BAO}
&\multicolumn{2}{c}{SNIa+BAO}\\
{} & $\chi^2_{\rm min}$ & B.F.P & $\chi^2_{\rm min}$ & B.F.P & $\chi^2_{\rm min}$ & B.F.P\\ 
\hline \hline 
$\Lambda$CDM & \multirow{2}{*}{$20.7$} & $\Omega_{m0}=0.31$ & \multirow{2}{*}{566} & $\Omega_{m0}=0.30$ & \multirow{4}{*}{$551$} & \multirow{2}{*}{$\Omega_{m0}=0.30$} \\ 
$h = 0.68 \pm 0.028$ &  & $\Omega_{\Lambda}=0.68$ &  & $\Omega_{\Lambda}=0.70$ &  & {} \\ 
\cline{1-5}

$\Lambda$CDM & \multirow{2}{*}{$21.0$} & $\Omega_{m0}=0.29$ & \multirow{2}{*}{567} & $\Omega_{m0}=0.30$ & \multirow{2}{*}{} & \multirow{2}{*}{$\Omega_{\Lambda}=0.73$} \\ 
$h = 0.738 \pm 0.024$ &  & $\Omega_{\Lambda}=0.79$ &  & $\Omega_{\Lambda}=0.76$ &  & \multirow{4}{*}{}\\ 
\hline
\hline

XCDM & \multirow{2}{*}{$20.7$} & $\Omega_{m0}=0.31$ & \multirow{2}{*}{566} & $\Omega_{m0}=0.31$ & \multirow{4}{*}{$551$} & \multirow{2}{*}{$\Omega_{m0}=0.30$} \\ 
$h = 0.68 \pm 0.028$ &  & $\omega_{X}=-0.99$ &  & $\omega_{X}=-1.02$ &  & {} \\
\cline{1-5}

XCDM & \multirow{2}{*}{$20.8$} & $\Omega_{m0}=0.28$ & \multirow{2}{*}{567} & $\Omega_{m0}=0.30$ & \multirow{2}{*}{} & \multirow{2}{*}{$\omega_{X}=-1.03$} \\ 
$h = 0.738 \pm 0.024$ &  & $\omega_{X}=-1.19$ &  & $\omega_{X}=-1.08$ &  & \multirow{4}{*}{}\\ 
\hline 
\hline

$\phi$CDM & \multirow{2}{*}{$20.7$} & $\Omega_{m0}=0.31$ & \multirow{2}{*}{566} & $\Omega_{m0}=0.30$ & \multirow{4}{*}{551} & \multirow{2}{*}{$\Omega_{m0}=0.30$} \\ 
$h = 0.68 \pm 0.028$ &  & $\alpha=0.05$ &  & $\alpha=0.00$ &  & {} \\
\cline{1-5}

$\phi$CDM & \multirow{2}{*}{$22.0$} & $\Omega_{m0}=0.29$ & \multirow{2}{*}{567} & $\Omega_{m0} =0.29$ & \multirow{2}{*}{} & \multirow{2}{*}{$\alpha =0.00$} \\ 
$h = 0.738 \pm 0.024$ &  & $\alpha =0.00$ &  & $\alpha =0.00$ &  & \multirow{4}{*}{}\\ 
\hline
\hline  
\end{tabular}

\caption{The minimum value of $\chi^2$ and the corresponding best fit 
points (B.F.P) which maximize the likelihood, for different combinations 
of data. The SNIa data values are for the case including systematic errors. }
\label{tab:results-2}
\end{center}
\end{table}

\begin{figure}[h!]
\centering
    \includegraphics[height=3.0in]{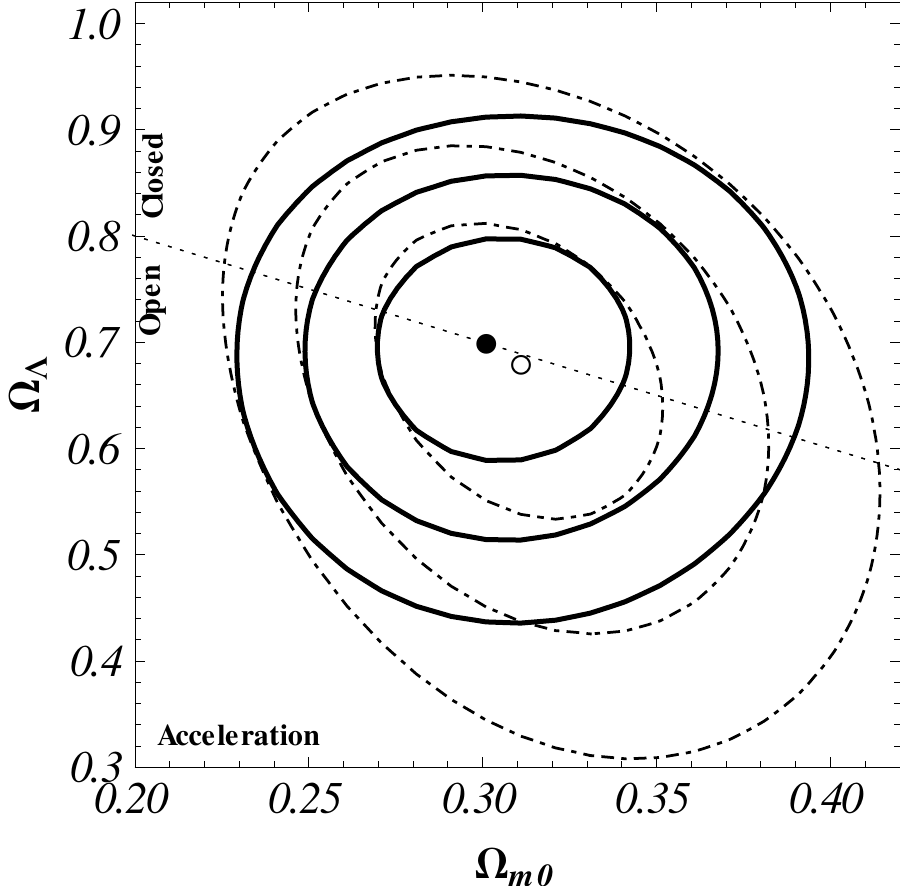}
    \includegraphics[height=3.0in]{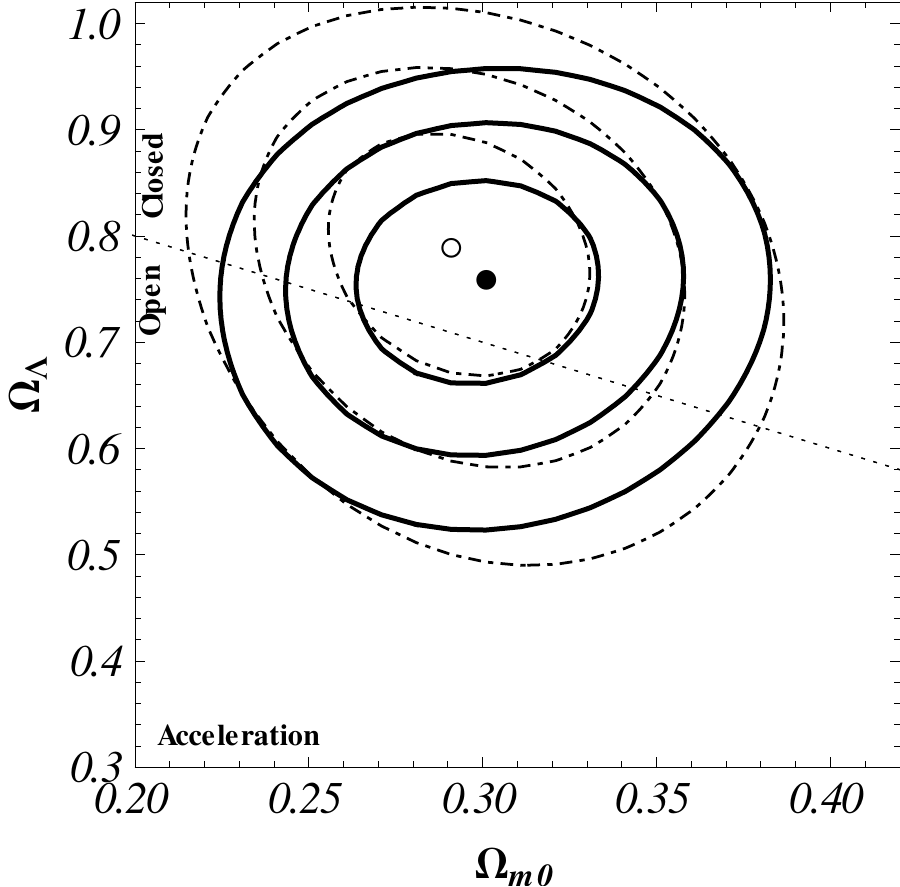}
\caption{
Thick solid (dot-dashed) lines are 1$\sigma$, 2$\sigma$, and 3$\sigma$ constraint
contours for the $\Lambda$CDM model from a joint analysis of the BAO
and $H(z)$ data, with (without) the SNIa data. 
The full (empty) circle marks the best-fit point determined from the 
joint analysis with (without) the SNIa data. The dotted sloping line 
corresponds to spatially-flat $\Lambda$CDM models. In the left panel 
we use the $H_0$ = 68 $\pm$ 2.8 km s$^{-1}$ Mpc$^{-1}$ prior while  
the right panel is for the $H_0$ = 73.8 $\pm$ 2.4 km s$^{-1}$ Mpc$^{-1}$ 
case. For quantitative details see Table (\ref{tab:results-2}).
}
\label{fig:LCDM_com2}
\end{figure}


\begin{figure}[h]
\centering
    \includegraphics[height=3.0in]{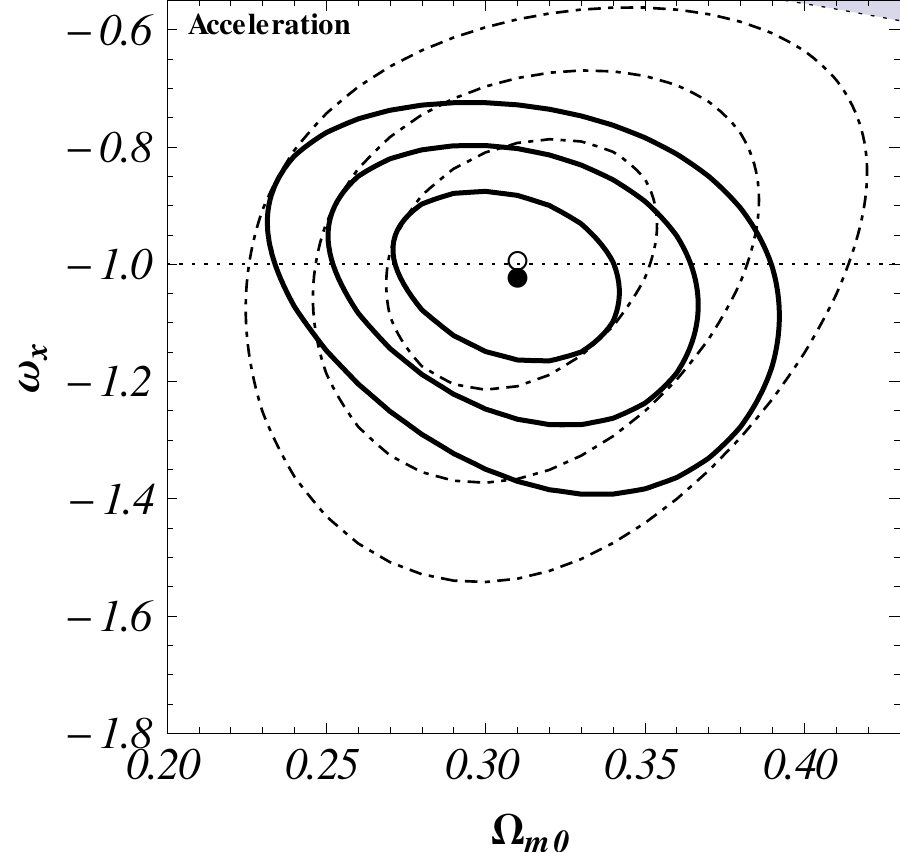}
    \includegraphics[height=3.0in]{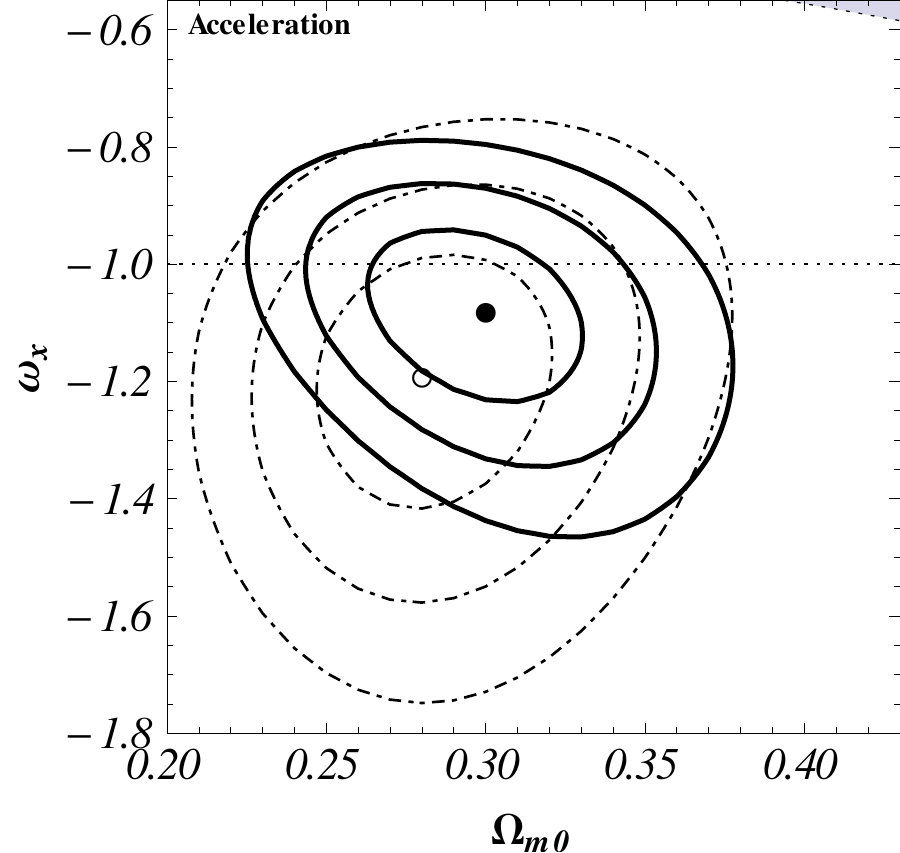}
 \caption{
Thick solid (dot-dashed) lines are 1$\sigma$, 2$\sigma$, and 3$\sigma$ constraint
contours for the XCDM parametrization from a joint analysis of the BAO
and $H(z)$ data, with (without) the SNIa data. 
The full (empty) circle marks the best-fit point determined from the 
joint analysis with (without) the SNIa data. The dotted horizontal 
line at $\omega_{\rm X} =-1$  corresponds to spatially-flat $\Lambda$CDM 
models. In the left panel we use the $H_0$ = 68 $\pm$ 2.8 km s$^{-1}$ 
Mpc$^{-1}$ prior while the right panel is for the $H_0$ = 73.8 $\pm$ 2.4 
km s$^{-1}$ Mpc$^{-1}$ case. The shaded area in the upper right corners 
are the region of decelerating expansion. For quantitative details see
Table (\ref{tab:results-2}).
} \label{fig:XCDM_com2}
\end{figure}


\begin{figure}[h!]
\centering
    \includegraphics[height=3.0in]{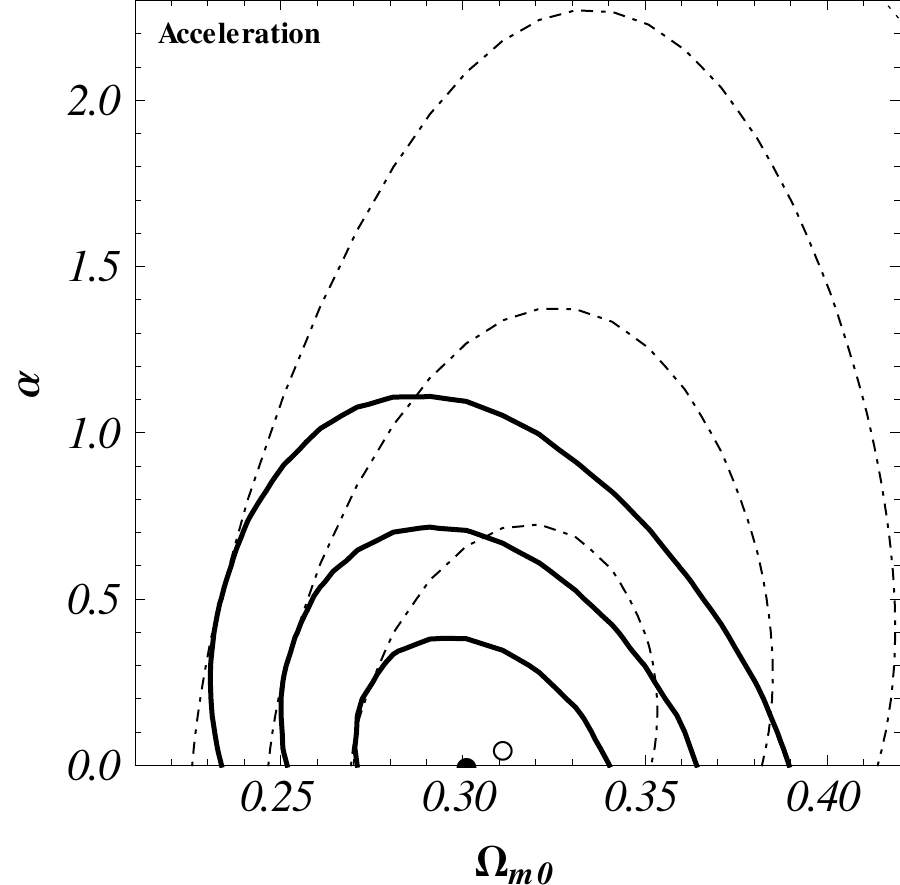}
    \includegraphics[height=3.0in]{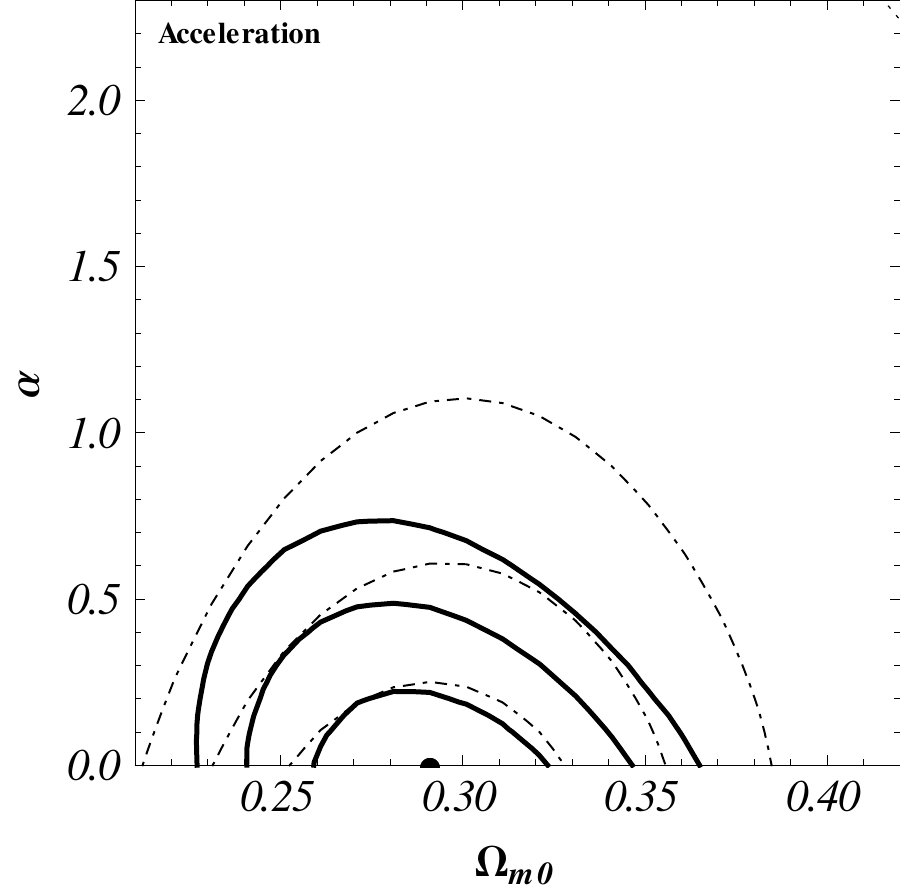}
 \caption{
Thick solid (dot-dashed) lines are 1, 2, and 3 $\sigma$ constraint
contours for the $\phi$CDM model from a joint analysis of the $H(z)$
and BAO data, with (without) the SNIa 
data. The full (empty) circle marks the best-fit point determined
from the joint analysis with (without) the SNIa data. The $\alpha = 0$
horizontal axes correspond to spatially-flat $\Lambda$CDM models.
In the left panel we use the $H_0$ = 68 $\pm$ 2.8 km s$^{-1}$ Mpc$^{-1}$ 
prior while the right panel is for the $H_0$ = 73.8 $\pm$ 2.4 km s$^{-1}$ 
Mpc$^{-1}$ case. For quantitative details see Table (\ref{tab:results-2}).
}
\label{fig:phiCDM_com2}
\end{figure}


Figures (\ref{fig:LCDM_com})---(\ref{fig:phiCDM_com}) show
constraints on the cosmological parameters for the $\Lambda$CDM and
$\phi$CDM models and the XCDM parametrization, from a joint analysis
of the BAO and SNIa data, as well as from a joint analysis of the
BAO, SNIa and $H(z)$ data. Table 3 lists information about best-fit
parameter values. Including the $H(z)$ data in the analysis 
tightens the constraints by more than one standard
deviation, in parts of the parameter spaces. 

\begin{table}[htb]
\begin{center}
\begin{tabular}{cccc}
\hline\hline
{Model and prior}	& {SNIa+BAO}	&{$H(z)$+BAO} & {$H(z)$+SNIa+BAO} \\
\hline
 {$\Lambda $CDM } & {0.25 $<$ }$\Omega_{m0}${ $<$ 0.36} &  {0.25 $<$ }$\Omega_{m0}${ $<$ 0.36} & {0.26 $<$ }$\Omega_{m0}${ $<$ 0.36} \\
 {$h=0.68 \pm 0.028$} & {0.53 $<$ }$\Omega _{\Lambda }${ $<$ 0.89} & {0.45 $<$ }$\Omega _{\Lambda }${ $<$ 0.85}  &{0.55 $<$ }$\Omega _{\Lambda }${ $<$ 0.83} \\
 \hline

 {$\Lambda $CDM} & {0.25 $<$ }$\Omega_{m0}${ $<$ 0.36} & {0.23 $<$ }$\Omega_{m0}${ $<$ 0.38} & {0.25 $<$ }$\Omega_{m0}${ $<$ 0.35} \\
 {$h = 0.738 \pm 0.024$} & {0.53 $<$ }$\Omega _{\Lambda }${ $<$ 0.89} & {0.60 $<$ }$\Omega _{\Lambda }${ $<$ 0.92} &{0.62 $<$ }$\Omega _{\Lambda }${ $<$ 0.88} \\
 \hline

 {XCDM} & {0.30 $<$ }$\Omega_{m0}${ $<$ 0.38} & {0.26 $<$ }$\Omega_{m0}${ $<$ 0.37} &  {0.29 $<$ }$\Omega_{m0}${ $<$ 0.37} \\
 {$h = 0.68 \pm 0.028$} & $-1.18 <  \omega_{\rm X} < -0.78$ & $-1.32 <  \omega_{\rm X} < -0.73$ & $-1.14 < \omega_{\rm X} < -0.78$ \\
 \hline 

 {XCDM} & {0.30 $<$ }$\Omega_{m0}${ $<$ 0.38} & {0.24 $<$ }$\Omega_{m0}${ $<$ 0.35} & {0.27 $<$ }$\Omega_{m0}${ $<$ 0.35} \\
 { $h = 0.738 \pm 0.024$} & $-1.18 < \omega_{\rm X} < -0.78$ & $-1.42 < \omega_{\rm X} < -0.88$ & $ -1.22 < \omega_{\rm X} < -0.86$ \\
  \hline

 {$\phi $CDM} & {0.25 $<$ }$\Omega_{m0}${ $<$ 0.35} & {0.25 $<$ }$\Omega_{m0}${ $<$ 0.36} & {0.26 $<$ }$\Omega_{m0}${ $<$ 0.35} \\
 {$h = 0.68 \pm 0.028$} & {0 $<$ $\alpha $ $<$ 0.54} & {0 $<$ $\alpha $ $<$ 1.01} & {0 $<$ $\alpha $ $<$ 0.54} \\
  \hline

 {$\phi $CDM} & {0.25 $<$ }$\Omega_{m0}${ $<$ 0.35} & {0.23 $<$ }$\Omega_{m0}${ $<$ 0.35} & {0.25 $<$ }$\Omega_{m0}${ $<$ 0.33} \\
 {$h = 0.738 \pm 0.024$} & {0 $<$ $\alpha $ $<$ 0.54} & {0 $<$ $\alpha $ $<$ 0.57} & {0 $<$ $\alpha $ $<$ 0.35} \\
\hline\hline
\end{tabular}

\caption{Two standard deviation bounds on cosmological
parameters using SNIa+BAO, $H(z)$+BAO and SNIa+BAO+$H(z)$ data, for three 
different models with two different $H_0$ priors.}
\label{tab:intervals}
\end{center}
\end{table}

Adding the $H(z)$ data for the $\bar{H_{0}}\pm\sigma_{H_{0}} 
= 68 \pm 2.8$ km s$^{-1}$ Mpc$^{-1}$ prior case improved the
constraints most significantly in the $\Lambda$CDM case (by more
than 1$\sigma$ on $\Omega_{\Lambda}$ in parts of parameter space), 
Fig.\ (\ref{fig:LCDM_com}), and least significantly for the $\phi$CDM model,
Fig.\ (\ref{fig:phiCDM_com}). For the case of the 
$\bar{H_{0}}\pm\sigma_{H_{0}} = 73.8 \pm 2.4$ km s$^{-1}$ Mpc$^{-1}$ prior, 
adding $H(z)$ again tightens up the constraints the most for the 
$\Lambda$CDM model (by more than 1$\sigma$ on $\Omega_{\Lambda}$), 
Fig.\ (\ref{fig:LCDM_com}), and least so for the 
XCDM parametrization, Fig.\ (\ref{fig:XCDM_com}).

Figures (\ref{fig:LCDM_com2})---(\ref{fig:phiCDM_com2}) show the constraints 
on the cosmological parameters of the three models, from a joint analysis 
of the BAO and $H(z)$ data, as well as from a joint analysis of the 
three data sets. Table (\ref{tab:results-2}) lists the best-fit parameter values. Comparing these 
figures to Figs.\ (\ref{fig:LCDM_com})---(\ref{fig:phiCDM_com}) allows for a
comparison between the discriminating power of the SNIa and $H(z)$ data.  

Figure (\ref{fig:LCDM_com2}) shows that adding SNIa data to the $H(z)$ and BAO 
data combination for the $\bar{H_{0}}\pm\sigma_{H_{0}} = 68 \pm 2.8$ 
km s$^{-1}$ Mpc$^{-1}$ prior case tightens up the constraints by more than 
1$\sigma$ on $\Omega_{\Lambda}$ from below, while addition of SNIa data 
for the $\bar{H_{0}}\pm\sigma_{H_{0}} = 73.8 \pm 2.4$ km s$^{-1}$ Mpc$^{-1}$ 
prior case tightens up the constraints by more than 1$\sigma$ on 
$\Omega_{\Lambda}$ from above. Addition of SNIa data to the $H(z)$ and BAO 
combination doesn't much improve the constraints on $\Omega_{m0}$ for 
either prior.

Figures (\ref{fig:LCDM_com2})---(\ref{fig:phiCDM_com2}) show that adding 
SNIa data to the $H(z)$ and BAO combination results in the most prominent 
effect for the XCDM case, Fig.\ (\ref{fig:XCDM_com2}). Here for the 
$\bar{H_{0}}\pm\sigma_{H_{0}} = 68 \pm 2.8$ km s$^{-1}$ Mpc$^{-1}$ 
prior it tightens up the constraints by more than 1$\sigma$ on 
$\omega_{X}$ from above and below while for the 
$\bar{H_{0}}\pm\sigma_{H_{0}} = 73.8 \pm 2.4$ km s$^{-1}$ Mpc$^{-1}$ 
prior it tightens up the constraints by more than 2 $\sigma$ on $\omega_{X}$
from below. Addition of SNIa data to the $H(z)$ and BAO combination 
doesn't  much improve the constraints on $\Omega_{m0}$ for either prior
in this case.

In the $\phi$CDM case, Fig.\ (\ref{fig:phiCDM_com2}), adding SNIa data to 
$H(z)$ and BAO combination affects the constraint on $\alpha$ the most
for the $\bar{H_{0}}\pm\sigma_{H_{0}} = 68 \pm 2.8$ km s$^{-1}$ Mpc$^{-1}$ 
prior case. The effect on $\Omega_{m0}$ is little stronger than what 
happens in the $\Lambda$CDM and XCDM cases but still less than 1$\sigma$.

Table\ ({\ref{tab:intervals}}) lists the two standard deviation bounds on the individual 
cosmological parameters, determined from their one-dimensional
posterior probability distributions  functions (which are derived
by marginalizing the two-dimensional likelihood over the other
cosmological parameter) for different combinations of data set. 

The constraints on the cosmological parameters that we derive from 
only the BAO and SNIa data are restrictive, but less so than those 
shown in Fig.\ 4 of Yun \& Ratra.\cite{Chen2011b} This is because the new 
Suzuki \textit{et al.}\cite{suzuki2012} SNIa compilation data we use here is based on a more
careful accounting of the systematic errors, which have increased.
Consequently, including the $H(z)$ data, in addition to the BAO and 
SNIa data, in the analysis, more significantly tightens 
the constraints: compare Figs.\ (\ref{fig:LCDM_com})---(\ref{fig:phiCDM_com})
here to Figs.\ 4---6 of  Yun \& Ratra.\cite{Chen2011b} We emphasize, however, that
this effect is prominent only in some parts of the parameter spaces. 

\section{Conclusion}
\label{summary}

In summary, the results of a joint analysis of the $H(z)$, BAO, and SNIa 
data are very consistent with the predictions of a spatially-flat 
cosmological model with energy budget dominated by a time-independent 
cosmological constant, the standard $\Lambda$CDM model. However, the 
data are not yet good enough to strongly rule out slowly-evolving dark 
energy density. More, and better quality, data are needed to discriminate
between constant and slowly-evolving dark energy density.

It is probably quite significant that current $H(z)$ data constraints
are almost as restrictive as those from SNIa data. The acquisition 
of $H(z)$ data has been an interesting backwater of cosmology for the 
last few years. We hope that our results will help promote more 
interest in this exciting area. Since the $H(z)$ technique has not 
been as much studied as, say, the SNIa apparent magnitude technique, 
a little more effort in the $H(z)$ area is likely to lead to very
useful results.

\section{Addition of ${z=2.3}$ Data Point}

At the same time when we were doing the above analysis Nicol\^{a}s Busca and his team\cite{busca12} did the measurements of Hubble parameter $z$ at the high redshift $(z=2.3)$ using the Lyman $\alpha$ forest baryon acoustic oscillations measurements. Here we want to briefly describe the Lyman $\alpha$ forest which is becoming a very useful source of information in physical cosmology.   

\begin{figure}[h]
\centering
    \includegraphics[height=4.0in]{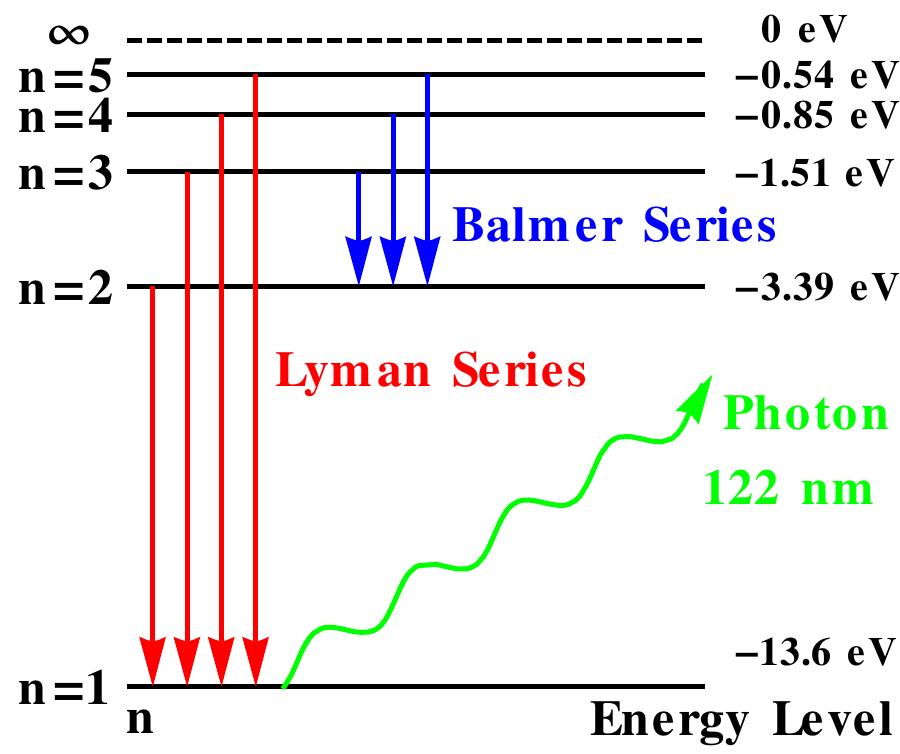}
 \caption{
Energy levels of neutral hydrogen atom. The transition to $n=1$ shell from any other shell
is called the Lyman series while the transition from any higher then $n=2$ shell to $n=2$ shell
is called the Balmer series that lies in visible (blue) range. When an electron jumps from 
any excited shell of the hydrogen atom to the ground level $(n=1)$ it emits a photon. If the transition is
from $n=2$ to $n=1$ then the wavelength of the photon is calculated to be 122 nm from Bohr's theory.
}
\label{fig:hydrogenenergy}
\end{figure}

The Lyman series is the series of energies required to excite an electron in the (neutral) hydrogen atom from its lowest energy state (n=1) to any higher energy state. The case of particular interest for cosmology is where a hydrogen atom with its electron in the lowest energy configuration n=1, gets hit by a photon (electromagnetic wave coming probably from a quasar behind the gas cloud) and is excited to the next lowest energy level n=2. The energy levels are given by $E_{\mathrm n} = -13.6 \mathrm{eV}/{\mathrm n}^2$ using Bohr's theory of atomic model, see Fig.\ (\ref{fig:hydrogenenergy}), and the energy difference between the lowest $(n=1)$ and second lowest $(n=2$) levels corresponds to a photon with wavelength 1216 $\AA$ $\approx$ 122 nm.\footnote{$\AA$ is called angstrom, and, 1 $\AA=10^{-10}$ m.} The reverse process i.e., emission of the photon and hence energy, after the electron jumps from $n=2$ to $n=1$ can and does occur as well.
\begin{figure}[h!]
\centering
    \includegraphics[height=7.0in]{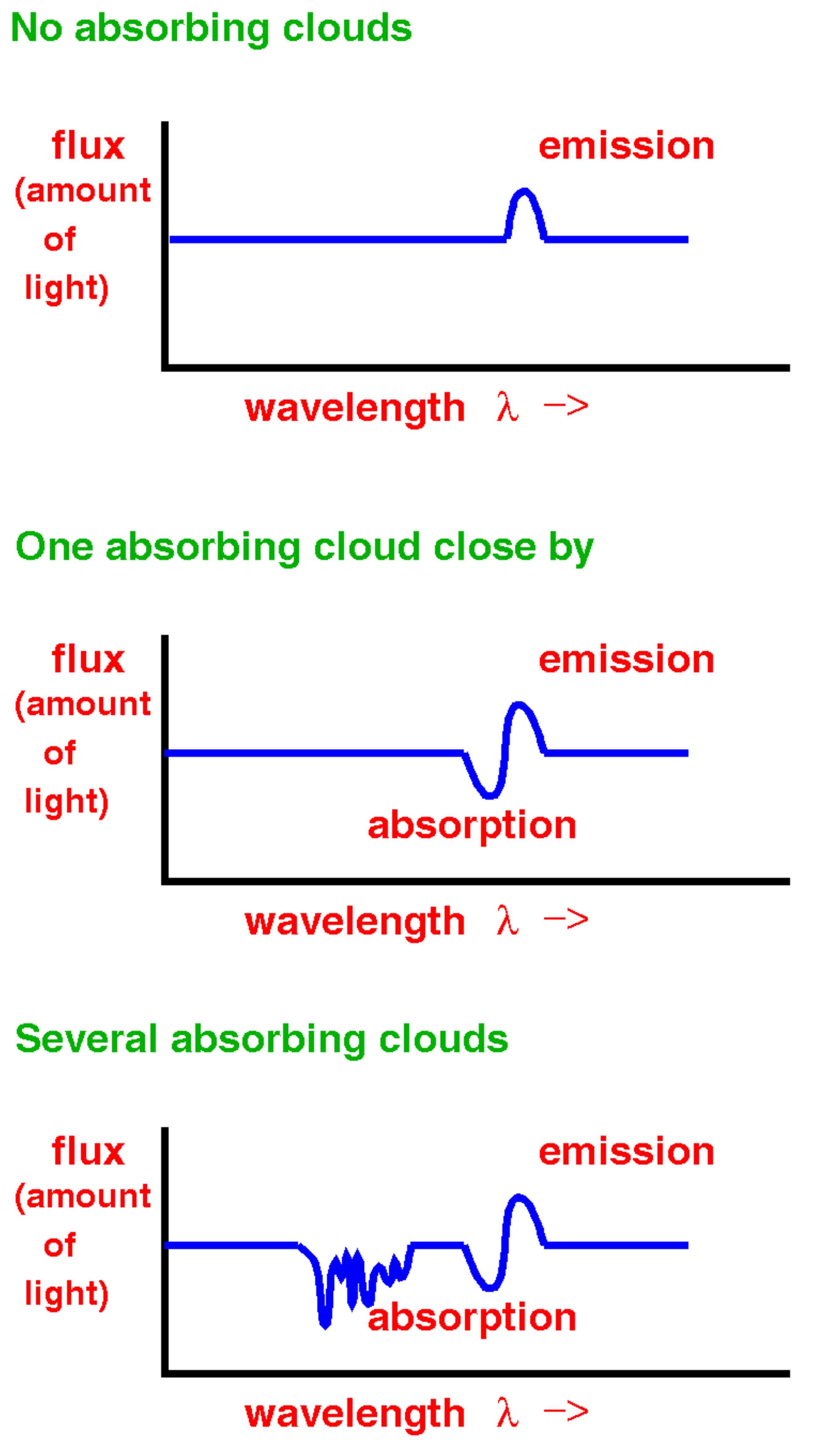}
 \caption{
Top picture shows a cartoon of how a quasar spectrum (the flux of light as a function of wavelength) might look if there were no intervening neutral hydrogen between the quasars and us. The middle picture shows the flux for one nearby region, while the bottom picture shows the case for several intervening regions.
}
\label{fig:lyman}
\end{figure}
The absorption or emission of photons with the correct wavelength can tell us something about the presence of hydrogen and free electrons in space. That is, if you shine a light with wavelength 122 nm at a bunch of neutral hydrogen atoms, in their ground state, the atoms will absorb the light, using it to boost the electron to a higher energy state. If there are a lot of neutral hydrogen atoms in their ground state n=1, they will absorb more and more of the light (photons of this particular wavelength). So if you look at the light you receive and view the intensity as a function of wavelength, you will see a dip in the intensity at 122 nm. This dip in the intensity is dependent on the amount of neutral hydrogen present in its ground state. The amount of light absorbed, the optical depth is proportional to the probability that the hydrogen will absorb the photon (cross-section) times the number of hydrogen atoms along its path. 

Since the Universe has many high energy photons and hydrogen atoms, both the absorption and emission of photons occurs frequently. In Lyman $\alpha$ systems, the hydrogen is found in regions in space, and the source for the photons are quasars (also called QSOs), very high energy light sources, shining at us from behind these regions. 

Because the Universe is expanding, one can learn more than just the number of neutral hydrogen atoms between us and the quasar. As these photons travel to us, the Universe expands, stretching out all the light waves. This increases the wavelength $\lambda$ and lowers the energies of the photons. 

Neutral hydrogen atoms in their lowest state will interact with whatever light has been red-shifted to a wavelength of 122 nm when it reaches them. The rest of the light will keep traveling to us. 

The quasar shines with a certain spectrum or distribution of energies, with a certain amount of power in each wavelength. In Fig.\ (\ref{fig:lyman})\footnote{This picture is taken from \textit{\textcolor{blue}{http://astro.berkeley.edu/~jcohn/lya.html}}} the top picture shows a cartoon of how a quasar spectrum (the flux of light as a function of wavelength) would look if there were no intervening neutral hydrogen between the quasar and us. In reality, gas around the quasar both emits and absorbs photons. With the presence of neutral hydrogen, including that near the quasar, the emitted flux is depleted for certain wavelengths, indicating the absorption by this intervening neutral hydrogen. As the 122 nm wavelength is preferably absorbed, we know that at the location the photon is absorbed, its wavelength is probably 122 nm. Its wavelength was stretched by the expansion of the Universe from what it was initially at the quasar, and, if it had continued to travel to us, it would have been stretched some more from the 122 nm wavelength it had at the absorber. Thus we see the dip in flux at the wavelength corresponding to that which the 122 nm (when it was absorbed) photon would have had if it had reached us. As we can calculate how the Universe is expanding, we can tell where the photons were absorbed in relation to us. Thus one can use the absorption map to plot the positions of region of intervening hydrogen between us and the quasar. The middle picture in Fig.\ (\ref{fig:lyman}) shows the flux for one nearby region while the bottom picture shows the case for several intervening regions. It is common to see absorption systems spread out into a `forest' of lines because each line is red-shifted by a different amount in proportion to the absorbing cloud's distance from us.

Hence, Busca \textit{et al.}\cite{busca12} reported the value of Hubble parameter $H$ at high redshift $z=2.3$ using the combined constraints from WMAP, CMB anisotropy, and baryon acoustic oscillations peak in the Ly$\alpha$ forest, when the Universe was matter dominated, as $H(z=2.3)=224 \pm 8$ kms$^{-1}$ Mpc$^{-1}$. The analysis was based on 48,600 quasars which are at $1.96\leq z\leq3.38$.

We decided to use this data point which has only 4\% error to get improved constraints on the dark energy model parameters. With the addition this high redshift $H(z)$ data point our whole $H(z)$ data look like Table\ (\ref{tab:Hz2}).

\section{Improved Constraints}

To constrain cosmological parameters $\textbf{p}$ of the models of 
interest we follow the procedure of discussed above in this 
chapter Farooq \textit{et al.}\cite{Farooq:2012ev}

\begin{figure}[h!]
\centering
    \includegraphics[height=3.0in]{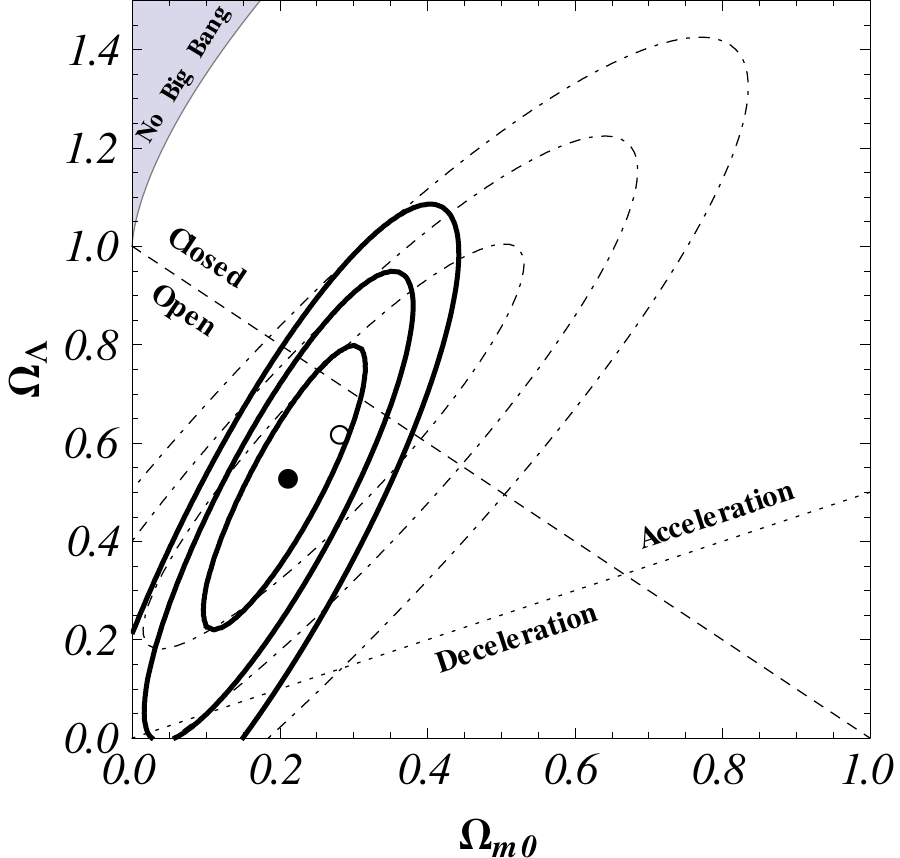}
    \includegraphics[height=3.0in]{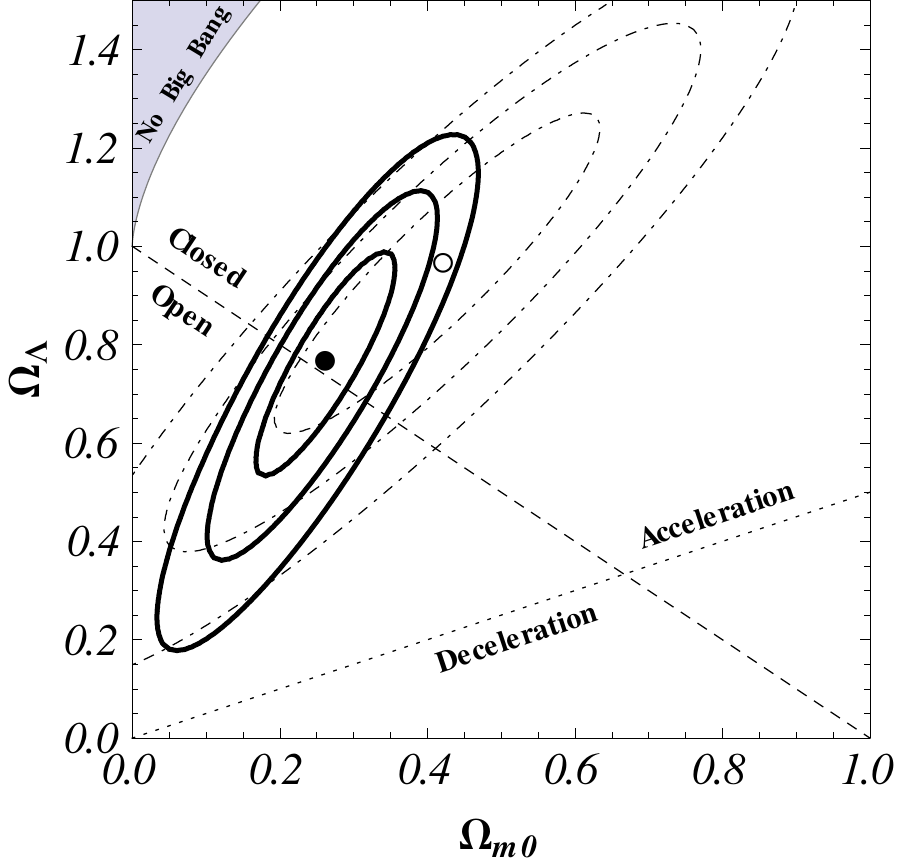}
 \caption{
Thick solid (thin dot-dashed) lines correspond to 1$\sigma$, 2$\sigma$, and 3$\sigma$ 
constraint contours from the new (old, Farooq \textit{et al.}\citep{Farooq:2012ev}) $H(z)$ data 
for the $\Lambda$CDM model. The filled (empty) circle is the best fit 
point from the new (old) $H(z)$ data. The left panel is for the 
$H_0 = 68 \pm 2.8$ km s$^{-1}$ Mpc$^{-1}$ prior and the right 
panel is for the $H_0 = 73.8 \pm 2.4$ km s$^{-1}$ Mpc$^{-1}$ one.
The dashed diagonal lines correspond to spatially-flat models, the 
dotted lines demarcate zero-acceleration models, and the shaded area 
in the upper left-hand corners are the region  for which there is no 
big bang. The filled circles correspond to best-fit pair 
$(\Omega_{m0}, \Omega_{\Lambda}) = (0.21, 0.53)$ with $\chi^2_{\rm min}
= 15.1$ (left panel) and best-fit pair $(\Omega_{m0}, \Omega_{\Lambda}) = 
(0.26, 0.77)$ with $\chi^2_{\rm min}=16.1$ (right panel). The empty 
circles correspond to best-fit pair 
$(\Omega_{m0}, \Omega_{\Lambda}) = (0.28, 0.62)$ with $\chi^2_{\rm min}
= 14.6$ (left panel) and best-fit pair $(\Omega_{m0}, \Omega_{\Lambda}) = 
(0.42, 0.97)$ with $\chi^2_{\rm min}=14.6$ (right panel). 
}
\label{fig:LCDMHz22}
\end{figure}

\begin{figure}[h!]
\centering
    \includegraphics[height=2.9in]{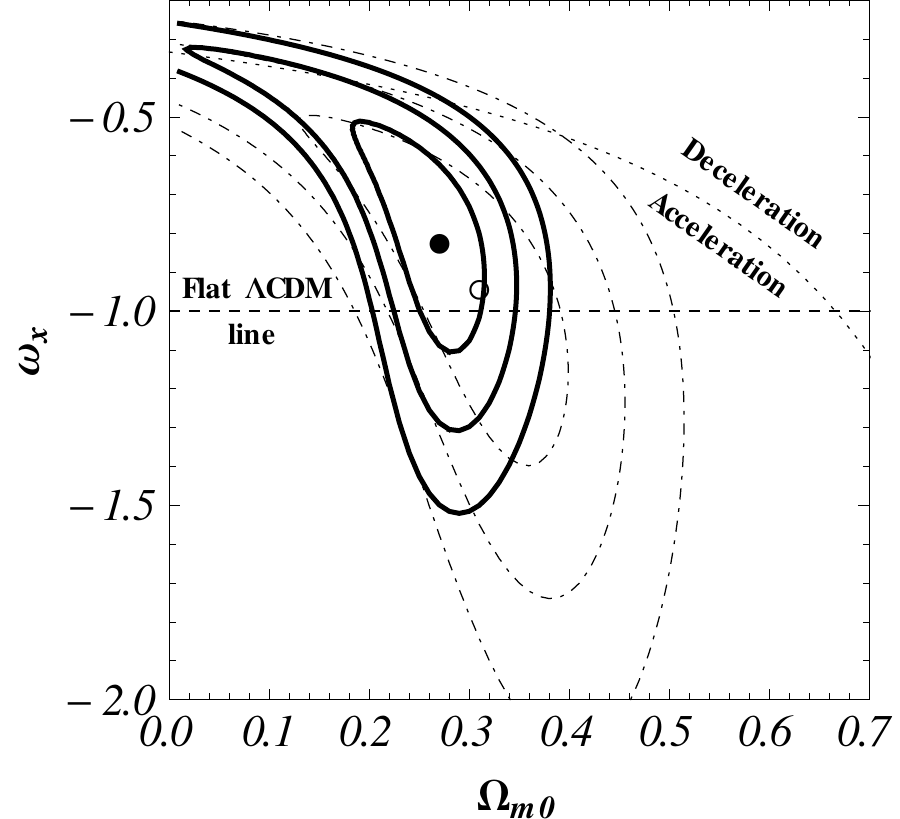}
    \includegraphics[height=2.9in]{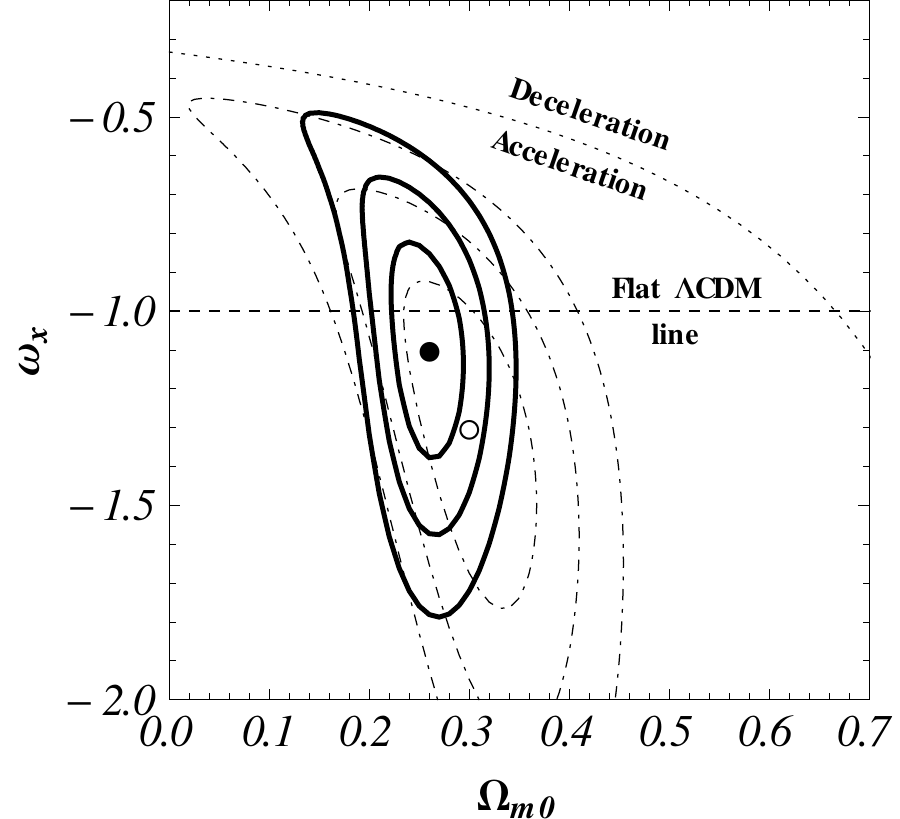}
 \caption{
Thick solid (thin dot-dashed) lines correspond to 1$\sigma$, 2$\sigma$, and 3$\sigma$ 
constraint contours from the new (old, Farooq \textit{et al.}\citep{Farooq:2012ev}) $H(z)$ data 
for the XCDM model. The filled (empty) circle is the best fit point from 
the new (old) $H(z)$ data. The left panel is for the 
$H_0 = 68 \pm 2.8$ km s$^{-1}$ Mpc$^{-1}$ prior and the right 
panel is for the $H_0 = 73.8 \pm 2.4$ km s$^{-1}$ Mpc$^{-1}$ one.
The dashed horizontal lines at $\omega_{\rm X} = -1$ correspond to 
spatially-flat $\Lambda$CDM models and the curved dotted lines demarcate 
zero-acceleration models. The filled circles correspond to best-fit 
pair $(\Omega_{m0}, \omega_{\rm X}) = (0.27, -0.82)$ with $\chi^2_{\rm min}
= 15.2$ (left panel) and best-fit pair $(\Omega_{m0}, \omega_{\rm X}) 
= (0.36, -1.1)$ with $\chi^2_{\rm min}=15.9$ (right panel).
The empty circles correspond to best-fit 
pair $(\Omega_{m0}, \omega_{\rm X}) = (0.31, -0.94)$ with $\chi^2_{\rm min}
= 14.6$ (left panel) and best-fit pair $(\Omega_{m0}, \omega_{\rm X}) 
= (0.30, -1.30)$ with $\chi^2_{\rm min}=14.6$ (right panel).
}
\label{fig:XCDMHz22}
\end{figure}

\begin{figure}[h!]
\centering
    \includegraphics[height=3.0in]{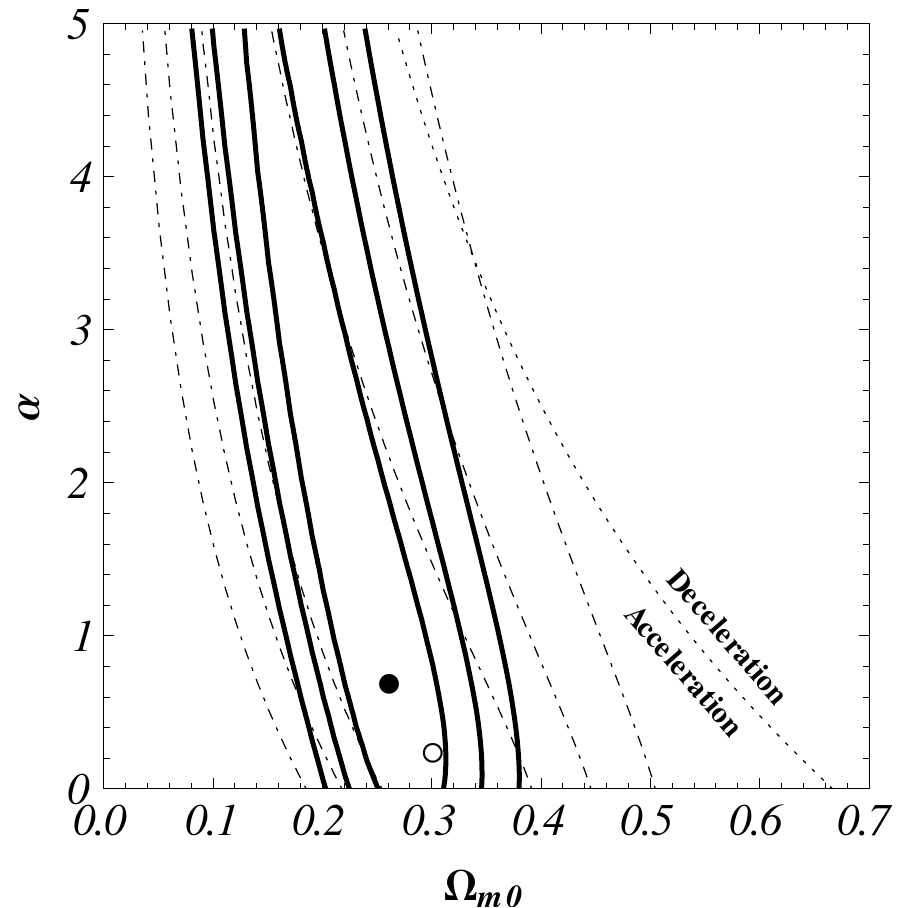}
    \includegraphics[height=3.0in]{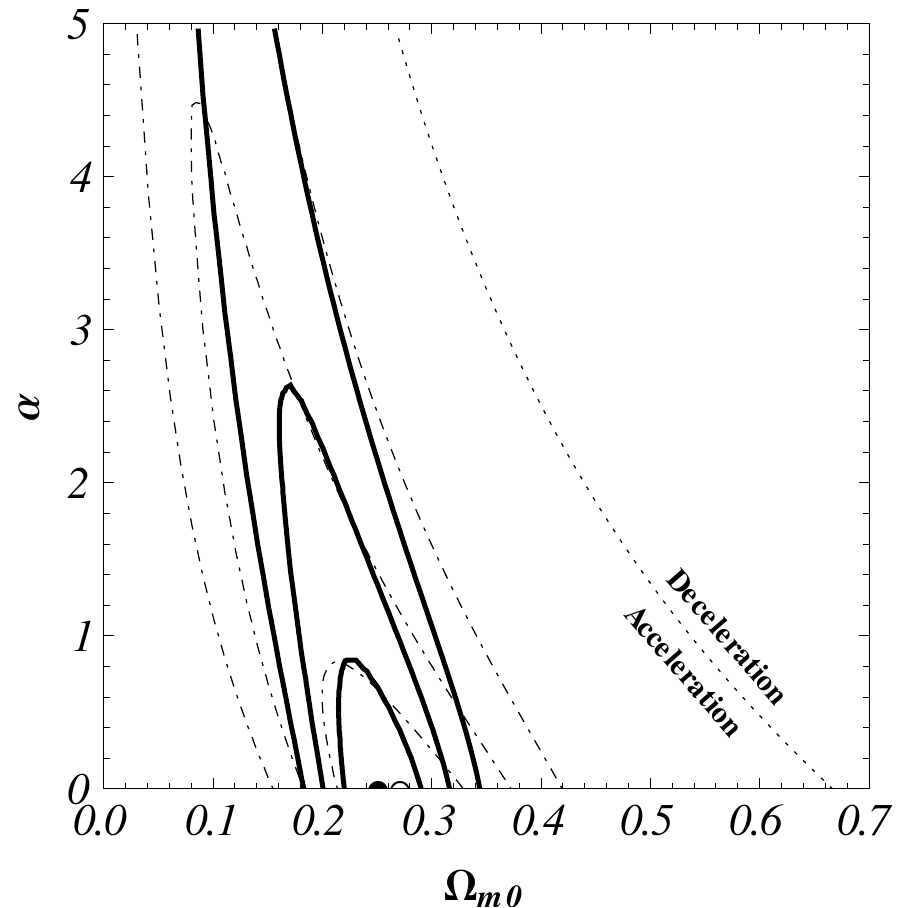}
 \caption{
Thick solid (thin dot-dashed) lines correspond to 1$\sigma$, 2$\sigma$, and 3$\sigma$ 
constraint contours from the new (old, Farooq \textit{et al.}\citep{Farooq:2012ev}) $H(z)$ data 
for the $\phi$CDM model. The filled (empty) circle is the best fit point 
from the new (old) $H(z)$ data. The left panel is for the 
$H_0 = 68 \pm 2.8$ km s$^{-1}$ Mpc$^{-1}$ prior and the right 
panel is for the $H_0 = 73.8 \pm 2.4$ km s$^{-1}$ Mpc$^{-1}$ one.
The horizontal axes at $\alpha = 0$ correspond to spatially-flat 
$\Lambda$CDM models and the curved dotted lines demarcate 
zero-acceleration models. The filled circles correspond to best-fit pair 
$(\Omega_{m0}, \alpha) = (0.36, 0.70)$ with $\chi^2_{\rm min}=15.2$
(left panel) and best-fit pair $(\Omega_{m0}, \alpha) = (0.25, 0)$ 
with $\chi^2_{\rm min}=16.1$ (right panel). The empty circles correspond to best-fit pair 
$(\Omega_{m0}, \alpha) = (0.30, 0.25)$ with $\chi^2_{\rm min}=14.6$
(left panel) and best-fit pair $(\Omega_{m0}, \alpha) = (0.27, 0)$ 
with $\chi^2_{\rm min}=15.6$ (right panel).
}
\label{fig:PCDMHz22}
\end{figure}
We again 
marginalize over the nuisance parameter $H_0$ using two different 
Gaussian priors with $\bar{H_0}\pm\sigma_{H_0}$=
68 $\pm$ 2.8 km s$^{-1}$ Mpc$^{-1}$ and with $\bar{H_0}\pm\sigma_{H_0}$=
73.8 $\pm$ 2.4 km s$^{-1}$ Mpc$^{-1}$. As discussed there, the Hubble 
constant measurement uncertainty can significantly affect 
cosmological parameter estimation for a recent example see, e.g.,
Calabrese \textit{et al.}\cite{calabrese12}
The lower of the two values we use is from a median 
statistics analysis Gott \textit{et al.}\citep{Gott2001} of 553 measurements of $H_0$
Chen \textit{et al.}\citep{Chen2011a}; this estimate has been stable for over 
a decade now \citep{Gott2001, chen03}. The other value is a recent, 
HST based one Riess \textit{et al.}\citep{Riess2011} Other recent estimates are 
compatible with at least one of the two values we use. See for example, 
Freedman \textit{et al.},\cite{Freedman2012} Sorce \textit{et al.},\cite{Sorce2012} and 
Tammann \text{et al.}\cite{Tammann2012}

We maximize the likelihood $\mathcal{L}_H(\textbf{p})$ with respect 
to the parameters $\textbf{p}$ to find the best-fit parameter values 
$\mathbf{p_0}$. In the models we consider $\chi_H^2=-2 {\rm ln}
\mathcal{L}_H(\textbf{p}) $ depends on 
two parameters. We define 1$\sigma$, 2$\sigma$, and 3$\sigma$ 
confidence intervals as two-dimensional parameter sets bounded by 
$\chi_H^2(\textbf{p}) = \chi_H^2(\mathbf{p_0})+2.3,~\chi_H^2(\textbf{p}) = 
\chi_H^2(\mathbf{p_0})+6.17$, and $\chi_H^2(\textbf{p}) = 
\chi_H^2(\mathbf{p_0})+11.8$, respectively.

Figures (\ref{fig:LCDMHz22})---(\ref{fig:PCDMHz22}) show the constraints
from the $H(z)$ data derived here, as well as those derived by Farooq \textit{et al.},
\cite{Farooq:2012ev} for the three dark energy models, and for the two 
different $H_0$ priors. Clearly, the $H(z = 2.3)$ measurement of Busca \textit{et al.}
\cite{busca12} significantly tightens the constrains. Given that the 
nonrelativistic matter density is larger relative to the dark energy density 
at $z = 2.3$, it is perhaps not unexpected that the Busca \textit{et al.}\cite{busca12} 
measurement tightens the constraints on $\Omega_{m0}$ much more 
significantly than it does for the constraints on the other 
parameter which more strongly affects the evolution of the dark energy
density, see Figs.\ (\ref{fig:XCDMHz22}) and (\ref{fig:PCDMHz22}).

\begin{figure}[h!]
\centering
    \includegraphics[height=3.0in]{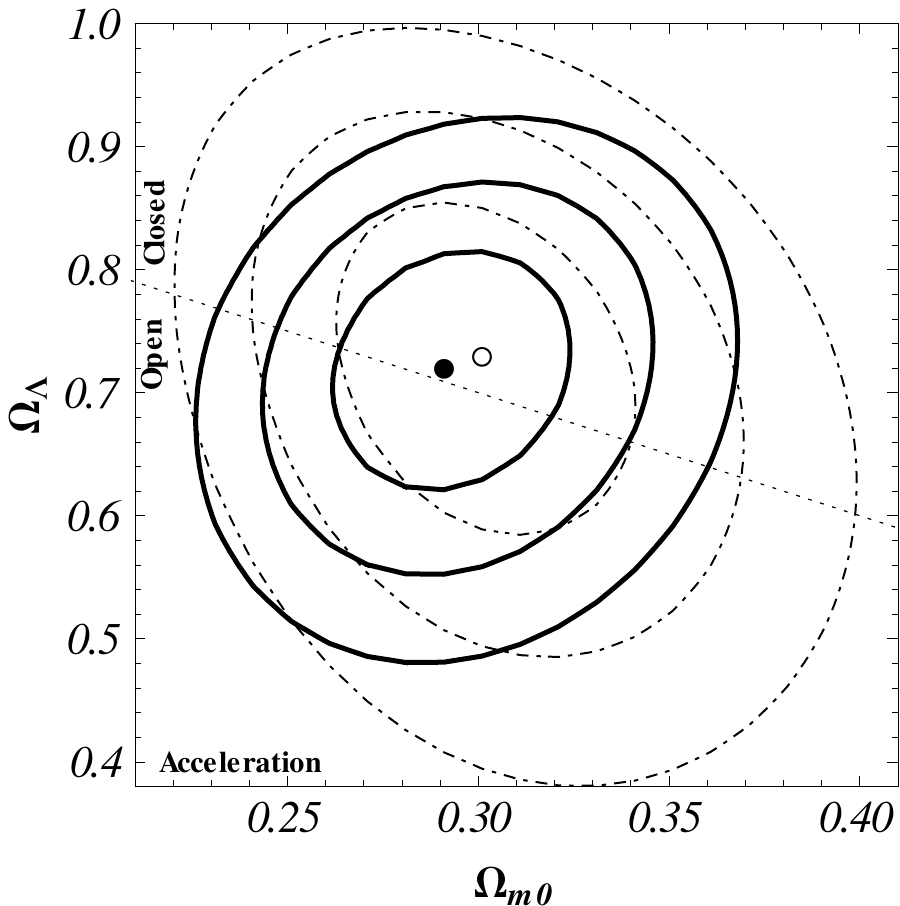}
    \includegraphics[height=3.0in]{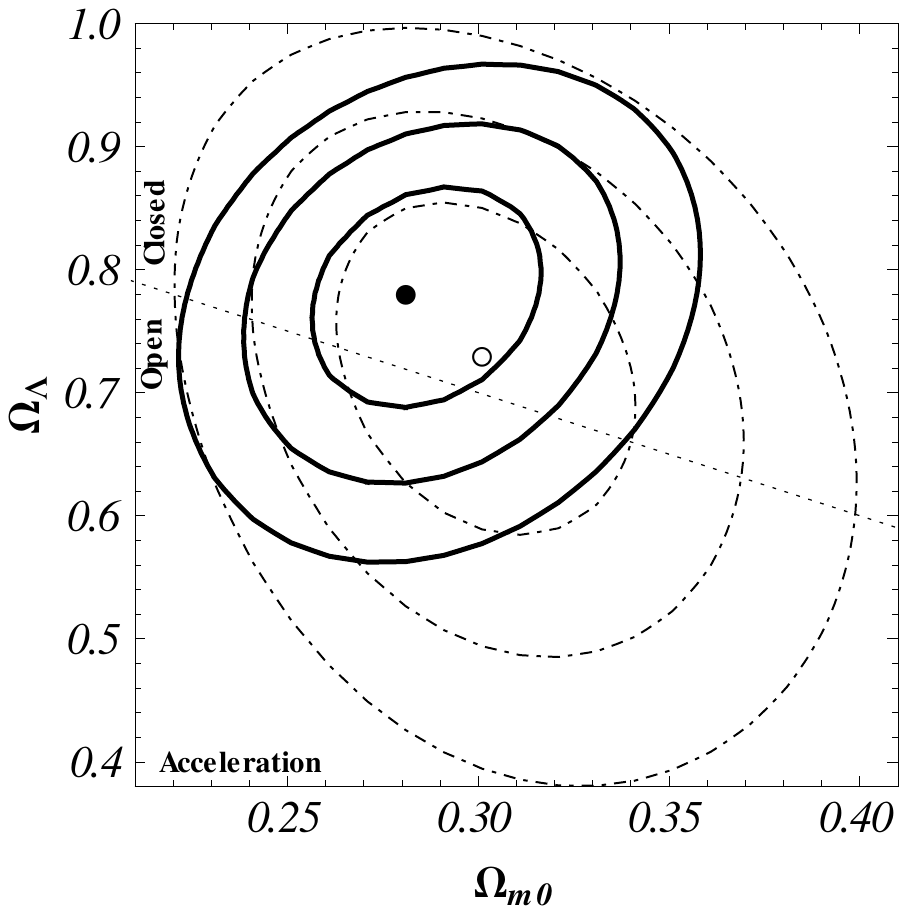}
 \caption{
Thick solid (thin dot-dashed) lines are 1$\sigma$, 2$\sigma$, and 3$\sigma$ constraint
contours for the $\Lambda$CDM model from a joint analysis of the BAO
and SNIa (with systematic errors) data, with (without) the $H(z)$ data. 
The full (empty) circle marks the best-fit point determined from the 
joint analysis with (without) the $H(z)$ data. The dotted sloping line 
corresponds to spatially-flat $\Lambda$CDM models. In the left panel 
we use the $H_0$ = 68 $\pm$ 2.8 km s$^{-1}$ Mpc$^{-1}$ prior. Here the
empty circle [no $H(z)$ data] corresponds to best-fit pair 
$(\Omega_{m0}, \Omega_{\Lambda}) = (0.30,0.73)$ with $\chi^2_{\rm min}=551$
while the full circle [with $H(z)$ data] indicates best-fit pair 
$(\Omega_{m0}, \Omega_{\Lambda}) = (0.29,0.72)$ with $\chi^2_{\rm min}=567$. 
In the right panel we use the $H_0$ = 73.8 $\pm$ 2.4 km s$^{-1}$ Mpc$^{-1}$ 
prior. Here the empty circle [no $H(z)$ data] corresponds to best-fit pair 
$(\Omega_{m0}, \Omega_{\Lambda}) = (0.30,0.73)$ with $\chi^2_{\rm min}=551$
while the full circle [with $H(z)$ data] demarcates best-fit pair 
$(\Omega_{m0}, \Omega_{\Lambda}) = (0.28,0.78)$ with $\chi^2_{\rm min}=568$.
}
\label{fig:LCDMall+snbao22}
\end{figure}

\begin{figure}[h!]
\centering
    \includegraphics[height=3.0in]{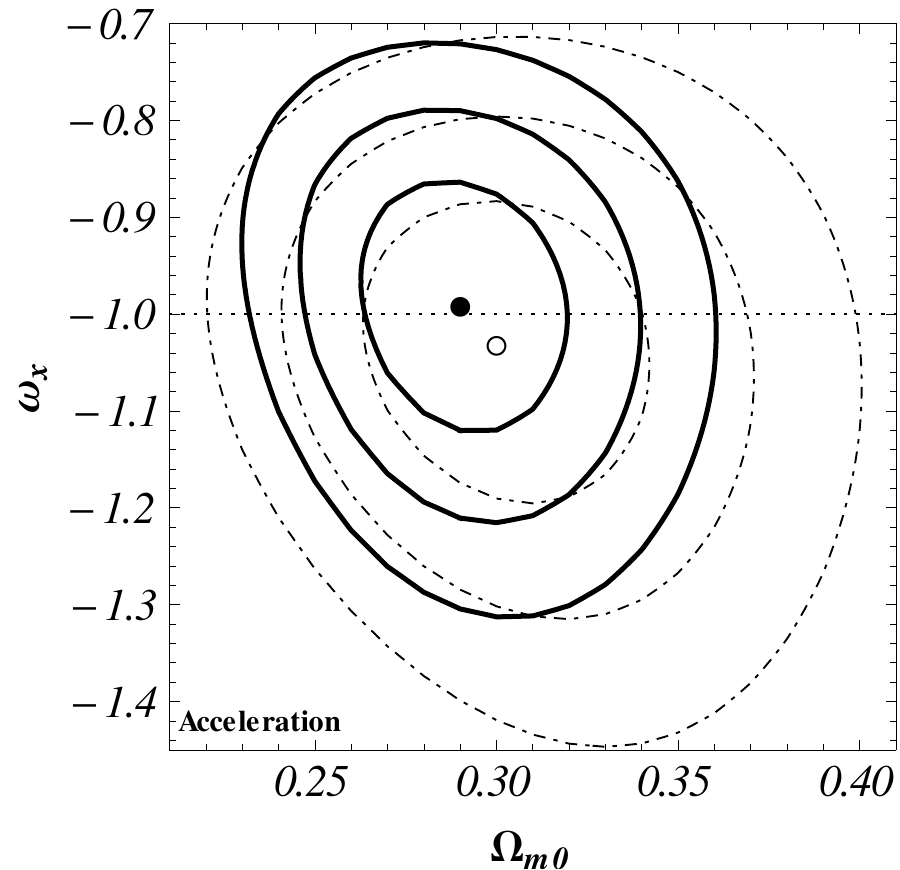}
    \includegraphics[height=3.0in]{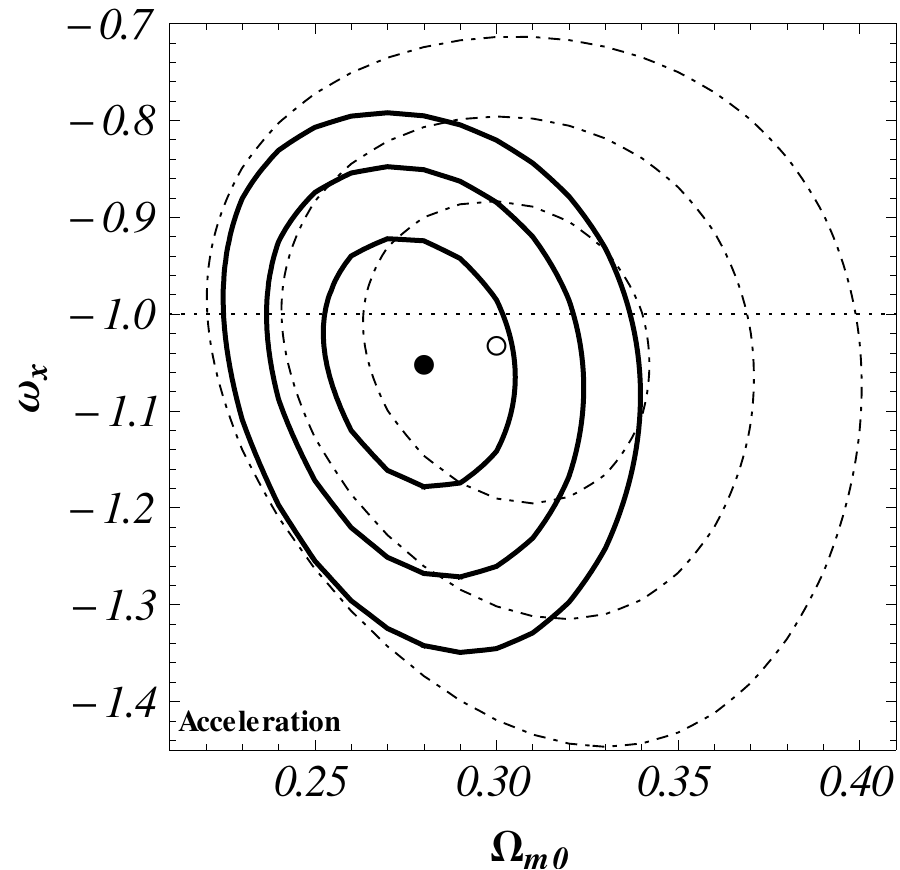}
 \caption{
Thick solid (thin dot-dashed) lines are 1$\sigma$, 2$\sigma$, and 3$\sigma$ constraint
contours for the XCDM parametrization from a joint analysis of the BAO
and SNIa (with systematic errors) data, with (without) the $H(z)$ data. 
The full (empty) circle marks the best-fit point determined from the 
joint analysis with (without) the $H(z)$ data. The dotted horizontal 
line at $\omega_{\rm X} =-1$  corresponds to spatially-flat $\Lambda$CDM 
models. In the left panel we use the 
$H_0$ = 68 $\pm$ 2.8 km s$^{-1}$ Mpc$^{-1}$ prior. Here the empty
circle [no $H(z)$ data] corresponds to best-fit pair $(\Omega_{m0}, 
\omega_{\rm X}) = (0.30,-1.03)$ with $\chi^2_{\rm min}=551$, while 
the full circle [with $H(z)$ data] demarcates best-fit pair 
$(\Omega_{m0}, \omega_{\rm X}) = (0.29,-0.99)$ with $\chi^2_{\rm min}=568$. 
In the right panel we use the $H_0$ = 73.8 $\pm$ 2.4 km s$^{-1}$ Mpc$^{-1}$ 
prior. Here the empty circle [no $H(z)$ data] corresponds to best-fit pair 
$(\Omega_{m0}, \omega_{\rm X}) = (0.30,-1.03)$ with $\chi^2_{\rm min}=551$
while the full circle [with $H(z)$ data] indicates best-fit pair 
$(\Omega_{m0}, \omega_{\rm X}) = (0.28,-1.05)$ with $\chi^2_{\rm min}=569$. 
}
\label{fig:XCDMall+snbao22}
\end{figure}

\begin{figure}[h!]
\centering
    \includegraphics[height=3.0in]{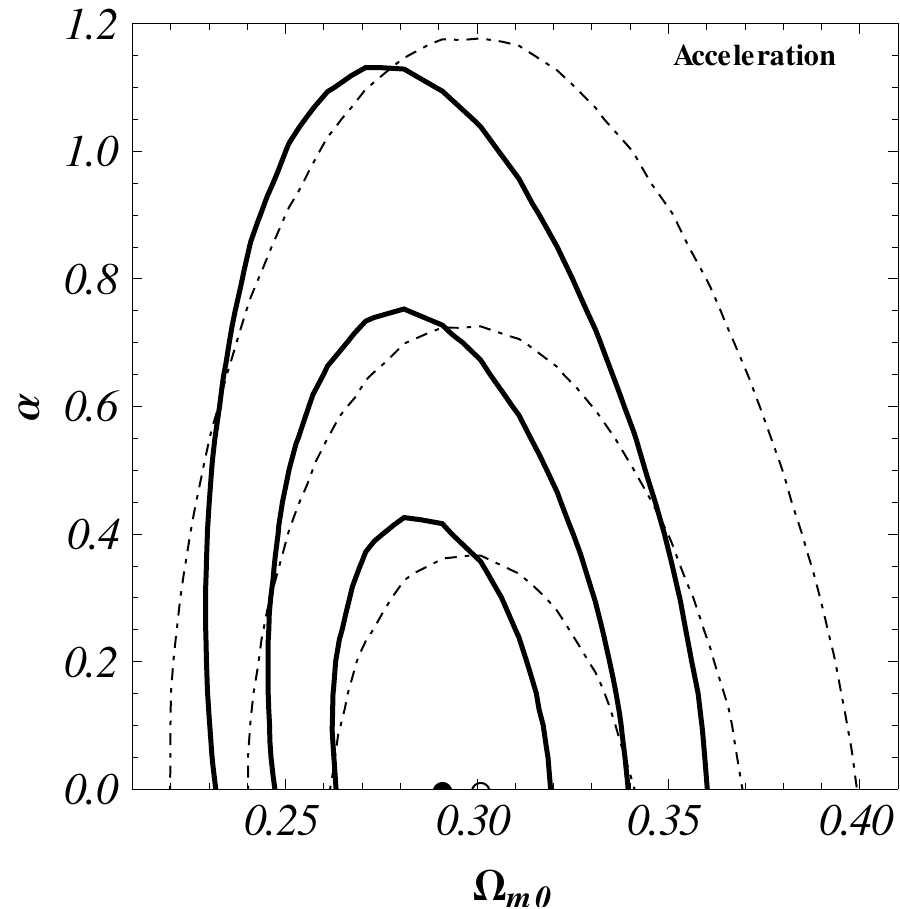}
    \includegraphics[height=3.0in]{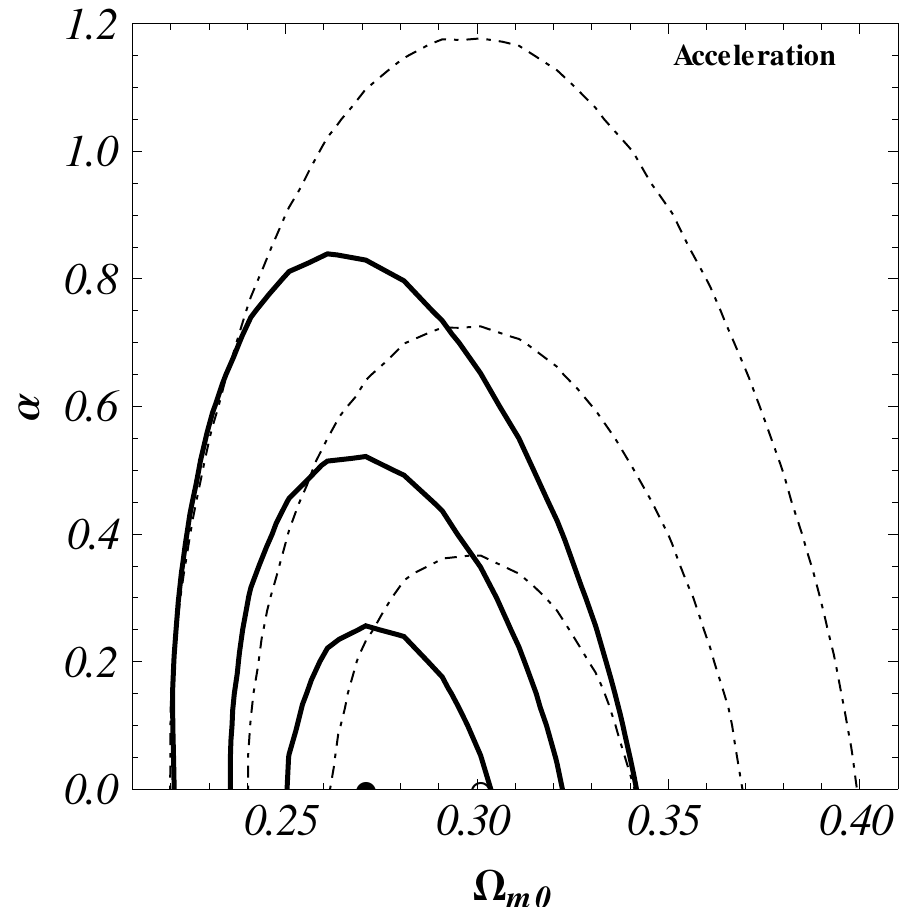}
 \caption{
Thick solid (thin dot-dashed) lines are 1$\sigma$, 2$\sigma$, and 3$\sigma$ constraint
contours for the $\phi$CDM model from a joint analysis of the BAO
and SNIa (with systematic errors) data, with (without) the $H(z)$ 
data. The full (empty) circle marks the best-fit point determined
from the joint analysis with (without) the $H(z)$ data. The $\alpha = 0$
horizontal axes correspond to spatially-flat $\Lambda$CDM models.
In the left panel we use the $H_0$ = 68 $\pm$ 2.8 km s$^{-1}$ Mpc$^{-1}$ 
prior. Here the empty circle corresponds to best-fit pair 
$(\Omega_{m0}, \alpha) = (0.30, 0)$ with $\chi^2_{\rm min}=551$
while the full circle indicates best-fit pair $(\Omega_{m0}, \alpha) 
= (0.29, 0)$ with $\chi^2_{\rm min}=567$. In the right panel we use 
the $H_0$ = 73.8 $\pm$ 2.4 km s$^{-1}$ Mpc$^{-1}$ prior. Here the empty
circle corresponds to best-fit pair $(\Omega_{m0}, \alpha) = (0.30, 0)$ 
with $\chi^2_{\rm min}=551$ while the full circle demarcates best-fit 
pair $(\Omega_{m0}, \alpha) = (0.27, 0)$ with $\chi^2_{\rm min}=569$.
}
\label{fig:PCDMall+snbao22}
\end{figure}

Comparing the $H(z)$ constraints derived here, and shown in Figs.\ 
(\ref{fig:LCDMHz22})---(\ref{fig:PCDMHz22}) here, to the SNIa constraints
shown in Fig.\ (\ref{fig:SNEIA-lxCDM}), we see that the new $H(z)$
data constraints are significantly more restrictive than those 
that follow on using the SNIa data. This is a remarkable result.
Qualitatively, because of the dependence on the $H_0$ prior 
and on the model used in the analysis, Figs.\ 
(\ref{fig:LCDMHz22})---(\ref{fig:PCDMHz22}) show that the $H(z)$
data alone require accelerated cosmological expansion at 
approximately the two standard deviation confidence level.

While the $H(z)$ data provide tight constraints on a linear
combination of cosmological parameters, the banana-like constraint
contours of Figs.\ (\ref{fig:LCDMHz22})---(\ref{fig:PCDMHz22})
imply that these data alone cannot significantly discriminate
between cosmological models. To tighten the constraints we must add other 
data to the mix. Following Farooq \textit{et al.}
\cite{Farooq:2012ev}, we derive constraints on 
cosmological parameters of the three models from a joint analysis of the $H(z)$
data with the 6 BAO peak length scale measurements of Percival \textit{et al.}\cite{Percival2010}, Beutler \textit{et al.}
\cite{Beutler11}, and Blake \textit{et al.}\cite{blake11}, and the Union2.1 compilation of
580 SNIa apparent magnitude measurements (covering a redshift range
$0.015<z <1.4$) from Suzuki \textit{et al.}\cite{suzuki2012}.

\begin{figure}[h!]
\centering
    \includegraphics[height=4.0in]{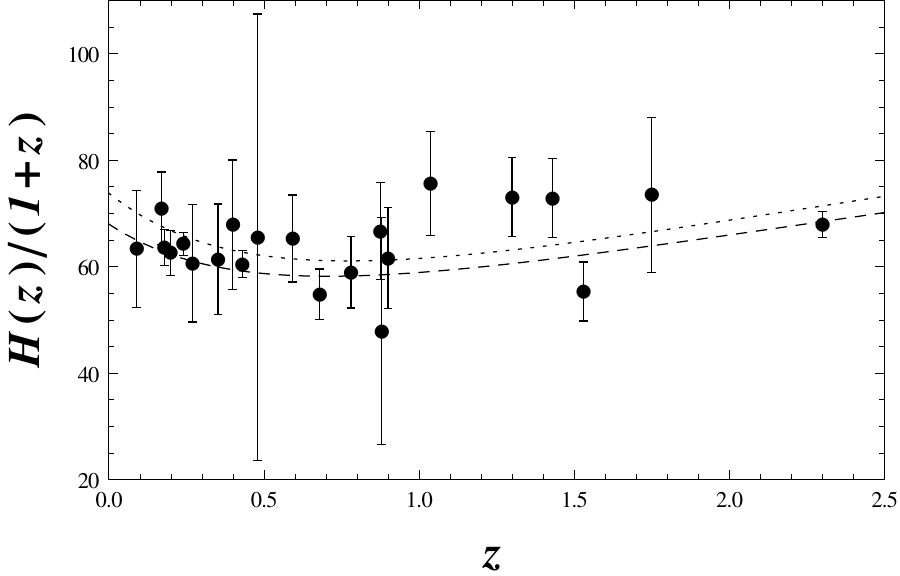}
 \caption{
Measurements and predictions for $H(z)/(1+z)$ as a function of $z$.
Dashed (dotted) lines show the predictions for the best-fit $\Lambda$CDM
model from the combined BAO, SNIa, and $H(z)$ data analyses, with
cosmological parameter values $(\Omega_{m0}, \Omega_\Lambda, h) = 
(0.29, 0.72, 0.68) [(0.28, 0.78, 0.738)]$.
}
\label{fig:ddota22}
\end{figure}

Figures\ (\ref{fig:LCDMall+snbao22})---(\ref{fig:PCDMall+snbao22}) show the
constraints on cosmological parameters for the $\Lambda$CDM and
$\phi$CDM models and the XCDM parametrization, from a joint analysis
of the BAO and SNIa data, as well as from a joint analysis of the
BAO, SNIa, and $H(z)$ data. Including the $H(z)$ data in the analysis 
tightens the constraints, somewhat significantly (sometimes by more 
than two standard deviations), in parts of the parameter spaces. 
Figure\ (\ref{fig:ddota22}) shows the $H(z)$ data and the two best-fit
$\Lambda$CDM models. The $H(z)$ data do support the idea of a 
deceleration to acceleration transition somewhere in the range
$0.5 < z < 1$.

\begin{table}[t]
\begin{center}
\begin{tabular}{ccc}
\hline\hline
{Model and prior} & {BAO+SNIa} & {BAO+SNIa+$H(z)$} \\
\hline
 {$\Lambda $CDM, $h=0.68 \pm 0.028$} & {0.25 $<$ }$\Omega_{m0}${ $<$ 0.36} & 
      {0.26 $<$ }$\Omega_{m0}${ $<$ 0.33} \\
 {} & {0.53 $<$ }$\Omega _{\Lambda }${ $<$ 0.89} & {0.60 $<$ }$\Omega _{\Lambda }${ $<$ 0.84} \\
 \hline
 {$\Lambda $CDM, $h = 0.738 \pm 0.024$} & {0.25 $<$ }$\Omega_{m0}${ $<$ 0.36} & 
      {0.25 $<$ }$\Omega_{m0}${ $<$ 0.32} \\
 {} & {0.53 $<$ }$\Omega _{\Lambda }${ $<$ 0.89} & {0.66 $<$ }$\Omega _{\Lambda }${ $<$ 0.89} \\
 \hline
 {XCDM, $h = 0.68 \pm 0.028$} & {0.30 $<$ }$\Omega_{m0}${ $<$ 0.38} & 
      {0.27 $<$ }$\Omega_{m0}${ $<$ 0.32} \\
 {} & $-1.18 <  \omega_{\rm X} < -0.78$ & $-1.03 < \omega_{\rm X} < -0.77$ \\
 \hline 
 {XCDM, $h = 0.738 \pm 0.024$} & {0.30 $<$ }$\Omega_{m0}${ $<$ 0.38} & 
      {0.25 $<$ }$\Omega_{m0}${ $<$ 0.30} \\
 {} & $-1.18 < \omega_{\rm X} < -0.78$ & $ -1.15 < \omega_{\rm X} < -0.90$ \\
  \hline
 {$\phi $CDM, $h = 0.68 \pm 0.028$} & {0.25 $<$ }$\Omega_{m0}${ $<$ 0.35} & 
      {0.25 $<$ }$\Omega_{m0}${ $<$ 0.32} \\
 {} & {0 $<$ $\alpha $ $<$ 0.54} & {0 $<$ $\alpha $ $<$ 0.56} \\
  \hline
 {$\phi $CDM, $h = 0.738 \pm 0.024$} & {0.25 $<$ }$\Omega_{m0}${ $<$ 0.35} & 
      {0.25 $<$ }$\Omega_{m0}${ $<$ 0.30} \\
 {} & {0 $<$ $\alpha $ $<$ 0.54} & {0 $<$ $\alpha $ $<$ 0.21} \\
\hline\hline
\end{tabular}

\caption{Two standard deviation bounds on cosmological
parameters using BAO+SNIa and BAO+SNIa+$H(z)$ data, for three 
models and two $H_0$ priors.}
\label{tab:intervalsfrom22data}
\end{center}
\end{table}

Table\ (\ref{tab:intervalsfrom22data}) lists the two standard deviation bounds on the individual 
cosmological parameters, determined from their one-dimensional
posterior probability distributions  functions (which are derived
by marginalizing the two-dimensional likelihood over the other
cosmological parameter). 

Adding the Busca \textit{et al.}\cite{busca12} $z = 2.3$ measurement of the Hubble parameter,
from BAO in the Ly$\alpha$ forest, to the 21 $H(z)$ data points tabulated
in Farooq \textit{et al.}\cite{Farooq:2012ev}, results in an $H(z)$ data set that provides quite
restrictive constraints on cosmological parameters. These constraints
are tighter than those that follow from the SNIa data of Suzuki \textit{et al.}\cite{suzuki2012},
which carefully accounts for all known systematic uncertainties. The $H(z)$
field is much less mature than the SNIa one, and there might be some as 
yet undetected $H(z)$ systematic errors that could broaden the $H(z)$ 
error bars, as has happened in the SNIa case. However, we emphasize that 
the observers have done a careful analysis and the error bars we have used 
in our analysis have been carefully estimated. In addition to providing
more restrictive constraints, the $H(z)$ data alone requires accelerated
cosmological expansion at the current epoch at approximately 2 $\sigma$
confidence level, depending on model and $H_0$ prior used in the analysis.  

In summary, the results of the joint analysis of the $H(z)$, BAO, and SNIa 
data are quite consistent with the predictions of the standard spatially-flat
$\Lambda$CDM cosmological model, with current energy budget dominated by 
a time-independent cosmological constant. However, currently-available 
data cannot rule out slowly-evolving dark energy density. We anticipate 
that, soon to be available, better quality data will more clearly discriminate
between constant and slowly-evolving dark energy density.


\cleardoublepage


\chapter{Constraints on Transition Red-Shift from $\boldsymbol {H(z)}$ Data}
\label{Chapter6}

This chapter is based on Farooq \& Ratra\cite{Farooq:2013hq}
\\
 
In order to put tighter constraints on the cosmological parameters of one time independent and two time-evolving dark energy models, and on the deceleration-acceleration transition redshift we compile a list of 28 independent measurements of the Hubble parameter between redshifts $0.07 \leq z \leq 2.3$, listed in Table\ (\ref{tab:Hz3}). These $H(z)$ measurements by themselves require a currently accelerating cosmological expansion at about, or better than, 3 $\sigma$ confidence, as will be explained latter.

\section{Introduction}

In the standard picture of cosmology, dark energy powers the current 
accelerating cosmological expansion, but it played a less significant role 
in the past when nonrelativistic (cold dark and baryonic) matter 
dominated and powered the then decelerating cosmological 
expansion.\footnote{
For reviews of dark energy see Bolotin \textit{et al.}\cite{Bolotin2011}, 
Martin \textit{et al.}\cite{Martin2012}, and references therein. The observed accelerating 
cosmological expansion has also be interpreted as indicating the
need to modify general relativity. In this paper we assume that general 
relativity provides an adequate description of gravitation on 
cosmological length scales. For reviews of modified gravity see 
Bolotin \textit{et al.}\cite{Bolotin2011}, Capozzielo \textit{et al.}\cite{Capozziello2011}, and references therein.}
It is of some interest to determine the redshift of the 
deceleration-acceleration transition predicted to exist in dark energy 
cosmological models. There have been a number of attempts to do
so, see, e.g., Lu \textit{et al.}\cite{Lu2011a}, Giostri \textit{et al.}\cite{Giostri2012}, Lima \textit{et al.},\cite{Lima2012}
and references therein. However, until very recently, this has not been 
possible because there has not been much high-quality data at high enough 
redshift (i.e., for $z$ above the transition redshift in standard dark 
energy cosmological models).

The recent Busca \textit{et al.}\cite{busca12} detection of the baryon acoustic oscillation 
(BAO) peak at $z = 2.3$ in the Ly$\alpha$ forest has dramatically 
changed the situation by allowing for a high precision measurement of the 
Hubble parameter $H(z)$ at $z = 2.3$, well in the matter dominated epoch
of the standard dark energy cosmological model. Busca \textit{et al.}\cite{busca12} use 
this and 10 other $H(z)$ measurements, largely based on BAO-like data,
and the Riess \textit{et al.}\cite{Riess2011} HST determination of the Hubble constant, in the 
context of the standard $\Lambda$CDM cosmological model, to estimate a 
deceleration-acceleration transition redshift of $z_{\rm da} = 0.82 \pm 0.08$.

We extend the analysis of Busca \textit{et al.}\cite{busca12} by first compiling
a list of 28 independent $H(z)$ measurements see Table\ (\ref{tab:Hz3}).\footnote{
It appears that some of the measurements listed in Table 2 of 
Busca \textit{et al.}\cite{busca12} might not be independent. For instance, the Chuang
\textit{et al.}\cite{Chuang2012a}
and the Xu \textit{et al.}\cite{Xu2012b} determinations of $H(z = 0.35)$ listed in the table 
are both based on the use of Sloan Digital Sky Survey Data Release 7 
measurements of luminous red galaxies.}
We then use these 28 measurements to constrain cosmological parameters 
in 3 different dark energy models and establish that the models are
a good fit to the data and that the data provide tight constraints on the 
model parameters. Finally, we use the models to estimate the redshift of the 
deceleration-acceleration transition. Busca \textit{et al.}\cite{busca12} have one measurement 
(of 11) above their estimated $z_{\rm da} = 0.82$, while we have 9 of 28 above this
(and 10 of 28 above our estimated redshift $z_{\rm da} = 0.74$). 
Granted, the Busca \textit{et al.}\cite{busca12} $z$ = 2.3 measurement carries great weight because 
of the small, 3.6\%, uncertainty, but 9 of our 10 high redshift
measurements, from Simon \textit{et al.},\cite{simon05} Stern \textit{et al.},\cite{Stern2010}
and Moresco \textit{et al.},\cite{moresco12} 
include 3 11\%, 13\%, and 14\% measurements from Moresco \textit{et al.}\cite{moresco12} and
3 10\% measurements from Simon \textit{et al.},\cite{simon05} all 6 of which carry significant 
weight. 

We only include
independent measurements of $H(z)$, listing only the most recent 
result from analyses of a given data set. The values in Table\ (\ref{tab:Hz3}) have been 
determined using a number of different techniques; for details see 
the papers listed in the table caption. Table\ (\ref{tab:Hz3}) is the largest set of 
independent $H(z)$ measurements considered to date. 

We first use these data to derive constraints on cosmological parameters
of the 3 models described in Chapter (\ref{Chapter3}). The constraints derived here are compatible 
with cosmological parameter constraints determined by other techniques. 
These constraints are more restrictive than those derived by Farooq \&\ Ratra \cite{Farooq20131}
using the previous largest set of $H(z)$ measurements, as well as those
derived from the recent SNIa data compilation of Suzuki \textit{et al.}\cite{suzuki2012}. 
The $H(z)$ data considered here require accelerated cosmological expansion 
at the current epoch at about or more than 3 $\sigma$ confidence.

\section{Constraints on Parameters and Transition Redshift}

Following Farooq \textit{et al.},\cite{Farooq:2012ev} we use the 
28 independent $H(z)$ data points listed in Table (\ref{tab:Hz3}) 
to constrain cosmological model parameters. 
The observational data consist of measurements of the 
Hubble parameter $H_{\rm obs}(z_i)$ at redshifts $z_i$, with the 
corresponding one standard deviation uncertainties $\sigma_i$.
To constrain cosmological parameters $\textbf{p}$ of the models of 
interest we build the posterior likelihood function 
$\mathcal{L}_{H}(\textbf{p})$ that depends only on the 
$\textbf{p}$ by integrating the product of exp$(-\chi_H^2 /2)$ and 
the $H_0$ prior likelihood function 
exp$[-(H_0-\bar H_0)^2/(2\sigma^2_{H_0})]$, as in Eq.\ 18 of 
Farooq \textit{et al.}\cite{Farooq:2012ev} We marginalize over the nuisance parameter $H_0$ 
using two different Gaussian priors with $\bar{H_0}\pm\sigma_{H_0}$=
68 $\pm$ 2.8 km s$^{-1}$ Mpc$^{-1}$ \citep{chen03, Chen2011a}
and with $\bar{H_0}\pm\sigma_{H_0}$ = 73.8 
$\pm$ 2.4 km s$^{-1}$ Mpc$^{-1}$ Riess \textit{et al.}\cite{Riess2011} As discussed there, 
the Hubble constant measurement uncertainty can significantly affect 
cosmological parameter estimation for a recent example see, e.g., Calabrese
\textit{et al.}\cite{calabrese12} We determine the parameter values that maximize 
the likelihood function and find 1$\sigma$, 2$\sigma$, and 3$\sigma$ constraint 
contours by integrating the likelihood function, starting from the 
maximum and including 68.27 \%, 95.45 \%, and 99.73 \% of the probability.

\begin{figure}[t!]
\centering
    \includegraphics[height=3.0in]{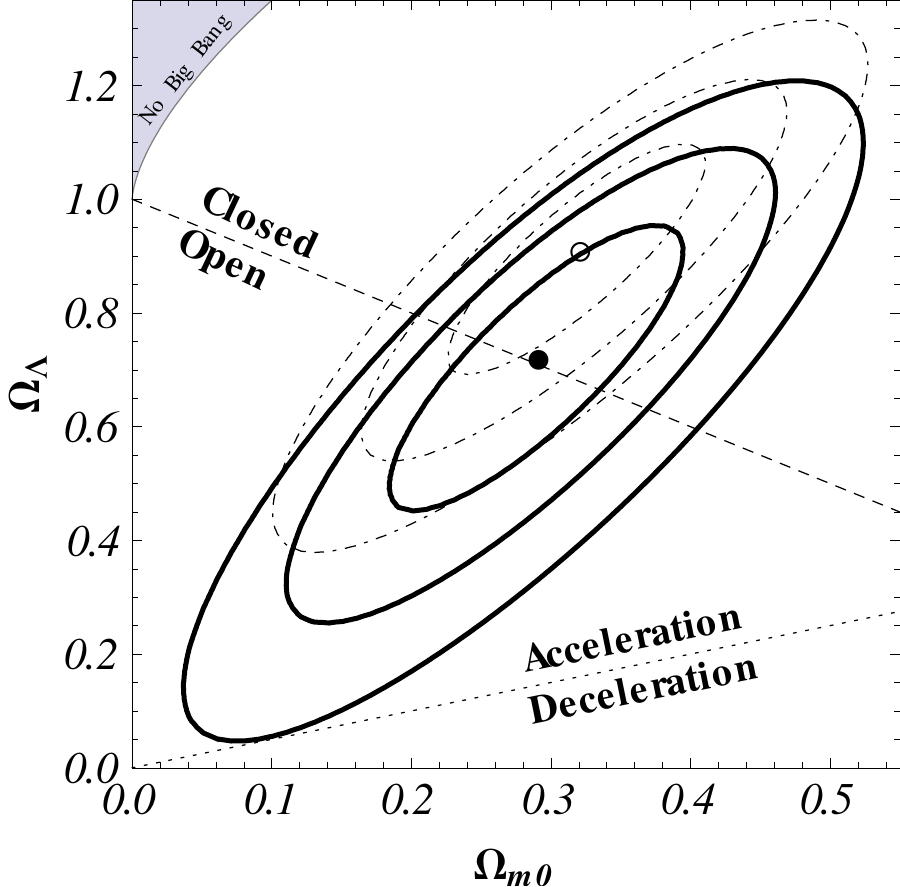}
 \caption{
Solid [dot-dashed] lines show 1$\sigma$, 2$\sigma$, and 3$\sigma$ constraint contours for 
the $\Lambda$CDM model from the $H(z)$ data given in Table (\ref{tab:Hz3}) for 
the prior $\bar{H}_0 \pm \sigma_{H_0} = 68 \pm 2.8$ km s$^{-1}$ Mpc$^{-1}$
[$\bar{H}_0 \pm \sigma_{H_0} = 73.8 \pm 2.4$ km s$^{-1}$ Mpc$^{-1}$].
The filled [empty] circle best-fit point is at 
$(\Omega_{m0},\Omega_{\Lambda})=(0.29,0.72)$ 
[$(0.32,0.91)$] with $\chi^2_{\rm min}=18.24$ [$19.30$].
The dashed diagonal line corresponds to spatially-flat models, the 
dotted line demarcates zero-acceleration models, and the area 
in the upper left-hand corner is the region for which there is no 
big bang. The 2 $\sigma$ intervals from the one-dimensional marginalized 
probability distributions are $0.15\leqslant \Omega_{m0}\leqslant 0.42$, 
$0.35\leqslant \Omega_{\Lambda}\leqslant 1.02$ 
[$0.20\leqslant \Omega_{m0}\leqslant 0.44$, 
$0.62\leqslant \Omega_{\Lambda}\leqslant 1.14$].
}
\label{fig:LCDMHz28}
\end{figure}

\begin{figure}[t!]
\centering
    \includegraphics[height=3.0in]{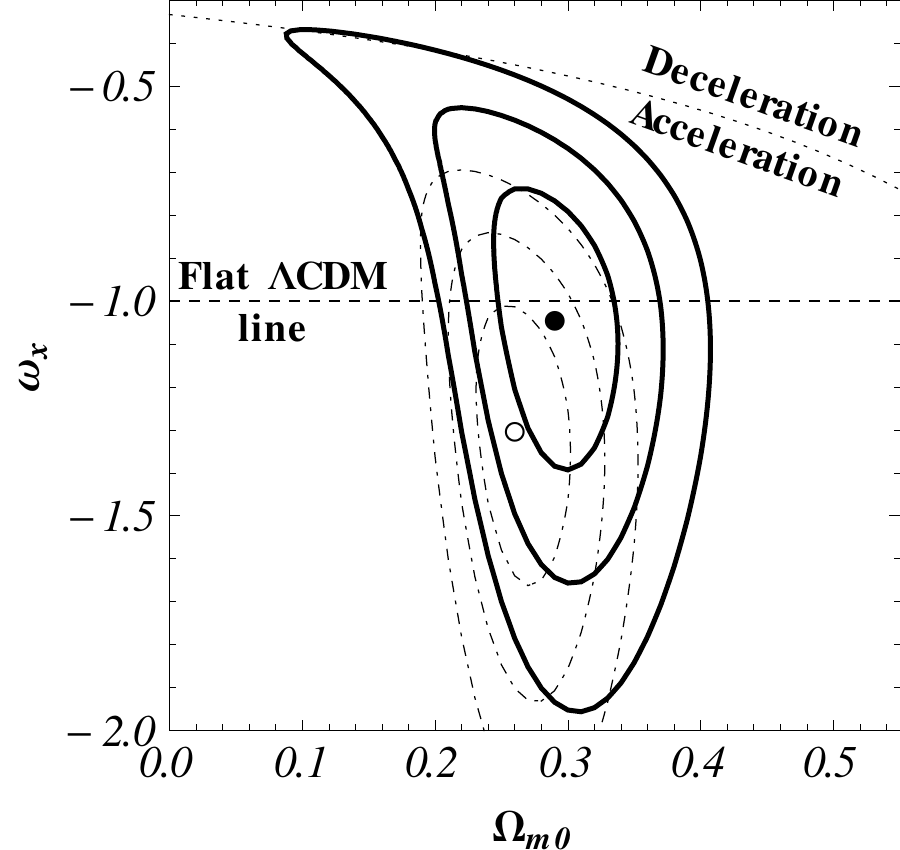}
 \caption{
Solid [dot-dashed] lines show 1$\sigma$, 2$\sigma$, and 3$\sigma$ constraint contours for 
the XCDM parametrization from the $H(z)$ data given in Table (\ref{tab:Hz3}) 
for the prior $\bar{H}_0 \pm \sigma_{H_0} = 68 \pm 2.8$ km s$^{-1}$ Mpc$^{-1}$
[$\bar{H}_0 \pm \sigma_{H_0} = 73.8 \pm 2.4$ km s$^{-1}$ Mpc$^{-1}$].
The filled [empty] circle is the best-fit point at
$(\Omega_{m0},\omega_{X})=(0.29,-1.04)$  [$(0.26,-1.30)$] 
with $\chi^2_{\rm min}=18.18$ [$18.15$].
The dashed horizontal line at $\omega_{\rm X} = -1$ corresponds to 
spatially-flat $\Lambda$CDM models and the curved dotted line demarcates 
zero-acceleration models. The 2 $\sigma$ intervals from the one-dimensional 
marginalized probability distributions are 
$0.23\leqslant \Omega_{m0}\leqslant 0.35$, 
$-1.51\leqslant \omega_{X}\leqslant -0.64$ 
[$0.22\leqslant \Omega_{m0}\leqslant 0.31$, 
$-1.78\leqslant \omega_{X}\leqslant -0.92$].
}
\label{fig:XCDMHz28}
\end{figure}

\begin{figure}[t!]
\centering
    \includegraphics[height=3.0in]{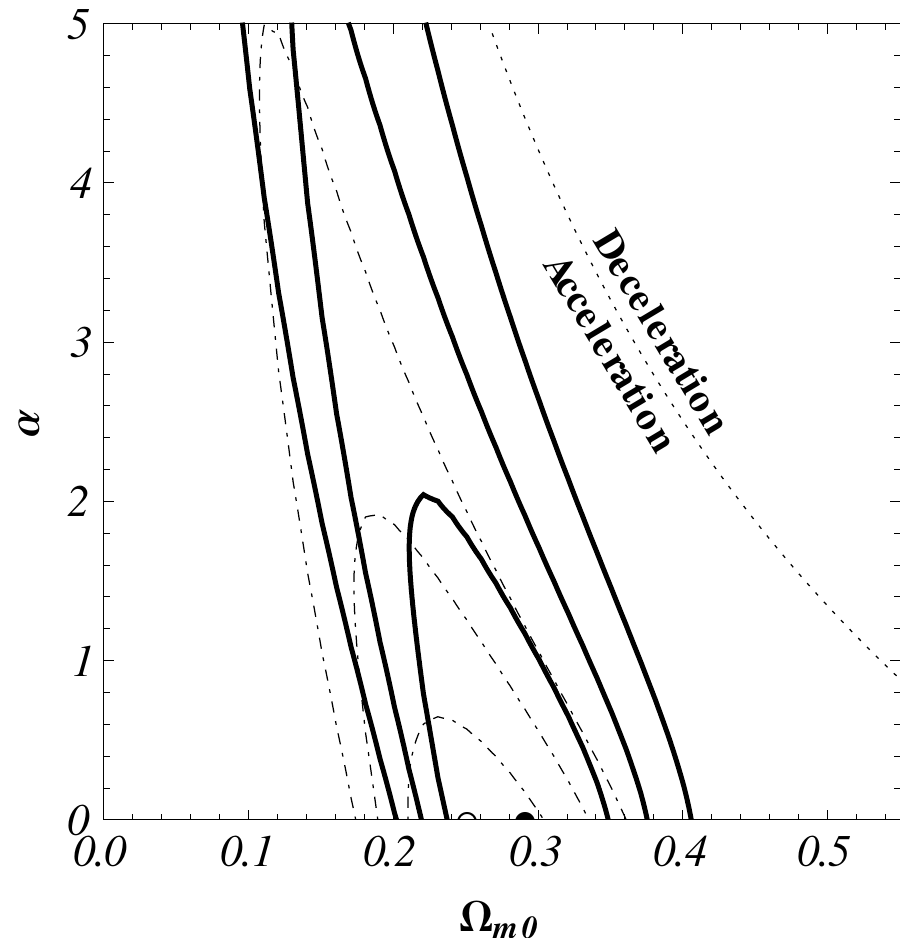}
 \caption{
Solid [dot-dashed] lines show 1$\sigma$, 2$\sigma$, and 3$\sigma$ constraint contours for 
the $\phi$CDM model from the $H(z)$ data given in Table (\ref{tab:Hz3}) for 
the prior $\bar{H}_0 \pm \sigma_{H_0} = 68 \pm 2.8$ km s$^{-1}$ Mpc$^{-1}$
[$\bar{H}_0 \pm \sigma_{H_0} = 73.8 \pm 2.4$ km s$^{-1}$ Mpc$^{-1}$].
The filled [empty] circle best-fit point is at $(\Omega_{m0},\alpha)=(0.29,0)$ 
[$(0.25,0)$] with $\chi^2_{\rm min}=18.24$ [$20.64$].
The horizontal axis at $\alpha = 0$ corresponds to spatially-flat 
$\Lambda$CDM models and the curved dotted line demarcates 
zero-acceleration models. The 2 $\sigma$ intervals from the one-dimensional 
marginalized probability distributions are 
$0.17\leqslant \Omega_{m0}\leqslant 0.34$, 
$\alpha \leqslant 2.2$ [$0.16\leqslant \Omega_{m0}\leqslant 0.34$, 
$\alpha \leqslant 0.7$].
}
\label{fig:PCDMHz28}
\end{figure}

\begin{figure}[t!]
\centering
    \includegraphics[height=4.0in]{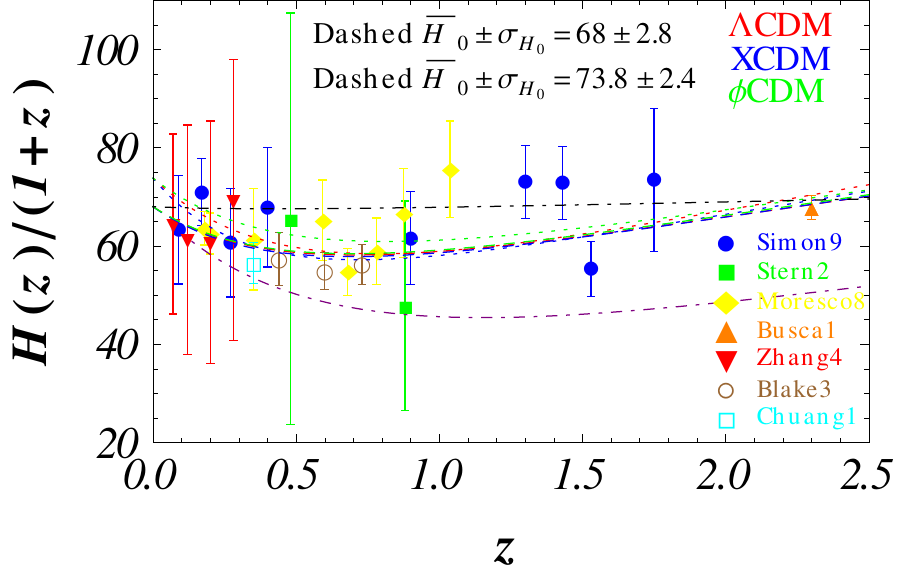}
 \caption{
$H(z)/(1+z)$ data (28 points) and model predictions (lines for 6 
best-fit models) as a function of redshift. The dashed [dotted] lines 
are for the prior $\bar{H}_0 \pm \sigma_{H_0} = 
68 \pm 2.8$ km s$^{-1}$ Mpc$^{-1}$ 
[$\bar{H}_0 \pm \sigma_{H_0} = 73.8 \pm 2.4$ km s$^{-1}$ Mpc$^{-1}$], with
red, blue, and green lines corresponding to the $\Lambda$CDM, XCDM, and 
$\phi$CDM cases. The black (purple) dot-dashed lines correspond to 
two models that are 3 $\sigma$ away from best-fit $\Lambda$CDM ($\phi$CDM) 
and have parameters $\Omega_m = 0.04$, $\Omega_\Lambda = 0.06$
($\Omega_m = 0.2$, $\alpha = 0$), both for the lower value of $\bar{H}_0$.
}
\label{fig:ddota28}
\end{figure}

Figures (\ref{fig:LCDMHz28})---(\ref{fig:PCDMHz28}) show the constraints
from the $H(z)$ data for the three dark energy models we consider, and
for the two different $H_0$ priors. In all 6 cases the $H(z)$ data of
Table \ref{tab:Hz3} require accelerated cosmological expansion at the current epoch,
at, or better than, 3 $\sigma$ confidence. The previous largest 
$H(z)$ data set used, that in Chapter\ (\ref{Chapter5}), required this
accelerated expansion at, or better than, 2 $\sigma$ confidence.
Comparing Figs.\ (\ref{fig:LCDMHz28})---(\ref{fig:PCDMHz28}) here to Figs.\ (\ref{fig:LCDMHz22})---(\ref{fig:PCDMHz22}), we
see that in the XCDM and $\phi$CDM cases the $H(z)$ data we use in
this paper significantly tightens up the constraints on 
$w_{\rm X}$ and $\alpha$, but does not much affect the $\Omega_{m0}$
constraints. However, in the $\Lambda$CDM case the $H(z)$ data used here
tightens up constraints on both $\Omega_\Lambda$ and $\Omega_{m0}$. 
We found that as we increase the value of the nuisance parameter 
$H_0$ the best-fit point for $\Lambda$CDM moves from the spatially-flat 
case to the closed case, and for XCDM the best-fit point moves almost 
orthogonally to the flat $\Lambda$CDM line, towards more negative values 
of $\omega_{X}$. 

As indicated by the $\chi^2_{\rm min}$ values listed in the captions of 
Figs.\ (\ref{fig:LCDMHz28})---(\ref{fig:PCDMHz28}), all 6 best-fit models are very consistent with the $H(z)$
data listed in Table\ (\ref{tab:Hz3}). It is straightforward to compute the cosmological
deceleration-acceleration transition redshift in these cases. Those are given in Table\ (\ref{tab:dars1})

The mean and standard deviation gives $z_{\rm da} = 0.74 \pm 0.05$, which is in good
agreement with the recent Busca \textit{et al.}\cite{busca12} determination of 
$z_{\rm da} = 0.82 \pm 0.08$ based on less data, possibly not all 
independent. Figure (\ref{fig:ddota28}) shows $H(z)/(1+z)$ data from 
Table (\ref{tab:Hz3}) and the 6 best-fit model predictions as a function of 
redshift. The deceleration-acceleration transition is not impossible
to discern in the data.

\begin{table}
\begin{center}
\begin{tabular}{ ccc }
\hline\hline
Model& Prior $H_0$ & Transition Redshift\\
used& (km s$^{-1}$ Mpc $^{-1}$) & $z_{da}$\\
\hline
\multirow{2}{*}{$\Lambda$CDM}& $68 \pm 2.8$   & 0.706 \\
\cline{2-3}
                           {}& $73.8 \pm 2.4$ & 0.785 \\
\hline
\multirow{2}{*}{$\Lambda$CDM}& $68 \pm 2.8$   & 0.706 \\
\cline{2-3}
                           {}& $73.8 \pm 2.4$ & 0.785 \\
\hline
\multirow{2}{*}{$\Lambda$CDM}& $68 \pm 2.8$   & 0.706 \\
\cline{2-3}
                           {}& $73.8 \pm 2.4$ & 0.785 \\
\hline
\multicolumn{2}{c}{\textbf{Average}}                   & $\boldsymbol{0.74 \pm 0.05}$ \\
\hline\hline
\end{tabular}
\end{center}
\caption{Results of the transition redshifts in different models.
}\label{tab:dars1}
\end{table}

From Fig.\ \ref{fig:ddota28} one sees that there are only 6 data 
points for $z>1$, but 22 data points for $z<1$. The larger errors of
some of the $z<1$ data, as compared to those of the $z>1$ measurements, 
are likely responsible for the excellent reduced $\chi^2$ values 
of the best-fit models. 

\section{Conclusion}
\label{summary}

In summary, we have extended the analysis of Busca \textit{et al.}\cite{busca12} to a larger
independent set of 28 $H(z)$ measurements and determined the 
cosmological deceleration-acceleration transition redshift 
$z_{\rm da} = 0.74 \pm 0.05$. These $H(z)$ data are well-described by all
6 best-fit models, and provide tight constraints on the model
parameters. The $H(z)$ data require accelerated cosmological expansion 
at the current epoch, and are consistent with the decelerated 
cosmological expansion at earlier times predicted and required in standard 
dark energy models. While the standard spatially-flat 
$\Lambda$CDM model is very consistent with the $H(z)$ data, current
$H(z)$ data are not able to rule out slowly evolving dark energy.
More, and better quality, data are needed to better discriminate
between constant and slowly-evolving dark energy density.


\cleardoublepage


\chapter{Binned Hubble Parameter Measurements and the Cosmological Deceleration-Acceleration Transition}
\label{Chapter7}

This chapter is based on Farooq \textit{et al.}\cite{farooq4}
\\

\section{Introduction}

In the standard cosmological model dark energy dominates the current epoch 
energy budget, but was less important in the past when non-relativistic (cold dark 
and baryonic) matter dominated. The transition from non-relativistic matter 
dominance to dark energy dominance results in a transition from decelerated 
to accelerated cosmological expansion. The existence of this transition is a strong 
prediction of the standard cosmological model and attempts have been made to
measure the transition redshift.\footnote{
For different attempts to measure deceleration-acceleration transition redshift see  
Lu \textit{et al.},\cite{Lu2011a} Giostri \textit{et al.},\cite{Giostri2012} 
Lima \textit{et al.}\cite{Lima2012} and references therein.}

 However, only very recently has this become possible, due to high redshift 
(i.e., $z$ above the deceleration-acceleration transition) data that recently became
available, with the most striking being the Busca \textit{et al.}\cite{busca12} measurement of the 
Hubble parameter $H(z=2.3)=224 \pm 8$ km s$^{-1}$ Mpc$^{-1}$, well in the matter 
dominated epoch of the standard $\Lambda$CDM model. 

From a compilation of 28 independent $H(z)$
measurements over $0.07\leqslant  z \leqslant 2.3$ Table\ (\ref{tab:Hz3}), 
the transition redshift was found in Chapter (\ref{Chapter6}) to be $z_{\rm da} = 0.74 \pm 0.05$. 
This was determined from the 6 best-fit transition redshifts measured in three 
different cosmological models, $\Lambda$CDM, XCDM, and $\phi$CDM, for two different 
Hubble constant priors.

The spatially-flat $\Lambda$CDM model\cite{peebles84} is the reigning standard 
cosmological model. Here we consider the more general $\Lambda$CDM model that 
allows for non-zero space curvature. In the standard model 
the cosmological constant, $\Lambda$, contributes around 70$\%$ of the 
present cosmological energy budget, non-relativistic, pressure-less, cold
dark matter (CDM) contributes a little more than 20$\%$, and non-relativistic baryonic matter 
makes up the remaining 5$\%$ or so.\footnote{According to the new data from \textit{Planck Collaboration} 
(see Ade \textit{et al.}\cite{Planckdata}) the dark energy density corresponding to cosmological
constant $\Lambda$, contributes 68.5\% of the current energy density. The next largest contributor is non-relativistic
cold dark matter at 26.7\%, and almost all the rest is cold baryonic matter which makes up to 4.9\% of the current 
energy budget.} In the $\Lambda$CDM model time-independent dark energy, 
$\Lambda$, is modeled as a spatially homogeneous 
fluid with equation of state $p_{\rm \Lambda} = -\rho_{\rm \Lambda}$ where 
$p_{\rm \Lambda}$ and $\rho_{\rm \Lambda}$ are the fluid pressure and energy 
density respectively. It has been known for a while now that the spatially-flat 
$\Lambda$CDM model is consistent with most observational data.\footnote{For 
early indications see, e.g.,Jassal \textit{et al.},\cite{jassal10} Wilson 
\textit{et al.},\cite{wilson06} Davis \textit{et al.},\cite{Davis2007} and 
Allen \textit{et al.}\cite{allen08} 
Note, however, there are some preliminary observational hints that the standard CDM 
structure formation model, assumed in the flat $\Lambda$CDM cosmological model, 
might need to be improved upon 
\citep[][and references therein]{Peebles&Ratra2003, Perivolaropoulos2010}.} 
It is also well known that if, instead of staying constant
like $\Lambda$, the dark energy density gradually decreased 
in time (and correspondingly slowly varied in
space), it would alleviate a conceptual coincidence 
problem associated with the $\Lambda$CDM model.\footnote{For recent discussions 
of time-varying dark energy models see 
Guendelman \textit{et al.},\cite{Guendelman2012} Wang,\cite{Wang12} De-Santiago
\textit{et al.},\cite{De-Santiago12} Lima \textit{et al.},\cite{lima13} 
Capozziello \textit{et al.},\cite{Capozziello13} Adak Lima \textit{et al.},\cite{Adak2012}  
and references therein.}  

Here we not only constraints the parameters of the two models and a dark energy 
parametrization, which are discussed in Chapter (\ref{Chapter3}),
namely non-flat $\Lambda$CDM, flat $\phi$CDM, and XCDM respectively but also find the 
deceleration-acceleration best-fit value of transition redshift using these models.

In addition to being affected by the cosmological model used in the analysis, the 
measured deceleration-acceleration transition redshift $z_{\rm da}$ depends on the assumed 
value of the Hubble constant. Consequently, to quantify the effect, we use
two Gaussian $H_0$ priors in the analyses. The first prior is 
$\overline{H}_{0}$ $\pm$ $\sigma_{H_{0}}$ = 68 $\pm$ 2.8 km s$^{-1}$ Mpc$^{-1}$. 
This comes from a median statistics analysis of 553 $H_0$ measurements Chen \textit{et al.}\citep{Chen2011a} 
and is consistent with the earlier estimates of Gott \textit{et al.}\cite{Gott2001} and Chen \textit{et al.}\cite{chen03}. The
second prior of  
$\overline{H}_{0}$ $\pm$ $\sigma_{H_{0}}$ = 73.8 $\pm$ 2.4 km s$^{-1}$ Mpc$^{-1}$ comes 
from recent Hubble Space Telescope measurements Riess \textit{et al.}\citep{Riess2011}.\footnote{Other
recent measurements are consistent with either the smaller or larger $H_0$ value
we consider, see, e.g., Freedman \textit{et al.},\cite{Freedman2012} Sorce \textit{et al.},\cite{Sorce2012},
and Tammann \textit{et al.}\cite{Tammann2012}, although it might now be significant that both BAO 
see, e.g., Colless \textit{et al.},\cite{colless12} and \textit{Planck} CMB anisotropy Ade \textit{et al.},\citep{Planckdata} 
measurements favor the lower $H_0$ value we use. It might also be significant that the
lower value of $H_0$ does not require the presence of dark radiation
\citep[][and references there in]{calabrese12}.}
 
In Chapter (\ref{Chapter6}) we determined the redshift of the deceleration-acceleration
transition by finding the mean and standard deviation of the six best-fit $z_{\rm da}$ values
in the 3 models (with 2 different $H_0$ priors). Here we use a different technique to measure
 $z_{\rm da}$ and the related uncertainty in each of these six cases. We then determine 
 summary estimates of $z_{\rm da}$ by considering various weighted mean combinations of 
 these six estimates. The transition redshifts take the forms:
\begin{eqnarray}
&z_{\rm da}=\left({\frac{2 \Omega_{\Lambda}}{\Omega_{m0}}}\right)^{1/3}-1,
\label{Eq:6-1}
\end{eqnarray}
\begin{eqnarray}
&z_{\rm da}=\left({\frac{\Omega_{m0}}{(\Omega_{m0}-1)(1+3 \omega_X)}}\right)^{1/{3\omega_{X}}}-1,
\label{Eq:6-2}
\end{eqnarray}
for the $\Lambda$CDM and XCDM cases where $\Omega_{\Lambda}$ and $\Omega_{m0}$ 
are the cosmological constant and non-relativistic matter density parameters,
and $\omega_X$ is the equation-of-state-parameter of dark energy. 
As for $\phi$CDM, from Eqs.\ (3) of Peebles \& Ratra\cite{Peebles&Ratra1988} we first derive:
\begin{eqnarray}
\frac{\ddot{a}}{a}&=&-\frac{4 \pi G}{3} \left[\rho_{m}+\rho_{\phi}(1+3\omega_{\phi})\right] \nonumber \\
				&=& -\frac{1}{2} H_0^2 \left[\Omega_{m0} (1+z)^3+\Omega_{\phi}(z,\alpha)(1+3\omega_{\phi}(z))\right],
\label{eq:friedman}
\end{eqnarray}
where $\Omega_{\phi}(z)$ is the scalar field energy density parameter and:
\begin{align}
\omega_{\phi}(z)=\frac{\frac{1}{2}\dot{\phi^2}-V(\phi)}{\frac{1}{2}\dot{\phi^2}+V(\phi)}.
\end{align}
The redshift $z_{\rm da}$ is determined by requiring that the right hand side 
of Eq.\ (\ref{eq:friedman}) vanish:
\begin{align}
&\Omega_{m0} (1+z_{\rm da})^3+\Omega_{\phi}(z_{\rm da},\alpha)\left[1+3~\omega_{\phi}(z_{\rm da})\right]=0.
\label{eq:zdaphi}
\end{align}
To determine $z_{\rm da}$ we numerically integrate the $\phi$CDM 
model equations of motion, Eqs. (3) of Peebles \& Ratra\cite{Peebles&Ratra1988},
using the initial conditions described there.\footnote{These initial conditions are 
discussed in detail in Chapter (\ref{Chapter3}).} These solutions 
determine the needed functions in Eq. (\ref{eq:zdaphi}), which we then numerically
solve for $z_{\rm da}(\Omega_{m0},\alpha)$. 

To find the expected values
$\langle z_{\rm da} \rangle$ and $\langle z_{\rm da}^2 \rangle$ 
we use:
\begin{eqnarray}
\label{expected zda}
\langle z_{\rm da} \rangle=\frac{\iint z_{\rm da}(\textbf{p}) \mathcal{L}(\textbf{p}) d\textbf{p}}{\iint \mathcal{L}(\textbf{p}) d\textbf{p}},~~~~~~~~~~~~
\langle z_{\rm da}^2 \rangle=\frac{\iint z_{\rm da}^2(\textbf{p}) \mathcal{L}(\textbf{p}) d\textbf{p}}{\iint \mathcal{L}(\textbf{p}) d\textbf{p}}.
\label{Eq:6-6}
\end{eqnarray} 
Here $\mathcal{L}(\textbf{p})$ is the $H(z)$ data likelihood function after marginalization
over the $H_0$ prior in the model under consideration. It depends only on the model 
parameters $\textbf{p}=(\Omega_{m0},\Omega_{\Lambda})$ for $\Lambda$CDM,
$=(\Omega_{m0},\omega_{X})$ for XCDM, and 
$=(\Omega_{m0},\alpha)$ for $\phi$CDM. 
The standard deviation in $z_{\rm da}$ is calculated from the standard formula
$\sigma_{z_{\rm da}}=\sqrt{\langle z_{\rm da}^2 \rangle-\langle z_{\rm da} \rangle^2}$.
The results of this computation are summarized in Table\ (\ref{table:Un-binned data details}).

It is reassuring that the results of the penultimate and the last columns of Table\ (\ref{table:Un-binned data details}) are 
very consistent. In Chapter\ (\ref{Chapter6}) we determined a summary estimate of $z_{\rm da}=0.74 \pm 0.05$ by 
computing the mean and standard deviation of the six values in the last column of Table\ (\ref{table:Un-binned data details}). It is 
of interest to estimate similar summary values for each of the two $H_{0}$ priors. We find that 
$z_{\rm da}=0.70 \pm 0.05$ ($z_{\rm da}= 0.77 \pm 0.04$) for 
$H_{0}\pm\sigma_{H_0}$ = 68 $\pm$ 2.8 (73.8 $\pm$ 2.4) km s$^{-1}$ Mpc$^{-1}$. 
Perhaps  more realistic summary estimates are determined by the weighted means
of the two sets of 3 values in the penultimate column of Table\ (\ref{table:Un-binned data details}): 
$z_{\rm da}=0.69 \pm 0.06$ ($z_{\rm da}=0.76 \pm 0.05$) for 
$H_{0}\pm\sigma_{H_0}$ = 68 $\pm$ 2.8 (73.8 $\pm$ 2.4) km s$^{-1}$ Mpc$^{-1}$, 
and $z_{\rm da}=0.74 \pm 0.04$ are the results if all six values are used.

More conventionally, cosmological data are used to constrain model parameters 
values such as $\Omega_{m0}$ and $\Omega_{\Lambda}$ for the $\Lambda$CDM model. 
A number of different data sets have been used for this purpose. 
These include Type Ia supernova (SNIa) apparent magnitude verses redshift data e.g.,Ruiz \textit{et al.},\cite{Ruiz2012}
Chiba \textit{et al.},\cite{chiba2013} Cardenas \textit{et al.},\cite{cardenas2011} 
Liao \textit{et al.},\cite{Liao2013} Farooq \textit{et al.},\cite{Farooq:2012ev} 
Campbell \textit{et al.},\cite{Campbell2013} cosmic microwave 
background (CMB) anisotropy measurements Ade \textit{et al.}\citep[][and references therein]{Planckdata}, 
baryonic acoustic oscillation (BAO) peak length scale data Mehta \textit{et al.},\cite{Mehta2012}
Anderson \textit{et al.},\cite{Anderson12} Li \textit{et al.},\cite{Li2012a} Scovacricchi \textit{et al.},\cite{Scovacricchi2012} 
Farooq \& Ratra\cite{Farooq:2013hq} and references therein, galaxy cluster gas mass fraction
as a function of redshift e.g.,
Allen \textit{et al.},\cite{allen08}
Samushia \& Ratra,\cite{Samushia&Ratra2008}
Tong \& Noh \textit{et al.},\cite{tongnoh11}
Lu \textit{et al.},\cite{Lu2011b}
Solano \textit{et al.},\cite{solano12}
Landry \textit{et al.},\cite{landry12}
and, of special interest here, measurement of the Hubble
parameter as a function of redshift 
Jimenezeta \textit{et al.},\cite{Jimenezetal2003}
Samushia \& Ratra,\cite{Samushia&Ratra2006}
Samushia \textit{et al.},\cite{samushia07}
Sen \textit{et al.},\cite{Sen&Scherrer2008}
Yun \& Ratra,\cite{Chen2011b}
Aviles\textit{et al.},\cite{Aviles2012}
Wang \textit{et al.},\cite{wangjcap12}
Campos \textit{et al.},\cite{campos12}
Chimento \textit{et al.},\cite{Chimento13}
and references therein. These data,
separately and in combination, provide strong evidence for accelerated cosmological
expansion at the current epoch.\footnote{Other data, with larger error
bars, support these results. See, e.g., Chae \textit{et al.},\cite{chae04}
Cao \textit{et al.},\cite{Cao2012}, Yun \& Ratra\cite{Chen2012}, 
Jackson\cite{Jackson2012}, Campanelli \textit{et al.},\cite{campanelli11}, 
Mania \& Ratra,\cite{maniaratra12} 
Poitras,\cite{Poitras2012} and Pan \textit{et al.}\cite{pan13}} However their error bars are still
too large to allow for a discrimination between constant and time-varying 
dark energy densities.

Of course, both methods are equivalent, since they make use of the same data,
but each has it own advantages and disadvantages. In particular, it is of
some interest to actually discern the deceleration-acceleration transition
in the $H(z)$ data. While the data does indicate the transition, see Fig.\ (\ref{fig:ddota28}),
the data points bounce around quite a bit. Given the low reduced $\chi^2$
for the best-fit models (see Chapter (\ref{Chapter6}) and Table\ (\ref{table:Un-binned data details}) here), all of which show significant 
evidence for a deceleration-acceleration transition. We investigate different data binning
techniques here, to see if binned versions of the $H(z)$ measurements more 
clearly illustrate the presence of a deceleration-acceleration transition.

\section{Binning the Data}
\label{Binning the data}

The 28 individual $H(z)$ measurements bounce around on the $H(z)/(1+z)$ plot,
Fig.\ 4 of Farooq \& Ratra.\cite{Farooq:2013hq} To try to get a smoother observed $H(z)/(1+z)$ function we form
bins in redshift and then combine the data points in each bin to give
a single observed value of $z$, $H(z)$, and $\sigma$ for that bin. The measurements
in each bin are combined using two different statistical techniques, weighted 
mean and median statistics. 

Table (\ref{table:WA}) lists  the weighted mean results. These results were computed using the 
standard formulae  see, e.g., Podariu \textit{et al.}\citep[][]{Podariu2001}. That is:
\begin{equation}  
\overline{H}(z)=\frac{\sum_{i=1}^{N} H(z_i)/\sigma_{i}^{2}}{\sum_{i=1}^{N}1/\sigma_{i}^{2}},
\end{equation}
where $N$ is the number of data points in the bin under consideration, $\overline{H}(z)$ is the weighted 
mean of the Hubble parameter in that bin, $H(z_i)$ is the value of the Hubble parameter measured
at redshift $z_i$ and $\sigma_i$ is the corresponding uncertainty. Weighted mean redshifts, denoted 
by $\overline{z}$, were similarly computed:
\begin{equation}
\overline{z}=\frac{\sum_{i=1}^{N} z_{i}/\sigma_{i}^{2}}{\sum_{i=1}^{N}1/\sigma_{i}^{2}}.
\end{equation}
The weighted mean standard deviation, denoted by $\overline{\sigma}$, for each bin was found from:
\begin{equation}
\overline{\sigma}=\left(\sum_{i=1}^{N} 1/\sigma_i^{2}\right)^{-1/2}.
\end{equation}
The assumptions underlying use of weighted mean statistics are that the measurement errors
are Gaussian, and there are no systematic errors. Hence, one can compute $\chi^2$, the
goodness-of-fit parameter, for each bin:
\begin{equation}
\chi^2=\frac{1}{N-1}\sum_{i=1}^{N}\frac{[H(z_i)-\overline{H}(z)]^2}{\sigma_{i}^2},
\label{eq.chi}
\end{equation}
which has expected value unity and error $1/\sqrt{2(N-1)}$, so we can use this to determine 
the number of standard deviations that $\chi$ deviates from unity for each bin:
\begin{equation}
N_{\overline\sigma}=|\chi-1|\sqrt{2(N-1)}.
\end{equation}
An unaccounted for systematic error, the 
presence of significant correlations between the measurements, and breakdown of the Gaussian error
assumption for each measurement, are the three factors that can make $N_{\overline\sigma}$ 
much greater than unity.

The second technique 
we use to combine measurements in a bin is median statistics, 
as developed in\footnote{For other applications of median
statistics see, e.g., Sereno,\cite{sereno03} Chen \textit{et al.},\cite{chen03b} Richards \textit{et al.},\cite{Richards2009}
and Shafieloo \textit{et al.}\cite{Shafieloo11}} Gott \textit{et al.}\cite{Gott2001} Table\ (\ref{table:Med}) lists the median 
statistics results. The median is the value for which there is a $50\%$ 
chance of finding a measurement below or above it. 
It is fair to use median statistics to combine the $H(z)$ data of Table\ (\ref{tab:Hz3}) 
since we assume that all the measurements are independent and there is no over-all systematic 
error in the $H(z)$ data as a whole. \citep[Individual measurements can have individual systematic errors,
 for discussion see Chen \& Ratra.][]{Chen2011a}
The median will be revealed as the true value 
as the number of measurements grow to infinity, and this technique reduces the effect of outliers 
of a set of measurements on the estimate of a true value. 
If $N$ measurements are considered, the probability of 
finding the true value between values $N_{i}$ and $N_{i+1}$ (where $i=1,2,...N$) is 
(see Gott \textit{et al.}\citep{Gott2001}):
\begin{equation}
P_{i}=\frac{2^{-N}N!}{i!(N-1)!}
\end{equation}
This process of finding a median value was used for the redshift and the Hubble parameter,
and the Hubble parameter probability distribution was used to determine  $\sigma$ for each bin. 

We would like to have as many measurements as possible in each bin, 
as well as bins that are as narrow as possible in redshift space. 
Obviously, since these requirements are contradictory, compromise 
is necessary. In addition, we require roughly the same number of 
measurements per bin, so as to have approximately similar errors 
on the binned measurements. As indicated in Table\ (\ref{table:WA}) 
and Table\ (\ref{table:Med}) we consider four different binnings 
of the 23 lower redshift, $z<1.04$, measurements; the five higher 
redshift measurements are sparsely spread over  too large a 
redshift range to allow for a useful binning.

The last column of Table\ (\ref{table:WA}) shows that the first 
two binnings, with approximately 3 and 5 measurements per bin, do 
not show any deviation from what is expected from Gaussian 
errors. On the other hand, the last binning, with about 8 
measurements per bin, appears to be not so consistent with 
the assumption of Gaussian errors. This is likely a  consequence 
of the large width in redshift of these bins, so the measurements 
at the low $z$ end and at the high $z$ end of each bin differ too much to
be combined together. Median statistics does not make
use of the error bars of the individual measurements. As a result, 
it is a more conservative technique, and when used to combine data in bins
it results in larger error bars. A comparison of the results in Tables 
(\ref{table:WA}) and (\ref{table:Med}) clearly illustrates this point. Fortunately the weighted mean
results we have found show that the individual lower redshift data points have 
reasonable error bars and so there is no obvious danger in 
using the more constraining weighted mean results to draw 
physical conclusions.

The weighted-mean and median 
statistics binned results of Tables (\ref{table:WA}) and (\ref{table:Med}) are plotted in the 
top panels of Figs.\ (\ref{fig:For table 2,3})---(\ref{fig:For table 8,9})
(in purple).
These figures also show the 5 higher $z$ unbinned measurements listed 
in Table\ (\ref{tab:Hz3}) (in cyan). Both sets of observations show $1$
and $2$ $\sigma$ error bars. Also shown are the unbinned data (Table\ (\ref{tab:Hz3}))
best-fit predictions for $\Lambda$CDM (red), XCDM (blue), and $\phi$CDM (green)
for the two priors, $\overline{H}_0\pm\sigma_{H_0}$= 68 
$\pm$ 2.8 km s$^{-1}$ Mpc$^{-1}$ (dashed lines) and $\overline{H}_0\pm\sigma_{H_0}$= 7.8 
$\pm$ 2.4 km s$^{-1}$ Mpc$^{-1}$ (dotted lines), from Farooq \& Ratra.\cite{Farooq:2013hq} Focusing on the 
weighted-mean panels in each of these plots, and comparing to Fig.\ (\ref{fig:ddota28}),
we see that the binned data of Figs.\ 1---3 clearly demarcates a declaration-acceleration
transition.

\section{Constraints from the binned data}
\label{constraints}

In this section we use the weighted-mean and median statistics binned data to derive constraints 
on cosmological parameters of $\Lambda$CDM, XCDM, and $\phi$CDM, and compare these constraints
to those that follow from the unbinned data of Table\ (\ref{tab:Hz3}).  

In order to derive constraints on the parameters $\textbf{p}$ of the dark energy models 
discussed above, using the binned 
data from Tables\ (\ref{table:WA}) and (\ref{table:Med}), we follow the procedure of Chapter (\ref{Chapter5}).
The observational data consist of measurements 
of the Hubble parameter $H_{\rm obs}(z_i)$ at redshifts $z_i$, with the corresponding 
one standard deviation uncertainties $\sigma_i$. To constrain parameters 
of cosmological models, we define the posterior likelihood function 
$\mathcal{L}_{H}(\textbf{p})$, that depends only on the model parameters $\textbf{p}$, 
by integrating the product of the $H_0$ prior likelihood function 
$\propto$ exp$[-(H_0-\bar H_0)^2/(2\sigma^2_{H_0})]$ and the usual likelihood function exp$(-\chi_H^2 /2)$, 
as in Eq.\ (18) of \cite{Farooq:2012ev}. 
Two different Gaussian priors, $\overline{H}_0\pm\sigma_{H_0}$= 68 
$\pm$ 2.8 km s$^{-1}$ Mpc$^{-1}$ 
\citep{Chen2011a} and 
$\overline{H}_0\pm\sigma_{H_0}$ = 73.8 $\pm$ 2.4 km s$^{-1}$ Mpc$^{-1}$
\citep{Riess2011} are used in the marginalization of the likelihood function over the nuisance 
parameter $H_0$. 

The best-fit point (BFP) $\mathbf{p_0}$ are those parameter values that
maximize the likelihood function $\mathcal{L}_H(\mathbf{p})$.
To find the 1, 2, and 3 $\sigma$ confidence intervals as two-dimensional 
parameter sets, we start from the BFP and integrate the volume under $\mathcal{L}_H(\mathbf{p})$  
until we include  68.27, 95.45, and 99.73 \% of the probability.

The lower 3 rows of panels in Figs.\ (\ref{fig:For table 2,3})---(\ref{fig:For table 8,9}) show the 
constraints (1$\sigma$, 2$\sigma$, and 3$\sigma$ contours)
from the unbinned $H(z)$ data of Table\ 1 of Farooq \& Ratra,\cite{Farooq:2013hq} (in blue dot-dashed contours) and from the binned $H(z)$ 
data of Tables (\ref{table:WA}) and (\ref{table:Med}) here (in red solid contours), for the three dark energy 
models we consider, and for the two different $H_0$ priors mentioned above. 
The red filled circles and the blue empty circles are the best fit points for the binned and 
unbinned data respectively. Some relevant results are listed in Tables\ 4---7.
Comparing the weighted-mean BFP cosmological parameter values listed in these tables, to those 
listed in the captions of Figs.\ (\ref{fig:LCDMHz28})---(\ref{fig:PCDMHz28}), establishes the very good 
agreement between the values derived here using the binned data (especially
for fewer measurements per bin) and the Farooq \& Ratra,\cite{Farooq:2013hq} values derived
using the unbinned data.

It is clear from the left two columns of the lower three rows of 
Figs.\ (\ref{fig:For table 2,3})---(\ref{fig:For table 8,9}) that the weighted-mean
binning of the first 23 data points in Table\ (\ref{tab:Hz3}) give almost exactly the 
same constraints on model parameters $\textbf{p}$ for the three cosmological 
models as do the unbinned data of Table\ (\ref{tab:Hz3}).
Since the weighted-mean technique
assumes that the error in the measurements has a Gaussian distribution
and that the measurements are uncorrelated, this 
result is consistant with this assumption that the 
$H(z)$ data in Table\ (\ref{tab:Hz3}) have Gaussian errors. Consequently, 
the best way to combine the measurements in a bin is to use the weighted-mean method. 
It is also useful to note that when there are fewer data points in a narrower bin, 
the constraints from the binned data matches better with the constraints derived 
from the unbinned data. This is not unexpected. In the case of median 
statistics, however, the constraints on model parameters for all three models 
from the binned data are much less restrictive than those 
derived from the unbinned data, see the right hand column of panels in the lower three rows of 
Figs\ (\ref{fig:For table 2,3})---(\ref{fig:For table 8,9}). This is because median statistics is a more
conservative technique and so, in this case, is not the
best way of combining $H(z)$ measurements in bins. 
It is also interesting to note, from Tables 
(\ref{table:fig1 details})---(\ref{table:fig4 details}), that $\chi^2_{\rm min}$ for the case of median 
statistics is significantly smaller than $\chi^2_{\rm min}$ for 
the weighted mean case. This is a direct 
consequence of the larger error bars estimated by the more conservative median statistics approach.

\section{Conclusion}
\label{summary}

We have shown that the weighted-mean combinations of the lower redshift $H(z)$
measurements by bins in redshift provide close to identical constraints on 
cosmological model parameters as do the unbinned $H(z)$ data tabulated in Farooq \& Ratra.\cite{Farooq:2013hq}
This is consistent with the $H(z)$ measurements errors being Gaussian.

When plotted against $z$, the weighted-mean binned $H(z)/(1+z)$ measurements
bounce around much less than the individual measurements considered in Farooq \& Ratra,\cite{Farooq:2013hq} 
and now much more clearly show the presence of a cosmological 
deceleration-acceleration transition, consistent with the new summary 
redshift $z_{\rm da}=0.74 \pm 0.04$ estimated here and consistent with that 
estimated in Farooq \& Ratra,\cite{Farooq:2013hq} This result is also consistent with what is
expected in the standard spatially-flat $\Lambda$CDM model, and in other cosmological
models with present-epoch energy budget dominated by dark energy.

More, and more precise, measurements of $H(z)$ in the redshift range 
$1\lesssim  z \lesssim 2.5$ will allow for a clearer demarcation of 
the cosmological deceleration-acceleration transition. We anticipate 
that such data will soon become available.

\begin{figure}[h!]
    \includegraphics[height=2.1in]{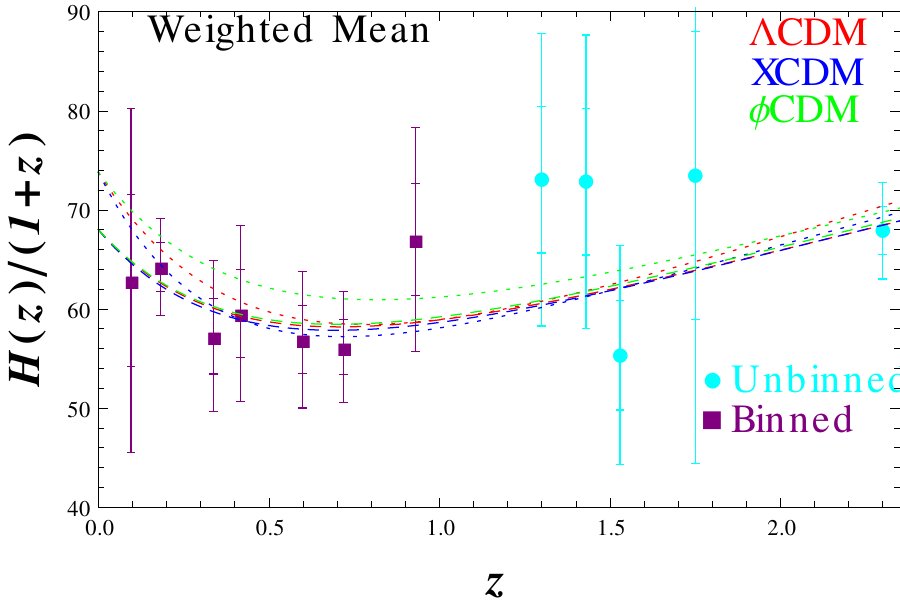}
    \includegraphics[height=2.1in]{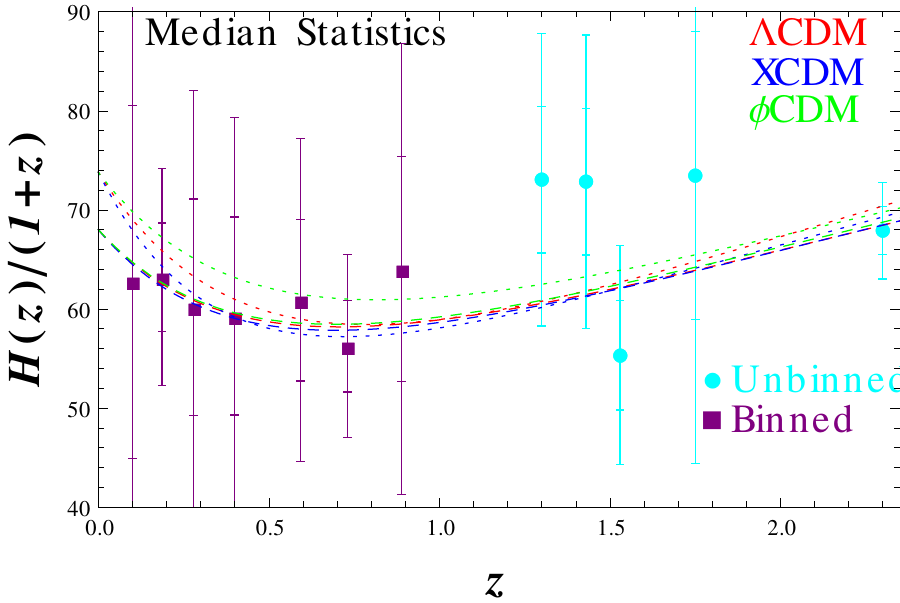}
    \includegraphics[height=1.5in]{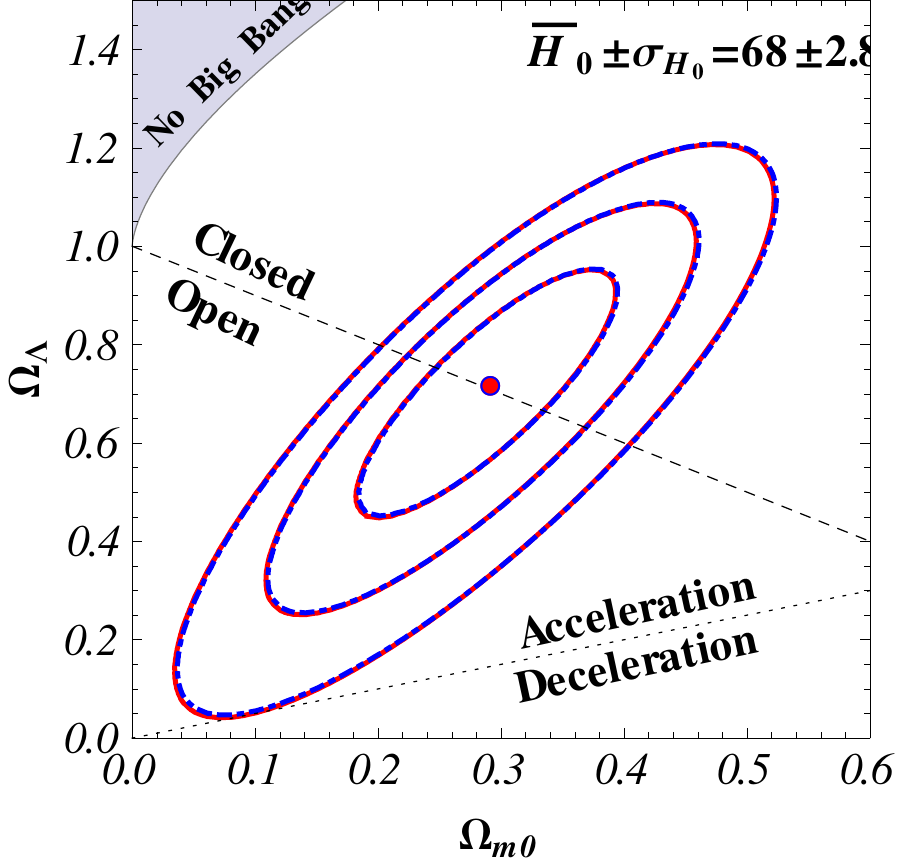}
    \includegraphics[height=1.5in]{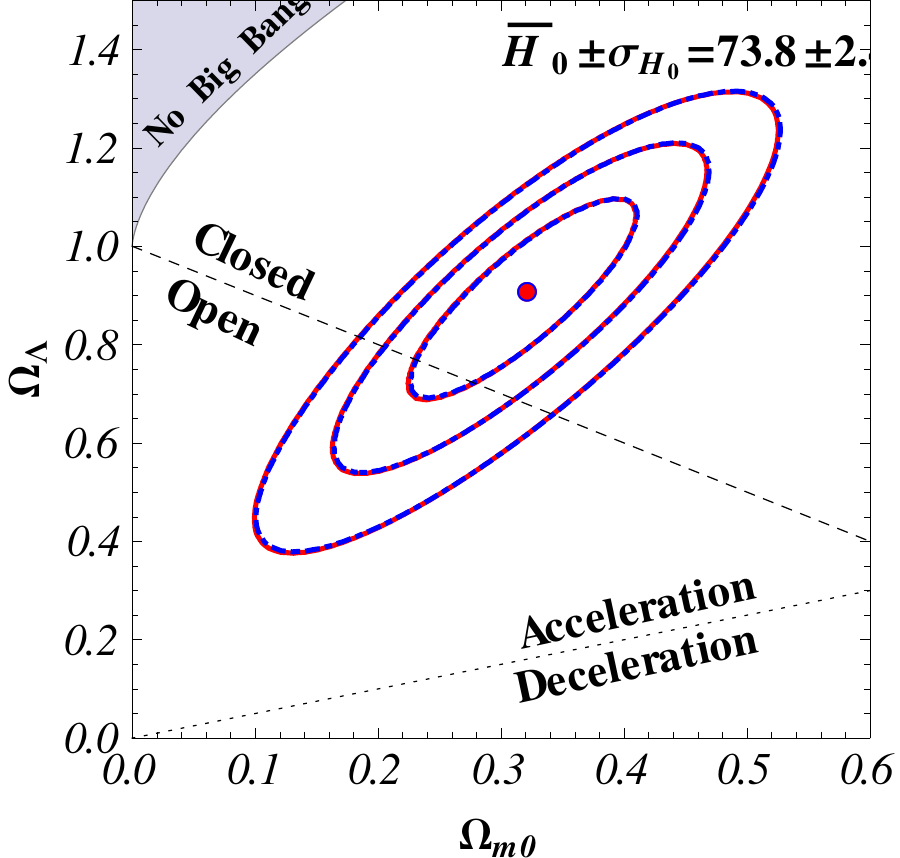}
    \includegraphics[height=1.5in]{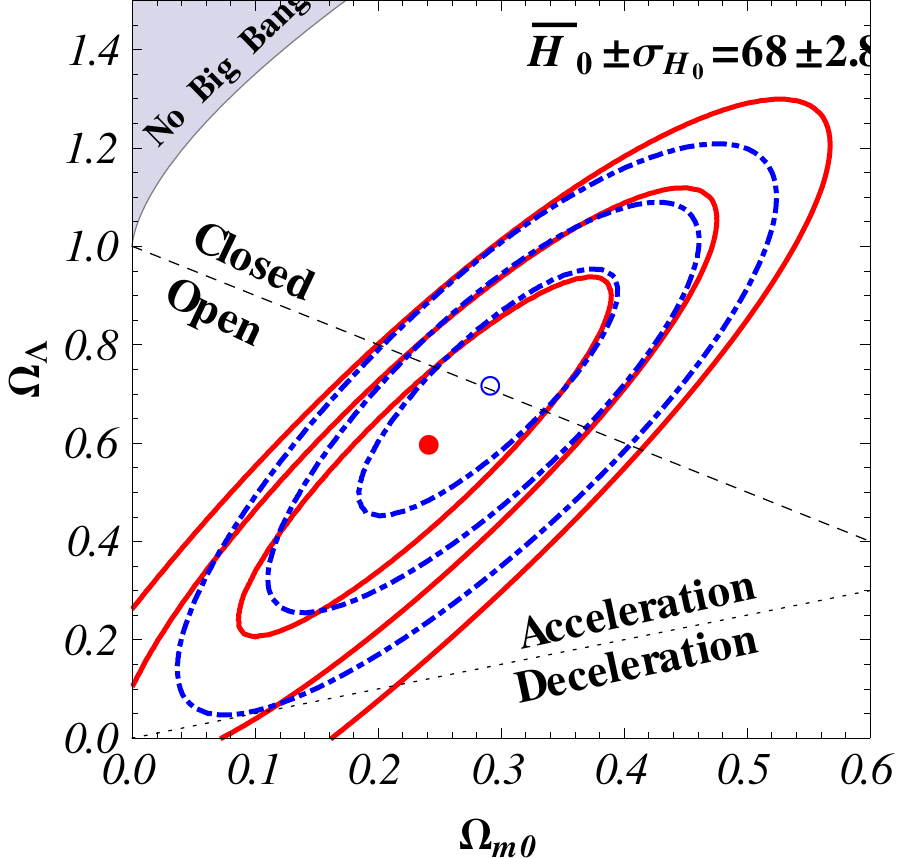}
    \includegraphics[height=1.5in]{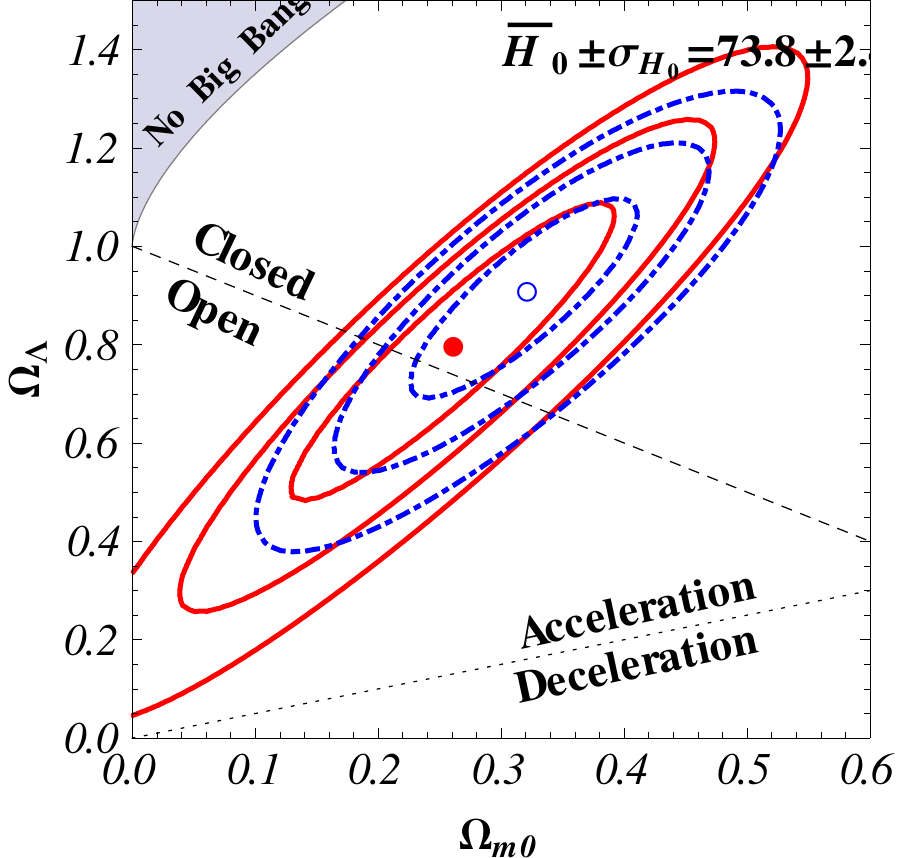}
    \includegraphics[height=1.45in]{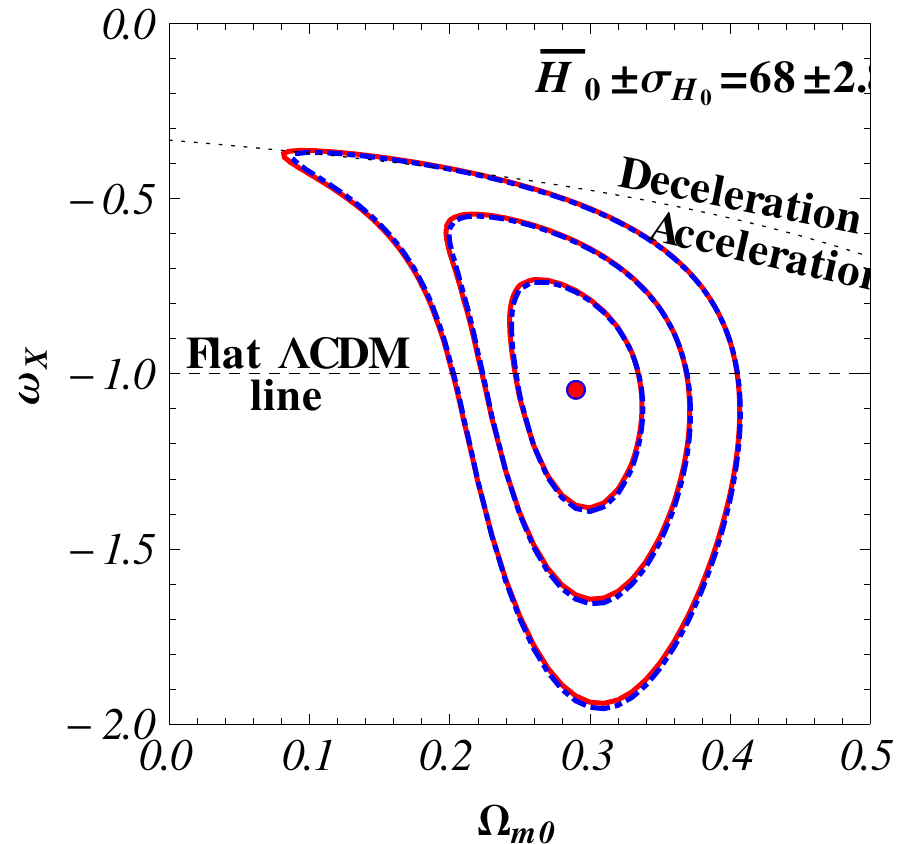}
    \includegraphics[height=1.45in]{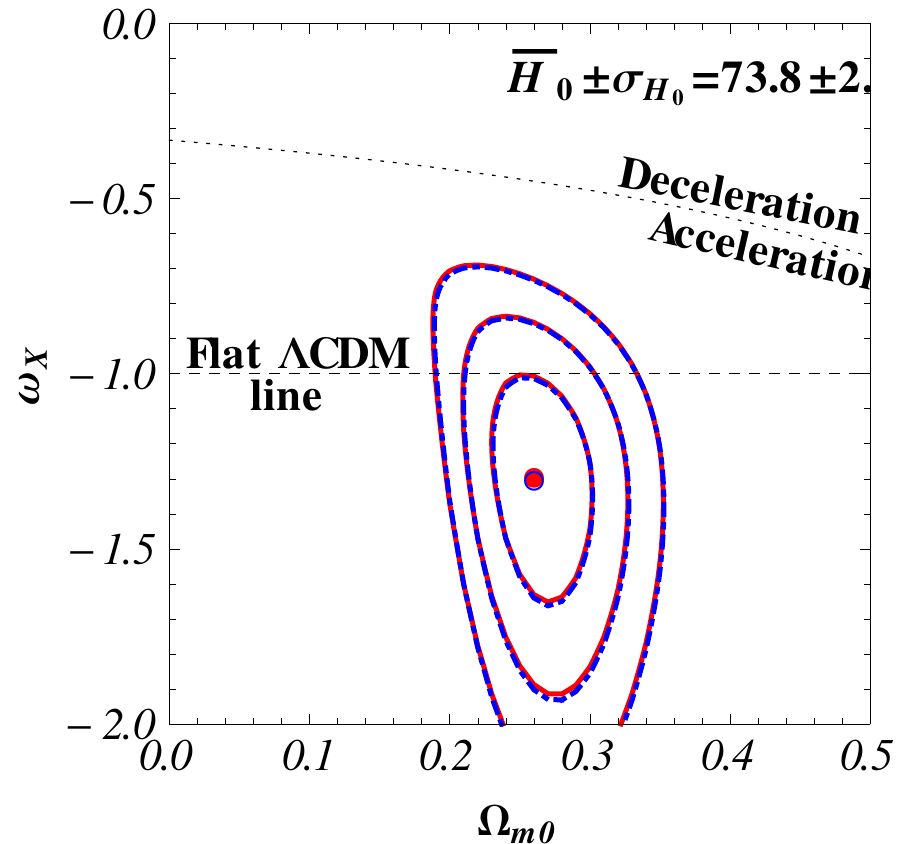}
    \includegraphics[height=1.45in]{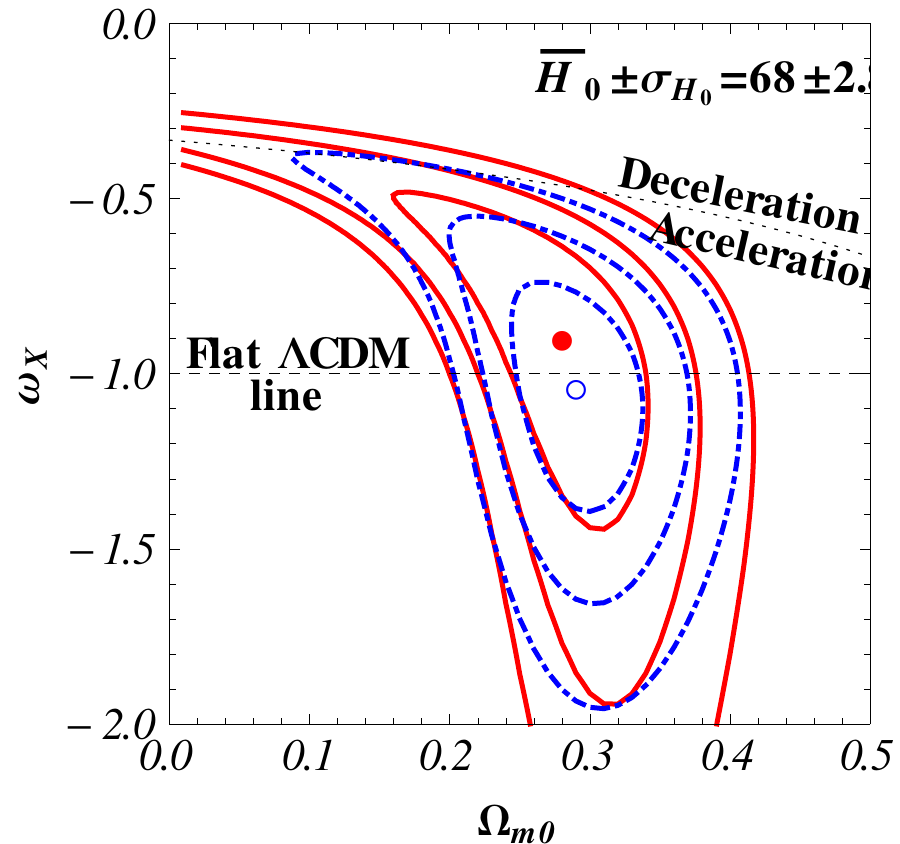}
    \includegraphics[height=1.5in]{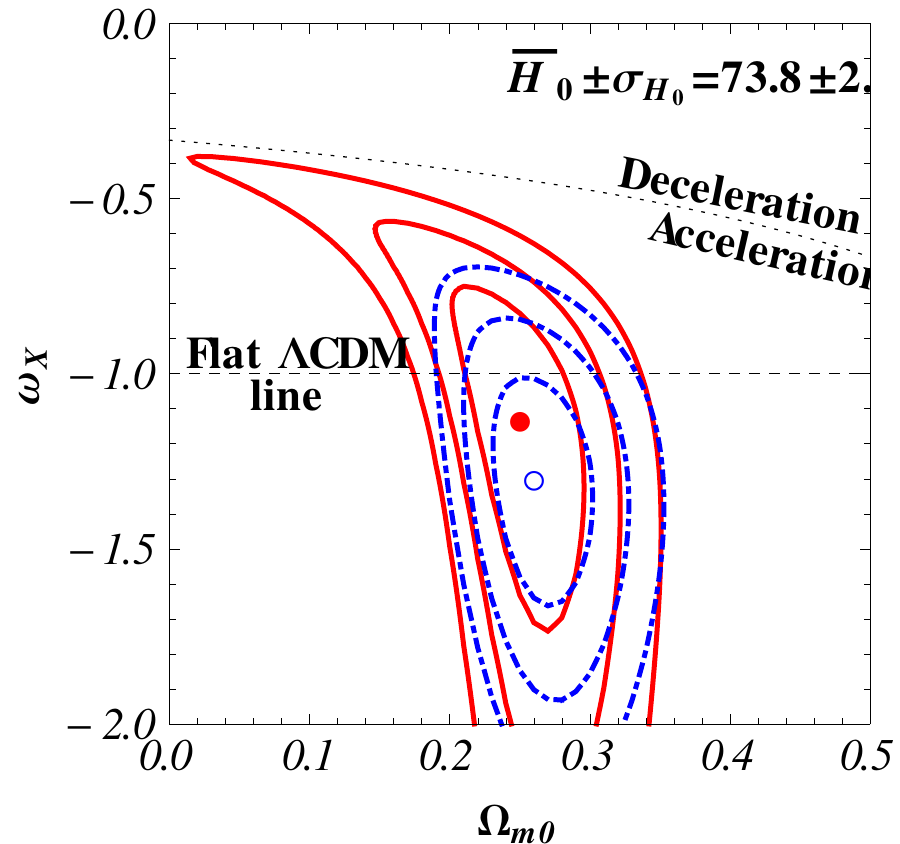}
    \includegraphics[height=1.55in]{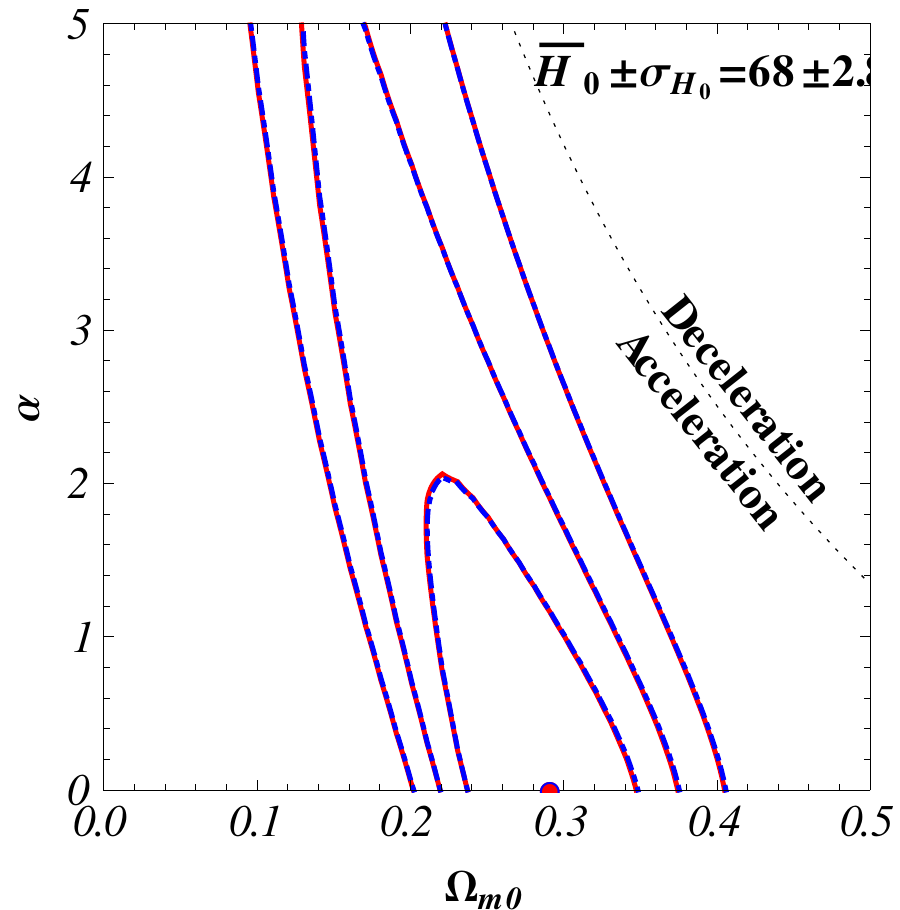}
    \includegraphics[height=1.55in]{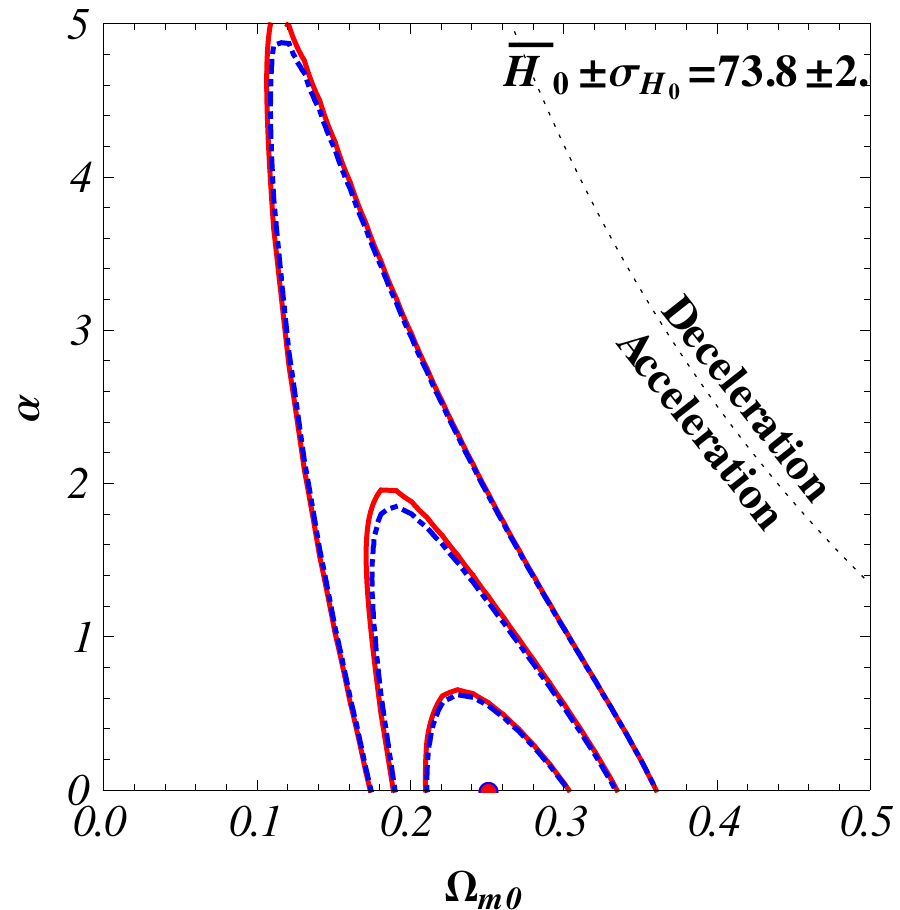}
    \includegraphics[height=1.55in]{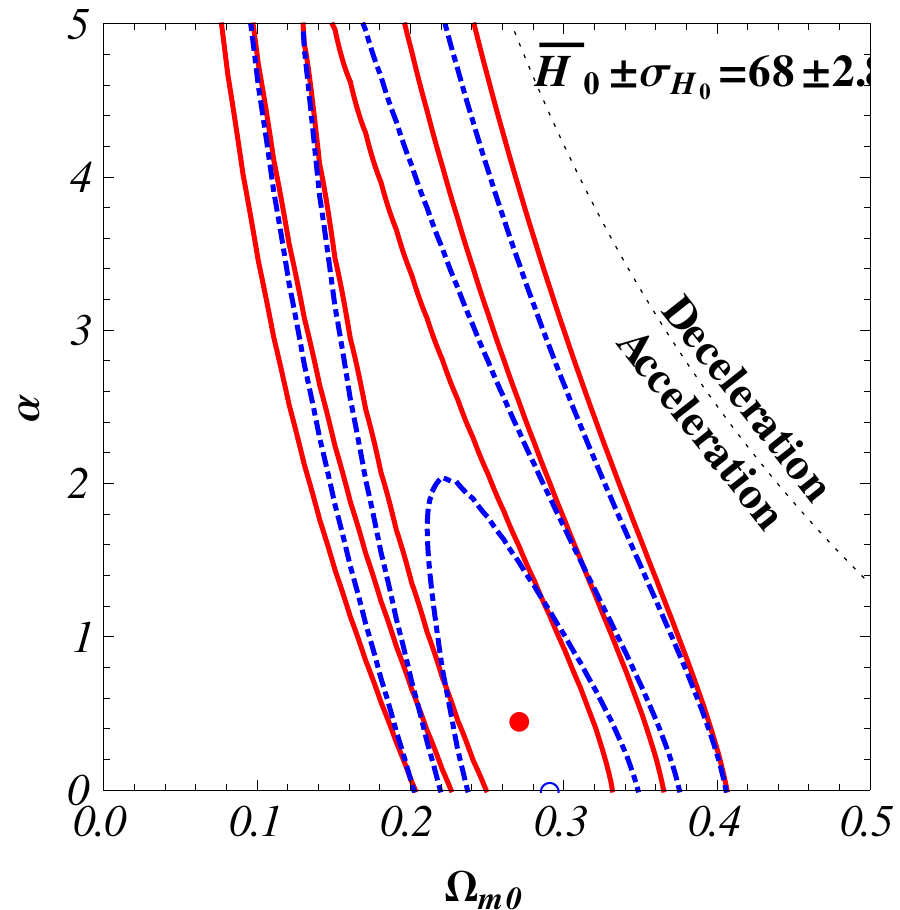}
    \includegraphics[height=1.55in]{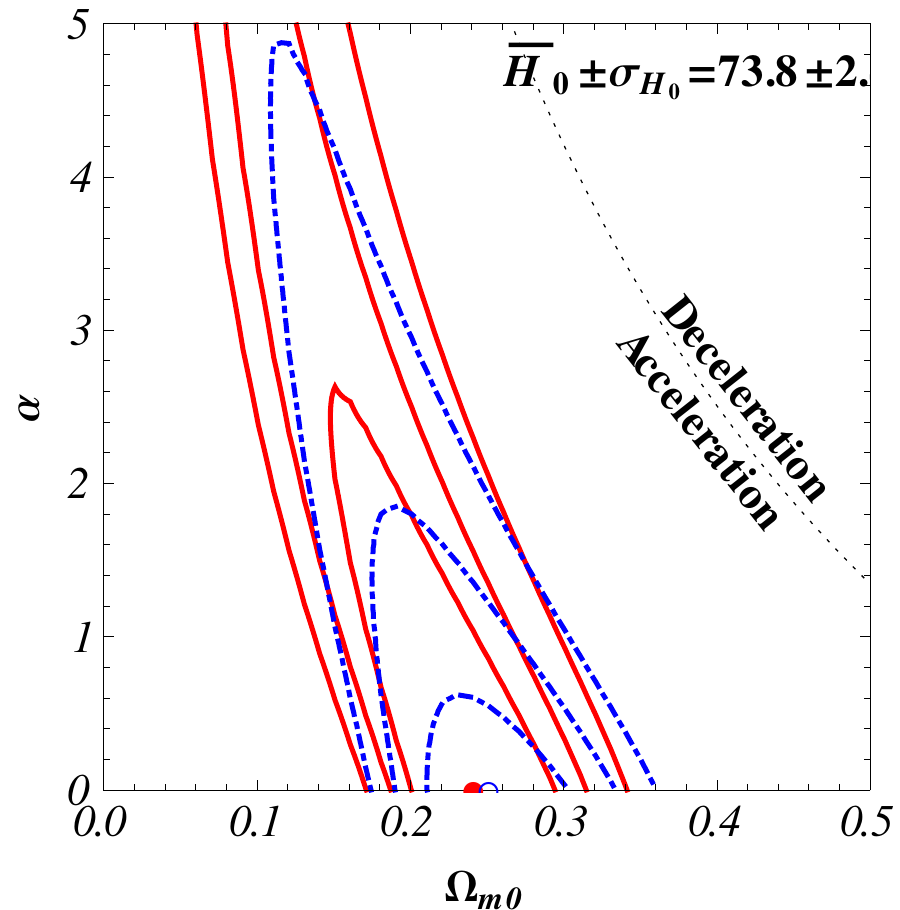}
 \caption{
Top left (right) panel shows the $H(z)/(1+z)$ data, binned with 3 or 4 
measurements per bin,  as well as 5 higher $z$ measurements, and the Farooq \& Ratra\cite{Farooq:2013hq} 
best-fit model predictions, dashed (dotted) for lower (higher) $H_0$ prior. 
The 2nd through 4th rows show the $H(z)$ constraints for $\Lambda$CDM, 
XCDM, and $\phi$CDM.
Red (blue dot-dashed) contours are 1$\sigma$, 2$\sigma$, and 3$\sigma$ confidence interval results from 3 or 4
measurements per bin (unbinned Table\ (\ref{tab:Hz3})) data. 
In these three rows, the first two plots 
include red weighted-mean constraints while the second two include red median statistics ones. The 
filled red (empty blue) circle is the corresponding best-fit point.
Dashed diagonal lines show spatially-flat models, and dotted
lines indicate zero-acceleration models. For quantitative details see Table (\ref{table:fig1 details}).
}
\label{fig:For table 2,3}
\end{figure}

\begin{figure}[h!]
    \includegraphics[height=2.1in]{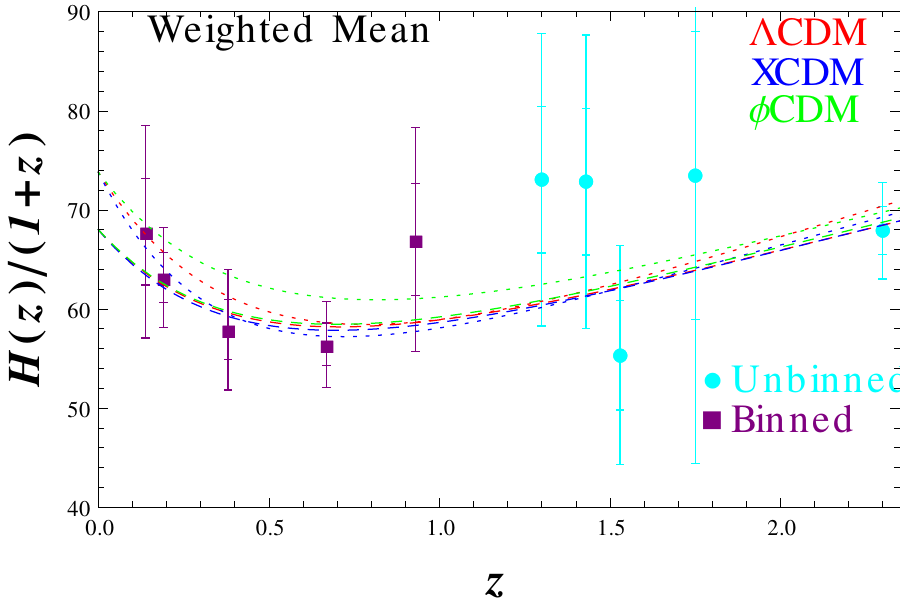}
    \includegraphics[height=2.1in]{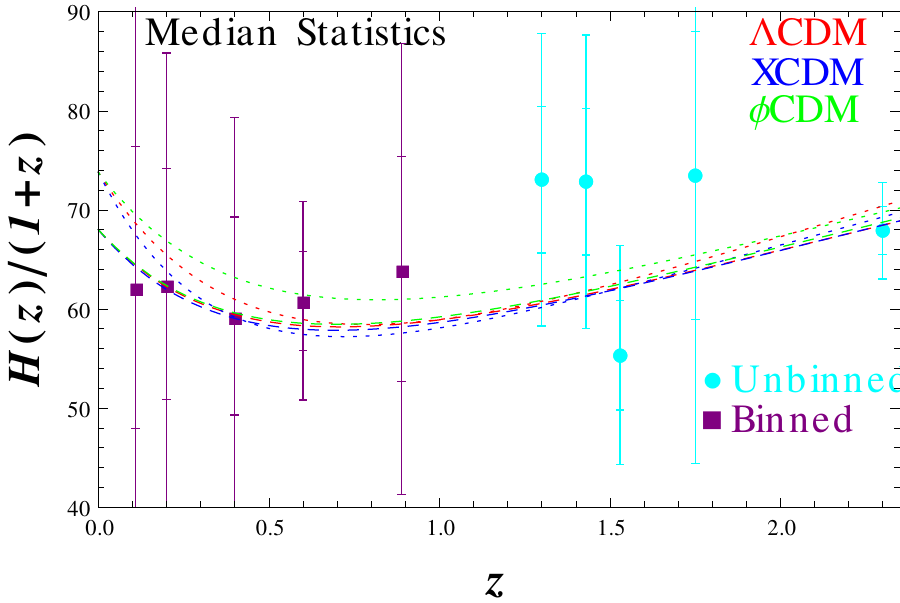}
    \includegraphics[height=1.5in]{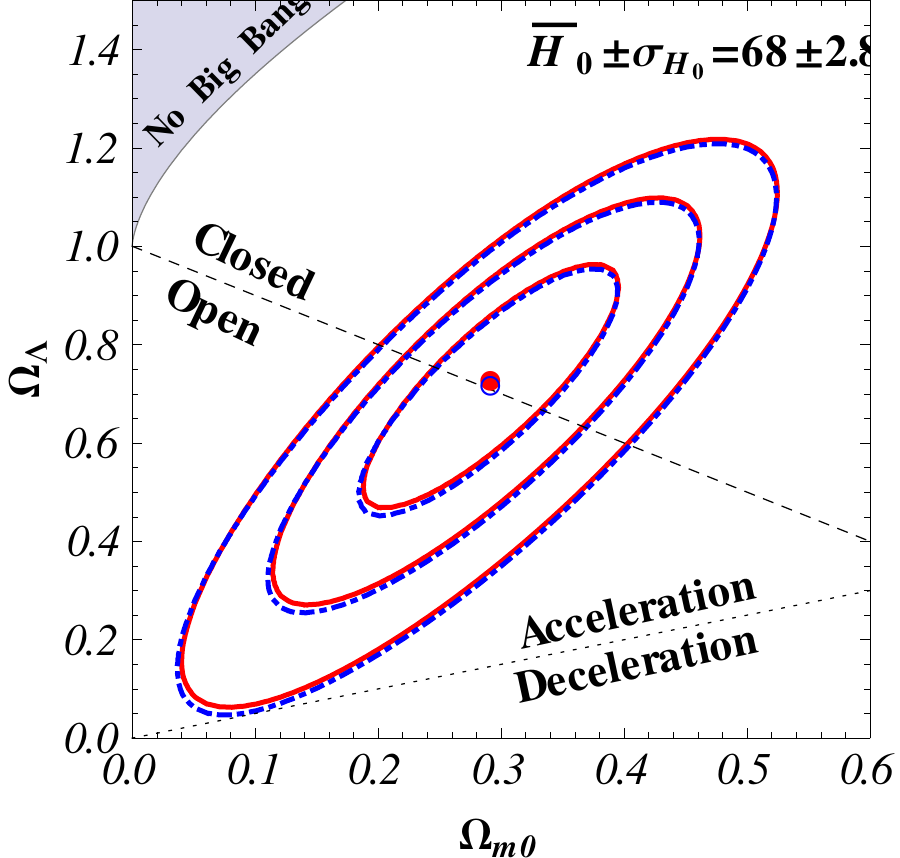}
    \includegraphics[height=1.5in]{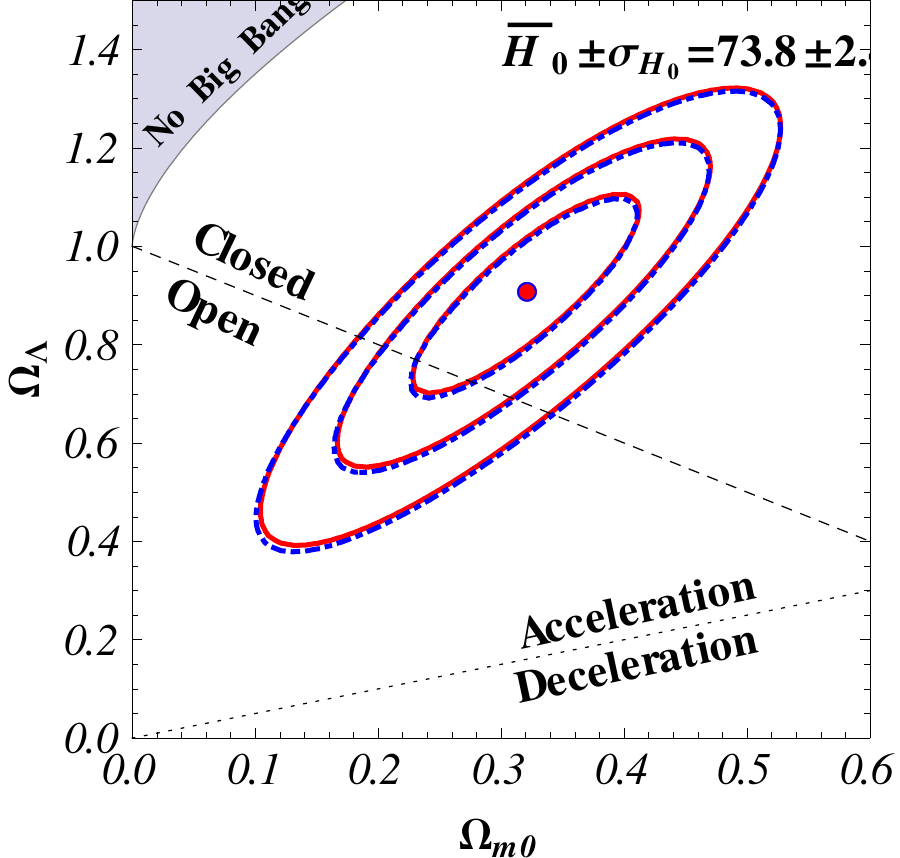}
    \includegraphics[height=1.5in]{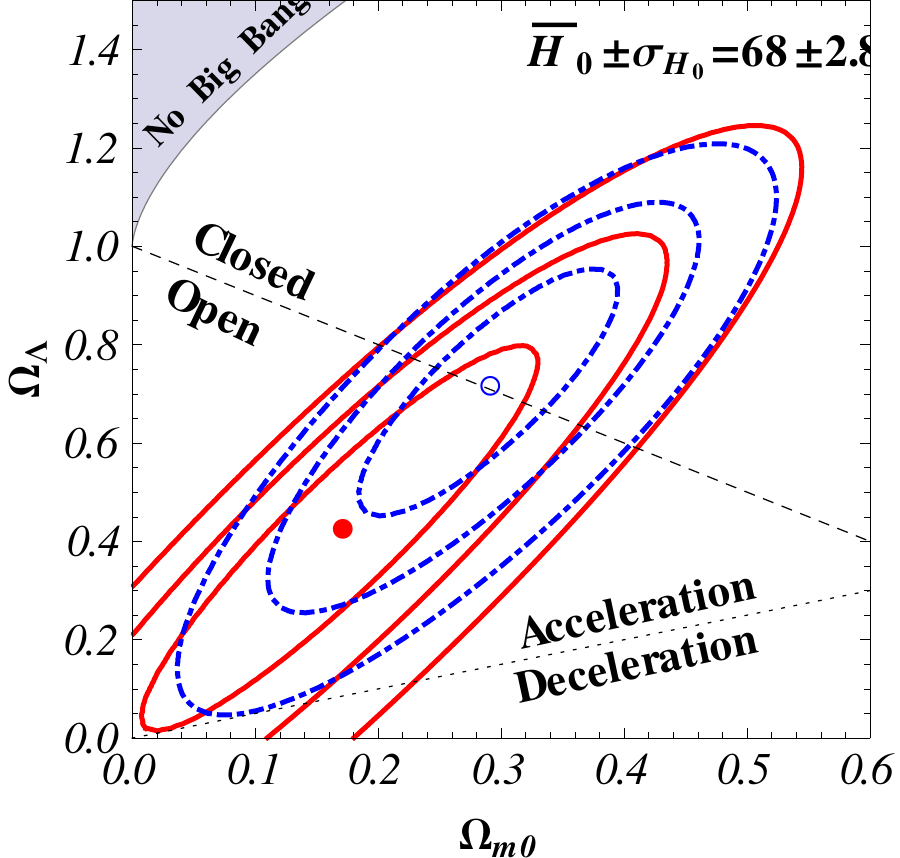}
    \includegraphics[height=1.5in]{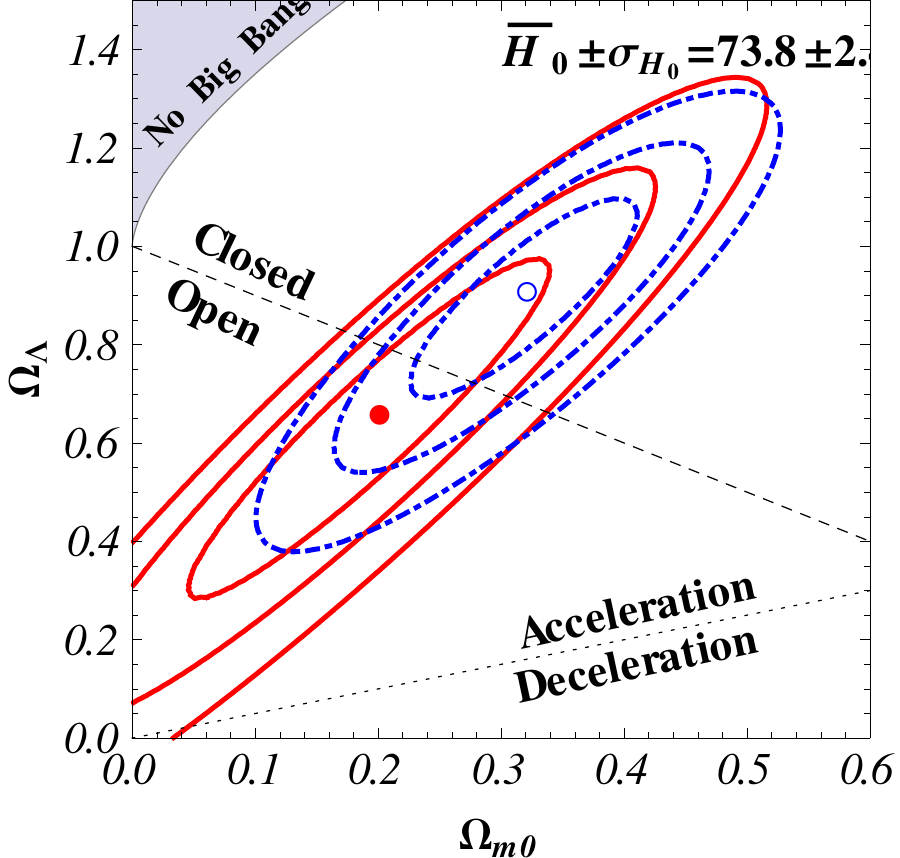}
    \includegraphics[height=1.45in]{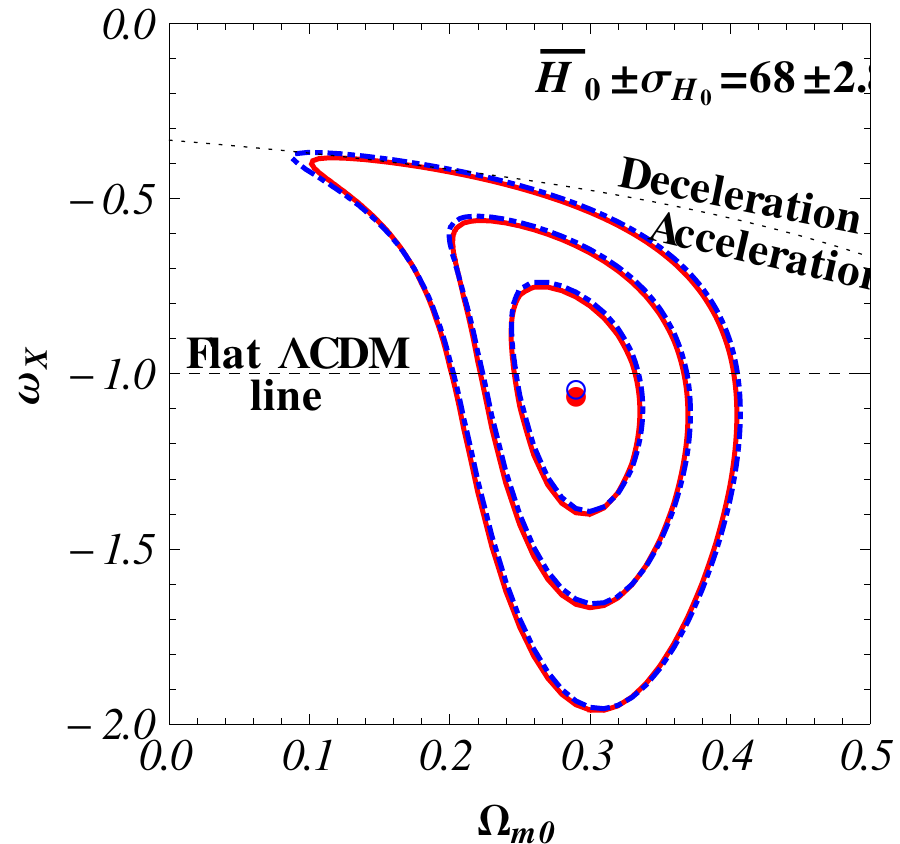}
    \includegraphics[height=1.45in]{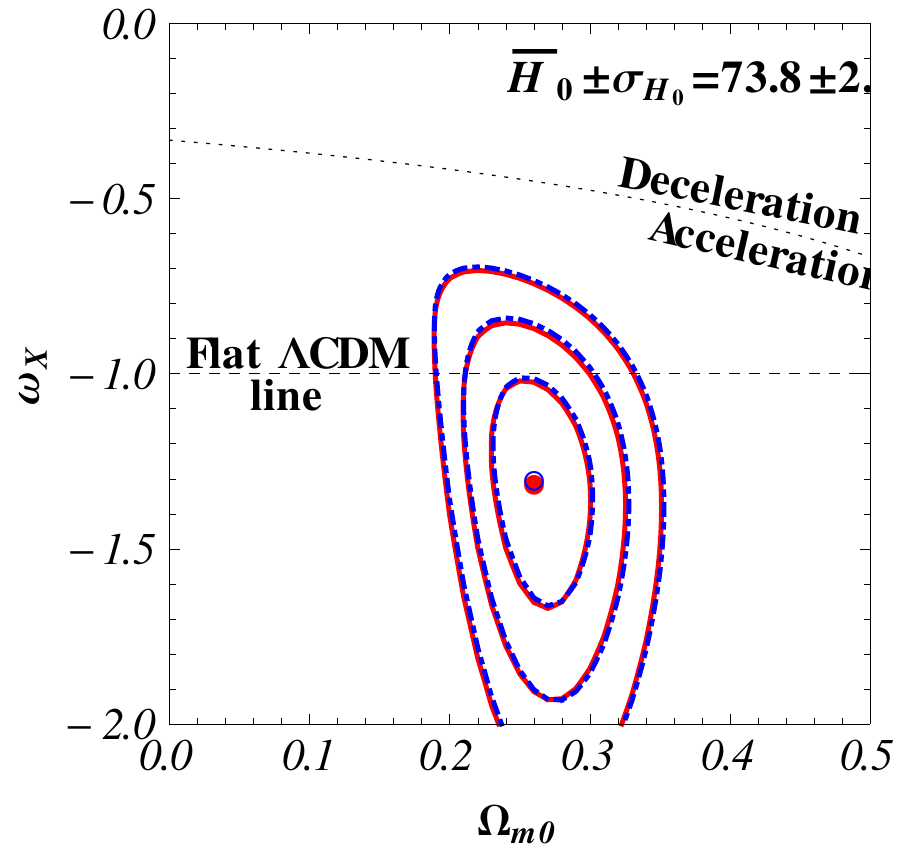}
    \includegraphics[height=1.45in]{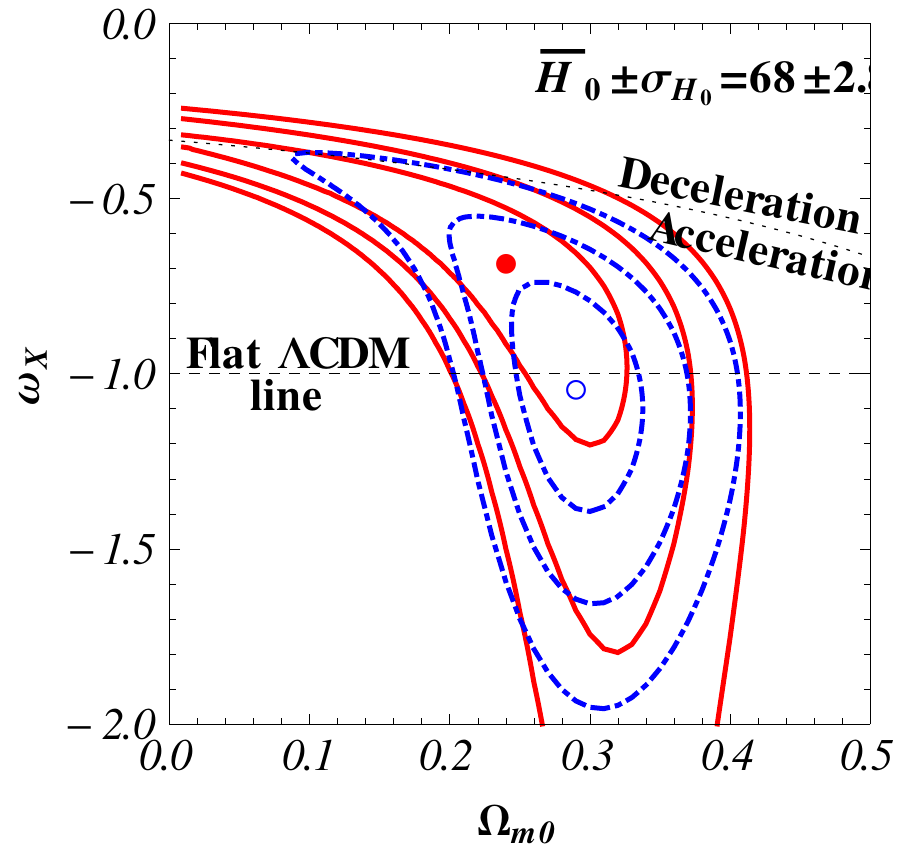}
    \includegraphics[height=1.5in]{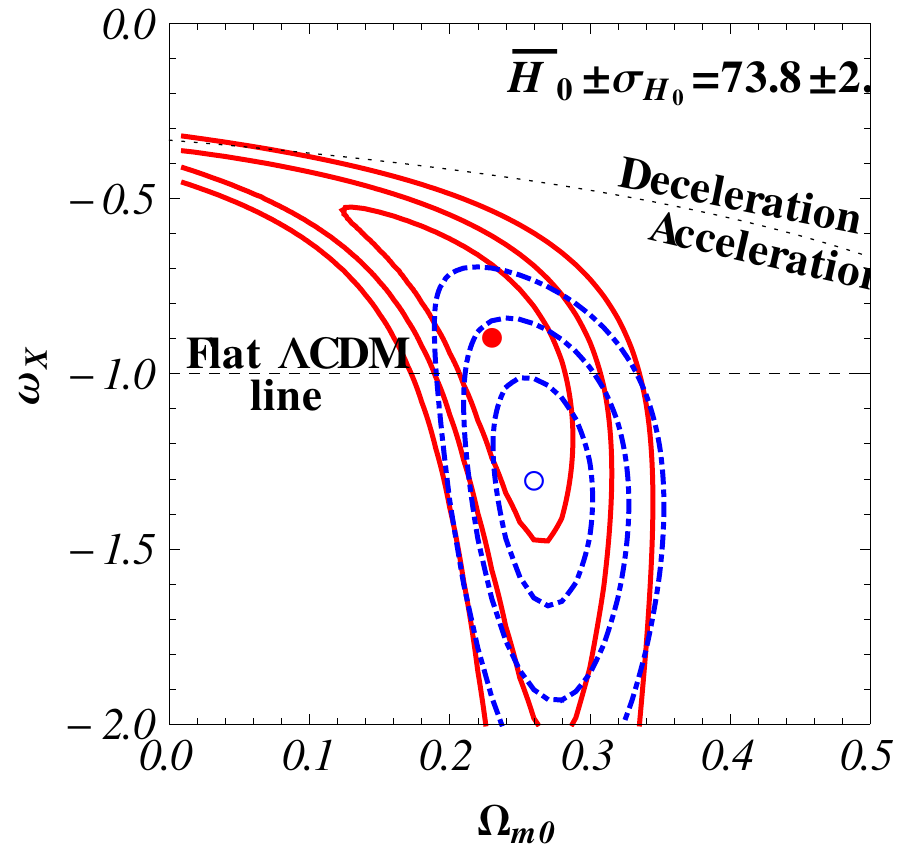}
    \includegraphics[height=1.55in]{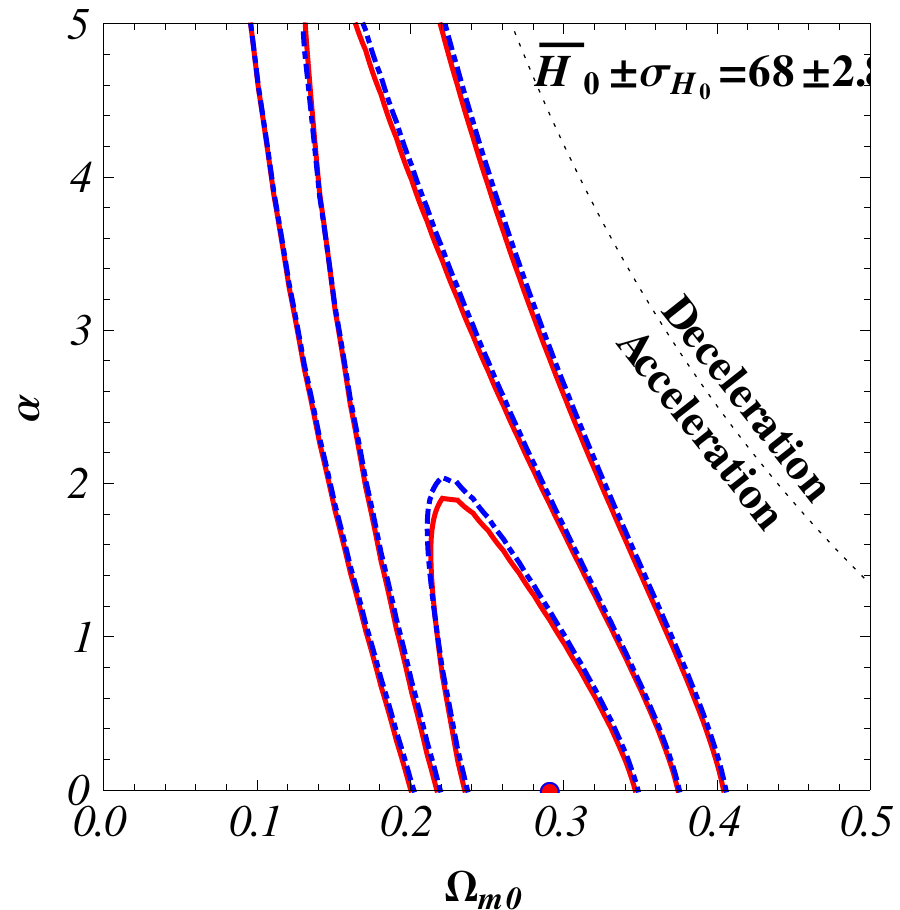}
    \includegraphics[height=1.55in]{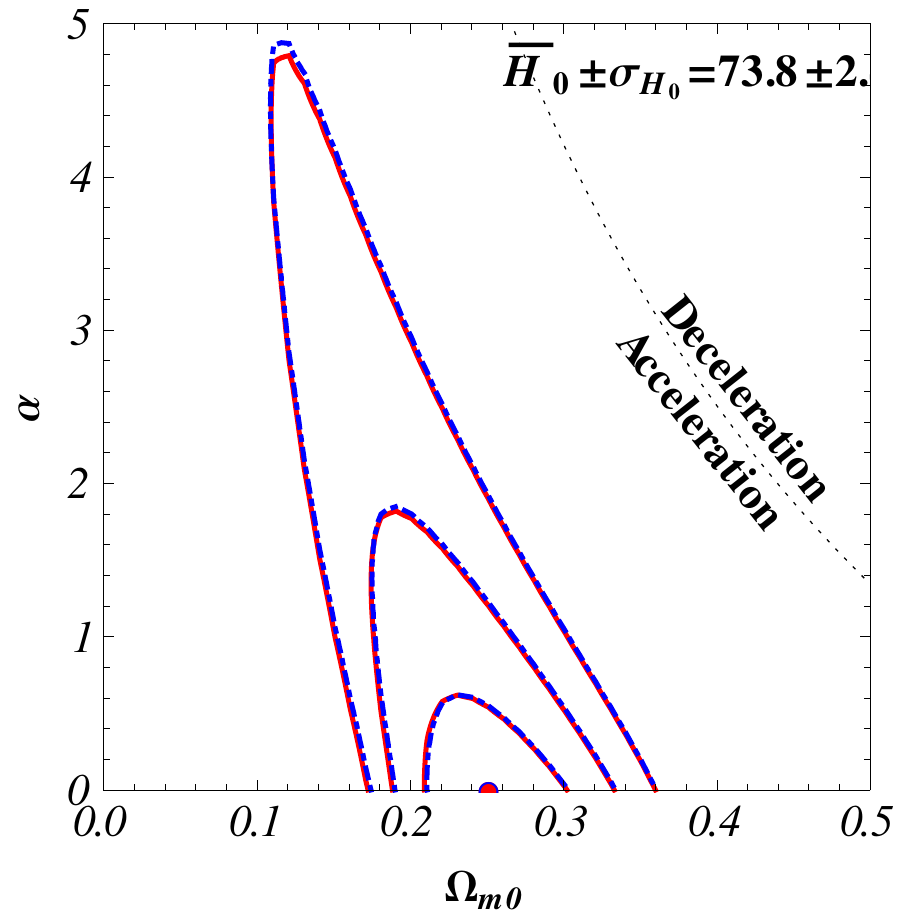}
    \includegraphics[height=1.55in]{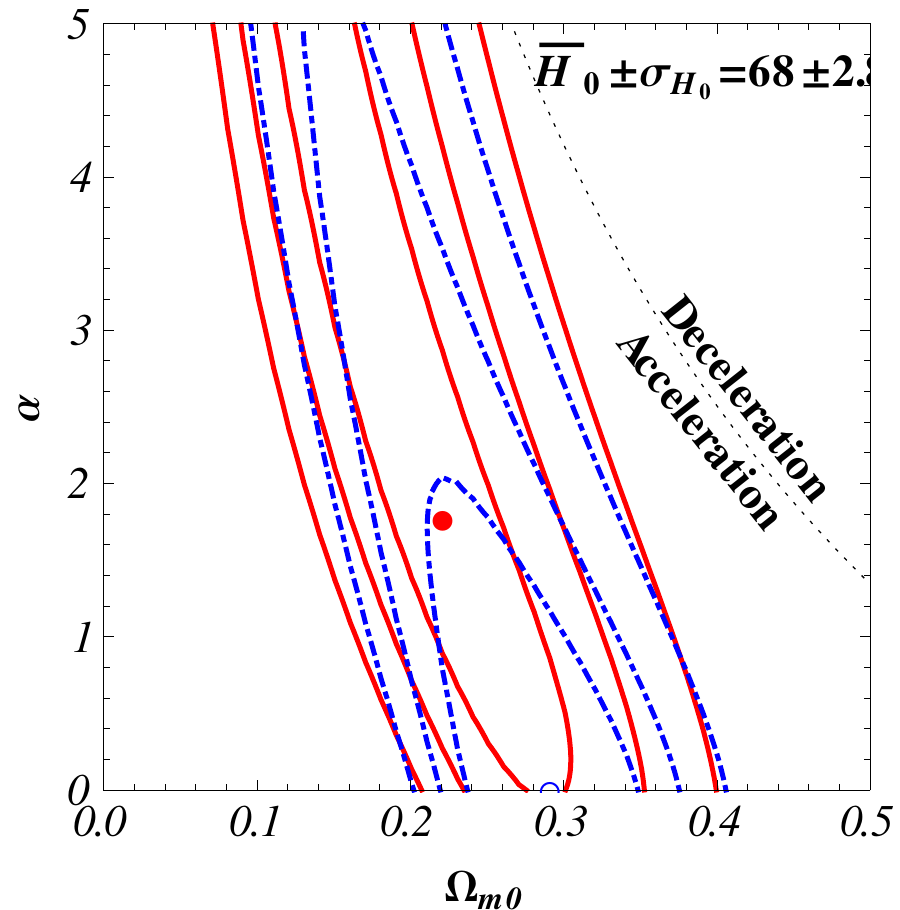}
    \includegraphics[height=1.55in]{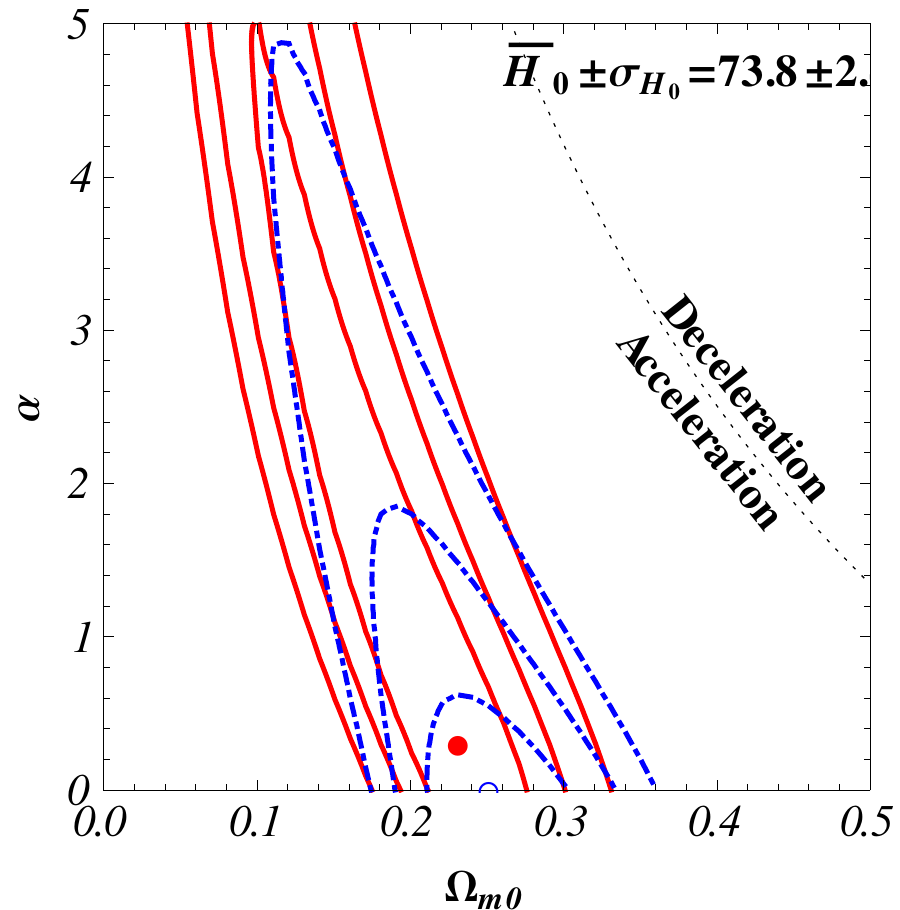}
 \caption{
Top left (right) panel shows the $H(z)/(1+z)$ data, binned with 4 or 5 
measurements per bin,  as well as 5 higher $z$ measurements, and the Farooq \& Ratra\cite{Farooq:2013hq} 
best-fit model predictions, dashed (dotted) for lower (higher) $H_0$ prior. 
The 2nd through 4th rows show the $H(z)$ constraints for $\Lambda$CDM, 
XCDM, and $\phi$CDM.
Red (blue dot-dashed) contours are 1$\sigma$, 2$\sigma$, and 3$\sigma$ confidence interval results from 4 or 5
measurements per bin (unbinned Farooq \& Ratra,\cite{Farooq:2013hq} Table\ 1) data. 
In these three rows, the first two plots 
include red weighted-mean constraints while the second two include red median statistics ones. The 
filled red (empty blue) circle is the corresponding best-fit point.
Dashed diagonal lines show spatially-flat models, and dotted
lines indicate zero-acceleration models. For quantitative details see Table (\ref{table:fig2 details}).
}
\label{fig:For table 4,5}
\end{figure}

\begin{figure}[h!]
    \includegraphics[height=2.1in]{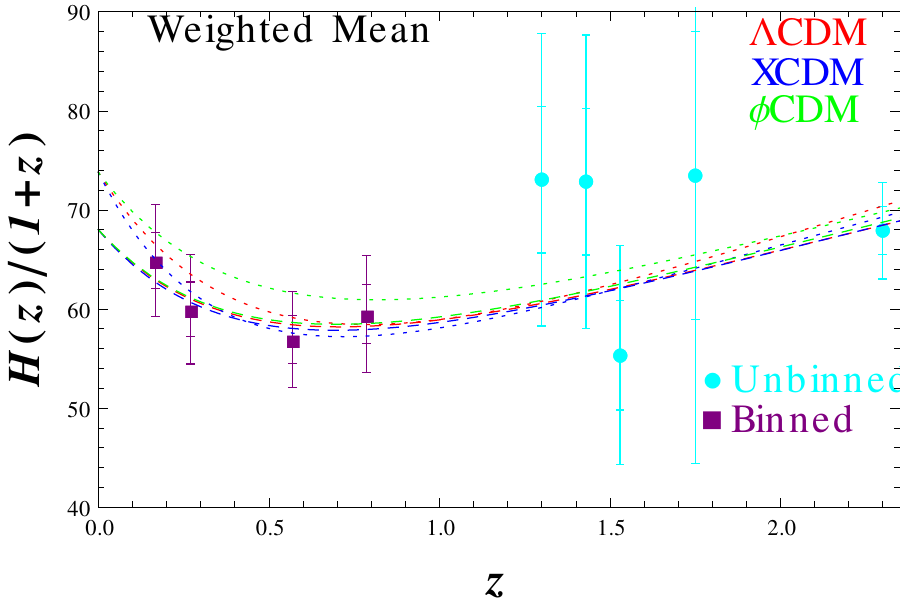}
    \includegraphics[height=2.1in]{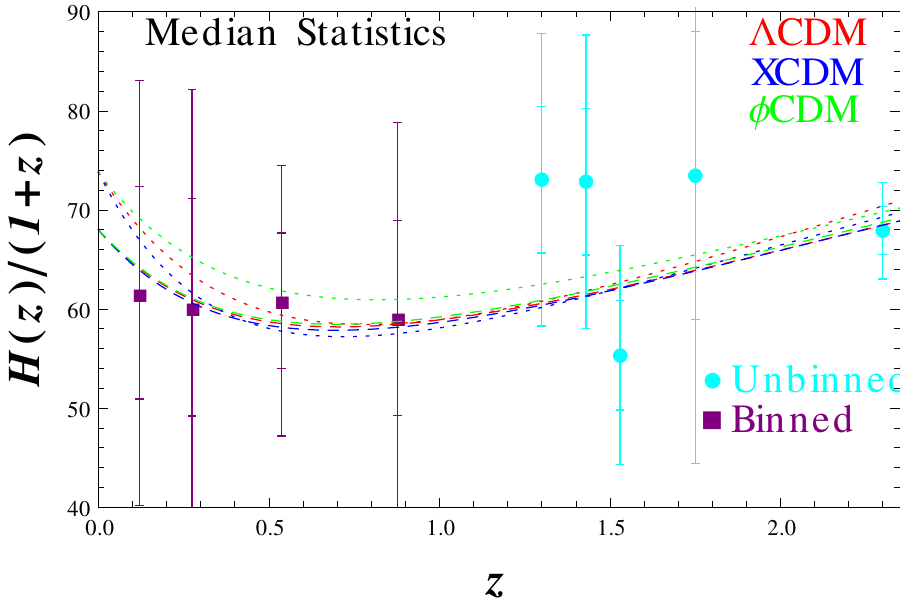}
    \includegraphics[height=1.5in]{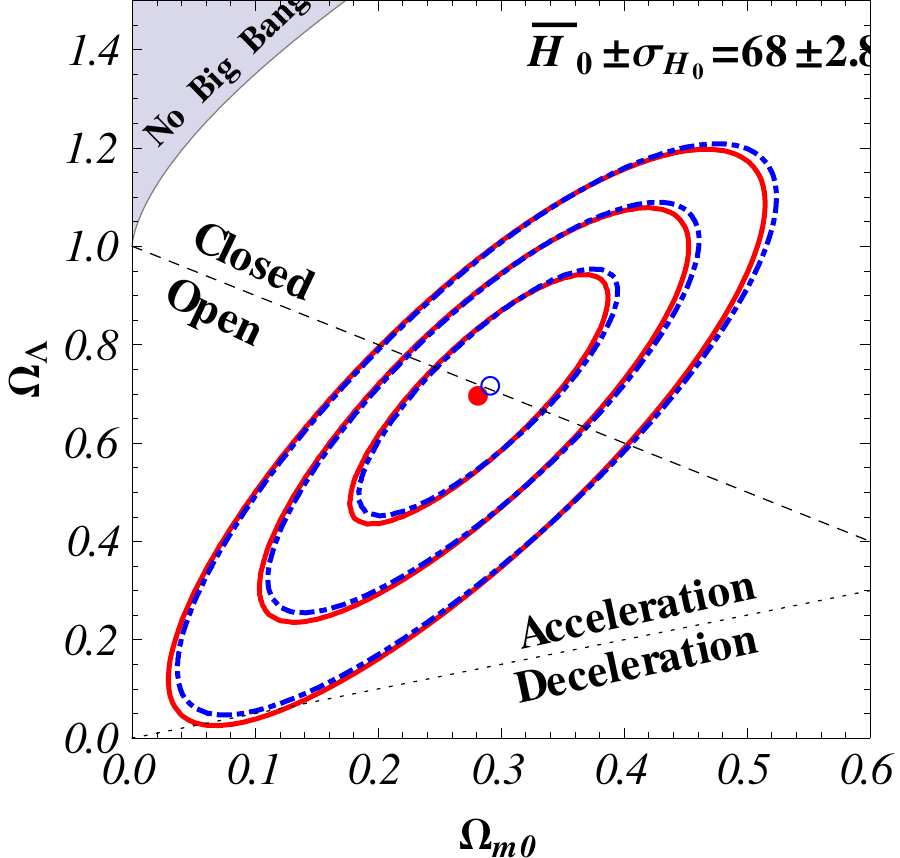}
    \includegraphics[height=1.5in]{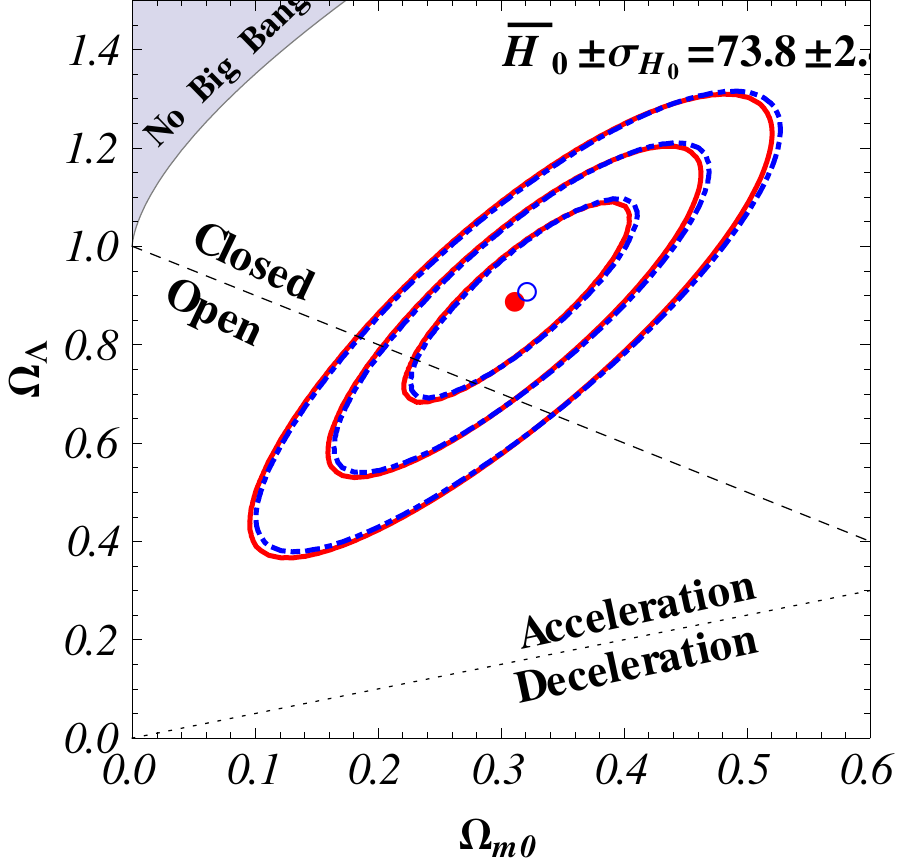}
    \includegraphics[height=1.5in]{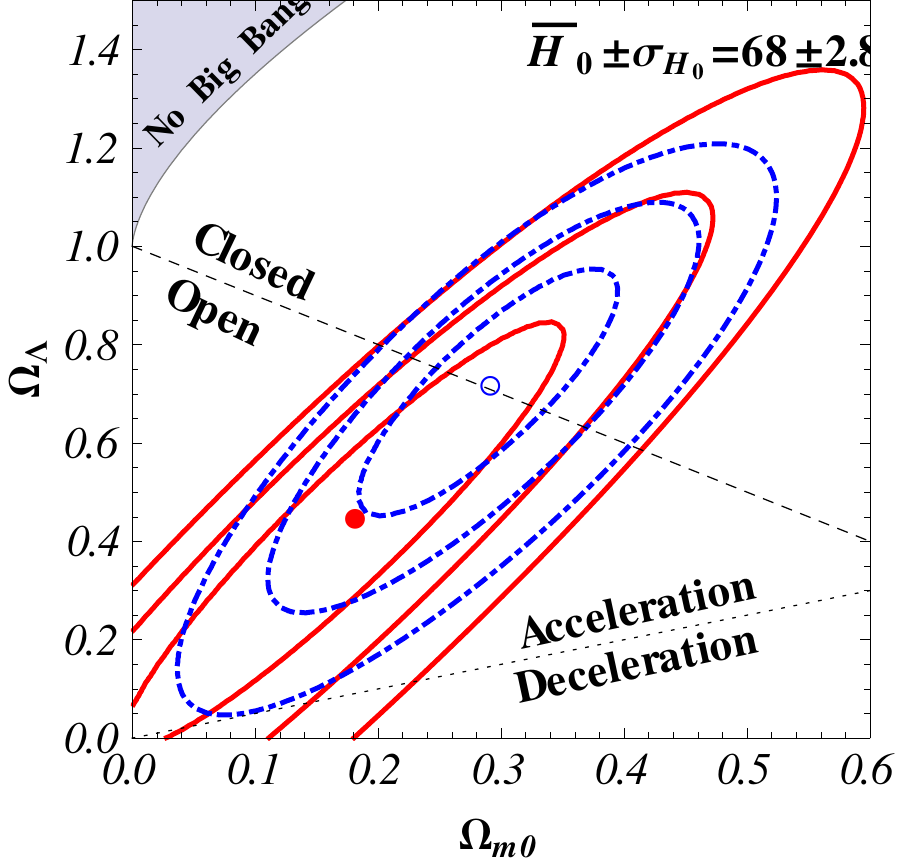}
    \includegraphics[height=1.5in]{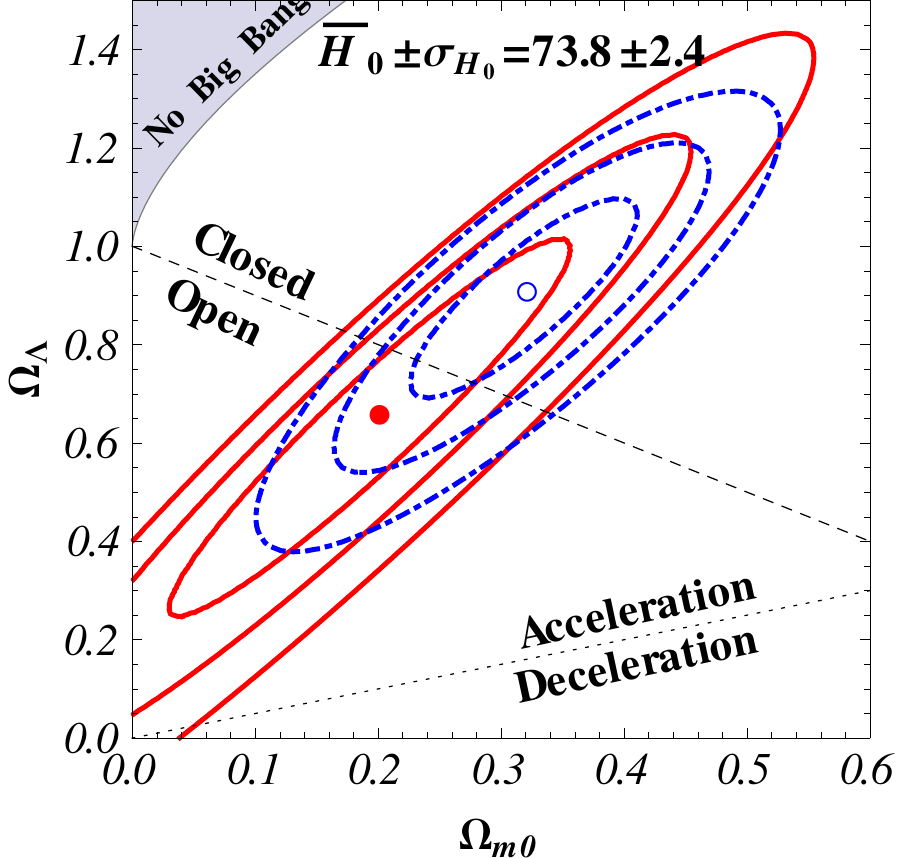}
    \includegraphics[height=1.45in]{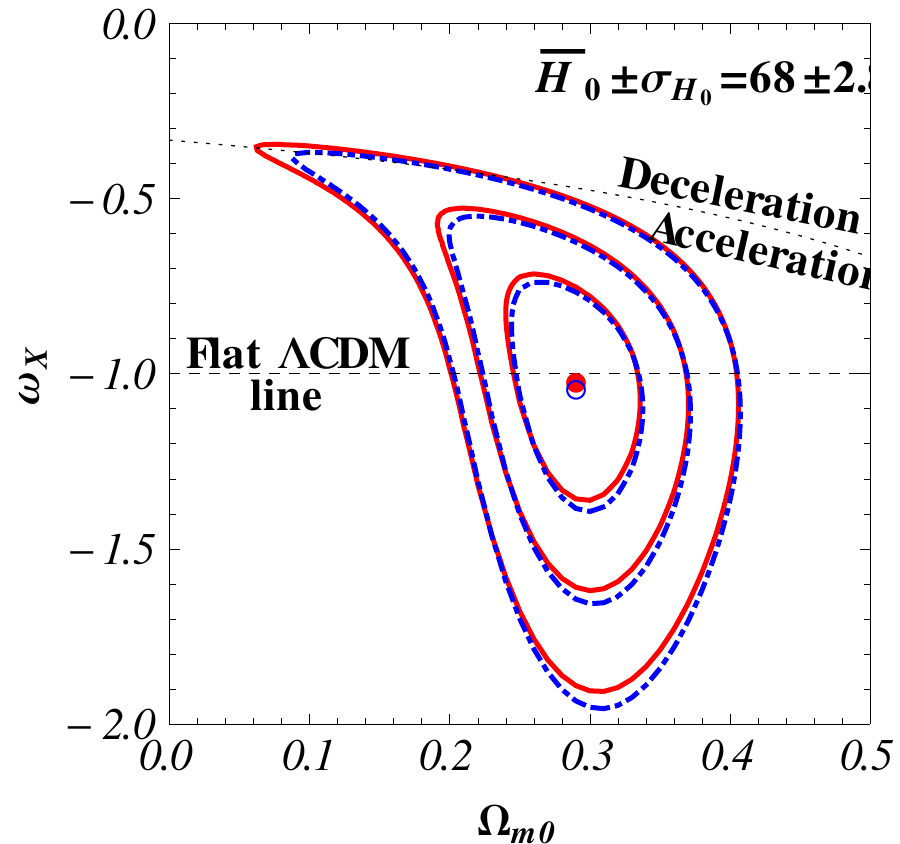}
    \includegraphics[height=1.45in]{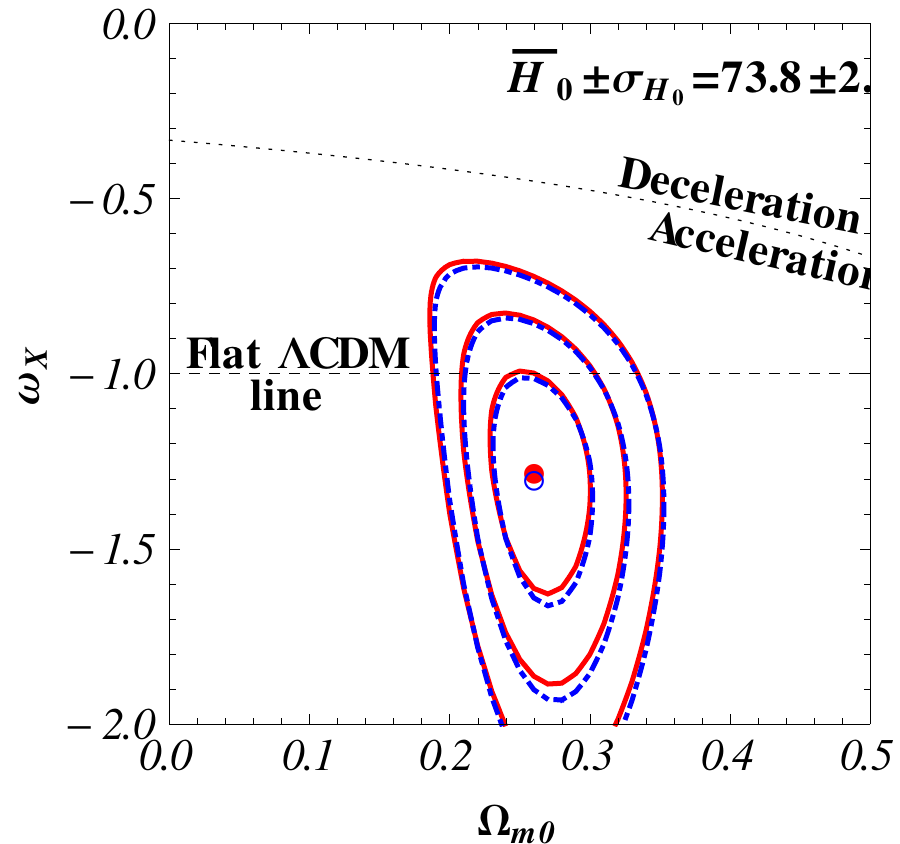}
    \includegraphics[height=1.45in]{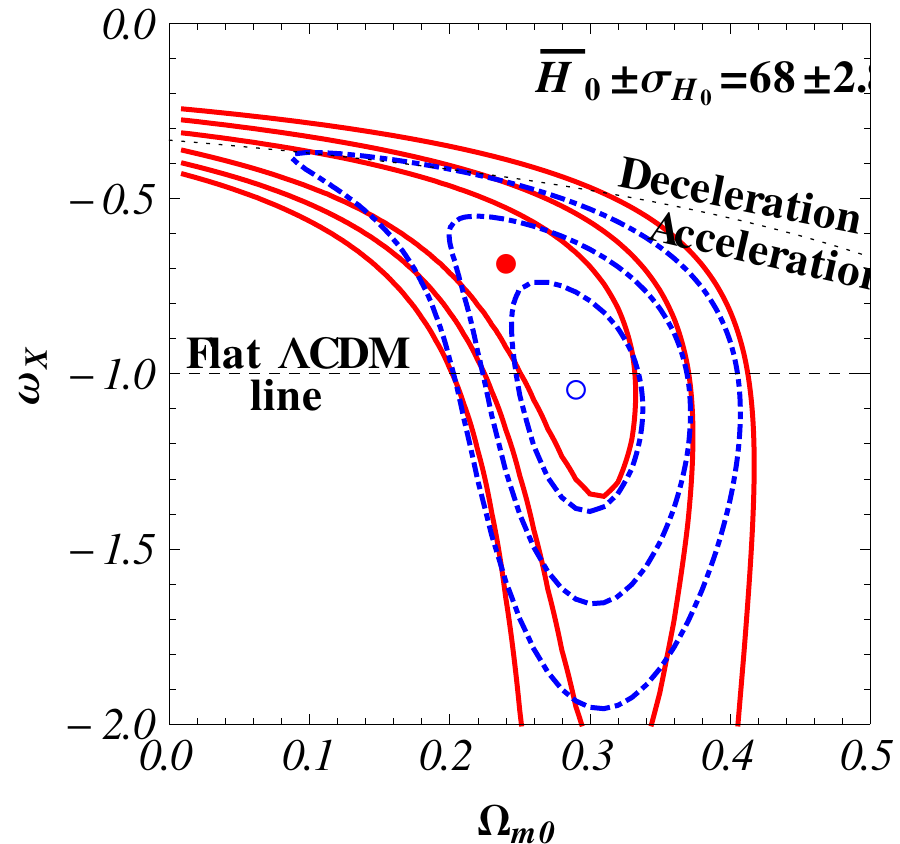}
    \includegraphics[height=1.5in]{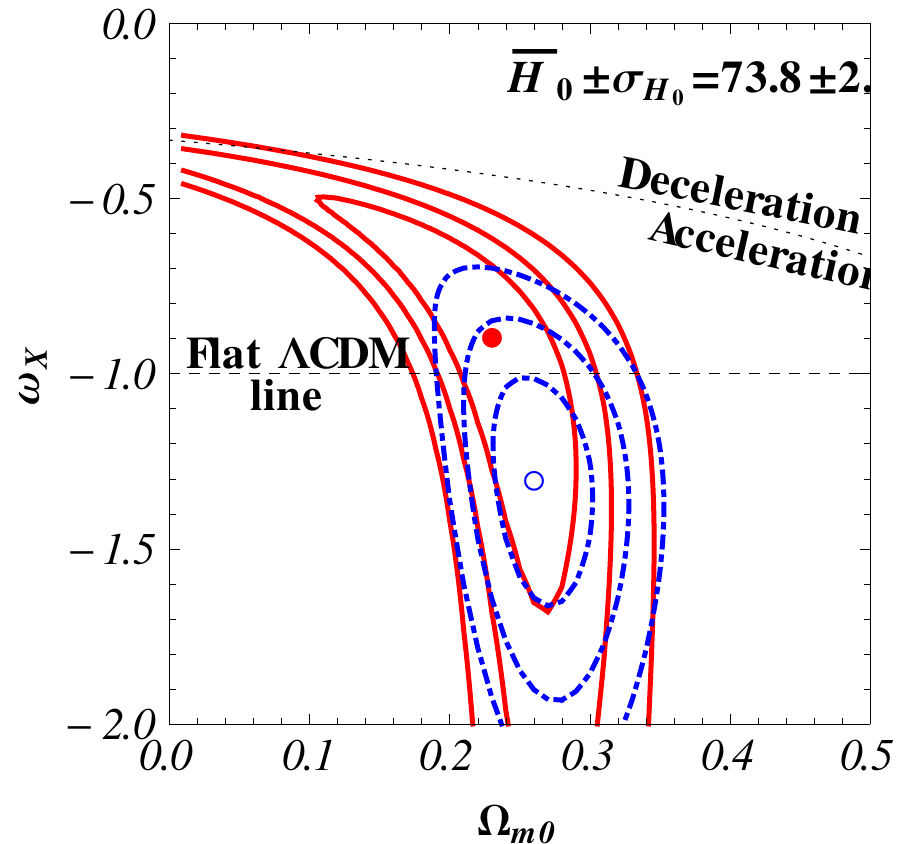}
    \includegraphics[height=1.55in]{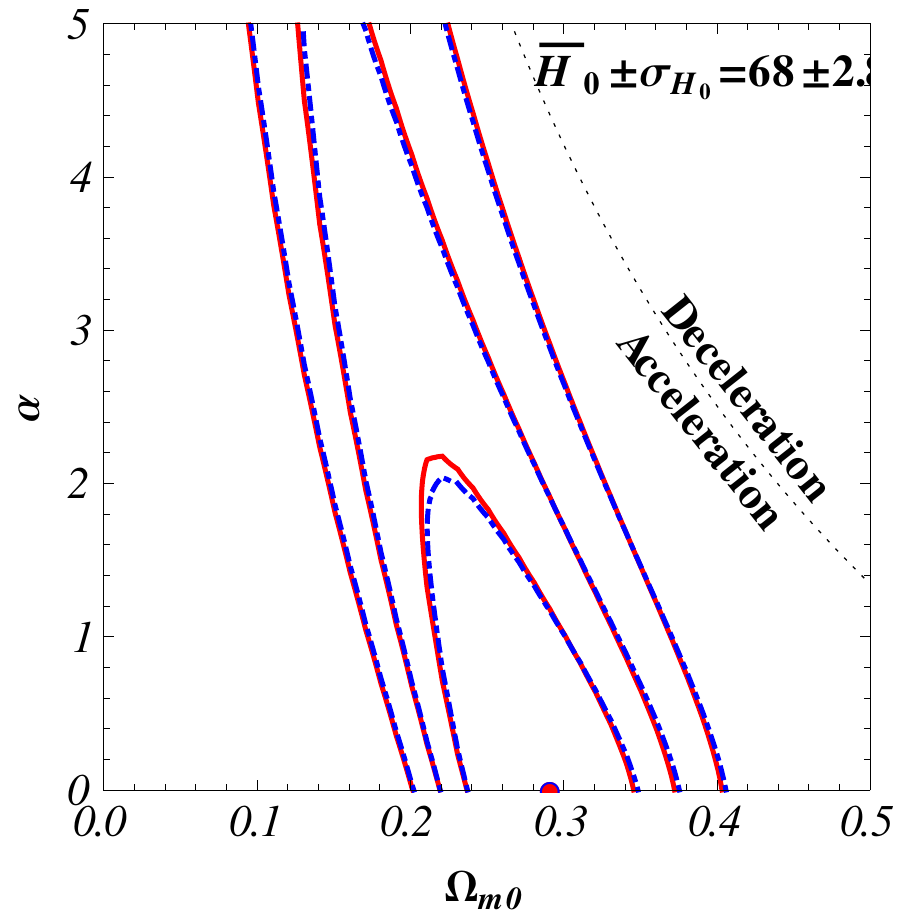}
    \includegraphics[height=1.55in]{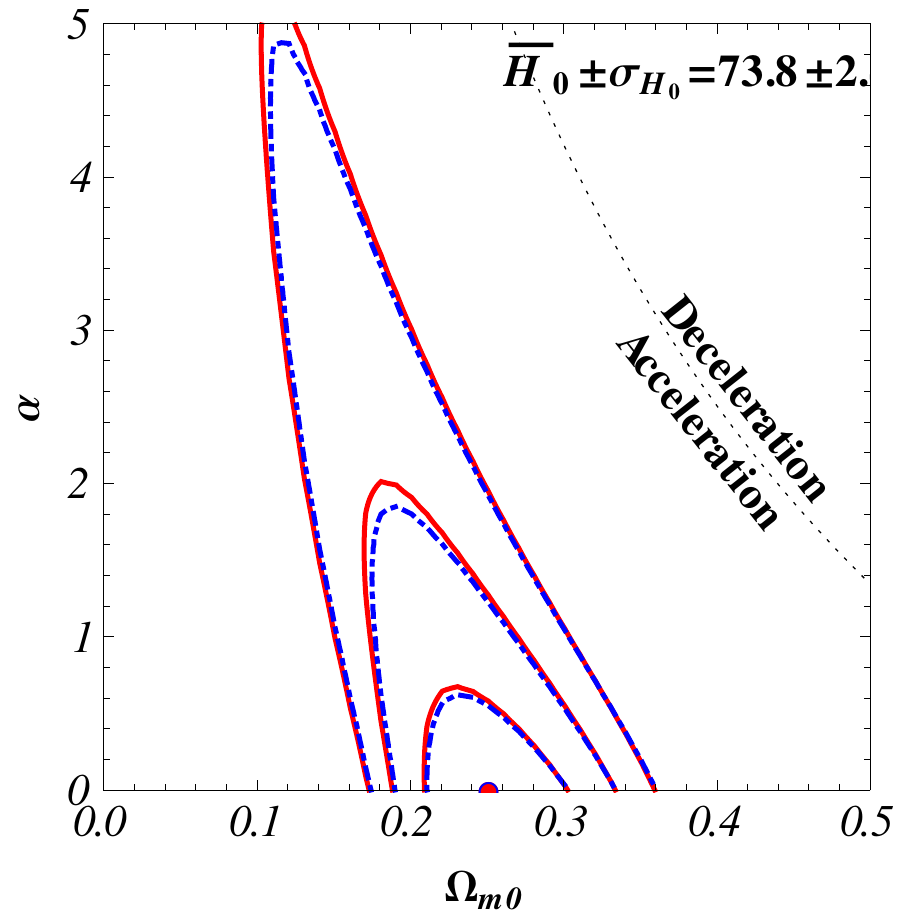}
    \includegraphics[height=1.55in]{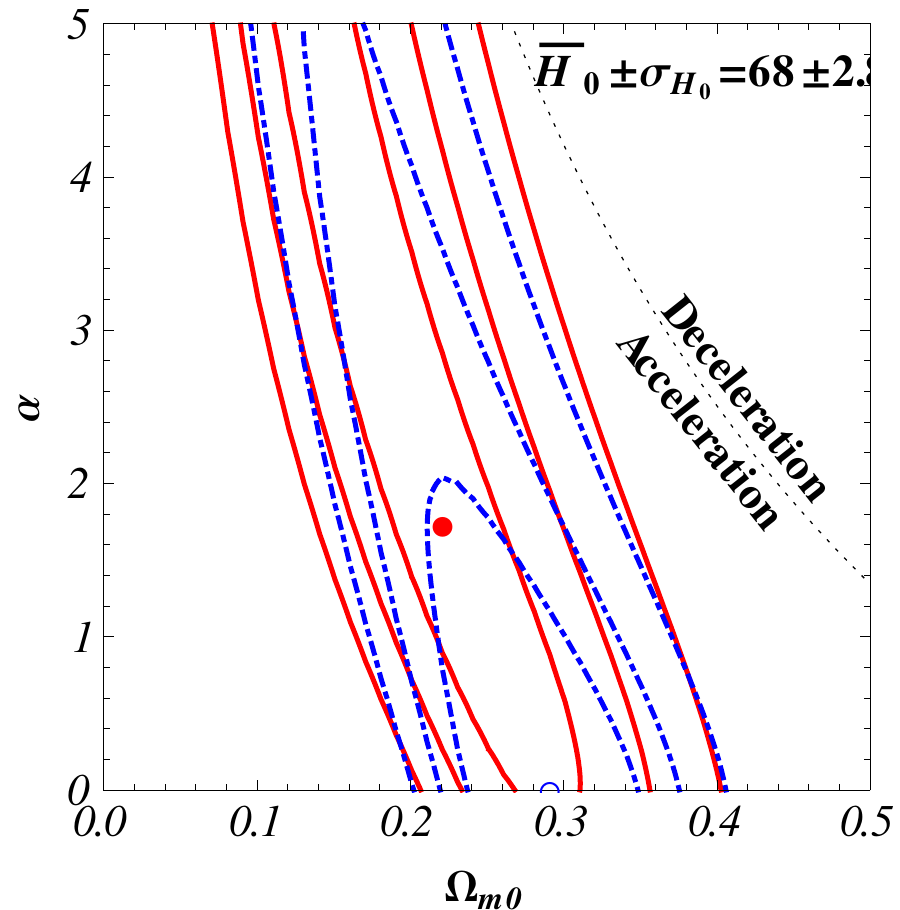}
    \includegraphics[height=1.55in]{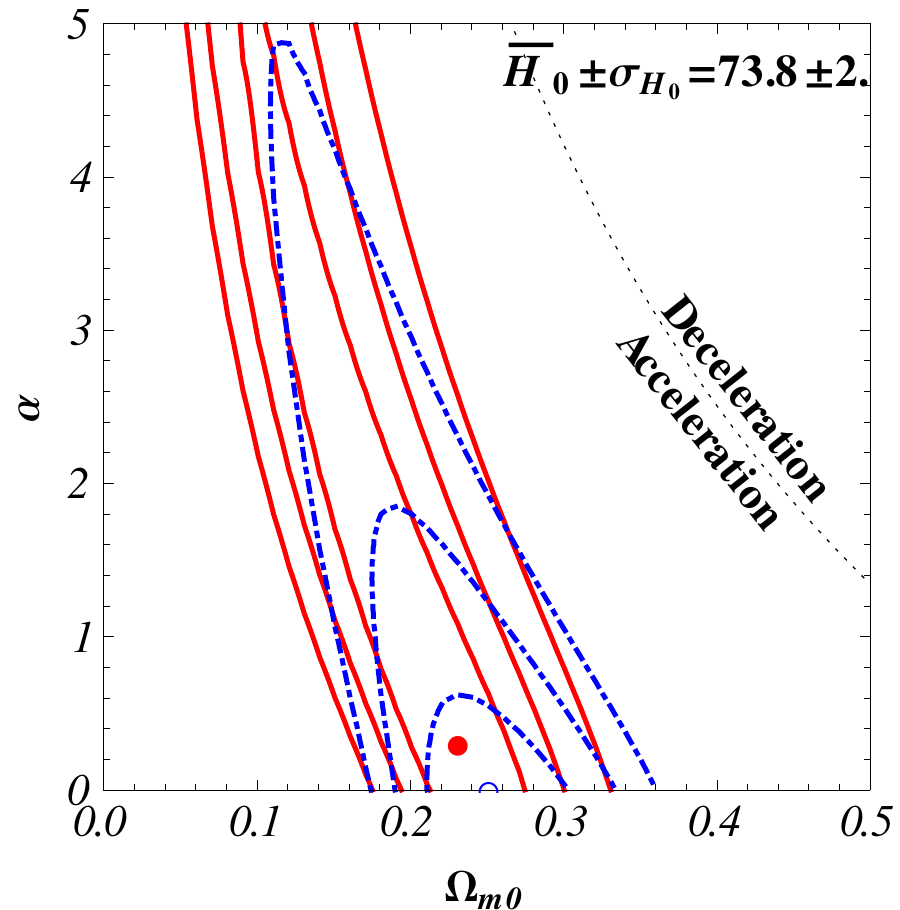}
 \caption{
Top left (right) panel shows the $H(z)/(1+z)$ data, binned with 5 or 6 
measurements per bin,  as well as 5 higher $z$ measurements, and the Farooq \& Ratra\cite{Farooq:2013hq}
best-fit model predictions, dashed (dotted) for lower (higher) $H_0$ prior. 
The 2nd through 4th rows show the $H(z)$ constraints for $\Lambda$CDM, 
XCDM, and $\phi$CDM.
Red (blue dot-dashed) contours are 1$\sigma$, 2$\sigma$, and 3$\sigma$ confidence interval results from 5 or 6
measurements per bin (unbinned Farooq \& Ratra,\cite{Farooq:2013hq} Table\ 1) data. 
In these three rows, the first two plots 
include red weighted-mean constraints while the second two include red median statistics ones. The 
filled red (empty blue) circle is the corresponding best-fit point.
Dashed diagonal lines show spatially-flat models, and dotted
lines indicate zero-acceleration models. For quantitative details see Table (\ref{table:fig3 details}).
}
\label{fig:For table 6,7}
\end{figure}

\begin{figure}[h!]
    \includegraphics[height=2.1in]{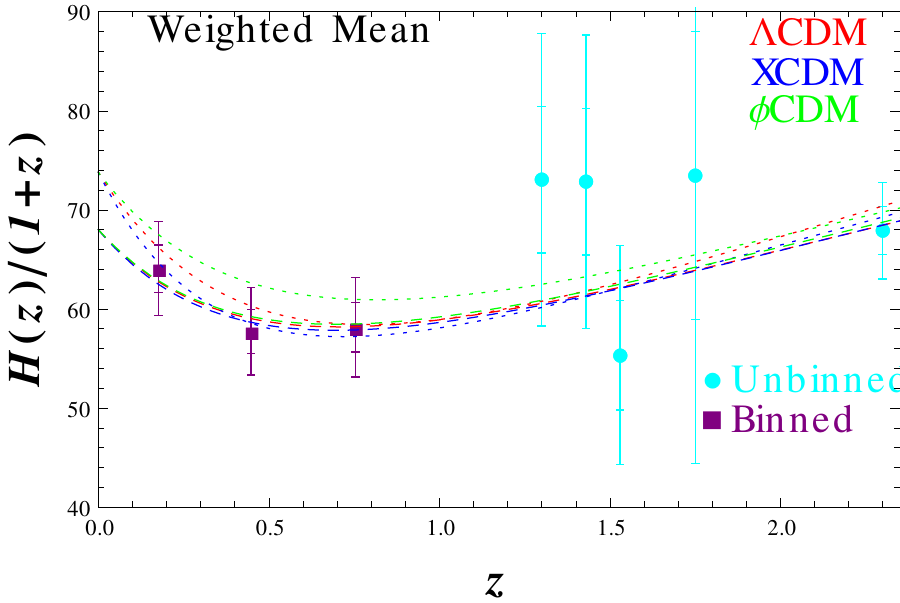}
    \includegraphics[height=2.1in]{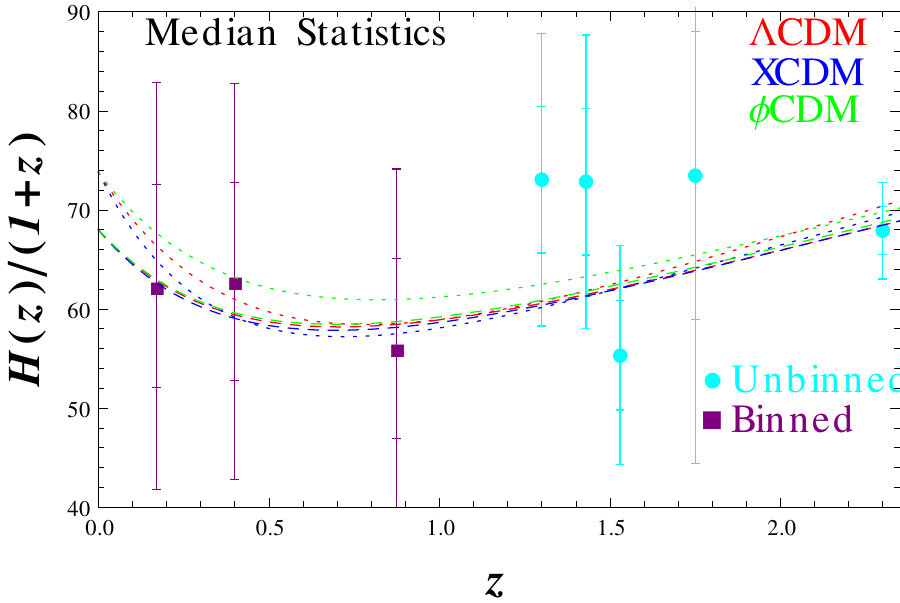}
    \includegraphics[height=1.5in]{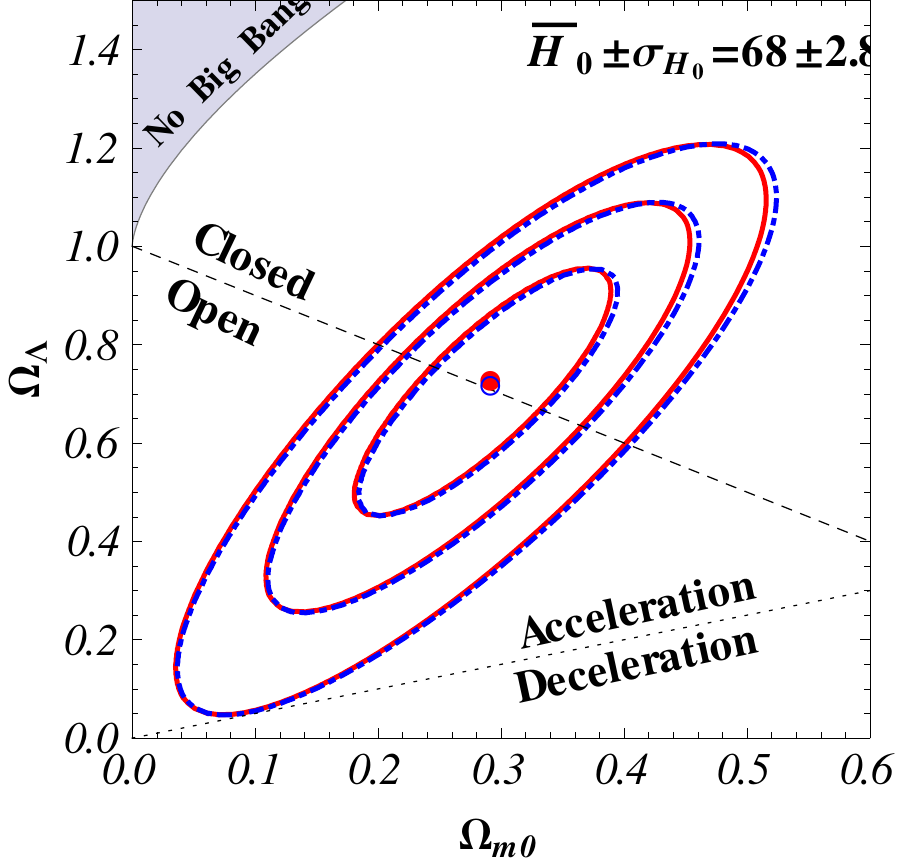}
    \includegraphics[height=1.5in]{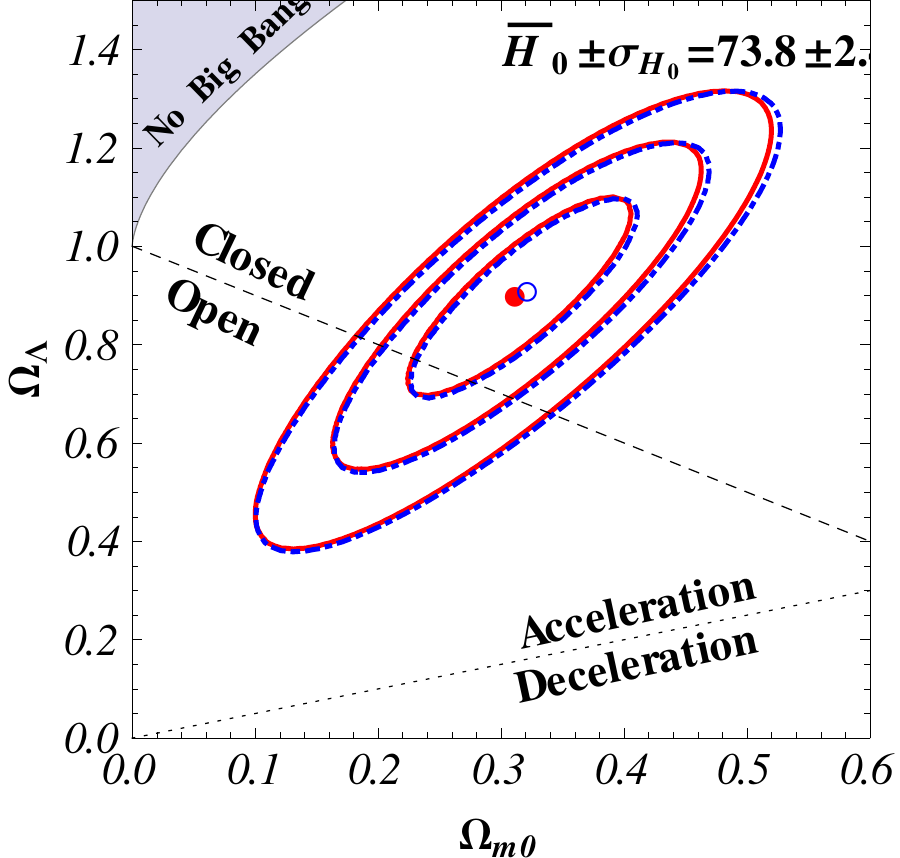}
    \includegraphics[height=1.5in]{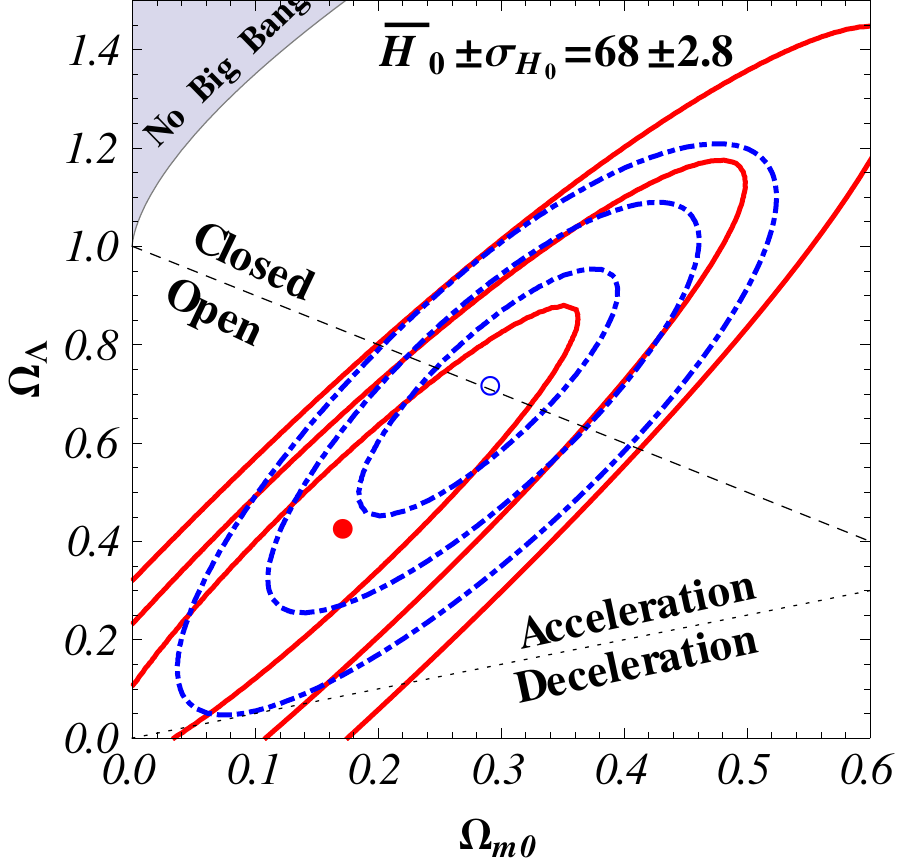}
    \includegraphics[height=1.5in]{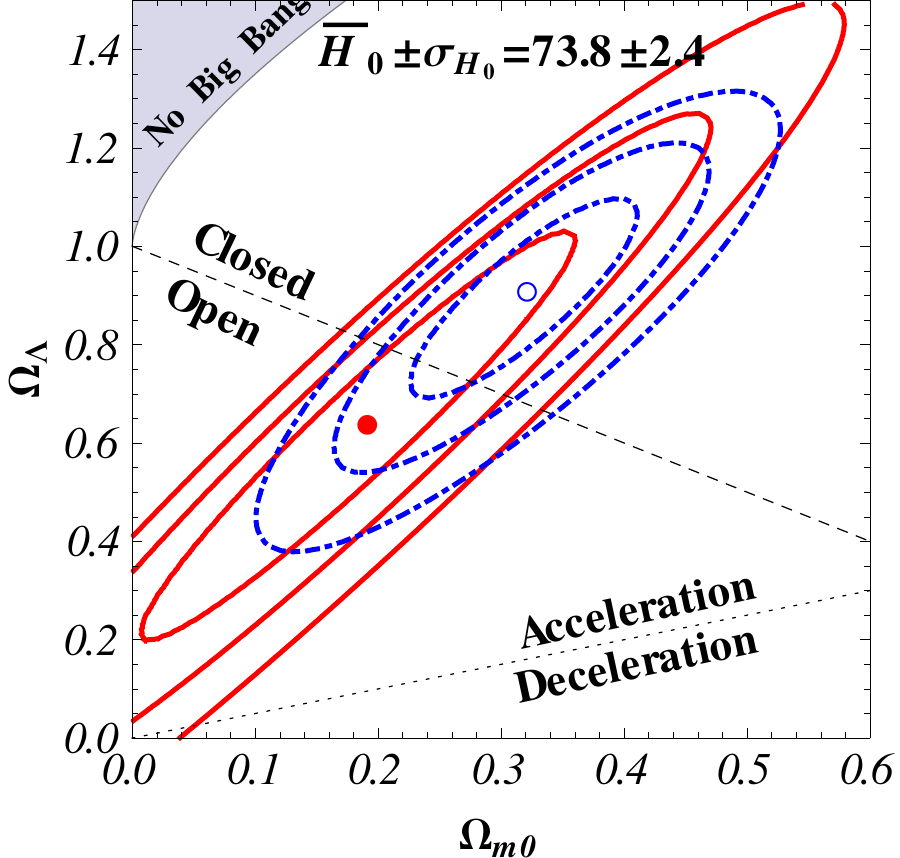}
    \includegraphics[height=1.45in]{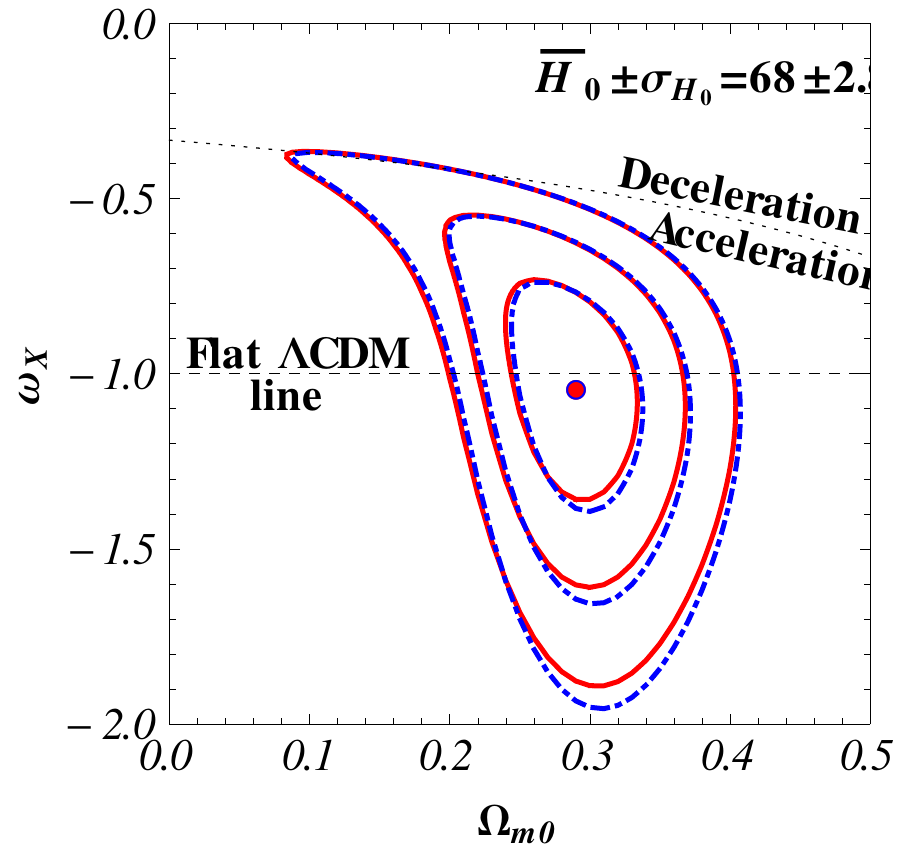}
    \includegraphics[height=1.45in]{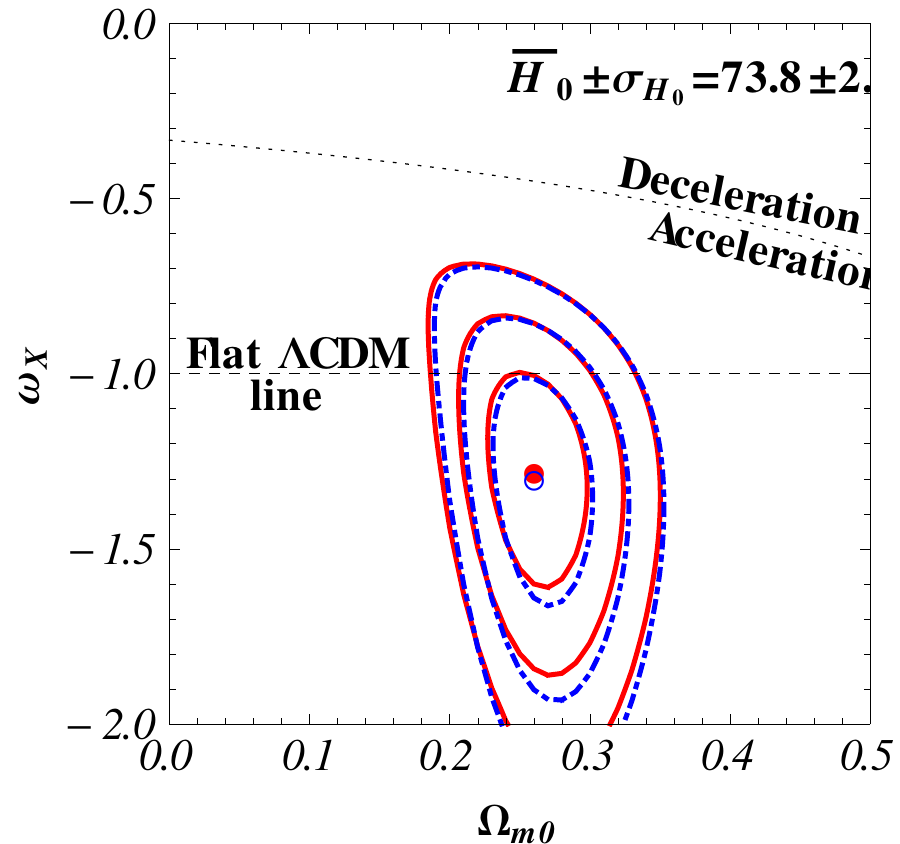}
    \includegraphics[height=1.45in]{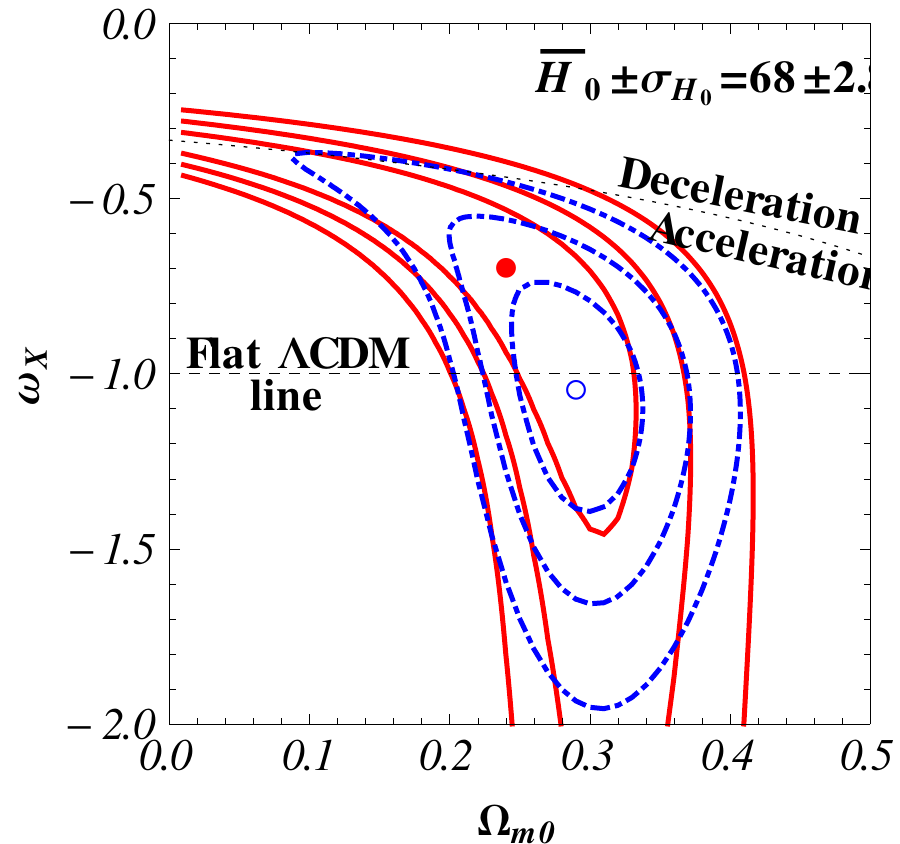}
    \includegraphics[height=1.5in]{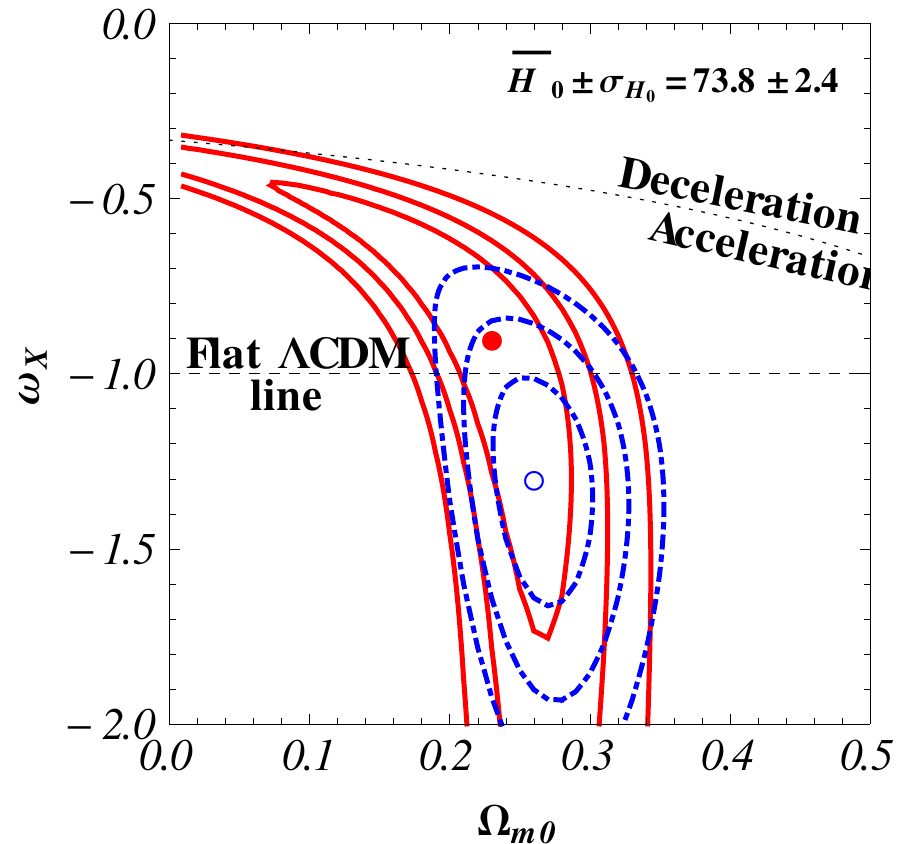}
    \includegraphics[height=1.55in]{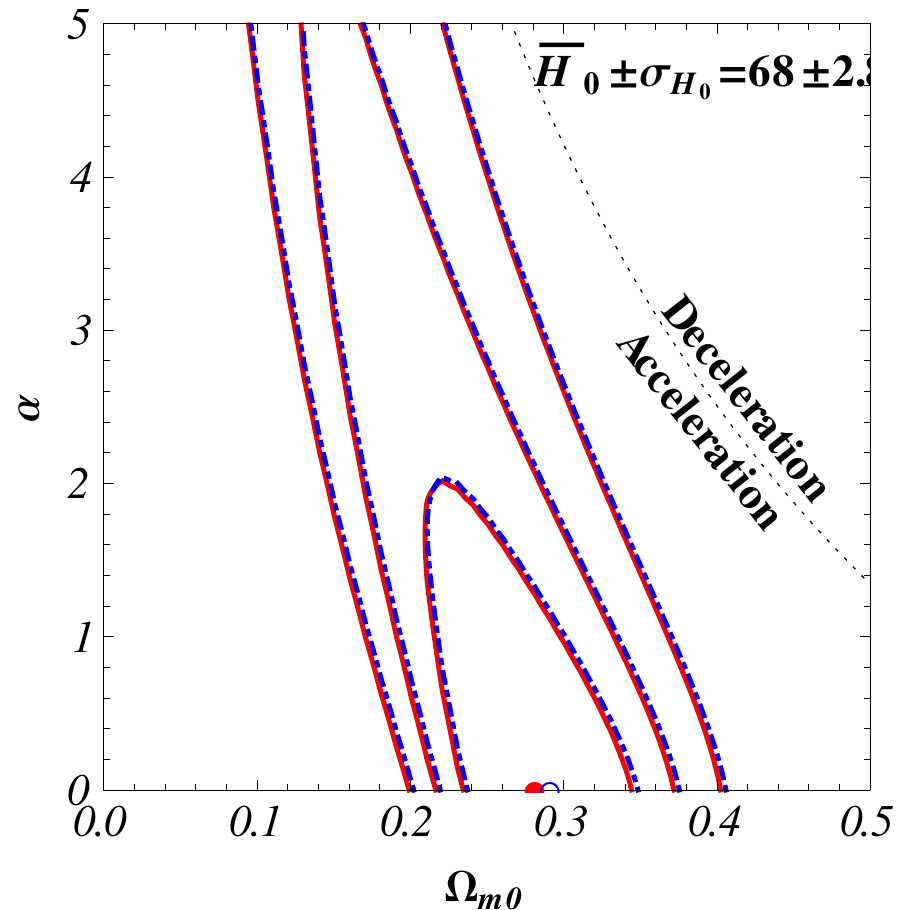}
    \includegraphics[height=1.55in]{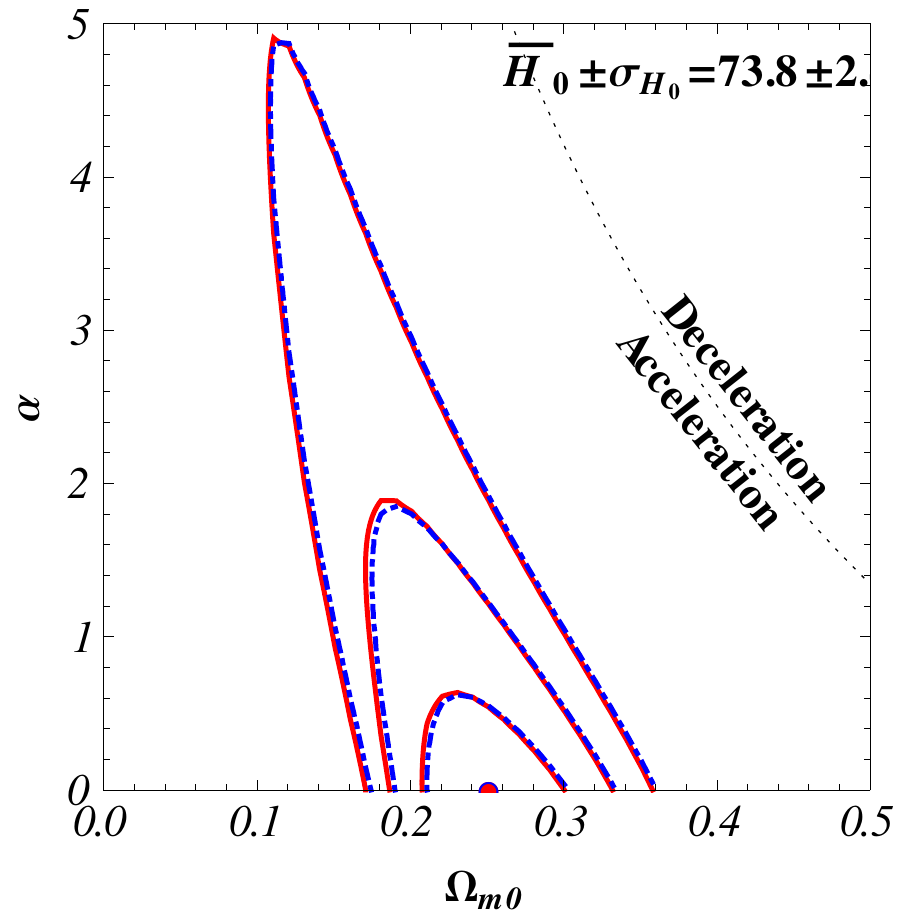}
    \includegraphics[height=1.55in]{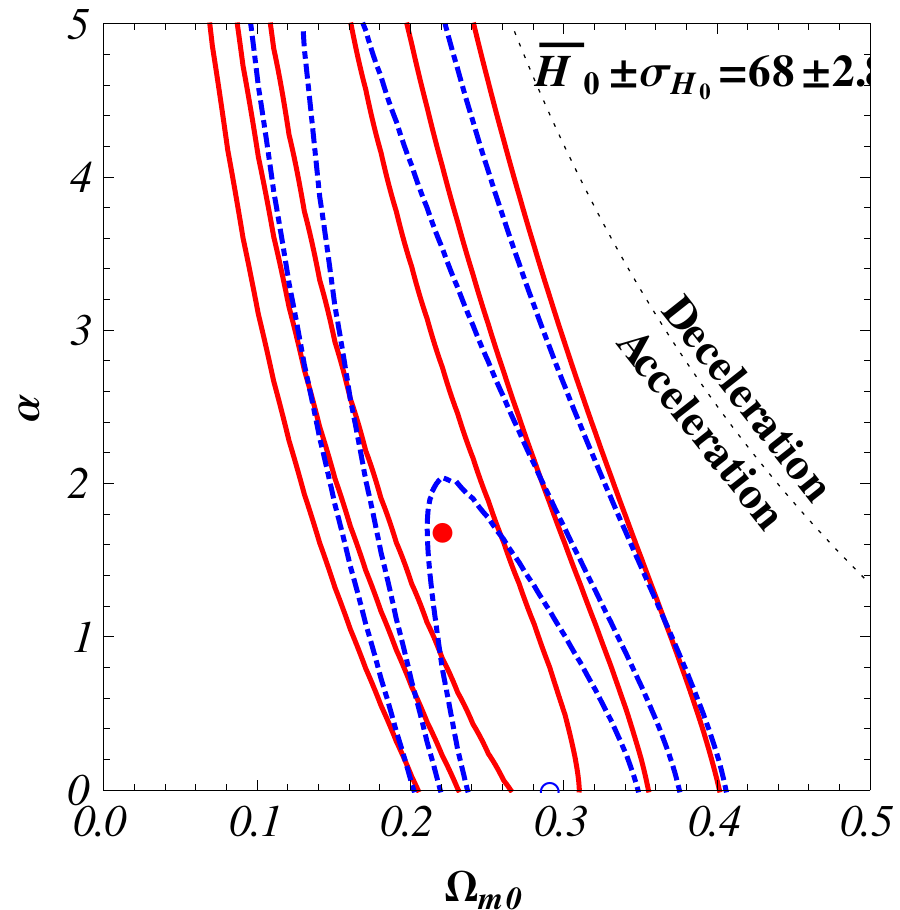}
    \includegraphics[height=1.55in]{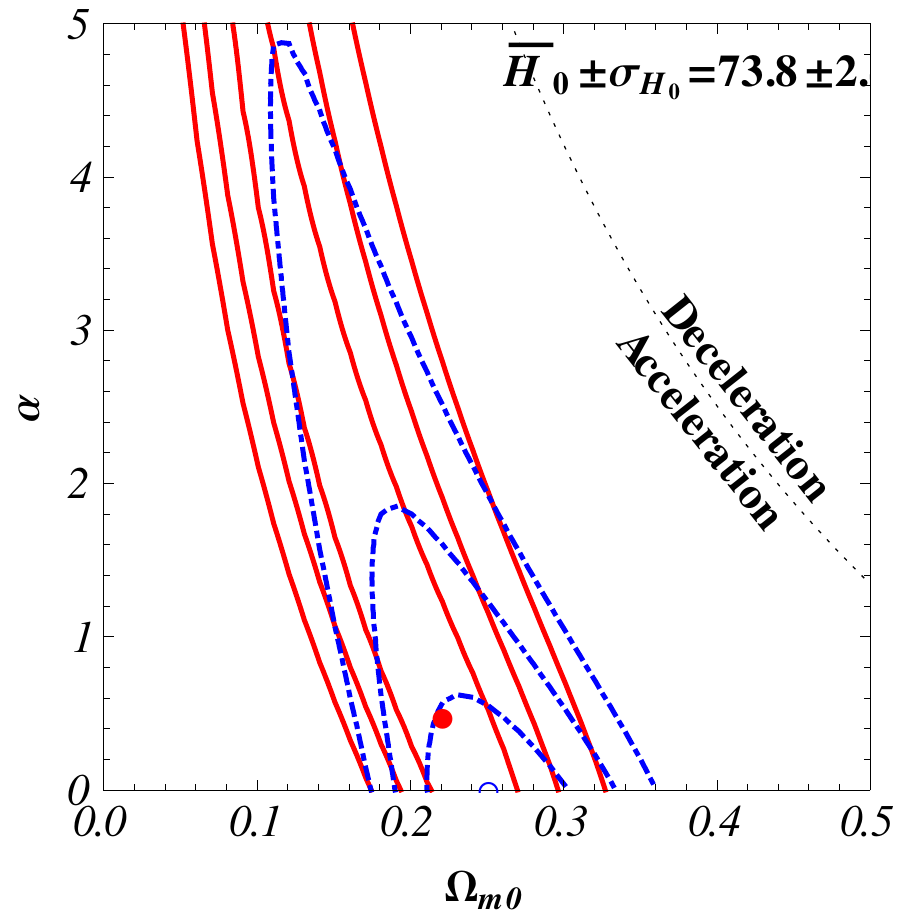}
 \caption{
Top left (right) panel shows the $H(z)/(1+z)$ data, binned with 7 or 9 
measurements per bin,  as well as 5 higher $z$ measurements, and the Farooq \& Ratra\cite{Farooq:2013hq} 
best-fit model predictions, dashed (dotted) for lower (higher) $H_0$ prior. 
The 2nd through 4th rows show the $H(z)$ constraints for $\Lambda$CDM, 
XCDM, and $\phi$CDM.
Red (blue dot-dashed) contours are 1$\sigma$, 2$\sigma$, and 3$\sigma$ confidence interval results from 7 or 9
measurements per bin (unbinned Farooq \& Ratra,\cite{Farooq:2013hq} Table\ 1) data. 
In these three rows, the first two plots 
include red weighted-mean constraints while the second two include red median statistics ones. The 
filled red (empty blue) circle is the corresponding best-fit point.
Dashed diagonal lines show spatially-flat models, and dotted
lines indicate zero-acceleration models. For quantitative details see Table (\ref{table:fig4 details}).
}
\label{fig:For table 8,9}
\end{figure}

\begin{center}
\begin{threeparttable}[h!]
\caption{Deceleration-Acceleration Transition Redshifts\tnote{a}}
\vspace{5 mm}
\begin{tabular}{ccccc}

\hline\hline

$h$ Prior\tnote{b} & Best-Fit Values & $\chi_{\rm min}^2$ & $z_{da} \pm \sigma_{z_{da}}$\tnote{c} & $z_{da}$\tnote{d}\\
\hline

\multicolumn{5}{c}{$\Lambda$CDM}\\

\hline

0.68 $\pm$ 0.028&	$(\Omega_{m0},\Omega_{\Lambda})=(0.29,0.72)$&	18.2&	0.690 $\pm$ 0.096&	   0.706\\

0.738 $\pm$ 0.024& $(\Omega_{m0},\Omega_{\Lambda})=(0.32,0.91$)&	19.3&	0.781 $\pm$ 0.067&	   0.785\\

\hline

\multicolumn{5}{c}{XCDM}\\

\hline

0.68 $\pm$ 0.028&	$(\Omega_{m0},\omega_{X})=(0.29,-1.04)$&	18.2&	0.677 $\pm$ 0.097&	   0.695\\

0.738 $\pm$ 0.024&	 $(\Omega_{m0},\omega_{X})=(0.26,-1.30)$&	18.2&	0.696 $\pm$ 0.082&	   0.718\\

\hline

\multicolumn{5}{c}{$\phi$CDM}\\

\hline

0.68 $\pm$ 0.028&	$(\Omega_{m0},\alpha)=(0.29,0.00)$&	18.2&	0.724 $\pm$ 0.148&	   0.698\\

0.738 $\pm$ 0.024&	 $(\Omega_{m0},\alpha)=(0.25,0.00)$&	20.7&	0.850 $\pm$ 0.116&	   0.817\\
\hline
\hline
\footnotetext{Estimated using the unbinned data in Table\ (\ref{tab:Hz3}).}
\end{tabular}
\begin{tablenotes}
\item[a]{Estimated using the unbinned data in Table\ (\ref{tab:Hz3}).}
\item[b]{Hubble constant in units of 100 km s$^{-1}$ Mpc$^{-1}$.}
\item[c]{Computed using Eqs.\ (\ref{Eq:6-1})---(\ref{Eq:6-6}).}
\item[d]{The deceleration-acceleration transition redshift in the model
with the best-fit values of the cosmological parameters, as computed in Farooq \& Ratra\cite{Farooq:2013hq}.}
\end{tablenotes}
\label{table:Un-binned data details}
\end{threeparttable}
\end{center}

\begin{center}
\begin{threeparttable}[h!]
\caption{Weighted Mean Results For 23 Lower Redshift Measurements}
\vspace{5 mm}
\begin{tabular}{ccccccc}
\hline\hline

\multirow{2}{*}{Bin} & \multirow{2}{*}{$N$} & \multirow{2}{*}{$z\tnote{a}$} & \multicolumn{1}{c}{$H(z)$} &  \multicolumn{1}{c}{$H(z)$ (1 $\sigma$ range)} & \multicolumn{1}{c}{$H(z)$ (2 $\sigma$ range)} & \multirow{2}{*}{$N_{\sigma}$}\\

{} & {} & {} & (km s$^{-1}$ Mpc $^{-1}$) & (km s$^{-1}$ Mpc $^{-1}$) & (km s$^{-1}$ Mpc $^{-1}$) & {}\\

\hline

\multicolumn{7}{c}{3 or 4 measurements per bin}\\

\hline

1&	3&	0.096&	69.0&	59.4$-$78.5&   49.9$-$88.0&	2.00\\

2&	4&	0.185&	76.0&	73.1$-$78.9&   70.2$-$81.8&	1.73\\

3&	3&  	0.338&	76.6&	71.5$-$81.8&   66.4$-$86.9&   1.89\\

4&	3&	0.417&	84.4&	78.1$-$90.7&	71.8$-$97.0&	1.55\\

5&	3&	0.598&	90.9&	85.4$-$96.4&	  79.9$-$102&	0.73\\

6&	3&	0.720&	96.6&         91.8$-$101&  87.0$-$106&	1.17\\

7&	4&	0.929&	129&	118$-$140&   107$-$151&	0.13\\

\hline

\multicolumn{7}{c}{4 or 5 measurements per bin}\\

\hline

1&	4&	0.139&	77.2&	71.1$-$83.3&   64.9$-$89.5&	1.41\\

2&	5&	0.191&	75.2&	72.1$-$78.2&   69.1$-$81.2&	2.71\\

3&	5&      0.380&	79.9&	75.8$-$84.1&   71.6$-$88.3&  1.81\\

4&	5&	0.668&	94.1&	90.5$-$97.7&   86.8$-$101&	0.91\\

5&	4&	0.929&	129&	118$-$140&   107$-$151&  0.13\\

\hline

\multicolumn{7}{c}{5 or 6 measurements per bin}\\

\hline

1&	5&	0.167&	75.7&	72.3$-$79.0&   69.0$-$82.3&	1.86\\

2&	6&	0.271&	76.2&	72.7$-$79.7&   69.3$-$83.1&	2.89\\

3&	6&      0.569&	89.4&	85.5$-$93.2&   81.7$-$97.0&   1.70\\

4&	6&	0.787&	106&	101$-$112&   95.8$-$117&	2.50\\

\hline

\multicolumn{7}{c}{7 or 9 measurements per bin}\\

\hline

1&	7&	0.177&	75.4&	72.7$-$78.2&   69.9$-$81.0&	2.66\\

2&	9&	0.448&	83.6&	80.3$-$86.8&   77.1$-$90.0&	1.29\\

3&	7&	0.754&	102&	97.6$-$106&   93.2$-$111&   2.99\\
\hline
\hline
\end{tabular}
\begin{tablenotes}
\item[a]{Weighted mean of $z$ values of measurements in the bin.}
\end{tablenotes}
\label{table:WA}
\end{threeparttable}
\end{center}

\begin{center}
\begin{threeparttable}[h!]
\caption{Median Statistics Results For 23 Lower Redshift Measurements}
\vspace{0 mm}
\begin{tabular}{cccccc}
\hline\hline

\multirow{2}{*}{Bin} & \multirow{2}{*}{$N$} & \multirow{2}{*}{$z\tnote{a}$} & \multicolumn{1}{c}{$H(z)$} &  \multicolumn{1}{c}{$H(z)$ (1 $\sigma$ range)} & \multicolumn{1}{c}{$H(z)$ (2 $\sigma$ range)}\\

{}& {} & {} & (km s$^{-1}$ Mpc $^{-1}$) & (km s$^{-1}$ Mpc $^{-1}$) & (km s$^{-1}$ Mpc $^{-1}$)\\

\hline

\multicolumn{6}{c}{3 or 4 measurements per bin}\\

\hline

1&	3&	0.100&	69.0&	49.4$-$88.6&   29.8$-$108\\

2&	4&	0.189&	75.0&	68.5$-$81.5&    62.0$-$88.0\\

3&	3&      0.280&	77.0&	63.0$-$91.0&   49.0$-$105\\

4&	3&	0.400&	83.0&	69.0$-$97.0&   55.0$-$111\\

5&	3&	0.593&	97.0&	84.0$-$110&   71.0$-$123\\

6&	3&	0.730&	97.3&  89.3$-$105&  81.3$-$113\\

7&	4&	0.890&	121&	99.5$-$143&  78.0$-$164\\

\hline

\multicolumn{6}{c}{4 or 5 measurements per bin}\\

\hline

1&	4&	0.110&	69.0&	53.2$-$84.8&    37.4$-$101\\

2&	5&	0.200&	75.0&	61.0$-$89.0&    47.0$-$103\\

3&	5&      0.400&	83.0&	69.0$-$97.0&    55.0$-$111\\

4&	5&	0.680&	97.3&	89.3$-$105&   81.3$-$113\\

5&	4&	0.890&	121&	99.5$-$143&    78.0$-$164\\

\hline

\multicolumn{6}{c}{5 or 6 measurements per bin}\\

\hline

1&	5&	0.120&	69.0&	57.0$-$81.0&   45.0$-$93.0\\

2&	6&	0.275&	76.7&	62.7$-$90.7&   48.7$-$105\\

3&	6&      0.537&	93.5&	83.0$-$104&   72.5$-$115\\

4&	6&	0.878&	111&	92.5$-$130&   74.0$-$148\\

\hline

\multicolumn{6}{c}{7 or 9 measurements per bin}\\

\hline

1&	7&	0.170&	72.9&	60.9$-$84.9&   48.9$-$96.9\\

2&	9&	0.400&	87.9&	73.9$-$102&   59.9$-$116\\

3&	7&	0.875&	105&	88.0$-$122&   71.0$-$139\\

\hline
\hline
\end{tabular}
\begin{tablenotes}
\item[a]{Median of $z$ values of measurements in the bin.}
\end{tablenotes}
\label{table:Med}
\end{threeparttable}
\end{center}
\newpage
~
~

\begin{table}[h!]
\begin{center}
\caption{Best-Fit Points And Minimum $\chi^2$s For 3 Or 4 Measurements Per Bin}
\vspace{2 mm}
\begin{tabular}{c c c c c c} 
\hline\hline 
{} & {} & \multicolumn{2}{c}{Weighted Mean} & \multicolumn{2}{c}{Median}\\

Model & $h$ Prior &  BFP & $\chi^2_{\rm min}$ &  BFP & $\chi^2_{\rm min}$\\

\hline

\multirow{4}{*}{$\Lambda$CDM} & \multirow{2}{*}{$0.68 \pm 0.028$} & $\Omega_{m0}=0.29$ &\multirow{2}{*}{13.0} & $\Omega_{m0}=0.24$ & \multirow{2}{*}{8.75}\\

& {} & $\Omega_{\Lambda}=0.72$ &{} & $\Omega_{\Lambda}=0.60$ \\

\cline{2-6}

& \multirow{2}{*}{$0.738 \pm 0.024$} & $\Omega_{m0}=0.32$ & \multirow{2}{*}{ 14.1} & $\Omega_{m0}=0.26$ & \multirow{2}{*}{9.37}\\

&{}& $\Omega_{\Lambda}=0.91$ &{}& $\Omega_{\Lambda}=0.80$\\ 

\hline

\multirow{4}{*}{XCDM} & \multirow{2}{*}{$0.68 \pm 0.028$} & $\Omega_{m0}=0.29$ &\multirow{2}{*}{13.0} & $\Omega_{m0}=0.28$ & \multirow{2}{*}{8.85}\\

& {} & $\omega_{X}=-1.04$ &{} & $\omega_{X}=-0.90$ \\

\cline{2-6}

& \multirow{2}{*}{$0.738 \pm 0.024$} & $\Omega_{m0}=0.26$ & \multirow{2}{*}{ 13.0} & $\Omega_{m0}=0.25$ & \multirow{2}{*}{9.18}\\

&{}& $\omega_{X}=-1.29$ &{}& $\omega_{X}=-1.13$\\ 

\hline

\multirow{4}{*}{$\phi$CDM} & \multirow{2}{*}{$0.68 \pm 0.028$} & $\Omega_{m0}=0.29$ &\multirow{2}{*}{13.0} & $\Omega_{m0}=0.27$ & \multirow{2}{*}{8.82}\\

& {} & $\alpha=0.00$ &{} & $\alpha=0.46$ \\

\cline{2-6}

& \multirow{2}{*}{$0.738 \pm 0.024$} & $\Omega_{m0}=0.25$ & \multirow{2}{*}{ 15.4} & $\Omega_{m0}=0.24$ & \multirow{2}{*}{9.43}\\

&{}& $\alpha=0.00$ &{}& $\alpha=0.00$\\ 

\hline  
\hline
\end{tabular}
\label{table:fig1 details}
\end{center}
\end{table}



\begin{table}[h!]
\begin{center}
\caption{Best-Fit Points And Minimum $\chi^2$s For 4 Or 5 Measurements Per Bin}
\vspace{2 mm}
\begin{tabular}{c c c c c c} 
\hline\hline

{} & {} & \multicolumn{2}{c}{Weighted Mean} & \multicolumn{2}{c}{Median}\\

Model & $h$ Prior &  BFP & $\chi^2_{\rm min}$ &  BFP & $\chi^2_{\rm min}$\\

\hline

\multirow{4}{*}{$\Lambda$CDM} & \multirow{2}{*}{$0.68 \pm 0.028$} & $\Omega_{m0}=0.29$ &\multirow{2}{*}{12.9} & $\Omega_{m0}=0.17$ & \multirow{2}{*}{7.62}\\

& {} & $\Omega_{\Lambda}=0.73$ &{} & $\Omega_{\Lambda}=0.43$ \\

\cline{2-6}

& \multirow{2}{*}{$0.738 \pm 0.024$} & $\Omega_{m0}=0.32$ & \multirow{2}{*}{ 13.7} & $\Omega_{m0}=0.20$ & \multirow{2}{*}{8.04}\\

&{}& $\Omega_{\Lambda}=0.91$ &{}& $\Omega_{\Lambda}=0.66$\\ 

\hline

\multirow{4}{*}{XCDM} & \multirow{2}{*}{$0.68 \pm 0.028$} & $\Omega_{m0}=0.29$ &\multirow{2}{*}{12.9} & $\Omega_{m0}=0.24$ & \multirow{2}{*}{7.75}\\

& {} & $\omega_{X}=-1.06$ &{} & $\omega_{X}=-0.68$ \\

\cline{2-6}

& \multirow{2}{*}{$0.738 \pm 0.024$} & $\Omega_{m0}=0.26$ & \multirow{2}{*}{ 12.5} & $\Omega_{m0}=0.23$ & \multirow{2}{*}{8.17}\\

&{}& $\omega_{X}=-1.31$ &{}& $\omega_{X}=-0.89$\\ 

\hline

\multirow{4}{*}{$\phi$CDM} & \multirow{2}{*}{$0.68 \pm 0.028$} & $\Omega_{m0}=0.29$ &\multirow{2}{*}{13.0} & $\Omega_{m0}=0.22$ & \multirow{2}{*}{7.70}\\

& {} & $\alpha=0.00$ &{} & $\alpha=1.77$ \\

\cline{2-6}

& \multirow{2}{*}{$0.738 \pm 0.024$} & $\Omega_{m0}=0.25$ & \multirow{2}{*}{ 15.2} & $\Omega_{m0}=0.23$ & \multirow{2}{*}{8.13}\\

&{}& $\alpha=0.00$ &{}& $\alpha=0.30$\\ 

\hline  
\hline
\end{tabular}
\label{table:fig2 details}
\end{center}
\end{table}



\begin{table}[h!]
\begin{center}
\caption{Best-Fit Points And Minimum $\chi^2$s For 5 Or 6 Measurements Per Bin}
\vspace{2 mm}
\begin{tabular}{c c c c c c} 
\hline\hline 

{} & {} & \multicolumn{2}{c}{Weighted Mean} & \multicolumn{2}{c}{Median}\\

Model & $h$ Prior &  BFP & $\chi^2_{\rm min}$ &  BFP & $\chi^2_{\rm min}$\\

\hline

\multirow{4}{*}{$\Lambda$CDM} & \multirow{2}{*}{$0.68 \pm 0.028$} & $\Omega_{m0}=0.28$ &\multirow{2}{*}{10.2} & $\Omega_{m0}=0.18$ & \multirow{2}{*}{7.65}\\

& {} & $\Omega_{\Lambda}=0.70$ &{} & $\Omega_{\Lambda}=0.45$ \\

\cline{2-6}

& \multirow{2}{*}{$0.738 \pm 0.024$} & $\Omega_{m0}=0.31$ & \multirow{2}{*}{ 11.2} & $\Omega_{m0}=0.20$ & \multirow{2}{*}{8.13}\\

&{}& $\Omega_{\Lambda}=0.89$ &{}& $\Omega_{\Lambda}=0.66$\\ 

\hline

\multirow{4}{*}{XCDM} & \multirow{2}{*}{$0.68 \pm 0.028$} & $\Omega_{m0}=0.29$ &\multirow{2}{*}{10.2} & $\Omega_{m0}=0.24$ & \multirow{2}{*}{7.77}\\

& {} & $\omega_{X}=-1.01$ &{} & $\omega_{X}=-0.68$ \\

\cline{2-6}

& \multirow{2}{*}{$0.738 \pm 0.024$} & $\Omega_{m0}=0.26$ & \multirow{2}{*}{ 10.2} & $\Omega_{m0}=0.23$ & \multirow{2}{*}{8.25}\\

&{}& $\omega_{X}=-1.28$ &{}& $\omega_{X}=-0.89$\\ 

\hline

\multirow{4}{*}{$\phi$CDM} & \multirow{2}{*}{$0.68 \pm 0.028$} & $\Omega_{m0}=0.29$ &\multirow{2}{*}{10.2} & $\Omega_{m0}=0.22$ & \multirow{2}{*}{7.72}\\

& {} & $\alpha=0.00$ &{} & $\alpha=1.73$ \\

\cline{2-6}

& \multirow{2}{*}{$0.738 \pm 0.024$} & $\Omega_{m0}=0.25$ & \multirow{2}{*}{ 12.4} & $\Omega_{m0}=0.23$ & \multirow{2}{*}{8.21}\\

&{}& $\alpha=0.00$ &{}& $\alpha=0.30$\\ 

\hline  
\hline
\end{tabular}
\label{table:fig3 details}
\end{center}
\end{table}

\begin{table}[h!]
\begin{center}
\caption{Best-Fit Points And Minimum $\chi^2$s For 7 Or 9 Measurements Per Bin}
\vspace{2 mm}
\begin{tabular}{c c c c c c} 
\hline\hline 
{} & {} & \multicolumn{2}{c}{Weighted Mean} & \multicolumn{2}{c}{Median}\\
Model & $h$ Prior &  BFP & $\chi^2_{\rm min}$ &  BFP & $\chi^2_{\rm min}$\\
\hline
\multirow{4}{*}{$\Lambda$CDM} & \multirow{2}{*}{$0.68 \pm 0.028$} & $\Omega_{m0}=0.29$ &\multirow{2}{*}{9.7} & $\Omega_{m0}=0.17$ & \multirow{2}{*}{7.76}\\
& {} & $\Omega_{\Lambda}=0.73$ &{} & $\Omega_{\Lambda}=0.43$ \\
\cline{2-6}
& \multirow{2}{*}{$0.738 \pm 0.024$} & $\Omega_{m0}=0.31$ & \multirow{2}{*}{ 10.4} & $\Omega_{m0}=0.19$ & \multirow{2}{*}{7.88}\\
&{}& $\Omega_{\Lambda}=0.90$ &{}& $\Omega_{\Lambda}=0.64$\\ 
\hline
\multirow{4}{*}{XCDM} & \multirow{2}{*}{$0.68 \pm 0.028$} & $\Omega_{m0}=0.29$ &\multirow{2}{*}{9.7} & $\Omega_{m0}=0.24$ & \multirow{2}{*}{7.82}\\
& {} & $\omega_{X}=-1.04$ &{} & $\omega_{X}=-0.69$ \\
\cline{2-6}
& \multirow{2}{*}{$0.738 \pm 0.024$} & $\Omega_{m0}=0.26$ & \multirow{2}{*}{ 9.5} & $\Omega_{m0}=0.23$ & \multirow{2}{*}{7.97}\\
&{}& $\omega_{X}=-1.28$ &{}& $\omega_{X}=-0.90$\\ 
\hline
\multirow{4}{*}{$\phi$CDM} & \multirow{2}{*}{$0.68 \pm 0.028$} & $\Omega_{m0}=0.28$ &\multirow{2}{*}{9.7} & $\Omega_{m0}=0.22$ & \multirow{2}{*}{7.80}\\
& {} & $\alpha=0.00$ &{} & $\alpha=1.69$ \\
\cline{2-6}
& \multirow{2}{*}{$0.738 \pm 0.024$} & $\Omega_{m0}=0.25$ & \multirow{2}{*}{ 11.8} & $\Omega_{m0}=0.22$ & \multirow{2}{*}{7.93}\\
&{}& $\alpha=0.00$ &{}& $\alpha=0.48$\\ 
\hline  
\hline
\end{tabular}
\label{table:fig4 details}
\end{center}
\end{table}


\cleardoublepage


\chapter{Observational Constraints on Non-Flat Dynamical Dark Energy Cosmological Models}
\label{Chapter8}

This chapter is based on Farooq \textit{et al.}\cite{farooq5}
\\

We constrain two non-flat time-evolving dark energy cosmological models
by using Hubble parameter data, Type Ia supernova apparent magnitude
measurements, and baryonic acoustic oscillation peak length scale observations. The
inclusion of space curvature as a free parameter in the analysis results 
in a significant broadening of the allowed range of values of the parameter
that governs the time evolution of the dark energy density in these models.
While consistent with the ``standard" spatially-flat $\Lambda$CDM cosmological model,
these data are also consistent with a range of mildly non-flat, slowly time-varying dark 
energy models. After marginalizing over all other parameters, these data require the
averaged magnitude of the curvature density parameter $|\Omega_{k0}| \lesssim 0.15$ 
at 1$\sigma$ confidence.

\section{Introduction}

There is significant observational evidence that the Universe
is currently undergoing accelerated expansion.
Most cosmologists believe that dark energy dominates the current cosmological
energy budget and is responsible for this accelerated 
expansion (for reviews of dark energy see Li \text{et al.},\cite{li2012} Tsujikawa,\cite{tsujikawa2013} Sol\`{a},\cite{sola2013}
and references therein).\footnote{
Some instead argue that these observations should be viewed as an indication 
that general relativity needs to be modified on these large cosmological
length scales. For recent reviews of modified gravity see Capozziello \& De Laurentis,\cite{Capozziello2011} 
Trodden, \cite{trodden2012} and
references therein. We assume here that general relativity provides an accurate description
of gravitation on cosmological length scales.} In addition, if one assumes that the dark energy density is close to or time independent, 
cosmic microwave background (CMB) anisotropy measurements indicate that the Universe must be close to 
or spatially flat (Ade \textit{et al.},\cite{Planckdata} and references therein; for an early indication see Podariu \textit{et al.},\cite{Podariu2001}).
Conversely, if one assumes that
the space sections are flat, the data favor a time-independent cosmological constant. However,
as far as we are aware, there has not been an analysis of observational data based 
on a consistent non-spatially-flat dynamical dark energy model. In this chapter we present the first
such analysis.

As a warm-up exercise, we consider the popular XCDM parameterization of dynamical dark energy.\footnote{Here
dark energy is taken to be a spatially-homogeneous $X$-fluid with a time-evolving energy density that 
dominates the current cosmological energy budget, with non-relativistic cold dark matter (CDM) being the next
largest contributor.} The XCDM parameterization is a generalization of the standard $\Lambda$CDM cosmological 
model from Peebles.\cite{peebles84} In the $\Lambda$CDM, case the current energy budget is dominated by a time-independent 
cosmological constant $\Lambda$. It is well-known that the $\Lambda$CDM model has some puzzling 
features which are more easily understood if, instead of remaining constant like $\Lambda$, the 
dark energy density gradually decreases with time.\footnote{Note
that there also are tentative observational indications that the standard CDM structure formation model, which is assumed in
the $\Lambda$CDM cosmological model, might need to be improved upon. For details see,
Peebles \& Ratra,\cite{Peebles&Ratra2003} Weinberg,\cite{Weinberg2013} and references therein.}

The simplest, complete and consistent time-varying dark energy 
model is $\phi$CDM (see Peebles \& Ratra,\cite{Peebles&Ratra1988} and 
Ratra \& Peebles,\cite{Ratra&Peebles1988} for detail).\footnote{For recent discussions of other time-varying 
dark energy models see Lui \textit{et al.},\cite{liu2012} Garcia \textit{et al.},\cite{Garcia-Salcedo2012} Benaoum,\cite{Benaoum2012}
Ayaita \textit{et al.},\cite{Ayaita2012} Ferreira \textit{et al.},\cite{Ferreira2013} 
Bezrukov \textit{et al.},\cite{Bezrukov2013} Liao \& Zhu,\cite{Liao2013} and references therein.}
Here the dark energy is modeled as a scalar 
field, $\phi$, with a gradually decreasing (in $\phi$) potential energy 
density $V(\phi)$. In this paper we assume an inverse-power-law potential energy 
density, $V(\phi) \propto \phi^{-\alpha}$, where $\alpha$ is a nonnegative 
constant Peebles \& Ratra.\cite{Peebles&Ratra1988} At $\alpha = 0$ the
$\phi$CDM model reduces to the corresponding $\Lambda$CDM case. 
The $\phi$CDM model was originally formulated in a spatially-flat cosmological model. In this chapter we consider 
the \cite{Anatoly2} generalization of the $\phi$CDM model to non-flat space.\footnote{Curved-space scalar field dark energy models have been studied in the past see e.g., Aurich \& Steiner,\cite{Aurich2002,Aurich2003,Aurich2004} Thepsuriya \& Gumjudpai,\cite{Thepsuriya2009} 
Chen \& Guo,\cite{Chen&Guo2012} Gumjudpai \& Thepsuriya.\cite{Gumjudpai2012}
However, as far as we are aware, Anatoly \textit{et al.},\cite{Anatoly2} were the first to establish that the scalar field solution is a time-dependent fixed point or attractor even in the curvature dominated epoch.}

For some time now, most observational constraints have 
been reasonably consistent with the predictions of the ``standard'' 
spatially-flat 
$\Lambda$CDM model \citep[for early indications see e.g.,][]{jassal10,
wilson06,Davis2007,allen08}. The big four, CMB anisotropy e.g., Plank results\cite{Planckdata}, supernova Type Ia (SNIa)
apparent magnitude versus redshift e.g., Suzuki \textit{et al.},\cite{suzuki2012} Salzano \textit{et al.},\cite{Salzano2012}
Campbell \textit{et al.},\cite{Campbell2013} baryonic acoustic oscillation (BAO) peak
length scale \citep[e.g.,][]{Percival2010,Beutler11,blake11,Mehta2012}, and Hubble parameter 
as a function of redshift \citep[e.g.,][]{Chen2011b,moresco12,busca12,Farooq:2013hq} measurements
provide the strongest support for this conclusion. 

Other measurements that have been used to constrain cosmological 
parameters include, for example, galaxy cluster gas mass fraction as a function of redshift 
e.g., Allen \textit{et al.},\cite{allen08} Samushia \& Ratra,\cite{Samushia&Ratra2008} 
Tong \textit{et al.},\cite{tongnoh11} Lu \textit{et al.},\cite{Lu2011b} Landry \textit{et al.},\cite{landry12} 
galaxy cluster and other large-scale structure properties 
Mortonson \textit{et al.},\cite{mortonson2011} Devi \textit{et al.},\cite{Devi2011} 
Wang,\cite{Wang2012} De Boni \textit{et al.},\cite{deboni13} Batista \textit{et al.},\cite{Batista2013} 
and references therein,
gamma-ray burst luminosity distance as a function of redshift e.g.,
Samushia \& Ratra, \cite{Samushia&Ratra2010} Wang \& Dai,\cite{Wang2011} Busti \textit{et al.},\cite{Busti2012} Pan \textit{et al.}\cite{pan13}, 
HII starburst galaxy 
apparent magnitude as a function of redshift e.g., Plionis \textit{et al.},\cite{plionisetal10,plionisetal11} Mania \& Ratra,\cite{maniaratra12}, 
angular size as a function of redshift e.g.,
Guerra \textit{et al.},\cite{guerra00} Bonamente \textit{et al.},\cite{Bonamente2006} Chen \& Ratra,\cite{Chen2012}, 
and strong gravitational lensing Chae \textit{et al.},\cite{chae04} Lee \& Ng,\cite{lee07} 
Biesiada \textit{et al.},\cite{biesiada10} Suyu \textit{et al.},\cite{Suyu2013} and references therein.\footnote{
Future space-based SNIa and BAO-like measurements 
\citep[e.g.,][]{podariu01a, Samushia2011, Sartoris2012, Basse2012, Pavlov2012},
as well as measurements based on new techniques 
\citep[][and references therein]{Appleby2013, Arabsalmani2013}
 should soon provide interesting constraints on cosmological
parameters.}
 While the constraints from 
these data are typically less restrictive than
those derived from the $H(z)$, SNIa, CMB anisotropy, and BAO data, both types of
measurements result in largely compatible constraints that generally 
support a currently accelerating cosmological expansion. This provides
confidence that the broad outlines of a ``standard'' cosmological
model are now in place.

In this chapter we consider an extension of this ``standard" cosmological 
model by allowing for the possibility of non-zero space curvature. As mentioned
above we consider two possibilities, a generalization of the XCDM parameterization
as well as the Pavlov \textit{et al.}\cite{Anatoly2} generalization of the $\phi$CDM model. In this 
chapter we derive constraints on the parameters of these options by using $H(z)$,
SNIa, and BAO data.

Here, we do not make use of the last of the big four data, that of CMB anisotropy.
While these data are widely credited with providing the strongest evidence 
for a very small contribution to the current energy budget from spatial
curvature (see discussion above), it is not straightforward to include them 
in the analyses because they require an analysis of the 
evolution of spatial inhomogeneities. In the case of the XCDM
parametrization this is not possible without an additional ad hoc extension. In the $\phi$CDM
case this requires a detailed computation, including the assumption of an early epoch of inflation
in non-flat space and a derivation of the concomitant power spectrum needed for the CMB anisotropy
computation. It is well known that the CMB anisotropy is almost certainly a consequence of 
quantum-mechanical zero-point fluctuations generated during inflation \citep[see e.g.,][]{Fischler1985,ratra92}.
While conceptually similar, the computation of the primordial spectrum is somewhat 
more involved in the spatially curved case \cite[see e.g.,][]{Gott1982,Ratra&Peebles1994,Ratra&Peebles1995,
Kamionkowski1994,Bucher1995,Lyth1995,Yamamoto1995,Ganga1997,gorshi1998}. Consequently, the computation
of CMB anisotropy constraints is beyond the scope of this initial paper. Even though we do not
use CMB anisotropy constraints here, a combination of the other three of the big four data--- $H(z)$,
SNIa, and BAO--- results in reasonably tight constraints on space curvature. For technical
computational reasons we believe our XCDM parametrization constraints are more reflective of the true
constraints on space curvature.\footnote{It is much more time consuming
 to do the $\phi$CDM computation, so we assumed a
narrower prior on space curvature in this case, which we suspect leads to slightly tighter but less reliable
constraints.} In the XCDM case, marginalizing over all other parameters, the $H(z)$, SNIa, and BAO data
require $|\Omega_{k0}|\leq 0.15$ and 0.3 at about 1$\sigma$ and 2$\sigma$ confidence.

\section{Time Varying Dark Energy Models in Curved Space}

In this section we summarize the two models we constrain. These are the \cite{Anatoly2}
generalization to curved space of the time-evolving dark energy $\phi$CDM model 
\citep{Peebles&Ratra1988, Ratra&Peebles1988}, as well as the curved space
generalization of the widely-used XCDM
dynamic dark energy parameterization in which dark energy is 
modeled as a spatially-homogeneous time-dependent $X$-fluid. 

We assume that general relativity provides an accurate description
of gravitation on cosmological scales. The equations of motion are 
Einstein's field equations,
\begin{equation}
R_{\mu\nu} - \frac{1}{2}Rg_{\mu\nu} = 8 \pi G T_{\mu\nu} - \Lambda g_{\mu\nu}.
\end{equation} 
Here $R_{\mu\nu}$ and $R$ are the Ricci tensor and scalar, 
$g_{\mu\nu}$ is the metric tensor, $\Lambda$ is the cosmological constant,
$T_{\mu\nu}$ is the energy-momentum tensor of the matter present,  and 
$G$ is the Newtonian gravitational constant.

At late times, we can ignore radiation and model non-relativistic (cold dark and
baryonic) matter as a perfect fluid with 
energy-momentum tensor $T_{\mu\nu} = 
{\rm diag} (\rho, p, p, p)$ where $\rho$ and $p$ are the energy density 
and the pressure of the fluid. Assuming the cosmological principle of
spatial homogeneity, Einstein's equations reduce to 
the two independent Friedmann equations:
\begin{equation}
\label{eq:F1}
\left(\frac{\dot{a}}{a}\right)^2 = \frac{8\pi G}{3}\rho-\frac{k}{a^2}+\frac{\Lambda}{3},
\end{equation}
\begin{equation}
\label{eq:F2}
\hspace{1 cm} \frac{\ddot{a}}{a}\ \ =-\frac{4\pi G}{3} (\rho +3p)+\frac{\Lambda}{3}.
\end{equation}
Here, $a(t)$ is the cosmological scale factor which is the ratio of the physical 
distance to the co-moving distance of a sufficiently distant object (so that 
the spatial homogeneity assumption is valid), an overdot denotes a 
derivative with respect to cosmological time, $k$ represents the curvature of 
spatial hypersurfaces (and can have three discrete values $-1$, $0$, or $+1$,
corresponding to hyperbolic, flat, and spherical geometry respectively), and $\rho$ and $p$
are the sums of all (time-dependent) densities and pressures of the various forms of matter
present. 

With a single type of matter, the Friedmann equations\ (\ref{eq:F1})---(\ref{eq:F2}) are two equations with 
three time-dependent unknowns: $a(t)$, $\rho(t)$, and $p(t)$. We can complete the system 
of equations with an equation of state for each type of matter. This
is a relation between pressure and energy
density for each type of matter:
\begin{equation}
\label{eq:EoS}
   p= p(\rho) = \omega \rho, 
\end{equation}
where $\omega$ is the dimensionless equation-of-state parameter.
For non-relativistic matter $\omega=0$, while
$\omega=-1$ corresponds to a standard cosmological constant $\Lambda$, and $\omega 
< -1/3$ corresponds to the XCDM parameterization.
 
Equations\ (\ref{eq:F1})---(\ref{eq:EoS}) form a closed set and can be used to derive 
the energy conservation equation:
\begin{equation}
\label{eq:EC}
   \frac{\dot{\rho}}{\rho} =-3\ \frac{\dot{a}}{a}\ (1+\omega).
\end{equation} 
This first-order linear differential equation can be solved 
with the boundary condition $\rho(t_0)=\rho_0$, where $t_0$ is the current time
and $\rho_0$ is the current value of the energy density of the particular type of
matter under consideration. The solution is:
\begin{equation}
\label{eq:EC1}
   \rho(t)=\rho_0 \left(\frac{a_0}{a}\right)^{3(1+\omega)},
\end{equation}
where $a_0$ is the current value of the scale factor. If there 
are a number of different species of non-interacting fluids, 
then Eq.\ (\ref{eq:EC1}) holds separately for each of them 
with the corresponding $\omega$ and $\rho_0$. For a non-relativistic 
gas (cold matter) $\omega=\omega_m=0$ and $\rho_m \propto a^{-3}$, 
for a homogeneous $X$-fluid $\omega=\omega_{X}<-1/3$ and $\rho_X\propto a^{-3(1+\omega_X)}$, 
and for spatial curvature $\omega=\omega_{k}=-1/3$ and $\rho_k\propto a^{-2}$. 

The ratio $\dot{a}(t)/a(t)$ in Eq.\ (\ref{eq:F1}) is the Hubble parameter
$H(t)$. The present value of the Hubble parameter is the Hubble constant 
$H_0$. To rewrite the Friedmann equation\ (\ref{eq:F1}) in
terms of observable parameters, we define the dimensionless redshift $z = a_0/a - 1 $ and 
the present value of the density parameters:
\begin{equation}
\label{eq:density parameters}
   \Omega_{m0} = \frac{8\pi G\rho_{m0}}{3H_0^2}, \ \ \ \ \ \ \
   \Omega_{k0}=\frac{-k}{(H_0 a_0)^2}, \ \ \ \ \ \ \
   \Omega_{X0} = \frac{8\pi G\rho_{X0}}{3H_0^2}.
\end{equation}
Here we have parameterized dark energy as a spatially homogeneous $X$-fluid with 
current density parameter value $\Omega_{X0}$, $\Omega_{m0}$ is the current non-relativistic
(baryonic and cold dark) matter density parameter, and $\Omega_{k0}$ is that of
spatial curvature (with $\Omega_{k0}>0$ corresponding to an open or hyperbolic spatial geometry). 
With these definitions Eq.\ (\ref{eq:F1}) becomes:
\begin{eqnarray}
\label{eq:XCDMwithc}
H^2(z; H_0, \textbf{p})\hspace{-2mm}&=&\hspace{-2mm}H_0^2 \left[\Omega_{m0} (1+z)^3 + (1-\Omega_{m0}-\Omega_{k0})(1+z)^{3(1+\omega_{X})}
                          + \Omega_{k0} (1+z)^2 \right], 
\end{eqnarray}
where we have made use of $\Omega_{X0} = 1-\Omega_{m0}-\Omega_{k0}$.
This is the Friedmann equation for the XCDM parameterization with non-zero 
spatial curvature. In this case, the cosmological parameters are taken to be
$\textbf{p}=({\Omega_{m0},\omega_{X},\Omega_{k0}})$. The XCDM parameterization is incomplete, as it
cannot describe the evolution of energy density inhomogeneities see e.g. Ratra,\cite{ratra91} and Podariu \textit{et al.}\cite{podariu2000}

The second model we consider is the simplest, complete and consistent dynamical dark energy model, 
$\phi$CDM, generalized to include non-zero spatial curvature Pavlov \textit{et al.}\cite{Anatoly2} 
In this case, dark energy is modeled as a slowly-rolling scalar field 
$\phi$ with an, e.g., inverse-power-law potential energy density 
$V(\phi)=\kappa m_p^2 \phi^{-\alpha}/2 $ where $m_p=1/\sqrt{G}$ is 
the Planck mass and $\alpha$ is a non-negative parameter that 
determines the coefficient $\kappa$.\citep{Peebles&Ratra1988} 
The scalar field part of the $\phi$CDM model action is:
\begin{equation}
   S=\frac{m_p^2}{32\pi}\int{\sqrt{-g}\left( ~g^{\mu \nu}
   \partial_\mu \phi \partial_\nu \phi - \kappa m_p^2 \phi^{-\alpha} 
   \right) d^4x},
\end{equation}
where the parameter $\kappa$ is: \citep{Peebles&Ratra1988,Anatoly2}
\begin{equation}
\kappa=\frac{8}{3}\left(\frac{2\alpha}{3}\right)^{\alpha /2}(\alpha +4)(\alpha +2)^{(\alpha -2)/2}.
\end{equation}

In this model, at the current
epoch, scalar field dark energy dominates the cosmological energy budget
and fuels the accelerating cosmological expansion. Prior to 
that, space curvature dominated and at even earlier times, non-relativistic
matter powered the decelerating cosmological expansion.
In the matter dominated epoch at
$a \ll a_0$, $\rho_\phi \ll \rho_{m}$ and $\rho_k \ll \rho_{m}$, the Einstein-de Sitter model applies, and the
initial conditions are that the cosmological scale factor evolves as $a(t)\propto t^{2/3}$,
the scalar field $\phi(t) \propto t^{2/(\alpha+2)}$, 
and the scalar field energy density evolves as 
$\rho_{\phi}\propto a^{-3\alpha/(\alpha+2)}\propto t^{-2\alpha/(\alpha+2)}$, as described in 
\cite{Peebles&Ratra1988}. In the space curvature dominated epoch $\rho_{\phi} \ll \rho_{k}$ and 
$\rho_{m} \ll \rho_{k}$ and $a(t) \propto t$, the scalar field $\phi (t) \propto t^{2/(2+\alpha)}$,
and the scalar field energy density evolves as $\rho_{\phi}\propto a^{-2\alpha/(2+\alpha)} \propto t^{-2\alpha/(2+\alpha)}$,\footnote{As
long as the scalar field energy density does not dominate, the scalar field energy density $\rho_{\phi}\propto
t^{-2\alpha/(2+\alpha)}$, independent of the type of matter that dominates.} as determined in Pavlov \textit{et al.}\cite{Anatoly2} Hence, for positive
values of $\alpha$, the scalar field energy decreases, but less
rapidly than that of space curvature in the space curvature 
dominated epoch ($\rho_k \propto a^{-2}\propto t^{-2}$)
and less rapidly than that of non-relativistic matter in the matter dominated epoch 
($\rho_{m}\propto a^{-3} \propto t^{-2}$). So at late times 
the Universe will become dark energy dominated \citep{Anatoly2,Ratra&Peebles1988}. As in the radiation and matter
dominated epochs, Peebles \& Ratra,\cite{Peebles&Ratra1988} Ratra \& Peebles,\cite{Ratra&Peebles1988} and Pavlov \textit{et al.},\cite{Anatoly2} show that in the curvature 
dominated epoch the solution for the scalar field is a time-dependent fixed point or attractor. This means that for a wide 
range of initial conditions the solution will approach this special time-dependent fixed point solution. 

The equation of motion of the scalar field is:
\begin{equation}
\label{eq:dotphi} 
    \ddot{\phi} + 3 \frac{\dot{a}}{a}\dot{\phi} -\frac{1}{2}
    \kappa \alpha m_p^2 \phi^{-(\alpha+1)} = 0.
\end{equation}
In the presence of spatial curvature the $\phi$CDM model 
Friedmann equation takes the form:
\begin{eqnarray}
\label{eq:phicdmfriedman}
   H^2(z; H_0, \textbf{p})\hspace{-2mm}
   &=&\hspace{-2mm} H_0^2[\Omega_{m0}(1+z)^3+\Omega_\phi (z,\alpha)+\Omega_{k0}(1+z)^2],
\end{eqnarray}
where the time-dependent scalar field density parameter $\Omega_{\phi}$ is defined as: 
\begin{eqnarray}
\label{eq:Omegaphi} 
   \Omega_\phi(z,\alpha) \equiv \frac{8 \pi G}{3H^{2}_{0}}\rho_\phi= 
   \frac{1}{12H^{2}_{0}}\left(\dot\phi^2+\kappa m^{2}_{p}\phi^{-\alpha}\right).
\end{eqnarray}
In the limit $\alpha=0$, the $\phi$CDM model
is equivalent to the ordinary time-independent 
cosmological constant $\Lambda$ model. 
This makes the $\phi$CDM model a generalization of the 
standard $\Lambda$CDM model of cosmology.

Solving the coupled differential 
equations\ ({\ref{eq:dotphi}})---({\ref{eq:Omegaphi}}), 
with the initial conditions described in Peebles \& Ratra,\cite{Peebles&Ratra1988} 
and Pavlov \textit{et al.},\cite{Anatoly2} allows for a numerical computation of 
the Hubble parameter $H(z; H_0, \textbf{p})$, as well
as the other functions needed for applications of the cosmological
tests. In this case the model parameters 
are taken to be $\textbf{p}=(\Omega_{m0},\alpha, \Omega_{k0})$.

\section{Observational Constraints}
\label{sec:Constraints}

To constrain cosmological parameters \textbf{p},
we generalize the technique described in Farooq \& Ratra,\cite{Farooq20131} to models
with three free parameters, $\textbf{p}=(\Omega_{m0},\omega_X, \Omega_{k0})$ 
for the XCDM parameterization and $\textbf{p}=(\Omega_{m0},\alpha, \Omega_{k0})$ for
$\phi$CDM. Following Farooq \textit{et al.},\cite{Farooq:2012ev} we compute a likelihood function 
$\mathcal{L}(\textbf{p})$ that depends on the three \textbf{p} parameters. We
compute these likelihood functions over the parameter ranges
$-0.7\leq \Omega_{k0} \leq 0.7$, $-2.0 \leq \omega_{X} \leq 0$, and  $ 0\leq \Omega_{m0} \leq 1.0$ 
for the XCDM parameterization, and $-0.2\leq \Omega_{k0} \leq 0.2$, $0 \leq \alpha \leq 5$, 
and  $ 0\leq \Omega_{m0} \leq 1.0$ for the $\phi$CDM model. For the sake of 
computational tractability the $\Omega_{k0}$ range considered in the case of $\phi$CDM
is much smaller than that used in the XCDM parameterization computation. 

To get two-dimensional likelihood functions $\mathcal{L}(\boldsymbol\theta)$, we
marginalize the three-dimensional likelihood function $\mathcal{L}(\textbf{p})$ over 
each of the three model parameters in turn, with flat priors.  Here
\begin{eqnarray}
\mathcal{L}(\boldsymbol\theta) \equiv \int \limits^{\beta_2}_{\beta_1}\mathcal{L}(\textbf{p})d\beta=\int \limits^{\beta_2}_{\beta_1}\mathcal{L}(\boldsymbol\theta,\beta)d\beta,
\label{eq:marg3to2}
\end{eqnarray}
where $\boldsymbol\theta$ is the set of two parameters at a time and $\beta$ is the
third parameter with marginalization limits of $\beta_1$ and $\beta_2$.

To maximize the two-dimensional likelihood function $\mathcal{L}(\boldsymbol\theta)$
we minimize $\chi^{2}(\boldsymbol\theta)\equiv
-2\mathrm{ln}\mathcal{L}(\boldsymbol\theta)$ with respect to model 
parameters $\boldsymbol\theta$ to find the best-fit 
parameter values $\boldsymbol\theta_0$. We define $1\sigma$, 
$2\sigma$, and $3\sigma$ confidence contours as two-dimensional 
parameter sets bounded by $\chi^2(\boldsymbol\theta) =
\chi^2(\boldsymbol\theta_0)+2.3,~\chi^2(\boldsymbol\theta) = 
\chi^2(\boldsymbol\theta_0)+6.17$, and $\chi^2(\boldsymbol\theta) = 
\chi^2(\boldsymbol\theta_0)+11.8$, respectively.\footnote{Farooq \& Ratra,\cite{Farooq:2013hq} 
found that the two-dimensional
contours obtained from integrating the likelihood function and
those obtained using the $\chi^2$ prescription described here hardly differ. 
To save computational time we use the $\chi^2$ prescription in this paper.}

\subsection{Constraints from $H(z)$, SNIa, and BAO data sets, one at a time}

We first consider $H(z)$ data constraints. For this we use 22 independent $H(z_i)$ 
measurements and one standard deviation uncertainties at redshift $z_i$, $i=1,2,\ldots,22$ (covering the redshift range of 0.09 to 2.3),
listed in Table\ (\ref{tab:Hz2})\footnote{In Farooq \& Ratra,\cite{Farooq:2013hq} 
we found that an augmented set of $H(z)$ measurements shows
clear evidence for the cosmological deceleration-acceleration transition predicted to 
occur in cosmological models dominated by dark energy at the current epoch. 
Farooq \textit{et al.},\cite{farooq4} more clearly illustrate the presence of this transition in the data by binning 
and combining the $H(z)$ data.} to constrain cosmological 
model parameters \textbf{p}. Using Eq.\ (18) of Farooq \textit{et al.},\cite{Farooq:2012ev}
which is obtained after marginalizing over the nuisance 
parameter $H_0$ using a Gaussian prior with
$H_0 = 68 \pm 2.8$ km s$^{-1}$ Mpc$^{-1}$,\footnote{As
discussed in \cite{Farooq:2012ev}, the constraint contours are sensitive
to the $H_0$ prior used. The $H_0$ prior we use is obtained from a 
median statistics analysis \citep{Gott2001} of 553 $H_0$ measurements \cite{Chen2011a},
and has been stable now for more than a decade \cite{Gott2001,chen03}.
Recent measurements of $H_0$ are consistent with this value \cite[see e.g.,][]{colless12,Planckdata}
although some suggest slightly larger or smaller values \cite[see e.g., ][]{Freedman2012,
Sorce2012,Tammann2012}. It may be significant that the value of $H_0$ we use does
not demand the presence of dark radiation calabrese \textit{et al.}\cite{calabrese12}}
we get a likelihood function $\mathcal{L}_H(\textbf{p})$ 
that depends only on model parameters \textbf{p} $=(\Omega_{m0}, 
\omega_X, \Omega_{k0})$ for the XCDM parameterization and $(\Omega_{m0}, \alpha, \Omega_{k0})$ for the 
$\phi$CDM model. Then using 
Eq.\ (\ref{eq:marg3to2}), we compute $\mathcal{L}_{H}(\boldsymbol\theta)$
from $\mathcal{L}_{H}(\textbf{p})$, and the two-dimensional confidence 
contours are obtained following the procedure discussed above.

To tighten constraints on model parameters we also
use a second data
set,  the Suzuki \textit{et al.},\cite{suzuki2012} Union2.1 compilation of 580 SNIa distance 
modulus measurements at measured redshifts
(covering the redshift range of 0.015 to 1.414) with corresponding
one standard deviation uncertainties including systematic uncertainties. To constrain 
cosmological model parameters using this data the three-dimensional 
likelihood function $\mathcal{L}_{SN}(\textbf{p})$ is defined by
generalizing Eq.\ (26) of Farooq \textit{et al.},\cite{Farooq:2012ev} and marginalizing over a 
flat $H_0$ prior (for these SNIa data). Then using 
Eq.\ (\ref{eq:marg3to2}) we determine $\mathcal{L}_{SN}(\boldsymbol\theta)$
from $\mathcal{L}_{SN}(\textbf{p})$, and the two-dimensional confidence 
contours are obtained as discussed above.
 
The third set of data we consider are the 6 BAO peak length scale
measurements (covering the redshift range of 0.1 to 0.75) with corresponding one standard deviation uncertainties 
from Percival \textit{et al.},\cite{Percival2010} Beutler \textit{et al.},\cite{Beutler11} and Blake \textit{et al.}\cite{blake11}. 
To constrain model parameters \textbf{p}
we compute the three-dimensional likelihood function $\mathcal{L}_{BAO}(\textbf{p})$
by again marginalizing over a flat $H_0$ prior (for this BAO data), as discussed in
Sec.\ 5 of Farooq \textit{et al.}\cite{Farooq:2012ev} 
Then using Eq.\ (\ref{eq:marg3to2}) we compute $\mathcal{L}_{BAO}(\boldsymbol\theta)$
from $\mathcal{L}_{BAO}(\textbf{p})$, and the two-dimensional confidence 
contours are obtained using the procedure discussed above.

\begin{figure}[h!]
\centering
    \includegraphics[height=1.95in]{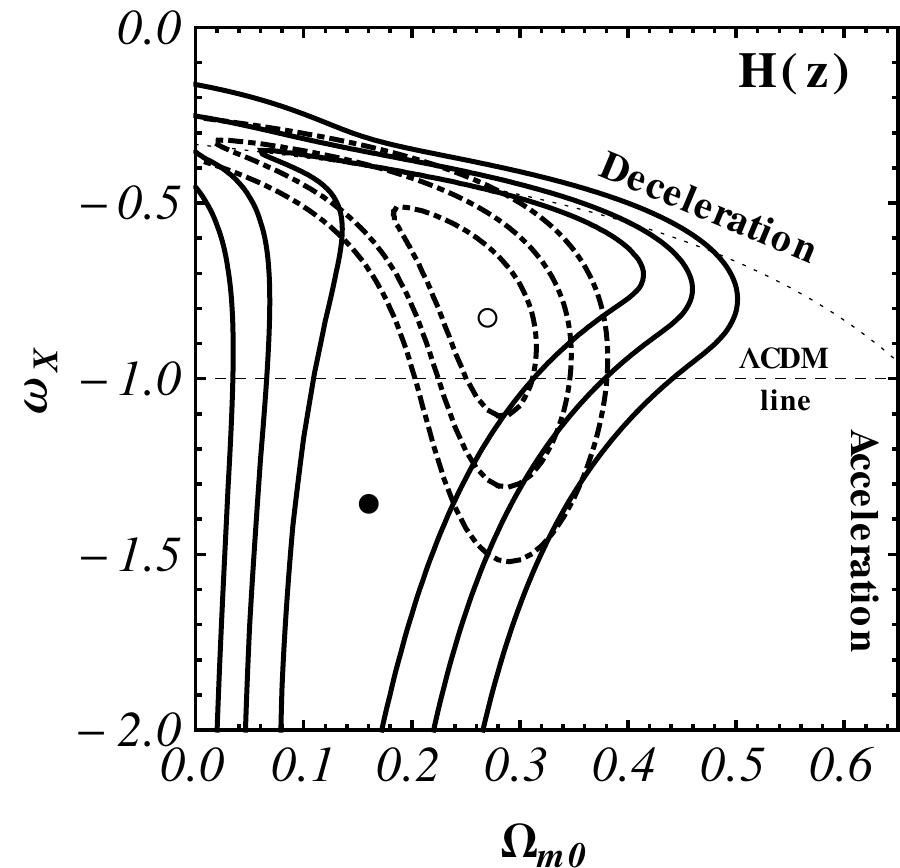}
    \includegraphics[height=1.95in]{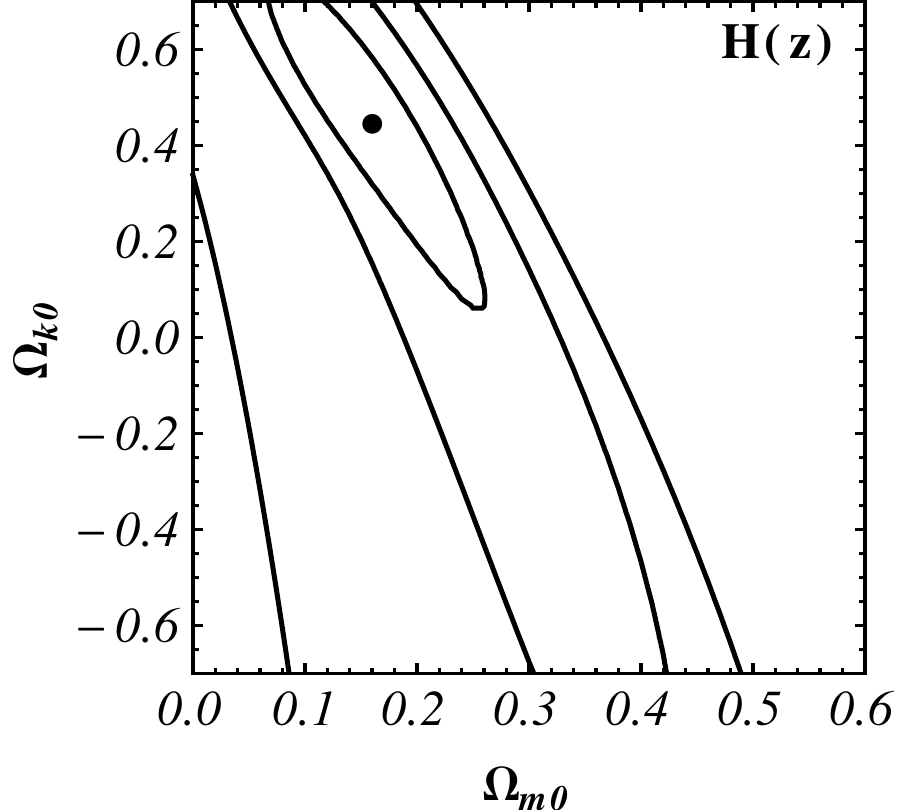}
    \includegraphics[height=1.95in]{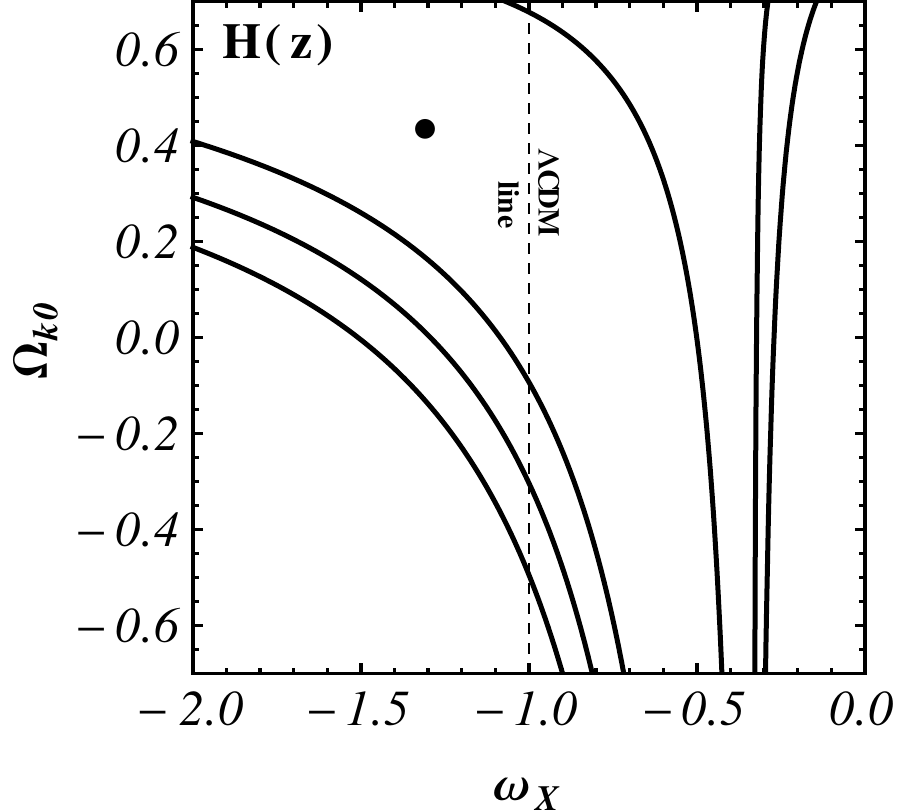}
    \includegraphics[height=1.95in]{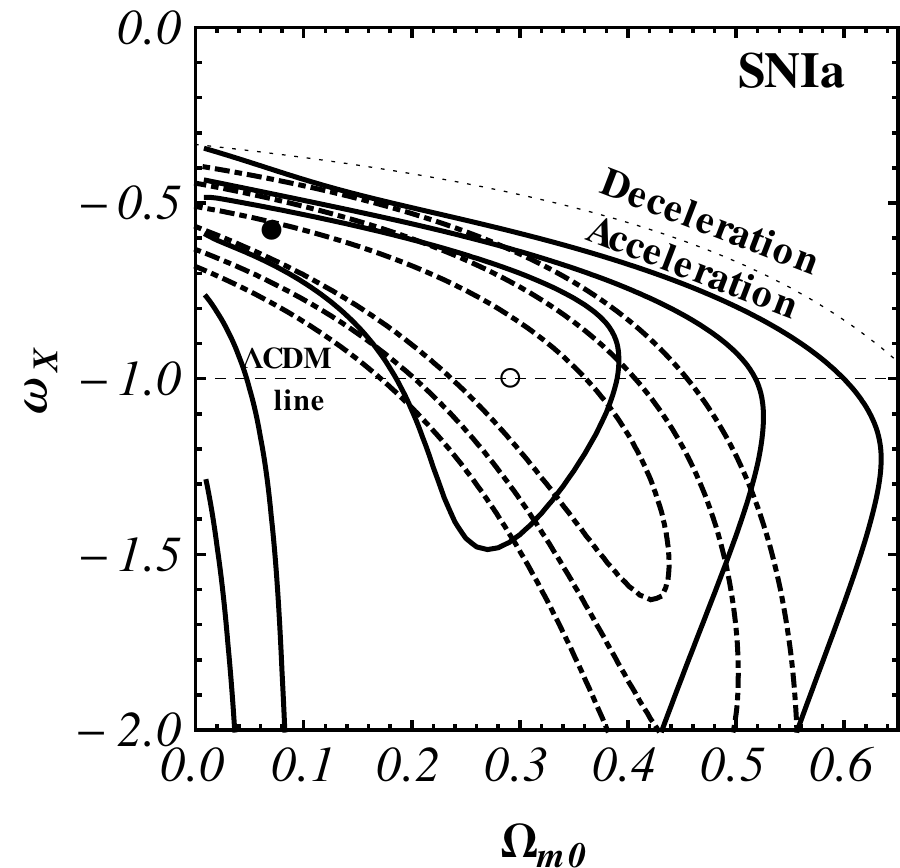}
    \includegraphics[height=1.95in]{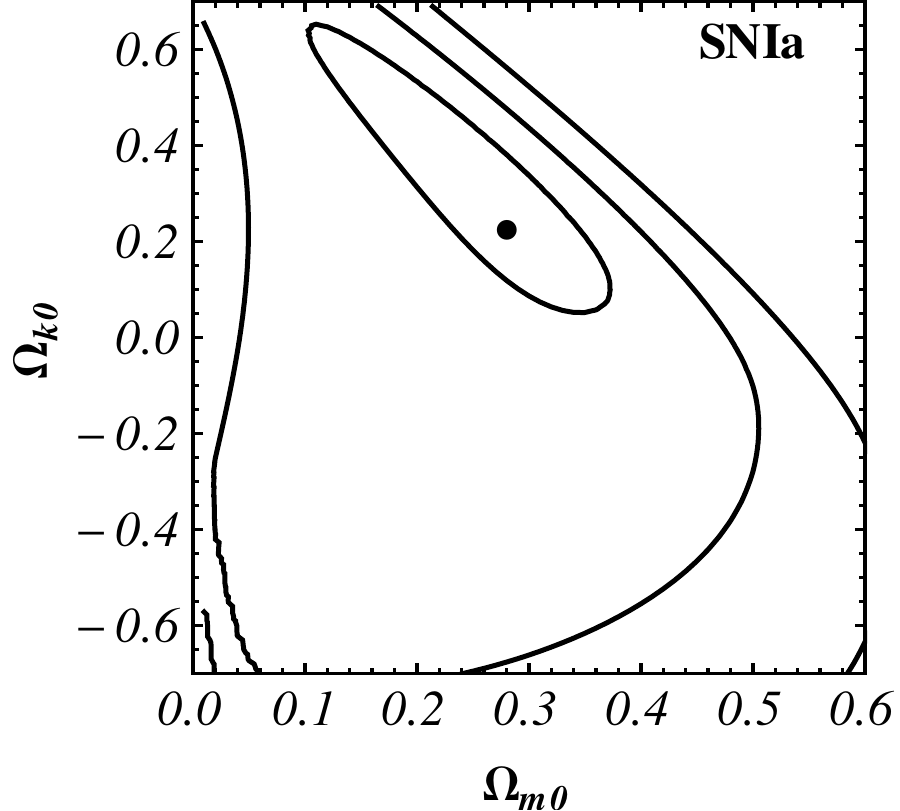}
    \includegraphics[height=1.95in]{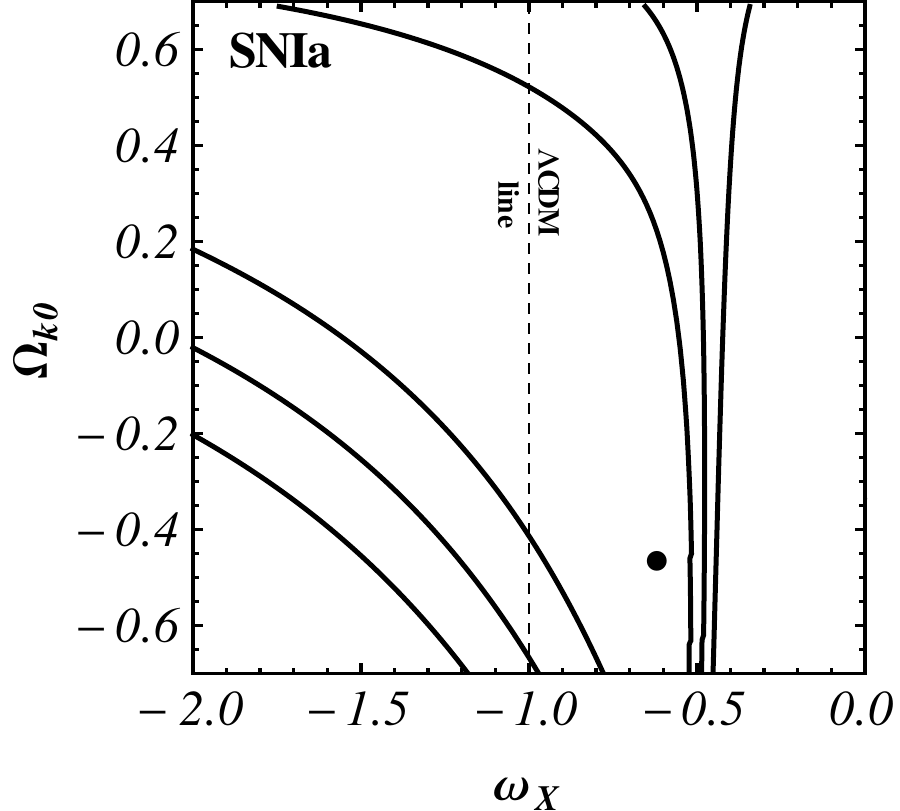}
    \includegraphics[height=1.95in]{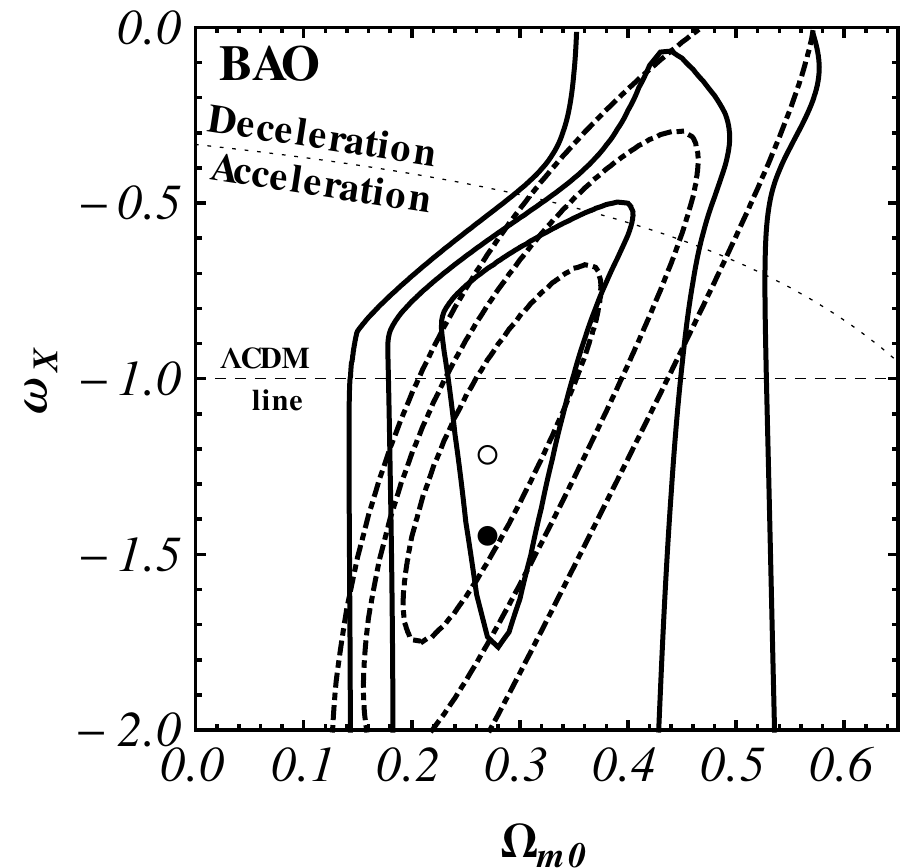}
    \includegraphics[height=1.95in]{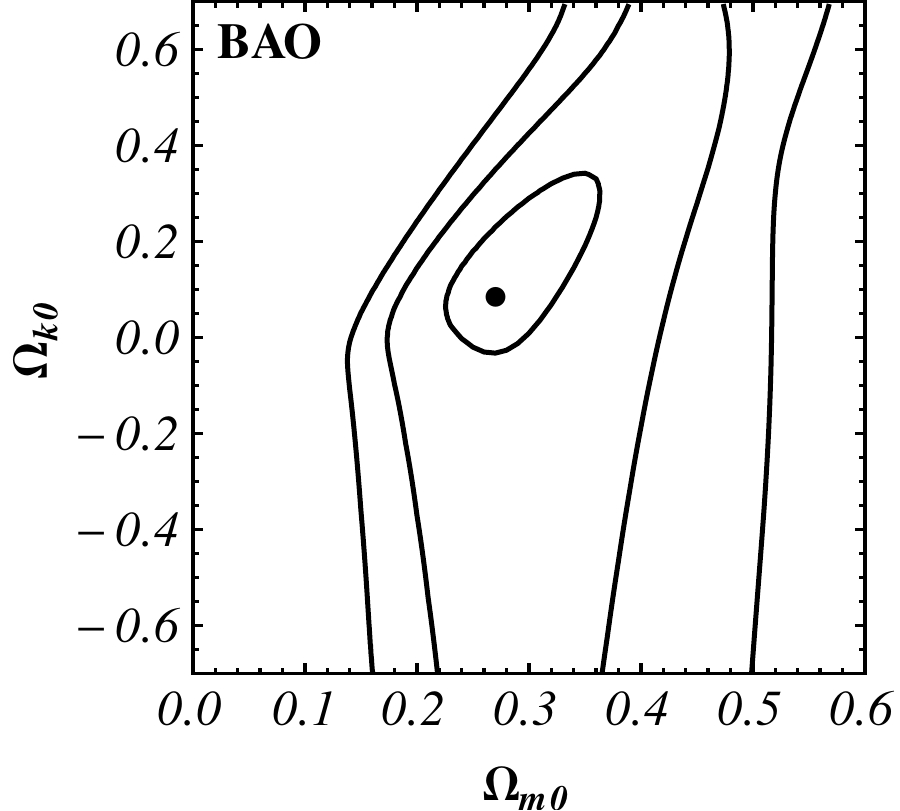}
    \includegraphics[height=1.95in]{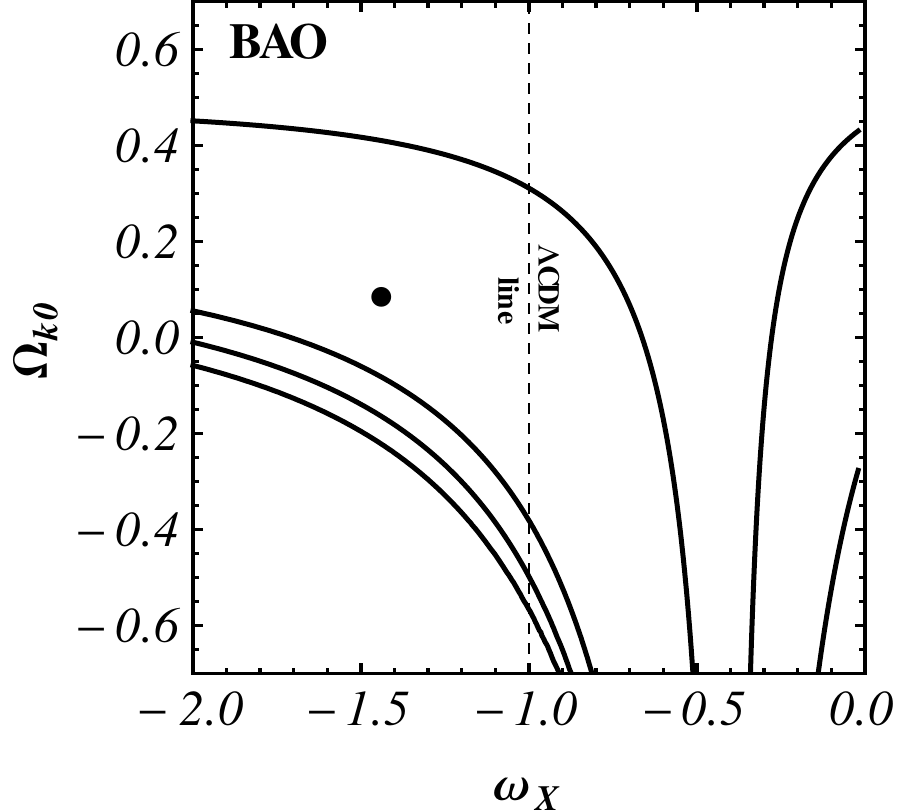}
\caption{%
$1\sigma$, $2\sigma$, and $3\sigma$ constraint contours (solid lines) for parameters of the non-flat XCDM dark energy parameterization from 
$H(z)$ (first row), SNIa (second row), and BAO (third row) measurements; filled circles show best-fit points. The dot-dashed lines in the first column panels are $1\sigma$, $2\sigma$, and $3\sigma$ constraint contours
derived by Farooq \& Ratra,\cite{Farooq20131} using the spatially-flat XCDM parameterization (open circles show best-fit points); here dotted lines distinguish between
accelerating and decelerating models (at zero space curvature) and dashed lines (here and in the third column) correspond to the 
$\Lambda$CDM model. The first, second, and third columns correspond to marginalizing over $\Omega_{k0}$, 
$\omega_X$, and $\Omega_{m0}$ respectively.
}
\label{fig:XCDM_S}
\end{figure}

\begin{figure}[h!]
\centering
    \includegraphics[height=1.95in]{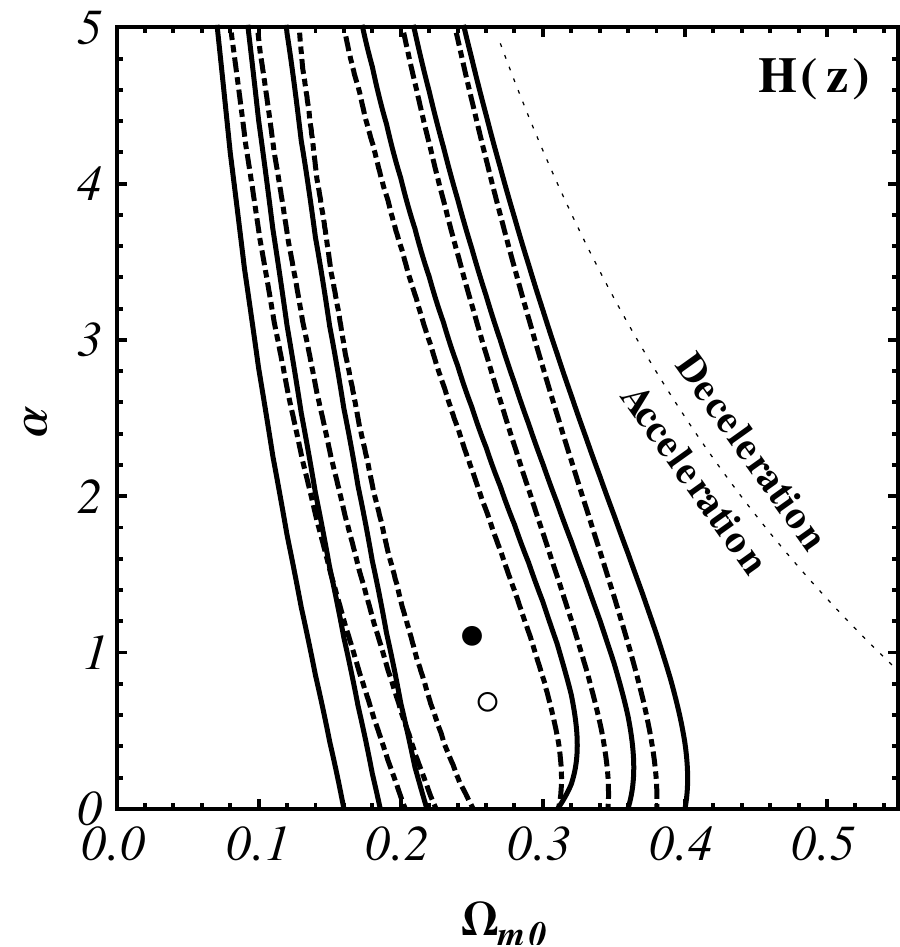}
    \includegraphics[height=1.95in]{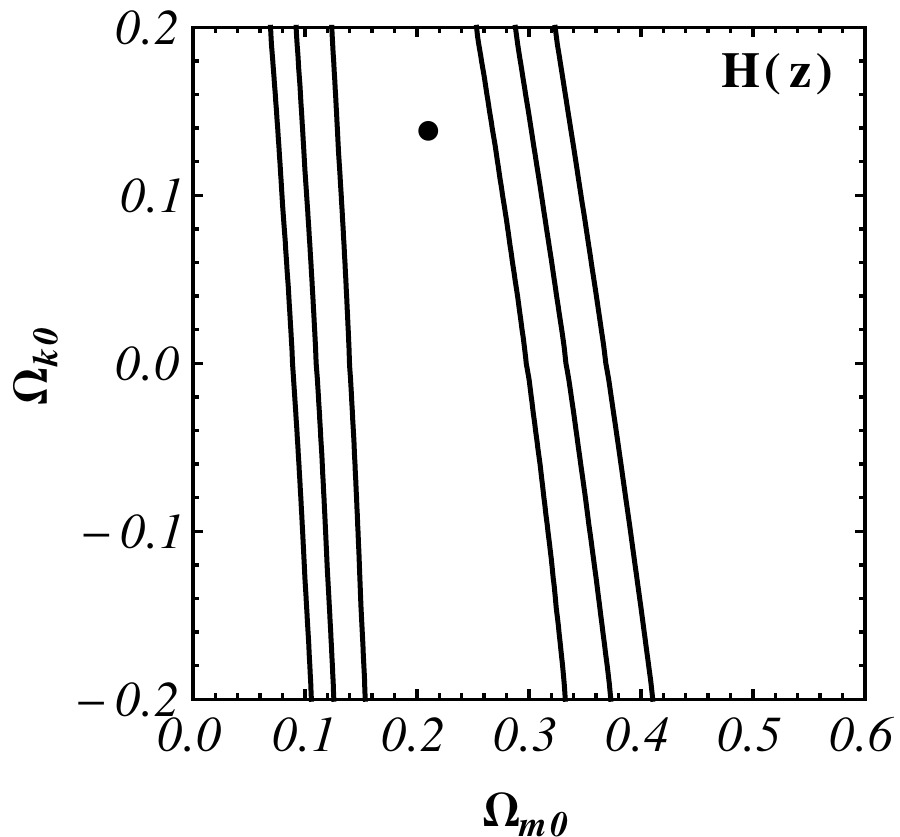}
    \includegraphics[height=1.95in]{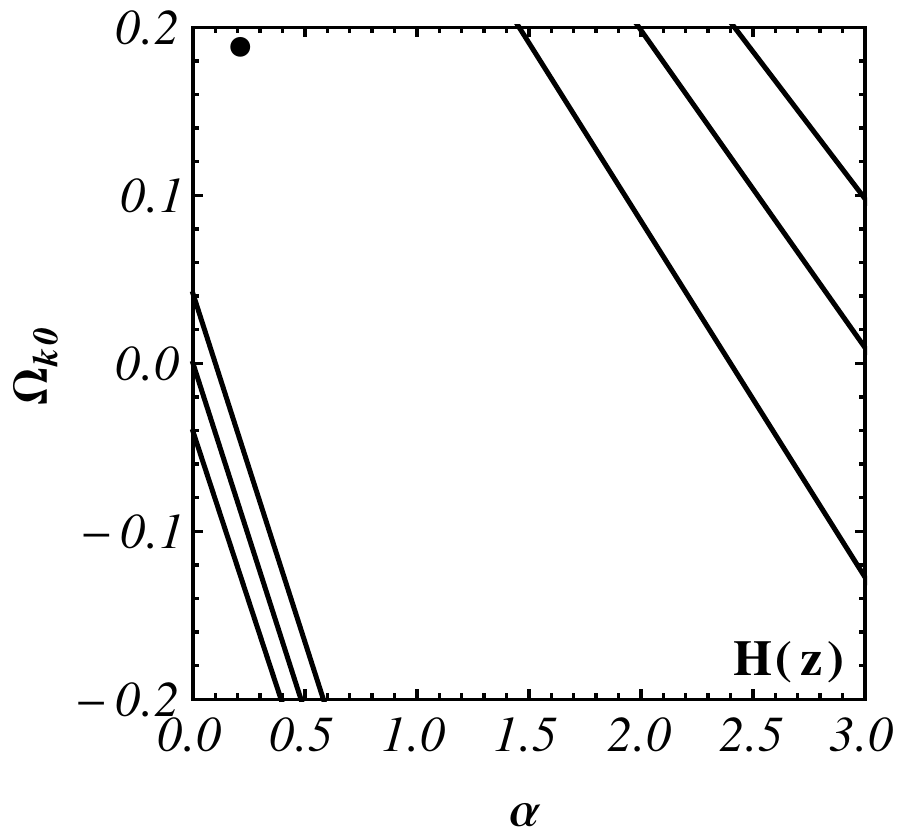}
    \includegraphics[height=1.95in]{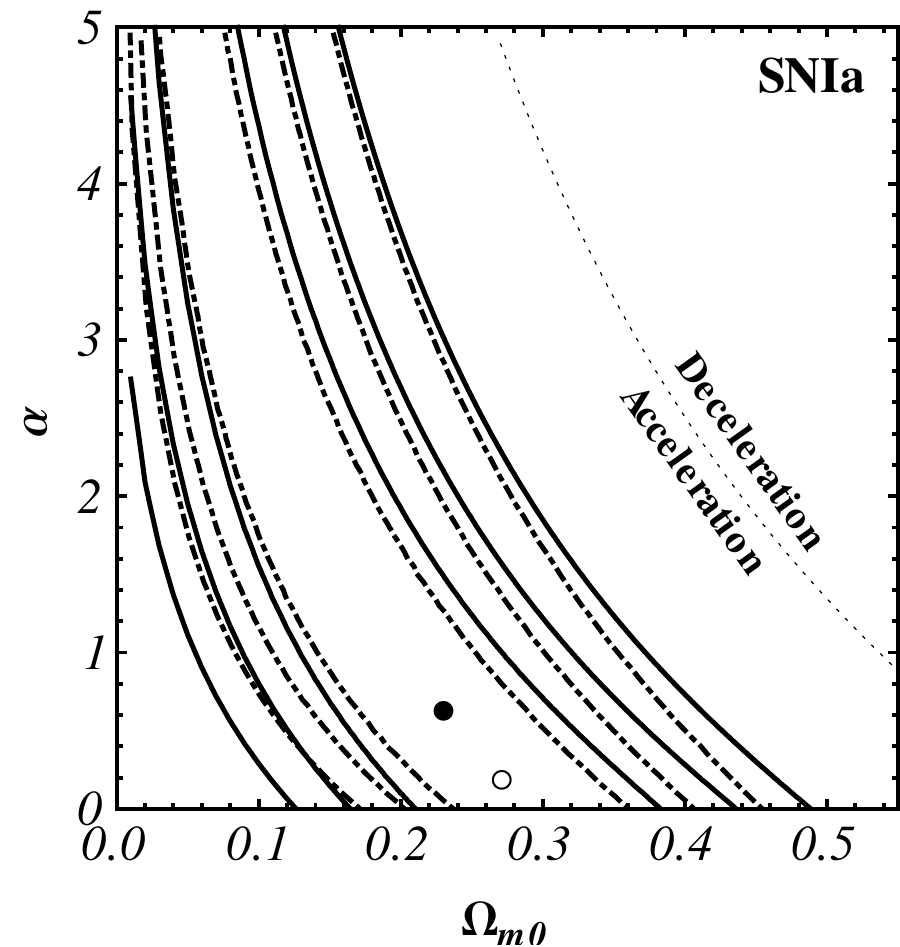}
    \includegraphics[height=1.95in]{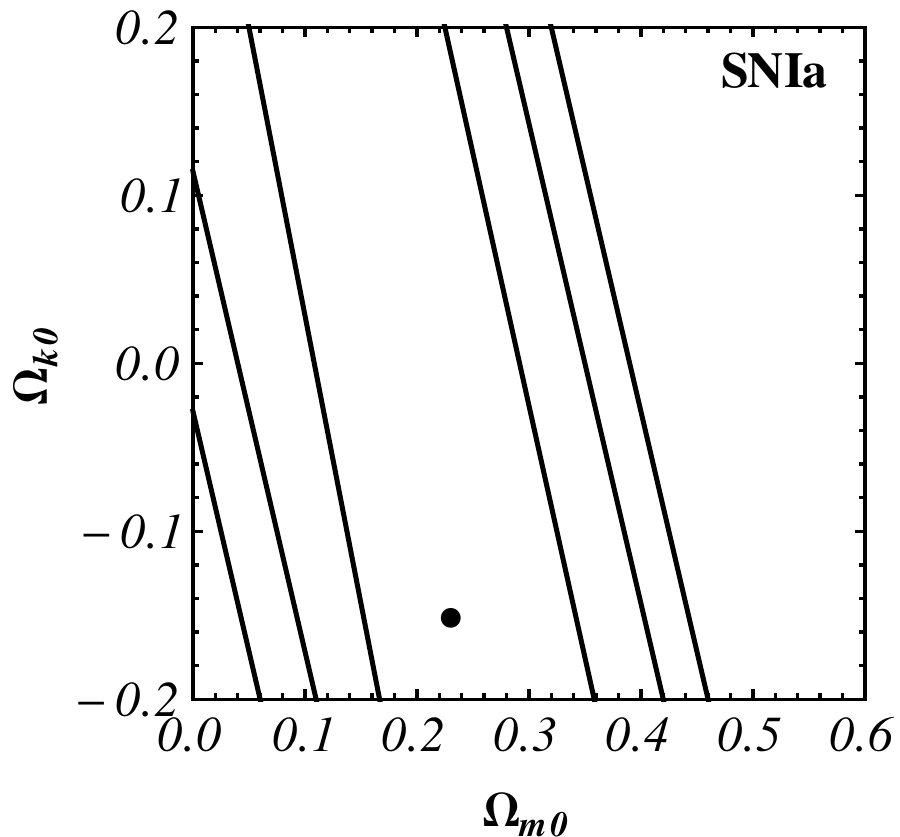}
    \includegraphics[height=1.95in]{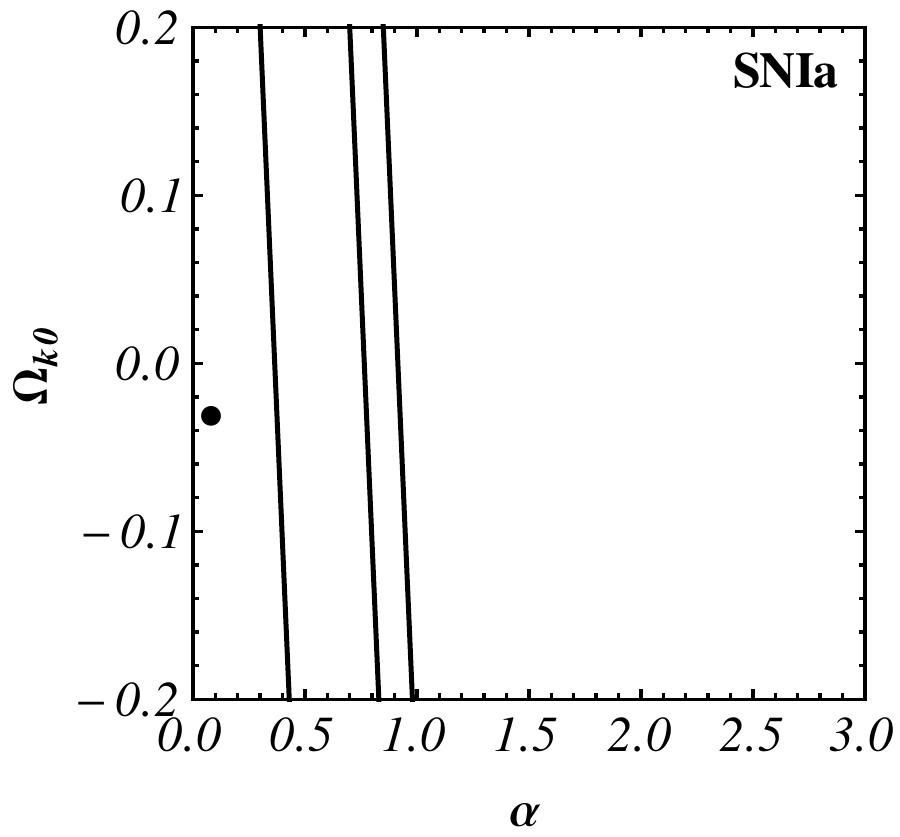}
    \includegraphics[height=1.95in]{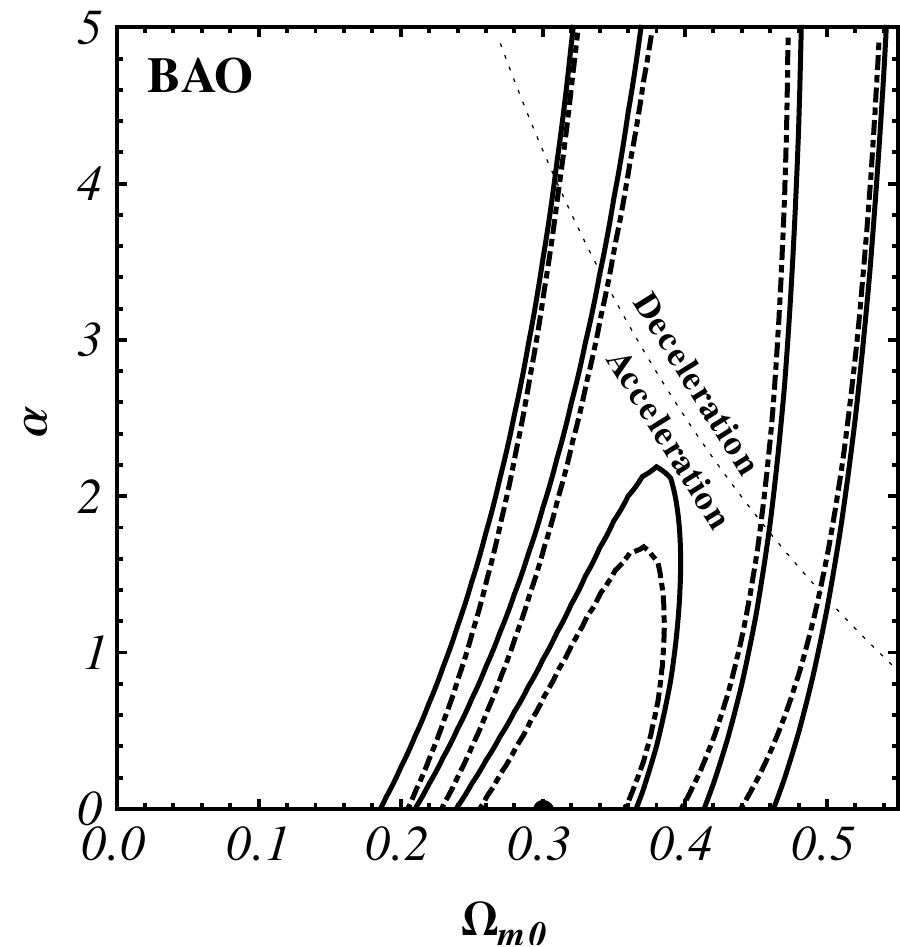}
    \includegraphics[height=1.95in]{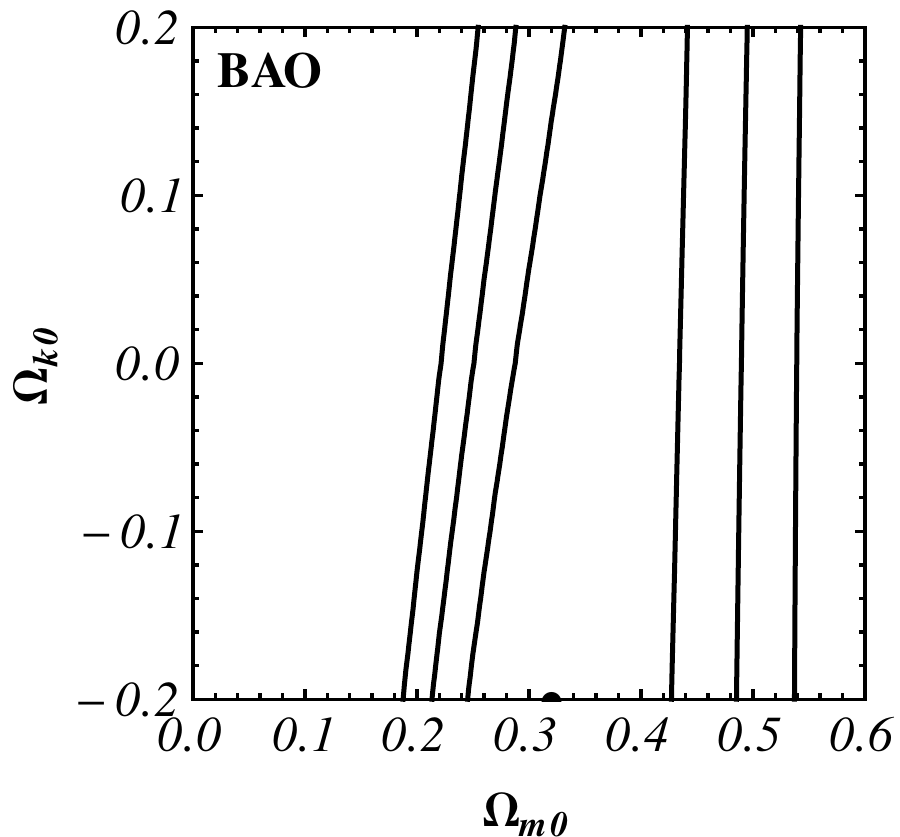}
    \includegraphics[height=1.95in]{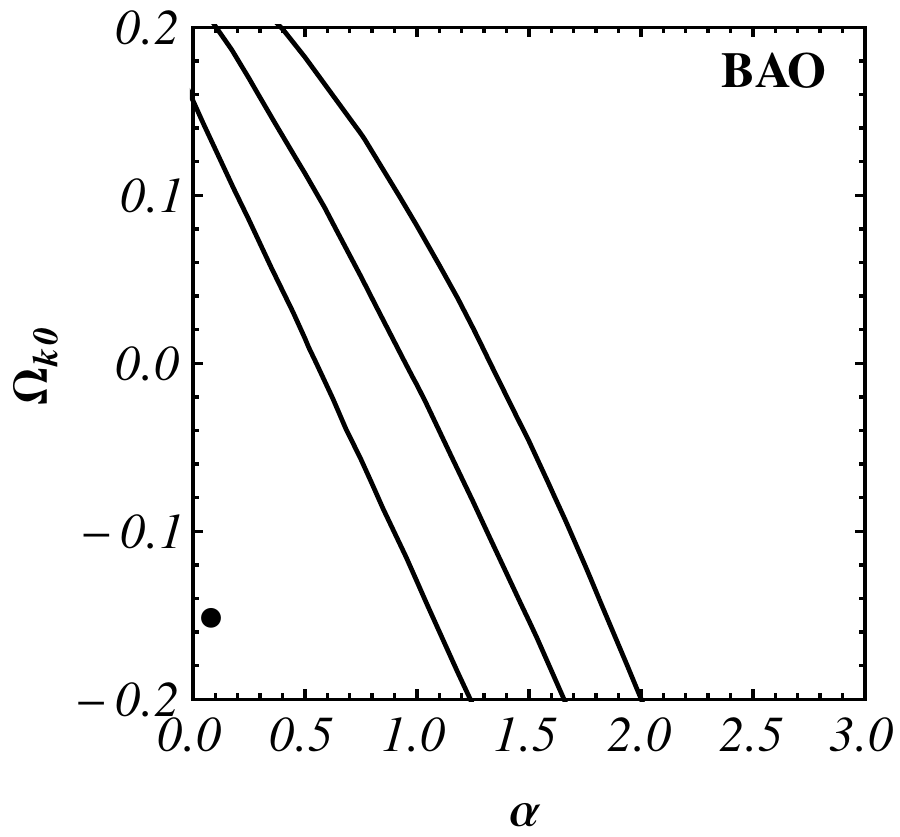}
\caption{
$1\sigma$, $2\sigma$, and $3\sigma$ constraint contours (solid lines) for parameters of the non-flat $\phi$CDM dark energy model from 
$H(z)$ (first row), SNIa (second row), and BAO (third row) measurements; filled circles show best-fit points. The dot-dashed lines in the first column panels are $1\sigma$, $2\sigma$, and $3\sigma$ constraint contours
derived by Farooq \& Ratra,\cite{Farooq20131} using the spatially-flat $\phi$CDM model (open circles show best-fit points); here dotted lines distinguish between
accelerating and decelerating models (at zero space curvature) and  the $\alpha=0$ axes (here and in the third column) correspond to the $\Lambda$CDM model. The first, second, and third columns correspond to marginalizing over $\Omega_{k0}$, 
$\alpha$, and $\Omega_{m0}$ respectively.
} 
\label{fig:phiCDM_S}
\end{figure}

\renewcommand{\arraystretch}{1.1}
\begin{deluxetable}{cccc}
\tablecaption{XCDM Parameterization Results}
\tablewidth{0pt}
\tabletypesize{\small}
\tablehead{
\colhead{\multirow{1}{*}{Data Set}}& 
\colhead{\multirow{1}{*}{Marginalization Range}}& 
\colhead{\multirow{1}{*}{Best-Fit Point}}& 
\colhead{$\chi^2_{\mathrm{min}}$}\\
\vspace{-4 mm}
}
\startdata
\noalign{\vskip -1mm}
\multirow{4}{*}{$H(z)$}& $ \Omega_{k0}=0$\tablenotemark{a} & $(\Omega_{m0}, \omega_X)=(0.27,-0.82)$ & 15.2\\
\cline{2-4} 
{}& $-0.7\leqslant \Omega_{k0} \leqslant 0.7$ & $(\Omega_{m0}, \omega_X)=(0.16,-1.35)$ & 17.8\\ 
 {}& $-2\leqslant \omega_{X} \leqslant 0$ & $(\Omega_{m0}, \Omega_{k0})=(0.16,0.45)$ & 14.5\\
 {}& $0\leqslant \Omega_{m0} \leqslant 1$ & $(\omega_X, \Omega_{k0})=(-1.31,0.44)$ & 20.4\\
\hline
\multirow{4}{*}{$\mathrm{SNIa}$} & $ \Omega_{k0}=0$\tablenotemark{a} & $(\Omega_{m0}, \omega_X)=(0.29,-0.99)$ & 545\\
\cline{2-4} 
{} & $-0.7\leqslant \Omega_{k0} \leqslant 0.7$ & $(\Omega_{m0}, \omega_X)=(0.07,-0.57)$ & 546\\
{} & $-2\leqslant \omega_{X} \leqslant 0$ & $(\Omega_{m0}, \Omega_{k0})=(0.28,0.23)$ & 545\\
{} & $0\leqslant \Omega_{m0} \leqslant 1$ & $(\omega_X, \Omega_{k0})=(-0.62,-0.46)$ & 549\\
\hline
\multirow{4}{*}{$\mathrm{BAO}$} & $ \Omega_{k0}=0$\tablenotemark{a} & $(\Omega_{m0}, \omega_X)=(0.27,-1.21)$ & 5.50\\
\cline{2-4} 
{} & $-0.7\leqslant \Omega_{k0} \leqslant 0.7$ & $(\Omega_{m0}, \omega_X)=(0.27,-1.44)$ & 6.50\\
{} & $-2\leqslant \omega_{X} \leqslant 0$ & $(\Omega_{m0}, \Omega_{k0})=(0.27,0.09)$ & 4.90\\
{} & $0\leqslant \Omega_{m0} \leqslant 1$ & $(\omega_X, \Omega_{k0})=(-1.44,-0.09)$ & 10.4\\
\hline
\multirow{4}{*}{$H(z)+\mathrm{SNIa}$} & $ \Omega_{k0}=0$\tablenotemark{a} & $(\Omega_{m0}, \omega_X)=(0.27,-0.90)$ & 561\\
\cline{2-4}
{} & $-0.7\leqslant \Omega_{k0} \leqslant 0.7$ & $(\Omega_{m0}, \omega_X)=(0.24,-0.97)$ & 562\\
{} & $-2\leqslant \omega_{X} \leqslant 0$ & $(\Omega_{m0}, \Omega_{k0})=(0.18,0.41)$ & 561\\
{} & $0\leqslant \Omega_{m0} \leqslant 1$ & $(\omega_X, \Omega_{k0})=(-0.98,0.15)$ & 566\\
\hline
\multirow{4}{*}{$H(z)+\mathrm{BAO}$} & $ \Omega_{k0}=0$\tablenotemark{a} & $(\Omega_{m0}, \omega_X)=(0.29,-0.99)$ & 22.4\\
\cline{2-4}
{} & $-0.7\leqslant \Omega_{k0} \leqslant 0.7$ & $(\Omega_{m0}, \omega_X)=(0.31,-0.79)$ & 24.2\\
{} & $-2\leqslant \omega_{X} \leqslant 0$ & $(\Omega_{m0}, \Omega_{k0})=(0.31,-0.19)$ & 26.9\\
{} & $0\leqslant \Omega_{m0} \leqslant 1$ & $(\omega_X, \Omega_{k0})=(-0.78,-0.19)$ & 27.5\\
\hline
\multirow{4}{*}{$\mathrm{SNIa}+\mathrm{BAO}$} & $ \Omega_{k0}=0$\tablenotemark{a} & $(\Omega_{m0}, \omega_X)=(0.30,-1.03)$ & 551\\
\cline{2-4}
{} & $-0.7\leqslant \Omega_{k0} \leqslant 0.7$ & $(\Omega_{m0}, \omega_X)=(0.29,-0.77)$ & 553\\
{} & $-2\leqslant \omega_{X} \leqslant 0$ & $(\Omega_{m0}, \Omega_{k0})=(0.31,0.22)$ & 552\\
{} & $0\leqslant \Omega_{m0} \leqslant 1$ & $(\omega_X, \Omega_{k0})=(-0.93,-0.10)$ & 556\\
\hline
\multirow{4}{*}{$H(z)+\mathrm{SNIa}+\mathrm{BAO}$} & $ \Omega_{k0}=0$\tablenotemark{a} & $(\Omega_{m0}, \omega_X)=(0.31,-1.02)$ & 566\\
\cline{2-4}
{}& $-0.7\leqslant \Omega_{k0} \leqslant 0.7$ & $(\Omega_{m0}, \omega_X)=(0.30,-0.88)$ & 571\\
{} & $-2\leqslant \omega_{X} \leqslant 0$ & $(\Omega_{m0}, \Omega_{k0})=(0.29,-0.15)$ & 582\\
{} & $0\leqslant \Omega_{m0} \leqslant 1$ & $(\omega_X, \Omega_{k0})=(-0.90,-0.10)$ & 573\\
\enddata
\tablenotetext{a}{From Farooq \& Ratra.\cite{Farooq20131}}
\label{table:XCDMresults}
\end{deluxetable}


\renewcommand{\arraystretch}{1.1}
\begin{deluxetable}{cccc}
\tablecaption{$\phi$CDM Model Results}
\tablewidth{0pt}
\tabletypesize{\small}
\tablehead{
\colhead{\multirow{1}{*}{Data Set}}& 
\colhead{\multirow{1}{*}{Marginalization Range}}& 
\colhead{\multirow{1}{*}{Best-Fit Point}}& 
\colhead{$\chi^2_{\mathrm{min}}$}\\
\vspace{-4 mm}
}
\startdata
\noalign{\vskip -1mm}
\multirow{4}{*}{$H(z)$} & $ \Omega_{k0} =0$\tablenotemark{a} & $(\Omega_{m0}, \alpha)=(0.26,0.70)$ & 15.2\\
\cline{2-4}
{} & $-0.2\leqslant \Omega_{k0} \leqslant 0.2$ & $(\Omega_{m0}, \alpha)=(0.25,1.12)$ & 17.8\\
{} & $0\leqslant \alpha \leqslant 5$ & $(\Omega_{m0}, \Omega_{k0})=(0.21,0.14)$ & 13.9\\
{} & $0\leqslant \Omega_{m0} \leqslant 1$ & $(\alpha, \Omega_{k0})=(0.21,0.19)$ & 20.4\\
\hline
\multirow{4}{*}{$\mathrm{SNIa}$} & $ \Omega_{k0} =0$\tablenotemark{a} & $(\Omega_{m0}, \alpha)=(0.27,0.20)$ & 545\\
\cline{2-4}
{} & $-0.2\leqslant \Omega_{k0} \leqslant 0.2$ & $(\Omega_{m0}, \alpha)=(0.23,0.64)$ & 548\\
{} & $0\leqslant \alpha \leqslant 5$ & $(\Omega_{m0}, \Omega_{k0})=(0.23,-0.15)$ & 547\\
{} & $0\leqslant \Omega_{m0} \leqslant 1$ & $(\alpha, \Omega_{k0})=(0.08,-0.03)$ & 550\\
\hline
\multirow{4}{*}{$\mathrm{BAO}$} & $ \Omega_{k0} =0$\tablenotemark{a} & $(\Omega_{m0}, \alpha)=(0.30,0.00)$ & 5.9\\
\cline{2-4}
{} & $-0.2\leqslant \Omega_{k0} \leqslant 0.2$ & $(\Omega_{m0}, \alpha)=(0.30,0.01)$ & 8.30\\
{} &  $0\leqslant \alpha \leqslant 5$ & $(\Omega_{m0}, \Omega_{k0})=(0.32,-0.20)$ & 5.50\\
{} & $0\leqslant \Omega_{m0} \leqslant 1$ & $(\alpha, \Omega_{k0})=(0.08,-0.15)$ & 10.6\\
\hline
\multirow{4}{*}{$H(z)+\mathrm{SNIa}$} & $ \Omega_{k0} =0$\tablenotemark{a} & $(\Omega_{m0}, \alpha)=(0.26,0.35)$ & 561\\
\cline{2-4}
{} & $-0.2\leqslant \Omega_{k0} \leqslant 0.2$ & $(\Omega_{m0}, \alpha)=(0.26,0.31)$ & 564\\
{} & $0\leqslant \alpha \leqslant 5$ & $(\Omega_{m0}, \Omega_{k0})=(0.25,0.11)$ & 562\\
{} & $0\leqslant \Omega_{m0} \leqslant 1$ & $(\alpha, \Omega_{k0})=(0.09,0.08)$ & 567\\
\hline
\multirow{4}{*}{$H(z)+\mathrm{BAO}$} & $ \Omega_{k0} =0$\tablenotemark{a} & $(\Omega_{m0}, \alpha)=(0.29,0.00)$ & 22.4\\
\cline{2-4}
{} & $-0.2\leqslant \Omega_{k0} \leqslant 0.2$ & $(\Omega_{m0}, \alpha)=(0.30,0.34)$ & 25.2\\
{} & $0\leqslant \alpha \leqslant 5$ & $(\Omega_{m0}, \Omega_{k0})=(0.31,-0.20)$ & 21.9\\
{} & $0\leqslant \Omega_{m0} \leqslant 1$ & $(\alpha, \Omega_{k0})=(0.77,-0.20)$ & 27.5\\
\hline
\multirow{4}{*}{$\mathrm{SNIa}+\mathrm{BAO}$} & $ \Omega_{k0} =0$\tablenotemark{a} & $(\Omega_{m0}, \alpha)=(0.30,0.00)$ & 551\\
\cline{2-4}
{} & $-0.2\leqslant \Omega_{k0} \leqslant 0.2$ & $(\Omega_{m0}, \alpha)=(0.30,0.08)$ & 554\\
{} & $0\leqslant \alpha \leqslant 5$ & $(\Omega_{m0}, \Omega_{k0})=(0.30,-0.05)$ & 553\\
{} & $0\leqslant \Omega_{m0} \leqslant 1$ & $(\alpha, \Omega_{k0})=(0.02,-0.03)$ & 557\\
\hline
\multirow{4}{*}{$H(z)+\mathrm{SNIa}+\mathrm{BAO}$} & $ \Omega_{k0} =0$\tablenotemark{a} & $(\Omega_{m0}, \alpha)=(0.29,0.00)$ & 567\\
\cline{2-4}
{} & $-0.2\leqslant \Omega_{k0} \leqslant 0.2$ & $(\Omega_{m0}, \alpha)=(0.30,0.46)$ & 571\\
{} & $0\leqslant \alpha \leqslant 5$ & $(\Omega_{m0}, \Omega_{k0})=(0.30,-0.05)$ & 569\\
{} & $0\leqslant \Omega_{m0} \leqslant 1$ & $(\alpha, \Omega_{k0})=(0.01,0.00)$ & 573\\
\enddata
\tablenotetext{a}{From Farooq \& Ratra.\cite{Farooq20131}}
\label{table:phiCDMresults}
\end{deluxetable}


Figures\ (\ref{fig:XCDM_S}) and
(\ref{fig:phiCDM_S}) show the constraints on parameters of the XCDM parameterization
and the $\phi$CDM model from the $H(z)$ (top row), SNIa (middle row), 
and BAO (bottom row) measurements. In these figures the panels in the first, second, and
third columns show the two-dimensional probability density 
constraint contours (solid lines) from $\mathcal{L}(\Omega_{m0},\omega_X)
[\mathcal{L}(\Omega_{m0},\alpha)]$, $\mathcal{L}(\Omega_{m0},\Omega_{k0})$, and
$\mathcal{L}(\omega_X,\Omega_{k0})[\mathcal{L}(\alpha,\Omega_{k0})]$ for the
XCDM parameterization [the $\phi$CDM model]. The dot-dashed contours in the panels of
the first columns of Figs.\ (\ref{fig:XCDM_S}) and 
(\ref{fig:phiCDM_S}) are $1\sigma$, $2\sigma$, and $3\sigma$ confidence contours
corresponding to spatially-flat models, reproduced from Farooq \& Ratra.\cite{Farooq20131}
Tables\ (\ref{table:XCDMresults})
and (\ref{table:phiCDMresults}) list best-fit points and $\chi^2_{\mathrm{min}}$ values.

Comparing the solid contours to the dot-dashed contours in the panels in the first columns of 
Figs.\ (\ref{fig:XCDM_S}) and (\ref{fig:phiCDM_S}), we see that the addition of space curvature
as a third free parameter results in a fairly significant broadening of the 
constraint contours, as might have been anticipated.

For the XCDM parameterization (first column of Fig.\ (\ref{fig:XCDM_S})), since
the data constrain $\omega_X$ reasonably well in the spatially-flat case, the
inclusion of space curvature as a free parameter significantly weakens the bounds on
$\omega_X$. For the $\phi$CDM model (first column of Fig.\ (\ref{fig:phiCDM_S}))
the data do not constrain $\alpha$ (the corresponding parameter that governs
the time-variability of dark energy in this case) as tightly in the 
spatially-flat case, so inclusion of space curvature appears to have a relatively
less significant effect (this is probably also a consequence of the significantly smaller
$\Omega_{k0}$ range considered, $-0.2 \leq \Omega_{k0} \leq 0.2$, for
computational tractability). This interplay between space curvature and the parameter that 
governs the time-variability of dark energy is also evident in the second
and third columns of panels of Figs.\ (\ref{fig:XCDM_S}) and (\ref{fig:phiCDM_S}).
Clearly, for the single data sets, including space curvature in the analysis
significantly weakens the support for a constant cosmological constant $\Lambda$,
while allowing dark energy density to be dynamical significantly weakens 
support for a spatially-flat model.

These results show very clearly that when spatial curvature is a free parameter a 
single data set cannot significantly constrain cosmological parameters of
the dynamical dark energy models considered here. 
To tighten constraints on cosmological parameters, we next consider combinations of data sets.

\subsection{Constraints from Combinations of Data Sets}

\begin{figure}[h!]
\centering
    \includegraphics[height=1.95in]{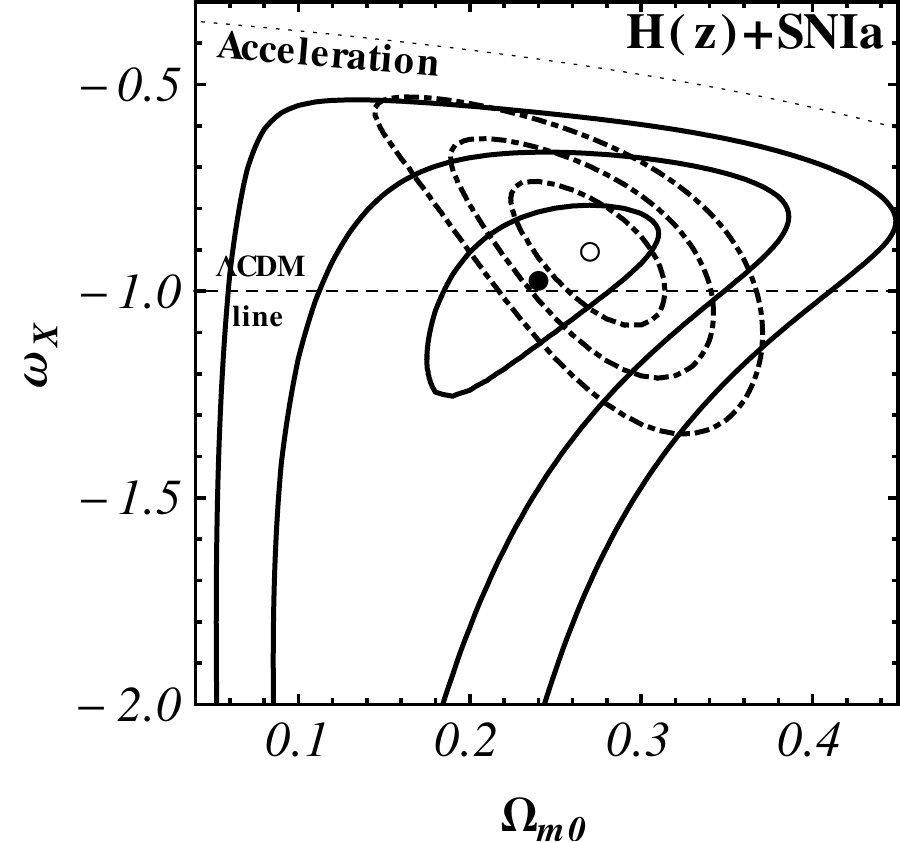}
    \includegraphics[height=1.95in]{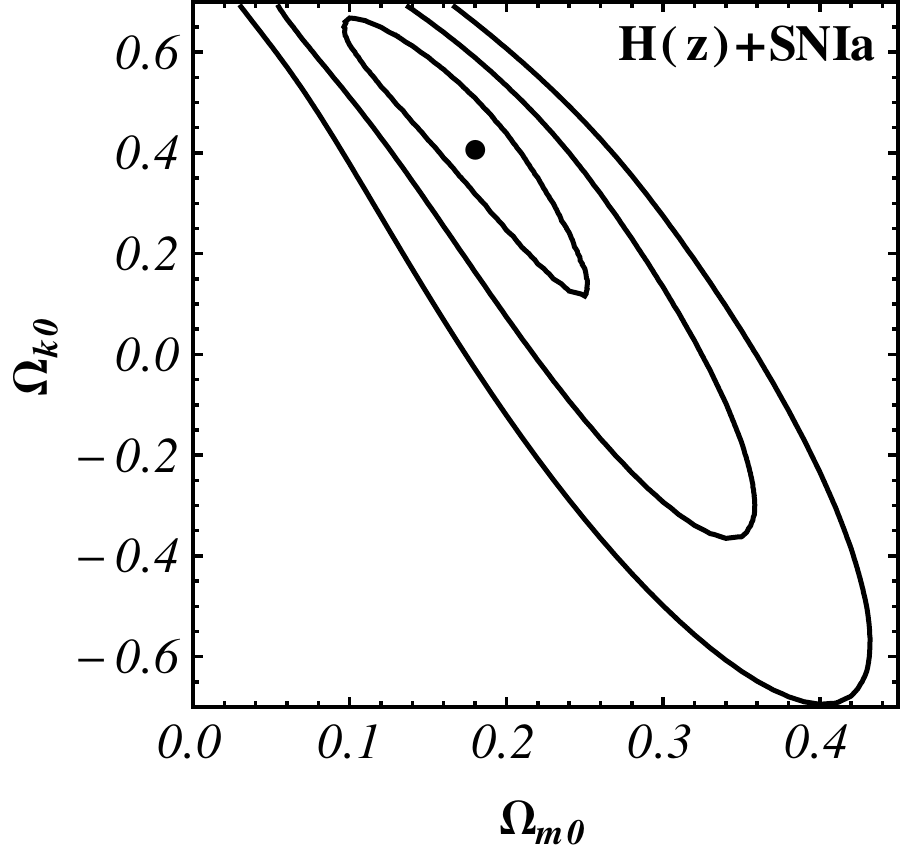}
    \includegraphics[height=1.95in]{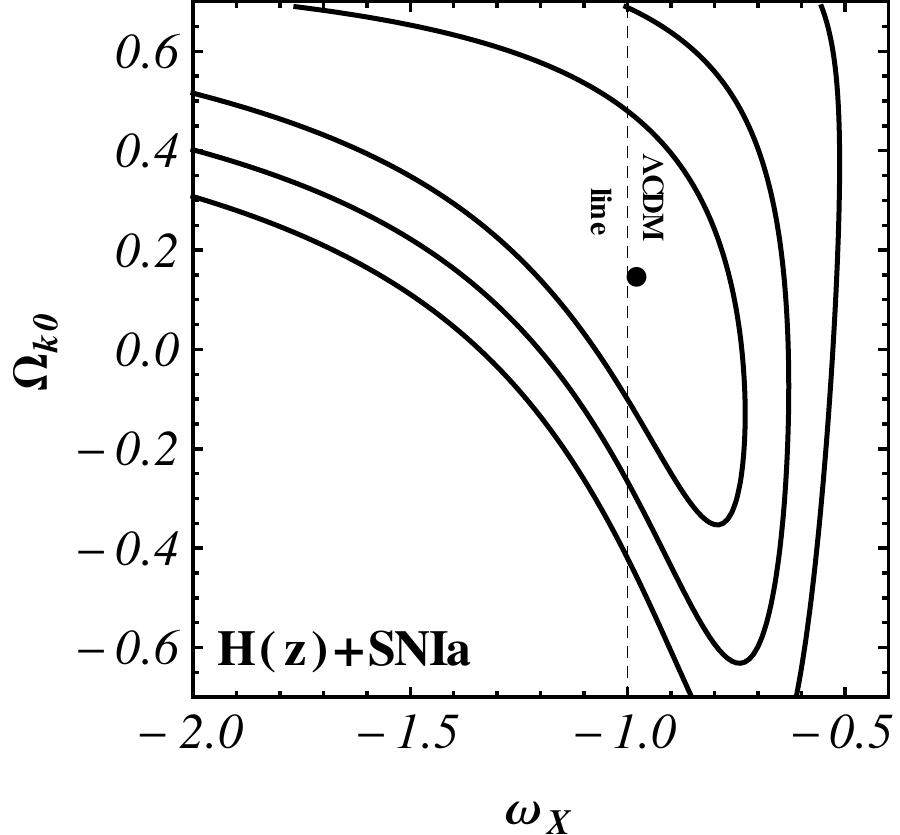}
    \includegraphics[height=1.95in]{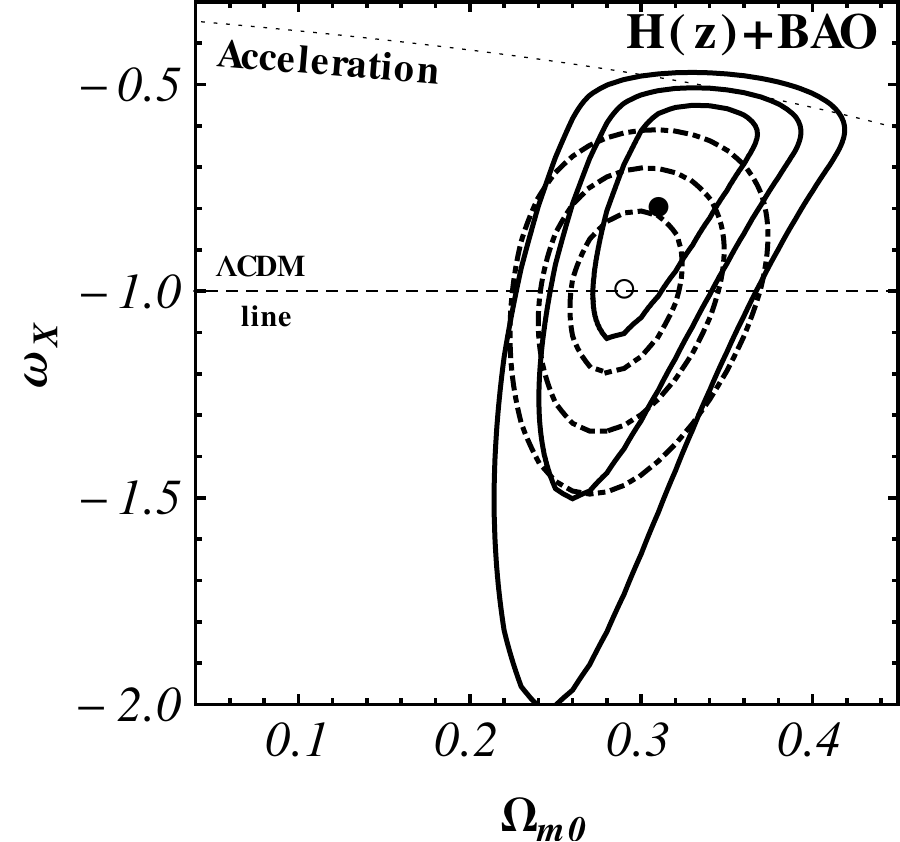}
    \includegraphics[height=1.95in]{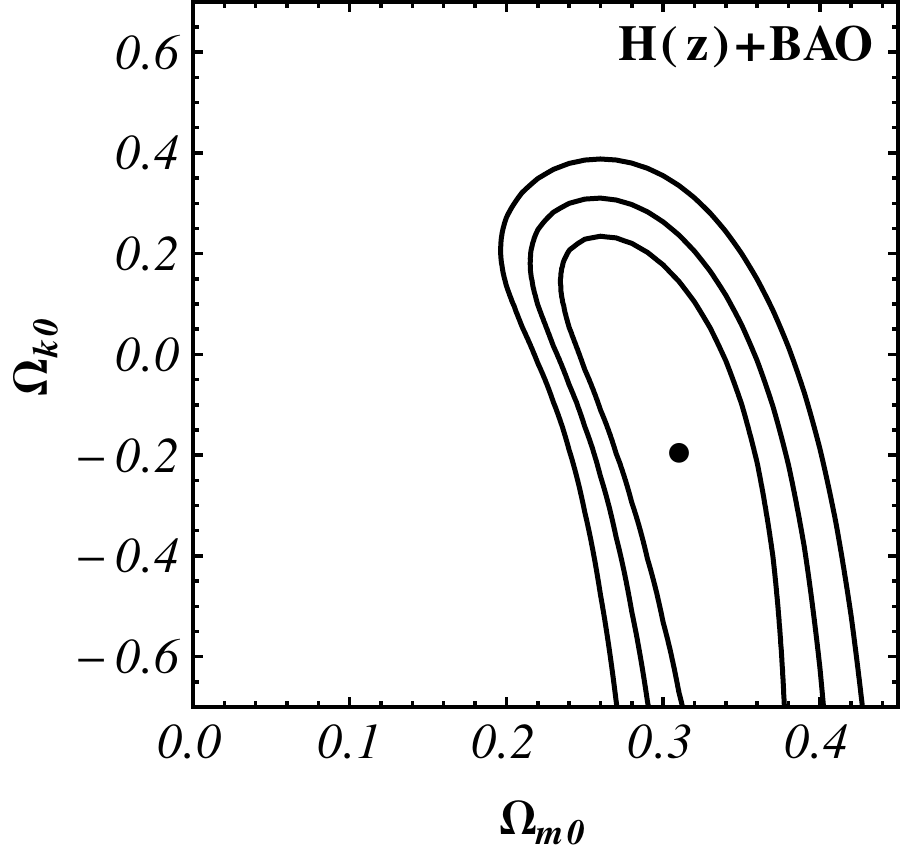}
    \includegraphics[height=1.95in]{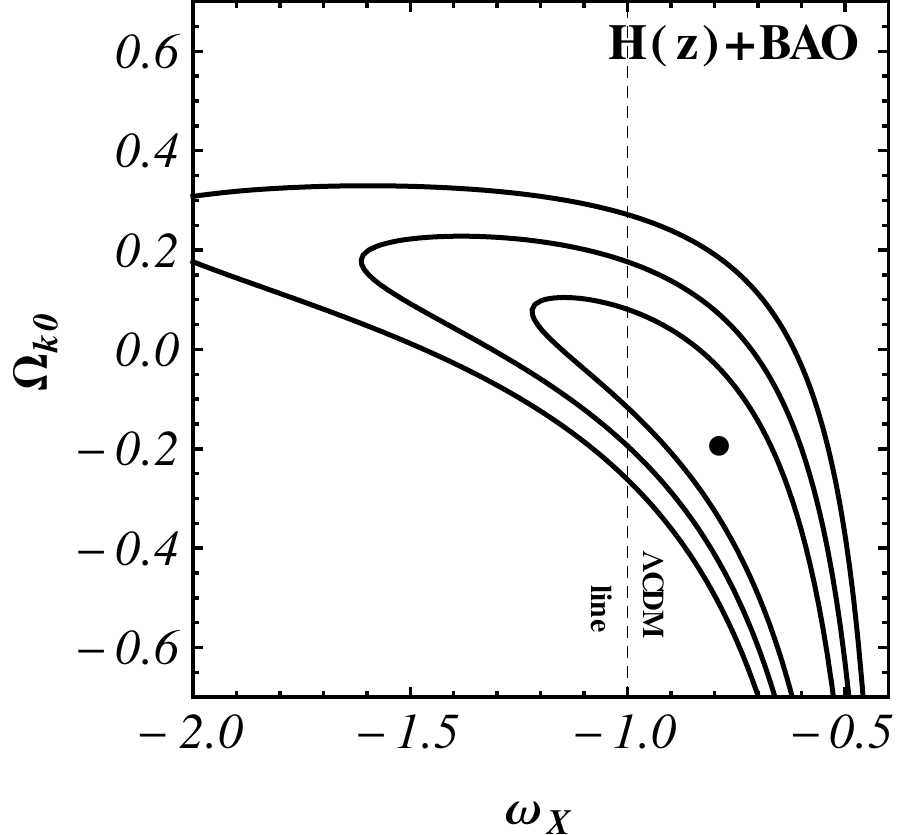}
    \includegraphics[height=1.95in]{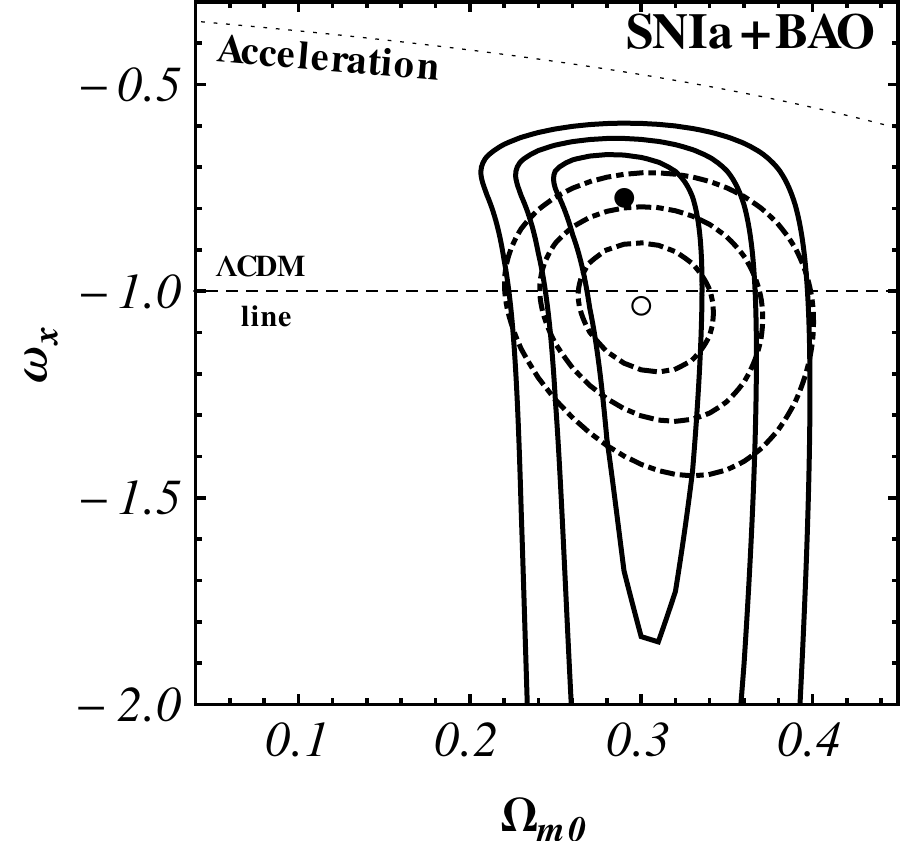}
    \includegraphics[height=1.95in]{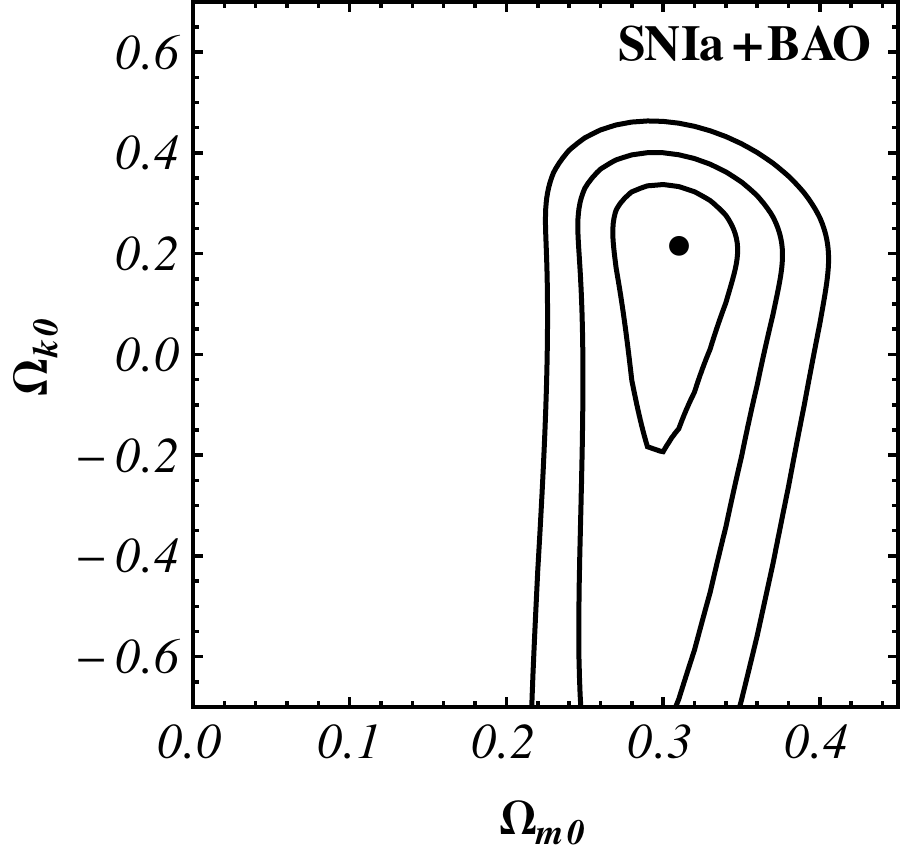}
    \includegraphics[height=1.95in]{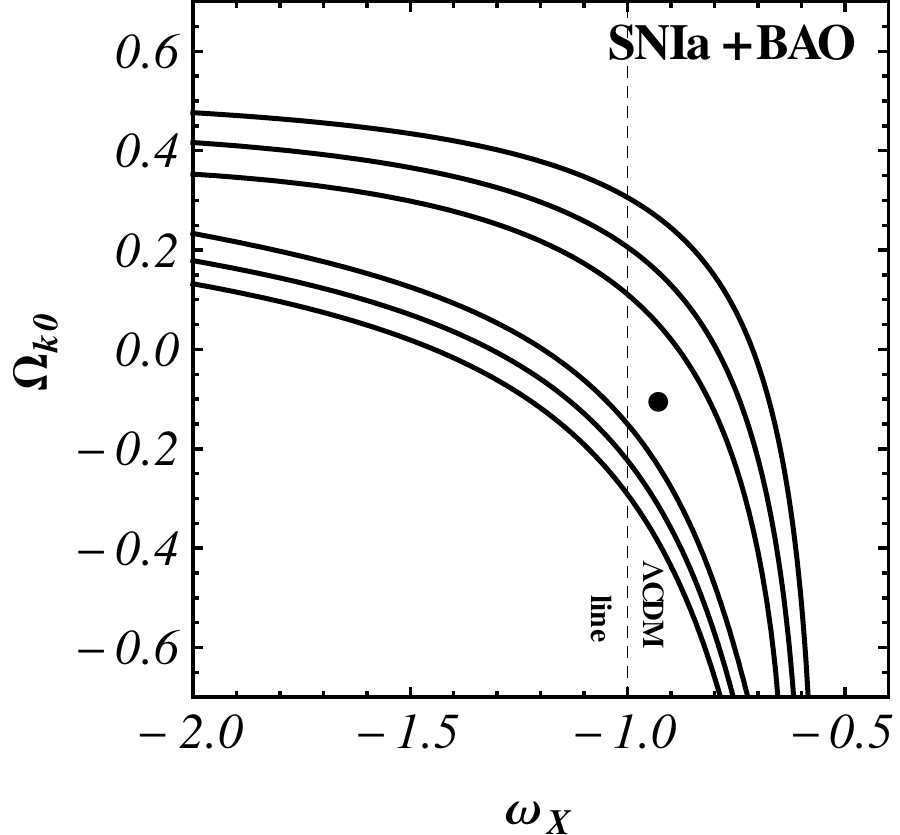}
\caption{
$1\sigma$, $2\sigma$, and $3\sigma$ constraint contours (solid lines) for parameters of the non-flat XCDM dark energy parameterization from 
$H(z)+$SNIa (first row), $H(z)+$BAO (second row), and SNIa$+$BAO (third row) measurements; filled circles show best-fit points. The dot-dashed lines in the first column panels are $1\sigma$, $2\sigma$, and $3\sigma$ constraint contours
derived by Farooq \& Ratra\cite{Farooq20131} using the spatially-flat XCDM dark energy parameterization (open circles show best-fit points); here dotted lines distinguish between
accelerating and decelerating models (at zero space curvature) and dashed lines (here and in the third column) correspond to the $\Lambda$CDM model. The first, second, and third columns correspond to marginalizing over $\Omega_{k0}$, 
$\omega_X$, and $\Omega_{m0}$ respectively.
} 
\label{fig:XCDM_D}
\end{figure}

\begin{figure}[h!]
\centering
    \includegraphics[height=1.95in]{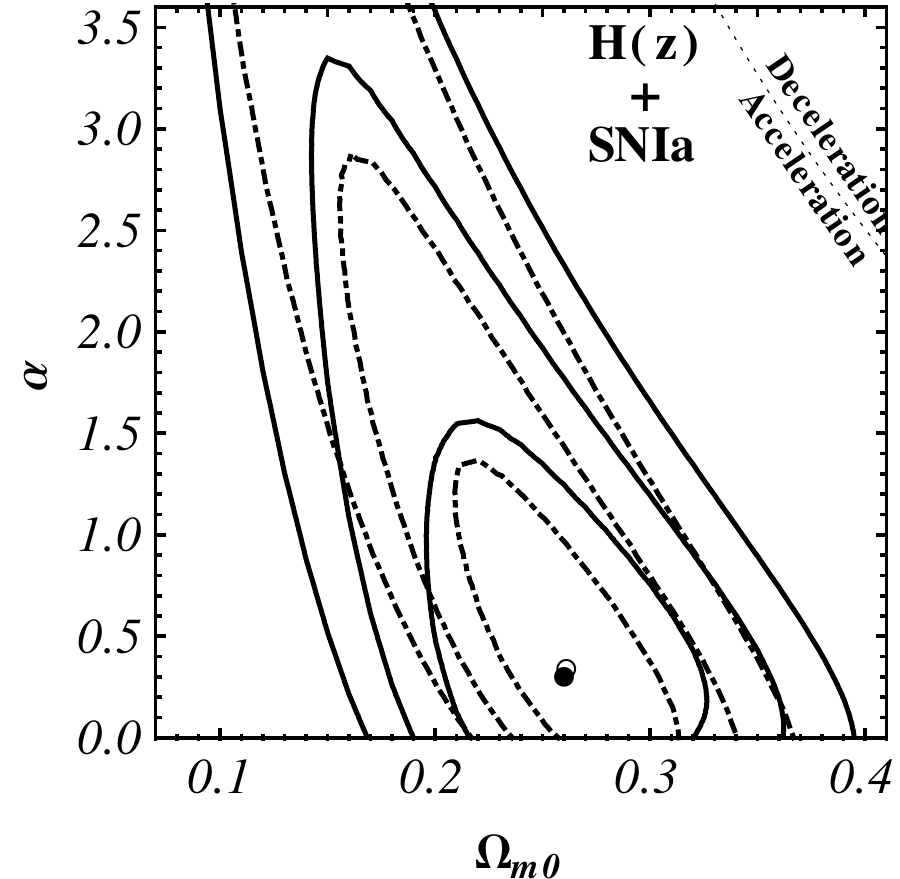}
    \includegraphics[height=1.95in]{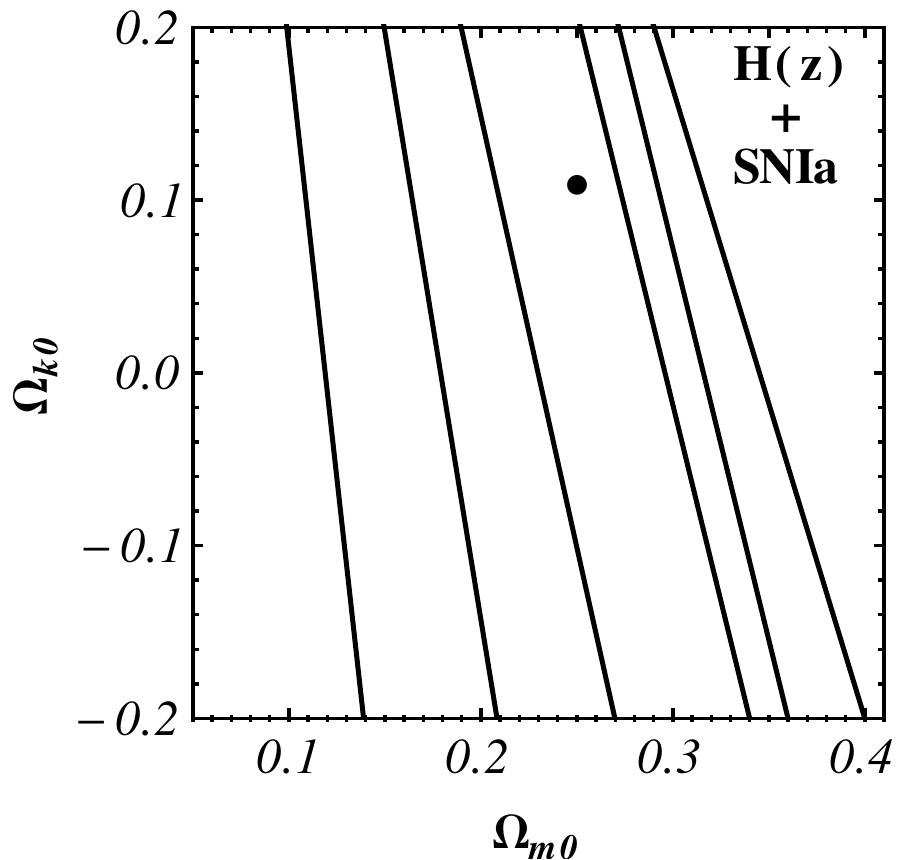}
    \includegraphics[height=1.95in]{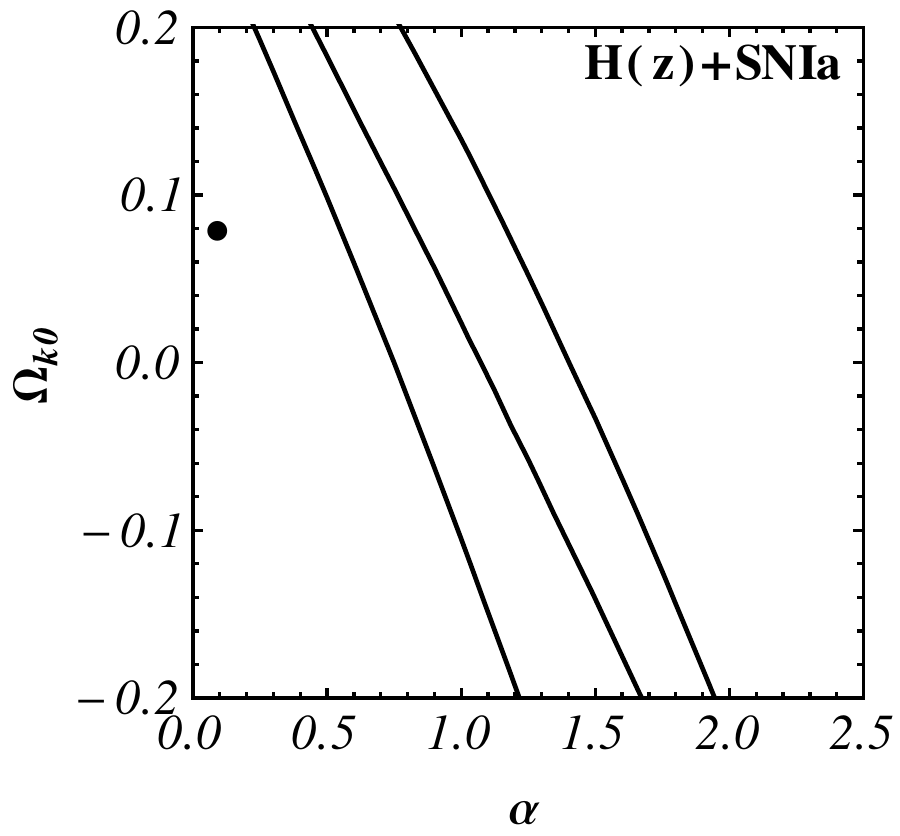}
    \includegraphics[height=1.95in]{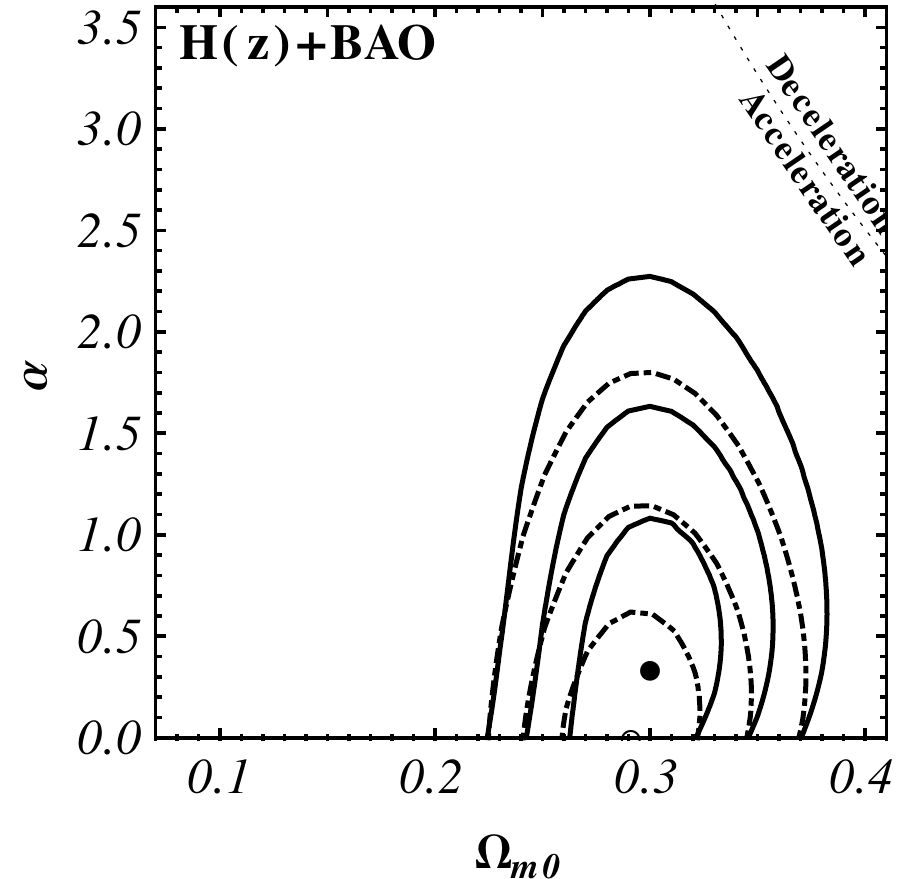}
    \includegraphics[height=1.95in]{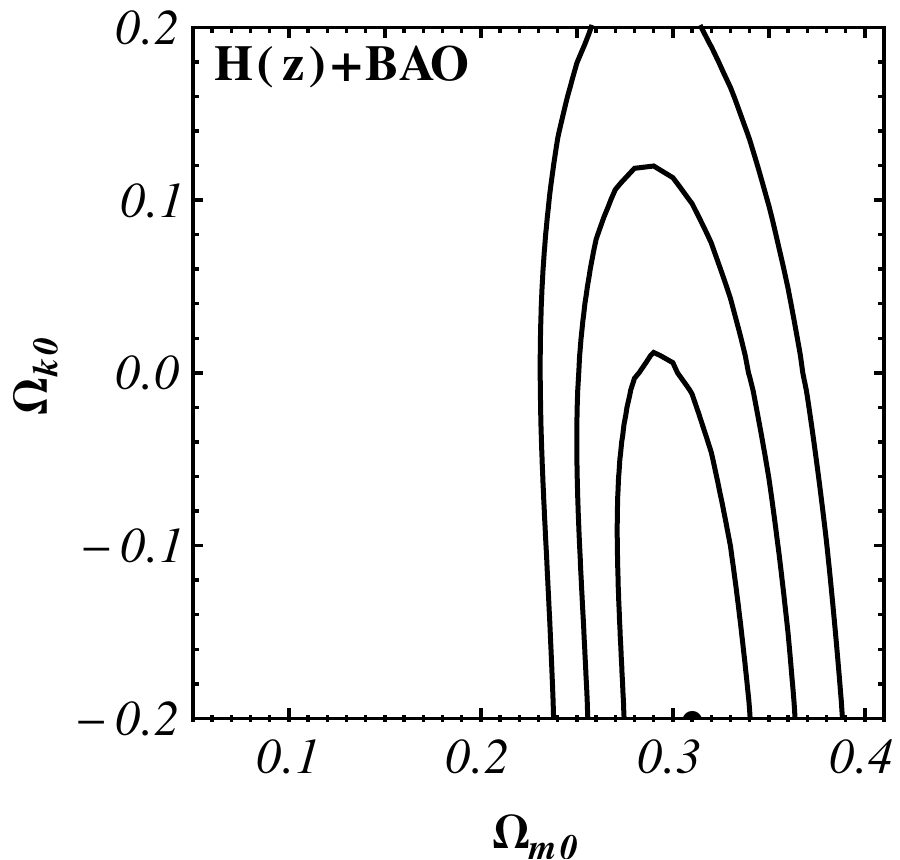}
    \includegraphics[height=1.95in]{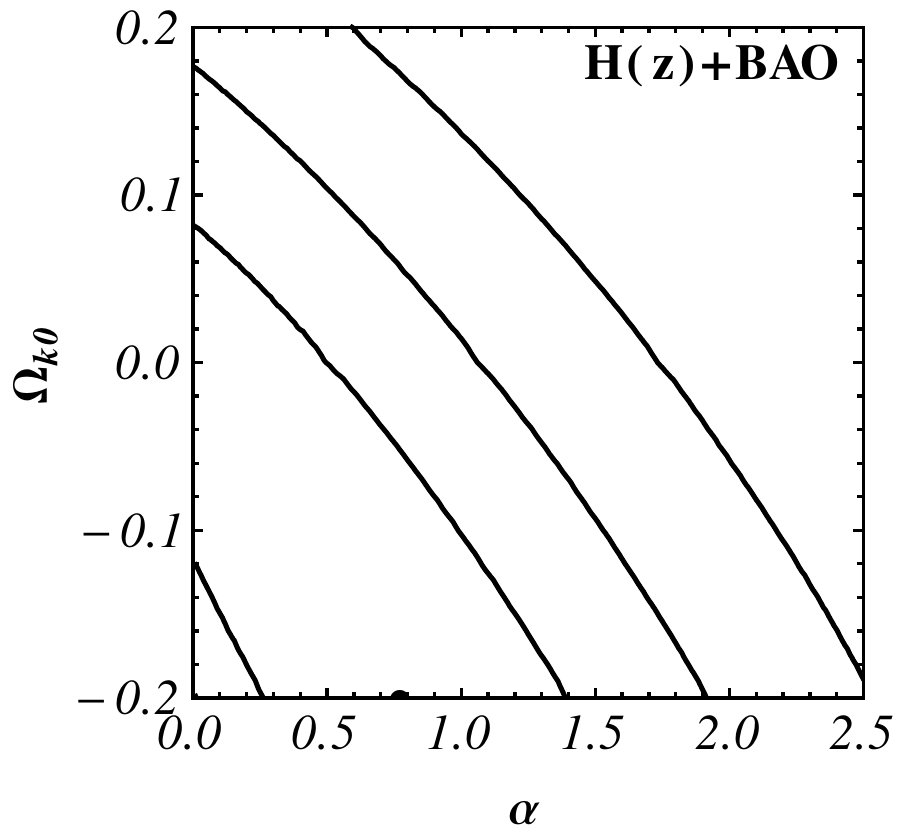}
    \includegraphics[height=1.95in]{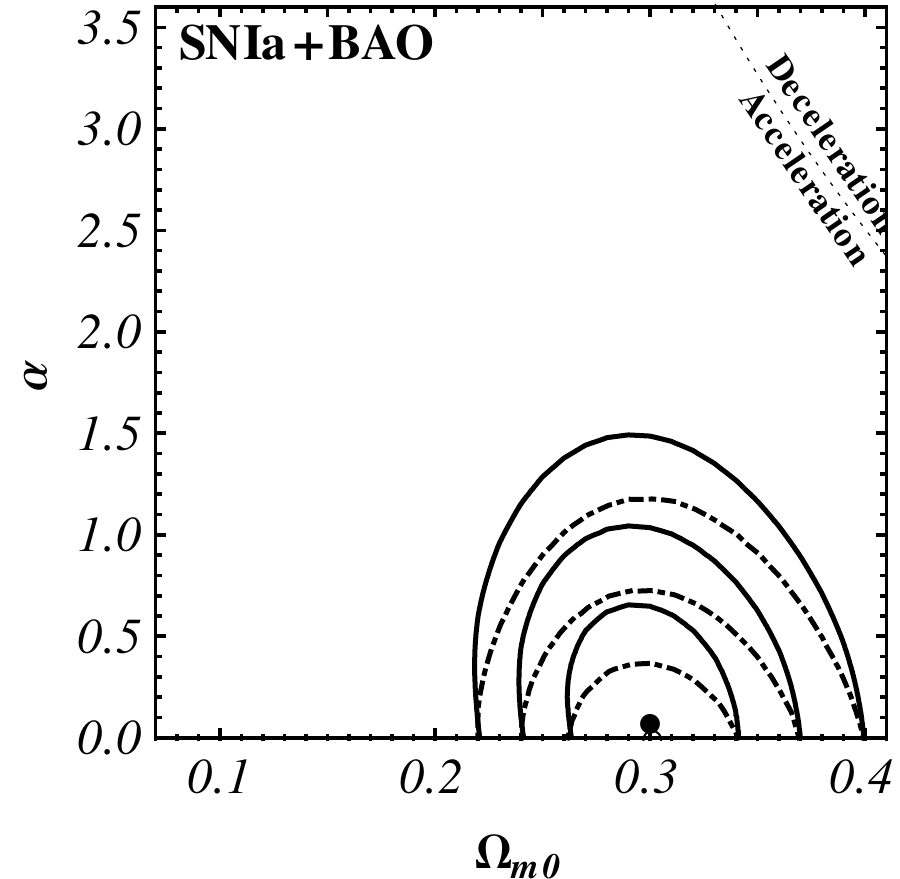}
    \includegraphics[height=1.95in]{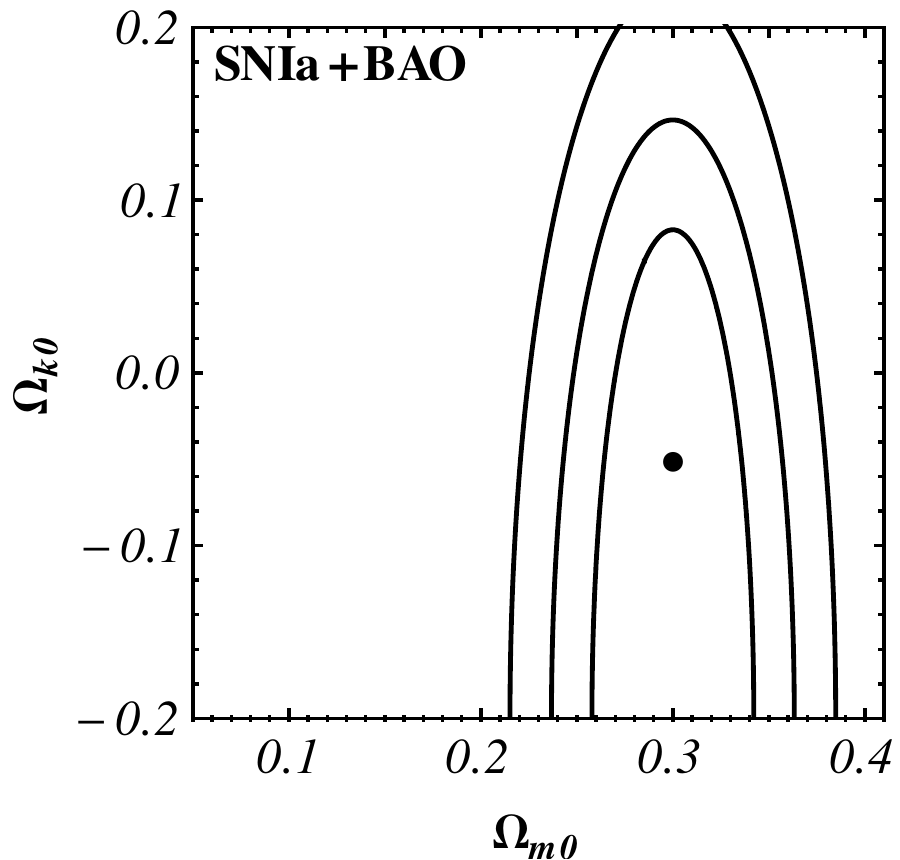}
    \includegraphics[height=1.95in]{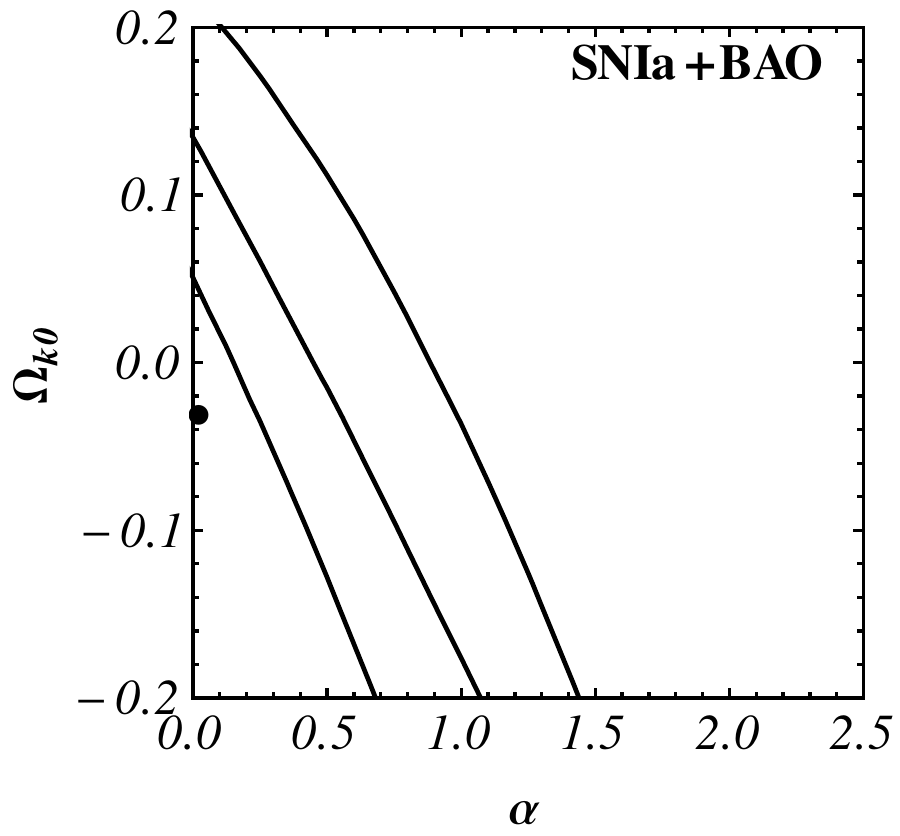}
\caption{
$1\sigma$, $2\sigma$, and $3\sigma$ constraints contour (solid lines) for parameters of the non-flat $\phi$CDM dark energy model from 
$H(z)+$SNIa (first row), $H(z)+$BAO (second row), and BAO$+$SNIa (third row) measurements; filled circles show best-fit points. The dot-dashed lines in the first column panels are $1\sigma$, $2\sigma$, and $3\sigma$ constraints contours
derived by Farooq \& Ratra\cite{Farooq20131} using the spatially-flat $\phi$CDM model (open circles show best-fit points); here dotted line distinguish between
accelerating and decelerating models (at zero space curvature) and the $\alpha=0$ axes (here and in the third column) correspond to the $\Lambda$CDM model. The first, second, and third columns correspond to marginalizing over $\Omega_{k0}$, 
$\alpha$, and $\Omega_{m0}$ respectively. 
}
\label{fig:phiCDM_D}
\end{figure}

Figures\ (\ref{fig:XCDM_D}) and
(\ref{fig:phiCDM_D}) show constraints on the parameters of the XCDM parameterization
and the $\phi$CDM model from the $H(z)$+SNIa (top row), $H(z)$+BAO (middle row), 
and SNIa+BAO (bottom row) measurements. In these figures the panels in the first, second, and
third columns show the two-dimensional probability density 
constraint contours (solid lines) from $\mathcal{L}(\Omega_{m0},\omega_X)
[\mathcal{L}(\Omega_{m0},\alpha)]$, $\mathcal{L}(\Omega_{m0},\Omega_{k0})$, and
$\mathcal{L}(\omega_X,\Omega_{k0})[\mathcal{L}(\alpha,\Omega_{k0})]$ for the
XCDM parameterization [the $\phi$CDM model]. The dot-dashed contours in the panels of
the first columns of Figs.\ (\ref{fig:XCDM_D}) and 
(\ref{fig:phiCDM_D}) are $1\sigma$, $2\sigma$, and $3\sigma$ confidence contours
corresponding to spatially-flat models, reproduced from Farooq \& Ratra.\cite{Farooq20131}
Tables\ (\ref{table:XCDMresults})
and (\ref{table:phiCDMresults}) list best-fit points and $\chi^2_{\mathrm{min}}$ values.

Comparing the solid contours of Figs.\ (\ref{fig:XCDM_D}) and (\ref{fig:phiCDM_D}) to those derived from the
single data sets in Figs.\ (\ref{fig:XCDM_S}) and (\ref{fig:phiCDM_S}), we see that combinations of pairs of data
sets result in a significant tightening of constraints, especially on $\Omega_{m0}$, and less so on 
$\Omega_{k0}$, $\omega_{X}$, and $\alpha$. 

Comparing the solid contours to the dot-dashed contours in the panels in the first columns of 
Figs.\ (\ref{fig:XCDM_D}) and (\ref{fig:phiCDM_D}) we see that the addition of space curvature
as a third free parameter results in a fairly significant broadening of the 
constraint contours, even using two data sets at a time, particularly in the direction along the parameter
that governs the time evolution of the dark energy density ($\omega_{X}$ for the XCDM parameterization and $\alpha$ for the $\phi$CDM
model). Again, when space curvature is included as a free parameter the constraint contours broaden more significantly for the
XCDM parameterization than for the $\phi$CDM model: compare the solid and dot-dashed contours in the first columns 
of Figs.\ (\ref{fig:XCDM_D}) and (\ref{fig:phiCDM_D}) (this is probably partially a consequence of the 
smaller range of space curvature, $-0.2 \leq \Omega_{k0} \leq 0.2$, considered for
computational tractability in the $\phi$CDM case).

\begin{figure}[h!]
\centering
    \includegraphics[height=2.8in]{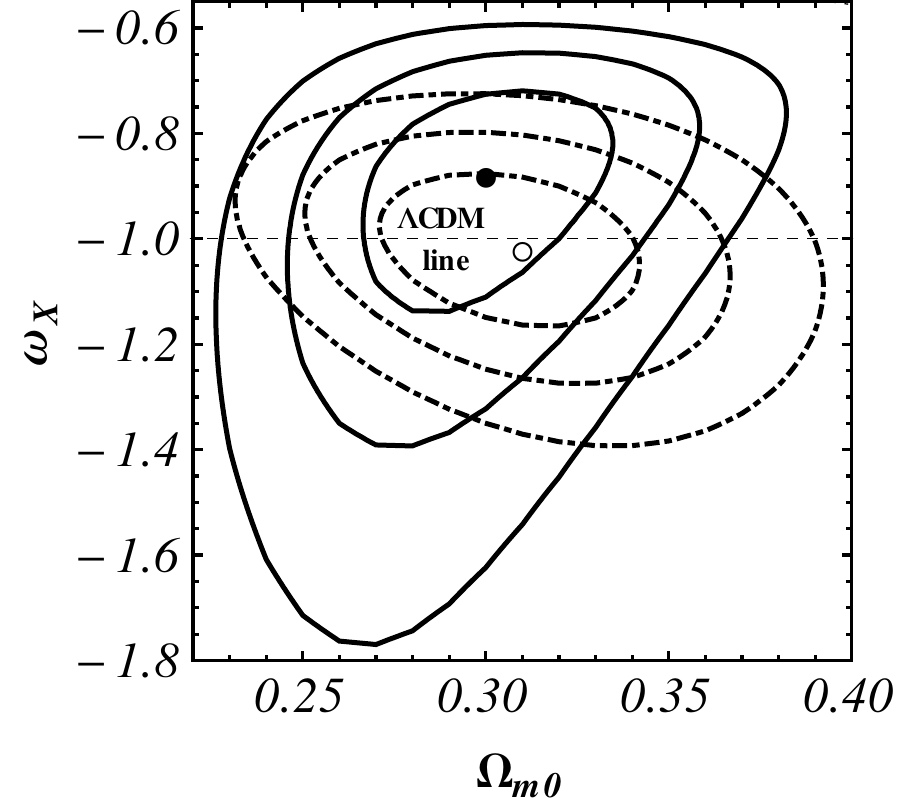}
    \includegraphics[height=2.8in]{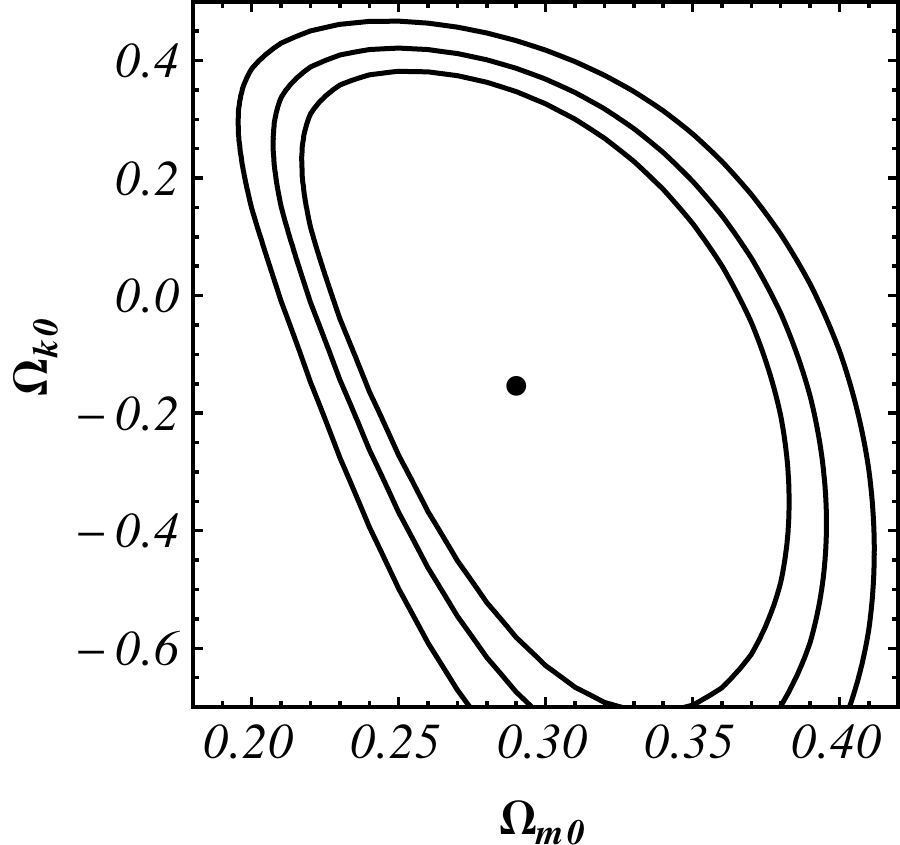}
    \includegraphics[height=2.8in]{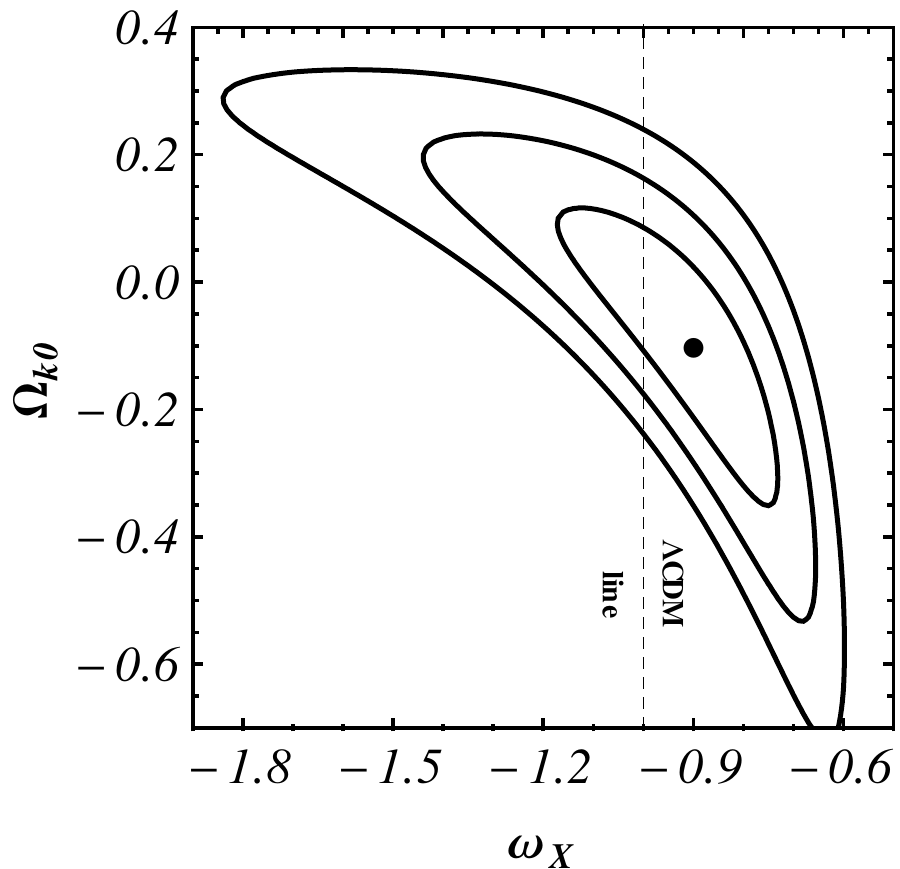}
\caption{
$1\sigma$, $2\sigma$, and $3\sigma$ constraint contours (solid lines) for parameters of the non-flat XCDM dark energy parameterization from 
$H(z)+$SNIa$+$BAO measurements; filled circles show best-fit points. The dot-dashed lines in the top left panel are $1\sigma$, $2\sigma$, and $3\sigma$ constraint contours
derived by Farooq \& Ratra\cite{Farooq20131} using the spatially-flat XCDM parameterization (open circle shows best-fit point); here dashed lines (in the top left and bottom panels) correspond to the $\Lambda$CDM model. The top left, top right and bottom panel correspond to marginalizing over $\Omega_{k0}$, 
$\omega_X$, and $\Omega_{m0}$ respectively.
}
\label{fig:XCDM_A}
\end{figure}

\begin{figure}[h!]
\centering
    \includegraphics[height=2.95in]{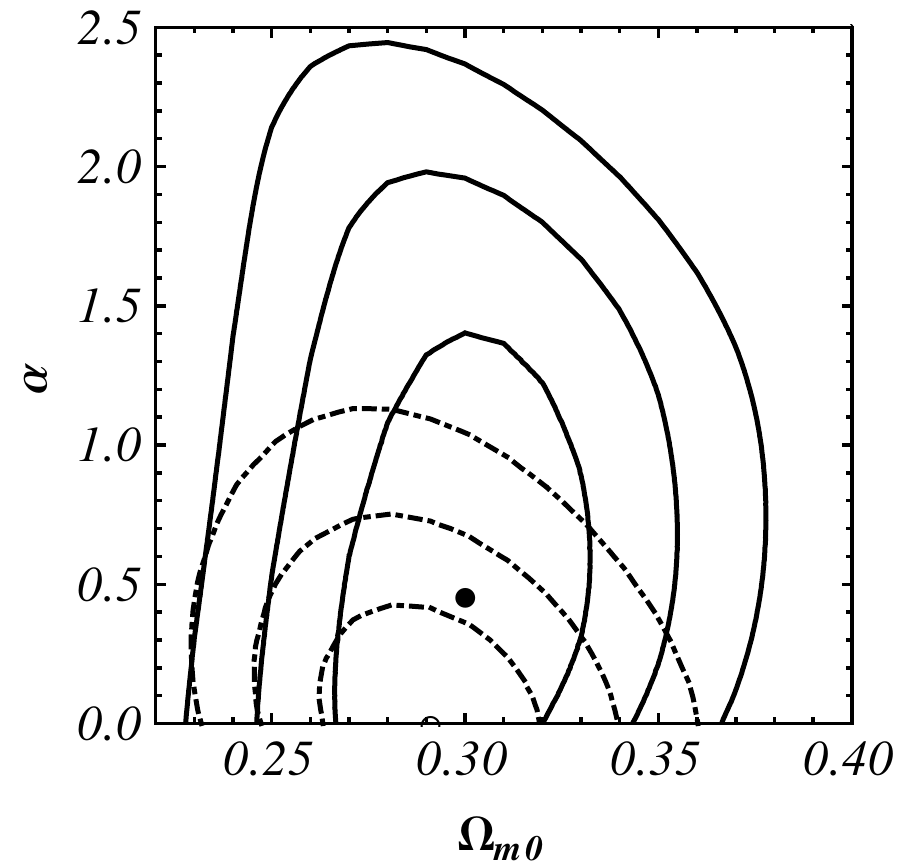}
    \includegraphics[height=2.95in]{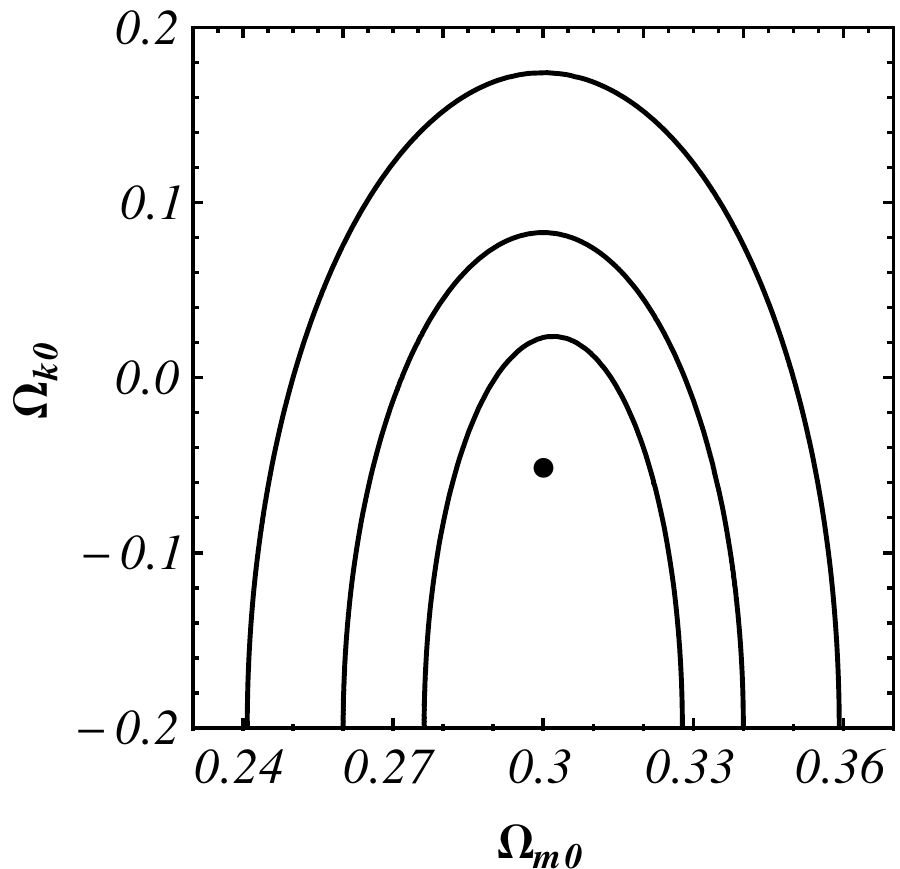}
    \includegraphics[height=2.95in]{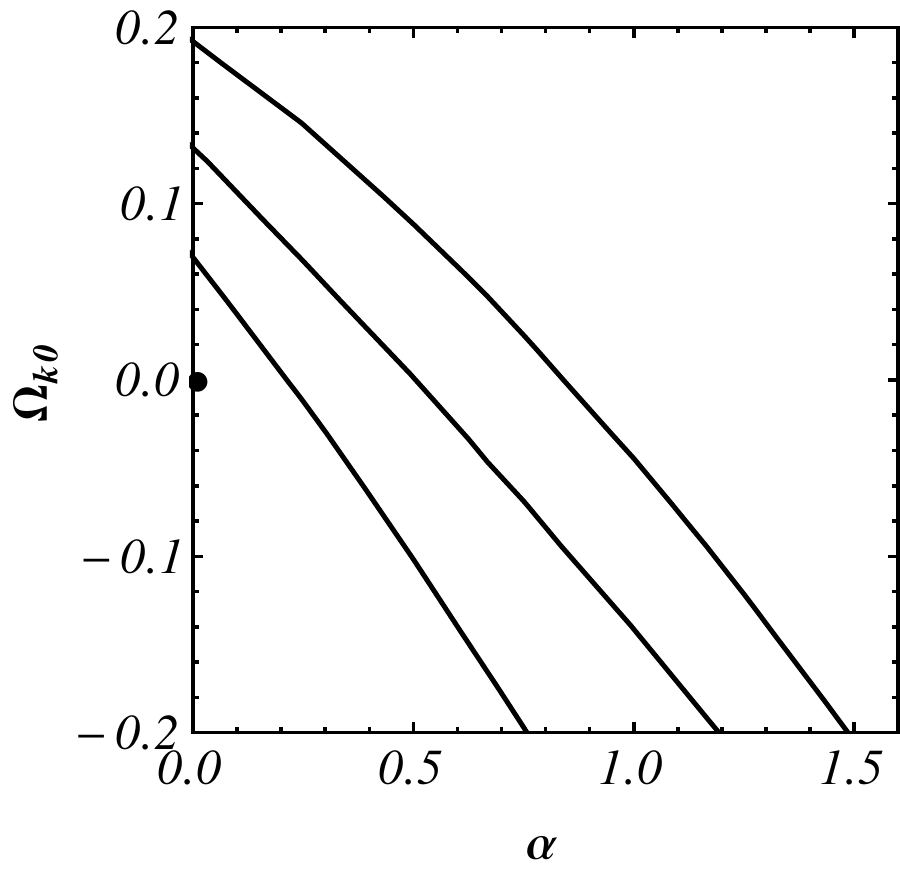}
\caption{
$1\sigma$, $2\sigma$, and $3\sigma$ constraint contours (solid lines) for parameters of the non-flat $\phi$CDM dark energy model from
$H(z)+$SNIa$+$BAO measurements; filled circles show best-fit points. The dot-dashed lines are $1\sigma$, $2\sigma$, and $3\sigma$ constraint contours
derived by Farooq \& Ratra\cite{Farooq20131} using the spatially-flat $\phi$CDM model (open circle shows best-fit point); here the  $\alpha=0$ axes in the top left and bottom panels correspond to the $\Lambda$CDM model. The top left, top right and bottom panel correspond to marginalizing over $\Omega_{k0}$, 
$\alpha$, and $\Omega_{m0}$ respectively.
}
\label{fig:phiCDM_A}
\end{figure}

Encouraged by the tightening of the constraint contours when two data sets are analyzed together, we now discuss the result of a joint analysis of the $H(z)$, SNIa, and BAO data. 
Figures.\ (\ref{fig:XCDM_A}) and
(\ref{fig:phiCDM_A}) show constraints on the parameters of the XCDM parameterization
and the $\phi$CDM model from the $H(z)$+SNIa+BAO measurements. 
In these figures the top left panels, top right panels, and the bottom panels
show the two-dimensional probability density 
constraint contours (solid lines) from $\mathcal{L}(\Omega_{m0},\omega_X)
[\mathcal{L}(\Omega_{m0},\alpha)]$, $\mathcal{L}(\Omega_{m0},\Omega_{k0})$, and
$\mathcal{L}(\omega_X,\Omega_{k0})[\mathcal{L}(\alpha,\Omega_{k0})]$ for the
XCDM parameterization [the $\phi$CDM model]. The dot-dashed contours in the left top panels of
Figs.\ (\ref{fig:XCDM_A}) and 
(\ref{fig:phiCDM_A}) are $1\sigma$, $2\sigma$, and $3\sigma$ confidence contours
corresponding to spatially-flat models, reproduced from Farooq \& Ratra.\cite{Farooq20131}
Tables\ (\ref{table:XCDMresults})
and (\ref{table:phiCDMresults}) list best-fit points and $\chi^2_{\mathrm{min}}$ values.

Comparing the solid contours of Figs.\ (\ref{fig:XCDM_A}) and (\ref{fig:phiCDM_A}) to those derived from the
data set pairs of Figs.\ (\ref{fig:XCDM_D}) and (\ref{fig:phiCDM_D}), we see that the joint analysis of all three data sets results in a significant tightening of constraints.

Comparing the solid contours to the dot-dashed contours in the left top panels in 
Figs.\ (\ref{fig:XCDM_A}) and (\ref{fig:phiCDM_A}) we see that the addition of space curvature
as a third free parameter results in a significant broadening of the 
constraint contours, but this time less than when only two data sets were used in Fig.\ (\ref{fig:XCDM_D}) 
and (\ref{fig:phiCDM_D}). The broadening is more significant in the direction along the parameter
that governs the time evolution of the dark energy density ($\omega_{X}$ for the XCDM parameterization and $\alpha$ for the $\phi$CDM
model).

We also computed the $1\sigma$ and $2\sigma$ bounds on model parameters that follow from the joint analysis 
of $H(z)$, SNIa, and BAO measurements. Tables\ (\ref{table:XCDMintervals}) and (\ref{table:phiCDMintervals}) 
list these bounds on
individual cosmological parameters, determined from their one-dimensional posterior probability distribution 
functions (which we obtained by marginalizing the three-dimensional likelihood over the other two cosmological parameters). The numerical values listed in these tables confirm the results described in the discussion above of Figs.\ (\ref{fig:XCDM_A}) and (\ref{fig:phiCDM_A}).

\renewcommand{\arraystretch}{1.1}
\begin{deluxetable}{ccc}
\tablecaption{XCDM Parametrization Results From $H(z)+$SNIa$+$BAO Data}
\tablewidth{0pt}
\tabletypesize{\small}
\tablehead{
\colhead{\multirow{1}{*}{Marginalization Range}}& 
\colhead{\multirow{1}{*}{1$\sigma$ intervals}}& 
\colhead{2$\sigma$ intervals}\\
\vspace{-4 mm}
}
\startdata
\noalign{\vskip -1mm}
 \multirow{2}{*}{$\Omega_{k0} =0$\tablenotemark{a}} & $0.29\leqslant \Omega_{m0} \leqslant 0.31$ & $0.27\leqslant \Omega_{m0} \leqslant 0.32$\\
{}& $-1.01\leqslant \omega_{X} \leqslant -0.83$ & $-1.03 \leqslant \omega_{X} \leqslant -0.77$\\
\cline{1-3}
\multirow{2}{*}{$-0.7 \leq \Omega_{k0}\leq -0.7$} & $0.27\leqslant \Omega_{m0} \leqslant 0.32$ & $0.25\leqslant \Omega_{m0} \leqslant 0.34$\\
{}& $-1.03\leqslant \omega_{X} \leqslant -0.77$ & $-1.25 \leqslant \omega_{X} \leqslant -0.69$\\
\cline{1-3}
\multirow{2}{*}{$-2 \leq \omega_{X}\leq 0$} & $0.27\leqslant \Omega_{m0} \leqslant 0.32$ & $0.25\leqslant \Omega_{m0} \leqslant 0.34$\\
{}& $-0.21\leqslant \Omega_{k0} \leqslant 0.10$ & $-0.39 \leqslant \Omega_{k0} \leqslant 0.22$\\
\cline{1-3}
\multirow{2}{*}{$0 \leq \Omega_{m0}\leq 1$} & $-1.03\leqslant \omega_{X} \leqslant -0.77$ & $-1.30\leqslant \omega_{X} \leqslant -0.69$\\
{}& $-0.21\leqslant \Omega_{k0} \leqslant 0.10$ & $-0.39 \leqslant \Omega_{k0} \leqslant 0.22$\\
\enddata
\tablenotetext{a}{From Farooq \& Ratra.\cite{Farooq20131}}
\label{table:XCDMintervals}
\end{deluxetable}


\renewcommand{\arraystretch}{1.1}
\begin{deluxetable}{ccc}
\tablecaption{$\phi$CDM Model Results From $H(z)+$SNIa$+$BAO Data}
\tablewidth{0pt}
\tabletypesize{\small}
\tablehead{
\colhead{\multirow{1}{*}{Marginalization Range}}& 
\colhead{\multirow{1}{*}{1 $\sigma$ intervals}}& 
\colhead{2 $\sigma$ intervals}\\
\vspace{-4 mm}
}
\startdata
\noalign{\vskip -1mm}
\multirow{2}{*}{$\Omega_{k0} =0$\tablenotemark{a}} & $0.27\leqslant \Omega_{m0} \leqslant 0.29$ & $0.25\leqslant \Omega_{m0} \leqslant 0.30$\\
{}& $0.00 \leqslant \alpha \leqslant 0.31$ & $0.00 \leqslant \alpha \leqslant 0.56$\\
\cline{1-3}
\multirow{2}{*}{$-0.2 \leq \Omega_{k0}\leq -0.2$} & $0.28\leqslant \Omega_{m0} \leqslant 0.32$ & $0.26\leqslant \Omega_{m0} \leqslant 0.34$\\
{}& $0.00\leqslant \alpha \leqslant 1.03$ & $0.00 \leqslant \alpha \leqslant 1.64$\\
\cline{1-3}
\multirow{2}{*}{$0 \leq \alpha\leq 5$} & $0.28\leqslant \Omega_{m0} \leqslant 0.31$ & $0.26\leqslant \Omega_{m0} \leqslant 0.33$\\
{}& $-0.2\leqslant \Omega_{k0} \leqslant 0.09$ & $-0.2 \leqslant \Omega_{k0} \leqslant 0.12$\\
\cline{1-3}
\multirow{2}{*}{$0 \leq \Omega_{m0}\leq 1$} & $0.00\leqslant \alpha \leqslant 1.03$ & $ 0.00 \leqslant \alpha \leqslant 1.64$\\
{}& $-0.2\leqslant \Omega_{k0} \leqslant 0.09$ & $-0.2 \leqslant \Omega_{k0} \leqslant 0.12$\\
\enddata
\tablenotetext{a}{From Farooq \& Ratra.\cite{Farooq20131}}
\label{table:phiCDMintervals}
\end{deluxetable}


Of some interest are the bounds on the curvature density parameter
$\Omega_{k0}$ perhaps the useful summary is that 1$\sigma$ limit 
$|\Omega_{k0}|\lesssim  0.15$ derived by symmetrizing about $\Omega_{k0}=0$
the 1$\sigma$ range from the central columns of Table (\ref{table:XCDMintervals}) and
(\ref{table:phiCDMintervals}). Note that the possible 2$\sigma$ of $\Omega_{k0}$ is
significantly smaller for $\phi$CDM than for XCDM (compare the relevant entries in 
the last columns of Table (\ref{table:XCDMintervals}) and
(\ref{table:phiCDMintervals}). This is almost certainly a consequence of the
smaller range of $\Omega_{k0}$, $-0.2 \leq \Omega_{k0} \leq 0.2$ we have used in
the $\phi$CDM computation; the 2$\sigma$ XCDM bound on $|\Omega_{k0}|$ is the
more reliable one.

\section{Conclusion}
\label{summary}

A joint analysis of $H(z)$, SNIa, and BAO data using the XCDM parametrization and the $\phi$CDM model
of time evolving dark energy density
in a non-flat geometry leads to the conclusion that more, and more precise, data are required to tightly
pin down the spatial curvature of the Universe in dynamical dark energy models. These data require 
$|\Omega_{k0}|\lesssim  0.15$ at 1$\sigma$ confidence. It would be of interest
to determine the constraints on space curvature in the non-flat $\phi$CDM model
from CMB anisotropy measurements. Such an analysis, possibly in combination with 
that of other data of the kind considered here, and extended over a wider range of
$\Omega_{k0}$ than we have considered, could go a long way towards establishing whether
space curvature contributes significantly to the current cosmological energy budget.


\cleardoublepage


\chapter{Conclusions}
\label{Chapter9}

In this thesis we used three different probes to constrain a number of spatially-flat and non-flat cosmological models. One of the key features of this work is that we used $H(z)$ data which had not previously been used as much as the other probes, like SNIa apparent magnitude verses redshift measurements, and BAO peak length scale observations. From different combinations of data sets, different constraints have been obtained for the different models we consider. 

We found that using 22 $H(z)$ data points [given in Table\ (\ref{tab:Hz2})] can constrain the flat cosmological models better than the SNIa data set that has 580 data points [given in Table\ (\ref{tab:SNeIaData})], which is an amazing result and is largely a consequence of large systematic error bars currently associated with SNIa data. Considering spatially-flat models, the data fits the best with the standard $\Lambda$CDM model, even in the case of the inverse power-law-potential energy density scalar field model [see Fig.\ (\ref{fig:phiCDM_com2})].

We also consider the models with space curvature in which case dark energy is dynamical. These data require $|\Omega_{K0}|\lesssim  0.15$ at 1$\sigma$ confidence. It would be of significant interest to determine the constraints on space curvature in the non-flat $\phi$CDM model from CMB anisotropy measurements.





\def\mnras{MNRAS}
\def\aapr{A\&A~Rev.}
\def\jcap{J. Cosmology Astropart. Phys.}
\def\apj{ApJ}
\def\apjl{ApJL}
\def\pasp{PASP}
\def\aap{A\&A}
\def\apss{Ap\&SS}
\def\apjs{ApJS}
\def\prd{Phys. Rev. D}
\def\prl{Phys. Rev. Lett.}
\def\physrep{Phys. Reports}
\def\nat{Nature}



\appendix

\cleardoublepage

\chapter{Derivation of Scalar Field Equation of Motion}
\label{Appendix:A}


The general scalar field action in a Riemann spacetime is [see Eq.\ (\ref{eq:EL})]
\begin{eqnarray}
S_{\phi}=\int\sqrt{-g}\mathcal{L}_{\phi}(\phi,\partial_{\alpha}\phi)d^4x.
\label{eq:A0.5}
\end{eqnarray}
In spacetime with signature $(-,+,+,+)$, the Lagrangian density of the $\phi$ field is
\begin{eqnarray}
\mathcal{L}_{\phi}=-\frac{1}{2}g^{\mu \nu}\partial_{\mu}\phi \partial_{\nu}\phi-V(\phi).
\label{eq:A1}
\end{eqnarray}
Here, we have neglected the constant in front of $\mathcal{L}_{\phi}$ as it does not effect the equation of motion.

The scalar field equation of motion is the Euler-Lagrange equation\ (\ref{eq:ELE1})  
\begin{eqnarray}
\frac{\partial }{\partial \phi}\Big(\sqrt{-g}\mathcal{L}_{\phi}\Big)-
\frac{\partial}{\partial x^\lambda}\left[\frac{\partial }{\partial (\partial_\lambda \phi)}\left(\sqrt{-g}\mathcal{L}_{\phi}\right)\right]=0.
\label{eq:A2}
\end{eqnarray}
The first term of Eq.\ (\ref{eq:A2}) is

\begin{eqnarray}
&=&\frac{\partial}{\partial \phi}\big(\sqrt{-g}\mathcal{L}_{\phi}\big)=\sqrt{-g}\frac{\partial \mathcal{L}_{\phi}}{\partial \phi}, \nonumber\\
&=&\sqrt{-g}\frac{\partial}{\partial \phi}\left[-\frac{1}{2}g^{\mu \nu}\partial_{\mu}\phi \partial_{\nu}\phi-V(\phi)\right]=-\sqrt{-g}\frac{\partial V(\phi)}{\partial \phi}.
\label{eq:A3}
\end{eqnarray}

The second term of Eq.\ (\ref{eq:A2}) is

\begin{eqnarray}
&=&\partial_{\lambda}\left[\frac{\partial}{\partial(\partial_{\lambda} \phi)}\big(\sqrt{-g}\mathcal{L}_\phi\big)\right], \hspace{1 cm} \left(\mathrm{where}\ \partial_\lambda=\frac{\partial}{\partial x^{\lambda}}\right)\nonumber \\
&=&\partial_{\lambda}\left[\frac{\partial}{\partial(\partial_\lambda \phi)}\left(-\sqrt{-g}\frac{1}{2}g^{\mu \nu}\partial_{\mu}\phi \partial_{\nu}\phi-\sqrt{-g}V(\phi)\right)\right], \nonumber \\
&=&-\partial_{\lambda}\left[\sqrt{-g}\frac{1}{2}g^{\mu \nu}\Big(\partial_{\mu}\phi \delta_{\nu \lambda}+\partial_{\nu}\phi \delta_{\mu \lambda}\Big)\right], \hspace{3.2 cm} \left(\mathrm{Leibniz's\ Rule }\right)\nonumber \\
&=&-\partial_{\lambda}\left[\sqrt{-g}\frac{1}{2}g^{\mu \nu}\partial_{\mu}\phi \delta_{\nu \lambda}\right]-\partial_{\lambda}\left[\sqrt{-g}\frac{1}{2}g^{\mu \nu}\partial_{\nu}\phi \delta_{\mu \lambda}\right], \nonumber \\
&=&-\partial_{\nu}\left[\sqrt{-g}\frac{1}{2}g^{\mu \nu}\partial_{\mu}\phi\right]-\partial_{\mu}\left[\sqrt{-g}\frac{1}{2}g^{\mu \nu}\partial_{\nu}\phi\right], \nonumber \\
&=&-2\partial_{\mu}\left[\sqrt{-g}\frac{1}{2}g^{\mu \nu}\partial_{\nu}\phi\right], \hspace{3.7cm} \left(\mu\ \mathrm{and}\ \nu\ \mathrm{are\ dummy\ indices}\right)\nonumber \\
&=&-\partial_{\mu}\left[\sqrt{-g}g^{\mu \nu}\partial_{\nu}\phi\right], \nonumber \\
&=&-\left(\frac{1}{2\sqrt{-g}}\partial_{\mu}(-g)\right)g^{\mu \nu}\partial_{\nu}\phi-\sqrt{-g}\partial_{\mu}(g^{\mu \nu})\partial_{\nu}\phi-\sqrt{-g}g^{\mu \nu}\partial_\mu \partial_\nu \phi, \nonumber \\
&=&-\sqrt{-g}\left[\left(\frac{1}{2g}\partial_{\mu}(g)\right)g^{\mu \nu}\partial_{\nu}\phi+\partial_{\mu}\big(g^{\mu \nu}\big)\partial_{\nu}\phi+g^{\mu \nu}\partial_\mu \partial_\nu \phi \right].
\label{eq:A4}
\end{eqnarray}
With the FLWR metric
\begin{equation} 
g_{\alpha \beta}=
            \left( {\begin{array}{cccc}
-1 & 0 & 0 & 0 \\

0 & a^2(t) & 0 & 0 \\

0 & 0 & a^2(t) & 0 \\

0 & 0 & 0 & a^2(t) \\
\end{array} } \right),\nonumber
\label{eq:A5}
\end{equation}
and a homogeneous scalar field $\phi = \phi(t)$ we have,
\begin{eqnarray}
g&=&-a^6,\ \nonumber \\
\partial_0 g=-6a^5\dot{a}&,&\ \partial_ig=0, \nonumber \\
\partial_0 \phi=\dot \phi&,&\ \partial_i \phi=0,\nonumber \\ 
\partial_\mu g^{\mu \nu}&=& 0.
\label{eq:A6}
\end{eqnarray}
Where $a$ is the scale factor, $\mu$, $\nu$ $\in\ \{0,1,2,3\}$ are spacetime indices, and $i\ \in\ \{1,2,3\}$ are spatial indices. Hence, the second term in Eq.\ (\ref{eq:A4}) is
\begin{eqnarray}
&=&-a^3\left[\frac{1}{2(-a^6)}\big(-6a^5\dot a\big)g^{00}+0+g^{00}\ddot\phi\right], \nonumber \\
&=&a^3\left[3\left(\frac{\dot a}{a}\right)\dot \phi-\ddot\phi\right].
\label{eq:A7}
\end{eqnarray}
Using Eq.\ (\ref{eq:A6}) in Eq.\ (\ref{eq:A3}), the first term is
\begin{eqnarray}
=\ -a^3\frac{\partial V(\phi)}{\partial \phi}.
\label{eq:A8}
\end{eqnarray} 
Now, using Eqs.\ (\ref{eq:A7}) and (\ref{eq:A8}) in Eq.\ (\ref{eq:A2}) gives
\begin{eqnarray}
\ddot \phi+3\left(\frac{\dot a}{a}\right)\dot \phi+\frac{\partial V(\phi)}{\partial \phi}=0.
\label{eq:A9}
\end{eqnarray}
Which is the required equation of motion for the $\phi$ field. The form of equation of motion of $\phi$ field (\ref{eq:A9}) does not depends upon even if non-flat space curvature is considered, since $\phi$ field is not directly coupled with the curvature.


\cleardoublepage

\chapter{Derivation of Scalar Field Stress-Energy Tensor}
\label{Appendix:B}
(Special thanks to Shawn Westmoreland for helpful discussion related to this appendix.)
\\

\section{Spcetime Signature}
In the literature there are two different sign conventions for the spacetime metric $g_{\mu \nu}$. These two signatures are $(+,-,-,-)$ and $(-,+,+,+)$. 
In the first case the spacetime metric for the flat space is  
\begin{equation} 
g_{\alpha \beta}=
            \left( {\begin{array}{cccc}
+1 & 0 & 0 & 0 \\

0 & -a^2(t) & 0 & 0 \\

0 & 0 & -a^2(t) & 0 \\

0 & 0 & 0 & -a^2(t) \\
                \end{array} } \right).
\label{eq:B10}
\end{equation} 
While in the second case, the flat space spacetime metric is   
\begin{equation} 
g_{\alpha \beta}=
            \left( {\begin{array}{cccc}
-1 & 0 & 0 & 0 \\

0 & a^2(t) & 0 & 0 \\

0 & 0 & a^2(t) & 0 \\

0 & 0 & 0 & a^2(t) \\
                \end{array} } \right).
\label{eq:B20}
\end{equation}

It is reasonable to expect that the energy-momentum tensor $T_{\mu\nu}$ for a given Lagrangian $\mathcal{L}$ (which is a physical quantity) will be of the same form and independent of the choice of spacetime signature, which is just a mathematical construct.
Here, we show this by deriving the energy-momentum tensor $T_{\mu \nu}$ for the $\phi$ field in flat expanding space in both signatures. 

The scalar field $\phi$ is taken to be spatially homogeneous so $\phi=\phi(t)$ and is independent of the space coordinates $x^i$, $i \in \{1,2,3\}$.

\section{Signature $(+,-,-,-)$}
If the signature is $(+,-,-,-)$ the corresponding stress energy tensor $T_{\mu \nu}$ that follows from the Lagrangian density $\mathcal{L}_{\phi}$ is
\begin{eqnarray}
T_{\mu\nu}&=&2\frac{\partial \mathcal{L}_{\phi}}{\partial g^{\mu\nu}} - g_{\mu\nu}\mathcal{L}_{\phi},
\label{eq:B30}
\end{eqnarray}
where $\mathcal{L}_{\phi}$ is given in Eq.\ (\ref{eq:A1}).
The variation is
\begin{eqnarray}
\frac{\partial\mathcal{L}_{\phi}}{\partial g^{\mu\nu}} &=&\frac{\partial}{\partial g^{\mu \nu}}\left[\frac{1}{2}g^{\alpha \beta}\partial_\alpha \phi \partial_\beta \phi-V(\phi)\right]=\frac{1}{2}\delta^\alpha_{\phantom{\alpha}\mu}\delta^\beta_{\phantom{\beta}\nu}\partial_\alpha \phi \partial_\beta\phi=\frac{1}{2}\partial_\mu \phi \partial_\nu\phi.
\label{eq:B40}
\end{eqnarray}
Using Eq.\ (\ref{eq:A1}) and Eq.\ (\ref{eq:B40}) in Eq.\ (\ref{eq:B30}) we have
\begin{eqnarray}\label{4}
T_{\mu\nu}&=& \partial_\mu \phi \partial_\nu\phi - g_{\mu\nu}\left[\frac{1}{2}g^{\alpha\beta}\partial_\alpha \phi \partial_\beta \phi - V(\phi)\right].
\label{eq:B50}
\end{eqnarray}
Using the metric of Eq.\ (\ref{eq:B10}) and the fact that $\phi$ only depends on time $x^0=t$ leads to
\begin{eqnarray}
T_{00}&=&\partial_0\phi \partial_0\phi-g_{00}\left[\frac{1}{2}g^{00}\partial_0\phi \partial_0\phi-V(\phi)\right]=\frac{1}{2}\dot\phi^2+V(\phi),
\label{eq:B60}
\end{eqnarray}
and
\begin{eqnarray}
T_{11}&=&\partial_1\phi \partial_1\phi-g_{11}\left[\frac{1}{2}g^{11}\partial_1\phi \partial_1\phi-V(\phi)\right]=a^2\left[\frac{1}{2}\dot\phi^2-V(\phi)\right]=T_{22}=T_{33}.
\label{eq:B70}
\end{eqnarray}

We can transform from the coordinate basis to an orthonormal basis as follows. Let us denote the coordinate basis ---  which we have already been using without writing it down explicitly --- as $({\bf e}_0, {\bf e}_1, {\bf e}_2, {\bf e}_3)$. The corresponding dual basis is denoted  by $({\bf e}^0, {\bf e}^1, {\bf e}^2, {\bf e}^3)$. With respect to this basis, the metric tensor ${\bf g}$ and the stress-energy tensor ${\bf T}$ are
\begin{eqnarray}
{\bf g} &=& g_{\mu\nu}{\bf e}^\mu\otimes{\bf e}^\nu = 1\cdot{\bf e}^0\otimes{\bf e}^0 -a^2\cdot{\bf e}^1\otimes{\bf e}^1 -a^2\cdot{\bf e}^2\otimes{\bf e}^2-a^2\cdot{\bf e}^3\otimes{\bf e}^3,
\label{eq:B80}
\end{eqnarray}
\begin{eqnarray}
{\bf T} &=& T_{\mu\nu}{\bf e}^\mu\otimes{\bf e}^\nu = T_{00}\cdot{\bf e}^0\otimes{\bf e}^0 +T_{11}\cdot{\bf e}^1\otimes{\bf e}^1 +T_{22}\cdot{\bf e}^2\otimes{\bf e}^2+T_{33}\cdot{\bf e}^3\otimes{\bf e}^3.
\end{eqnarray}
We can find a transformation to an orthonormal (dual) basis   by taking something like the square-root of the metric (this is easy when the metric is diagonal, and more difficult when it is not diagonal). There are of course infinitely orthonormal bases to write down, but the most straightforward one is
\begin{eqnarray}
{\bf e}^{\hat 0}&=&{\bf e}^{0},
\label{eq:B90}
\end{eqnarray}
\begin{eqnarray}
{\bf e}^{\hat 1}&=&a{\bf e}^{1},
\label{eq:B91}
\end{eqnarray}
\begin{eqnarray}
{\bf e}^{\hat 2}&=&a{\bf e}^{2},
\label{eq:B92}
\end{eqnarray}
\begin{eqnarray}
{\bf e}^{\hat 3}&=&a{\bf e}^{3}.
\label{eq:B93}
\end{eqnarray}
Note that $({\bf e}^{\hat 0}, {\bf e}^{\hat 1}, {\bf e}^{\hat 2}, {\bf e}^{\hat 3})$ is an orthonormal basis because with respect to this basis, ${\bf g}$ has the form of the Minkowski metric,\footnote{Locally (at a point)  every metric is Minkowskian --- note that all of this analysis that we are doing is purely local and we are not saying the metric is globally Minkowskian.} 
\begin{eqnarray}
{\bf g}&=&1\cdot{\bf e}^0\otimes{\bf e}^0 -a^2\cdot{\bf e}^1\otimes{\bf e}^1 -a^2\cdot{\bf e}^2\otimes{\bf e}^2-a^2\cdot{\bf e}^3\otimes{\bf e}^3\nonumber\\
&=&1\cdot{\bf e}^{\hat 0}\otimes{\bf e}^{\hat 0} -1\cdot{\bf e}^{\hat 1}\otimes{\bf e}^{\hat 1} -1\cdot{\bf e}^{\hat 2}\otimes{\bf e}^{\hat 2}-1\cdot{\bf e}^{\hat 3}\otimes{\bf e}^{\hat 3}.
\label{eq:B100}
\end{eqnarray}
As for the stress-energy tensor ${\bf T}$, it is
\begin{eqnarray}
{\bf T}&=&T_{00}\cdot{\bf e}^0\otimes{\bf e}^0 +T_{11}\cdot{\bf e}^1\otimes{\bf e}^1+T_{22}\cdot{\bf e}^2\otimes{\bf e}^2+T_{33}\cdot{\bf e}^3\otimes{\bf e}^3\nonumber\\
&=&T_{00}\cdot{\bf e}^{\hat 0}\otimes{\bf e}^{\hat 0}+\left(T_{11}/a^2\right)\cdot{\bf e}^{\hat 1}\otimes{\bf e}^{\hat 1} +\left(T_{22}/a^2\right)\cdot{\bf e}^{\hat 2}\otimes{\bf e}^{\hat 2}+\left(T_{33}/a^2\right)\cdot{\bf e}^{\hat 3}\otimes{\bf e}^{\hat 3}\nonumber\\
&=&T_{\hat 0\hat 0}\cdot{\bf e}^{\hat 0}\otimes{\bf e}^{\hat 0}+T_{\hat 1\hat 1}\cdot{\bf e}^{\hat 1}\otimes{\bf e}^{\hat 1}+T_{\hat 2\hat 2}\cdot{\bf e}^{\hat 2}\otimes{\bf e}^{\hat 2}+T_{\hat 3\hat 3}\cdot{\bf e}^{\hat 3}\otimes{\bf e}^{\hat 3}.
\label{eq:B110}
\end{eqnarray}
In the orthonormal basis, the energy-density and pressures can be read from the components.  

The energy density $\rho_\phi$ is
\begin{eqnarray}
\rho_\phi &= & T_{\hat 0\hat 0}\ = \ \frac{1}{2}\dot\phi^2 + V(\phi).
\label{eq:B120}
\end{eqnarray}
Since $T_{\hat1\hat1} = T_{\hat 2\hat 2} = T_{\hat 3\hat 3}$, these components represent the pressure $p_\phi$, which is
\begin{eqnarray}
p_\phi&=& \frac{1}{2}\dot\phi^2 - V(\phi)
\label{eq:B130}
\end{eqnarray}
\section{Signature $(-,+,+,+)$}

If the metric has signature $(-,+,+,+)$, then we can get $T_{\mu\nu}$ from the  Lagrangian density $\mathcal{L}_{\phi}$ using\footnote{See page 125 in \emph{A Relativist's Toolkit} by E. Poisson\cite{poissonrelativist}}
\begin{eqnarray}
T_{\mu\nu}&=&-2\frac{\partial \mathcal{L}_{\phi}}{\partial g^{\mu\nu}} + g_{\mu\nu}\mathcal{L}_{\phi},
\label{eq:B140}
\end{eqnarray}
where $\mathcal{L}_{\phi}$ is now given by
\begin{eqnarray}
\mathcal{L}_{\phi} &=&-\frac{1}{2}g^{\alpha\beta}\partial_\alpha \phi \partial_\beta \phi - V(\phi).
\label{eq:B150}
\end{eqnarray}
The important thing to note is that we need to introduce a minus sign in front of the kinetic and space gradient part of the Lagrangian density. To see why, remember that in the present problem, since $\phi$ is spatially homogeneous, the first term is just $\frac{1}{2}\dot\phi^2$ but since $g^{00} =-1$, we will need that minus sign. 

Varying, one finds 
\begin{eqnarray}
\frac{\partial\mathcal{L}_{\phi}}{\partial g^{\mu\nu}} &=&\frac{\partial}{\partial g^{\mu \nu}}\left[-\frac{1}{2}g^{\alpha \beta}\partial_\alpha \phi \partial_\beta \phi-V(\phi)\right]=-\frac{1}{2}\delta^\alpha_{\phantom{\alpha}\mu}\delta^\beta_{\phantom{\beta}\nu}\partial_\alpha \phi \partial_\beta\phi=-\frac{1}{2}\partial_\mu \phi \partial_\nu\phi,
\label{eq:B160}
\end{eqnarray}
so
\begin{eqnarray}\label{4}
T_{\mu\nu}&=& \partial_\mu \phi \partial_\nu\phi + g_{\mu\nu}\left[-\frac{1}{2}g^{\alpha\beta}\partial_\alpha \phi \partial_\beta \phi - V(\phi)\right].
\label{eq:B170}
\end{eqnarray}
Using the signature $(-,+,+,+)$ and the fact that $\phi$ only depends on time $x^0=t$ leads to

\begin{eqnarray}
T_{00}&=&\partial_0\phi \partial_0\phi+g_{00}\left[-\frac{1}{2}g^{00}\partial_0\phi \partial_0\phi-V(\phi)\right]=\frac{1}{2}\dot\phi^2+V(\phi),
\label{eq:B180}
\end{eqnarray}
and
\begin{eqnarray}
T_{11}&=&\partial_1\phi \partial_1\phi+g_{11}\left[-\frac{1}{2}g^{11}\partial_1\phi \partial_1\phi-V(\phi)\right]= a^2\left[\frac{1}{2}\dot\phi^2-V(\phi)\right]=T_{22}=T_{33}.
\label{eq:B190}
\end{eqnarray}

As discussed earlier, we can transform to an orthonormal basis. Denoting the coordinate basis  which we have been using as $({\bf e}_0, {\bf e}_1, {\bf e}_2, {\bf e}_3)$, and the corresponding dual basis is denoted  by $({\bf e}^0, {\bf e}^1, {\bf e}^2, {\bf e}^3)$, the metric tensor ${\bf g}$ and the stress-energy tensor ${\bf T}$ in the dual basis are
\begin{eqnarray}
{\bf g} &=& g_{\mu\nu}{\bf e}^\mu\otimes{\bf e}^\nu=-1\cdot{\bf e}^0\otimes{\bf e}^0 +a^2\cdot{\bf e}^1\otimes{\bf e}^1 +a^2\cdot{\bf e}^2\otimes{\bf e}^2+a^2\cdot{\bf e}^3\otimes{\bf e}^3,
\end{eqnarray}
\begin{eqnarray}
{\bf T} &=& T_{\mu\nu}{\bf e}^\mu\otimes{\bf e}^\nu=T_{00}\cdot{\bf e}^0\otimes{\bf e}^0 +T_{11}\cdot{\bf e}^1\otimes{\bf e}^1 +T_{22}\cdot{\bf e}^2\otimes{\bf e}^2+T_{33}\cdot{\bf e}^3\otimes{\bf e}^3.
\end{eqnarray}
The transformation to the most straightforward orthonormal (dual) basis $({\bf e}^{\hat 0}, {\bf e}^{\hat 1}, {\bf e}^{\hat 2}, {\bf e}^{\hat 3})$  is given in Eqs.\ (\ref{eq:B90})-(\ref{eq:B91}) and the metric and stress tensor in this basis are given in Eqs.\ (\ref{eq:B100})-(\ref{eq:B110}), so we can read of the  
energy density $\rho_\phi$,
\begin{eqnarray}
\rho_\phi &= &  \frac{1}{2}\dot\phi^2 + V(\phi),
\end{eqnarray}
and the pressure $p_\phi$,
\begin{eqnarray}
p_\phi&=& \frac{1}{2}\dot\phi^2 - V(\phi),
\end{eqnarray}
which are identical as Eqs.\ (\ref{eq:B120}) and (\ref{eq:B130}), as they must be.

\cleardoublepage

\chapter{Solution of Standard Cubic Equation}
\label{Appendix:C}

The general form of the cubic equation is:\footnote{Here we have taken the coefficient of the cubic term $\left(y^3\right)$ as unity without the loss of any generality.}
\begin{eqnarray}
y^3+\alpha y^2+\beta y+\gamma=0.
\label{eq:cubic1}
\end{eqnarray}
We begin with an observation: by a simple translation we can remove the ``square term" from \textit{any cubic equation we wish to solve}. Substitute
\begin{eqnarray}
x=y+\frac{\alpha}{3}, \hspace{1.cm} \mathrm{or} \hspace{1.5cm} y=x-\frac{\alpha}{3}, 
\label{eq:cubic2}
\end{eqnarray} 
in Eq.\ (\ref{eq:cubic1}):
\begin{eqnarray}
\left(x-\frac{\alpha}{3}\right)^3+\alpha \left(x-\frac{\alpha}{3}\right)^2+\beta \left(x-\frac{\alpha}{3}\right)+\gamma=0,
\label{eq:cubic3}
\end{eqnarray} 
which is simplified as:
\begin{eqnarray}
x^3+\left(\beta-\frac{\alpha^2}{3}\right)x+\left(\frac{2\alpha^3}{27}-\frac{\alpha\beta}{3}+\gamma\right)=0.
\label{eq:cubic4}
\end{eqnarray}
It follows that we can always reduce all cubic equations of the form Eq.\ (\ref{eq:cubic1}) to
\begin{eqnarray}
x^3+Px+Q=0.
\label{eq:cubic4}
\end{eqnarray}
The point to appreciate is that any solution, $x$, to Eq.\ (\ref{eq:cubic4}) will gives rise to a solution $y=x-\frac{\alpha}{3}$ to Eq.\ (\ref{eq:cubic1}), and of course, vise verse.
Since the special cases where either $P=0$ or $Q=0$ are trivial, henceforth, we assume that $P\neq 0$, and $Q\neq 0$. Our intention is to exploit the trigonometric identities:
\begin{eqnarray}
\mathrm{cos^3}\theta-\frac{3}{4}\mathrm{cos}\theta-\frac{1}{4}\mathrm{cos}3\theta=0.
\label{eq:cubic5}
\end{eqnarray}
Comparison of the Eqs.\ (\ref{eq:cubic4}) and (\ref{eq:cubic5}) we will get:
\begin{eqnarray}
x=\mathrm{cos}\theta,\hspace{1cm}  P=\frac{-3}{4}\hspace{1cm}  Q=-\frac{1}{4}\mathrm{cos}3\theta.
\label{eq:cubic6}
\end{eqnarray}
This means that our comparison is reasonable if:
\begin{eqnarray}
\Rightarrow |4Q|<1.
\label{eq:cubic7}
\end{eqnarray}
From the third equation in Eqs.\ (\ref{eq:cubic6}) 
\begin{eqnarray}
\Rightarrow \theta=\frac{1}{3}\mathrm{cos^{-1}}\left[-4Q\right]+\frac{2n\pi}{3}, \hspace{1cm} n \in \{0,1,2\}.
\label{eq:cubic8}
\end{eqnarray}
\begin{eqnarray}
\Rightarrow x=\mathrm{cos}\left[\frac{1}{3}\mathrm{cos^{-1}}\left[-4Q\right]+\frac{2n\pi}{3}\right], \hspace{1cm} n \in \{0,1,2\}.
\label{eq:cubic9}
\end{eqnarray}
which will be the solution of the cubic equation of the form given in Eq.\ (\ref{eq:cubic4}) for the case when $|4Q|<1$.\footnote{From here, we find the solution $y$ for the original cubic equation (\ref{eq:cubic1}) using Eqs.\ (\ref{eq:cubic2}), but since we do not need that in this work hence we will stop here.}

Now if $|4Q|>1$, then we have to use the following trigonometric identity for comparison:
\begin{eqnarray}
\mathrm{cosh^3}\eta-\frac{3}{4}\mathrm{cosh}\eta-\frac{1}{4}\mathrm{cosh}3\eta=0.
\label{eq:cubic10}
\end{eqnarray}
Comparing the coefficients of Eqs.\ (\ref{eq:cubic4}) and  (\ref{eq:cubic10}):
\begin{eqnarray}
x=\mathrm{cosh}\eta,\hspace{1cm}  P=\frac{-3}{4}\hspace{1cm}  Q=-\frac{1}{4}\mathrm{cosh}3\eta.
\label{eq:cubic11}
\end{eqnarray}
Here, our comparison works if $4Q<-1$. Using the last equation in Eq.\ (\ref{eq:cubic11}) results in:
\begin{eqnarray}
\eta=\frac{1}{3}\mathrm{cosh^{-1}}\left[-4Q\right].
\label{eq:cubic12}
\end{eqnarray}
\begin{eqnarray}
\Rightarrow x=\mathrm{cosh}\left[\frac{1}{3}\mathrm{cosh^{-1}}\left[-4Q\right]\right],
\label{eq:cubic13}
\end{eqnarray}
That completes the solution of cubic equation.


\cleardoublepage

\chapter{Different $\boldsymbol {H(z)}$ Data Sets}
\label{Appendix:D}

Here we compiled the $H(z)$ data sets that were used in this thesis.
The data is taken from different sources mentioned in the caption of the Table (\ref{tab:Hz1}). 
Here, there are 21 data points. After that, there was an addition of one higher $z$ 
data point from Busca \textit{et al.}\cite{busca12} $H(z=2.3)=224 \pm 8$ km s$^{-1}$ Mpc $^{-1}$ which gave us better constraints on the 
parameters of the dark energy models discussed in this thesis, as explained 
in chapter (\ref{Chapter5}). The $H(z)$ data with 22 data points are shown in Table.\ (\ref{tab:Hz2}). The more complete and the largest data set with 28 data points used in Farooq \textit{et al.},\cite{Farooq:2013hq,farooq4} is given in Table\ (\ref{tab:Hz3}).\footnote{ 
Simon \textit{et al.}\cite{simon05}, Stern \textit{et al.}\cite{Stern2010}, Moresco \textit{et al.}\cite{moresco12}, and Zhang \textit{et al.}\cite{Zhang2012}, 
estimate $H(z)$ from measurements of differential ages of
passively evolving galaxies. Busca \textit{et al.}\cite{busca12} use Ly$\alpha$ 
BAO and WMAP7 data while Blake \textit{et al.}\cite{Blake2012} and Chuang \textit{et al.}\cite{Chuang2012b} 
use galaxy clustering BAO data to estimate $H(z)$.}

\begin{table}
\begin{center}
\begin{tabular}{ cccc }
\hline\hline
$z$ & $H(z)$ &$\sigma_{H}$ & Reference\\
   & (km s$^{-1}$ Mpc $^{-1}$) &(km s$^{-1}$ Mpc $^{-1}$)& \\
\hline

0.090&	69&	12&	1\\
0.170&	83&	8&	1\\
0.179&	75&	4&	4\\
0.199&	75&	5&	4\\
0.240&	79.69&	2.65&	2\\
0.270&	77&	14&	1\\
0.352&	83&	14&	4\\
0.400&	95&	17&	1\\
0.430&	86.45&	3.68&	2\\
0.480&	97&	62&	3\\
0.593&	104&	13&	4\\
0.680&	92&	8&	4\\
0.781&	105&	12&	4\\
0.875&	125&	17&	4\\
0.880&	90&	40&	3\\
0.900&	117&	23&	1\\
1.037&	154&	20&	4\\
1.300&	168&	17&	1\\
1.430&	177&	18&	1\\
1.530&	140&	14&	1\\
1.750&	202&	40&	1\\
\hline\hline
\end{tabular}
\end{center}
\caption{Hubble parameter versus redshift data. The last column reference numbers are:
1. Simon \textit{et al.}\cite{simon05}, 2. Gazta\^{n}aga \textit{et al.}\cite{gaztanaga09}, 3. Stern \textit{et al.}\cite{Stern2010}, 
4. Moresco \textit{et al.}\cite{moresco12}.
}\label{tab:Hz1}
\end{table}

\begin{table}
\begin{center}
\begin{tabular}{ cccc }
\hline\hline
$z$ & $H(z)$ &$\sigma_{H}$ & Reference\\
   & (km s$^{-1}$ Mpc $^{-1}$) &(km s$^{-1}$ Mpc $^{-1}$)& \\
\hline

0.090&	69&	12&	1\\
0.170&	83&	8&	1\\
0.179&	75&	4&	4\\
0.199&	75&	5&	4\\
0.240&	79.69&	2.65&	2\\
0.270&	77&	14&	1\\
0.352&	83&	14&	4\\
0.400&	95&	17&	1\\
0.430&	86.45&	3.68&	2\\
0.480&	97&	62&	3\\
0.593&	104&	13&	4\\
0.680&	92&	8&	4\\
0.781&	105&	12&	4\\
0.875&	125&	17&	4\\
0.880&	90&	40&	3\\
0.900&	117&	23&	1\\
1.037&	154&	20&	4\\
1.300&	168&	17&	1\\
1.430&	177&	18&	1\\
1.530&	140&	14&	1\\
1.750&	202&	40&	1\\
2.300&	224&	08&	5\\
\hline\hline
\end{tabular}
\end{center}
\caption{Hubble parameter versus redshift data. The last column reference numbers are
1. Simon \textit{et al.}\cite{simon05}, 2. Gazta\^{n}aga \textit{et al.}\cite{gaztanaga09}, 3. Stern \textit{et al.}\cite{Stern2010}, 
4. Moresco \textit{et al.}\cite{moresco12}, 5. Busca \textit{et al.}\cite{busca12}.
}\label{tab:Hz2}
\end{table}

\begin{table}
\begin{center}
\begin{tabular}{cccc}
\hline\hline
\multirow{2}{*}{$z$} &\ \ $H(z)$ & \ \ $\sigma_{H}$ &\ \  \multirow{2}{*}{Reference}\\
~~~~~    & (km s$^{-1}$ Mpc $^{-1}$) &~~~~~~~ (km s$^{-1}$ Mpc $^{-1}$)& \\
\hline
0.070& \ \	69&\ \	19.6&\ \ 5\\
0.100& \ \	69&\ \	12&\ \	1\\
0.120& \ \	68.6&\ \	26.2&\ \	5\\
0.170& \ \	83&\ \	8&\ \	1\\
0.179& \ \	75&\ \	4&\ \	3\\
0.199& \ \	75&\ \	5&\ \	3\\
0.200& \ \	72.9&\ \	29.6&\ \	5\\
0.270& \ \	77&\ \	14&\ \	1\\
0.280& \ \	88.8&\ \	36.6&\ \	5\\
0.350& \ \	76.3&\ \	5.6&\ \	7\\
0.352& \ \	83&\ \	14&\ \	3\\
0.400& \ \	95&\ \	17&\ \	1\\
0.440& \ \	82.6&\ \	7.8&\ \	6\\
0.480& \ \	97&\ \	62&\ \	2\\
0.593& \ \	104&\ \	13&\ \	3\\
0.600& \ \	87.9&\ \	6.1&\ \	6\\
0.680& \ \	92&\ \	8&\ \	3\\
0.730& \ \	97.3&\ \	7.0&\ \	6\\
0.781& \ \	105&\ \	12&\ \	3\\
0.875& \ \	125&\ \	17&\ \	3\\
0.880& \ \	90&\ \	40&\ \	2\\
0.900& \ \	117&\ \	23&\ \	1\\
1.037& \ \	154&\ \	20&\ \	3\\
1.300& \ \	168&\ \	17&\ \	1\\
1.430& \ \	177&\ \	18&\ \	1\\
1.530& \ \	140&\ \	14&\ \	1\\
1.750& \ \	202&\ \	40&\ \	1\\
2.300& \ \	224&\ \	8&\ \	4\\

\hline\hline
\end{tabular}
\end{center}
\caption{Hubble parameter versus redshift data. The last column reference numbers are
1. Simon \textit{et al.}\cite{simon05}, 2. Stern \textit{et al.}\cite{Stern2010}, 3. Moresco \textit{et al.}\cite{moresco12}, 4. Busca  \textit{et al.}\cite{busca12}, 5. Zhang \textit{et al.}\cite{Zhang2012}, 6. Blake \textit{et al.}\cite{Blake2012}, 7. Chuang \textit{et al.}\cite{Chuang2012b}.
}
\label{tab:Hz3}
\end{table}



\cleardoublepage

\chapter{SNeIa ``Union 2.1" Compilation Data}
\label{Appendix:E}

This data is taken from Supernova Cosmology Project (SCP) ``Union2.1" SN Ia compilation.\footnote{http://supernova.lbl.gov/Union/}
There were 833 SNe data points which were drawn from 19 different datasets, but only 580 SNe pass usability cuts.\cite{suzuki2012}

\begin{center}
\begin{longtable}[c]{c c c }
\caption[SNeIa ``union'' data set.]{SNeIa ``union'' data set. The redshift $z$, distance modulii $\mu$, and 1$\sigma$ statistical measurement errors on the measurement of $\mu$.}\\
\hline 
$z$&
$\mu$&
$\sigma_\mu$\\
\hline 
\endfirsthead

\multicolumn{3}{c}%
{{\bfseries \tablename\ \thetable{} -- continued from previous page}} \\
\hline 
\hspace{0.1 cm}
$z$&
$\mu$&
$\sigma_\mu$\\
\hline 
\endhead

\hline \multicolumn{3}{r}{{Continued on next page}} \\ 
\endfoot

\hline \hline
\endlastfoot

0.028488 & 35.34658339 & 0.223905933	 \\
0.050043 & 36.68236792 & 0.166828851	 \\
0.052926 & 36.81769125 & 0.155755915	 \\
0.070086 & 37.44673654 & 0.158466934	 \\
0.062668 & 37.48340935 & 0.156099435	 \\
0.087589 & 38.22905705 & 0.187745679	 \\
0.078577 & 37.48816226 & 0.155635656	 \\
0.017227 & 34.65436995 & 0.19933718	 \\
0.042233 & 36.33645955 & 0.167174042	 \\
0.045295 & 36.64027218 & 0.164981249	 \\
0.03648 & 35.90532197 & 0.170174953	 \\
0.019599 & 34.58521743 & 0.18469122	 \\
0.100915 & 38.4567456 & 0.167333482	 \\
0.027342 & 35.08576569 & 0.175510836	 \\
0.074605 & 37.58811576 & 0.159770865	 \\
0.026489 & 35.4806852 & 0.19131227	 \\
0.049922 & 36.56697347 & 0.16230382	 \\
0.030604 & 35.55023776 & 0.173295444	 \\
0.016345641 & 34.04402778 & 0.142912931	 \\
0.0154363 & 33.9409484 & 0.14869411	 \\
0.030529 & 35.59924572 & 0.088750664	 \\
0.024525 & 35.05817066 & 0.102438504	 \\
0.023953 & 34.96871038 & 0.107041197	 \\
0.026038 & 35.36726207 & 0.108499792	 \\
0.048948 & 36.7315974 & 0.172547619	 \\
0.024314 & 35.10949506 & 0.181662706	 \\
0.015166 & 34.10166662 & 0.215239341	 \\
0.03572 & 35.96054064 & 0.171186987	 \\
0.048818 & 36.38201078 & 0.160299265	 \\
0.021980006 & 34.85297336 & 0.187544764	 \\
0.1244 & 39.04478851 & 0.164268688	 \\
0.036 & 35.8210171 & 0.16788525	 \\
0.016321 & 34.01742111 & 0.204965074	 \\
0.01673 & 34.22633717 & 0.20946467	 \\
0.0275 & 35.64970591 & 0.176364689	 \\
0.021793 & 34.97378687 & 0.232449383	 \\
0.01645 & 34.18129629 & 0.25089656	 \\
0.023208 & 35.08554272 & 0.231479605	 \\
0.036457 & 36.13423313 & 0.217628588	 \\
0.019264 & 34.95261373 & 0.240633817	 \\
0.017605 & 34.3437957 & 0.271650645	 \\
0.031528 & 35.72876878 & 0.225362489	 \\
0.023536 & 35.16959909 & 0.234698202	 \\
0.016743 & 34.0027278 & 0.248385126	 \\
0.05371 & 36.47643849 & 0.221700261	 \\
0.016991 & 34.37877181 & 0.309889811	 \\
0.027865 & 35.09337833 & 0.222426417	 \\
0.017173 & 34.26067146 & 0.248216009	 \\
0.029955 & 35.97225783 & 0.224368552	 \\
0.016559 & 34.34383381 & 0.251177496	 \\
0.015 & 34.16350389 & 0.161452857	 \\
0.0544 & 36.95443541 & 0.086095353	 \\
0.1561 & 39.22925402 & 0.084144124	 \\
0.0393 & 36.33439501 & 0.099100528	 \\
0.1241 & 38.8220334 & 0.111614247	 \\
0.1441 & 38.8360423 & 0.156777492	 \\
0.1299 & 38.97918547 & 0.129444952	 \\
0.0784 & 37.68224045 & 0.087216805	 \\
0.0583 & 37.03263017 & 0.206256669	 \\
0.0309 & 35.92947288 & 0.183689296	 \\
0.0406 & 36.36563513 & 0.172317345	 \\
0.0152 & 34.0169043 & 0.215071231	 \\
0.0224 & 34.9470872 & 0.239011369	 \\
0.016 & 34.17401539 & 0.221115375	 \\
0.0362 & 35.9868706 & 0.171742204	 \\
0.0173 & 34.2497348 & 0.2155085	 \\
0.0312 & 35.62680976 & 0.180216697	 \\
0.0221 & 34.91154976 & 0.189960683	 \\
0.016 & 33.82460889 & 0.207959703	 \\
0.0249 & 34.80370708 & 0.193523924	 \\
0.0303 & 35.62826135 & 0.176747386	 \\
0.0283 & 35.52026995 & 0.180124446	 \\
0.0152 & 34.25836979 & 0.241474862	 \\
0.0345 & 35.97824448 & 0.211562474	 \\
0.036 & 35.67925968 & 0.179022122	 \\
0.0248 & 35.25617751 & 0.184605303	 \\
0.0292 & 35.99256678 & 0.174820108	 \\
0.0163 & 34.45325291 & 0.212891043	 \\
0.0187 & 35.04829985 & 0.197196789	 \\
0.0195 & 34.75691168 & 0.197526188	 \\
0.0256 & 35.68472179 & 0.186173892	 \\
0.0337 & 35.84369332 & 0.179737454	 \\
0.0546 & 36.60955851 & 0.176655181	 \\
0.024 & 35.17618578 & 0.195512674	 \\
0.0336 & 36.00539455 & 0.188520474	 \\
0.0341 & 35.8419047 & 0.175222462	 \\
0.0261 & 35.36041779 & 0.192469562	 \\
0.0211 & 34.66071824 & 0.188393368	 \\
0.0321 & 35.89599457 & 0.17384516	 \\
0.0221 & 34.9221291 & 0.189014756	 \\
0.0334 & 35.8799403 & 0.175772733	 \\
0.0341 & 35.94254084 & 0.172834592	 \\
0.0421 & 36.40046503 & 0.169930417	 \\
0.0576 & 37.08025666 & 0.161852672	 \\
0.0205 & 34.61728007 & 0.191412743	 \\
0.0402 & 36.37453308 & 0.17063735	 \\
0.026 & 35.38129323 & 0.183933665	 \\
0.0259 & 35.41596869 & 0.178489659	 \\
0.0239 & 35.03497414 & 0.181392417	 \\
0.069 & 37.56604798 & 0.17616364	 \\
0.0651 & 37.3066908 & 0.162808584	 \\
0.0229 & 35.19686717 & 0.185459075	 \\
0.0315 & 35.65113634 & 0.170993524	 \\
0.0215 & 34.93273633 & 0.187040818	 \\
0.0255 & 35.72017314 & 0.196167302	 \\
0.0325 & 35.81309142 & 0.170394736	 \\
0.0843 & 38.05183112 & 0.200253645	 \\
0.0308 & 35.62898081 & 0.178370444	 \\
0.0327 & 36.09416721 & 0.169492305	 \\
0.0423 & 36.3928259 & 0.171582197	 \\
0.0684 & 37.73119698 & 0.170591643	 \\
0.0153 & 34.70718573 & 0.213242896	 \\
0.0233 & 34.88212867 & 0.183424685	 \\
0.0491 & 36.73013588 & 0.175843223	 \\
0.0425 & 35.92810181 & 0.195296013	 \\
0.0192 & 34.73667845 & 0.197842881	 \\
0.0308 & 35.77913624 & 0.173310005	 \\
0.0212 & 34.84713453 & 0.191529722	 \\
0.0277 & 35.70500377 & 0.183298735	 \\
0.0335 & 35.97383254 & 0.170490252	 \\
0.0208 & 34.79545637 & 0.201629405	 \\
0.0173 & 34.23006604 & 0.220994474	 \\
0.036 & 36.14629886 & 0.177000364	 \\
0.0233 & 35.19835742 & 0.185738575	 \\
0.0589 & 37.11164751 & 0.165682351	 \\
0.0583 & 37.05828815 & 0.168089681	 \\
0.0688 & 37.48663565 & 0.198025809	 \\
0.0321 & 35.64807683 & 0.175025423	 \\
0.0522 & 36.67431461 & 0.189117803	 \\
0.0308 & 35.59350422 & 0.172493084	 \\
0.0329 & 35.94176854 & 0.170393176	 \\
0.023 & 35.0706425 & 0.186848385	 \\
0.015 & 34.3797728 & 0.217348535	 \\
0.0321 & 35.87036183 & 0.173741008	 \\
0.0643 & 37.17550222 & 0.16600801	 \\
0.032 & 35.82669134 & 0.193104692	 \\
0.0209 & 34.7028985 & 0.194226935	 \\
0.0219 & 34.84899048 & 0.188387184	 \\
0.032 & 35.58849732 & 0.173446311	 \\
0.0151 & 34.52587529 & 0.213682161	 \\
0.0192 & 34.49156233 & 0.197091648	 \\
0.0266 & 35.32349359 & 0.177716836	 \\
0.0377 & 35.79521443 & 0.294106561	 \\
0.0247 & 34.91369283 & 0.181621012	 \\
0.0242 & 35.18938058 & 0.180174916	 \\
0.0366 & 35.97129065 & 0.168963688	 \\
0.0229 & 35.13862607 & 0.18301804	 \\
0.0312 & 35.88285856 & 0.173762463	 \\
0.015 & 34.11141096 & 0.21338887	 \\
0.0341 & 35.77047476 & 0.177569141	 \\
0.0251 & 34.94827524 & 0.1792227	 \\
0.0189 & 34.37470166 & 0.20957163	 \\
0.029802137 & 35.47092171 & 0.122226818	 \\
0.032134017 & 35.37812789 & 0.12751209	 \\
0.027568726 & 35.47547181 & 0.128117983	 \\
0.046967335 & 36.49441345 & 0.110309142	 \\
0.018315232 & 34.3717549 & 0.160464926	 \\
0.080048144 & 37.68570934 & 0.101968492	 \\
0.024185299 & 35.05232428 & 0.133368622	 \\
0.015027043 & 33.9501915 & 0.178117024	 \\
0.028396027 & 35.54814847 & 0.125198882	 \\
0.044976673 & 36.55039542 & 0.108342547	 \\
0.032912371 & 35.96785202 & 0.118114324	 \\
0.075350112 & 37.58008992 & 0.102274673	 \\
0.020374725 & 34.65644385 & 0.14560335	 \\
0.022971168 & 35.13388295 & 0.135640926	 \\
0.026809197 & 35.26015054 & 0.127879596	 \\
0.017931283 & 34.34982 & 0.155413941	 \\
0.048392195 & 36.68583438 & 0.119215628	 \\
0.056683367 & 36.96362307 & 0.110070938	 \\
0.063864084 & 37.31594161 & 0.119614965	 \\
0.146290296 & 39.55900955 & 0.120248735	 \\
0.129278207 & 38.91800095 & 0.134402616	 \\
0.102715034 & 38.48725959 & 0.119892508	 \\
0.24250468 & 40.13044655 & 0.150915057	 \\
0.298409274 & 41.06240023 & 0.215818596	 \\
0.043718911 & 36.38656602 & 0.127214293	 \\
0.113042645 & 38.55145413 & 0.119237438	 \\
0.256475743 & 40.6108439 & 0.156286225	 \\
0.29558555 & 41.1370597 & 0.215064876	 \\
0.380359487 & 41.65749866 & 0.217910528	 \\
0.145668547 & 39.0567157 & 0.125750126	 \\
0.273454769 & 40.72715824 & 0.163668271	 \\
0.297518834 & 40.77220501 & 0.205185827	 \\
0.378965802 & 41.5814889 & 0.197094008	 \\
0.380416514 & 41.27234053 & 0.198558434	 \\
0.30175503 & 41.4938105 & 0.246761481	 \\
0.348345021 & 41.30779153 & 0.217774889	 \\
0.085689459 & 37.99741341 & 0.11636528	 \\
0.260586108 & 40.52316768 & 0.141595052	 \\
0.215543321 & 40.29652942 & 0.140207657	 \\
0.117625329 & 38.57941851 & 0.113609963	 \\
0.18221824 & 39.59381884 & 0.119334464	 \\
0.357507357 & 41.3288394 & 0.202587682	 \\
0.141787999 & 39.27381133 & 0.119521983	 \\
0.260533477 & 40.80813414 & 0.148279349	 \\
0.232781107 & 40.20452909 & 0.151788368	 \\
0.151857895 & 39.15578836 & 0.113847312	 \\
0.093908632 & 38.17282051 & 0.117952906	 \\
0.286618707 & 41.03840765 & 0.133800276	 \\
0.194316512 & 39.96152004 & 0.126373845	 \\
0.147025138 & 39.30358397 & 0.111657297	 \\
0.211586982 & 40.56032369 & 0.159324734	 \\
0.18011978 & 39.6385777 & 0.115560255	 \\
0.263491027 & 40.76732261 & 0.12729308	 \\
0.19214998 & 40.03921788 & 0.121087443	 \\
0.338802609 & 41.30537188 & 0.170908368	 \\
0.117277363 & 38.74593382 & 0.113702596	 \\
0.142404652 & 39.11647725 & 0.113661194	 \\
0.160861855 & 39.31917589 & 0.1186449	 \\
0.288418345 & 40.84278536 & 0.149598782	 \\
0.1228289 & 38.7997897 & 0.121065439	 \\
0.263647951 & 40.54817567 & 0.13982362	 \\
0.126473162 & 38.71046918 & 0.119200795	 \\
0.172742231 & 39.50293614 & 0.113458016	 \\
0.163795895 & 39.39456306 & 0.113134872	 \\
0.249511055 & 40.78092433 & 0.132323047	 \\
0.257740304 & 40.6514784 & 0.135891056	 \\
0.10671234 & 38.6366669 & 0.117621429	 \\
0.159889938 & 39.35058453 & 0.114747058	 \\
0.204979685 & 40.02450023 & 0.123526943	 \\
0.244378877 & 40.20193039 & 0.124309901	 \\
0.248508131 & 40.27217468 & 0.125645589	 \\
0.228528474 & 40.25443364 & 0.122011371	 \\
0.085854644 & 37.95244412 & 0.112589853	 \\
0.061835765 & 37.13099448 & 0.115680071	 \\
0.277853423 & 40.83800845 & 0.168994208	 \\
0.275440197 & 40.7433918 & 0.142941616	 \\
0.155247328 & 39.31732041 & 0.115415756	 \\
0.330512449 & 41.227939 & 0.156723683	 \\
0.361934309 & 41.38445335 & 0.166441915	 \\
0.330634628 & 41.05225989 & 0.152699339	 \\
0.144621086 & 39.29082977 & 0.112727024	 \\
0.389288788 & 41.60262843 & 0.200867576	 \\
0.173910057 & 39.53194111 & 0.128967776	 \\
0.300312696 & 40.84684819 & 0.151500986	 \\
0.116348503 & 38.74613817 & 0.112296117	 \\
0.218585189 & 40.19909746 & 0.117113951	 \\
0.119671538 & 38.75553812 & 0.110740034	 \\
0.154632097 & 39.32395023 & 0.112047394	 \\
0.408319092 & 41.84029353 & 0.206369619	 \\
0.177600695 & 40.07067585 & 0.124814155	 \\
0.250667631 & 40.74560303 & 0.12097096	 \\
0.251740186 & 40.54172176 & 0.118440111	 \\
0.391599213 & 41.5087271 & 0.171564316	 \\
0.128726735 & 38.86422531 & 0.10951334	 \\
0.399601335 & 41.72880031 & 0.205132434	 \\
0.309492645 & 41.01819247 & 0.134960171	 \\
0.170628397 & 39.47490012 & 0.116376625	 \\
0.278924676 & 40.76782977 & 0.130157044	 \\
0.308580866 & 40.90762171 & 0.177652783	 \\
0.185812447 & 39.82409758 & 0.120667757	 \\
0.066440312 & 37.37210505 & 0.117582401	 \\
0.316429845 & 41.39507338 & 0.192866122	 \\
0.257497888 & 40.59416459 & 0.131393591	 \\
0.298777444 & 41.02935798 & 0.154536936	 \\
0.270434443 & 40.66898384 & 0.145477929	 \\
0.279454733 & 40.50128437 & 0.138724733	 \\
0.292469756 & 40.91904865 & 0.141400627	 \\
0.188853175 & 39.79203749 & 0.119741861	 \\
0.26576248 & 40.49814992 & 0.124803942	 \\
0.124273529 & 38.7199869 & 0.111134859	 \\
0.182548913 & 39.78402156 & 0.114083137	 \\
0.312883364 & 41.01444878 & 0.139506514	 \\
0.309547337 & 41.19255672 & 0.154012333	 \\
0.089019429 & 37.82845293 & 0.117432705	 \\
0.402459619 & 41.91114676 & 0.211278494	 \\
0.20061172 & 39.85239182 & 0.132867364	 \\
0.326396483 & 40.97230418 & 0.144388066	 \\
0.211629598 & 39.98010475 & 0.115235181	 \\
0.179685641 & 39.72095995 & 0.11698488	 \\
0.093149403 & 38.28091373 & 0.11445383	 \\
0.302401621 & 41.0332001 & 0.134910116	 \\
0.108638266 & 38.6505002 & 0.119489655	 \\
0.085696117 & 37.99379171 & 0.117971238	 \\
0.20260868 & 39.96775926 & 0.128807331	 \\
0.196716069 & 39.92441738 & 0.134983154	 \\
0.126687989 & 38.91520843 & 0.123937308	 \\
0.214568259 & 40.20184967 & 0.116987446	 \\
0.320446989 & 41.10278514 & 0.134380311	 \\
0.218347445 & 40.22084984 & 0.124421503	 \\
0.189706567 & 39.93956251 & 0.119682792	 \\
0.092936818 & 38.14583004 & 0.118950356	 \\
0.379662294 & 41.43306189 & 0.14721974	 \\
0.210938395 & 39.98393773 & 0.120818771	 \\
0.366602899 & 41.85985724 & 0.19840443	 \\
0.420926821 & 42.133548 & 0.202321001	 \\
0.258028271 & 40.64467887 & 0.137108499	 \\
0.183568405 & 39.7107958 & 0.123775735	 \\
0.212548765 & 40.08759328 & 0.131576963	 \\
0.360034212 & 41.15730044 & 0.175048187	 \\
0.393974478 & 41.78115342 & 0.15070732	 \\
0.114712621 & 38.66909512 & 0.113372328	 \\
0.143705907 & 39.18577065 & 0.116311968	 \\
0.25248606 & 40.51842108 & 0.145050196	 \\
0.387297107 & 41.89246533 & 0.186429866	 \\
0.348583858 & 41.30094104 & 0.177371888	 \\
0.25549062 & 40.48594463 & 0.150967373	 \\
0.216582822 & 40.3602717 & 0.152479839	 \\
0.43 & 41.31885817 & 0.357827626	 \\
0.62 & 43.22796202 & 0.390344663	 \\
0.57 & 42.48911036 & 0.388952633	 \\
0.3 & 40.96279515 & 0.314865572	 \\
0.38 & 42.06317653 & 0.327833114	 \\
0.43 & 42.39854151 & 0.458327874	 \\
0.24 & 40.74219158 & 0.40284416	 \\
0.44 & 42.04953327 & 0.319794879	 \\
0.5 & 42.36300326 & 0.317337167	 \\
0.97 & 42.82120383 & 0.805210303	 \\
0.479 & 42.35175776 & 0.3581363	 \\
0.83 & 43.54240344 & 0.471046843	 \\
0.416 & 42.42361704 & 0.556342333	 \\
0.581 & 42.06342328 & 0.50503396	 \\
0.45 & 41.83199185 & 0.453701963	 \\
0.579 & 43.18636334 & 0.643297807	 \\
0.32 & 41.24460345 & 0.420998113	 \\
0.657 & 42.97524872 & 0.658728481	 \\
0.43 & 41.76613683 & 0.608328916	 \\
0.472 & 41.96916326 & 0.51447285	 \\
0.374 & 43.18511468 & 0.923198714	 \\
0.18 & 40.18951732 & 0.445447561	 \\
0.55 & 44.34346754 & 1.00681218	 \\
0.592 & 44.15235483 & 0.717963493	 \\
0.172 & 39.30222601 & 0.429709121	 \\
0.526 & 41.95608086 & 0.513679185	 \\
0.763 & 44.47242459 & 0.898163121	 \\
0.58 & 43.30512501 & 0.519962251	 \\
0.43 & 41.8019361 & 0.458804991	 \\
0.45 & 42.27097318 & 0.589947887	 \\
0.656 & 43.14485911 & 0.627202506	 \\
0.495 & 42.11890815 & 0.443653473	 \\
0.49 & 41.78743494 & 0.446484506	 \\
0.57 & 42.67146671 & 0.470382089	 \\
0.388 & 42.20761404 & 0.465017827	 \\
0.45 & 42.40244773 & 0.508290618	 \\
0.48 & 42.15994859 & 0.521420423	 \\
0.615 & 42.54263661 & 0.556222743	 \\
0.4 & 42.31276716 & 0.477589021	 \\
0.655 & 42.31508525 & 0.497081625	 \\
0.498 & 42.98372024 & 0.642406504	 \\
0.465 & 41.8251633 & 0.600030636	 \\
0.453 & 42.82120848 & 0.524912952	 \\
0.425 & 41.20213122 & 0.489149083	 \\
0.514 & 42.78619362 & 0.493735911	 \\
0.423 & 41.56552588 & 0.243579001	 \\
0.859 & 44.0929026 & 0.296965045	 \\
0.936 & 43.30710586 & 0.7059814	 \\
0.528 & 42.45434957 & 0.244290675	 \\
0.978 & 43.50007216 & 0.292647302	 \\
0.885 & 44.18433331 & 0.286230184	 \\
0.815 & 44.0744016 & 0.72892173	 \\
0.698 & 43.76872869 & 0.433480218	 \\
0.568 & 42.70924962 & 0.289546897	 \\
0.711 & 43.56654956 & 0.366725181	 \\
0.3396 & 41.08397338 & 0.226527941	 \\
0.3965 & 41.48792079 & 0.202374999	 \\
0.812 & 43.65357328 & 0.379016342	 \\
0.799 & 43.3788294 & 0.233174228	 \\
0.882 & 43.37503893 & 0.583987119	 \\
0.833 & 43.68809785 & 0.520913246	 \\
0.874 & 43.29537924 & 0.388089303	 \\
0.772 & 43.50999383 & 0.21883477	 \\
0.178 & 39.45479082 & 0.235746048	 \\
0.26 & 40.82336577 & 0.199851736	 \\
0.186 & 39.71216793 & 0.189043139	 \\
0.269 & 40.77616173 & 0.256400639	 \\
0.215 & 40.37805767 & 0.193933998	 \\
0.543 & 42.47880585 & 0.098141434	 \\
0.75 & 43.24299869 & 0.138318458	 \\
0.64 & 42.76424727 & 0.188809377	 \\
0.43 & 42.18412421 & 0.148171058	 \\
0.64 & 43.16438069 & 0.193749017	 \\
0.497 & 42.32455309 & 0.167980194	 \\
0.44 & 42.01064753 & 0.107484808	 \\
0.355 & 41.34231158 & 0.201468434	 \\
0.78 & 43.60080717 & 0.171318833	 \\
0.54 & 42.42462127 & 0.111551881	 \\
0.86 & 43.92166067 & 0.172411965	 \\
0.468 & 42.54890374 & 0.163914771	 \\
0.84 & 43.87310319 & 0.222348502	 \\
0.96 & 43.61321161 & 0.275514631	 \\
0.8218 & 43.81536396 & 0.213206984	 \\
0.93 & 43.55242958 & 0.289918379	 \\
0.451 & 41.79667441 & 0.143055822	 \\
0.61 & 42.90496571 & 0.156248253	 \\
0.83 & 44.05065993 & 0.208835356	 \\
0.707 & 43.28242562 & 0.256253916	 \\
0.415 & 41.87677046 & 0.141093703	 \\
0.557 & 42.5570112 & 0.157456031	 \\
0.791 & 43.57688112 & 0.201343679	 \\
0.695 & 43.21170426 & 0.205449194	 \\
0.633 & 43.04986353 & 0.169957138	 \\
0.2486 & 40.61118534 & 0.154515955	 \\
0.532 & 42.56581566 & 0.166651053	 \\
0.331 & 41.07847005 & 0.137704767	 \\
0.346 & 41.36462686 & 0.152213246	 \\
0.961 & 44.26420162 & 0.340056561	 \\
0.613 & 42.99372466 & 0.16509672	 \\
0.3402 & 41.32809153 & 0.135546197	 \\
0.983 & 44.15728437 & 0.434997641	 \\
0.71 & 43.02203881 & 0.184280955	 \\
0.73 & 43.27699552 & 0.194894052	 \\
0.47 & 42.13455949 & 0.153239602	 \\
0.62 & 43.0092526 & 0.169564686	 \\
0.521 & 42.18012945 & 0.160996803	 \\
0.369 & 41.63402302 & 0.152644122	 \\
0.571 & 42.39916207 & 0.174223964	 \\
0.604 & 42.52697648 & 0.15968705	 \\
0.9271 & 43.94049787 & 0.286707555	 \\
0.285 & 40.85424347 & 0.13659413	 \\
0.2912 & 40.83906428 & 0.146624503	 \\
0.548 & 42.29629856 & 0.188370893	 \\
0.868 & 43.49328741 & 0.246095165	 \\
0.496 & 42.21467603 & 0.15281454	 \\
0.811 & 43.40453602 & 0.210619526	 \\
0.756 & 43.81399219 & 0.192775955	 \\
0.817 & 43.65156374 & 0.210286864	 \\
0.752 & 43.30182988 & 0.207071437	 \\
0.5516 & 42.30318373 & 0.149254873	 \\
0.3578 & 41.43490959 & 0.136096152	 \\
1.01 & 44.01247454 & 0.375672417	 \\
0.741 & 43.71725229 & 0.22328686	 \\
0.43 & 41.8109208 & 0.151082143	 \\
0.526 & 42.40021068 & 0.166943097	 \\
0.592 & 42.57732276 & 0.172394047	 \\
0.905 & 43.63874001 & 0.257467547	 \\
0.949 & 43.44490985 & 0.284630783	 \\
0.4607 & 42.07166655 & 0.181262455	 \\
0.3709 & 41.6692044 & 0.154834006	 \\
0.8 & 43.71406011 & 0.215165244	 \\
0.679 & 43.45740634 & 0.195969077	 \\
0.5817 & 42.67204061 & 0.159803369	 \\
0.55 & 42.27564705 & 0.155204642	 \\
0.81 & 43.36989249 & 0.212685692	 \\
0.95 & 43.97660751 & 0.291243769	 \\
0.3373 & 41.29449095 & 0.137240589	 \\
0.91 & 44.30551642 & 0.267241049	 \\
0.263 & 40.63467212 & 0.134567833	 \\
0.643 & 43.01126699 & 0.166369772	 \\
0.691 & 43.08876693 & 0.253842245	 \\
0.357 & 41.42552796 & 0.137721477	 \\
0.721 & 43.17567838 & 0.187179852	 \\
0.581 & 42.74232334 & 0.155462518	 \\
0.6268 & 42.75861262 & 0.157227805	 \\
0.818 & 43.39289417 & 0.267558806	 \\
0.4627 & 42.042285 & 0.147568802	 \\
0.449 & 42.02306835 & 0.157063021	 \\
0.688 & 43.05398603 & 0.165777265	 \\
0.87 & 44.2313042 & 0.27881729	 \\
0.5043 & 42.3382007 & 0.149577507	 \\
0.591 & 43.20910597 & 0.306006112	 \\
0.426 & 41.76157748 & 0.224920671	 \\
0.329 & 41.34025673 & 0.269807401	 \\
0.583 & 42.40112537 & 0.28648393	 \\
0.519 & 43.07150971 & 0.314462491	 \\
0.401 & 41.69542447 & 0.229308722	 \\
0.205 & 39.90376943 & 0.217272004	 \\
0.34 & 41.21317877 & 0.22819077	 \\
0.436 & 41.8800162 & 0.216533947	 \\
0.363 & 41.55197569 & 0.208173297	 \\
0.436 & 41.9257693 & 0.22826276	 \\
0.309 & 41.17379571 & 0.235823876	 \\
0.342 & 41.38459818 & 0.214810321	 \\
0.159 & 39.4163641 & 0.242628008	 \\
0.332 & 41.2554783 & 0.239903276	 \\
0.469 & 42.35448121 & 0.283152406	 \\
0.239 & 40.25912878 & 0.20416793	 \\
0.352 & 41.42377935 & 0.224191541	 \\
0.612 & 42.81394094 & 0.283023415	 \\
0.631 & 42.37738312 & 0.244974123	 \\
0.645 & 42.81697317 & 0.23427313	 \\
0.429 & 41.89003881 & 0.21547139	 \\
0.497 & 42.07577382 & 0.221637713	 \\
0.539 & 42.22666321 & 0.241797816	 \\
0.561 & 42.8719504 & 0.306521909	 \\
0.41 & 41.34749333 & 0.256981142	 \\
0.412 & 41.42371219 & 0.312056382	 \\
0.599 & 42.7434649 & 0.382165112	 \\
0.619 & 43.05601864 & 0.247822482	 \\
0.422 & 41.72830946 & 0.230713846	 \\
0.54 & 42.51116352 & 0.282945442	 \\
0.401 & 42.55462036 & 0.369893982	 \\
0.218 & 40.07540742 & 0.222222738	 \\
0.633 & 42.20167248 & 0.259877853	 \\
0.383 & 41.6420339 & 0.251010175	 \\
0.302 & 41.31079828 & 0.27234655	 \\
0.34 & 41.08004325 & 0.231193988	 \\
0.51 & 41.88068537 & 0.226762462	 \\
0.421 & 42.13603163 & 0.322004955	 \\
0.399 & 41.48808176 & 0.292952212	 \\
0.493 & 42.14691467 & 0.263957547	 \\
0.687 & 42.99638508 & 0.285188799	 \\
0.687 & 42.83483512 & 0.272537782	 \\
0.495 & 42.24866337 & 0.232347966	 \\
0.603 & 42.64620471 & 0.278628903	 \\
0.421 & 42.17962775 & 0.233773507	 \\
0.348 & 41.5896262 & 0.217425302	 \\
0.213 & 40.1090412 & 0.223352611	 \\
0.344 & 41.17287112 & 0.209772839	 \\
0.271 & 40.53230207 & 0.215267001	 \\
0.564 & 42.37289175 & 0.292022401	 \\
0.274 & 40.72465065 & 0.208670921	 \\
0.181 & 39.68288852 & 0.227088514	 \\
0.582 & 43.16391026 & 0.325207918	 \\
0.68 & 42.90403446 & 0.290951926	 \\
0.401 & 41.93563231 & 0.332838743	 \\
0.416 & 41.55753695 & 0.303252122	 \\
0.286 & 41.21113084 & 0.258144621	 \\
0.562 & 43.05052259 & 0.333215442	 \\
0.266 & 40.39070739 & 0.250227167	 \\
0.314 & 41.23196804 & 0.242363894	 \\
0.581 & 43.66867977 & 0.39123693	 \\
0.463 & 41.94775375 & 0.266464454	 \\
0.341 & 40.99151386 & 0.22450432	 \\
0.631 & 42.88182529 & 0.23436967	 \\
0.522 & 42.67733035 & 0.247150869	 \\
0.368 & 41.46769524 & 0.207180967	 \\
0.309 & 40.86077465 & 0.214863256	 \\
0.528 & 42.36969184 & 0.302654151	 \\
0.216 & 40.40490894 & 0.215211775	 \\
0.284 & 40.83538258 & 0.206616021	 \\
0.508 & 42.19427073 & 0.231767203	 \\
0.781 & 43.43722955 & 0.321348684	 \\
0.613 & 42.61670959 & 0.26021447	 \\
0.278 & 40.56624057 & 0.198452988	 \\
0.477 & 42.05433516 & 0.183599934	 \\
0.95 & 43.63831761 & 0.297754711	 \\
1.057 & 44.1506301 & 0.238584685	 \\
0.816 & 43.69108288 & 0.457863422	 \\
0.455 & 42.32393227 & 0.22962754	 \\
1.02 & 44.3629247 & 0.220963762	 \\
1.14 & 44.32295709 & 0.22870103	 \\
0.854 & 43.60846429 & 0.225569906	 \\
1.37 & 45.04971335 & 0.26274845	 \\
0.975 & 44.3337255 & 0.212780603	 \\
0.97 & 44.4626017 & 0.291454456	 \\
0.74 & 43.2875467 & 0.198402487	 \\
1.39 & 44.87623434 & 0.250631965	 \\
0.46 & 42.14844961 & 0.210095498	 \\
1.02 & 44.16406553 & 0.232339527	 \\
1.12 & 44.51440549 & 0.224079095	 \\
1.23 & 45.02067572 & 0.235028401	 \\
1.19 & 44.36283166 & 0.247108114	 \\
0.839 & 43.39806572 & 0.222124147	 \\
1.01 & 44.91213783 & 0.333748617	 \\
0.521 & 42.37824456 & 0.199332155	 \\
0.475 & 42.1048814 & 0.255958317	 \\
0.95 & 43.88415029 & 0.235845124	 \\
1.3 & 45.0162581 & 0.242719168	 \\
1.305 & 44.74016933 & 0.259540115	 \\
0.216 & 40.5560466 & 0.244099861	 \\
0.735 & 43.09184035 & 0.200575809	 \\
1.14 & 44.19695213 & 0.368403294	 \\
1.307 & 45.41074411 & 0.31469671	 \\
1.265 & 44.94411084 & 0.23546996	 \\
0.67 & 43.14320955 & 0.209654285	 \\
0.64 & 42.92427543 & 0.280114332	 \\
1.34 & 45.06750558 & 0.275015093	 \\
0.84 & 43.51430438 & 0.208697053	 \\
0.935 & 43.54019724 & 0.227616538	 \\
0.953 & 44.27362023 & 0.954717441	 \\
1.124 & 44.56751918 & 0.197263048	 \\
0.552 & 42.51093428 & 0.103414503	 \\
0.671 & 42.982007 & 0.120545528	 \\
0.511 & 42.37366047 & 0.089061755	 \\
1.03 & 44.24009336 & 0.141254203	 \\
1.192 & 44.4587516 & 0.200258072	 \\
1.092 & 44.00784772 & 0.246511865	 \\
0.974 & 43.83414326 & 0.17423894	 \\
1.11 & 44.62533585 & 0.447832633	 \\
1.35 & 44.82706548 & 0.185705412	 \\
0.85 & 43.49425735 & 0.171796431	 \\
1.241 & 44.58170296 & 0.478371053	 \\
1.414 & 44.80376614 & 0.346181483	 \\
1.188 & 44.60764255 & 0.500544949	 \\
1.017 & 44.29397077 & 0.171046154	 \\
1.315 & 44.97135777 & 0.187508839	 \\
0.821 & 43.64093879 & 0.194013592	 \\
1.215 & 45.24652095 & 0.560317445	 \\
0.623 & 42.514524 & 0.241428135	 \\
\hline
\end{longtable}
\label{tab:SNeIaData}
\end{center}

\cleardoublepage

\chapter{Baryonic Acoustic Oscillation (BAO) Data}
\label{Appendix:F}

\begin{equation}
D_V(z)= \left[(1 + z)^2 d_A(z)^2 \frac{c~z}{H(z)}\right]^{1/3}.
\end{equation}
Here, $d_A(z)$ is the angular diameter distance from Eq.\ (\ref{eq:ada3}):
\begin{equation}
d_A(z)=\frac{y(z)}{H_0(1+z)},
\end{equation}
where $y(z)$ is the dimensionless coordinate distance given in Eq.\ 
({\ref{eq:Angular size distance}}):
\begin{equation}
A_{\rm th}(z)=\frac{100~D_V(z)~\sqrt{\Omega_mh^2}}{z}.
\end{equation}
Using Eqs.\ ({\ref{eq:D_V}})---({\ref{eq:A_z}}) we have: 
\begin{equation}
A_{\rm th}(z)=\sqrt{\Omega_m} \left[\frac{y^2(z)}{z^2 E(z)}\right]^{1/3},
\end{equation}
which is $h$ independent and where $E(z)$ is defined in 
Chapter\ (\ref{Chapter2}). \begin{equation}
d_{\rm th}(z)=\frac{r_s(z_d)}{D_V(z)},
\end{equation}
where $r_s(z_d)$ is the sound horizon at the drag epoch, is given in 
Eq.\ (6) of Eisenstein \textit{et al.}\cite{Eisenstein1998}
\begin{deluxetable}{cccccc}
\tablecaption{BAO Data--- Distilled and Acoustic Parameters measuremnts.\tablenotemark{a}}
\tablewidth{0pc}
\tablehead{
\colhead{\multirow{1}{*}{Sample}}& 
\colhead{\multirow{1}{*}{$z$}}& 
\colhead{\multirow{1}{*}{$d_z$}}& 
\colhead{$\sigma_{d_z}$}&
\colhead{\multirow{1}{*}{$A(z)$}}& 
\colhead{$\sigma_{A(z)}$}\\
\vspace{-5 mm}
}
\startdata
\noalign{\vskip -1mm}
6dFGS   & 0.106 &      \textbf{0.336} &   \textbf{0.015} &           0.526&          0.028 \\

SDSS    & 0.2   &      \textbf{0.1905}&   \textbf{0.0061}&           0.488&          0.016 \\

SDSS    & 0.35  &      \textbf{0.1097}&   \textbf{0.0036}&           0.484&          0.016 \\

WiggleZ & 0.44  &               0.0916&            0.0071&  \textbf{0.474}& \textbf{0.034} \\

WiggleZ & 0.6   &               0.0726&            0.0034&  \textbf{0.442}& \textbf{0.020} \\

WiggleZ & 0.73  &               0.0592&            0.0032&  \textbf{0.424}& \textbf{0.021} \\
\enddata
\tablenotetext{a}{From Blake \textit{et al.}\cite{blake11}}
\label{tab:BAODATA}
\end{deluxetable}


\cleardoublepage
\phantomsection
\addcontentsline{toc}{chapter}{Bibliography}

\bibdata{references}

\bibliography{references}

\end{document}